\pdfoutput=1

\documentclass[a4paper,eqsecnum,secnumarabic,11pt,twoside]{revtex4}
\usepackage{url}
\usepackage{cleveref}

\usepackage{graphicx}
\usepackage{tabularx}
\usepackage{booktabs}
\usepackage{xspace}
\usepackage{lscape}
\usepackage{subfigure} 
\usepackage{setspace}
\usepackage{epsfig}
\usepackage{units}
\usepackage{threeparttable}
\usepackage{multirow}
\usepackage{comment}
\usepackage{float}

\usepackage[intlimits]{amsmath}
 \usepackage{MnSymbol} 

\usepackage{hhline}
\usepackage{array}

\DeclareMathAlphabet{\mathscr}{OT1}{pzc}{m}{it} 

\usepackage{titlesec}
\titleformat{\chapter}[hang]{\large\bfseries}{\thechapter\quad}{0pt}{}
\titleformat{\section}[hang]{\Large\bfseries}{\thesection\quad}{0pt}{}
\titleformat{\subsection}[hang]{\large\bfseries}{\thesubsection\quad}{0pt}{}
\titleformat{\subsubsection}[hang]{\bfseries}{\thesubsubsection\quad}{0pt}{}
\titleformat{\paragraph}[hang]{\bfseries}{\theparagraph)~}{0pt}{}

\titlespacing{\chapter}{0pt}{-3em}{6pt}
\titlespacing{\section}{0pt}{6pt}{6pt}
\titlespacing{\subsection}{0pt}{18pt}{6pt}
\titlespacing{\subsubsection}{0pt}{12pt}{6pt}
\titlespacing{\paragraph}{0pt}{9pt}{3pt}

\newcommand{\six}{\ensuremath{6\times6\times6 \mbox{m}^3}\xspace}
\newcommand{\three}{\ensuremath{3\times1\times1 \mbox{m}^3}\xspace}

\begin{document}

\centerline{EUROPEAN ORGANIZATION FOR NUCLEAR RESEARCH}

\vspace{0.75cm}
\begin{flushright}
CERN-SPSC-2014-013 \\
SPSC-TDR-004 \\
March 31st, 2014
\end{flushright}

\vspace{0.3cm}

\title{\Large LBNO-DEMO \\ Large-scale neutrino detector demonstrators for 
phased performance assessment 
in view of a long-baseline oscillation experiment \\
\vspace{0.6cm}
\large The LBNO-DEMO (WA105) Collaboration
\vspace{0.3cm}}



\newcommand{\INSTA}{\affiliation{\bf ETH Zurich, Institute for Particle Physics, Zurich, Switzerland}}
\newcommand{\INSTB}{\affiliation{\bf University of Bern, Albert Einstein Center for Fundamental Physics, Laboratory for High Energy Physics (LHEP), Bern, Switzerland}}
\newcommand{\INSTC}{\affiliation{\bf University of Sheffield, Department of Physics and Astronomy, Sheffield, United Kingdom}}
\newcommand{\INSTD}{\affiliation{\bf University of Bucharest, Faculty of Physics, PO Box MG-11, Bucharest-Magurele, Romania}}
\newcommand{\INSTE}{\affiliation{\bf Universit\'e de Lyon, Universit\'e Claude Bernard Lyon 1, IPN Lyon (IN2P3), Villeurbanne, France}}
\newcommand{\INSTF}{\affiliation{\bf IFIC (CSIC \& University of Valencia), Valencia, Spain}}
\newcommand{\INSTG}{\affiliation{\bf Queen Mary University of London, School of Physics, London, United Kingdom}}
\newcommand{\INSTH}{\affiliation{\bf Institute of Nuclear Technology-Radiation Protection, National Centre
for Scientific Research ''Demokritos'', Athens, Greece}}

\newcommand{\INSTI}{\affiliation{\bf Institute for Nuclear Research of the Russian Academy of Sciences, Moscow, Russia}}
\newcommand{\INSTDUBNA}{\affiliation{\bf Joint Institute for Nuclear Research, Dubna, Moscow Region, Russia}}

\newcommand{\INSTJ}{\affiliation{\bf IRFU, CEA Saclay, Gif-sur-Yvette, France}}

\newcommand{\INSTPA}{\affiliation{\bf INFN Sezione di Padova and Universit\`a di Padova, Dipartimento di Fisica, Padova, Italy}}

\newcommand{\INSTRASC}{\affiliation{\bf INFN Laboratori Nazionali di Frascati, Frascati, Italy}}

\newcommand{\INSTJY}{\affiliation{\bf Department of Physics, University of Jyv\"askyl\"a, Finland}}
\newcommand{\INSTBU}{\affiliation{\bf Faculty of Physics, University of Bucharest, Bucharest, Romania}}
\newcommand{\INSTDF}{\affiliation{\bf National Centre for Nuclear Research (NCBJ), Warsaw, Poland}}

\newcommand{\INSTDH}{\affiliation{\bf Institute of Experimental Physics, Warsaw University (IFD UW), Warsaw, Poland}}

\newcommand{\INSTFD}{\affiliation{\bf University of Warwick, Department of Physics, Coventry, United Kingdom}}
\newcommand{\INSTGG}{\affiliation{\bf Oxford University, Department of Physics, Oxford, United Kingdom}}

\newcommand{\INSTFC}{\affiliation{\bf University of Liverpool, Department of Physics, Liverpool, United Kingdom}}
\newcommand{\INSTEG}{\affiliation{\bf University of Geneva, Section de Physique, DPNC, Geneva, Switzerland}}
\newcommand{\INSTCE}{\affiliation{\bf CERN, Geneva, Switzerland}}

\newcommand{\INSTDUR}{\affiliation{\bf Institute for Particle Physics Phenomenology, Durham University, United Kingdom}}

\newcommand{\INSTRO}{\affiliation{\bf Rockplan Ltd., Helsinki, Finland}}
\newcommand{\INSTTECHNO}{\affiliation{\bf Technodyne International Limited, Eastleigh, Hampshire, United Kingdom}}

\newcommand{\INSTTUA}{\affiliation{\bf Middle East Technical University (METU), Ankara, Turkey}}
\newcommand{\INSTTUB}{\affiliation{\bf Ankara University, Ankara, Turkey}}

\newcommand{\INSTTBU}{\affiliation{\bf Department of Atomic Physics, Faculty of Physics, St.\,Kliment Ohridski University of Sofia, Sofia, Bulgaria}}

\newcommand{\INSTNA}{\affiliation{\bf INFN Sezione di Napoli and Universit\`a di Napoli, Dipartimento di Fisica, Napoli, Italy}}

\newcommand{\INSTIMPER}{\affiliation{\bf Imperial College, London, United Kingdom}}

\newcommand{\INSTOULU}{\affiliation{\bf University of Oulu, Oulu, Finland}}

\newcommand{\INSTMILBIC}{\affiliation{\bf INFN Milano Bicocca, Milano, Italy}}

\newcommand{\INSTROMATRE}{\affiliation{\bf INFN Roma Tre, Roma, Italy}}
\newcommand{\INSTROMATREUNI}{\affiliation{\bf Universita' and INFN Roma Tre, Roma, Italy}}

\newcommand{\INSTBARI}{\affiliation{\bf INFN and Dipartimento interateneo di Fisica di Bari, Bari, Italy}}

\newcommand{\INSTLAPP}{\affiliation{\bf LAPP, Universit\'e de Savoie, CNRS/IN2P3, Annecy-le-Vieux, France}}

\newcommand{\INSTHELSINK}{\affiliation{\bf University of Helsinki, Helsinki, Finland}}

\newcommand{\INSTLANCAS}{\affiliation{\bf Physics Department, Lancaster University, Lancaster, United Kingdom}}

\newcommand{\INSTRAL}{\affiliation{\bf STFC, Rutherford Appleton Laboratory, Harwell Oxford, United Kingdom}}

\newcommand{\INSTPARIS}{\affiliation{\bf UPMC, Universit\'e Paris Diderot, CNRS/IN2P3, Laboratoire de Physique Nucl\'eaire et de Hautes Energies (LPNHE), Paris, France}}

\newcommand{\INSTAPC}{\affiliation{\bf APC, AstroParticule et Cosmologie, Universit\'e Paris Diderot, CNRS/IN2P3, CEA/Irfu, Observatoire de Paris, Sorbonne Paris Cit\'e, 10, rue Alice Domon et L\'eonie Duquet, 75205 Paris Cedex 13, France}}

\newcommand{\INSTPNPI}{\affiliation{\bf Petersburg Nuclear Physics Institute (PNPI), St-Petersburg, Russia}}

\newcommand{\INSTINFNTRIESTE}{\affiliation{\bf INFN Trieste, Trieste, Italy}}

\newcommand{\INSTINSUBRIA}{\affiliation{\bf Universita` dell'Insubria, sede di Como/ INFN Milano Bicocca, Como, Italy}}

\newcommand{\INSTGLAS}{\affiliation{\bf University of Glasgow, Glasgow, United Kingdom}}

\newcommand{\INSTUCL}{\affiliation{\bf Dept. of Physics and Astronomy, University College London, London, United Kingdom}}

\newcommand{\INSTMANCHEST}{\affiliation{\bf University of Manchester, Manchester, United Kingdom}}

\newcommand{\INSTCAG}{\affiliation{\bf INFN Sezione di Cagliari, Cagliari, Italy}}

\newcommand{\INSTUCAG}{\affiliation{\bf INFN Sezione di Cagliari and Universit\`a di Cagliari, Cagliari, Italy}}

\newcommand{\INSTCAMB}{\affiliation{\bf University of Cambridge, Cambridge, United Kingdom}}

\newcommand{\INSTSTRASS}{\affiliation{\bf IPHC, Universit\'e de Strasbourg, CNRS/IN2P3, Strasbourg, France}}

\newcommand{\INSTAACHEN}{\affiliation{\bf III. Physikalisches Institut, RWTH Aachen, Aachen, Germany}}

\newcommand{\INSTAGT}{\affiliation{\bf AGT Ingegneria S.r.l., Perugia, Italy}}

\newcommand{\INSTAAD}{\affiliation{\bf Alan Auld Engineering, Doncaster, United Kingdom}}

\newcommand{\INSTSOF}{\affiliation{\bf SOFREGAZ SA, Saint-Denis, France}}

\newcommand{\INSTCPPM}{\affiliation{\bf CPPM, Marseille, France}}

\newcommand{\INSTCIEMAT}{\affiliation{\bf Centro de Investigaciones Energ\'eticas, Medioambientales y Tecnol\'ogicas (CIEMAT), Madrid, Spain}}

\newcommand{\INSTBARCEL}{\affiliation{\bf Institut de Fisica d'Altes Energies (IFAE), Bellaterra (Barcelona), Spain}}

\newcommand{\INSTINFNPISA}{\affiliation{\bf INFN-Sezione di Pisa, Italy}}

\newcommand{\INSTKEK}{\affiliation{\bf High Energy Accelerator Research Organization (KEK), Tsukuba,  Ibaraki, Japan}}
\newcommand{\INSTIWATE}{\affiliation{\bf Iwate University, Department of Electrical Engineering and Computer Science, Morioka, Iwate, Japan}}

\newcommand{\INSTIFIN}{\affiliation{\bf Horia Hulubei National Institute of R\&D for
Physics and Nuclear Engineering - IFIN-HH, Magurele, Romania}}

\newcommand{\INSTOMEGA}{\affiliation{\bf OMEGA Ecole Polytechnique/CNRS-IN2P3,
route de Saclay,  Palaiseau, France}}

\author{I.\,De Bonis}\INSTLAPP
\author{P.\,Del Amo Sanchez}\INSTLAPP
\author{D.\,Duchesneau}\INSTLAPP
\author{H.\,Pessard}\INSTLAPP

\author{S.\,Bordoni}\INSTBARCEL
\author{M.\,Ieva}\INSTBARCEL
\author{T.\,Lux}\INSTBARCEL
\author{F.\,Sanchez}\INSTBARCEL

\author{A.\,Jipa}\INSTBU
\author{I.\,Lazanu}\INSTBU
% ...

\maketitle



\newpage

\centerline{\Large \bf Executive Summary}
\vspace{0.5cm}
In June 2012, an Expression of Interest for a long-baseline
experiment (LBNO)~\cite{Stahl:2012exa} has been submitted to the CERN SPSC and is presently under review.
LBNO considers three types of neutrino detector technologies: a double-phase liquid argon (LAr) TPC and a magnetised iron detector as far detectors. For the near detector, a high-pressure gas TPC embedded in a calorimeter and a magnet is the baseline design.

A mandatory milestone in view of any future long baseline experiment is a concrete prototyping effort towards the envisioned large-scale detectors, and an accompanying campaign of measurements aimed at assessing the systematic errors that will be affecting their intended physics programme. 
Following an encouraging feedback from 108th SPSC on the technology choices, we have defined as priority the construction and operation of a $6\times 6\times 6$m$^3$ (active volume) double-phase liquid argon (DLAr) demonstrator,
and a parallel development of the technologies necessary for large magnetised MIND detectors. 

The \six DLAr is an industrial prototype of the design proposed in
the EoI and scalable to 20 kton, 50~kton or more. It is to be
constructed and operated in a controlled laboratory and surface
environment with test beam access, such as the CERN North Area (NA). 
Its successful operation and full characterisation will be a
fundamental milestone, likely opening the path to an underground
deployment of larger detectors.
The response of the DLAr demonstrator will be measured and understood
with an unprecedented precision in a charged particle test beam (0.5-20
GeV/c). The exposure will certify the assumptions and calibrate the
response of the detector, and allow to develop and to benchmark
sophisticated reconstruction algorithms, such as those of 3-dimensional
tracking, particle ID and energy flow in liquid argon. All these steps
are fundamental for
validating the correctness of the physics performance described in the
LBNO EoI.

We anticipate that a successful operation of the double-phase \six
DLAr demonstrator and its campaign exposure to a charged particle beam, will provide very important and vital feedback for long baseline programmes, and in general for the field. It will represent a never-achieved milestone for LAr  detectors.  Its design specifically addresses and represents a concrete step towards
an extrapolation of the technology
to very large masses in the tens of kton range, such as the one 
considered and studied for several years within
the EU FP7 funded LAGUNA/LAGUNA-LBNO design studies. 
The parameters of the demonstrator will be directly scalable and
the components mass-produceable. Long drift paths will be 
assessed on a large scale.

As requested by SPSC, we submit a Technical Design Report,  in view of a realisation 
of the facility and an exposure to the charged particle beam before the LHC LS2.
%

\newpage
\tableofcontents

\newpage


\pagestyle{headings}

\section{Introduction}
\subsection{Main goals of the demonstrators}
The main objectives  towards long baseline neutrino oscillation experiment
(LBNO) in the coming years are to develop demonstrators of the neutrino detector
technologies considered in the Expression of Interest (EoI) CERN-SPSC-2012-021 (SPSC-EOI-007)~\cite{Stahl:2012exa}
submitted to the SPSC in 2012. Such demonstrators are needed to leverage large risks associated to the
extrapolation from existing experience to the huge mass required for far detectors. 

Already since almost a decade we have worked towards the realisation of a giant liquid argon underground 
detector for neutrino mass hierarchy (MH) and leptonic CP-violation (CPV) discovery, and neutrino astrophysics. In the recent years, 
two consecutive FP7 Design Studies (LAGUNA/LAGUNA-LBNO) have led to the development
of a conceptual design (fully engineered and costed) for a 20kton/50kton GLACIER-type~\cite{Rubbia:2004tz}
underground neutrino detector. In these studies, an underground implementation has been assumed {\it ab initio}
and such constraints have been important and taken into account in design choices.

In 2013, the SPSC has endorsed the physics case for
CP-violation and neutrino mass hierarchy determination described in the
EoI, the choice of the Liquid Argon (LAr) detector technology, and encourages the
development of the kind of activities described in this document, requesting
more detailed information (108th SPSC minutes):
\begin{quote}
{\it ``Concerning LAGUNA-LBNO, the SPSC supports the double-phase LAr TPC option as a promising technique to instrument very large LAr neutrino detectors in the future. The SPSC therefore encourages the LBNO consortium to proceed with the R\&D necessary to validate the technology on a large scale. For further review of the project, the SPSC requests a technical proposal describing the R\&D programme to be led at CERN''.}
\end{quote}

The successful operation of the \six DLAr demonstrator presented in this Technical Report will open the way to the construction of large and affordable liquid argon underground detectors addressing the complete investigation of 3-flavours neutrino oscillations and the determination of their still unknown parameters. These detectors will  as well be very powerful for important non-beam studies such as proton decay, atmospheric neutrinos and supernovae neutrinos. The \six dimensions are motivated by the fact that the 
basic readout component of the large-scale LAGUNA/LBNO 20-50 kton detectors are $4\times 4$~m$^2$ units. 
The $6\times6$~m$^2$ is consistent with having a fiducial volume corresponding to that readout unit. Surface
operation prohibits drift lengths above 6~m.
In parallel, the MIND detector will allow developing 
further this well-understood technology in particular for what concerns the optical readout and the large scale integration.

More generally, the technologies addressed and the results that will be obtained during
the commissioning and operation of the facility, are relevant to a broader range of experiments, which 
are presently being considered and could go online in the next decade. In the following we list
the main experiments where our proposal would have a potentially important impact:

\begin{itemize}

\item The vision of the long baseline neutrino experiment (LBNE) in the USA \cite{Adams:2013qkq}
is based on a prospective 1.2~MW proton beam power at FNAL and a far detector
at Homestake of 35~kton mass. The detectors proposed in this document and their successful operation could,
if performed on a timescale of less than 3~years, have a positive potential impact on the design of
the LBNE far detector, and represent the basis for a
significant European contribution to the project.

\item The Hyper-Kamiokande Detector in Japan (LOI September 2011) \cite{Abe:2011ts}
has been proposed as a next generation underground Water Cherenkov Detector (WCD). It will serve 
as a far detector of a long baseline neutrino oscillation experiment envisioned for the  J-PARC neutrino
complex upgraded up to 750~kW, 
and as a detector capable of observing -- far beyond the sensitivity of the Super-Kamiokande (Super-K) detector -- 
proton decays, atmospheric neutrinos, and neutrinos from astronomical origins. 
The baseline design of Hyper-K is based on the highly successful Super-K, taking full advantage of a well-proven technology.
However, in order to exploit the full statistical power, systematic errors have to be controlled by a set
of near detectors. The technologies considered in this proposal are directly relevant for improved near
detectors at J-PARC. In particular, a muon ranger behind a new WCD detector at the new 2~km position
is being discussed.

\item In the proposed short baseline neutrino experiment at CERN, CERN-SPSC-2012-010 (SPSC-P-347) \cite{Antonello:2012hf},
a new conventional neutrino beam (CENF) has been discussed. The location of the DLAr and MIND demonstrators
described in this proposal is made compatible with the ``near'' station of the eventual CENF neutrino beam. Exposure to the CENF
would allow to collect a large number of neutrino interactions 
in the range of 500~k/year, as illustrated in Appendix \ref{sec:appendixA}.

\item NUSTORM~ \cite{Kyberd:2012iz} (``Neutrinos from STORed Muons,") is a proposed facility,  
which, with an appropriate far detector for neutrino oscillation searches at short baseline, could
provide a confirmation or rejection of the LSND/MinBooNE results. The DLAr and MIND detectors
proposed in this document exposed to a NUSTORM-like beam,
could be used to make precision neutrino interaction cross section measurements 
important to the next generation of long-baseline neutrino oscillation experiments.
This would result in the detection of 100~k/year neutrino interactions, as illustrated
in Appendix \ref{sec:appendixA}.

\item In the recent proposal to Search for Heavy Neutral Leptons at the SPS, CERN-SPSC-2013-024 
(SPSC-EOI-010) \cite{Bonivento:2013jag}, the neutrinos produced in the dump are a source
of background, whose rate and spectra are not very well known. A recent calculation finds
a discrepancy of a factor four with CHARM data. The DLAr and MIND
demonstrators would directly measure these backgrounds, which could be used to predict
the rates and features of these interactions in the muon shield and decay volume of
the SHIP experiment.

\item The Neutrino Factory (see e.g. \cite{Bross:2013oua}) is conceived as the facility
for the ultimate long baseline neutrino oscillation experiment. As explained in the LBNO
EOI, the choice of baseline 2300~km is optimised for the neutrino factory, and the
LBNO far detector infrastructure is a first milestone towards such a facility. In this proposal,
the DLAr and MIND demonstrators will assess the scalability and costs of the far detectors
of the neutrino factory.

\end{itemize}

\subsection{The DLAr demonstrator}
\begin{figure}[hbtp]
\begin{center}
\begin{minipage}[t]{\columnwidth}
\begin{center}
\includegraphics[width=\columnwidth]{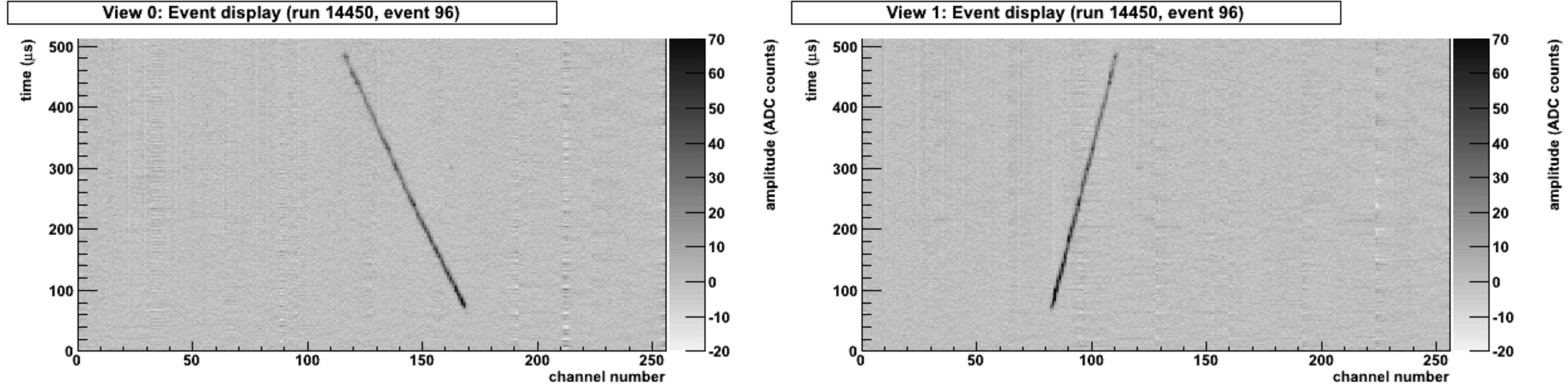}
\end{center}
\end{minipage}
\begin{minipage}[t]{\columnwidth}
\begin{center}
\includegraphics[width=\columnwidth]{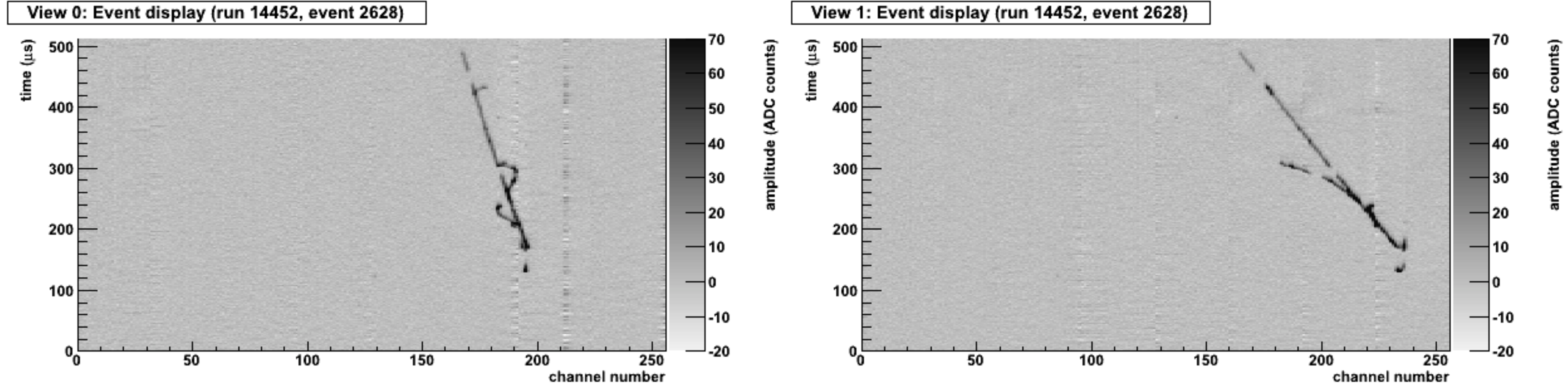}
\end{center}
\end{minipage}
\begin{minipage}[t]{\columnwidth}
\begin{center}
\includegraphics[width=\columnwidth]{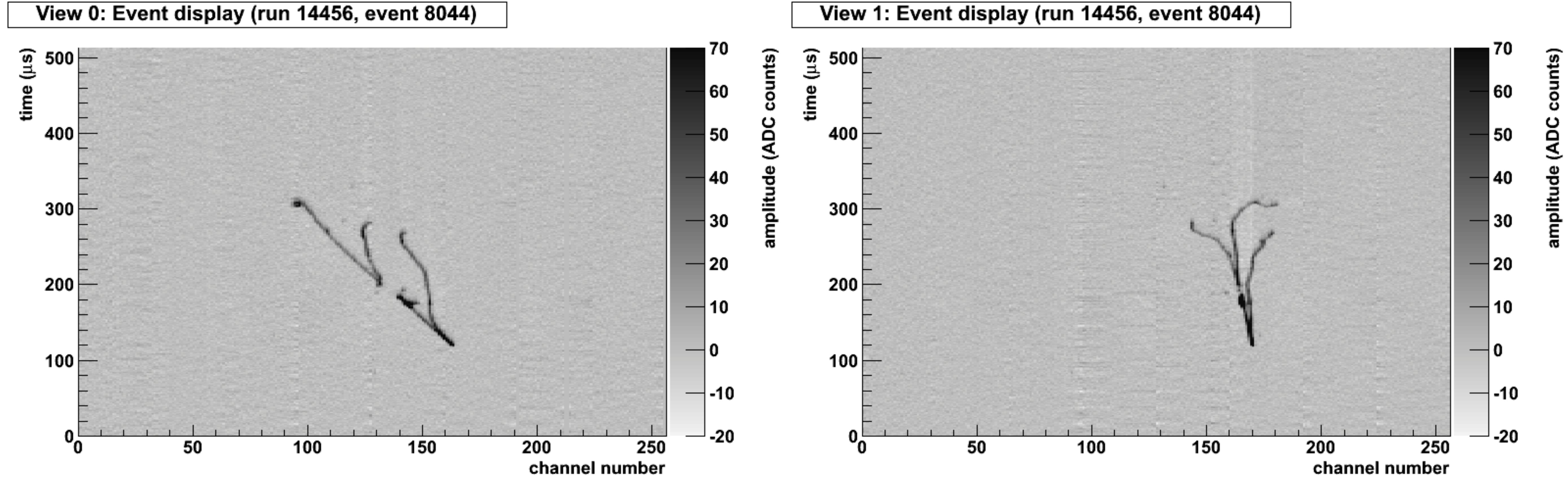}
\end{center}
\end{minipage}
\caption{Event display for three cosmic events recorded using the
40~$\times$~80~cm$^2$ LAr LEM-TPC~\protect\cite{Badertscher:2012dq}.
The left and right columns correspond to the $xt$- and $yt$-projections, respectively.
Top: a straight muon track crossing from top to bottom of the field cage. The time of the start ($t_0 \sim 60$ $\mu$s) and the end ($t_{\rm max} \sim 490$ $\mu$s) of the extraction are clearly visible. Middle: an example of a ``grazing'' muon track with delta rays, where the extraction starts later at $t \sim 150$ $\mu$s. 
Bottom: interacting event.}
\label{fig:fig15lemgallery}
\end{center}
\end{figure}

The concept of the LAr Time Projection
Chamber (LAr TPC)~\cite{Rubbia:1977zz} and its excellent tracking-calorimeter performances
allow for massive neutrino detectors with higher signal efficiency and 
effective background discrimination compared to other techniques.
The pioneering work of ICARUS on prototypes of ever increasing mass (0.003~\cite{Aprile:1985xz}, 
0.05~\cite{Arneodo:2006ug}, 
3~\cite{Cennini:1994ha}, and 10~tons~\cite{Arneodo:2003vh})
has culminated in the construction and operation  of the T600 detector (478~tons) on 
surface~\cite{Amerio:2004ze}, its characterization~\cite{Arneodo:2003rr,Amoruso:2003sw,Amoruso:2004ti,
Amoruso:2004dy,Ankowski:2006ts,Ankowski:2008aa}, 
and eventually to its underground 
commissioning~\cite{Rubbia:2011ft}. The so-called GLACIER design~\cite{Rubbia:2004tz,Rubbia:2009md},
based on the double phase Large Electron Multiplier (LEM) readout ``sandwich'' 
with charge extraction and amplification,
coupled to a very long drift paths in a single non-evacuated LNG-type tank has
been advocated as an attractive solution to reach very large detector masses,
beyond what is realistically achievable 
by linear extrapolation of the ICARUS design.
The successful phases of R\&D and prototyping on small-scale double phase
LAr LEM TPC  setups~\cite{Badertscher:2013wm,Badertscher:2012dq,Badertscher:2010zg,Badertscher:2009av}, 
benefitting from the  worldwide effort on the general development
of Micro Pattern Gas Detectors (MPGD)~\cite{Chefdeville:2011zz} 
-- in particular GEMs~\cite{Sauli:1997qp}, MicroMegas~\cite{Giomataris:1995fq} 
and LEMs/THGEMs~\cite{Breskin:2008cb} --
 already demonstrated the significantly improved  
performance of this novel concept.
The collection-only 
readout mode (avoiding the use of induction planes) is also an important asset
in the case of complicated topologies, like e.g. in electromagnetic or hadronic showers.
Three examples of cosmic events obtained with typical electric field 
configurations (the drift field of $\approx$0.4 kV/cm)
and 35 kV/cm in the LEM holes are presented in Figure~\ref{fig:fig15lemgallery}. 
Using a simple event display images for two projections, i.e. $xt$- and $yt$-projections, 
can be obtained for each event: $x$- and $y$-coordinates are two horizontal 
coordinates orthogonal to each other and $t$-coordinate corresponds to the vertical coordinate $z$. 

The LAGUNA and LAGUNA-LBNO EU FP7 design studies have been focussed on the GLACIER concept, a large double phase Liquid Argon TPC with a long drift and charge-sensitive detectors in the gas phase. 
According to these studies it is technically feasible to build a large (20-100 kt) underground tank, based on the LNG industrial technology,
in which the detector is composed of a large number of independent basic readout units of $4\times 4$~m$^2$~\cite{Rubbia:2013tpa}. In a more recent related development these studies have also considered a novel technique for the tank construction involving the membrane technology for the tank. In this technology, the functions of structural support, insulation and liquid containment are all realised by different components, namely an outer concrete structure, specially designed insulating panels and a thin layer of steel plates. These development are very promising for the realisation of a large underground detector, however several areas need to be verified  on a large scale prototype. 

In order to measure and assess the detector performance for LBNO, we consider dedicated test beam campaigns, to test and optimise the readout methods and the calorimetric performance of such detectors. 
The proposed test beam will address the following points:
\begin{enumerate}
\item Electron, neutral pion, charged pion, muon reconstruction: A crucial feature of the LAr TPC is the possibility for a very fine sampling, which should deliver unmatched performances in particle identification and reconstruction. 
\item Electron/$\pi^0$ separation: Another central feature of the LAr TPC is the possibility to precisely measure and identify electrons from neutral pion backgrounds. 
\item Calorimetry: A specific feature of the LAr TPC is its 100\% homogeneity and full sampling capabilities. As an extension of the measurements performed in above, more refined measurements with low energy particles will yield actual calorimetric performance and determine the ability to reconstruct full neutrino events in the GeV-range. 
\item Hadronic secondary interactions: With the large statistics expected, 
an exclusive final state study of pion secondary interactions will be attempted. Comparison of the data obtained with MC (e.g. GEANT4) will allow to cross-check and eventually tune these models. 

\end{enumerate}

The reconstruction of electrons, neutral and charged pions and muons will be demonstrated in the dedicated test beam
campaign. The obtained results will optimise the readout parameters to be used in the far detector.
They will represent a major milestone in the definition of LAGUNA/LBNO, which relies
on low energy neutrino beams and sensitive searches for proton decay.
They will complement direct measurements in a low energy neutrino beam.
The reconstruction software with a fully automatised reconstruction will be developed. Samples
of hadronic interactions produced by particles of well-known momentum will allow benchmarking these
software tools and determining their performance with high precision.

\subsection{The MIND demonstrator}
\begin{figure}[hbt]
\centering
\includegraphics*[width=0.50\textwidth]{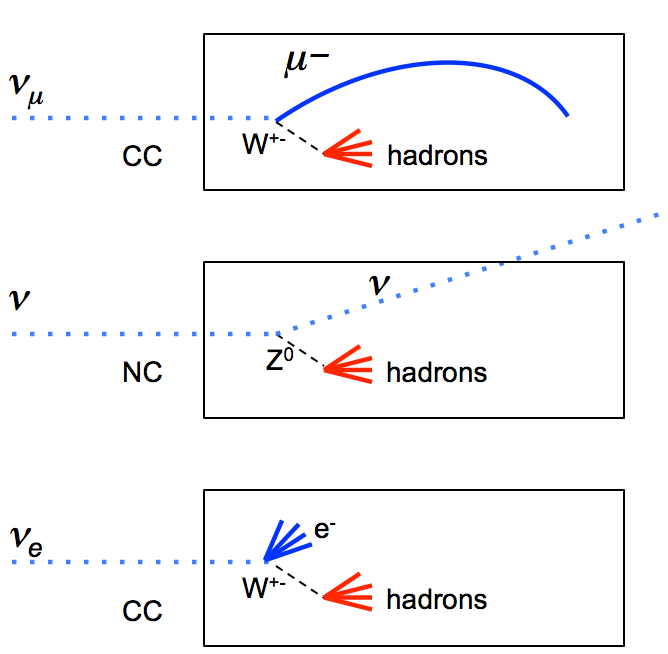}
\caption{Sketch of event topologies in a MIND. The dashed lines with the W/Z bosons are for illustration only and do not represent tracks.}
\label{event_topo}
\end{figure}

MIND-type detectors have been successfully designed, built and operated over several decades. They are particularly well suited for charge current (CC) neutrino interactions with an outgoing muon in the final state, see Figure \ref{event_topo}.  A MIND is the baseline detector for a Neutrino Factory, where a muon of opposite charge to that expected from the neutrino content of the beam, is detected through CC interactions of the oscillated $\nu_\mu$ \cite{Cervera:2010rz}. Along with the performance of detectors obtained online, dedicated test runs have provided valuable information, for example the calibration detector tests or CalDet carried out for MINOS at the CERN-PS, which measured the energy resolution for the setup to be $21.42\%/$$\sqrt{E}$ (GeV), $56.6\%/$$\sqrt{E}$ and $56.1\%/$$\sqrt{E}$ (GeV) for electrons, protons and pions respectively. 

Despite the considerable amount of experience gained on this detector technology, there are several gaps in our knowledge which merit further detailed studies. A MIND prototype located downstream of the DLAr demonstrator would enable the following studies:
\begin{itemize}
\item{Charge ID efficiencies specifically for low energy interactions ($<1$ GeV/c): Reconstruction tools such as the Kalman filter, work very well for high energy muons, where a typical event has several hits through several planes well away from the initial hadronic activity. When the muon range is comparable to the range of the hadronic shower, we have more of a challenge. Cellular automaton techniques are a possibility. We would benchmark new reconstruction techniques against test beam data. This will include testing how fast reconstruction and charge ID efficiencies drop as a function of decreasing momentum in a MIND, and establishing at which point an alternative detector (e.g. air core magnet) would be better suited.}
\item{Backgrounds in MIND analyses: backgrounds to wrong sign muon events come from secondary muons arising from decays of the initial hadronic activity (e.g pions, kaons). A charged particle beam incident first on a 6 m DLAr would generate several scenarios in the MIND, especially given the beam is typically composed of several particles (e.g. $>10$\% pions in a muon beam). Simulations can be benchmarked against these.}
\item{Momentum reconstruction for charged particles crossing the DLAr: the DLAr cannot reconstruct muons above a few GeV/c. Secondary particles arising from interactions of the primary charged particle beam in the DLAr may not stop in the DLAr, so event reconstruction would be complemented by the MIND.}
\end{itemize}
MIND-type detectors are planned downstream of a liquid argon detector at LBNO but also downstream of a water cherenkov at a near detector facility for Hyper-Kamiokande. 

\subsection{Location of the prototype and needed infrastructure}
The cryogenic vessel required for a detector of the size of \six is hardly transportable, and is 
most conveniently constructed {\it in situ}. On the other hand, the detector components
will be assembled and tested elsewhere, transported sealed onsite, and installed inside the
vessel, equipped during the installation phase as a clean-room. 

\begin{figure}[htb]
\begin{center}
\includegraphics[width=0.85\textwidth]{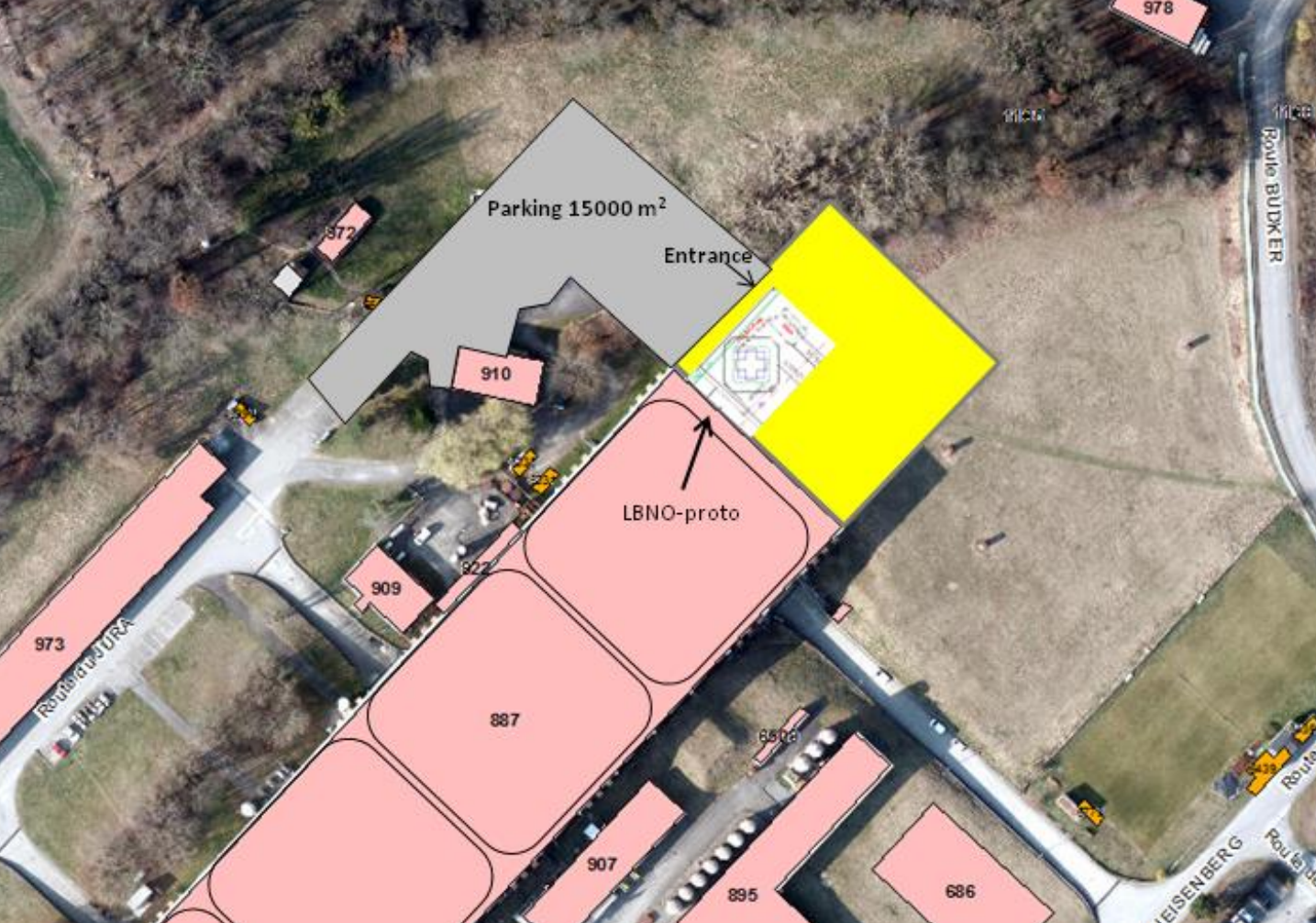} 
\caption{ Areal view of the foreseen extension of the CERN EHN1 (Bat 887) building to 
host the LAGUNA/LBNO detector demonstrators.}
\label{fig:laguna_lardetector_bat887_extension}
\end{center}
\end{figure}

An extension of the
CERN EHN1 experimental hall (See Figure~\ref{fig:laguna_lardetector_bat887_extension})
would provide an appropriate location to host the \six prototype.
This extension fulfils the following requirements: 
\begin{itemize}
\item possibility to host the infrastructure needed during the construction and installation phase, such
as crane, access, electricity, services, by extending the presently existing services in EHN1;
\item controlled area for safe cryogenic operation;
\item charge particle test beam by extension of the existing lines in EHN1.
\end{itemize}
The infrastructure should be provided by CERN, via their technical experimental area support teams.
Refer to \Cref{sec:genrequirements} for a general list of space and infrastructure requirements.

\clearpage
\section{Scientific and technical motivations}

\graphicspath{{./Section-Motivation/figs/}}

\subsection{Calorimetry in charge particle beams}
\label{sec:calochargebeam}

Besides its tracking capabilities, the LAr TPC can also be 
considered as a fully homogeneous
liquid argon ionization chambers, allowing
the energy measurement by total absorption, in the spirit
of the liquid argon calorimeter proposed
by Willis and Radeka~\cite{Willis:1974gi}. 
As was pointed out in their pioneering
work, liquid argon is a medium that satisfies 
all requirements better than many other materials.
All the energy is ideally 
converted into ionization through the development of
the cascade showers, which is then drifted across a gap 
and readout electronically. The integral charge,
once calibrated, yields the incoming particle energy.
In practice, the precision of the ionization measurement
is limited by (i) particle leakage out of the surface of the detector, such as high energy
neutrons or neutral strange kaons,
(ii) energy carried off by neutrinos, (iii) nuclear interactions
releasing or absorbing binding energies,  (iv) charge recombination
(v) charge quenching of
heavy nuclear fragments, alpha particles and nuclear evaporation products
(saturation of response on densely ionizing particles) and (vi) electronic noise.
The energy resolution of the calorimeter will be determined by the 
fluctuations in the above effects. 

Just as in the case of the liquid argon
ionization chamber, the LAr TPC is a single-carrier device 
as far as charge collection is concerned, since 
only the drifting electrons induce visible signals as they
arrive nearby the readout electrodes in the LAr TPC, which is picked
up by a charge sensing preamplifier, and then sampled
and converted to digital information. Offline the integrals of the digitized
signal waveforms yield the full amount of charge of the ionization 
electrons.
This situation is similar to that of calorimeters where
positive ions, due  to their very low mobility, contribute little to the signal 
charge in the short time 
electrons take to drift across the gap (which for
uniformly distributed ionization across the gap yields half the charge).

In order to study the calorimetric response of the LAr TPC, Monte
Carlo simulations with the GEANT3/GFLUKA and
GEANT4 (QGSP\_BERT physics list) were performed in the full geometry of the LAr detector. 
Charge recombination $R$ was taken into account according to the Birks law:
\begin{equation}
  R=\frac{A}{1+k/E_{drift} \left(dE/dx\right)}
 \end{equation}
where $A=0.8$, $k=0.0486$~kV/cm$\frac{\mathrm{g/cm}^2}{\mathrm{MeV}}$ and $E_{drift}$
is the drift field~\cite{Amoruso:2004dy}.
For 1~kV/cm, about 30\% of the produced ionization charge recombines in the
case of minimum ionizing particles.  The digitization of the charge is 
performed using the expected response of the charge preamplifier. The
charge is reconstructed using the same automatic reconstruction algorithm
as used for data.

\begin{figure}[tbh]
\begin{center}
\includegraphics[width=0.495\textwidth]{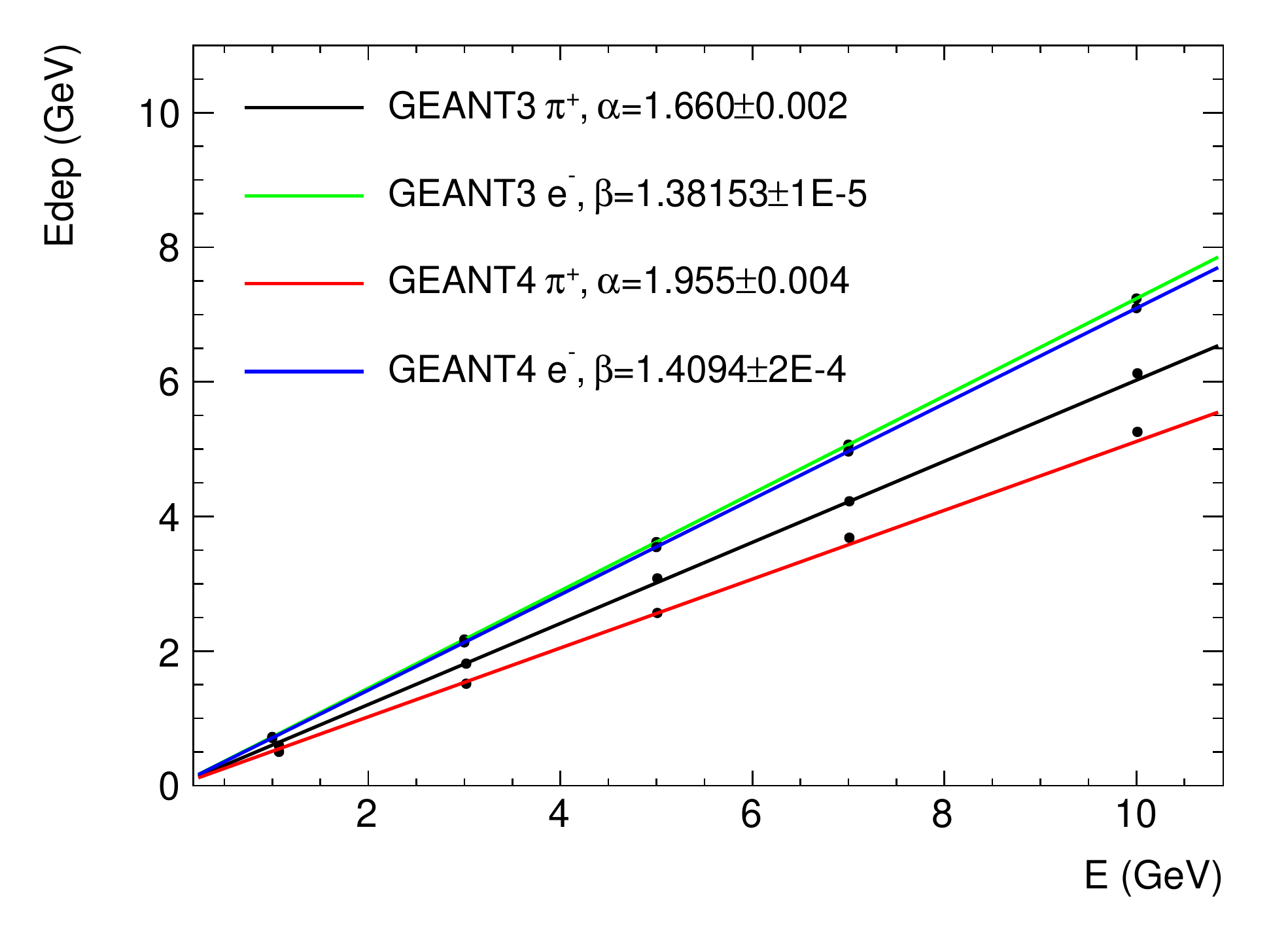}
\includegraphics[width=0.495\textwidth]{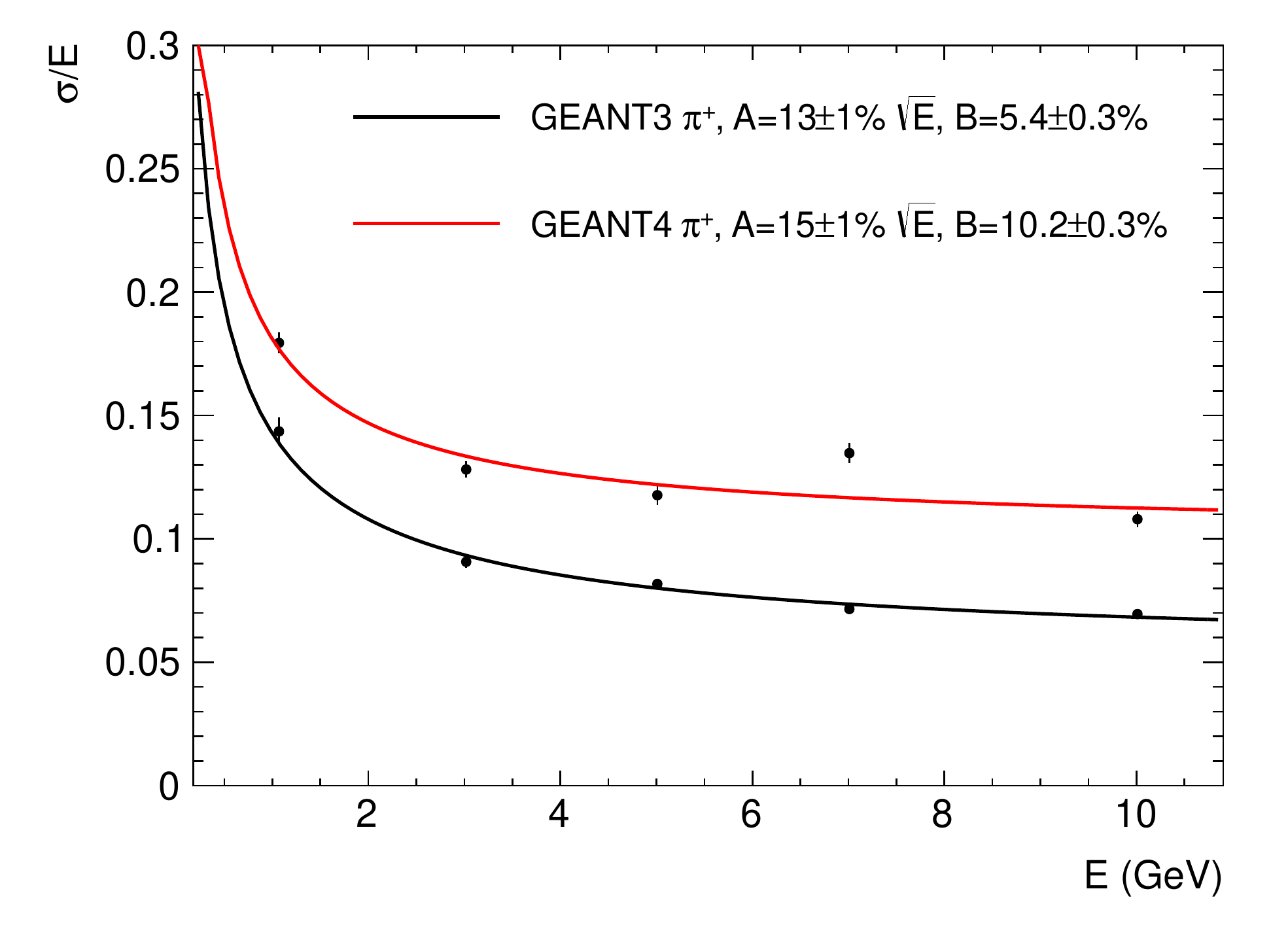}
\caption{LAr energy response of MC generated (GEANT3/GEANT4) $e^-$ and
  $\pi^+$ events for energies up to 10~GeV. The left plot shows the
  obtained results for the correction factors $\alpha$ and
  $\beta$. On the right the reconstructed energy resolution for $\pi^+$
  particles is shown.}
\label{fig:calibration}
\end{center}
\end{figure}

The electromagnetic and hadronic responses were studied with
1, 3, 5, 7 and 10~GeV/c electrons and charged pions.
Since pure liquid argon is a non-compensating medium, 
electromagnetic and hadronic showers require different calibrations.
The $e/\pi$ compensation is performed offline and the reconstructed energy is
therefore defined as
\begin{equation}
  E_{reco}=\alpha E_{had}+\beta E_{em}
\label{eq:ereco}
\end{equation}
where $E_{had}$ is the reconstructed hadronic component,
$E_{em}$ is the reconstruction electromagnetic component
and $\alpha$, $\beta$ are tunable parameters.
Neglecting the electromagnetic components from the charged pions, 
the parameters $\alpha$ and $\beta$ are measured from the hadronic and
electromagnetic energy response. The results are shown for the case of GEANT3
and GEANT4 in Figure~\ref{fig:calibration} on the left. 
Both simulations are in agreement for the electron showers, with a consistent
value for $\beta\simeq 1.4$, as expected from the charge recombination effect.
On the other hand, the results differ for hadronic showers, with GEANT4 predicting about
30\% less ionization than GEANT3. This difference is also reflected in the predicted
hadronic energy resolutions. The right plot
shows the energy dependence of the hadronic shower resolution, superimposed with the
fit $\sigma/E=A/E\oplus B$. The stochastic terms are similar at the level of 15\%, while
the predicted constant term is significantly larger with GEANT4 which predicts 10\%
than with GEANT3/GFLUKA which gives 5\%. The study of calorimetry will be performed
with the LAGUNA/LBNO prototype with unprecedented statistics of charged particles
crossing the detector of well-defined energy and direction. The understanding of
calorimetry is a fundamental milestone to achieve the required level of precision
in the reconstruction of the neutrino energy, fundamental in the future oscillation
experiments at long baselines.

%

As specified above, the calorimetric resolution of the chamber is
limited by particle leakage out of the surface of the detector. The
best energy resolution is obtained if the resulting cascade shower is
fully contained within the fiducial volume of the detector. The
containment of the showers in the \six were studied by comparing the
reconstructed energy\footnote{see Equation 2.2 for a definition of
  reconstructed energy} of $\pi^+$ events with two other simulated
detector geometries: (i) LBNO/GLACIER 20kton, which is the far detector
described in the LBNO expression of interest \cite{Stahl:2012exa}. Here the
event is generated at the center of the detector and the volume is
large enough that the event is fully contained (ii) \three which is a
chamber of 3 m along the beam direction 1m transverse and 1 m drift.

\begin{figure}[htb]
  \centering
  \includegraphics[width=.32\textwidth,height=0.24\textheight]{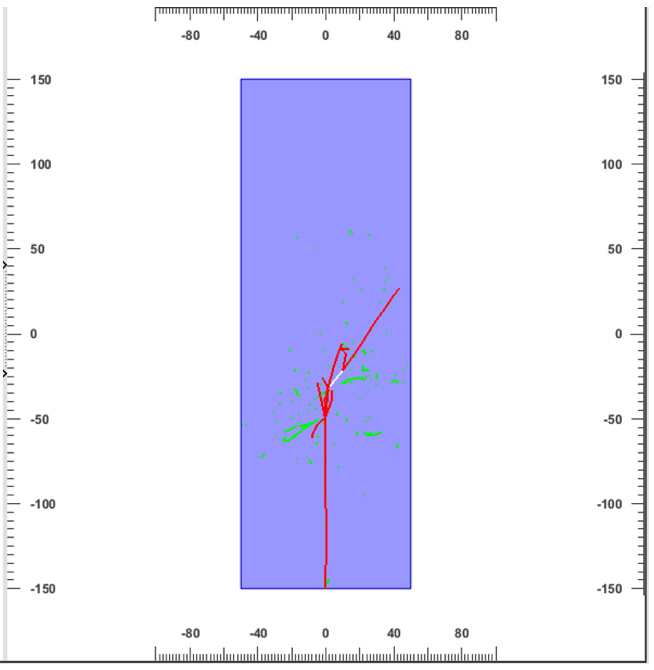}  
  \includegraphics[width=.32\textwidth,height=0.24\textheight]{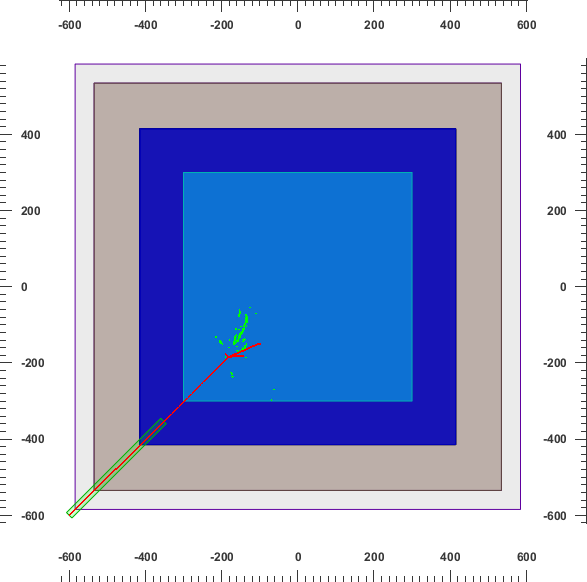}  
  \includegraphics[width=.32\textwidth,height=0.24\textheight]{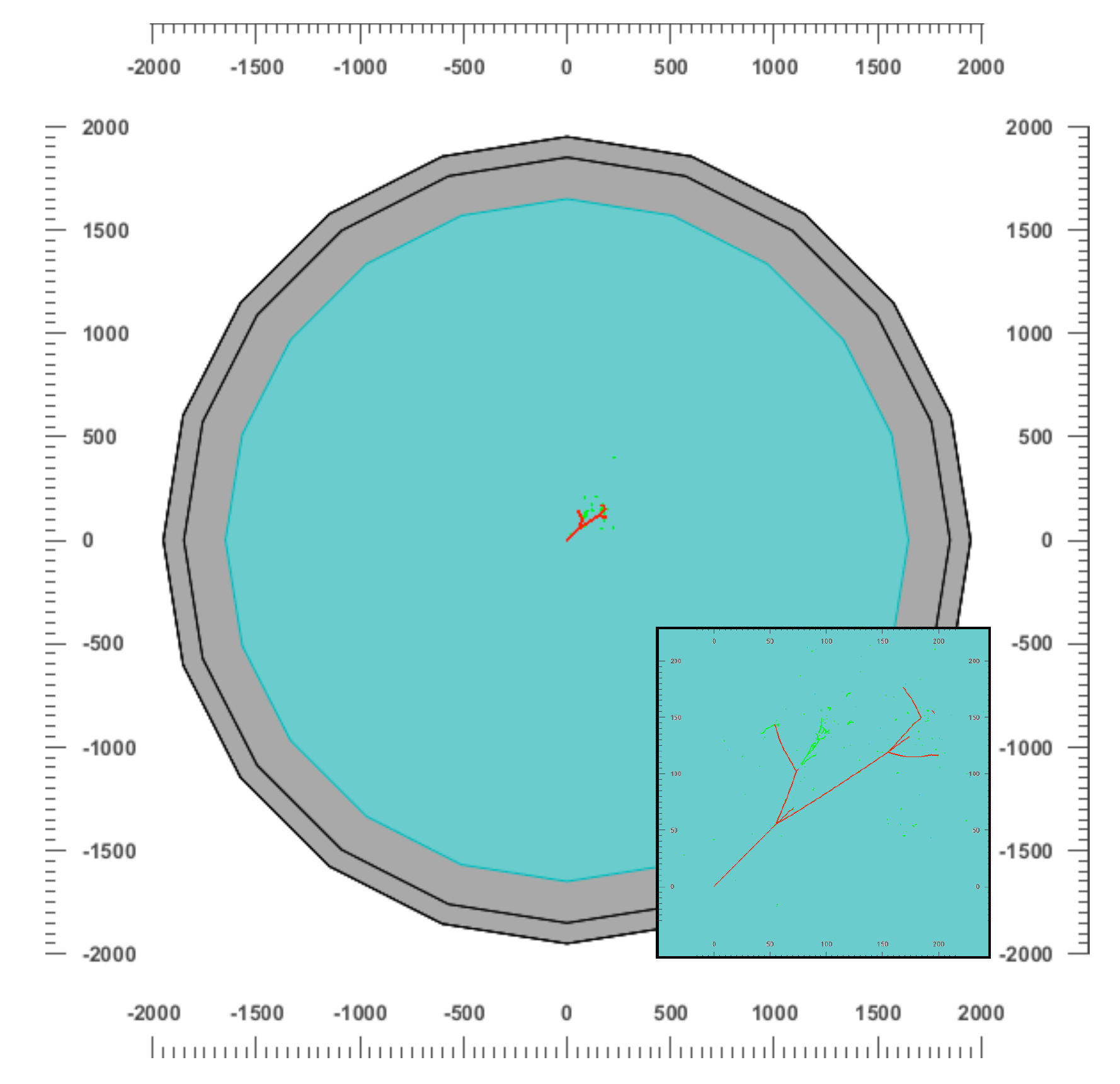}  
  \caption{Topview of a simulated 3 GeV $\pi^+$ in the \three (left),
    the \six (middle) and the LBNO/GLACIER 20kt with a zoom on the shower (right).}
    \label{fig:pip_events_geo}
    \end{figure}

\begin{figure}[htb]
  \centering
  \includegraphics[width=.49\textwidth]{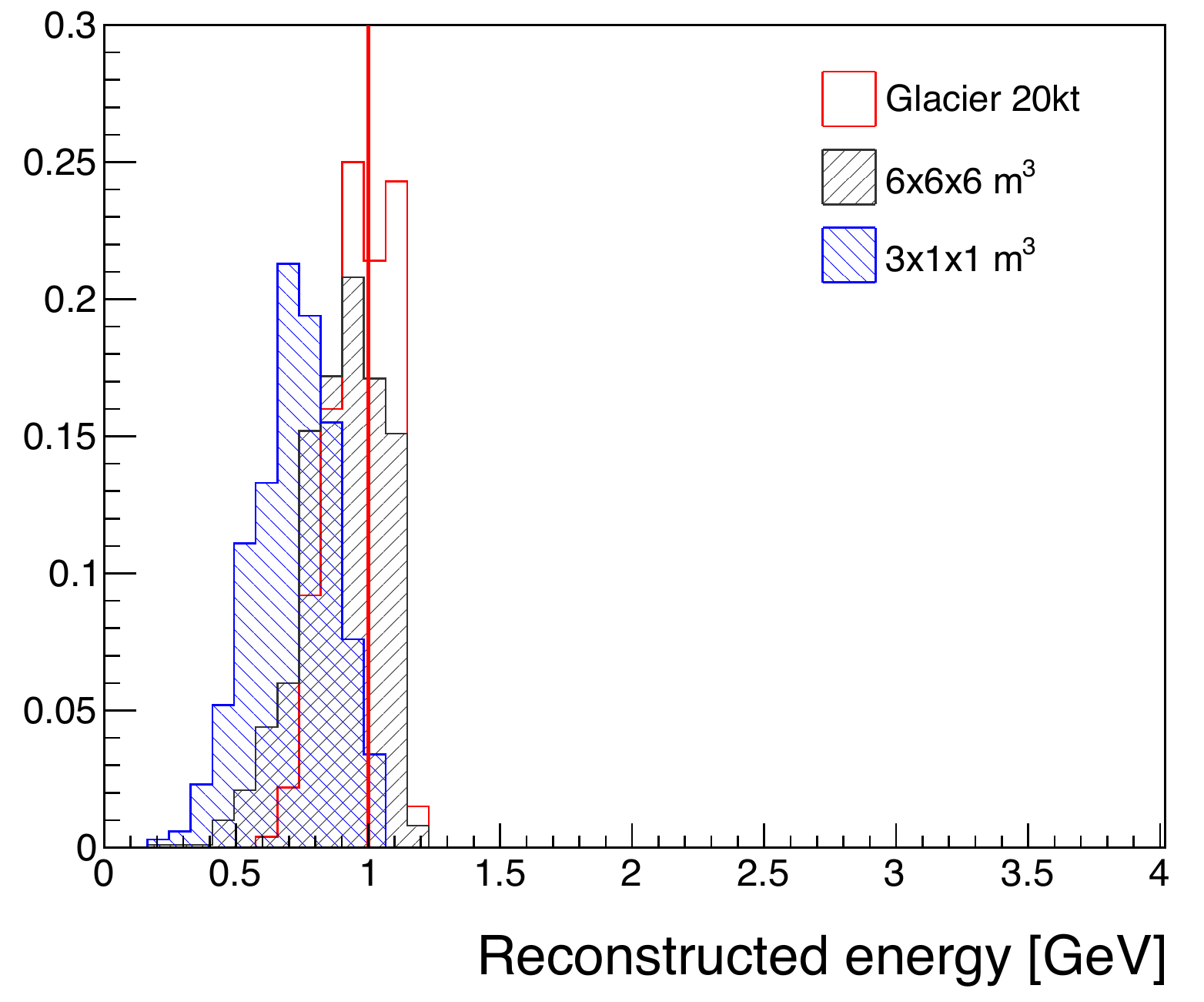}  
  \includegraphics[width=.49\textwidth]{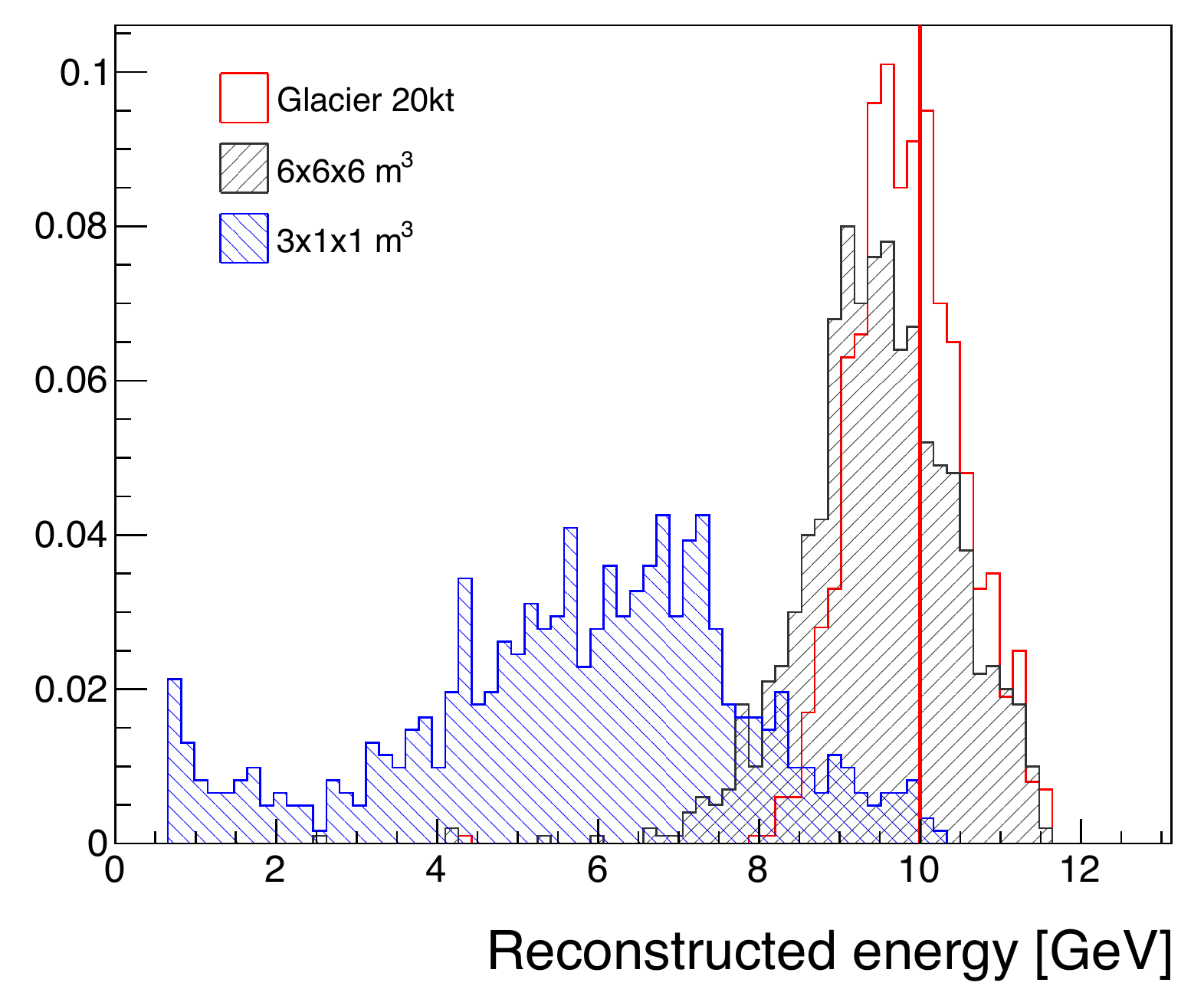}  
  \caption{ Distribution of reconstructed energy for a $\pi^+$ event
    of 1 GeV/c (left) and 10 GeV/c (right) for three different
    detector geometries. The initial energy of the particle is
    indicated by a red line. }
    \label{fig:ereco_dist}
    \end{figure}

\begin{figure}[htb]
  \centering
  \includegraphics[width=.9\textwidth]{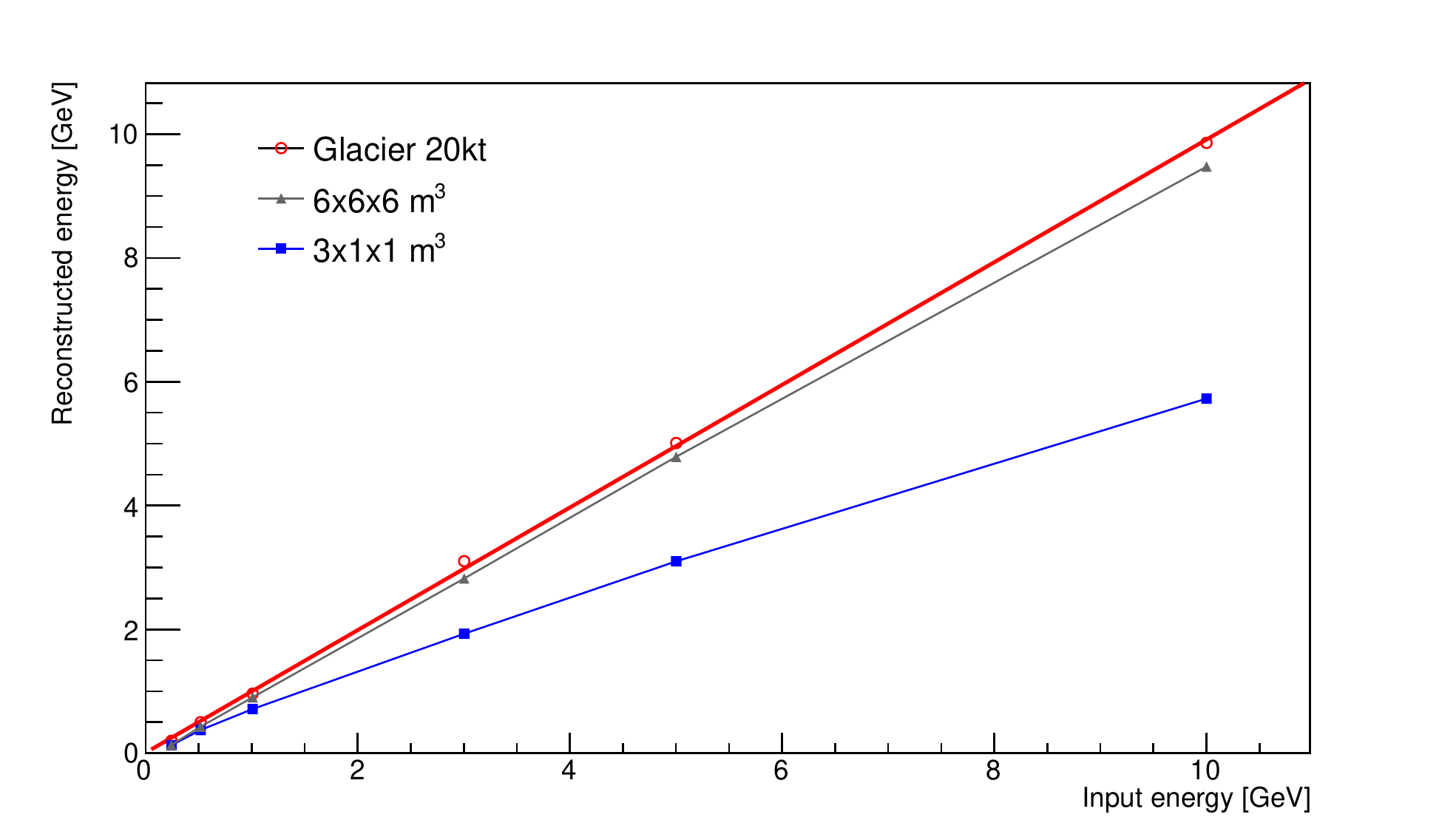}  
  \caption{Mean of the reconstructed energy as a function of the
    $\pi^+$ initial energy for the three considered geometries. }
    \label{fig:ereco_ein}
    \end{figure}

Figure~\ref{fig:pip_events_geo} shows an event display of a shower initiated
by a 3 GeV $\pi^+$ in each of those detectors. For a detailed
discussion on the simulation of detector geometries and particle
transport please refer to \Cref{sec:reco-mc-sim}. The distributions of
reconstructed energies were compared for generated $\pi^+$ of 0.2,
0.5, 1, 3, 5 and 10 GeV in the three detector geometries. The
distributions of reconstructed energies are shown in Figure
\ref{fig:ereco_dist} for pions of 1 and 10 GeV. If the event is fully
contained, as is the case with the LBNO/GLACIER 20kton geometry, the
distribution of reconstructed energy is centered on the initial energy
of the particle. On the other hand if the fiducial volume is not large
enough (e.g in the \three), the distribution is shifted to lower
energies. The mean of the distribution of reconstructed energy versus
the initial energy of the pion is plotted in Figure
\ref{fig:ereco_ein}. It shows that the \six detector is large enough
to contain pion showers at least up to 10 GeV, and is adequate
for energies up to 20~GeV as envisioned. On the other hand,
the smaller \three is not adequate to contain hadronic showers.

Beyond neutrino physics, the high statistics data sample of hadronic interactions collected by the DLAr demonstrator
and its detailed comparison with the predictions of the hadronic showers models will be of general interest, in particular
for the next generation collider experiments. 
These experiments aim at very accurate hadronic calorimetry, based on particle showers imaging and on the application 
of the particle flow approach (PFA) \cite{Schuwalow:2009zz}.

The recent R\&D activity on calorimetry for the future Linear Collider brought to the construction of prototypes 
with unprecedented granularity, bringing detailed information on the properties of hadronic showers. 
The CALICE collaboration constructed an analog hadronic calorimeter prototype (AHCAL) \cite{Simon:2010yh} 
(iron/scintillator with SiPM readout) with a maximal granularity of $3\times 3 cm^2$. The high granularity allowed 
studying, with high precision, properties of the hadronic showers, such as the longitudinal and transverse profiles, 
tracks multiplicities and angular distributions. These data were compared to Monte Carlo models and used to 
develop software e/h compensation techniques.

The DLAr demonstrator is
a completely homogeneous calorimeter with granularity of $3\times 3$~mm$^2$ granularity with full containment 
of hadronic showers. Its data sample of interactions from the charged particle beam will represent the 
ultimate tool for the high accuracy studies of the hadronic showers. They could bring a real breakthrough in the 
improvement of the showers simulation models useful, for a large number of applications in high energy physics.
The granularity is two orders of magnitude finer than the one achieved in the AHCAL prototype~\cite{Simon:2010yh} 
and it is also accompanied by other very powerful handles: 

\begin{itemize}
\item the $dE/dx$ measurements in LAr allow also for the identification of the secondary hadrons produced in the showers. 
This information will complement the multiplicity and angular distribution measurements;
\item the electromagnetic component of the shower can be identified and measured with very high resolution;
\item the energy and transverse momentum $P_t$ balance of the secondary vertices can be measured as well and compared to simulations bringing 
to improved comprehension of the sources of invisible energy in the hadronic showers;
\end{itemize}

All these features will allow for unprecedented studies of the hadronic showers development and very thorough data 
comparisons and validation of the simulation models.

\subsection{Development of automatic event reconstruction}
The LAr TPC technology provides an ``imaging'' of the events with a bubble-chamber-like quality.
Due to the large amount of information, the automatic reconstruction of such complex events is
an important task that needs to be addressed, in order to move towards an unbiased reconstruction
and analysis, not required an intense human intervention (in the spirit of the analysis of the large
sets of photographs collected in the bubble chambers).

While we have already shown that minimum ionising particles and tracking are well understood
and reproduced by our Qscan reconstruction~\cite{Lussi2012}, and that electromagnetic showers are rather
well modelled by Monte-Carlo codes such as GEANT, the main difficulty lies in the reconstruction
of hadronic showers. Indeed, the response is affected by nuclear debris which are highly quenched,
and offline reconstruction algorithm that can automatically handle  the e/pi compensation  must be 
developed.

\begin{figure}[htbp]
\begin{center}
  \includegraphics[width=0.95\textwidth]{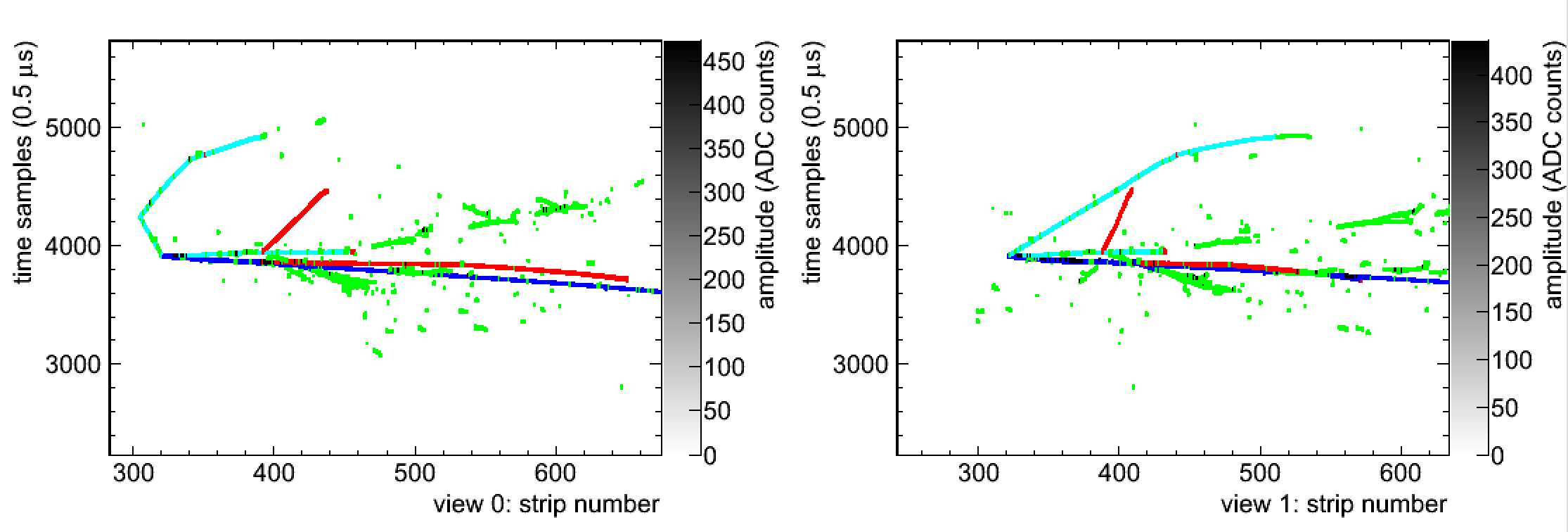}
  \includegraphics[width=0.95\textwidth]{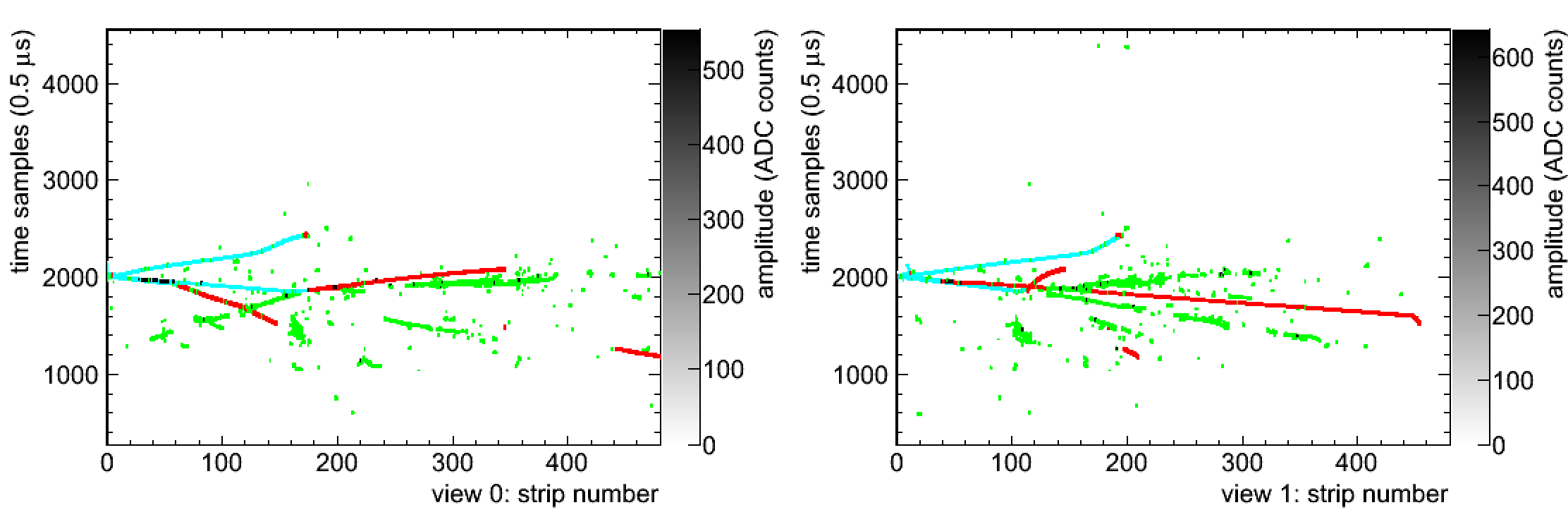}
    \caption{MC comparison of neutrino and pion interactions:
    (top) 5~GeV muon neutrino interaction (bottom) 5~GeV pion interaction. The secondary particles produced
    in the two interactions (blue=muon, green=electron, red=proton, cyan=pion)
    have very similar characteristics (except for the muon in the case of the neutrino charged current (CC) interactions)
    and can be used to assess the performance of the detector and test the energy flow reconstruction algorithms.
    A precise knowledge of the incoming particle energy (not possible in a neutrino beam) is necessary to 
    calibrate and check the linearity of the the response.}
    \label{fig:comparemupi}
\end{center}
\end{figure}

The collection of charged particles in a test beam configuration, as opposed to a neutrino beam,
is very important (and sufficient) to assess the performance and response
of the detectors. \Cref{fig:comparemupi}\,Êcompares the event produced by a muon neutrino interaction
of 5~GeV with the one of a 5~GeV pion interaction. The MC tracks are coloured according to their
type (blue=muon, green=electron, red=proton, cyan=pion).
The secondary particles produced
    in the two interactions have very similar characteristics and topologies (except for the muon in the case of the neutrino CC interactions)
    and can be used to assess the performance of the detector and test the energy flow reconstruction algorithms.

In the case of the charged particles, the well-known momentum of the incoming particle (not accessible on an event-by-event
basis in a neutrino beam) is necessary to
precisely calibrate the energy  and to check the linearity of the response.


\subsection{Charged pions and proton cross-section on Argon nuclei}

The neutrino energy reconstruction is a challenge for current and future long-baseline neutrino oscillation experiments. Several elements contribute to the neutrino energy reconstruction, such as the detector response (see \Cref{sec:calochargebeam}), the interaction of charged particles emitted in the interaction with matter and the re-interaction of hadrons within the nucleus. Concerning
this latter, the high density in the nucleus changes both the nature and the energy of the hadrons produced in the interaction affecting the total energy reconstruction of the event. Traditionally neutrino interaction cross-section Monte Carlos simulate these re-interactions base on a cascade model tuned by external pion-Nucleus cross-section data (see 
\footnote{P. de Peiro, (Neutrino 2012 poster), http://t2k.org/docs/poster/034.} for the description of the implementation for T2K). 

There are three type of interactions considered in the cascade models: 

\begin{itemize}

\item Elastic scattering, the particle changes momentum and direction. 

\item Absorption, the particles are absorbed emitting low energy nucleons. 

\item Charge exchange, the charged particle is absorbed and reemitted with different charge (i.e. proton $\rightarrow$ neutron, $ \pi^+ \rightarrow \pi^0 $,...)

\end{itemize}

\begin{figure}[htbp]
        \begin{center}
                \includegraphics[keepaspectratio=true,height=80mm]{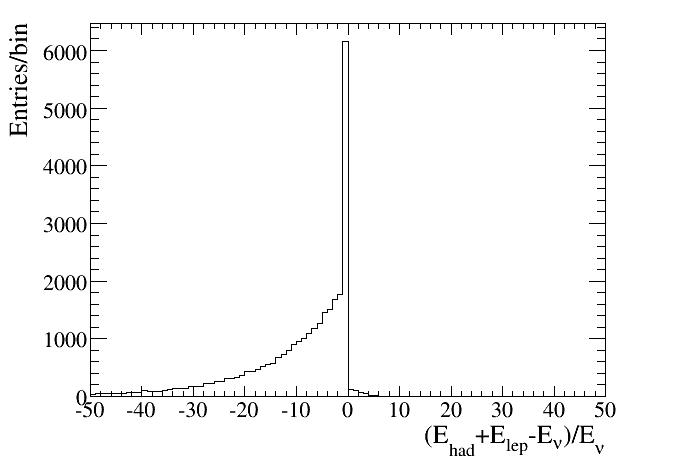}
                        \caption{ Fraction of the neutrino energy visible at the detector following a charge current neutrino nucleus interaction as predicted by GENIE Monte Carlo. }  
                        \label{fig:FSIEffect}
        \end{center}
\end{figure}

\begin{figure}[htbp]
        \begin{center}
                \includegraphics[keepaspectratio=true,height=80mm]{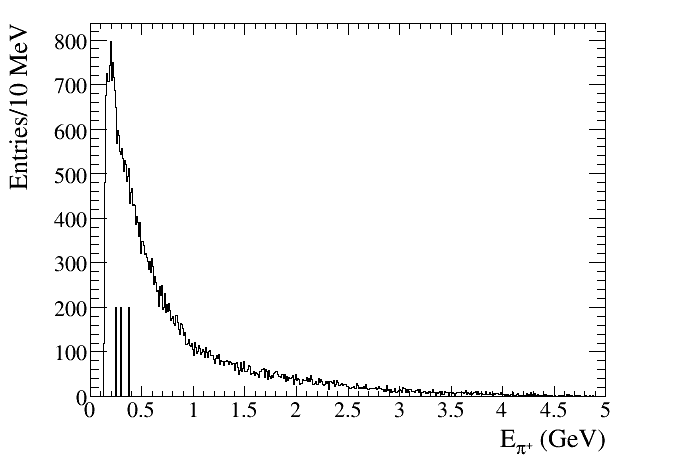}                 
                        \caption{ Pion momentum emitted in neutrino argon interactions as predicted by GENIE Monte Carlo.}  
                        \label{fig:PionE}
        \end{center}
\end{figure}

The last two processes affect the neutrino energy reconstruction by smearing or removing energy from the final state. Figure \ref{fig:FSIEffect}  shows the fraction of the neutrino energy  visible at the detector after a charge current neutrino interaction as simulated by the GENIE MC \cite{Andreopoulos:2009zz} for the typical
LBNO beam neutrino spectrum.  The reconstructed neutrino energy shows lower energy than the original one producing an asymmetric distribution. 
This distribution needs to be known to reconstruct precisely the oscillations parameters such as the mass difference squared.

The current knowledge about $\pi^+$ argon cross-section at low energy is limited. The most recent result \cite{Rowntree:1999dp}  provides the total cross-section and the final state multiplicity for three values of the pion kinetic energy (118, 162 and 239 MeV). These values do not cover, see \Cref{fig:PionE}, the full spectrum of $\pi^+$ emitted by the neutrino interactions. Another limitation is that the results in \cite{Rowntree:1999dp} do not distinguish between the three interactions modes described above, each one with a different contribution to the energy reconstruction bias. 

These interactions modify the reconstruction of the event worsening the neutrino energy reconstruction and increasing the probability of lower energy reconstruction. This effect is much stronger for the interactions happening inside the nucleus. The direct measurement of the hadron-nucleus cross sections in liquid argon will be helpful in the improvement of the nuclear cascade models which are based on these data.

\begin{figure}[htbp]
\begin{center}
  \includegraphics[width=0.95\textwidth]{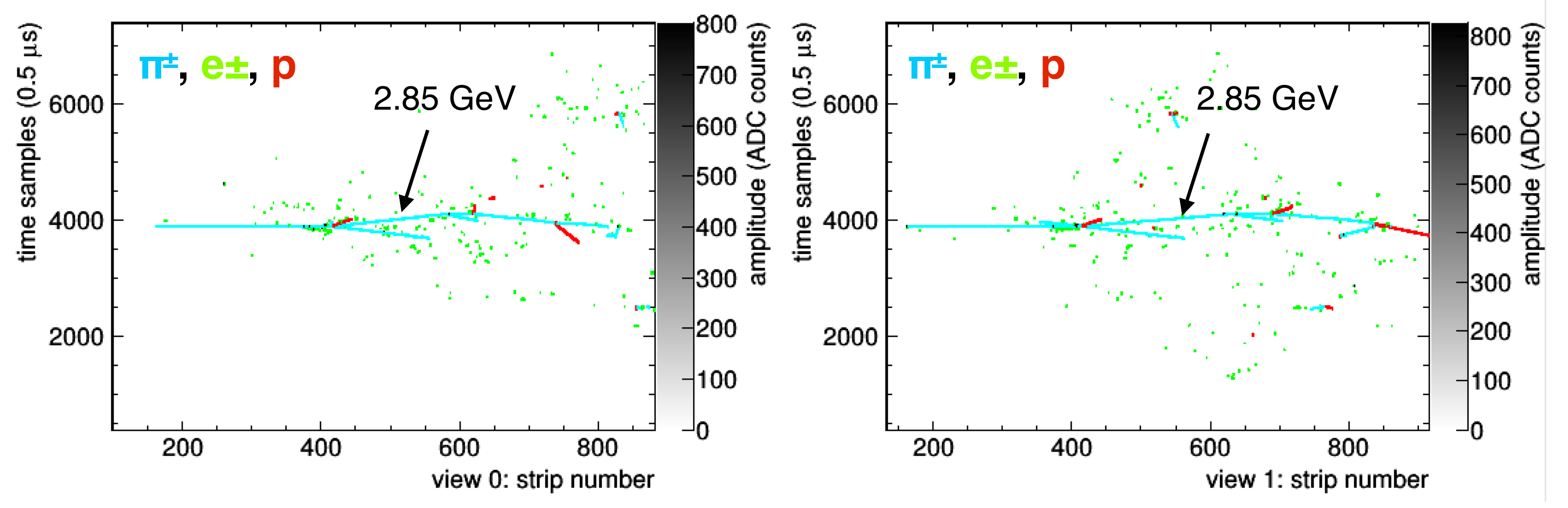}
    \caption{A 5 GeV pion interaction, where a secondary pion of 2.85~GeV reinteracts in the detector.}
    \label{fig:pireinteraction}
\end{center}
\end{figure}

The DLAr will help in reducing these uncertainties for future experiments by means of measuring the strength of the exclusive interaction channels and the properties of the final state particles. The cross-section can be studied via the response to beam particles. 
An alternative method is to study the low energy products of the interaction of high-energy  hadrons inside 
the detector volume by reconstructing the full cascade history of the event. 
This is illustrated in \Cref{fig:pireinteraction} where the event display of a 5~GeV pion interaction is displayed, in which
a secondary pion of 2.58~GeV has re-interacted.
The MC tracks are coloured according to their
type (blue=muon, green=electron, red=proton, cyan=pion). A full kinematical reconstruction (fit) of such events will yield
information on the exclusive interactions cross-sections of secondary particles of lower momenta.


\subsection{Development and proof-check of industrial solutions}

Several years of studies and engineering efforts in the context of the LAGUNA/LBNO design studies (funded by EC)
have led to the conceptual design of the 20 and 50~kton GLACIER detectors. Further steps beyond these efforts
require the actual demonstration of the techniques on a smaller scale, ideally in an optimal laboratory environment
as CERN.
The technical goals are to develop and assess the design in view of an extrapolation
to tens of kton mass scales and a long-term stable operation ($>$10~years), as
necessary to accomplish the LBNO physics programme~\cite{Stahl:2012exa}. One 
fundamental issue is the drift length, whose optimisation involves the interplay
of several technical components, such as the high-voltage (for the drift field), the
purity, the mechanical aspects of the field cage, etc. It
can only be addressed with a prototype with a drift length of several meters.
To test the feasibility and the performance of an ultimate drift length of 20~m~\cite{Rubbia:2004tz}, a minimum
length of 6~m is required, where the longer drift conditions expected at a drift
field of 1~kV/cm can be reproduced by the shorter drift length of 6~m at a field of
about 0.3~kV/cm.
From the technological point of view the most urgent questions to be addressed are
summarised in the following:

\begin{itemize}
\item \textbf{Very high purity}. The drift of ionization electrons over a distance of 20 m requires a very clean environment, with impurities at the level of 100 ppt O$_2$ equivalent for an electron lifetime of 3-10 ms. While this has been achieved on small prototypes, this will be the first test with a large scale non-evacuable prototype and the same tank construction technique foreseen for the far detector.  
\item \textbf{Large field cage.} This is a large and hanging structure with demanding requirements on its mechanical precision and capable of sustaining a large potential difference (up to 500 kV).    
\item \textbf{Very high voltage generation.} A very low noise and stable power supply able to reach 600~kV to generate an uniform drift field of 1~kV/cm (300~kV power supplies with the required specification are commercially available).   
\item \textbf{Large area micropattern charge readout.} A large 36 m$^2$ surface will be instrumented with a charge sensitive device providing gas amplification in ultra pure argon vapour.  
\item \textbf{Cold front-end charge read-out electronics.} A good S/N is crucial to reach the required physics performances, especially for the low energy neutrino physics. An innovative solution with preamplifiers located as close as possible to the charge-sensitive anode, but yet accessible without opening the inner vessel, will be tested. 
\item \textbf{Long term WLS coating.} A  method based on WLS deposition with very long stability ($>10$~years) will be implemented and tested.
\item \textbf{Integrated light readout electronics.} New integrated devices will be developed for the digitisation of argon scintillation light, scalable to very large detectors.
\end{itemize}

\clearpage
\section{DLAr Detector overview}

\graphicspath{{./Section-DetOverview/figs/}}

\subsection{Design concept of the $6\times 6\times 6$m$^3$ prototype}
\begin{figure}[htb]
\begin{center}
\includegraphics[width=0.975\textwidth]{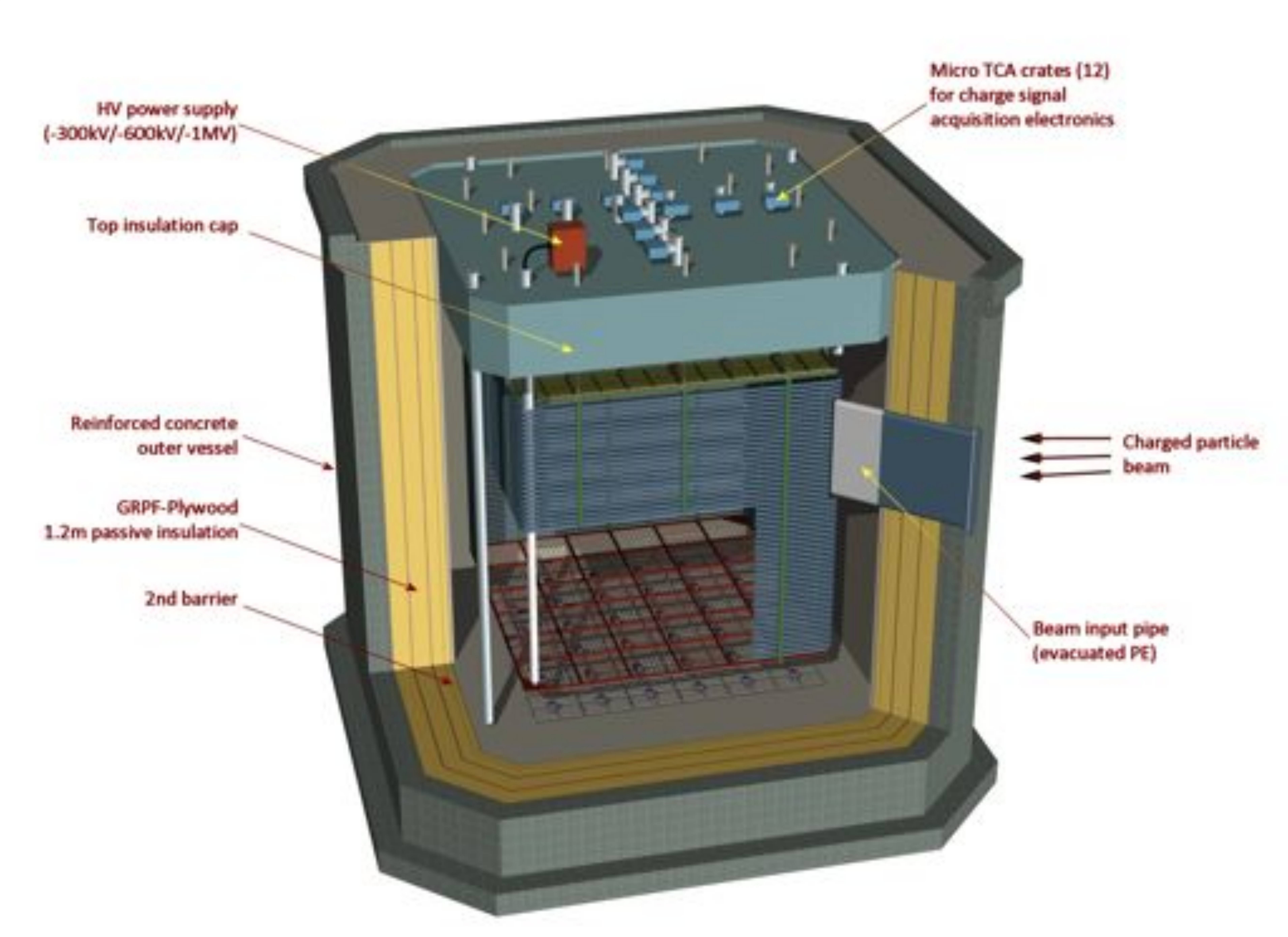}
\caption{\small Illustration of the \six with the inner detector inside the cryostat.}
\label{fig:laguna_innerdet_cryo}
\end{center}
\end{figure}

\begin{figure}[tbp]
\begin{center}
\includegraphics[width=0.95\textwidth]{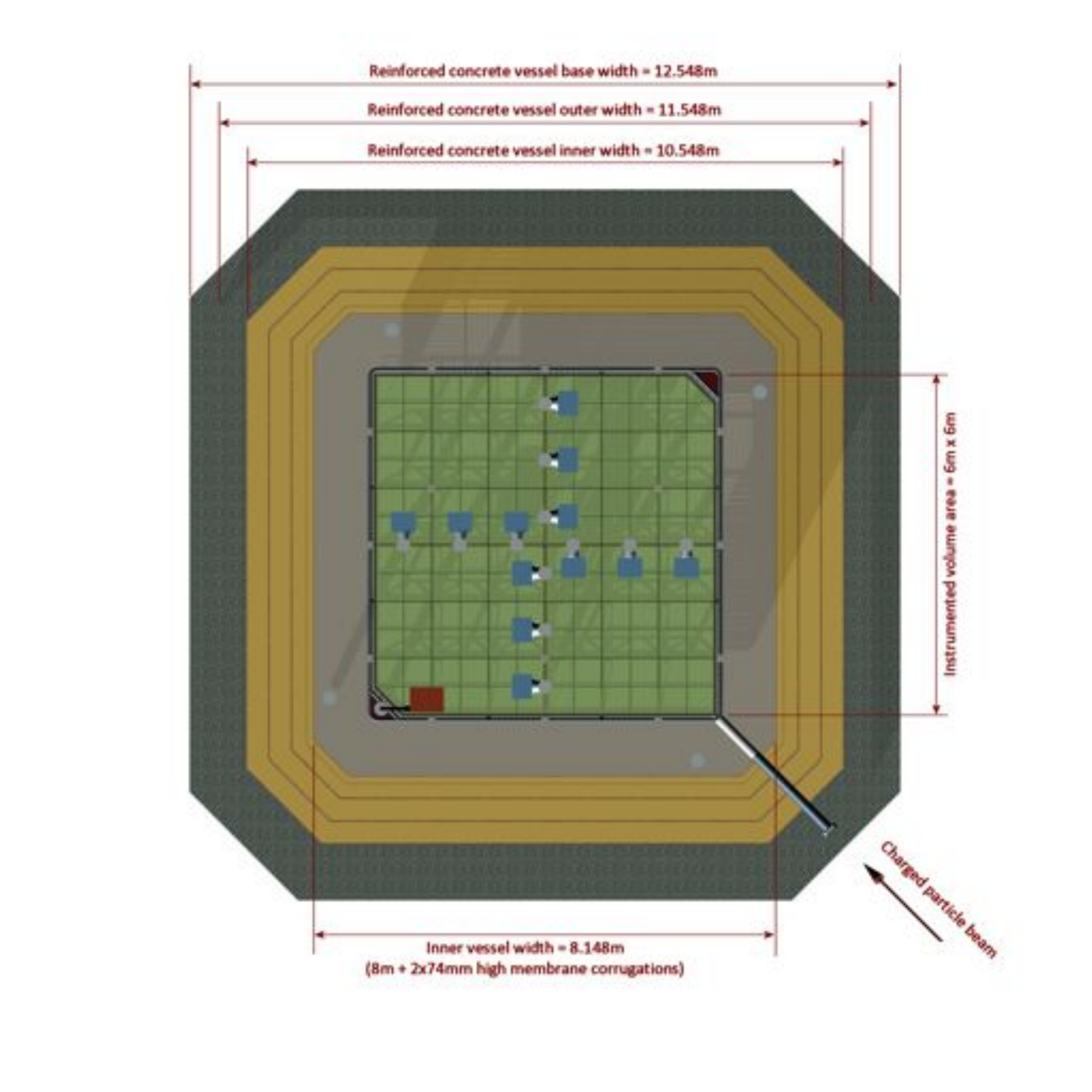}
\vspace*{-1.5cm}
\caption{\small Plan view section of the \six.}
\label{fig:laguna_horizsection}
\end{center}
\end{figure}

\begin{figure}[htp]
\begin{center}
\includegraphics[width=0.9\textwidth]{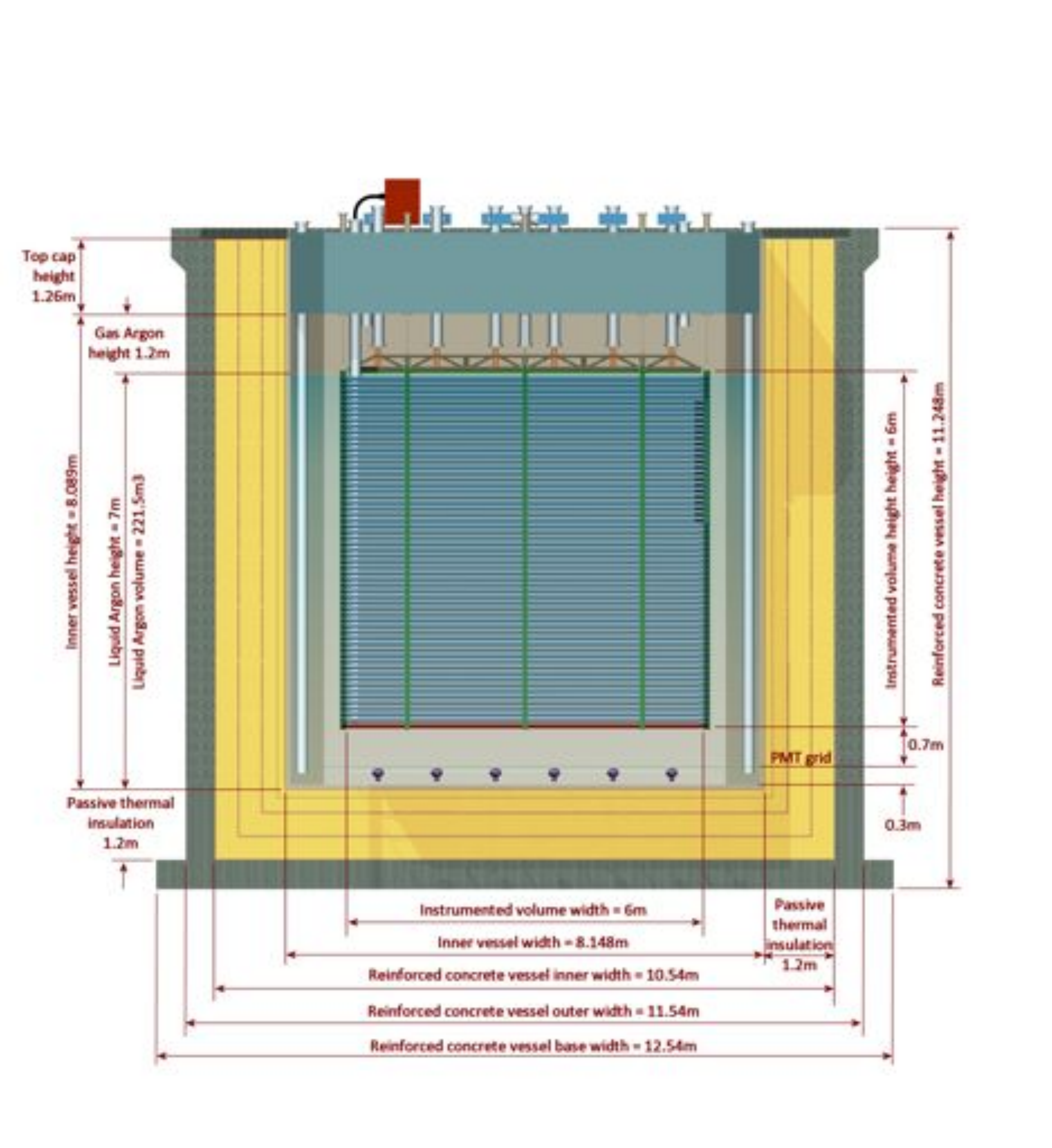}
\vspace*{-1.cm}
\caption{\small Vertical cross section of the \six.}
\label{fig:laguna_vertsection}
\end{center}
\end{figure}

The \six prototype is illustrated in \Cref{fig:laguna_innerdet_cryo}.
Following the GLACIER concept,
the LAr detector has the shape of a vertically standing volume, where
electrons are drifted vertically towards the liquid-vapor interface, extracted
from the liquid into the gas phase, amplified and collected
at a segmented anode~\cite{Badertscher:2013wm,Badertscher:2012dq,Badertscher:2010zg}.
The main parameters are summarised in Table~\ref{tab:params}.
The horizontal and vertical sections are shown in \Cref{fig:laguna_horizsection,fig:laguna_vertsection}.
\begin{table}[tbp]
  \centering
 \begin{tabular}{|l|c|c|}
\hline
Liquid argon density & T/m$^3$& 1.38 \\
Liquid argon volume height & m& 7.6 \\
Active liquid argon height& m  & 5.99 \\
Hydrostatic pressure at the bottom& bar & 1.03 \\
Inner vessel size (WxLxH) &m$^3$ & 8.3 $\times$ 8.3 $\times$ 8.1\\
Inner vessel base surface& m$^2$& 67.6 \\
Total liquid argon volume& m$^3$ & 509.6 \\
Total liquid argon mass & t & 705 \\
Active LAr area & m$^2$& 36 \\
Charge readout module (0.5 x0.5 m$^2$) & & 36\\
N of signal feedthrough & & 12 \\
N of readout channels & & 7680\\
N of PMT & & 36 \\
\hline 
\end{tabular}
 \caption{Main parameters of the LBNO prototype.}
 \label{tab:params}
\end{table}

The uniform drift field is created by a field cage composed
of several equally-spaced stainless-steel tubes, held in place
by insulating mechanical structures which are hanged from
the top cap of the vessel. 
The anode deck is also
suspended with stainless-steel
ropes linked to the top roof. 
The bottom field is 
closed by a transparent cathode and the top field by an anode, which also 
serves as the charge readout.
The light readout consists of PMTs uniformly
distributed below the cathode.

The detector is configured as a $6\times 6 \times 6$~m\textsuperscript{3} liquid argon
TPC with liquid-to-gas ionisation electron extraction
and multiplication before collection. The ionisation charge is
collected in a 2-dimensional readout plane on the top of the volume
with an area of $6\times 6$~m$^2$ and finely segmented with a 3~mm strip pitch.
The full active volume of $\sim$216 m\textsuperscript{3} is under an
uniform electric field  $E{\simeq} 0.5-1.0$~kV/cm generated from a bottom
cathode plane (also $6\times 6$~m\textsuperscript{2}) operated at $\simeq$300-600~kV and
kept uniform by a stack of field shaping electrodes (round pipes along
a square path, with rounded corners) polarised at linearly decreasing
voltage from the cathode voltage to ground.
%
%

The cathode plane is gridded and transparent to light
to allow the detection of the scintillation
light by an array of photomultipliers located at a distance of $\sim$1m
under it.
Ionization charge signals are sent to a set of signal feedthroughs (12),
located on the top face of the hosting LAr vessel and hosting the cold
readout electronics. Other chimneys/feedthroughs are foreseen for HV (1
or 2), top readout plane suspension and level regulation (3), PMT high
voltage and signal readout (4), monitoring instrumentation (level,
temperature, 1 or 2).
The front-end electronics is based on analog preamplifiers implemented in
CMOS ASIC circuits for high integration and large scale affordable production. The
baseline design is to integrate this electronics on the feed-through flange terminating
the chimneys on the roof of the tank, under the insulation layer in order to be cooled to a temperature near that of liquid argon.
Cathode and field shaping electrodes are kept in their position by a set
of insulating supports/spacers resting on the inner vessel floor.
The inner vessel has a cubic shape with chamfered vertical edges. 
Its walls are built with the so-called corrugate membrane
technique, to compensate thermal shrinkage. A manhole and a
detail-introduction hole are located in its top face. Thermal
insulation is passive, based on GRPF (glass reinforced polyurethane
foam) layers, interspersed with pressure distributing layers of
plywood. Its thickness and composition are such to reach a residual heat
input of 5~W/m\textsuperscript{2}. The estimated total heat input including
also the losses through the feed-throughs, cables, etc., and introduced by the LAr process, 
is $\sim$4~kW at liquid argon temperature, to be
dissipated by cryocooler(s).
The passive insulation is contained in a reinforced concrete
{\textquotedblleft}vessel{\textquotedblright} with $\simeq$~0.5~m thick
walls. The top outer ceiling is made by a framework reinforced
stainless steel plane, able to support the inner anode and outer
instrumentation (electronics, cryogenics, control). 

A charged beam pipe (evacuated) is indicated as crossing the concrete
outer vessel and the thermal insulation layers. Its vertical
orientation is adapted to the charged beam vertical axis in its last
section. 
%
%


\subsection{Liquid argon as detector medium and electron drift}

Liquid argon, and liquid noble gases in general, have interesting properties for detecting particles, what makes them a good choice as target material. A summary of the most important argon properties is given in \Cref{t_argon_specifications}.
\begin{table}[tb]
\begin{center}
\begin{threeparttable}[b]
\begin{tabular}{|c|c|}
\hline
\multicolumn{2}{|c|}{\textbf{General properties}} \\
\hline Atomic number & 18 \\
\hline Molecular weight & $39.948 \unit{~g/mol}$ \\
\hline & $^{36}$Ar (0.34\%)\\
Most important stable isotopes &$^{38}$Ar (0.06\%)\\
&$^{40}$Ar (99.60\%) \\
\hline Concentration in air & $0.934\%$ \\
\hline Melting point ($1 \unit{~atm}$) & $83.8 \unit{~K}$ \\
\hline Boiling point ($1 \unit{~atm}$) & $87.3 \unit{~K}$ \\
\hline Triple point & $83.8 \unit{~K}$ and $0.687 \unit{~bar}$\\
\hline Ratio LAr/GAr ($1 \unit{~atm}$)& $835 \unit{~vol/vol}$ \\
\hline \multicolumn{2}{|c|}{\textbf{Gaseous phase properties}} \\
\hline Gas density (at boiling point resp. $15\unit{~^{\circ}C}$) & $5.85 \mbox{ resp. }1.67 \unit{~kg/m^{3}}$ \\
\hline Heat capacity at constant pressure $C_{p}$ ($1 \unit{~bar}$ and $25 \unit{~C\,^{\circ}}$) & $0.02 \unit{~kJ/mol \cdot K}$ \\
\hline Heat capacity at constant volume $C_{v}$($1 \unit{~bar}$ and $25 \unit{~C\,^{\circ}}$) & $0.012 \unit{~kJ/mol \cdot K}$ \\
\hline Thermal conductivity of GAr ($1 \unit{~atm}$ and $273 \unit{~K}$)& $16.36 \unit{~mW/m \cdot K}$ \\
\hline Thermal conductivity of GAr (boiling point and $1 \unit{~atm}$)\tnote{$\ast$}& $5.66 \unit{~mW/m \cdot K}$ \\
\hline Ionization energy in gas $W_{ion}$& $26.4 \unit{~eV}$ \\
\hline\multicolumn{2}{|c|}{\textbf{Liquid phase properties}} \\
\hline Liquid density (at $87.3 \unit{~K}$)& $1392.8 \unit{~kg/m^{3}}$\\
\hline Latent heat of vaporization $L_{v}$ ($1 \unit{~atm}$) & $160.81 \unit{~kJ/kg}$ \\
\hline Thermal conductivity of LAr ($87.3 \unit{~K}$)\tnote{$\ast$}& $127 \unit{~mW/m \cdot K}$ \\
\hline $dE/dx_{min}$ for mip & $2.12 \unit{~MeV/cm}$ \\
\hline Ionization energy in liquid $W_{ion}$ & $23.6 \unit{~eV}$ \\
\hline Excitation energy $W_{\gamma}$ & $19.5 \unit{~eV}$ \\
\hline Maximum of emission spectrum & $\sim 128 \unit{~nm}$\\
\hline Radiation length $X_{0}$ & $ 14 \unit{~cm}$ \\
\hline Moli\`ere radius & $ 9.28\unit{~cm}$ \\
\hline Nuclear interaction length & $84 \unit{~cm}$ \\
\hline Maximal breakdown strength & $ 1.1-1.4\unit{~MV/cm}$ \\
\hline 
\end{tabular} 
 \begin{tablenotes}
 \tiny{\item[$\ast$] Value linear extrapolated from \cite{Younglove1986}}
 \end{tablenotes}
\caption{Physical and chemical properties of argon.\label{t_argon_specifications}}
 \end{threeparttable}
 \end{center}
\end{table}
Because of its density of $1.4 \unit{~g/cm^{3}}$, liquid argon is a good stopping material with an average energy loss $dE/dx$ of $2.12 \unit{~MeV/cm}$ for a minimum ionizing particle (mip). Also, an interaction of a particle with an argon atom excites it and creates scintillation light, or the atom is ionized. The mean excitation energy of $W_{\gamma}=19.5 \unit{~eV}$, and the mean ionization energy $W_{ion}=23.6 \unit{~eV}$, give a very good energy resolution also for detecting low energetic particles. 
The possibility to drift charges within the liquid noble gas along the electric drift field, without big diffusion, gives the option to read out ionising tracks with a high spacial resolution. 
The basic principle of every time projection chamber is that free charge carriers are produced in a relatively large volume and then drifted towards an electrode. The time needed between the initial ionisation and the moment the signal is read out from the electrode defines, together with the drift velocity, the distance between the interaction and the electrode along the lines of the electric field. In case of the double
phase LAr TPC, the drift velocity in gas argon as also in liquid argon have to be taken in account. 

Argon is a byproduct from the air liquefaction process and therefore relatively cheap
compared to other noble gases. This makes it the only viable candidate 
for very large detectors of several tens of kilotons of active volume. 


\subsection{Drift velocity and diffusion in argon gas}

\begin{figure}[t!]
\begin{minipage}[t]{0.5\linewidth}
\centering
\includegraphics[width=70mm]{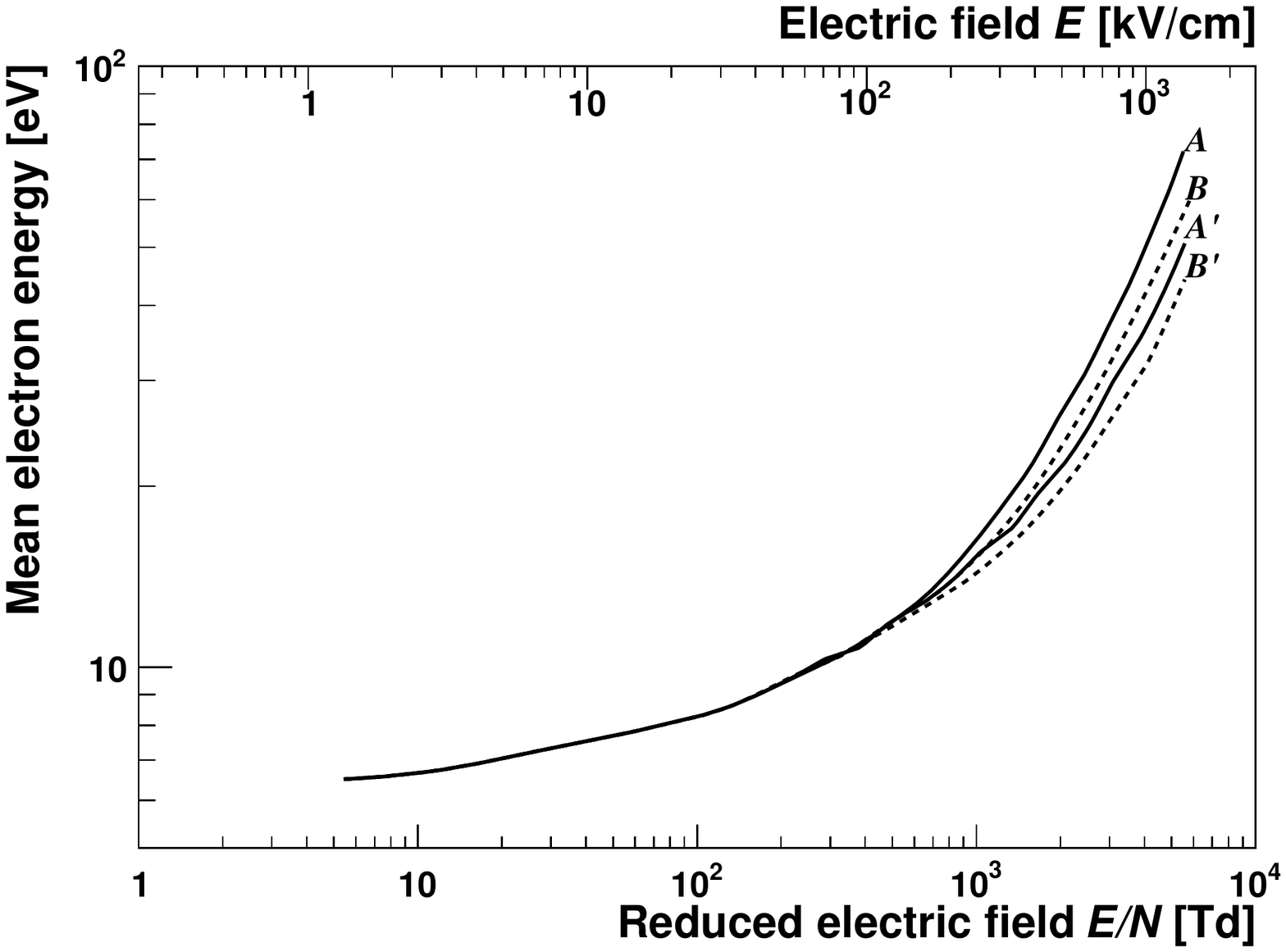}
\caption[The mean electron energy in argon gas as a function of the external electric field.]{The mean energy in argon gas as a function of the external (reduced) electric field. In case A, both the
elastic and inelastic scatterings are according to the elastic differential cross-section. In case B the
inelastic scattering is isotropic. $A$ and $B$ are the values before and $A'$ and $B'$ after scattering \cite{Kucukarpaci1981}. The scale of the electric field $E$ is given for $1 \unit{~bar}$ and $20 \unit{~^{\circ}C}$.}
\label{f_ElectronEnergy}
\end{minipage}
\hspace{0.5cm}
\begin{minipage}[t]{0.45\linewidth}
\centering
\includegraphics[width=70mm]{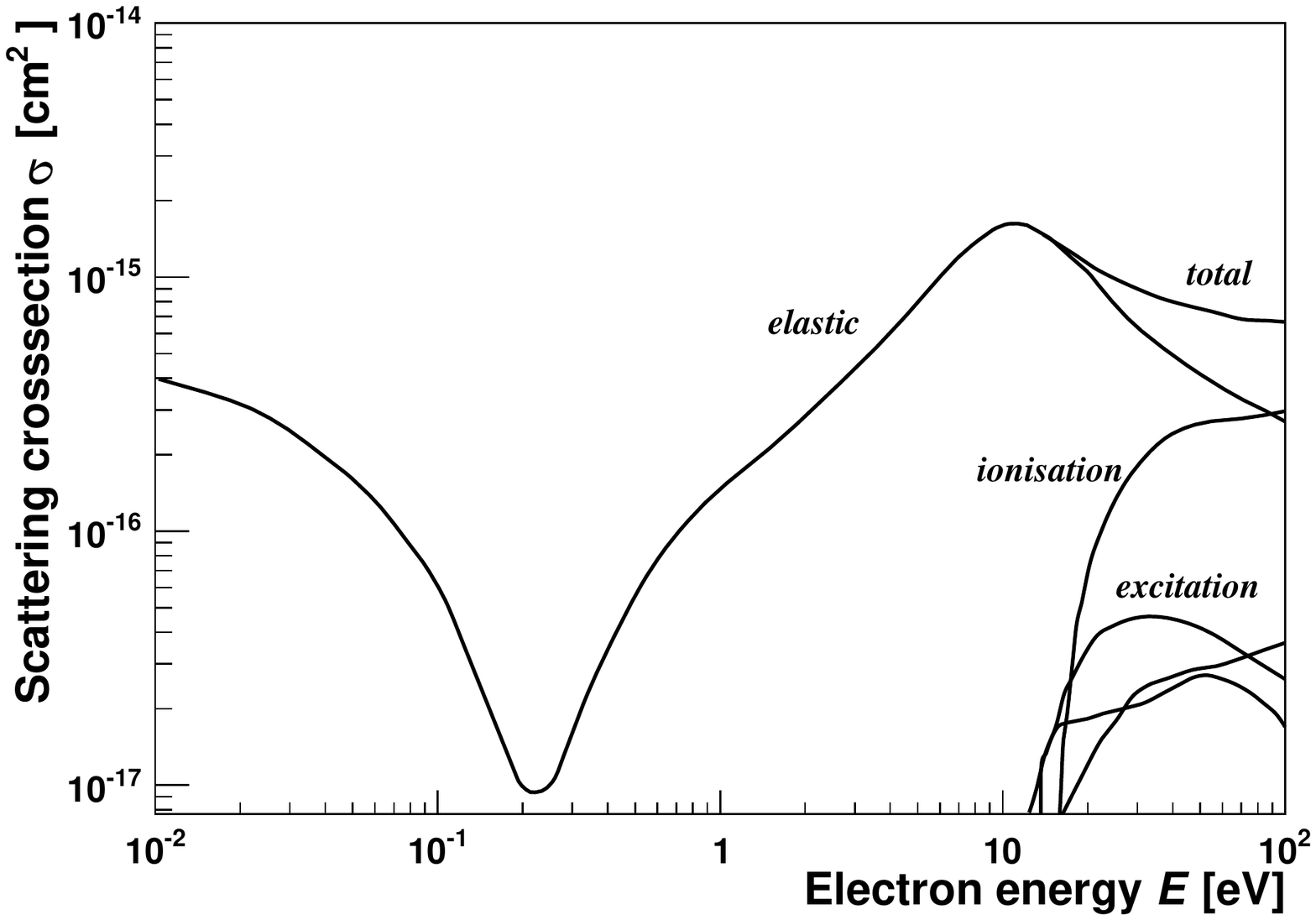}
\caption[The cross-section between electrons and argon gas as a function of the total energy calculated with Magboltz by S. Biagi]{The cross-section between electrons and argon gas as a function of the total energy of the electron, calculated with Magboltz \cite{Biagi1989}. Clearly visible is the Ramsauer-minimum around 0.2 eV.}
\label{f_CrosssectionArgon}
\end{minipage}
\end{figure}

Applying an electric field, the electron is drifted along the field line and the total energy depends on the electric field strength as shown in \Cref{f_ElectronEnergy}. 
The Figure is normalized to a reduced electric field $E/N$ with Townsend unit 
(\unit{Td}=$\unit{[E/N]}=10^{-17} \unit{~V\, cm^{2}}$),
where $E$ is the electric field and $N$ the molecular density.
The top label indicates the values for SATP conditions 
(Standard Ambient Temperature and Pressure: $20 \unit{~^{\circ}C}$ and $1 \unit{~bar}$).

\begin{figure}[t!]
\begin{minipage}[t]{0.5\linewidth}
\centering
\includegraphics[{width=70mm}]{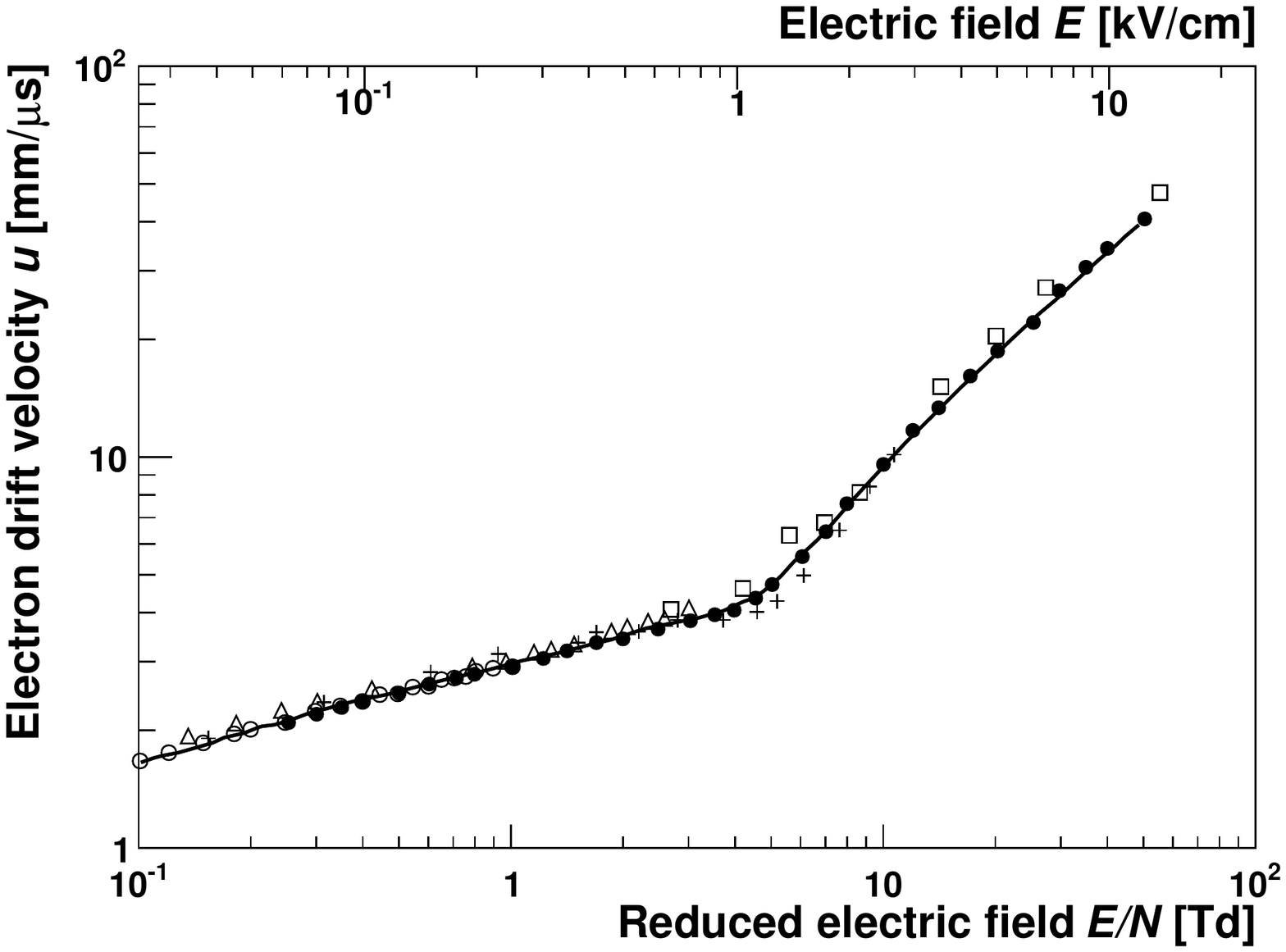}
\caption{The electron drift velocity $u$ in gaseous argon as a function of the normalized electric field ($\unit{[Td]=10^{-17} V cm^{2}}$) according to \cite{Nakamura1988}. The scale for the electric field $E$ is derived for SATP conditions.}
\label{f_ElectronDriftvelocity}
\end{minipage}
\hspace{0.5cm}
\begin{minipage}[t]{0.45\linewidth}
\centering
\includegraphics[width=70mm]{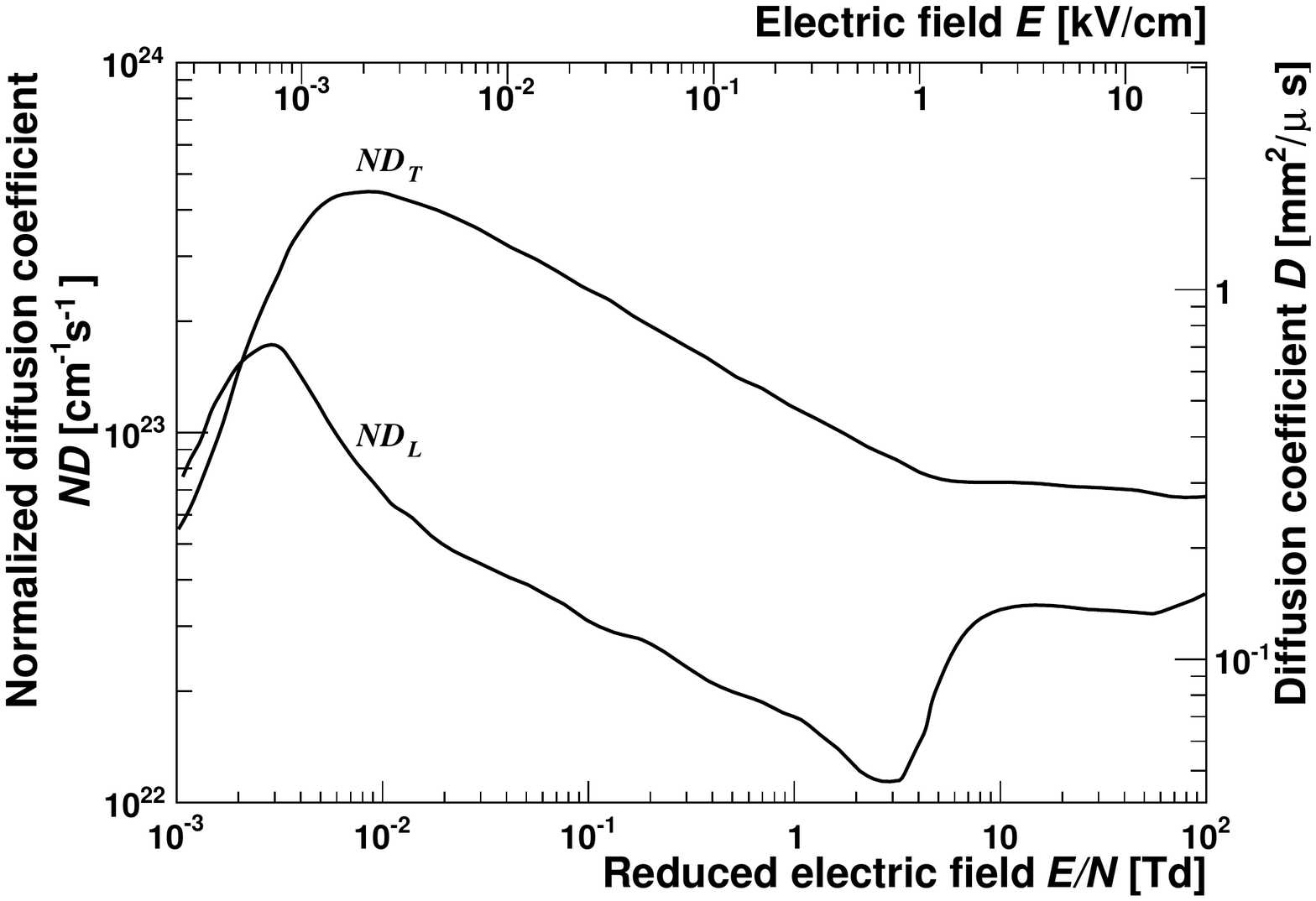}
\caption{The normalized longitudinal and transversal diffusion coefficient $ND$ for gaseous argon according to \cite{Suzuki1990}. The scale of the diffusion coefficient $D$ and the electric field $E$ are under SATP conditions.}
\label{f_diffusion_coefficient_GAr}
\end{minipage}
\end{figure}

For electrons drifting in gas, the energy gained between two collisions is, because of their small mass, rather large. 
The drift velocity  hence depends strongly on the total scattering cross-section between the electrons and the atoms of the medium, as shown in \Cref{f_CrosssectionArgon}. It is not uniform but has a minimum around $\sim 0.2\unit{~eV}$ caused from quantum mechanical effects (Ramsauer-minimum \cite{Ramsauer1921}). In the energy range relevant for the drifting electrons, there is mainly elastic scattering. 
In a first approximation the drift velocity $u$ is given by the acceleration in the electric field $E$ and the 
characteristic time $\tau_{c}$ between two collisions
\begin{equation}
u = \frac{ e\,E\, \tau_{c}}{m}\equiv \mu E
\label{e_driftvelocity1}
\end{equation}
where $\mu$ is the mobility.
The characteristic time $\tau_{c}$ is a function of the electric field $E$ \cite{Sauli1977}.
This is only an approximation since it is assumed that the electron completely stops after each interaction. An actual measurement of the drift velocity is reported in \cite{Nakamura1988} and shown in \Cref{f_ElectronDriftvelocity}.

An important property is the electron cloud diffusion. In case of an electric field, we have to distinguish between the diffusion in drift direction and the diffusion normal to it. The longitudinal and transversal diffusions are given by
\begin{equation}
\sigma_{l,t}^{2}=2 D_{l,t}\,t=2 D_{l,t}\,L/u
\label{e_transversal_diffusion_gas}
\end{equation}
where $L$ is the drift length, $u$ the drift velocity along the electric field lines and $D_{l,t}$ the diffusion coefficients in longitudinal and the transversal directions, respectively. As shown in \Cref{f_diffusion_coefficient_GAr}, the diffusion coefficients are depending on the electric field. Also it is shown that for low electric fields the diffusion in the transversal direction is almost one order of magnitude bigger than the one in longitudinal direction.

\subsection{Drift velocity and diffusion in liquid argon} \label{L_Driftvelocity}

The liquid phase changes the properties for drifting electrons. Electrons drift through a non-polar medium and locally, the argon atoms are polarised by the charge. The local field is no longer only given by the distance between the electron and the scattering atom but also by the sum of all other fields due to the induced dipoles in the neighbouring atoms. This interaction changes the scattering cross-section and the Ramsauer-minimum disappears \cite{Lekner1967}.
Other parameters besides the electric field affect the drift velocity. A relatively large effect is due to the temperature of the liquid argon. In \cite{Walkowiak2000} an average temperature gradient
of $u$ has been reported to be
\begin{equation}
\frac{\Delta u}{\Delta T \cdot \,u}= \left( - 1.72 \pm 0.08 \right) \unit{~\% \cdot K^{-1}}
\label{f_temperature_dependency}
\end{equation}

\begin{figure}[t!]
\begin{minipage}[t]{0.45\linewidth}
\centering
\includegraphics[height=80mm]{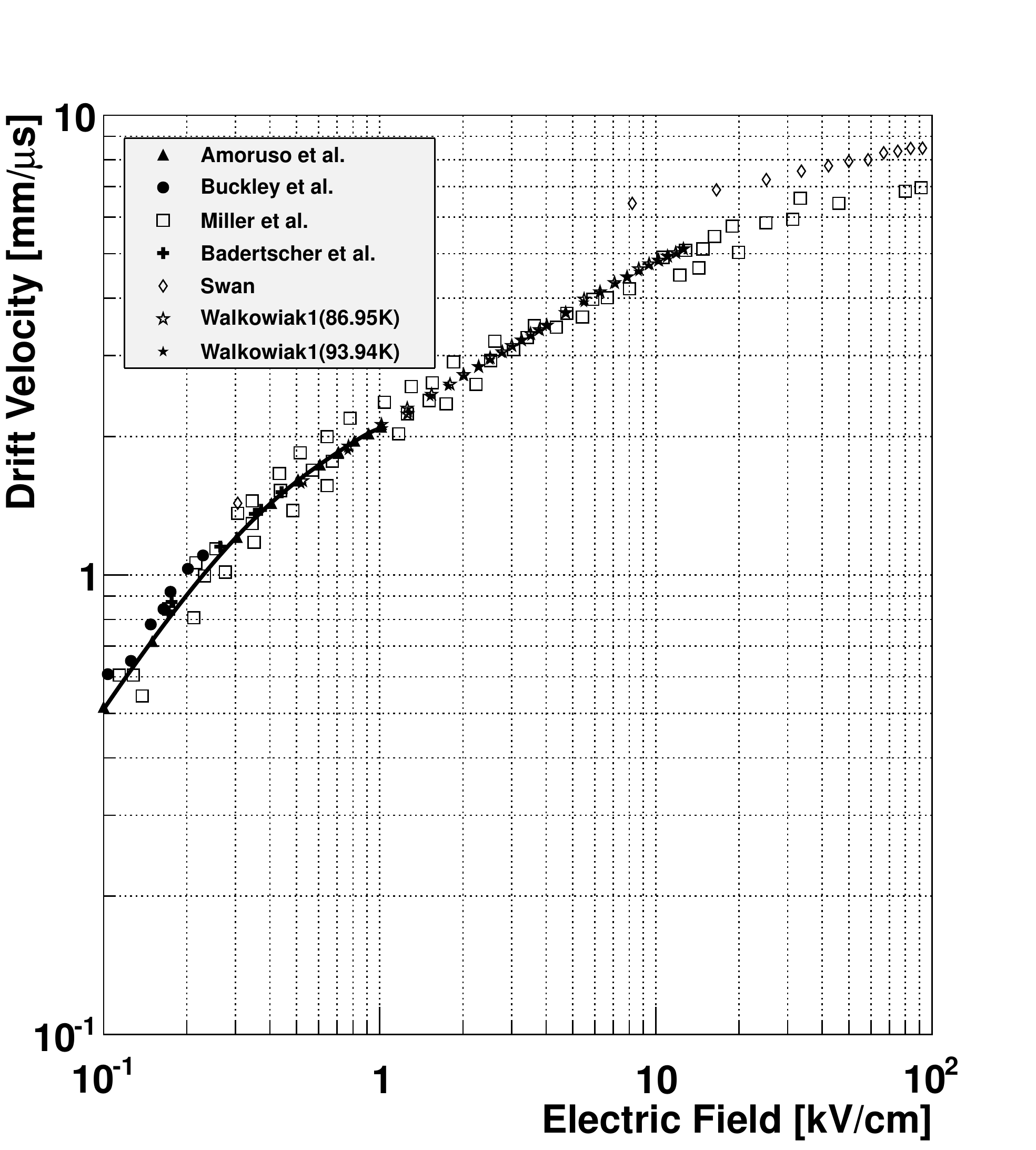}
\put(-138,25){\includegraphics[height=38mm]{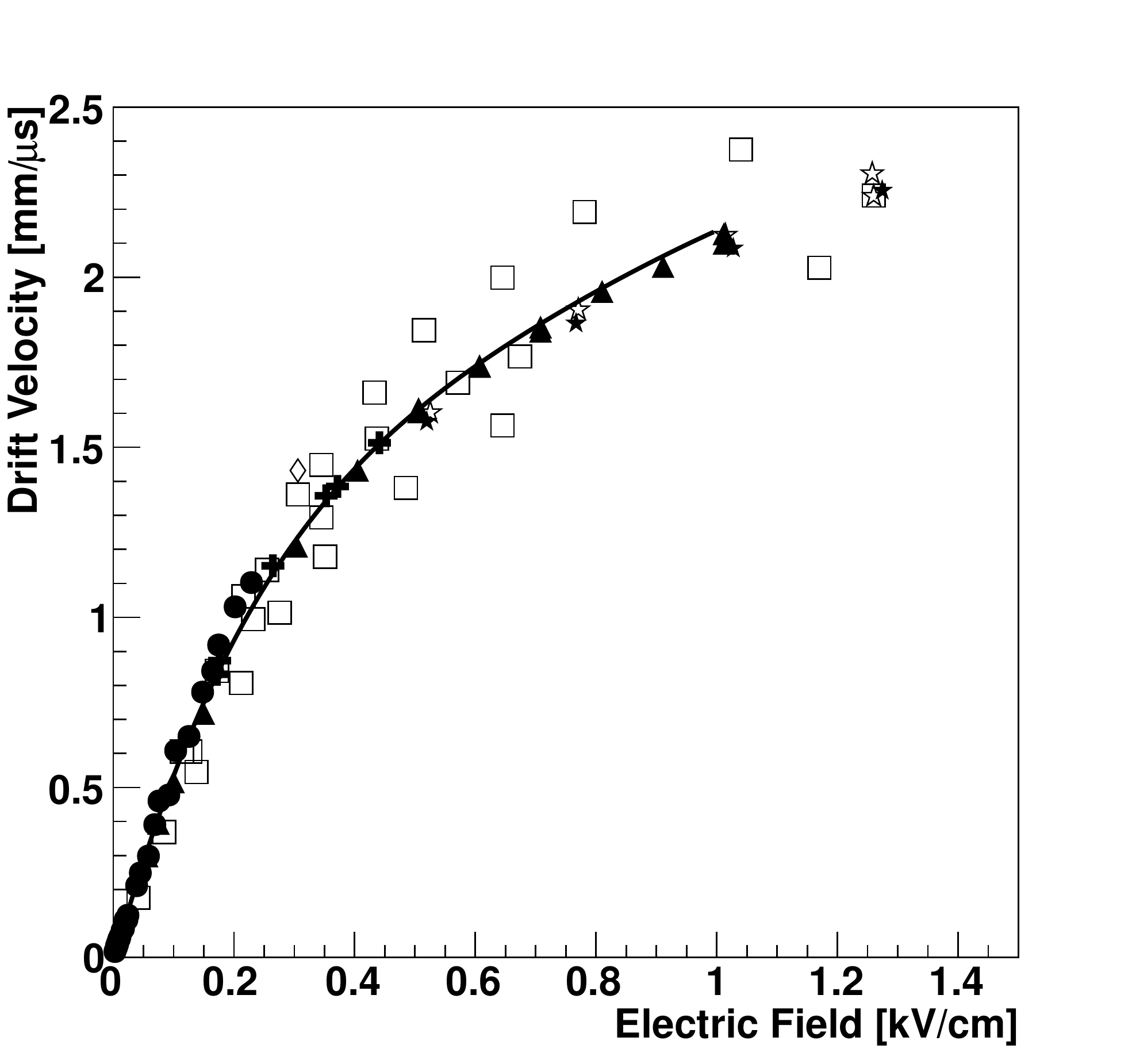}}
\caption{Drift velocity in liquid argon for fields up to $100 \unit{~kV/cm}$. The general plot (data collected from \cite{Amoruso2004,Buckley1989,Miller1967,Badertscher2012,Swan1964,Walkowiak2000}) is in log-scale to give a good overview, while the zoom-in, in the interesting region of the electric field for ArDM, is in linear scale. The solid line is a $5^{th}$ order polynomial proposed by \cite{Amoruso2004}. It is fitted between 0 and $2 \unit{~kV/cm} $ and includes all the presented data. All data are corrected according to Eq.~\ref{f_temperature_dependency} to a common temperature of 87 K.}
\label{f_driftvelocity_liquid}
\end{minipage}
\hspace{0.08\linewidth}
\begin{minipage}[t]{0.45\linewidth}
\centering
\includegraphics[height=80mm]{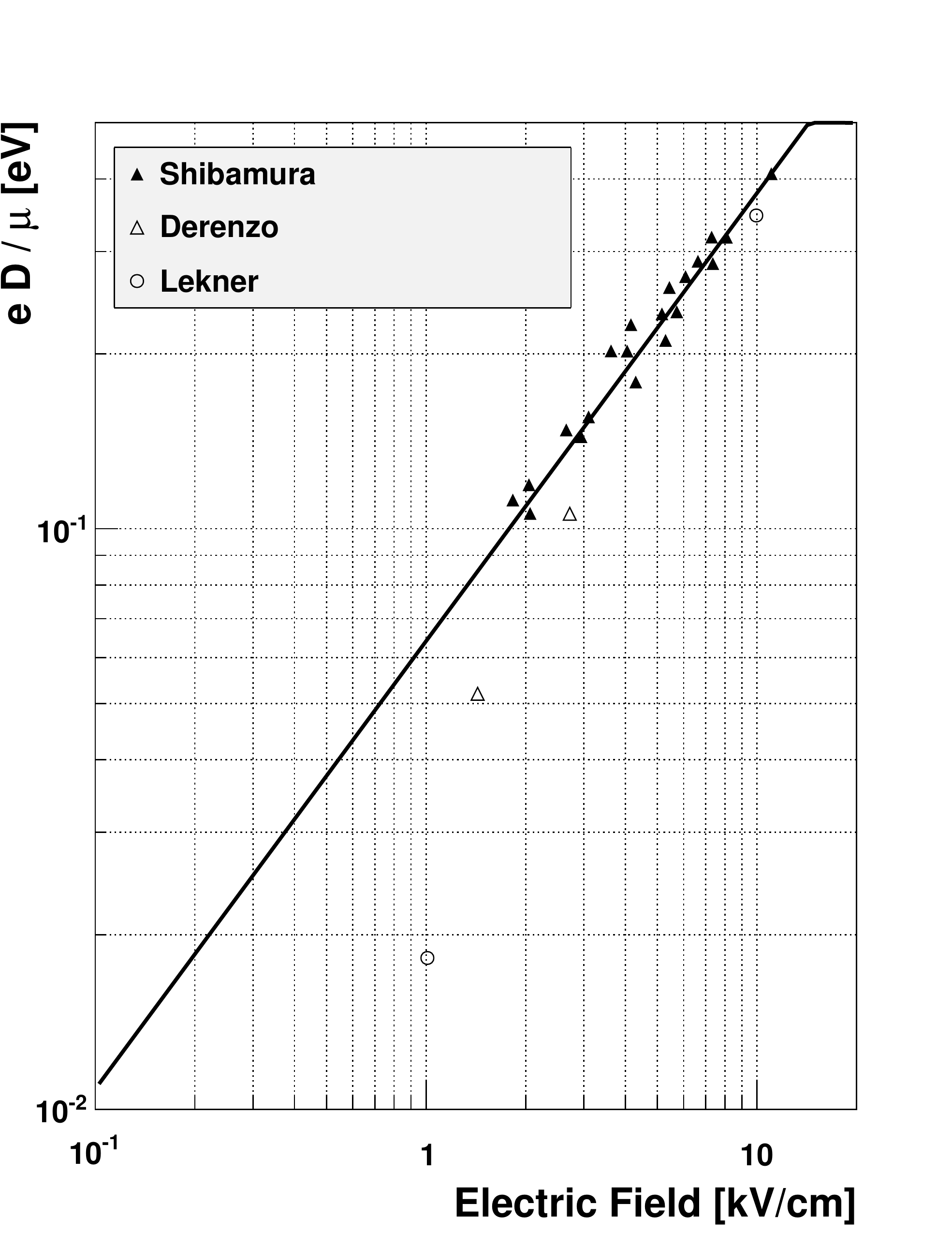}
\caption[Field dependence of the ratio $eD/\mu$ in liquid argon.]{Field dependence of the ratio $\frac{eD}{\mu}$ in liquid argon \cite{Shibamura1979}, where $D$ is the diffusion coefficient and $\mu$ the electron mobility. The different sets of data are described in \cite{Shibamura1979}.}
\label{f_DiffCoef}
\end{minipage}
\end{figure}

Experimentally, the drift velocity of electrons in liquid argon is measured by their drift time in a known electric field. One possibility for the measurement is to use crossing muons through the detector. There is a trigger on the primary scintillation light and the stopping trigger is either on secondary scintillation light or on the charge extracted and read out by an anode \cite{Badertscher2008}. Another option is to knock out the electrons from the HV cathode by the use of a laser \cite{Walkowiak2000} or a pulsed xenon lamp \cite{Amoruso2004}. The trigger is the pulse from the cathode when an electron is knocked out, and, as a stopping signal, again the arrival of the charge on the anode is used. \Cref{f_driftvelocity_liquid} shows the drift velocity as a function of the electric field, compiled from the literature \cite{Amoruso2004,Buckley1989,Miller1967,Badertscher2012,Swan1964,Walkowiak2000}. The data from the different sources are corrected to a common temperature of $87 \unit{~K}$ and fitted with a polynomial function\footnote{The actual function is $u(E)=1.004 \cdot E^{5}-5.083\cdot E^{4}+10.082\cdot E^{3}-10.280\cdot E^{2}+6.431\cdot E-0.016$ for $[E]=\unit{~kV/cm}$ and $[u]=\unit{~mm/\mu s}$}, as proposed by \cite{Amoruso2004}.

\begin{figure}[h!]
  \centering
  \includegraphics[width=\textwidth,height=0.32\textheight]{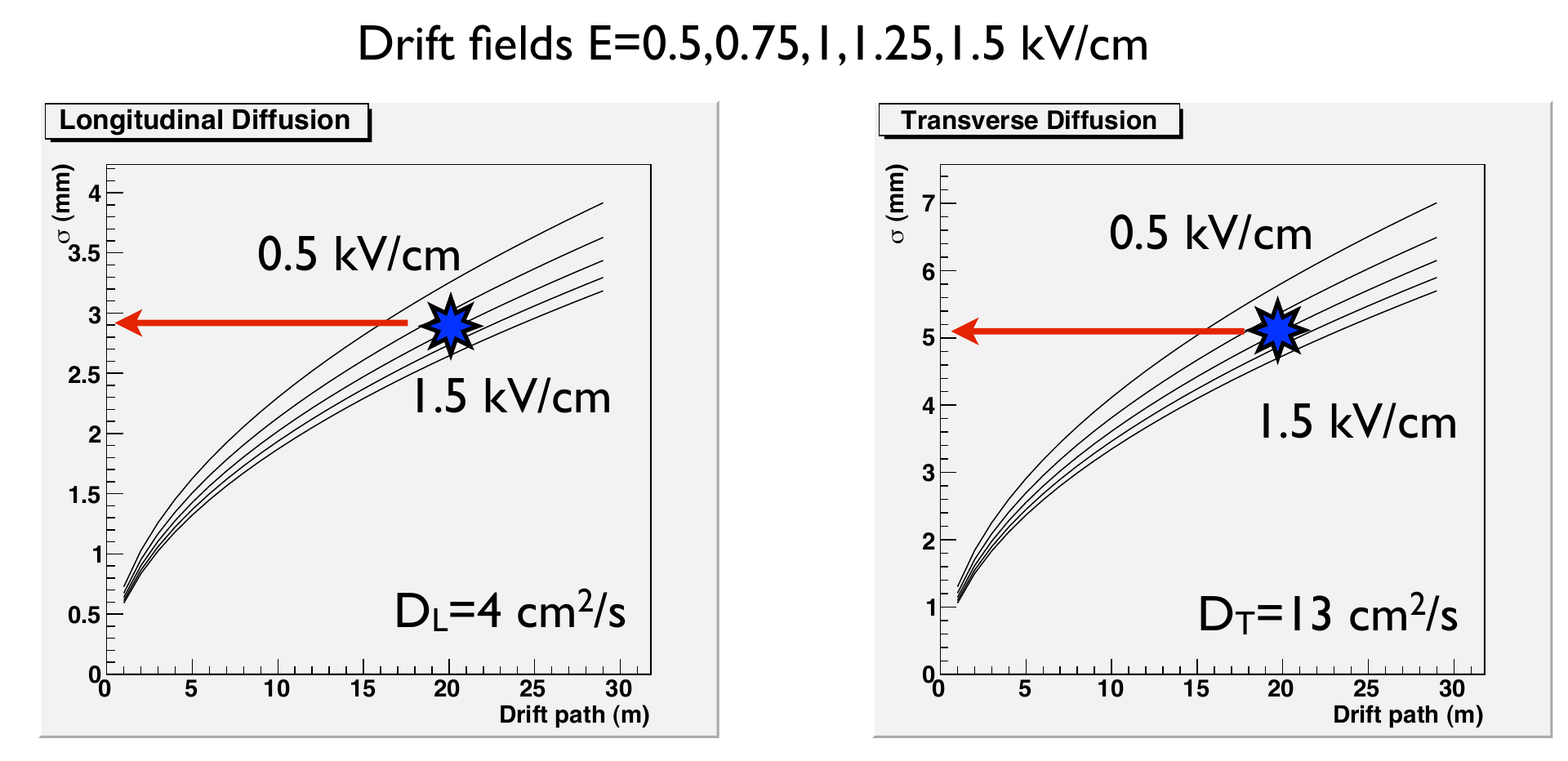}  
  \caption{Expected longitudinal and transversal diffusion in liquid argon as a function of the drift distance,
  for various drift fields.}
  \label{fig:diffusionexpected}
 \end{figure}

The electron cloud diffusion in the liquid is an important parameter to determine the maximum drift length of a detector,
which is presently not fully known experimentally.
The theoretical background for diffusion is similar to the diffusion in gas. It is derived from Einstein's theory about the molecular kinetic motion of particles suspended in a fluid \cite{Einstein1905}. Electrons in the liquid and with an electric field below $\sim 200 \unit{~V/cm}$ can be assumed to be thermal electrons, and
the ratio between diffusion coefficient and mobility is given by:
\begin{equation}
\frac{eD}{\mu}=kT=\frac{2}{3}<\! \epsilon \! >
\label{e_Diffusion}
\end{equation}
where $D$ is the diffusion coefficient. \Cref{f_DiffCoef} shows the ratio
$\frac{eD}{\mu}$ as a function of the electric field measured by \cite{Shibamura1979}. 
The  transversal diffusion $\sigma_{t}$ can be derived by multiplication of the coefficient with the electron mobility $\mu$ and dividing it by the drift velocity $u$ (Eq.~\ref{e_transversal_diffusion_gas}). 
The expected diffusion is shown as a function of the drift distance in
\Cref{fig:diffusionexpected}, assuming different electric drift fields ranging from 0.5 to 1.5~kV/cm. 
For the needed ratio of the diffusion coefficient and the mobility, the data presented in \cite{Shibamura1979} and shown in \Cref{f_DiffCoef} are fitted with a simple power law, 
${eD}/{\mu}=0.064\cdot E^{0.77}$
where $E$ is the electric field in kV/cm. 
This might not be true for low electric fields, as shown by the data points in \Cref{f_DiffCoef}, but the proposed function is giving an upper limit of the ratio. The points measured at lower electric field(empty triangle) might be too low due to space charge effects \cite{Shibamura1979}. The empty circles are theoretical values derived from Eq.~\ref{e_Diffusion} with $<\epsilon>$-values calculated in \cite{Lekner1967}. 
%
%

From \Cref{fig:diffusionexpected} it is clear that even for long drifts of 10 or 20 m the diffusion is still small and a charge readout  with a pitch of the readout of $\sim 3 \unit{~mm}$ can be used. 
Electric drift fields in the range 0.5-1~kV/cm are optimal from the point of view of the resolution and are sufficient for a wide range of detectors, up to a total drift length of up to 20~m~\cite{Rubbia:2004tz}). 

The \six will be a fundamental tool to investigate experimentally the effect of very long drift distances.
To reach a goal corresponding to drift times of about $10 \unit{~ms}$, one is considering a shorter drift distance with a reduced drift field in order to obtain similar drift times and therefore similar diffusions.


\subsection{Electrons extraction}
\label{section:extraction}

\begin{figure}[t]
\centering
\includegraphics[width=1.0\textwidth]{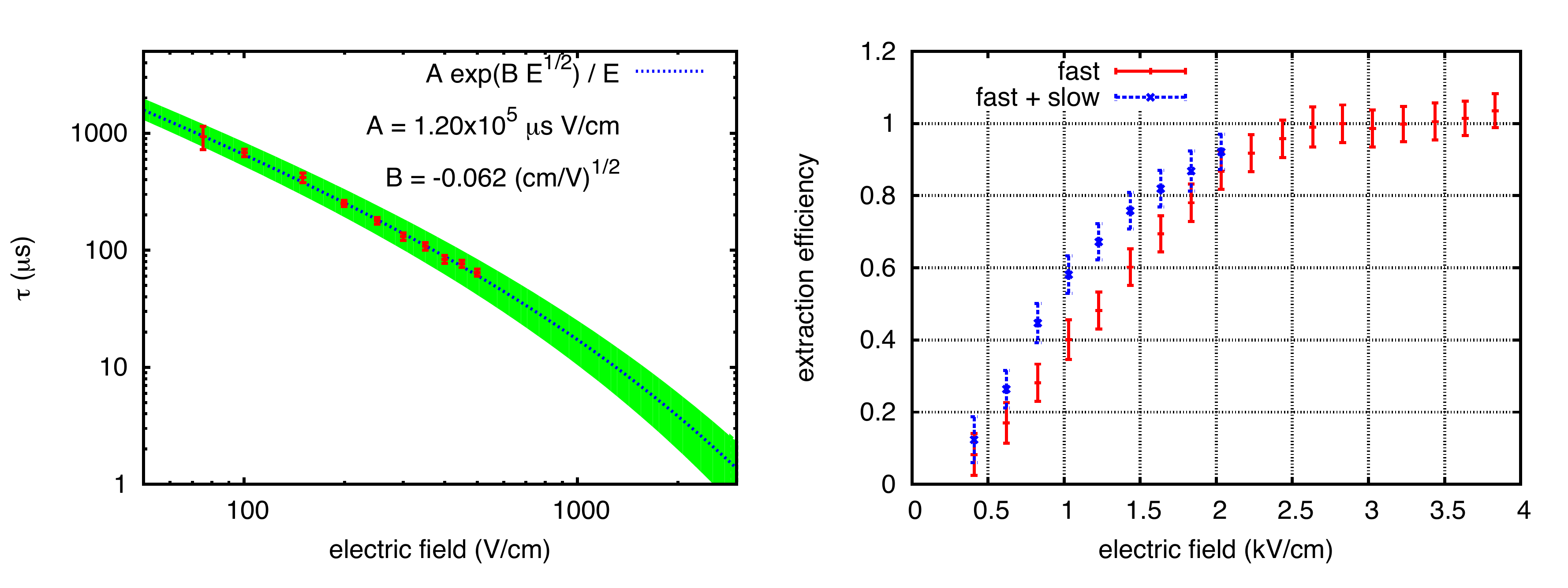}
\caption{The left picture shows the dependence of the extraction time on the electric field in liquid argon (T~$= 87.4$~K) as reported in~\cite{Borghesani1990}.
The picture on the right~\cite{Gushchin1982b} shows the extraction efficiency for fast and slow components as a function of the electric field in liquid argon (T~$= 90$~K).
As described in the paper, due to limitations of the electronics the measurement of the slow component has a semi-qualitative character.}
\label{figure:extractionVsElectricField}
\end{figure}

The \six will be operated in double-phase conditions~\cite{Badertscher:2013wm,Badertscher:2012dq,Badertscher:2010zg}, involving the transfer of the electron cloud from
liquid to the gas phase.
The transfer of electrons in excess from a condensed non-polar fluid to its saturated vapor using an electric field is a phenomenon investigated since the seventies~\cite{Dolgoshein1970}.
In particular, in argon it is experimentally shown that the electrons are extracted in two stages.
Near the triple point part of the charge is emitted on time scales that can be as high as 1~ms, strongly dependent on the electric field applied~\cite{Gushchin1982b, Borghesani1990}, while at larger temperatures the emission takes less than 100~ns.
At high electric fields the slow extraction time reduces, and the fraction of the slowly extracted electrons becomes negligible (see Figure~\ref{figure:extractionVsElectricField}).
This behavior can be understood in the framework of the Schottky model of electric field enhanced thermionic emission~\cite{Murphy1956}.

\begin{figure}[t]
\centering
\includegraphics[width=1.0\textwidth]{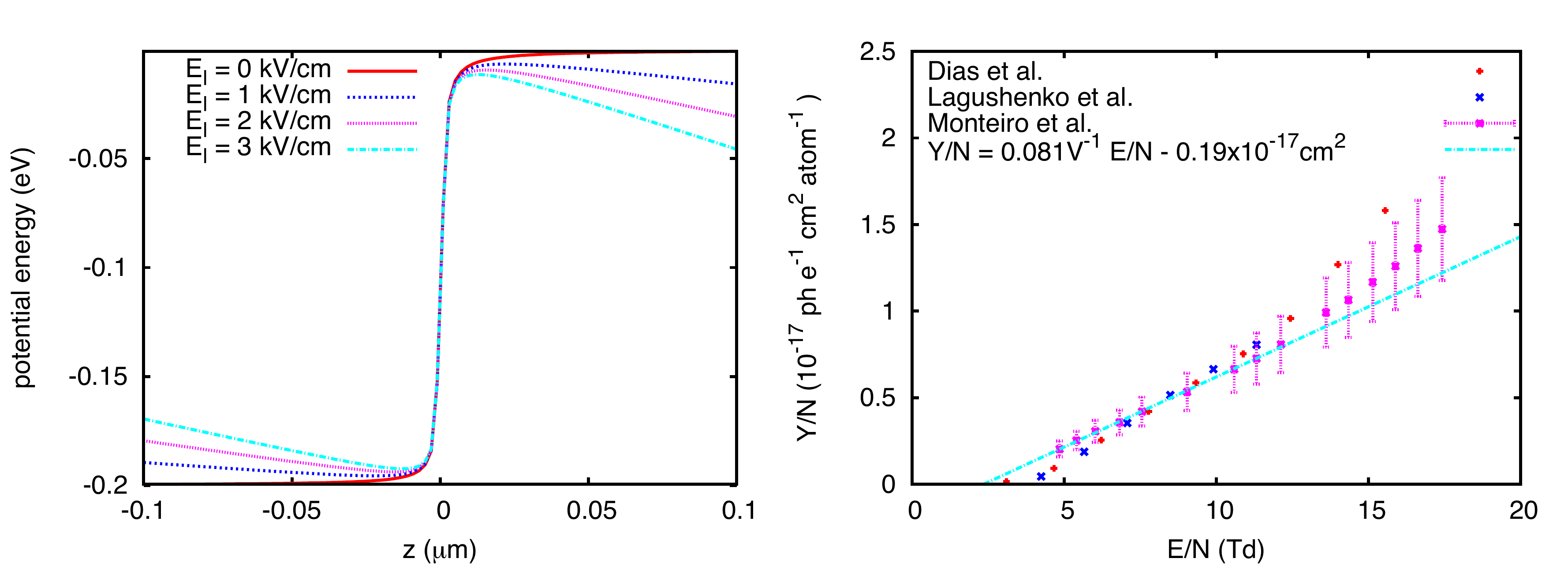}
\caption{Left: potential energy of an electron at the liquid vapor interface (at $z = 0$).
Right: reduced proportional scintillation yield in pure argon gas simulated by Lagushenko and Dias~\cite{Lagushenko1984, Dias1986} and measured by Monteiro and collaborators~\cite{Monteiro2008}.}
\label{figure:extraction}
\end{figure}

An electron in the vicinity of a dielectric surface feels the force of the charge induced on the surface by its presence.
In analogy to a conductor surface one can compute, with a method similar to the mirror charge method~\cite{Sometani1977}, the potential energy of the electron.
In the presence of an external electric field orthogonal to the liquid-vapor interface of argon, the energy potentials in the liquid ($\Phi_l$) and in the vapor ($\Phi_v$) as a function of the vertical position $z$ (the surface is set at $z = 0$) are~\cite{Gushchin1982b, Borghesani1990, Bolozdynya1999}:
\begin{align*}
\Phi_l = -V_0 - q_e \mathcal{E}_l z - A_l/z \mbox{ and } \Phi_v = - q_e \mathcal{E}_v z - A_v/z,
\end{align*}
with
\begin{align*}
A_l = \frac{q_e^2}{16 \pi \epsilon_0 \epsilon_l} \frac{\epsilon_l - \epsilon_v}{\epsilon_l + \epsilon_v} \mbox{ and } A_v = A_l \epsilon_l / \epsilon_v,
\end{align*}
where the terms inversely proportional to $z$ are due to the presence of the dielectric, $-q_e$ is the charge of the electron, $\mathcal{E}_l$ and $\mathcal{E}_v = \mathcal{E}_l \epsilon_l/\epsilon_v$ are the electric fields in the liquid and in the vapor respectively, $\epsilon_l$ and $\epsilon_v$ are the dielectric constants of the liquid and of the vapor respectively, $\epsilon_0$ is the permittivity of the vacuum, $-V_0$ is the minimum of the conduction band in the liquid with respect to the vapor (about $-0.2$~eV~\cite{Bolozdynya1999}).
The discontinuity of the potential on the surface is unphysical, but also not relevant for the discussion.
The regulated potential around $z = 0$ is shown in the plot on the left in Figure~\ref{figure:extraction} for different externally applied electric fields.

The minimum and the maximum of the potential in the liquid and in the vapor phase are:
\begin{align*}
\Phi_l^{min} & = 2\sqrt{A_l q \mathcal{E}_l} - V_0, & \mbox{ for } z_l^{min} & = -\sqrt{A_l/(q \mathcal{E}_l)},\\
\Phi_v^{max} & = -2\sqrt{A_v q \mathcal{E}_v}, & \mbox{ for } z_v^{max} & = \sqrt{A_v/(q \mathcal{E}_v)}.
\end{align*}
It is interesting to notice that the presence of the electric field reduces the potential barrier by:
\begin{align*}
\Delta \Phi = 2 (1 + \epsilon_l/\epsilon_v) \sqrt{A_l q_e \mathcal{E}_l},
\end{align*}
so that the gap becomes $V = V_0 - \Delta \Phi$.

An electron with momentum $p_z$ perpendicular to the liquid argon surface for which $p_z^2/(2m_e)>V$ (with $m_e$ the mass of the electron) is transferred to the vapor.
The electrons with kinetic energy $T > V$, but not satisfying the requirement on $p_z$, are reflected towards the liquid.
They undergo several elastic collisions that randomize the momentum direction almost without loss of energy, and they reach the liquid surface again.
If $p_z$ is not yet big enough, they repeat the process until all electrons are extracted in few tens of nanoseconds.
These are the so called \emph{hot} electrons (or fast component).

The electrons with $T < V$ are thermalized on the liquid surface and are emitted according to the thermionic emission with a characteristic time that depends on the energy gap~\cite{Borghesani1990}.
These electrons are the \emph{cold} ones (or slow component).

The electrons in liquid argon gain energy from the electric field.
At around 1~kV/cm, the electron average kinetic energy is of the order of 0.1~eV~\cite{Lekner1967, Gushchin1982}, larger than the thermal energy and comparable to the liquid-vapor interface energy gap.

The experimental facts can be summarized and explained:
\begin{enumerate}
\item at low temperature the electrons are emitted also slowly because their energy is not always larger than the potential barrier,
\item increasing the electric field the fraction of electrons with energy above the potential barrier increases,
\item the extraction time for the \emph{cold} electrons decreases at high field because the energy gap reduces.
\end{enumerate}

The electrons can be trapped by electronegative impurities diluted in the liquid argon and never be emitted in the vapor.
Similarly to what happens to the scintillation light in the presence of impurities, the amount of charge extracted decreases by a factor $\tau_{imp}/(\tau_{ext} + \tau_{imp})$, and the effective extraction time can be written as $\tau_{ext}^{eff} = (1/\tau_{ext}+1/\tau_{imp})^{-1}$, where $\tau_{ext}$ is the extraction time with no impurities and $\tau_{imp}$ is inversely proportional to the electronegative impurity concentration.

\subsection{Scintillation in liquid and gas phases}
Ionizing radiation in liquid noble gases leads to 
the formation of excimers in either singlet or triplet states~\cite{Doke1990617,0022-3719-11-12-024}, which decay radiatively
to the dissociative ground state with characteristic fast and slow lifetimes  ($\tau_{fast}\approx 6$~ns, $\tau_{slow}\approx 1.6\mu$s in liquid argon
 with the so-called second continua emission spectrum peaked at 
$128\pm10$~nm~\cite{larlight1}).
This prompt 128~nm  scintillation light is exploited 
in liquid Argon TPCs
to provide the absolute times ($T_0$) of the ionisation signals collected at the anode,
thereby after matching providing the absolute value of the drift coordinate of fully contained events,
as well as a prompt signal used for triggering purposes.

The proportional scintillation in argon, also referred to as secondary scintillation and luminescence, is the phenomenon of generating photons in gas or vapor in the presence of free charges and an electric field.
In a defined electric field window, that depends on the density of the argon, the amount of photons is proportional to the number of electrons, to the electric field and to the length of the path covered by the electrons.

The electric field range is defined such that between two successive collisions the drifting electrons, accelerated by the electric field, gain enough energy to excite argon atoms, but not ionize them.
In the case the electric field is lower, no photons are produced, in the case it is larger, because new charge is created, the amount of light grows nearly exponential with the field and the path length.
For the discussion on the charge amplification in gas, see \Cref{chapter:chargeAmplificationInGas}.

In order to take into account the argon density, the quantity used is the reduced electric field, defined as the electric field divided by the argon atomic density ($1\mbox{ Td} = 10^{-17}\mbox{ V cm}^2$), and the reduced light yield, defined as the number of photons produced per electron per unit path length divided by the argon density.
On the right of Figure~\ref{figure:extraction} the simulated~\cite{Lagushenko1984, Dias1986} and measured~\cite{Monteiro2008, Monteiro2001} reduced light yield at room temperature and pressure are 
plotted versus the reduced electric field.
The measured proportional scintillation threshold and ionization threshold are respectively 2.34~Td and 12.4~Td~\cite{Monteiro2001}, slightly lower than the simulated ones, but still in good agreement.

The electrons, when extracted from the liquid argon to the vapor, produce proportional scintillation.
From the data at room temperature and pressure one can extrapolate that in an electric field of 4.5~kV/cm, corresponding to an extraction field of 3~kV/cm in liquid, over 1~cm one electron generates about 200 photons.
Since the total amount of light produced is proportional to the charge extracted, this method is used by some double phase noble gas experiments for the direct Dark Matter searches~\cite{Aprile2010, Benetti2008,Rubbia:2005ge} to detect the ionisation charge. In the \six, they will be detected by the scintillation light system to provide a secondary trigger.

\subsection{Charge amplification in gas - Townsend avalanche}
\label{chapter:chargeAmplificationInGas}
\begin{figure}[t]
\centering
\includegraphics[width=1.0\textwidth]{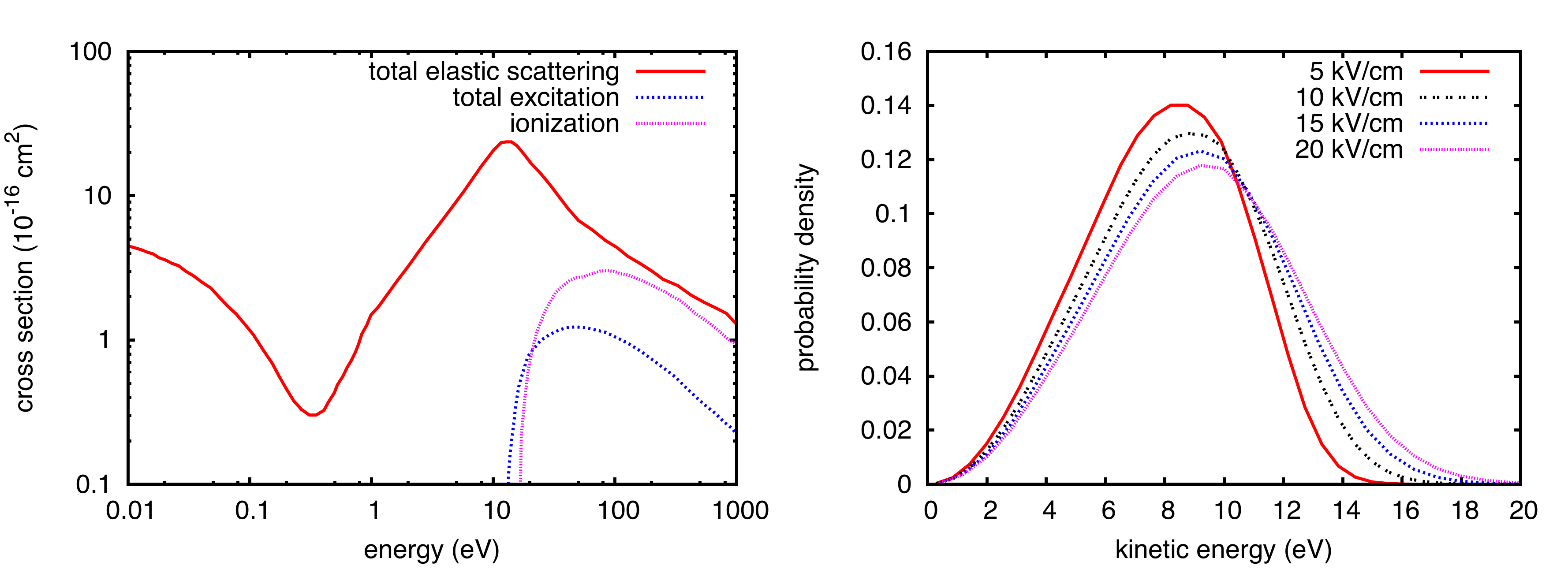}
\caption{Left: electron argon atom cross sections for different kind of interactions~\cite{garfieldWebSite}.
Right: energy distribution of free electrons under the action of an external electric field in pure argon gas at 1~atm and 20$^\circ$~C simulated with Magboltz~8.4~\cite{Biagi1999}.}
\label{figure:electronEnergyAndCrossSection}
\end{figure}

In a gas an electron accelerated under the action of an electric field ($E$) gains energy that is released in collisions against neutral atoms.
In addition to the diffusive random motion of the electron, the net drift velocity ($u$) in the direction of the field (but opposite) is obtained.
Under the assumption that the duration of the scattering is short compared to the average time between collisions $\tau_c$, and that the electrons undergo elastic collision only, so that the absolute magnitude of their speed does not change appreciably (the mass of the atom is much larger than the electron mass), the drift velocity can be 
written as in Eq.~\ref{e_driftvelocity1}~\cite{Raizer1991}.
$\tau_m$ is inversely proportional to the gas density $\rho$ and the momentum transfer cross-section $\sigma_m$~\cite{Raizer1991}.
In general, it depends on the energy of the electrons and therefore on $E$.
This makes $u$ proportional to $E$ only for low electric fields.
The kinetic energy due to the drift is much smaller than the kinetic energy due to thermal motion.
A more realistic computation should take into account the details of the electron velocity distribution.
The kinetic energy of the electrons increases by (1) decreasing the gas density and the electron-atom cross section and (2) increasing the electric field.
In other words, the peak of the electron energy distribution moves to higher energies with the increase of the electric field, as shown in the right plot of Figure~\ref{figure:electronEnergyAndCrossSection}.
It displays the results of a computation done using Magboltz~8.4~\cite{Biagi1999} in pure argon gas at 1~atm and 20$^\circ$~C.
Magboltz calculates diffusion coefficients, drift velocities and first Townsend coefficients (see later in the text) for gas mixtures in the presence of electric and magnetic fields.
The program inputs are tables of cross sections like the one reported on the left plot of Figure~\ref{figure:electronEnergyAndCrossSection}, provided in~\cite{garfieldWebSite}.

The high energy tail of the electron energy distribution extends beyond the first ionization potential, that for the argon is about 15.7~eV~\cite{Velchev1999}.
This means that a fraction of collisions gives rise to ionization of neutral atoms and to the production of new free electrons.
This process is called Townsend avalanche and is at the base of all the signal amplification techniques in gas chambers.
The larger the electric field is, the more populated becomes the tail above the ionization energy, and the larger the ionization rate is.
When the electric field is high enough that the majority of the electrons have an energy above the ionization potential, the ionization rate saturates, and its behavior reflects the ionization cross section dependence on the electron energy.
The first Townsend coefficient ($\alpha$) is the quantity that describes the number of ion-electron pairs created by an electron per unit drift length.
A very useful empirical approximation of the first Townsend Coefficient as a function of the electric field and the gas density is:
\begin{equation}
\alpha(\rho,E) = A \rho e^{-\frac{B\rho}{E}},
\label{equation:alphaVsE}
\end{equation}
where $A$ and $B$ are constants depending on the gas.
The expression $\alpha/\rho$ is a function of $E/\rho$ (also called reduced electric field).
The reason of this comes from the following considerations:
\begin{enumerate}
\item The electron mean free path ($\lambda$) can be written as $(\sigma \rho)^{-1}$, where $\sigma$ is the electron-atom total cross section (generally dependent on the electron energy, see Figure~\ref{figure:electronEnergyAndCrossSection}).
\item The energy gained by a free electron between two collisions is proportional to the product of the electric field $E$ and the mean free path $\lambda$, i.e.\ $E/(\sigma \rho)$.
\item $\alpha \lambda$, i.e.\ $\alpha/(\rho \sigma)$, can be regarded as a quantity proportional to the number of electron-ion pairs produced by an electron \emph{per collision}.
\end{enumerate}
Given a defined gas, in conditions where $E$ and $\rho$ are changed, but $E/\rho$ is kept constant, the electron-atom collisions are not distinguishable, i.e.\ same energy and, therefore, same ionization probability per collision (proportional to $\alpha/\rho$).
For the very weak assumption that the mean free path of the electrons is much smaller than the size of the avalanche, the total number of electrons $n_e$ created along a drifting
electron path $\cal S$ starting from $n^0$ electrons can be written as:
\begin{align}
n\equiv G n^0 =n_0\,\exp{\left[\int_{\cal S}\alpha(E(s)) ds\right]},
\label{equation:gainGeneralDefinition}
\end{align}
and  $G$ is the gain, 
under the assumption that
the gas amplification device is operated in proportional mode, meaning
that the detected charge is proportional to the initial charge.
In a parallel plate gap of length $x$ with a uniform electric field, where the Townsend coefficient is constant, 
the gain is:\begin{align}
G = e^{\alpha x}.
\label{equation:gainDefinition}
\end{align}

The above cited quantities must be considered as average values.
In fact, the Townsend avalanche is a stochastic process with the following behavior.
When $n_e < 10^5$, the number of electrons involved in an avalanche generated by a single electron follows the so called Furry distribution~\cite{Ficker2007}.
The probability that an electron frees a second electron traversing an infinitesimal distance $dx$ is $\alpha dx$.
As computed in~\cite{Furry1937}, given one initial particle entering into the uniform field of the multiplication region, the probability that $n$ particles emerge is
\begin{align*}
P(n) = (1-1/G)^{n-1}/G.
\end{align*}
The most probable number of particles at the end of the path is one, the average is $G = e^{\alpha x}$, and the variance is $G^2 (1 - 1/G)$.
For large gains the standard deviation of the distribution can be approximated with $G$, and for the central limit theorem, given a large number $ n_e^0$ of electrons at the beginning of the multiplication path, the distribution of the number of emerging electrons is a Gaussian with average $G n_e^0$ and standard deviation $G \sqrt{n_e^0}$.
To be noticed is that the relative width of the distribution is independent of the gain, and it decreases the larger the number of electrons at the beginning is.

While the gain is mainly due to the described Townsend mechanism, also
secondary effects, caused by the simultaneously
produced ions and photons have to be taken into account: although the energy
of the produced scintillation light is not sufficient to ionize directly
argon atoms, photons can extract electrons from electrodes by
photoelectric effect. A similar feedback, though much slower,  is due
to back drifting ions, which can eject further electrons when they
impinge on the collection electrode. 
Obviously any of the described feedback mechanisms can turn into a
continuous current, that may lead to the formation of an electron-ion
plasma or a \textit{streamer}. Due to the fact that such a streamer
persists until the amplification device is completely discharged, a
detector would be paralysed until the fields are back at the working point. As a consequence of the
heat production, heavy discharges may also damage the
amplification device. In order to decrease the rate of discharges or
to increase the maximum achievable stable gain,
typical gas detectors work with a gas mixture, including a so-called
quench gas that absorbs scintillation light and thus reduces the
feedback. However, since the LAr TPC requires extremely  
low levels of contaminations of less than 1~ppb in order to drift electrons
over large distances in LAr, chemical quenchers cannot be used. 
In the LAr LEM TPC option described in \Cref{sec:larlemoption}, mechanical quenching
is important to stabilise the system and allow high gains to be achieved.


\subsection{Primary and secondary scintillation light}
\label{sec:priseclight}
Primary scintillation light is produced promptly by the passage of ionising particles through the liquid argon volume.
The formation of excimers in either singlet or triplet states~\cite{Doke1990617,0022-3719-11-12-024}, which decay radiatively to the dissociative ground state with characteristic fast and slow lifetimes  ($\tau_{fast}\approx 6$~ns, $\tau_{slow}\approx 1.6\mu$s in liquid argon with the so-called second continua emission spectrum peaked at 
$128\pm10$~nm~\cite{larlight1}).
Secondary scintillation light is produced in the gas phase of the detector when electrons, extracted form the liquid, are accelerated in the electric field. Because the gas is less dense and the mean free path of the electron is longer
than in liquid, the extracted electrons can gain enough energy to excite argon atoms in collisions so that scintillation light appears. For appropriate electric fields, the amount of light is directly proportional to the amount of charge.
Measurements show a reduced secondary scintillation yield in pure argon per electron and per cm, according to \cite{Monteiro2008}:
\begin{equation}
\left(Y/p\right)=\left(81E/p\right)-47 \quad \unit{photons/(electron\, cm \, bar)}
\end{equation}
where $[E/p]=\unit{kV/(cm \, bar)}$. This formula is valid for room temperature. For the operation at cryogenic temperatures of about $90 \unit{~K}$ (e.g. argon vapor at 1 bar) the pressure has to be multiplied by a factor 3.36 due to the higher density. 

\begin{figure}[h]
\begin{center}
\fbox{\includegraphics[scale=0.5]{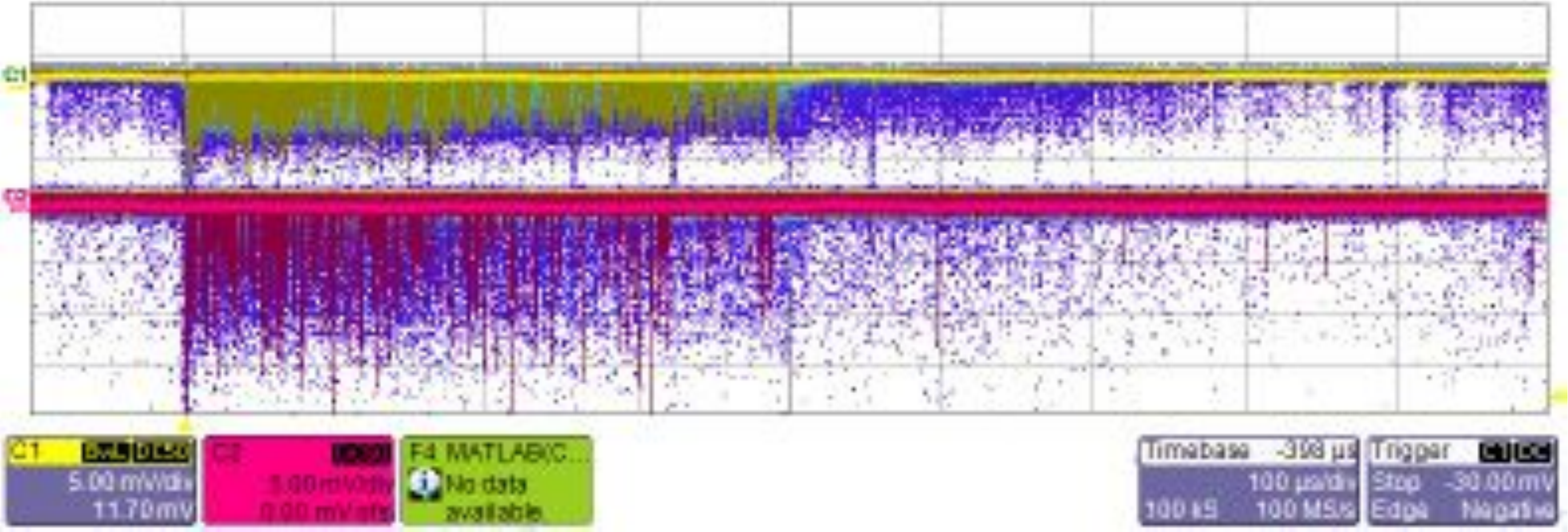}}
\caption{Oscilloscope screen shot of the primary and secondary scintillation light of a crossing track during the operation of the 40$\times$80~cm$^2$ LAr LEM TPC with 60 cm drift length\cite{Badertscher:2013wm} (see text).}
\label{f_screenshot_electronextraction}
\end{center}
\end{figure}

\Cref{f_screenshot_electronextraction} shows a screenshot from the oscilloscope of a cosmic track crossing the 40$\times$80~cm$^2$ LAr LEM TPC with 60 cm drift length\cite{Badertscher:2013wm}. 
The track ionised a straight track through the detector. As a result, 
there is a continuous extraction of electrons as long as the track drifts towards the liquid-gas interface. 
The time between the occurrence of the primary scintillation light an the secondary scintillation light is equivalent to the drift time of the electrons from the vertex to the liquid argon surface. 
The yellow and red traces correspond to two PMTs located below the cathode with a timebase of $100 \unit{~\mu s/div}$. They show an continuous extraction signal for a total duration of $400 \unit{~\mu s}$, which can be translated in a drift velocity of $\sim1.5 \unit{~mm/\mu s}$. With the given electric field of $\sim 480 \unit{~V/cm}$ the measurement agrees with values from literature as also the curve in \Cref{f_driftvelocity_liquid}.
The violet trace is the persistence of the signals from previous events and signals after $400 \unit{~\mu s}$ are from pile up events.

In the \six and
for an extraction field of 2.5~kV assuming the liquid level in the middle of the gap between
the extraction grid and the lower electrode of the CRP, the
secondary light yield is $Y\approx 75\,\gamma/e$. In addition, the avalanche inside the LEM holes
create light.


\subsection{Electron attachement to impurities}\label{c_attachement_impurity}
The amount of charge collected at the anode is affected by the amount of electronegative molecules (for example $\unit{O_{2}}$) dissolved in the liquid argon. By capturing free electrons these latter become negative ions which slowly drift towards the anode. Due to their larger mass and scattering cross-section with the argon atoms, they move much slower than  electrons. Attached electrons therefore do not contribute to the readout signal. 
Thus, purity is a fundamental requirement and a possible contamination  has to be kept at an extraordinary low level. This means the vessel needs to be sealed and absolutely vacuum tight, with a typical
leak rate upper limit of $10^{-9} \unit{~mbar ~ l ~ s^{-1}}$ to prevent molecule contaminants from entering
the argon volume. 

In practice, the most common impurities in the welding-grade liquid argon are oxygen, nitrogen and water. 
In absence of leaks to the outside, the majority of the electronegative impurities come from the outgassing of the detector material itself. Special attention has to be given to composite materials like glue, for they might have chlorine or fluorine compounds. These materials must be prevented from contaminating the liquid argon, since chlorine and fluorine are the most electronegative molecules and immediately capture the drifting electrons. 

\begin{figure}[t!]
\centering
\includegraphics[height=90mm]{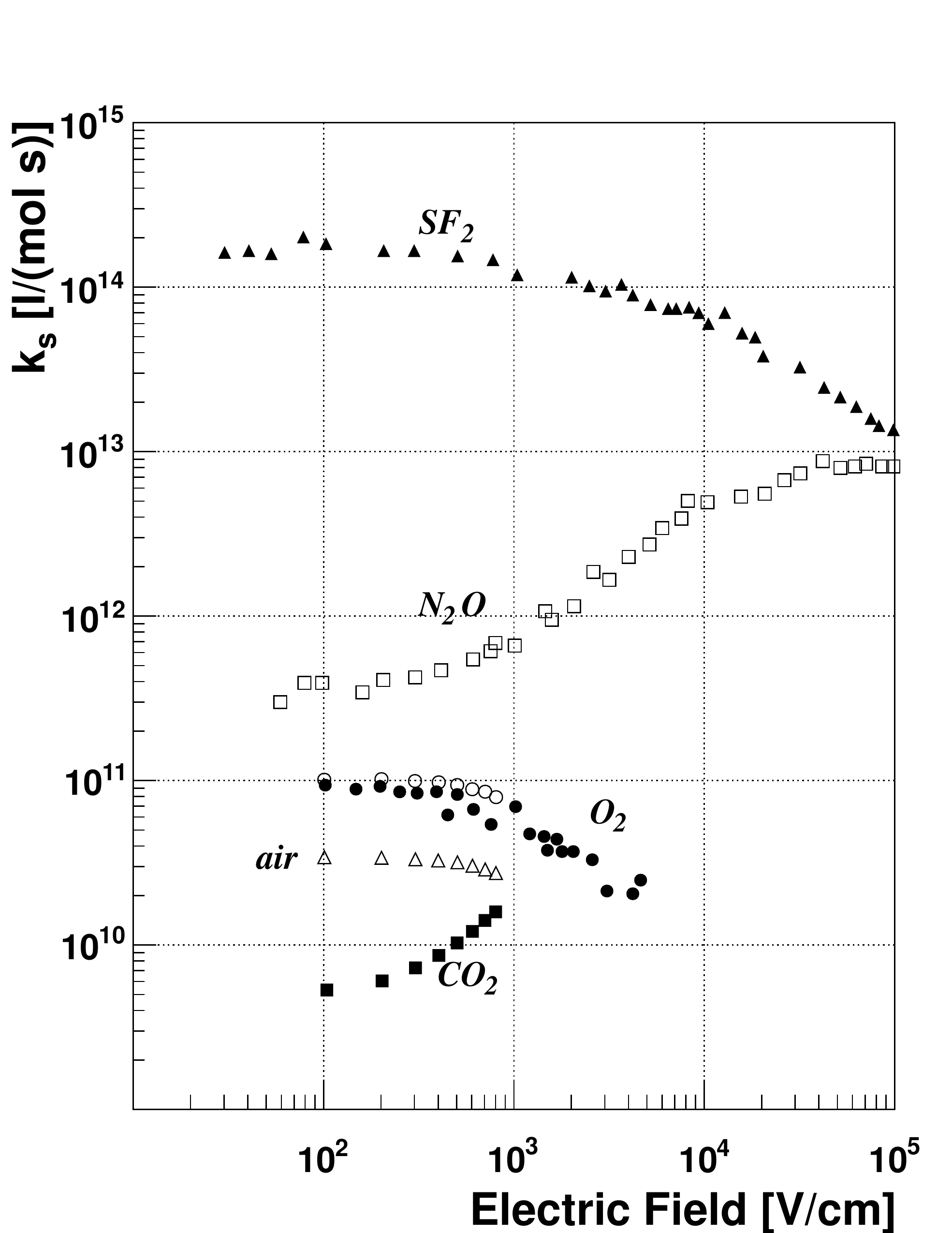}
\caption[Rate constant for the electron attachment of different molecules.]{Rate constant for the electron attachment to different molecules. The data points for $\unit{SF_{6}}$ $(\filledmedtriangleup)$, $\unit{N_{2}O}$ $(\medsquare)$ and $\unit{O_{2}}$ $(\bullet)$ are from \cite{Bakale1976}, the data points for $\unit{O_{2}}$ $(\circ)$, $\unit{air}$ $(\medtriangleup)$ and $\unit{CO_{2}}$ $(\filledmedsquare)$ from \cite{Suzuki1990_1}.}
\label{f_RateConst}
\end{figure}

In general, the attachment in liquid is a three-body process that involves the electron and the molecule to which it is attached. When attaching an electron, the molecular ion is excited and loses its energy to a third body through vibrational states \cite{Swan1963}.
\begin{equation}
e^{-}+M + X \stackrel{k_{s}}{\longrightarrow}M^{-}+X
\end{equation}
where $M$ is the electronegative molecule and $X$ the third body, in our case the argon atom. $k_{s}$ is the rate constant and it is given as a function of the electric field $E$ by
\begin{equation}
k_{s}(E)=\int{v \cdot \sigma(v) \cdot \mathrm{f}(v,E) \operatorname{d} \! v},
\end{equation}
with $v$ the electron velocity and $\mathrm{f}(\epsilon)$ the Maxwell distribution \cite{Bakale1976}. If the capture cross section $ \sigma(\epsilon)$ is a function of the energy $\epsilon$, the rate constant is given by
\begin{equation}
k_{s}(E)=\int{ \sigma(\epsilon) \cdot \mathrm{f}(\epsilon,E) \operatorname{d} \! \epsilon}
\end{equation}
For $E = 0$ the distribution $\operatorname{f(\epsilon)}$ is
\begin{equation}
\operatorname{f(\epsilon)}=\frac{2}{\sqrt{\pi} ~ k_{B} T} e^{-x} x^{1/2}
\end{equation}
with $x={\epsilon}/({k_{B}\,T})$.
The effects of an external electric field on the rate constant $k_{s}$ are described in \cite{Bakale1976} and shown in \Cref{f_RateConst} for different molecules.
Experimentally, the rate constant $k_{s}$ can be obtained by looking at the decay of the electron current after a controlled release of electrons from the cathode or inside the liquid argon by ionization. 

In case of a small number of released electrons $N_{e}$ with respect to the amount of 
impurities $N_{s}$ ($[ N_{s}]=mol/l$), the number of electrons captured per time interval is \begin{equation}
\frac{\operatorname{d} \! N_{e}}{\operatorname{d} \! t}=-k_{s}\cdot N_{s}\cdot N_{e}
\end{equation}
and therefore, for a homogeneous impurity contamination and drift field, the number of electrons $N_{e}$ as a function of time is
\begin{equation}
N_{e}(t)=N_{e\,0} ~ e^{-t/\tau}
\end{equation}
where $\tau$ is the mean electron lifetime and $N_{e\,0}$ the initial number of released electrons. The lifetime is directly related to the amount of impurities.
\begin{equation}
\tau = \frac{1}{k_{s}N_{s}}
\label{eq:e_lifetime}
\end{equation}

As shown in \Cref{f_RateConst}, the rate constant $k_{s}$ for attaching electrons is not only depending on the kind of molecule but also on the electric field strength. For electric fields up to $\sim1 \unit{~kV/cm}$, $k_{s}$ is approximately constant for oxygen and Eq.~\ref{eq:e_lifetime} can be written as \cite{Buckley1989}
\begin{equation}
\tau [\unit{\mu s}] \approx \frac{300}{\rho_{O_{2}}[ \unit{ppb}]}
\label{e_oxygen_impurity}
\end{equation}
where $\rho_{s}=N_{s}/N_{Ar}$ is the amount of impurities of the molecule $s$ with respect to the amount of argon.

Recent measurements of the electron lifetime in liquid argon have brought attention to water \cite{Andrews2009}. Derived from the presented data, a lifetime of 
\begin{equation}
\tau [\unit{\mu s}] = \frac{17.4 \pm 0.5}{\rho_{H_{2}O}[ \unit{ppb}]}
\label{e_water_impurity}
\end{equation}
has been found. This measurement has been presented without any information about the electric field, used for drifting the electrons.
In general, impurities are given as ``oxygen equivalent'' impurities. In that case, 
1~ppb of water corresponds to $\sim 17$~ppb of oxygen.


\subsection{Liquid argon purity requirements}
\begin{figure}
\begin{center}
	\resizebox{0.8\linewidth}{!}{\includegraphics{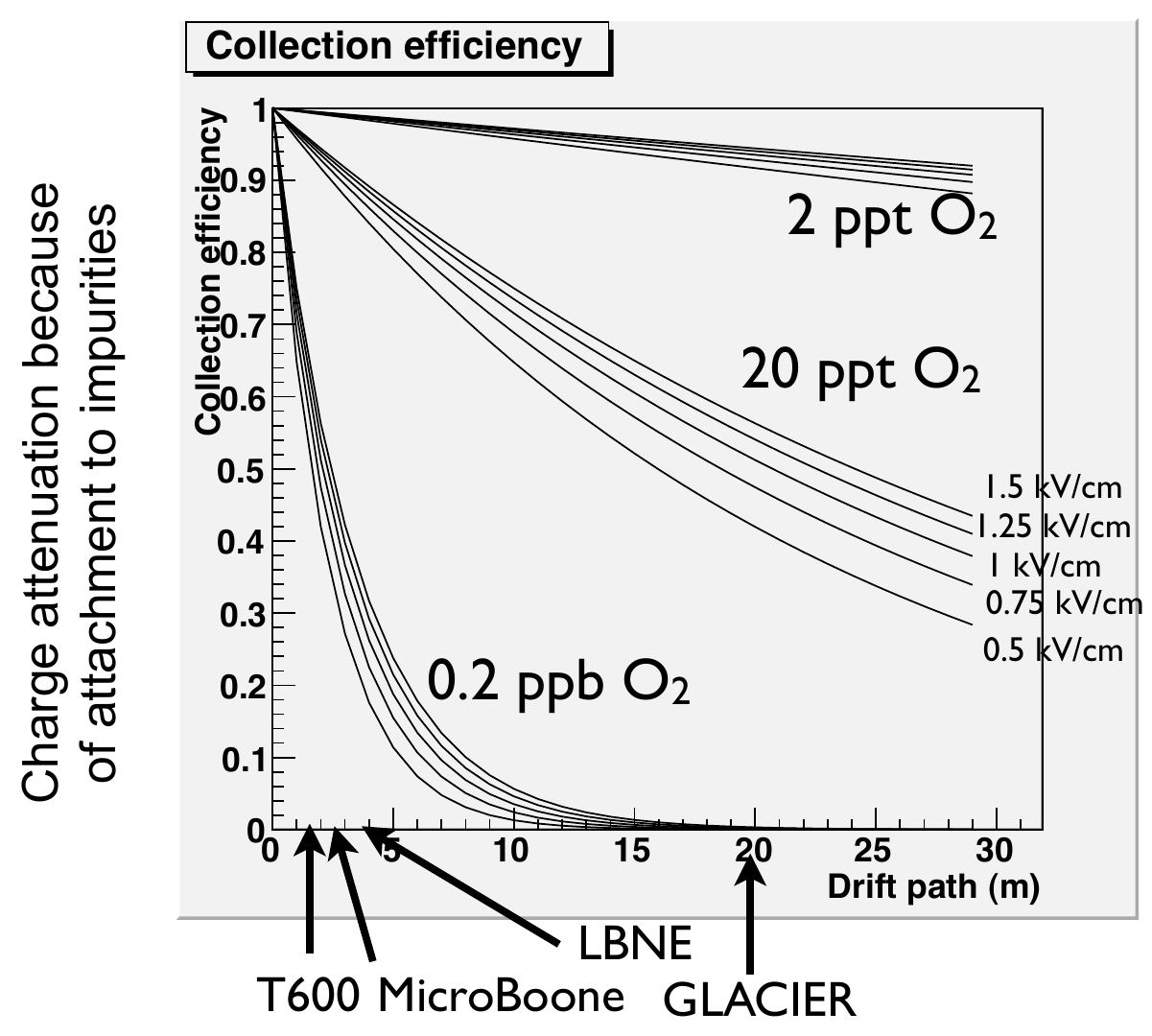}}
	\caption{Expected ionisation charge attenuation due to attachment
	to impurities as a function of the drift path for 0.2~ppb, 20~ppt and 
	2~ppt Oxygen-equivalent
	impurity levels and electric fields in the range 0.5--1.5~kV/cm. 
	The arrows indicate the drift length of
	the ICARUS T600, MicroBOONE, LBNE and LAGUNA-LBNO GLACIER.}
	\label{fig:larimpuritycomp}
\end{center}
\end{figure}

As discussed in the previous section, a long drift path requires an ultra-high level of purity in the medium.
The expected ionisation charge attenuation due to attachment
to impurities as a function of the drift path for various oxygen-equivalent
impurity levels and electric fields is shown in Figure~\ref{fig:larimpuritycomp}.
The arrows indicate the drift length of
the ICARUS T600~\cite{Amerio:2004ze}, 
MicroBOONE~\cite{Chen:2007ae}, LBNE~\cite{Akiri:2011dv} 
and LAGUNA-LBNO GLACIER~\cite{Rubbia:2009md,Rubbia:2010zz}.
The  free of electro-negative molecules (like $O_2$, $H_2O$, etc.) must
reach a level below 100~ppt $O_2$ level. An impurity level 
of $<30$~ppt $O_2$-equivalent is needed to obtain an electron lifetime
greater than 10~ms.
Compared to commercially
available bulk liquid argon deliveries which typically contain ppm-level
purities, the goal is to reduce those impurities by a factor $10^4$--$10^5$
before filling the main vessel tank. 
Excellent purity has been reproducibly achieved in various setups always relying on commercially available techniques, of various sizes and capacities, and should not pose a 
show-stopper for long drift paths.

Several independent groups performed numerical simulations and 
concluded that the vacuum evacuation of the main detector volume 
could be avoided for larger detectors,
thanks to 
(1) a more favourable surface / volume ratio for larger volume 
(also larger volumes are less sensitive to micro-leaks),
(2) a purification from ppm to $<<$ 1 ppb is anyhow needed
since the initial purity of argon when delivered is typ. ppm $O_2$ (see above),
and (3)  the outgassing of material is mostly from hot components
and impurities ÒfrozenÓ at low temperature.
GAr flushing and purging were shown to be effective ways to remove air and impurities.
Purging on 6~m$^3$ volume has been successfully demonstrated~\cite{Curioni:2010gd}.
The piston effect was seen in gas and the impurities
reached 3~ppm $O_2$ after several volumes exchange.

Although the \six adopts a drift length which is significantly longer than the others such
as ICARUS T600, MicroBOONE or LBNE as mentioned above, it should be noted
that all experiments require in order to collect efficiently ionisation charges a liquid argon purity
in the range below 0.1~ppb = 100~ppt of oxygen equivalent (See \Cref{fig:larimpuritycomp}). 
Hence, the challenge to reach the required level of purity starting from commercially available ppm-level
bulk argon is not mitigable by a shorter drift length in the meter range. It exists for all considered
detectors.


\subsection{Effect of impurities on scintillation light}\label{c_Impurities_Scintillation}
Electronegative impurities in the liquid argon do not only affect the lifetime of the free electrons
but  also quench the scintillation light, especially its slow component.  
In pure argon the total scintillation light emission rate $l(t)$ is given by the simple exponential equation
\begin{equation}
l(t)=\frac{A_{S}}{\tau_{S}}\,\exp{\left(-\frac{t}{\tau_{S}}\right)}+\frac{A_{T}}{\tau_{T}}\,\exp{\left(-\frac{t}{\tau_{T}}\right)},
\end{equation}
with $A_{S}$ the relative amplitude of the singlet and $A_{T}$ the relative amplitude of the triplet state. The sum of the two is normalized to $A_{S}+A_{T}=1$. 
In case of impurities, mainly O$_{2}$, N$_{2}$ and H$_{2}$O, the argon excimer Ar$^{\star}_{2}$ is de-excited by collisions with the impurity molecules and the average lifetime is therefore reduced \cite{Acciarri2009}. 
\begin{equation}
\tau^{\prime}_{T}=\frac{\tau_{T}}{1+k[\mathrm{X}_{2}]\tau_{T}}
\end{equation}
with $k[\mathrm{X}_{2}]$ the rate constant of the light quenching for different molecules $\mathrm{X}_{2}$. 
The measured values for oxygen and nitrogen are $k[\mathrm{O}_{2}]=0.54 \pm 0.03 \unit{~\mu s^{-1} ppm^{-1}}$ and $k[\mathrm{N}_{2}]=0.11 \unit{~\mu s^{-1} ppm^{-1}}$ respectively \cite{Acciarri2009}.
Comparing the actual measured lifetime and the maximum lifetime of the triplet state of $\sim 1.6 \unit{~\mu s}$, the purity of the liquid argon can be measured down to a contamination of $\sim 10- 100 \unit{~ppb}$.

In the case of a detector that reads out charge and light, the quenching of the light is not important, since the purity needed for drifting large distances is the limiting factor. It is much lower than the purity needed to recover all the light produced. For example, drifting over $1 \unit{~m}$ with an electric field of $1 \unit{~kV/cm}$ gives, according to \Cref{f_driftvelocity_liquid}, a total drift time of $\sim 500 \unit{~\mu s}$ and therefore, according to Eq.~\ref{e_oxygen_impurity} a maximum allowed $\unit{O_{2}}$ equivalent contamination in the order of $\sim 0.6 \unit{~ppb}$. 
This is too small to affect the lifetime of the scintillation light. The light quenching is useful for quickly checking the quality of the commercial argon when delivered. It also gives an immediate alarm in case of a leak in the cryostat, since the lifetime can be analyzed in every PMT signal instantaneously without time consuming processing of the data. 


\subsection{Ionisation space charge effects}
\label{sec:spacecharge}
Cosmic ray muons and other natural radiation create ionisation charge with a constant rate and uniformly distributed in the liquid argon.
Electron and positive argon ion pairs are separated by the applied electric field and drift towards the anode and the cathode respectively.

In a steady state, the charge concentration is proportional to the evacuation time, and, therefore, inversely proportional to the drift speed $v_x$.
The drift velocity of ions in liquid argon is known to be influenced by the self-motion of the liquid (in particular see convective motions
in the next Section~\ref{sec:convectivemotionlar}). 
Although the lowest measured mobility was about $3.6\times 10^{-4}$~cm$^2$V$^{-1}$s$^{-1}$, Dey and Lewis
have reported an extrapolation to zero current suggesting 
a true ion mobility, uninfluenced by liquid motion, of about $2\times 10^{-4}$~cm$^2$V$^{-1}$s$^{-1}$~\cite{DeyLewis:0022-3727-1-8-309}.
In practice, the velocity of the ions is influenced by the movement of the liquid, generally due to the convective motion (see \Cref{sec:convectivemotionlar}).

In a uniform electric field, the spatial charge distribution $\rho_x$ can written as a function of the drift length $z$ as
$$
\rho_x = \Phi \langle dQ/dl \rangle z / v_x,
$$
where $\Phi$ is the cosmic muon flux (considered the main contribution to the ionisation during surface operation), $\langle dQ/dl \rangle$ is the average charge per unit length created after the ion-electron recombination, and the subscript $x$ refers to the electron $e^-$ and positive ion $Ar^+$.
Assuming for the positive argon ions $v = 1$~cm/s and $E=1$kV/cm, $\Phi = 200$~m$^{-2}$/s and $\langle dQ/dl \rangle = 10$~fC/cm, for a drift of 6~m the ion concentration is of the order of $10^6$~ion/cm$^3$, which is enough to distort the drift field.
Due to the fact that electrons move much faster than ions, their contribution is negligible.

There are other processes that may have an impact on the final charge distribution.
Electrons and ions may recombine during the electron drift.
The recombination rate is proportional to the local densities of the two species
$$
d\rho_x/dt = -K \rho_{e^-}\rho_{Ar^+},
$$
with the constant $K = 10^{-4}$~cm$^3$/s~\cite{Bueno:2007um}.
This results in a decrease of the amount of charge in the volume and, consequently, in a decrease of the signals from the TPC.

Electronegative molecules diluted in the liquid argon may trap drifting electrons.
Since the amount of electronegative impurities is much greater than the amount of drifting electrons, the formation of the negative ions ($O_2^-$)\footnote{$O_2$ represents in this case any electronegative impurity, oxygen being a very common one.} can be expressed as
$$
d\rho_{O_2^-}/dt = \rho_{e^-}/\tau,
$$
where $\tau$ is the very well known drifting electron lifetime (assumed 5~ms for the computations performed here).
The negative ions have a drift velocity similar to the argon ions, but they move on the opposite direction, and they partially cancel the effects of the positive charges.

The TPC signal amplification and readout system (consisting of the extraction grid, the LEM and the two views anode) is based on the charge multiplication in argon vapour due to the Townsend effect, i.e.\ drifting electrons gain enough energy in a strong electric field to further ionise argon atoms.
In the process, there is a formation of positive argon ions, that drift back towards the active volume.
A significant part is collected on the bottom electrode of the LEM and on the extraction grid.
The ions that enter into the liquid phase contribute to the positive charge distribution.

\begin{figure}[htbp]
\centering
\includegraphics[width=0.9\textwidth]{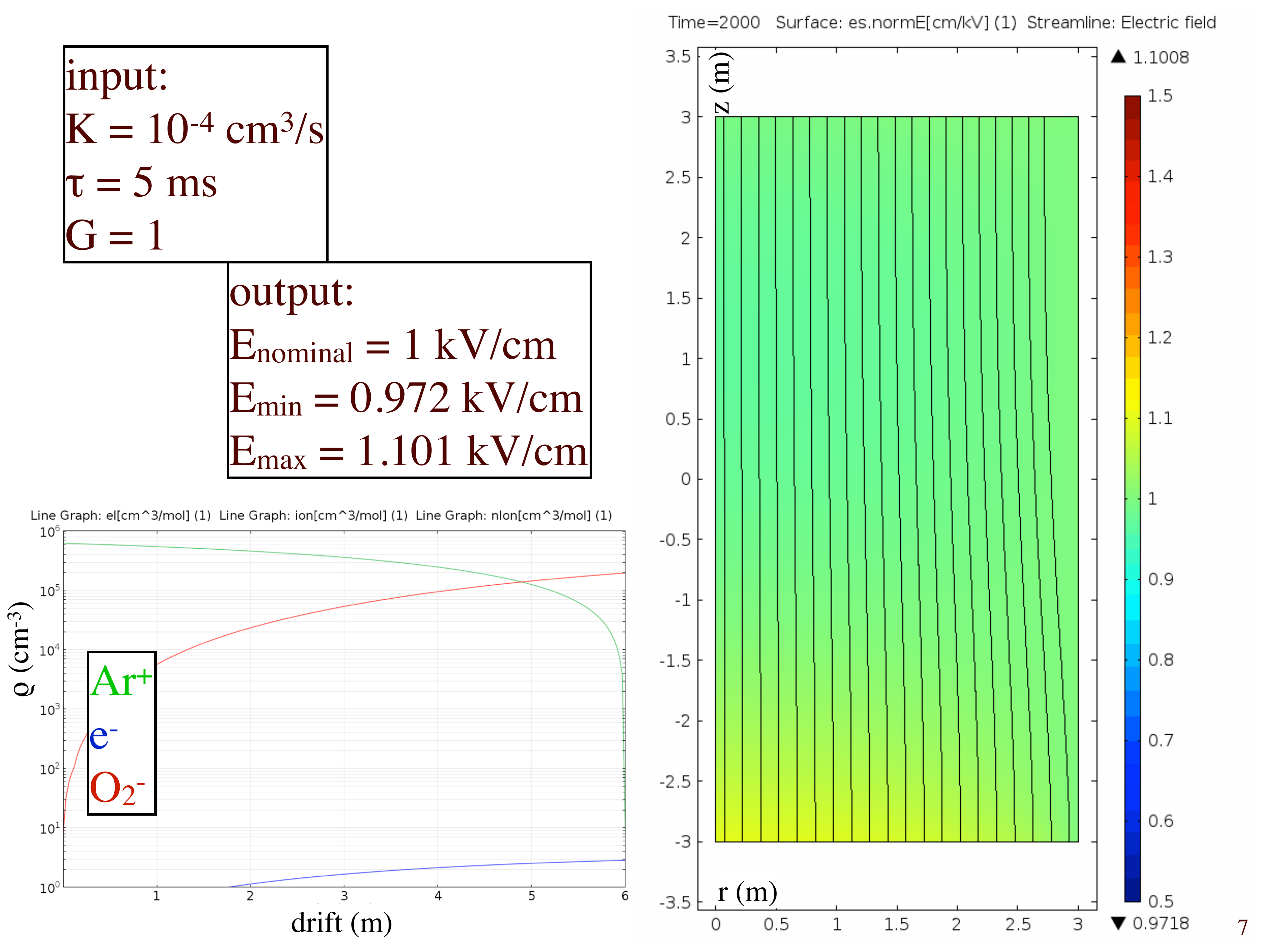}
\caption{Steady state computation in cylindrical coordinates of the electric field with space charges without (left) and with (right) argon ions from the signal amplification system. The colour scale represents the absolute value of the electric field and the curves are the field lines. See text for the details.}
\label{fig:spaceCharge}
\end{figure}

The solution to this problem can be computed with a Finite Element Analysis (FEA) approach.
The model was implemented in COMSOL Multiphysics~\footnote{COMSOL Multiphysics: \protect\url{http://www.comsol.com}}.
Figure~\ref{fig:spaceCharge} shows the computed absolute value of the electric field and the field lines in a cylindric volume of 6~m of diameter and 6~m long with a nominal electric field of 1~kV/cm.
For the computation the charge gain in the LEM was assumed to be 20, and the two images refer to the case where all the ions are collected on the extraction grid and on the LEM (left), and to the case where 20\% of the ions leaks into the liquid argon bulk.
In the model the $O_2^-$-$Ar^+$ recombination is neglected.
The nominal drift field is distorted: the value varies between -44\% and +27\% in the worst case, and the field lines followed by the drifting electrons are squeezed towards the centre of the volume.
This prevents the charge to be lost because exiting from the sensitive volume.
The drift field distortions can be mapped with straight cosmic muons, and at sea level the cosmic ray flux is large enough to provide the required statistics.
Another important aspect, neglected up to now, is described next.

\subsection{Convective motion of liquid argon}
\label{sec:convectivemotionlar}
\begin{figure}[htbp]
\centering
\includegraphics[width=0.3\textwidth]{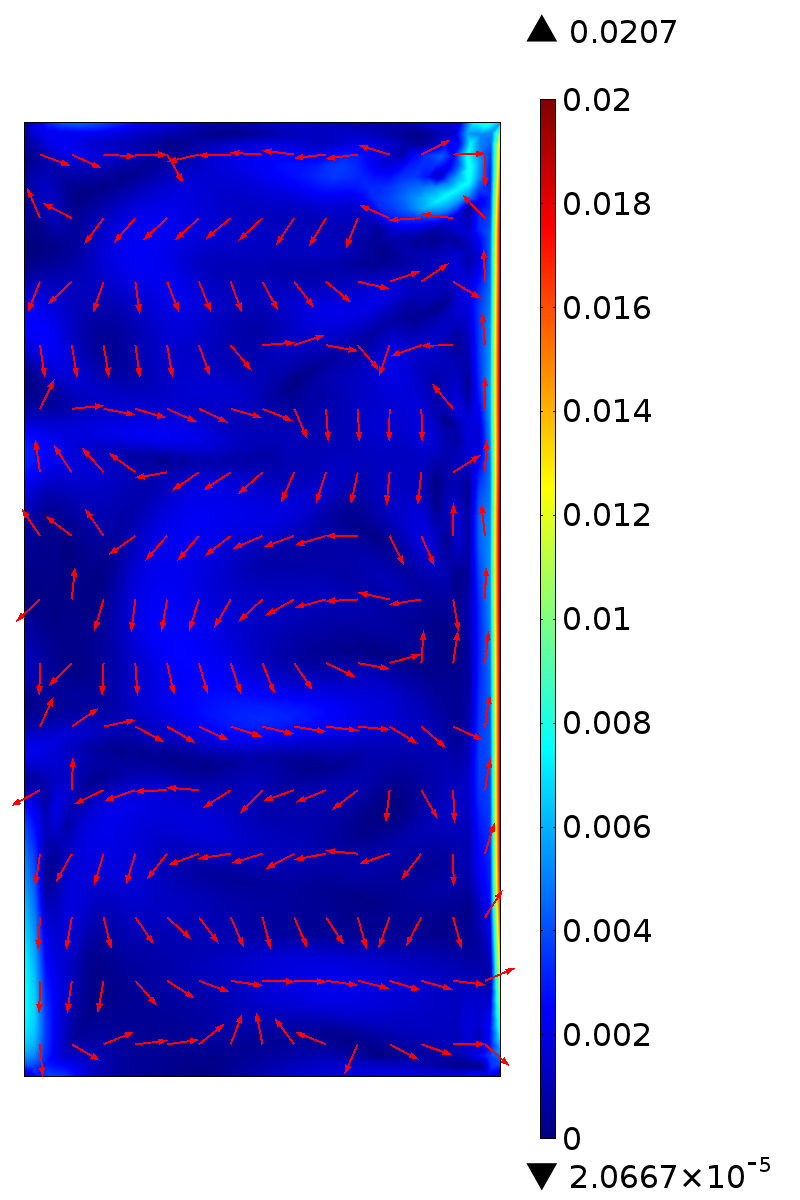}
\includegraphics[width=0.3\textwidth]{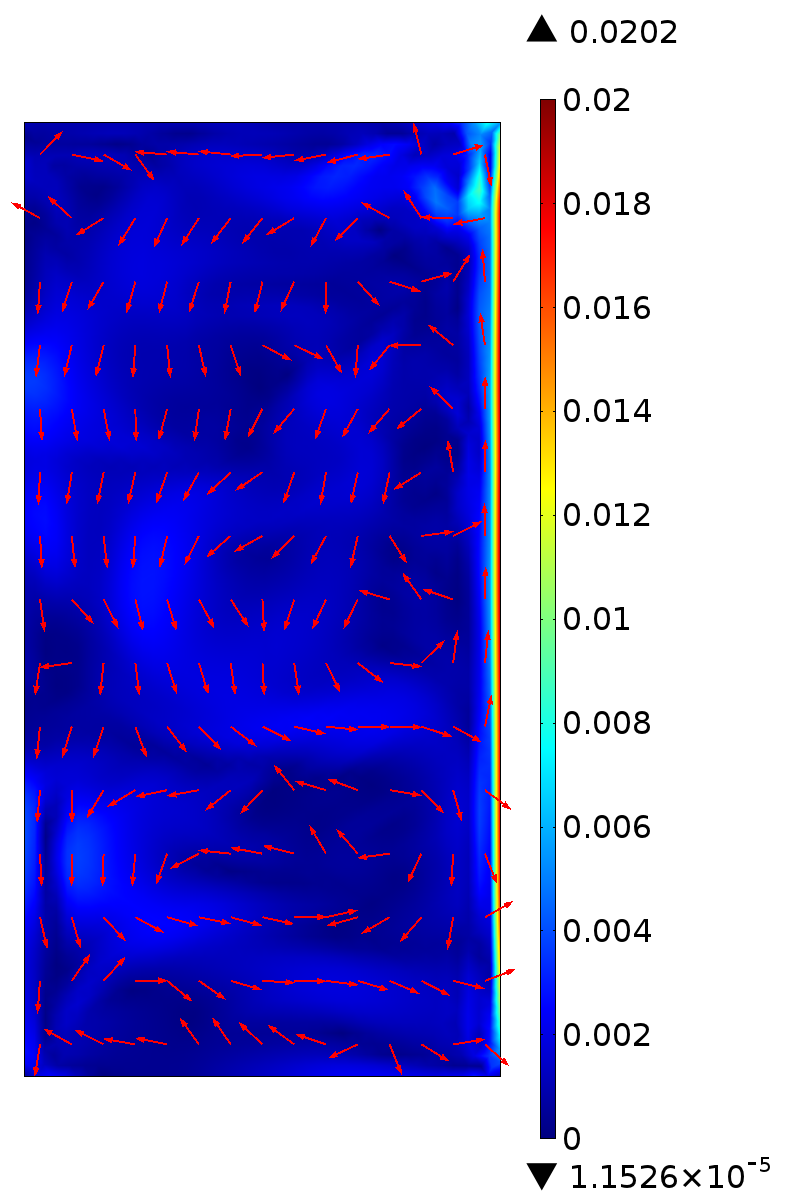}
\includegraphics[width=0.3\textwidth]{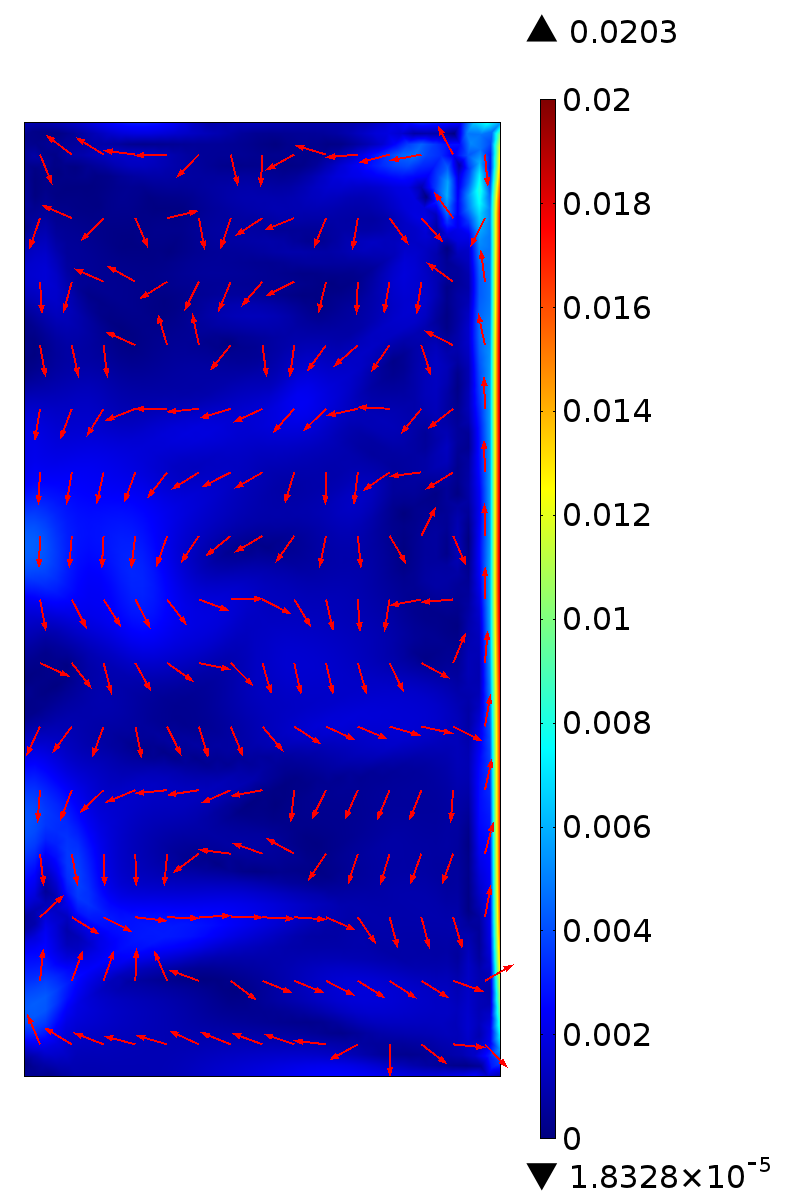}
\caption{Liquid argon velocity distribution at three times spaced by 100~s. The colour scale represents the magnitude of the speed and the arrows the direction.}
\label{fig:shuffle}
\end{figure}

Since liquid argon density decreases with its temperature, convective fluxes due to the buoyancy may occur in the presence of temperature gradient of liquid argon.
Thermal gradients are due to heat input and cooling power.
As an example, in \Cref{fig:shuffle} the convection of liquid argon in a 6~m diameter and 6~m long cylinder is computed with COMSOL, 
assuming a constant 5~W/m$^2$ heat input from the walls.
The three figures are snapshots of the convection velocity every 100~s at t=4000~s.

The liquid argon is mixed by convection and, since the ion drift speed is of the same order of magnitude of the argon convection speed, also the $Ar^+$ distribution is affected.
While in the space charge computation described before, the argon is supposed still, in this case the liquid argon motion must be considered to compute the ion motion.
The net effect of the convection is to even out the positive and negative ion distributions and to speed up the evacuation of the charges 
by neutralising them on the conducting surfaces.
This results in a decreases of the total amount of space charges in the volume and in a decrease of the electric field distortion.

\subsection{Electric breakdown in liquid argon}
\label{sec:electricbreakdowninlar}
The loss of the dielectric properties (of a gas, liquid or solid) as a result of application of a strong electric field greater
than a certain critical magnitude is called dielectric breakdown. The critical field at which the breakdown occurs
is the dielectric strength of the material (or breakdown voltage). It depends on the geometry, on the thickness,
the rate of the voltage increase, the shape of the voltage as a function of time, etc. In (pure) liquids, the breakdown
mechanism is similar to that in gases where current carriers are free electrons and ions generated by
external radiation. In the strong electric field, these particles acquire kinetic energy, large enough to ionise
further molecules or atoms, potentially leading to streamers. Compared to gases, the liquid has the advantage
of a much higher density, so in a first approximation one can expect the dielectric strength of liquids to be
much higher than those of gases (proportional to their relative densities). However, it is known that breakdown
in liquid can occur at much lower electric fields due to the formation of bubbles. The geometry of these
latter is altered by the field lines and ellipsoids can form and merge into larger high-conductivity channels
prone to discharges. A local increase of conductivity increases the temperature in the channel, the
liquid can boil, the vapour enlarging further the canal. Such phenomena have for example been observed
in liquid under the sudden application of strong electric fields, where electro-striction generates
by cavitation a gaseous phase prone to discharges~\cite{ushakov}. In general, the influence of a strong electric field
in a liquid is noticeable when the electric field density is comparable to the external pressure $p$:
\begin{equation}
\frac{1}{2}\epsilon \epsilon_0 E^2 \geq p
\end{equation}
where $E$ is the electric field. For liquid argon and in absence of bubbles, 
a field of 1~MV/cm$=$100~MV/m leads to $\approx$0.5~bar, which 
will have a noticeable effect. For comparison, the dielectric strength
of polyethylene is in the range 20--160~MV/m. 
A field of 100~kV/cm in liquid argon yields a pressure of about 5~mbar.
By assuming that the liquid argon in the \six will be sub-cooled by e.g. 10~mbar
via the liquid argon process system, the condition on the field strength to
prevent formation of bubbles is set to $E\leq 100$~kV/cm.

\begin{figure}[htbp]
\centering
\includegraphics[width=0.5\textwidth]{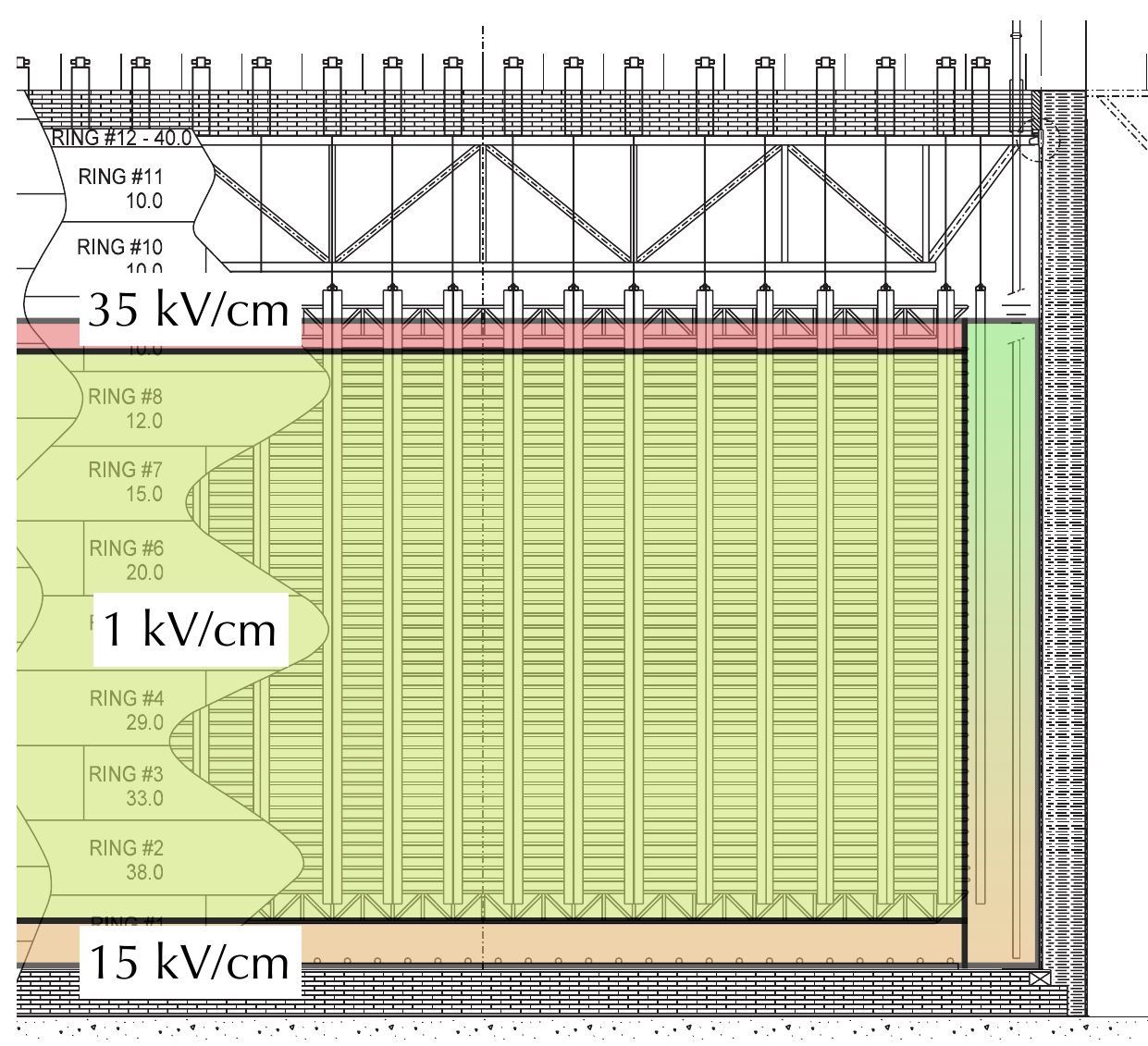}
\caption{Representation of the electric fields in a GLACIER-type detector with a 20~m drift distance.}
\label{fig:fieldglacier}
\end{figure}

The assumed electron drift length for GLACIER is 20~m, with an electric field of 1~kV/cm, corresponding to the path after which the spread due to the electron diffusion becomes larger than the readout pitch of 3~mm.
This requires a voltage at the cathode of -2~MV.
In the current design of GLACIER, a distance of 1.5~m is foreseen between the cathode structure and the bottom of the tank at ground, giving an average electric field of 13.3~kV/cm (See \Cref{fig:fieldglacier}).
In the vicinity of the cathode tubes the electric field reaches 50~kV/cm over distances of the order of centimetre.

\begin{figure}[htbp]
\centering
\includegraphics[width=0.9\textwidth]{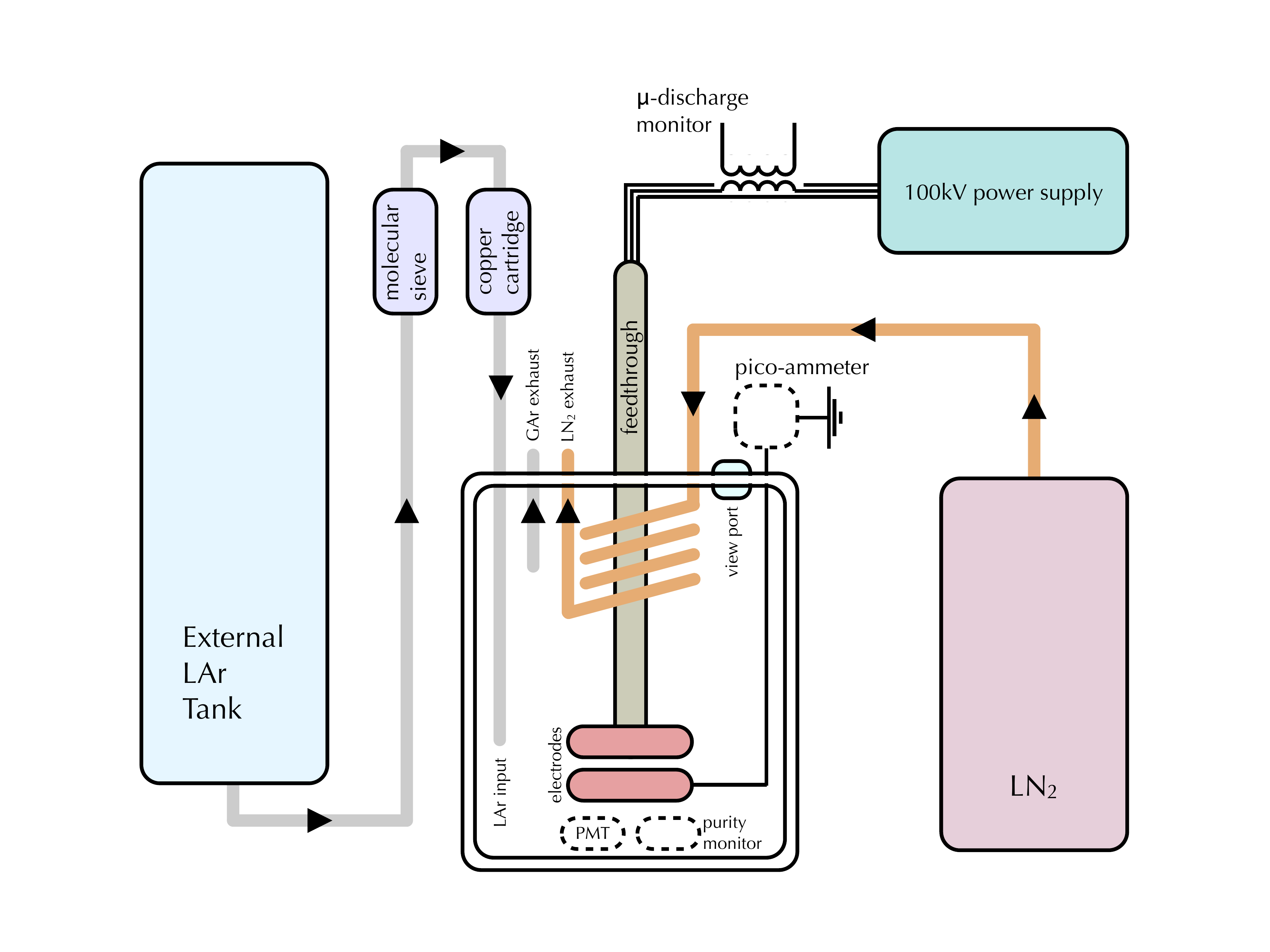}
\caption{Schematic representation of the apparatus. See text for the description.}
\label{fig:scheme}
\end{figure}

A dedicated test to measure the maximum electric field (up to 100~kV over 1~cm) that the liquid argon can sustain was performed~\cite{Bay:2014jwa}.
Figure~\ref{fig:scheme} shows the schematic representation of the test setup, which consists of a vacuum insulated dewar hosting a high voltage feedthrough and a couple of electrodes, in between which the high electric field is generated.

During operation, the dewar is filled with liquid argon purified through a molecular sieve (ZEOCHEM Z3-06), which blocks the water molecules, and a custom-made copper cartridge, which absorbs oxygen molecules.
Before the filling, the vessel is evacuated to residual pressure lower than $10^{-4}$~mbar in order to remove air traces, favour the outgassing of the materials in the dewar, and check the absence of leaks towards the atmosphere.
During the filling, the argon {\it boil off} is exhausted to control the pressure in the dewar.
The pressure is always kept at least 100~mbar above the atmospheric pressure, so that air contaminations are minimised.
Once the detector is completely full, the exhaust is closed, and the liquid argon is kept cold by means of liquid nitrogen flowing into a serpentine, which acts as heat exchanger that condenses the argon {\it boil off}.
The liquid argon level is visually checked through a vacuum sealed viewport.

The voltage is provided by a negative 100~kV power supply from Heinzinger~\footnote{Heinzinger PNC 100000. Heinzinger electronic GmbH, Rosenheim, Germany. \protect\url{http://www.heinzinger.com}.} though a HV coaxial cable modified to inductively couple the HV wire to an oscilloscope via a 1:1 transformer.

\begin{figure}[htbp]
\centering
\includegraphics[width=0.9\textwidth]{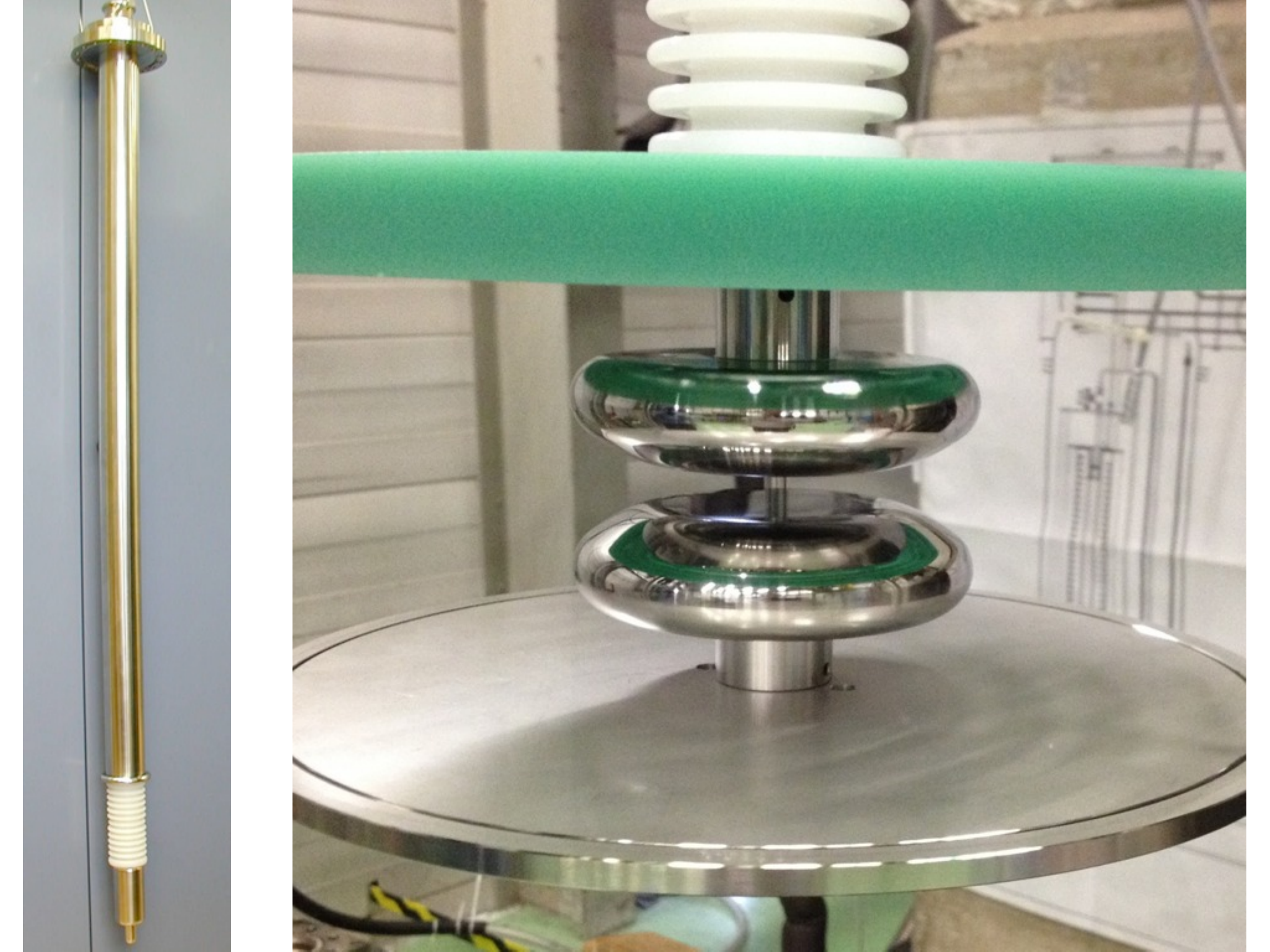}
\caption{Left: Image of the High Voltage feedthrough. Right: figure of the electrodes structure.}
\label{fig:feedthrough}
\end{figure}

When a pulsed current flows through the cable, it is detected on the scope.
This sensor is used to monitor the frequency of small discharges (not necessarily happening between the electrodes, e.g.\ the feedthrough and the cable may have leakage currents).
It is also used to monitor the current delivered by the power supply during the charging up.
The cable enters into a custom-made HV feedthrough (see figure~\ref{fig:feedthrough} left), which is vacuum tight and designed to sustain voltages larger than 150~kV.

\begin{figure}[htbp]
\centering
\includegraphics[width=0.45\textwidth]{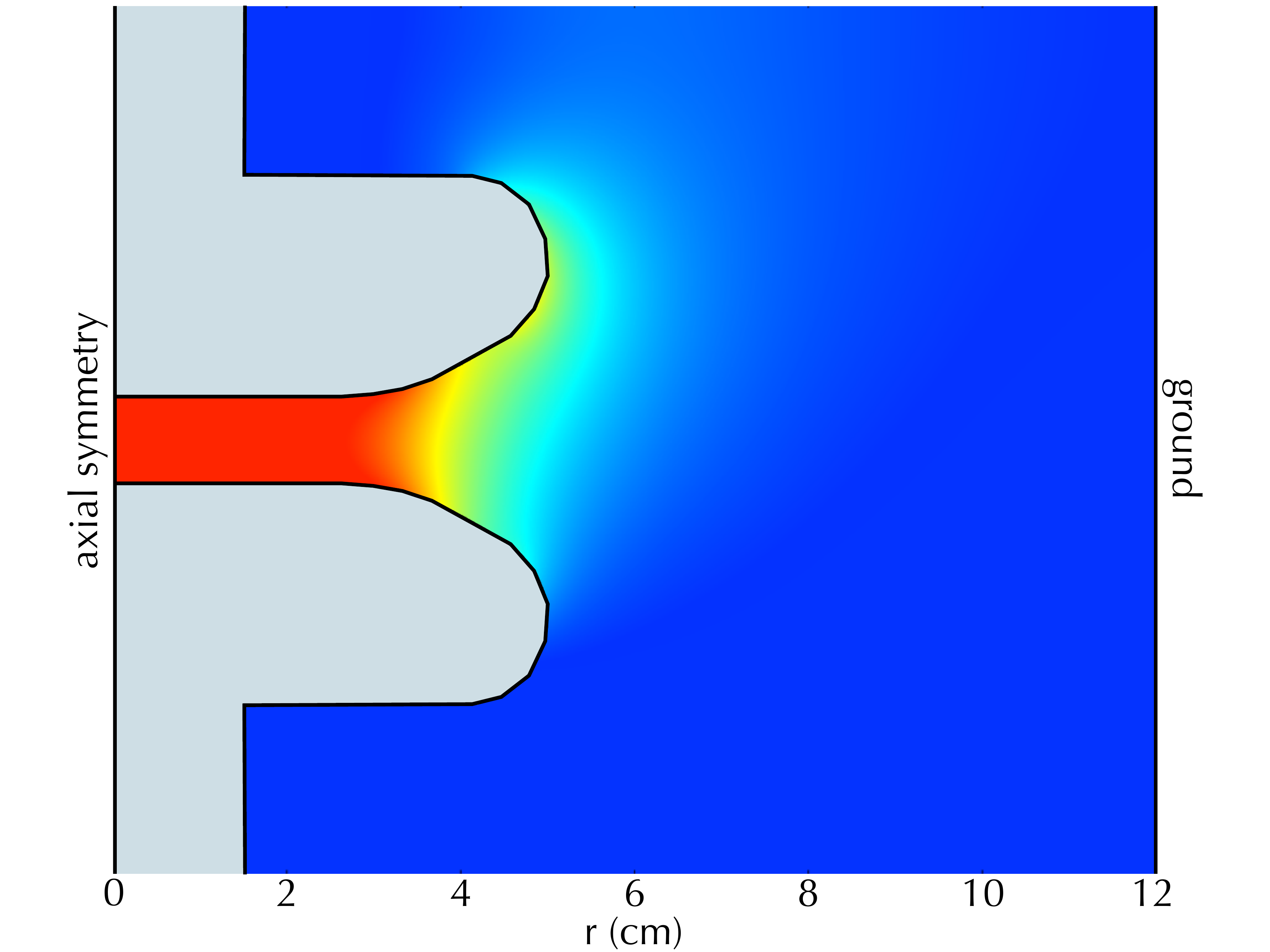}
\includegraphics[width=0.45\textwidth]{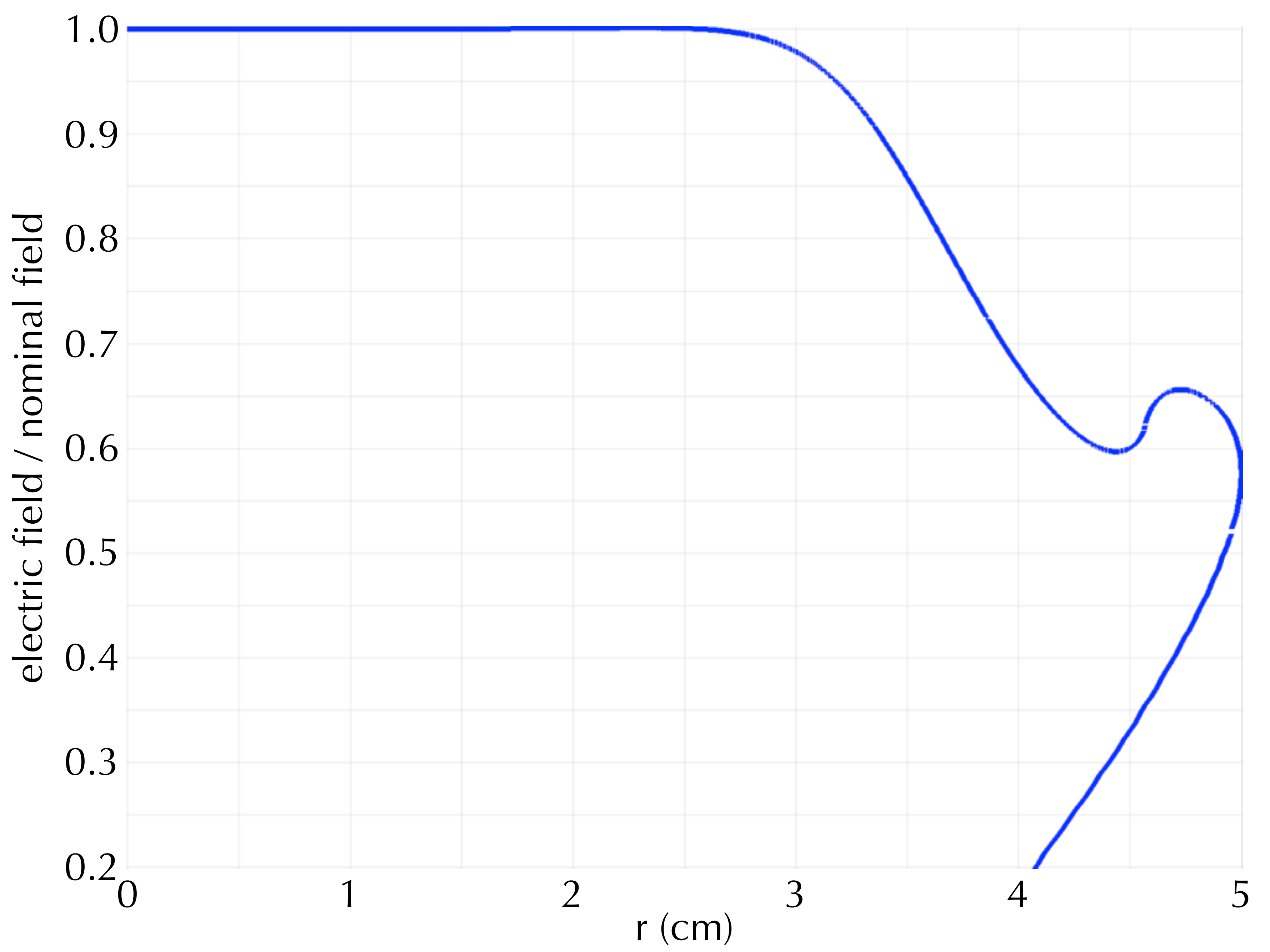}
\caption{Left: Computed electric field. The cross sections of the electrodes are shown in grey, and the colour pattern is proportional to the absolute value of the electric field. The electric field is essentially uniform in the central region. Right: Computed electric field on the profile of the top electrode as a function of the radius. The largest field is attained between the electrodes.}
\label{fig:field}
\end{figure}

High electric fields can be achieved with low potentials and electrodes with small curvature radii, but, since the breakdown is a random process, it is important to test a sizeable region of the electrodes.
For these reasons, a system that provides a uniform electric field over 20~cm$^2$ area was used.
A picture of the electrodes structure is shown in figure~\ref{fig:feedthrough} right.
The two 10~cm diameter electrodes have the same shape and are facing each other at a distance of 1~cm.
The top electrode is connected to the live contact of the HV feedthrough, and the bottom one is connected to ground through the vessel.
The electrodes, made out of mechanically polished stainless steel, are shaped according to the Rogowski profile~\cite{Rogowski:1923} that guarantees that the highest electric field is almost uniform (in a region of about 5~cm in diameter), and confined in between the two electrodes, as shown in figure~\ref{fig:field}.
The left image shows, in cylindrical coordinates, the absolute value of the electric field in the vicinity of the electrodes, computed with COMSOL.
On the right the electric field along the profile of the top electrode as a function of the radius is shown.

The two electrodes form a standalone structure, that is assembled first and then mounted.
By construction, the structure ensures the parallelism of the electrodes when cooled down to the liquid argon temperature.
The shrinkage of the materials in cold is computed to affect the distance between the electrodes less than 1\%.

\begin{figure}[htbp]
\centering
\includegraphics[width=0.5\textwidth]{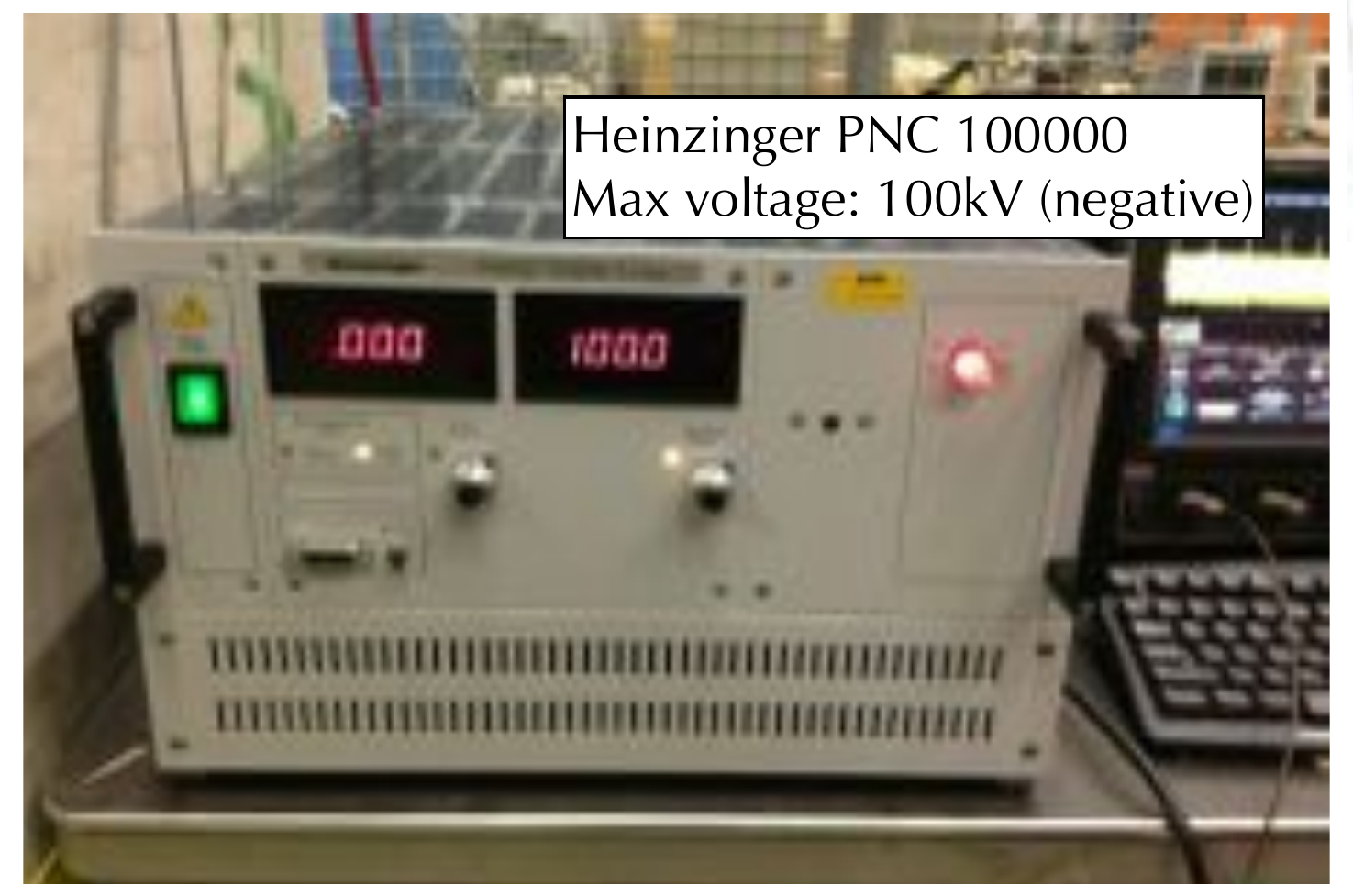}
\caption{Negative PNC 100000 power supply from Heinzinger in operation with 100~kV across a 1~cm liquid argon gap and 
no current. See text.}
\label{fig:field100kv}
\end{figure}

With the liquid argon temperature below the boiling point at a given pressure, a voltage of -100~kV was applied to the top electrode.
This value was limited by the maximum voltage of the power supply.
This configuration corresponds to a uniform electric field of 100~kV/cm in a region of about 20~cm$^2$ area between the electrodes.
Several cycles of discharging and charging up of the power supply were performed.
The system could also be stressed several times by ramping up the voltage from 0~V to -100~kV in about 20~s without provoking any breakdown.
At the centimetre scale, the dielectric rigidity of non-boiling liquid argon is larger than 100~kV/cm (See \Cref{fig:field100kv}).

A completely different behaviour was observed with boiling argon.
Several breakdowns between the electrodes at fields as low as 40~kV/cm occurred.
The stillness of the liquid argon was controlled by varying the pressure of the argon vapour and was monitored visually by looking through the viewport.
The pressure was regulated by acting on the flow of the liquid nitrogen passing through the heat exchanger.
In fact, the thermal inertia of the liquid argon bulk makes temperature variations very slow, hence increasing the cooling power translates in an rapid decrease of the vapour pressure.
When the pressure is above the boiling one, the liquid argon surface becomes flat and still.
On the contrary, the more the pressure decreases below the boiling point, the more the liquid argon boils.

The occurrence of breakdowns was found to be significantly sensitive to the presence of bubbles.
Above about 1000~mbar the argon was not boiling and 100~kV/cm could be achieved.
Around 930~mbar breakdowns occurred at 70~kV/cm and below 880~mbar at 40kV/cm.
This is interpreted as the evidence that breakdown can be triggered by bubbles in the liquid: the bubbles convey the breakdown forming a preferred channel for the discharge development, since argon gas has a much lower dielectric rigidity compared to the liquid.

Results from a similar test~\cite{Blatter:2014wua} with a different electrode geometry show that the maximum electric field (in a non uniform field) for a breakdown free operation is 40~kV/cm.
A strong dependence of the behaviour of the setup on the presence of bubbles is also observed.
In addition, the breakdown point is only weakly dependent on the liquid argon purity: the maximum electric field increases of 30\% changing the oxygen-equivalent contamination from 1~ppb to 20~ppm.

\clearpage
\section{DLAr detector components}

\subsection{Anode Charge Readout Plane}

\graphicspath{{./Section-AnodeCRP/figures_CRP/}}

\newcommand{\dqdxz}{\ensuremath{\Delta Q_0/\Delta s_0}\xspace}
\newcommand{\dqdxi}{\ensuremath{\Delta Q_i/\Delta s_i}\xspace}
\newcommand{\dqdxo}{\ensuremath{\Delta Q_1/\Delta s_1}\xspace}

\subsubsection{The top anode deck}
\label{sec:top_anode_deck}

%

The top anode deck is composed of a reinforced frame holding the independent
Charge Readout Planes (CRPs). The deck is suspended via ropes passing through dedicated
Feedthroughs as explained in \Cref{sec:supportingstruct}. The height of the deck is controlled externally and positioned precisely above the liquid level.  The 36 m2 surface is to be instrumented with a large number (144) of individual modules. This will require a semi-industrial production with external suppliers and a dedicated QA/QC validation and calibration effort. The anode modules provide two independent coordinate views, and are segmented with strips of 3~mm pitch. The baseline option for the CRP is based on the LAr LEM TPC technique, as described in the \Cref{sec:larlemoption}. In addition, we are also developing a second option based on the MicroMEGAS technique, as described in \Cref{sec:micromegas}.

\subsubsection{The supporting structure}
\label{sec:supportingstruct}

\begin{figure}[htb]
\begin{center}
\includegraphics[width=0.85\textwidth]{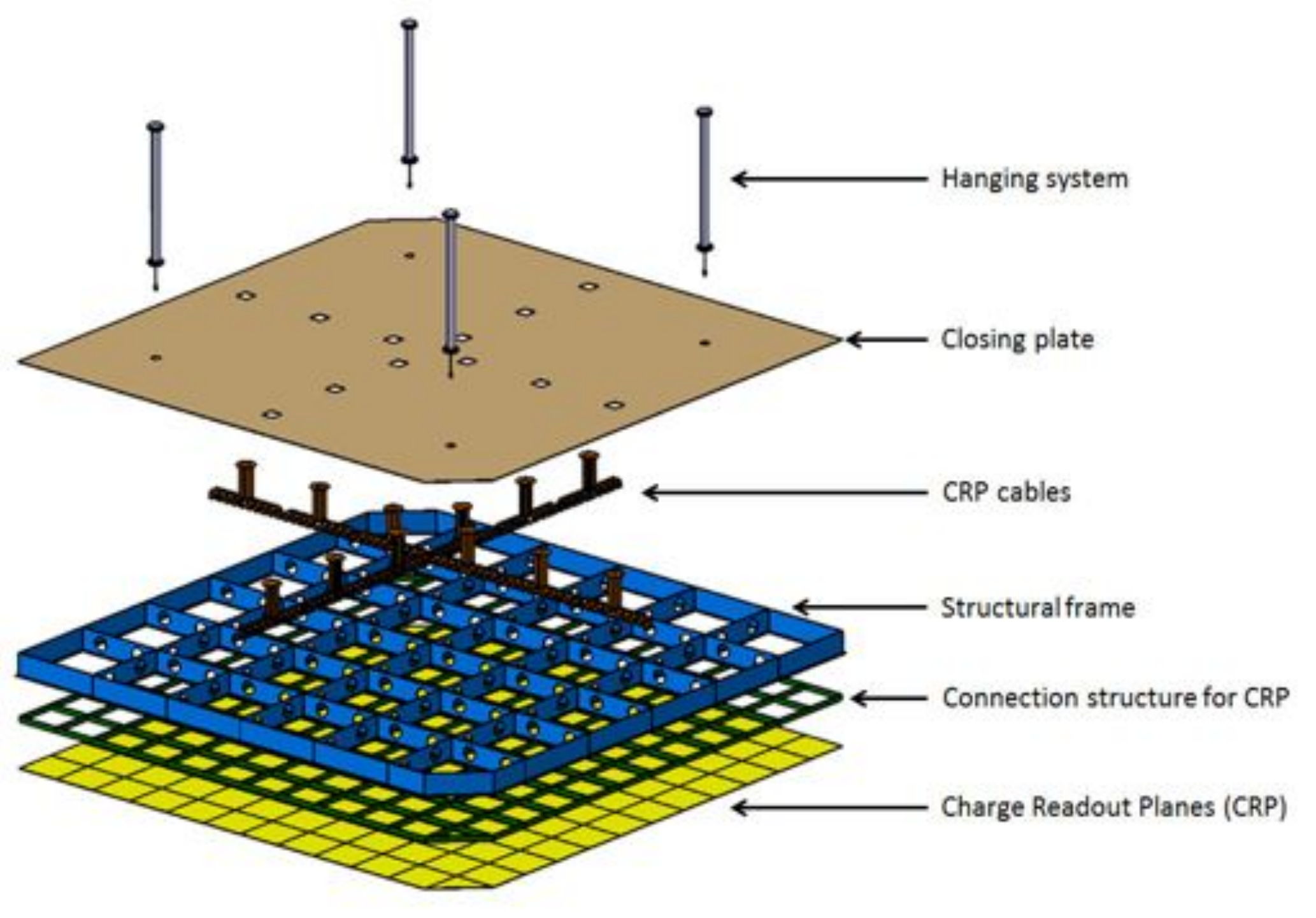} 
\caption{Exploded view of the anode deck structure and naming conventions.}
\label{fig:topdeckexploded}
\end{center}
\end{figure}

\begin{figure}[htb]
\begin{center}
\includegraphics[width=0.85\textwidth]{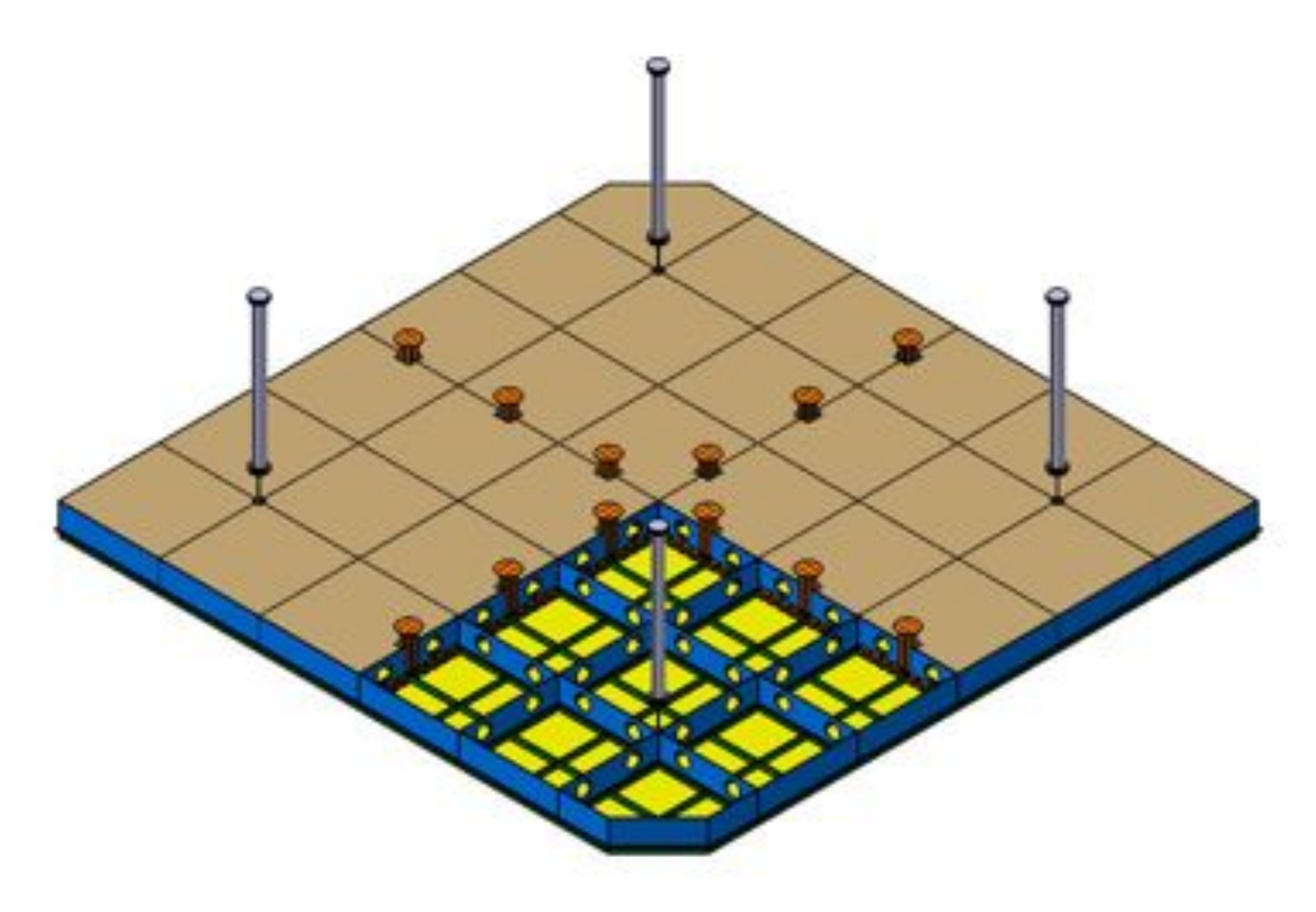} 
\caption{Assembled structure (with closing plate partially removed for viewing purpose).}
\label{fig:topdeckassembled}
\end{center}
\end{figure}

The anode deck structure is made of a sandwich of elements holding the independent Charge Readout Planes (CRP). \Cref{fig:topdeckexploded} shows an exploded view of the different components, whereas \Cref{fig:topdeckassembled} shows the assembled structure. Note that for clarity reason, the closing plate on the second figure has been partially removed to make the frame visible.

The structure is made of a connection structure for CRP on which is fixed a frame made of beams called structural frame. 
Then a plate is used to close this frame, playing the role of an upper skin. The combination of the frame, whose vertical 
dimension ($\sim$250mm) is clearly of prime importance, associated with a closing plate gives the desired stiffness of the 
final assembly. Moreover, the deck is suspended via four ropes fixed in the structural frame using dedicated feedthroughs. 
The use of four ropes makes the boundary conditions symmetric and is decisive to achieve the requirement in term of 
CRP deformations. Finally several feedthroughs for the CRP cables (electronics) are foreseen in the closing plate.
The mechanical support structure is designed to get a very small sag under gravity, but also to maintain the read out plane 
into positioning tolerance when going from ambient mounting temperature to liquid argon temperature ($+22^\circ$C to $-186^\circ$C). 
During this almost two hundred degree cooling phase, even small differences between materials thermal expansion coefficient could modify drastically the planarity of the structure because of bi-material thermal effects. Consequently, in order to avoid as much as possible the use of different material the retained proposal is to produce the whole mechanical structure in G10.

In order to minimize sagging, the design is based on a sandwich structure composed of two G10 plates separated by the G10 frame. The lower plate is composed of LEM detection plane and associated connection structure supporting frame. The upper one is purely dedicated to mechanical stiffness and structure symmetry. The later guarantees an in-plane thermal deformation of readout plane (displacements almost reduced to X and Y components), CRP and holding structure assembly.

\begin{figure}[htb]
\begin{center}
\includegraphics[width=0.85\textwidth]{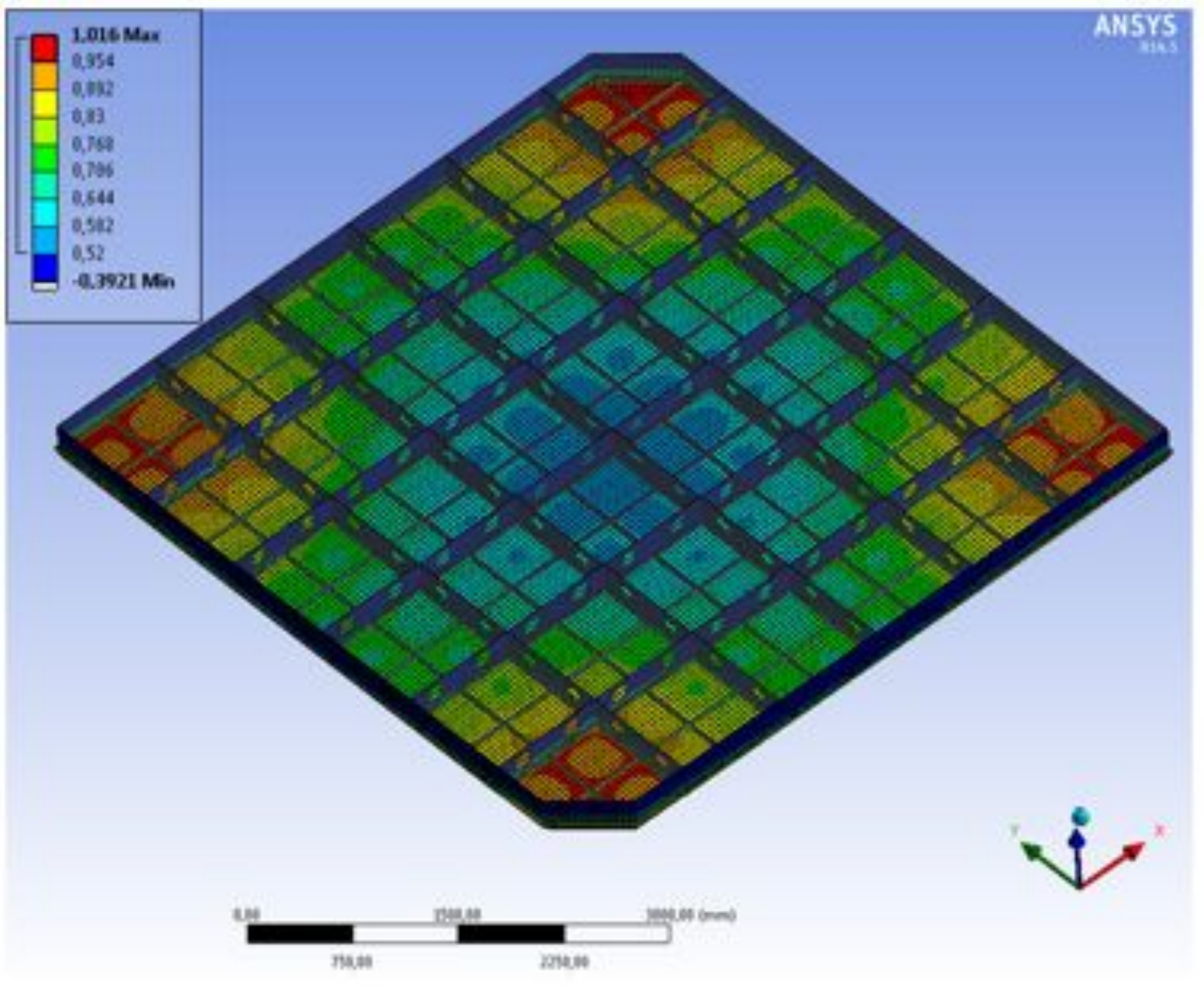} 
\caption{Vertical displacements under gravity and thermal loading.}
\label{fig:topdeckgravthermloading}
\end{center}
\end{figure}

A pre-sizing finite element study has been carried out to evaluate the thermo-mechanical behaviour of the proposed structure. The anode grid, tensioning system and electronic readout weight have been estimated to be about $10$kg/m$^2$. In this model a vertical zero displacement boundary condition has been applied to each hanging rope. The loading used consists of a gravity load and a uniform temperature change from $+22^\circ$C to $-186^\circ$C. As shown on \Cref{fig:topdeckgravthermloading}, the CRP vertical displacements are in the range $+0.52$~mm to $+1.016$mm equivalent to a peak to peak vertical displacement of $0.5$~mm. Moreover for better illustration of CRP vertical displacements, the upper closing plate has not been shown. These displacements are in agreement with the $+/- 0.5$~mm CRP plane planarity specification. 

\subsubsection{The position control system}
The aim of the control system is to manage the position of the anode plane with respect to the liquid argon surface. The system will have to cover a displacement range of the order of 100 mm and to keep the distance of the anode plane to the liquid equals to 10 mm at a sub mm precision. The speed and acceleration are not an issue due to the slow change of the argon liquid level with time. 

The real time monitoring and control will be carried out assuming that the position of each CRP pad (1 m$^2$ unit; total 36) is available thanks to capacitive measurement between the argon liquid level and the pads with an accuracy of $0.1$mm. The position is proportional to the capacitive measurement signal. 

The anode plane, weighting approximately 2 tons, will be hanged up by means of 4 rigid cables as described earlier and the action and control of the anode plane will be carried out outside of the tank to avoid the environment problems (temperature, vacuumÉ). The thermic dilatation of the cables is expected to be less than $10$mm. 

{\bf Positioning process:}

The positioning of the anode plane will be decomposed in two different steps.

\begin{figure}[htb]
\begin{center}
\includegraphics[width=0.85\textwidth]{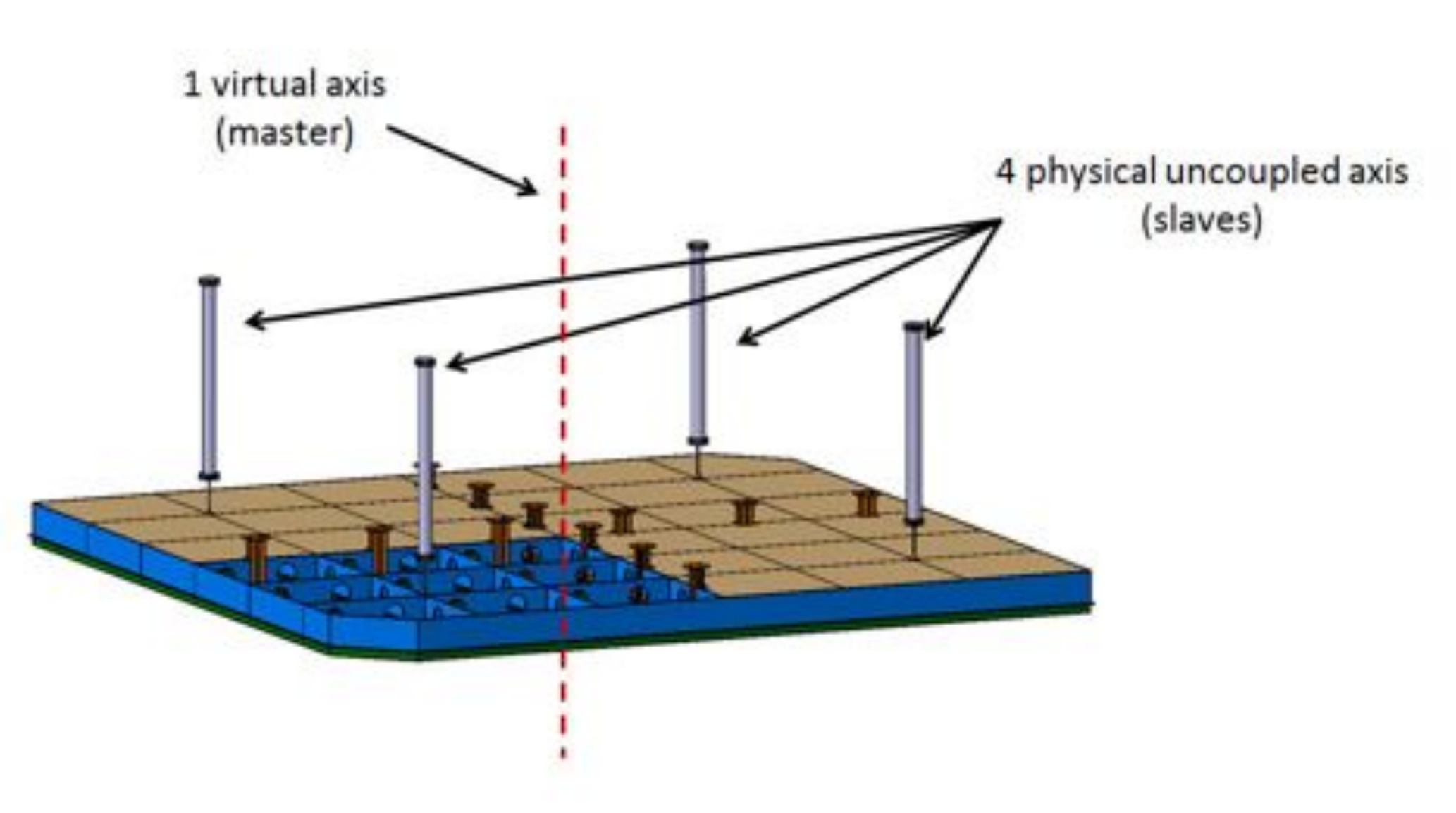} 
\caption{Anode plane controlled by four physical axes managed which are managed by a virtual axis.}
\label{fig:topdeckcontrol}
\end{center}
\end{figure}

The first one will be the displacement of the whole anode plane to move it as close as possible to the liquid argon level. This motion is named Translation Motion (TM) and will be based on a multi-axes coordinated motion. The control set point of each axis will be done via a virtual axis used as a master. The 4 physical axes will be the slaves (\Cref{fig:topdeckcontrol}). In such manner, the motion will be accurately and safely performed during the complete process. Note that this technology is commonly used in automation when requiring multi-axes motion control.

The second step, named Tilt Adjustment (TA), will be the correction of the anode plane tilt. In this case, the four axes will be decoupled and independently managed. The capacitive measurement of each pad will allow interpolating the shape and the planarity of the anode plane. A dedicated algorithm will be developed in order to generate the needed rotations and translations, and thus to obtain a charge readout plane as parallel as possible to the liquid argon level. In this way, the system will be able to manage rotation and deformation of the anode in accordance to the specifications.

The advantage of this setup is that the axes could be managed simultaneously like if they would be mechanically coupled for the whole plane positioning (TM) or completely independent for the tilt plane adjustment (TA). In parallel, constrain gauges could be installed all along the anode plane to guarantee that the applied forces wonÕt generate any damages.

{\bf Technology of the control system and actuators:}

The control architecture will be based on a Programmable Logic Controller (PLC) which is composed of two central processing units (one for the classical cycles and a second one for the drive part management), two fieldbus (one of which is dedicated to the drive) for the deported devices (input/output modules, touch panelÉ) and specific variators which are able to manage power cuts and synchronized displacements.

The actuators have to be able to manage accurately tiny displacements of a few millimeters with a relative heavy load. In this prospect, electric jacks are foreseen. This type of device coupled with an electrical gear-motor (brushless equipped by absolute encoders) thanks to an industrial integrated system (example: ESBF jacks models produced by Festo, EMC jacks models produced by BoschÉ with a range of 200 mm) will allow such linear displacement. Note that the jack supplier could be different from the motor supplier.

\subsubsection{The LAr LEM TPC option}
\label{sec:larlemoption}
The design concept of the novel Liquid Argon Large Electron Multiplier Time Projection Chamber (LAr LEM TPC)
and its promising performances have 
been demonstrated experimentally with a detector prototype having an active area of 10 $\times$ 10 cm$^2$ 
\cite{Badertscher:2010zg, Badertscher:2009av} and 40$\times$80~cm$^2$\cite{Badertscher:2013wm,Badertscher:2012dq}. 
This new type of LAr-TPC, operating in the double-phase (liquid-vapor) mode in pure argon,
is characterized by 
its charge
amplifying stage (LEM) 
and its projective anode charge readout system. 
Situated in the vapor phase on top of the active LAr volume, these latter provide an adjustable charge gain 
and two independent readout views, each with a canonical and optimal pitch of 3~mm, since 
the amplification stage produces sufficient charge to be shared among several readout electrodes
(smaller pitch sizes could be considered).
With $T_0$ time information provided by the LAr scintillation, they 
provide a real-time three-dimensional (3D) track imaging 
with $dE/dx$ information 
and the detector acts as a high resolution tracking-calorimeter. 
A  high signal-to-noise ratio  can be reached in the LAr LEM-TPC thanks to the gas amplification stage.
This significantly  improves the 
event reconstruction quality 
with a lower energy deposition threshold and a better resolution per volumetric pixel (voxel) 
compared to a conventional single-phase LAr-TPC 
\cite{Badertscher:2009av}.
In addition the charge amplification 
compensates for potential loss of signal-to-noise due to the charge diffusion and
attachment to electronegative impurities diluted in LAr, 
which both become more important as the drift length increases. The collection-only 
readout mode (avoiding the use of induction planes) is also an important asset
in the case of complicated topologies, like e.g. in electromagnetic or hadronic showers.


The extracted drift electrons in the vapor phase are finally drifted
towards a collection plane where they are detected.
Due to the applied electric field, electrons are accelerated
in between two collisions with argon atoms
and produce the \textit{Townsend
  avalanche} (see~\Cref{chapter:chargeAmplificationInGas}).
Being positioned on top of the LAr TPC in the vapor phase, the
LEM is responsible for the
multiplication of the ionization charges. 
\begin{figure*}[t]
\centering
\includegraphics[width=0.53\textwidth]{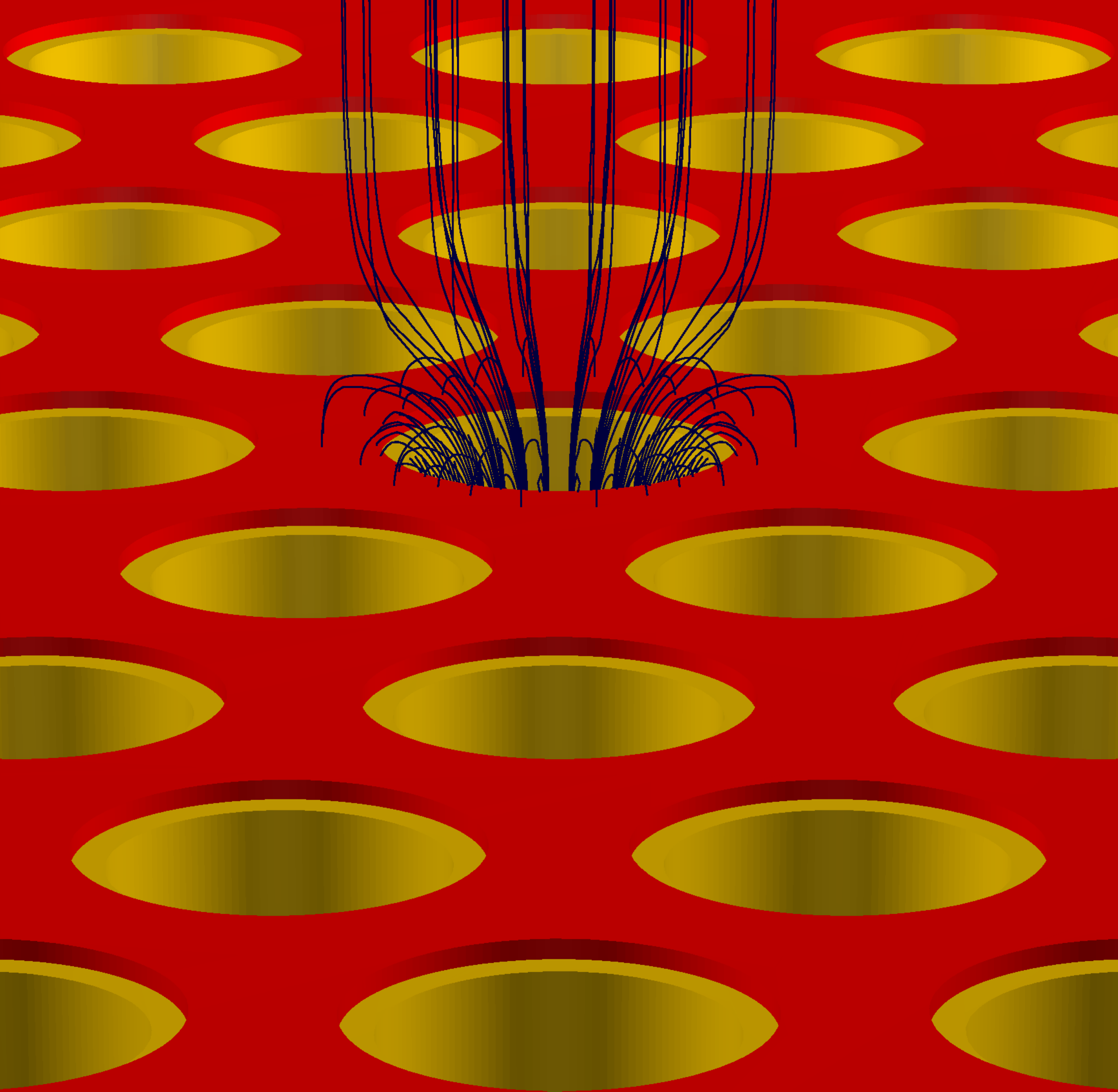}
\includegraphics[width=0.46\textwidth]{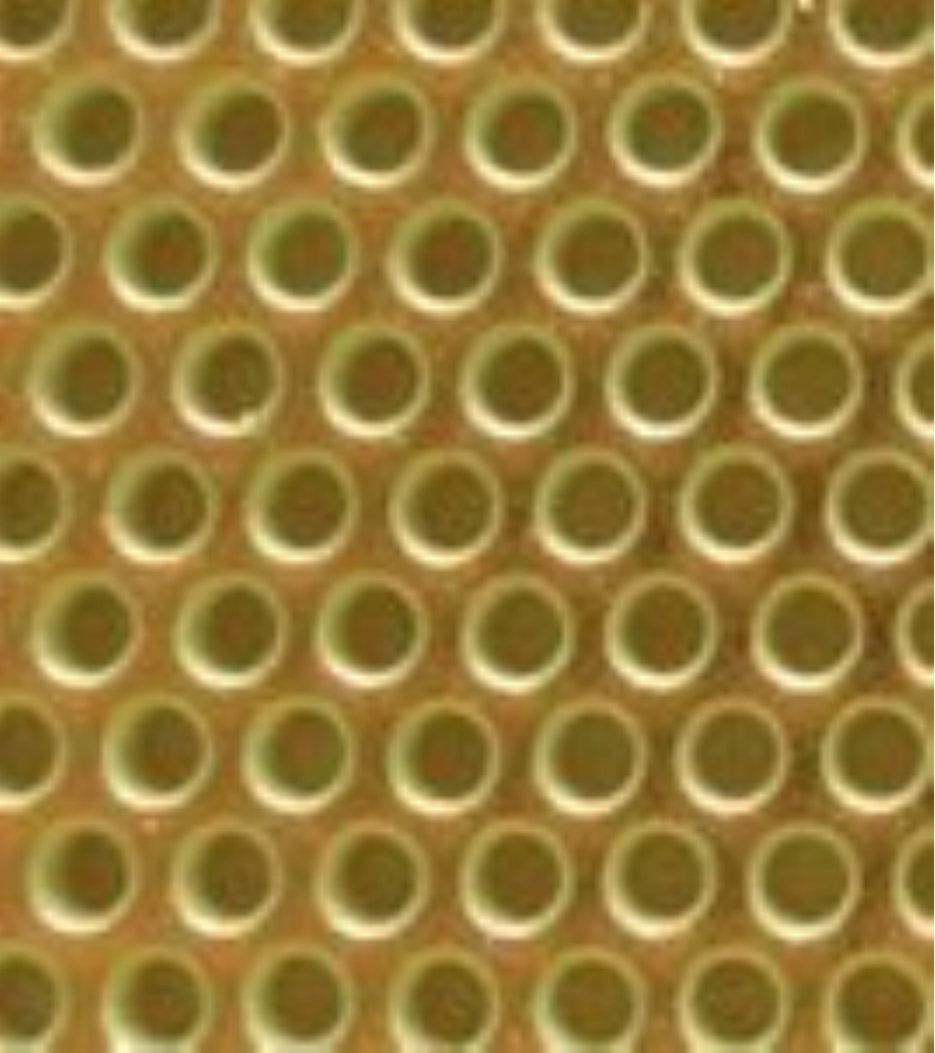}
\caption{Left: schematic drawing of the LEM, consisting of the
 copper electrode (red) and the insulating material FR4
 (yellow). The field lines passing through the central hole are drawn
 in black. Right: closeup picture of a manufactured LEM. As in the
 schematic drawing on the left, the dielectric rim is clearly
 visible. [See text for more details.]
}
\label{fig:detector-lem}
\end{figure*}
Similar to the more established Gaseous Electron Multipliers
(GEMs)~\cite{Sauli:1997qp}, the basic idea of the LEM is to create
electron avalanches in the gap of two holed metal planes, which are
spaced by an electrical insulator. While the GEMs are made of
50~$\mu$m thick Kapton foils that are likely not applicable to 
cryogenic double phase detectors, the LEM or THGEM (for THick GEM) is
a macroscopic, sturdy structure, made with standard PCB techniques. 
It consists of a 1~mm thick FR4 substrate with thin copper sheets 
laminated on both sides. The electrode-insulator-electrode sandwich is
perforated with 800~$\mu$m spaced cylindric holes with a diameter of
500~$\mu$m. The view from top is shown in the scheme on the left and the
picture on the right of \Cref{fig:detector-lem}. The scheme
shows that the hole diameter of the electrode (red) exceeds the
diameter of the hole through the insulator (yellow). As demonstrated
in~\cite{Breskin:2009aa} these so-called dielectric rims, typically 
about 50~$\mu$m thick, significantly reduce the occurrence of discharges
and therefore increase the maximum achievable gain. 
Besides the dielectric rims, 
the discharge probability is further reduced by the hole geometry of
the LEM. Since the avalanche is happening inside the 1~mm long and
0.5~mm wide holes that are only surrounded by insulating material and
no metal surfaces, secondary effects due to emitted scintillation light are
suppressed. This is sometimes called \textit{mechanical quenching}. A
very detailed discussion of the discharge mechanisms, including finite
element simulations, can be found in~\cite{Thesis_FilippoResnati}.

\begin{figure}[hbtp]
\begin{center}
\includegraphics[height=20pc]{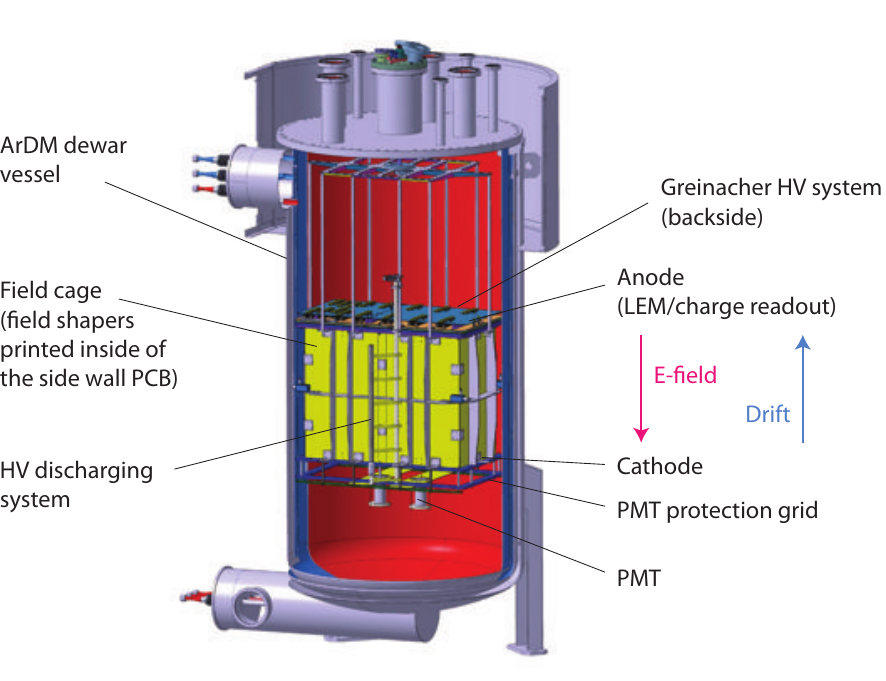}
\caption{3D CAD model of the LAr LEM-TPC prototype installed in the ArDM LAr vessel. The field cage has dimensions 76 cm (L) $\times$ 40 cm (W) $\times$ 60 cm (H). The LEM charge readout system is mounted on top of the field cage. Negative HVs linearly decreasing from bottom to top are generated by the built-in Greinacher HV multiplier and distributed to 31 field shapers formed by etching on the inner surface of the side wall PCBs. A vertical drift electric field thus is created uniformly over the full drift length of 60 cm.}
\label{fig:fig1}
\end{center}
\end{figure}

Figure \ref{fig:fig1} 
illustrates 
the structure of the 40 $\times$ 80 cm$^2$ LAr LEM-TPC hanging inside the ArDM detector vessel, that was successfully operated at CERN in 2011. 
The field cage, which defines the active LAr target and drift volume, is constructed with four PCB side walls in a shape of rectangular box with dimensions 
76 cm (L) $\times$ 40 cm (W) $\times$ 60 cm (H).
On the inner surface of each PCB 31 horizontal field shaping strips, which hereafter are called field shapers (FS), 
are formed by etching at a constant pitch of 20 mm.
The top (FS0) and the bottom (FS30) field shapers have a width of 9 mm while the ones in between have 18 mm, with a gap of 2~mm between neighboring ones. 
Negative high voltages linearly decreasing from bottom to top are distributed to the field shapers so that a vertical drift electric field is created uniformly over the full 
drift length of 
60 cm.

On top of the field cage two horizontal extraction grids are mounted with a gap of 10 mm in between. 
The grid is a 0.15-mm-thick 
stainless-steel mesh, where square holes with a size  2.85 $\times$ 2.85 mm$^2$ are etched at a pitch of 3 mm all over the active area. 
The lower grid is positioned at the top face of the field cage.
For the double-phase operation mode the LAr surface is adjusted at the middle of the two grids. 
The liquid level at each of the four corners can be monitored with a precision of $\sim$0.5 mm with the aid of four capacitive level meters. 
Ionization electrons produced by ionizing particles are drifted upwards to the liquid surface. 
These electrons are extracted across the liquid-vapor interface into the gas argon (GAr) phase 
with the aid of 
a strong extraction field of typically 3--4 kV/cm between the two grids. 
They are then collected by the charge readout 
system 
incorporating the LEM.
The bottom face of the field cage is covered by the cathode grid 
that is a stainless-steel mesh of the same type as used for the extraction grids.
The top (FS0) and the bottom (FS30) field shapers are electrically coupled to the lower extraction grid and respectively, to the cathode. 

\begin{figure}[hbtp]
\begin{center}
\includegraphics[height=17pc]{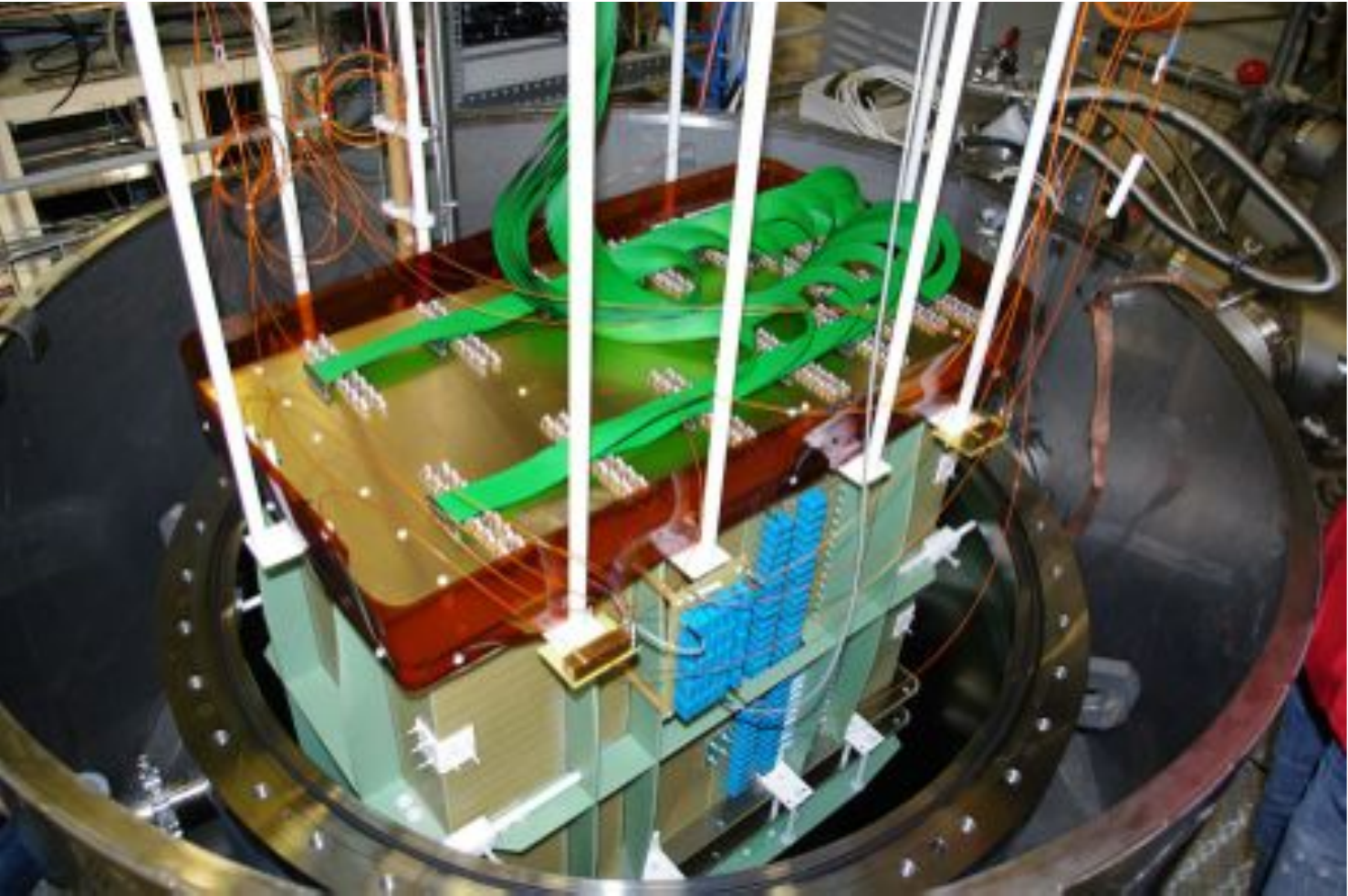}
\caption{Picture of the LAr LEM-TPC with readout sandwich
and drift field cage, which are about to be inserted into the ArDM vessel. 
The stack of blue capacitors on the front side wall shows the Greinacher HV multiplier.}
\label{fig:fig2}
\end{center}
\end{figure}

The HV for creating the drift field is generated using a built-in 30-stage Greinacher HV multiplier~\cite{Badertscher:2012dq}. 
The circuit is integrated in the design of one of the side-wall PCBs and the components are mounted directly on the outer surface of the PCB, 
as can be seen in Figure~\ref{fig:fig2}. 
As already mentioned the generator itself avoids the use of a voltage divider, since each multiplying stage provides
a characteristic DC voltage. The various multiplying stages generate a monotonously increasing potential to 
supply the electrodes surrounding the drift volume. In our setup, 
the DC output of each of the 30 Greinacher stages is connected to each field shaper via a wire through the PCB. 

\begin{figure}[t]
  \centering
  \includegraphics[width=1\textwidth]{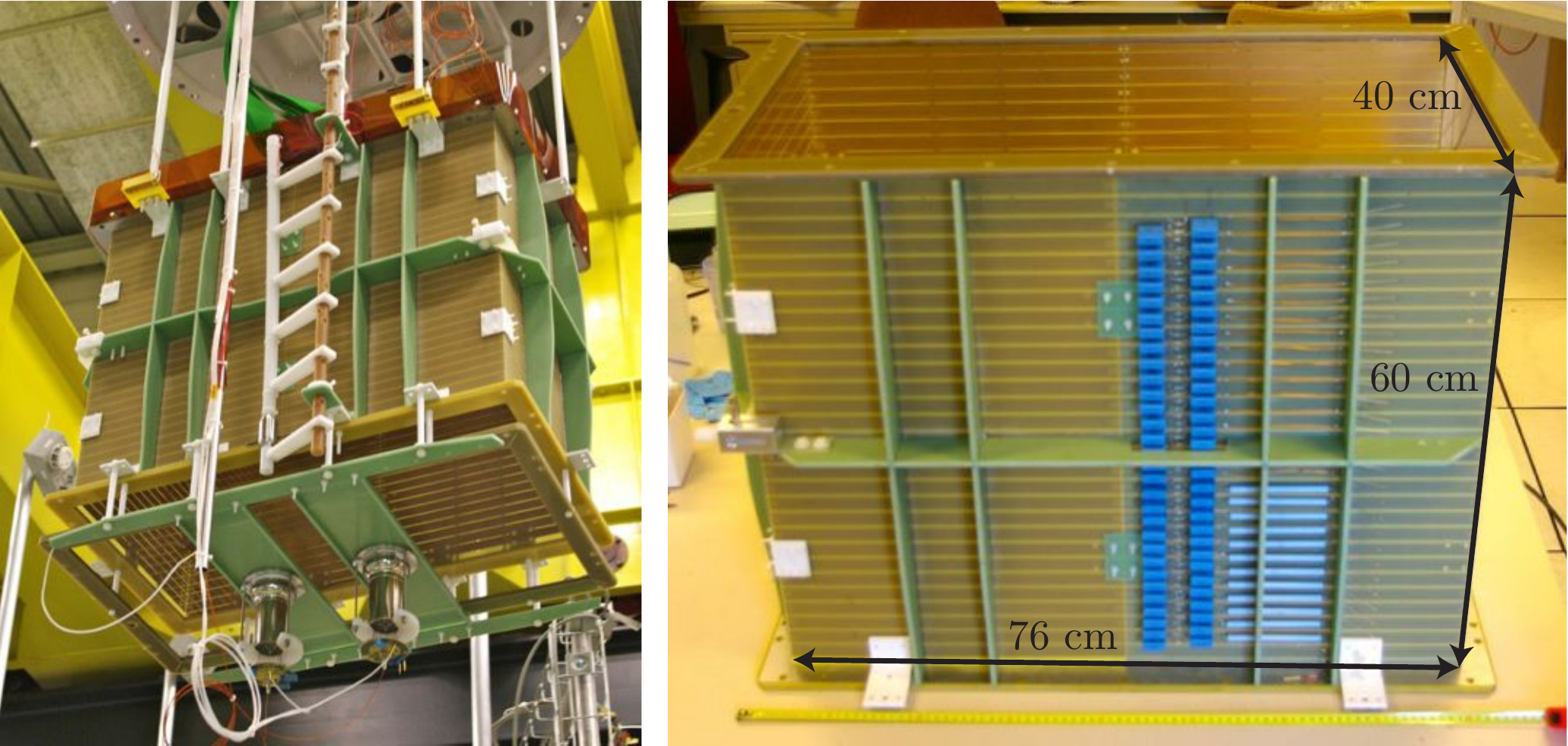}
  \caption{Pictures of the detector. Left: fully assembled detector
    hanging from the top flange of the ArDM vessel.  Right: TPC during
    the assemblage phase.} 
  \label{figure:250L}
\end{figure}

\begin{figure}[htb]
  \centering
  \includegraphics[width=1\textwidth]{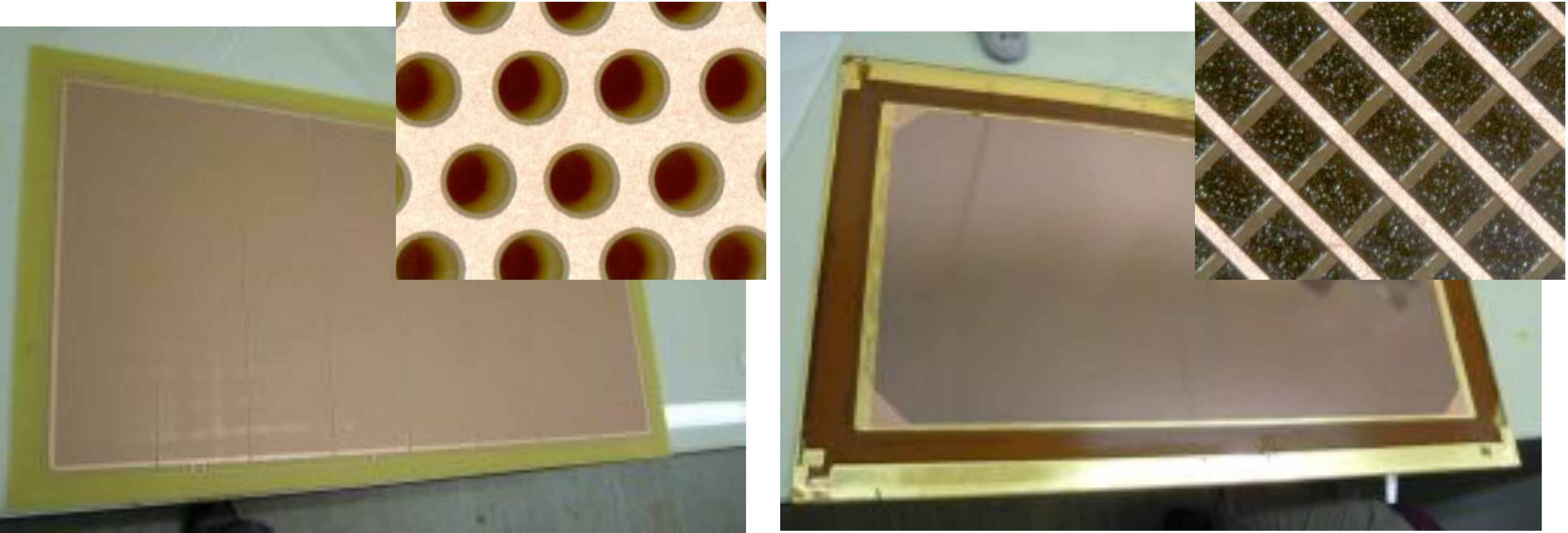}
  \caption{Picture of the $76 \times 40$~cm$^2$ LEM with close-up of
    its holes (left) and picture of the 2D anode with close-up of its strips (right).} 
  \label{figure:AnodeLEM}
\end{figure}

\begin{figure}[htb]
  \centering
  \includegraphics[width=1.0\textwidth]{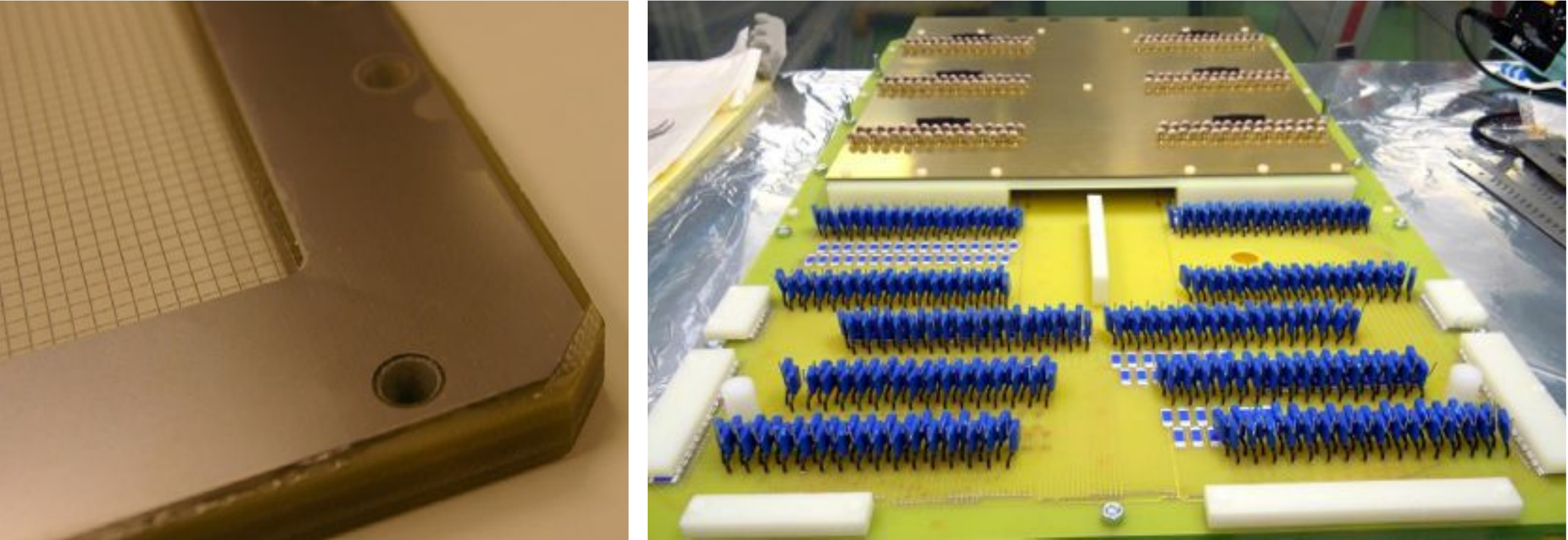}
  \caption{Pictures of the details of the detector: 
  one of the extraction grids (left) and the charge readout 
  sandwich during the assembling (right): the half in the back is already 
  covered with the signal plane, in the foreground the HV distribution plane 
  with the decoupling capacitors and the 500~M$\Omega$ resistors is seen.} 
  \label{figure:GridSignalPlane}
\end{figure}

\begin{figure}[t]
  \centering
  \includegraphics[width=1.0\textwidth]{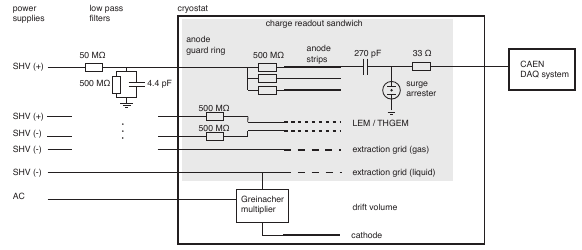}
  \caption{Electrical scheme of the TPC with the HV connections on the
    left and the data acquisition system on the right. The gray box
    shows the components which are embedded in the charge readout
    sandwich.} 
  \label{figure:electricScheme}
\end{figure}

As displayed in the scheme of Figure~\ref{figure:electricScheme}, each
electrode is connected independently via SHV feedthrough and a low pass
filter to a separate channel of the power supply. This configuration
allows to apply arbitrary field configurations. In order to be able to
operate the anode at positive voltages, each of the 512~readout strips
is connected via 500~M$\Omega$ resistors to the
guard ring of the anode.  They are mounted on a first \emph{signal
  routing plane}, which is positioned on top of the anode. The 270~pF
HV decoupling capacitors connect this
first plane to a second one, where a discharge protection circuit is
installed as well as the connectors for the sixteen signal
cables.  Finally, the signal cables are fed through the cryostat and
connected to the charge acquisition system. Both the discharge
protection circuit as well as the readout electronics and acquisition
system are detailed in the following section. 

Under the field cage two photomultiplier tubes (PMTs) are installed to detect scintillation light produced by charged particles crossing the LAr target (primary scintillation, S1), as well as the secondary scintillation light (S2) produced via electro-luminescence by the electrons extracted to the 
GAr phase 
\cite{Monteiro:2008zz}.
We employed Hamamatsu R11065 cryogenic 3'' PMTs having a high quantum efficiency of $\sim$30\%, specifically designed for the use in LAr 
\cite{Acciarri:2011qx}.
The quartz window of the PMT was coated with a wavelength shifter, 1,1,4,4-Tetraphenyl-1,3-butadiene (TPB) \cite{Boccone:2009kk}, 
in a Paraloid\texttrademark\, B-72 polymer matrix to detect 
deep ultra violet photons of the argon scintillation (peaked around 128~nm) . 
The PMT signal is used primarily for triggering and for determination of the event $T_0$ absolute time. 
To protect PMTs from possible discharges from the parts at HV, another mesh electrode, which is grounded to the detector vessel,  
is inserted between the cathode grid and the PMTs. 
The distance between the cathode and the protection grid was 9~cm and that between the protection grid and the PMTs 
was 1~cm. 

\begin{figure}[hbtp]
\begin{center}
\includegraphics[width=\columnwidth]{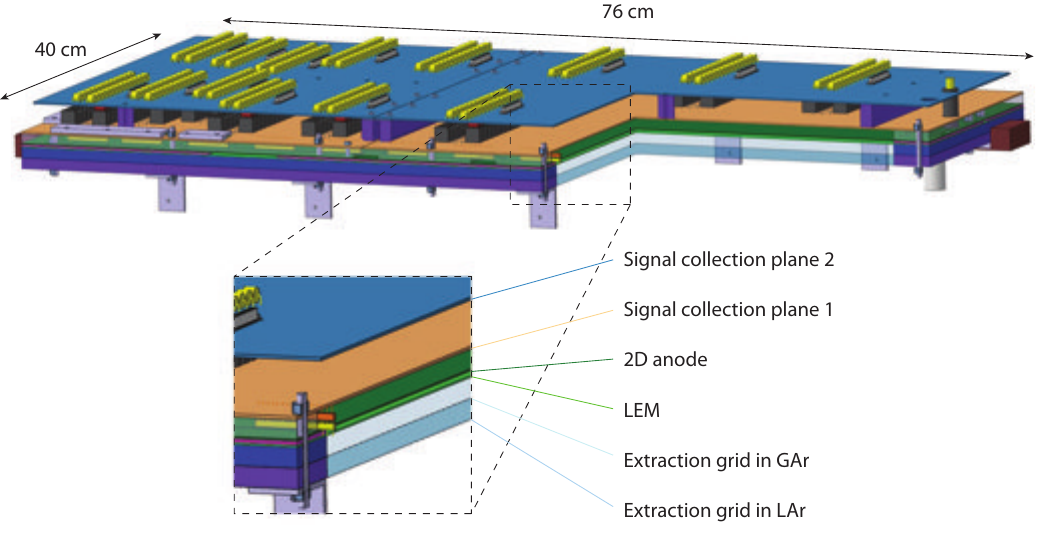}
\caption{Structure of the 40 $\times$ 80 cm$^2$ charge readout system incorporating a LEM (Large Electron Multiplier) and a 2D anode
successfully built and operated. Right-front corner is cut in the illustration to present different stages in the multi-stage structure.}
\label{fig:fig3}
\end{center}
\end{figure}

The  10~$\times$~10~cm$^2$ and 40 $\times$ 80 cm$^2$
are characterised by (1) a single 1-mm-thick LEM amplification stage and (2) a 2D readout anode realized on a single PCB. 
The system has a multi-stage structure consisting of two extraction grids, a LEM, a 2D readout anode and two signal collection planes as illustrated in 
Figure~\ref{fig:fig3}. 
The distances between the stages and typical configurations for the potentials and the inter-stage electric fields are summarized in Table~\ref{tab:table1}. 
In typical configurations a positive HV of $\sim$1 kV is applied to the anode while the extraction grid in liquid is operated at a negative HV of $\sim-7$ kV,
resulting in a virtual ground within the LEM plane. 
\begin{table}[hbtp]
\begin{tabularx}{\textwidth}{@{\extracolsep{\fill}}lccc} 
\hline
& Distance to the & Typical operating & Field to the stage \\
& stage above (cm) & potential (kV) & above (kV/cm) \\
\hline
\normalsize
Signal collection plane 2 & -- & 0 & -- \\
Signal collection plane 1 & 2 & $+$1 & -- \\
2D anode & 1 & $+$1 & -- \\
LEM (top electrode) & 0.2 & $+$0.5 & 2.5 \\
LEM (bottom electrode) & 0.1 & $-$3 & 35 \\
Extraction grid in GAr & 1 & $-$4 & 1 \\
Extraction grid in LAr & 1 & $-$7 & 3 \\
\hline
Cathode & 60 & $-$31 & 0.4 \\
\hline
\end{tabularx}
\caption{Baseline LAr LEM charge readout system configurations.}
\label{tab:table1}
\end{table}
The LEM is a macroscopic hole electron multiplier built with standard PCB techniques\footnote{It was manufactured at ELTOS Circuiti Stampati Professionali,
Italy ({http://www.eltos.com/}).}. 
It is a 1-mm-thick FR4 plate, double-side cladded with a passivated copper layer. 
Of the order of half a million holes 500 $\mu$m in diameter are CNC (Computer Numerical Control) drilled through the plate at a pitch of 800 $\mu$m between the centers of adjacent holes. After the PCB has been manufactured and drilled, the copper layers are further chemically etched to create 40~$\mu$m wide rims
around the holes.
A HV of typically 3.5 kV is applied across the two faces creating a strong electric field of 35 kV/cm in the LEM holes leading to multiplication of electrons by avalanches. 
The 2D anode plane has two orthogonal sets of readout strips on the bottom face providing two independent readout views. 
On top of the strips for one view, those for the other view are laid with a thin (50 $\mu$m) polyimide (Kapton) insulating bed underneath. 
For each view, copper strips are formed by etching at a pitch of 600 $\mu$m covering nearly the full area of 
76 $\times$ 40 cm$^2$.
The readout pitch of 3 mm is obtained by bridging five strips at one end. 
The strips on top are narrower (120~$\mu$m) than those lying underneath (500 $\mu$m) optimizing for an equal charge collection by the two views operating at the same potential. 
A more detailed description of the 2D anode can be found in Ref.~\cite{Badertscher:2010zg} and references
therein.
The strips for both of the views are tilted by 45$^{\circ}$ 
with respect to the sides of the rectangular plate,
to have symmetric conditions for both views.
We use two signal collection planes, between which 
HV decoupling capacitors (270 pF) are connected.
Each readout channel is routed to a collective 32-channel connector to a flat cable. 
A surge arrester  protecting the readout 
electronics from discharges is mounted for each channel. 
The cables bring the signals through a custom-made feedthrough on the upper flange, 
connected to the readout electronics placed outside of the vessel. 
The LEM charge readout system is assembled as a multi-layered ``sandwich'' unit with precisely defined inter-stage 
distances and inter-alignment. 
The multiplied charges coming from the ionization in LAr are finally
collected on a charge readout plane. The \textit{2D anode} basically consists of two
perpendicular sets of electrode strips (or: \textit{views}), which are separated by a very
thin insulating layer. Due to the semi-transparent design of the device,
arriving charges produce a similar response on both sets of strips,
finally delivering the two coordinates of the detected
charges. Although the concept was adapted from the readout for GEM
detectors~\cite{Bressan:1999aa}, the design parameters had to be
optimized for the application in a LAr TPC. In order to define an
optimal electrode geometry,
we have considered the response of a point-like energy deposition in
LAr. As a consequence of the electric 
field focussing and defocussing by the extraction grids and the LEM
but also the diffusion in gaseous argon, a $\delta$-function like charge distribution in LAr
arrives at the anode as a 1~mm extended charge cloud. Given this
assumption, the induced signals on both views of the anode have to be
mono-polar, fast and with similar amplitudes.
Mono-polarity is guaranteed, since all the electrodes act in
charge collection mode. As can be shown using the 
method of W. Shockley and S. Ramo~\cite{Shockley:1938aa,Ramo:1939aa},
electrons always drift against the weighting field of the
corresponding collection electrode, inducing an
entirely negative replacement current. Since the current is
proportional to the electron drift speed in gas, the expected signal
rise times are $<0.5~\mu$s.

\begin{figure*}[t]
\centering
\includegraphics[width=1\textwidth]{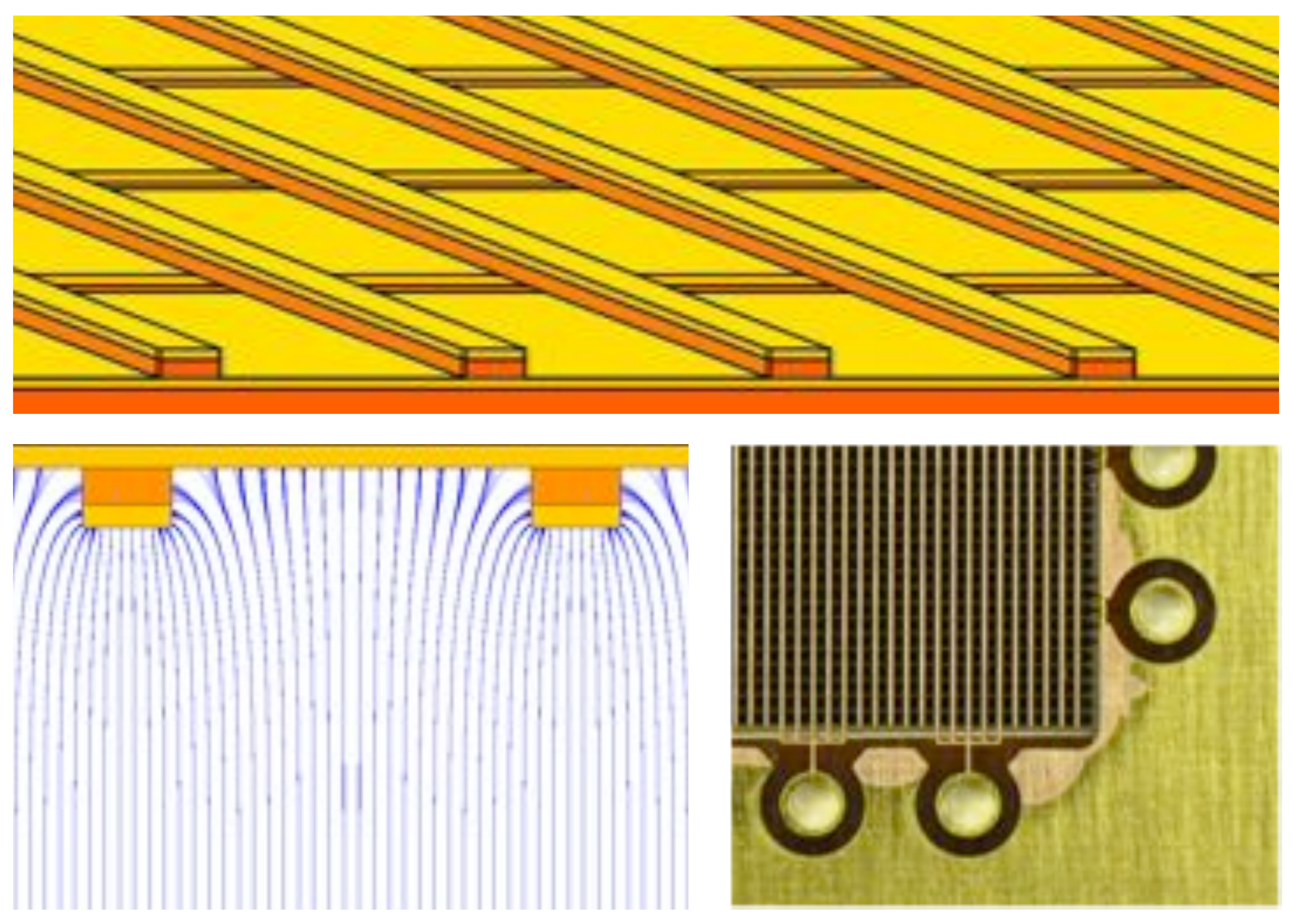}
\caption{Top: scheme of 2D anode (upside-down) with the FR4 substrate
  on the bottom (orange), followed by two perpendicular sets of electrodes
  (yellow), which are separated from each other by a thin Kapton layer
  (orange strips underneath the thin copper electrodes). Bottom left:
  electric field simulation to show that equal amount of charges are
  collected by both electrode sets. Bottom right:
  picture of manufactured 2D anode of the 3~L
  prototype~\cite{Badertscher:2010zg}.
}
\label{fig:detector-anode}
\end{figure*}

As sketched on top of \Cref{fig:detector-anode}, the readout consists
of a FR4 substrate with 500~$\mu$m wide 
copper strips (\textit{covered strips}), spaced by 600~$\mu$m. On top of them, thin
Kapton strips provide the electrical separation from the second set of
120~$\mu$m wide readout strips (\textit{exposed strips}). The
intrinsic readout pitch of 600~$\mu$s for each 
view ensures that even point-like charge deposits in the detector induce
signals on both views, guaranteeing a correct reconstruction of the
$(x,y)$ coordinates. In order to collect in average equal amounts of
charges on both views (exposed and covered strips), the local electric
field was simulated with finite element methods\footnote{COMSOL
  Multiphysics software, http://www.comsol.com}. 
It can be seen in the cut through the anode shown on the bottom left
plot of \Cref{fig:detector-anode} that the exposed electrodes have a
focusing effect on the electric field lines. Simulating different geometries,
we have found that roughly half of the field lines end up on
each of the two views if the width of the exposed strip equals
120~$\mu$m, as shown in the figure. The picture on
the bottom right of \Cref{fig:detector-anode} shows the anode of the
3L~prototype. It can be seen that the desired readout pitch of 3~mm is
obtained by connecting five consecutive strips with the intrinsic pitch
of 600~$\mu$m together.
Published performance obtained with these kind of detectors can be found
in Refs.~\cite{Badertscher:2013wm,Badertscher:2012dq,Badertscher:2010zg, Badertscher:2009av}.
The 40 $\times$ 80 cm$^2$, the largest detector operated so far, has been operated during
more than 1~month in 2011 under controlled pressure of $\pm$1~mbar. The effective gain
in stable condition was 14.6 with a $S/N=30$ for m.i.p. The charge sharing between the two
views was within 8\%.

\begin{figure}
\centering
\includegraphics[width=0.9\textwidth]{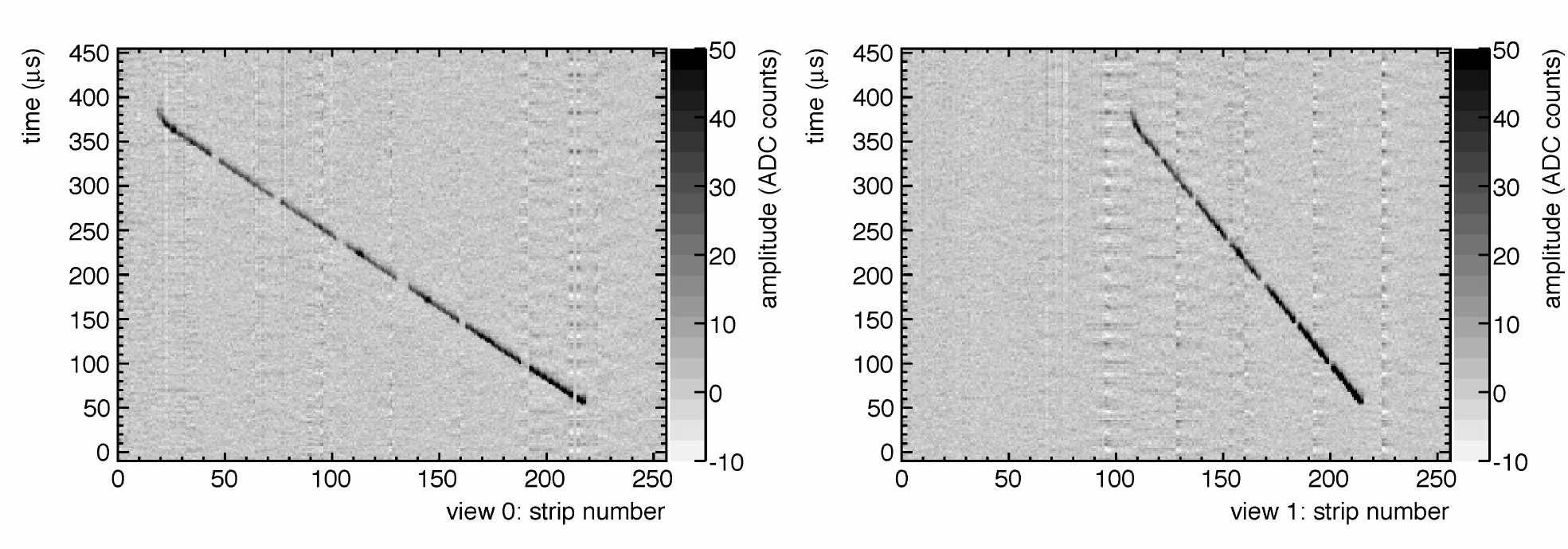}
\includegraphics[width=0.9\textwidth]{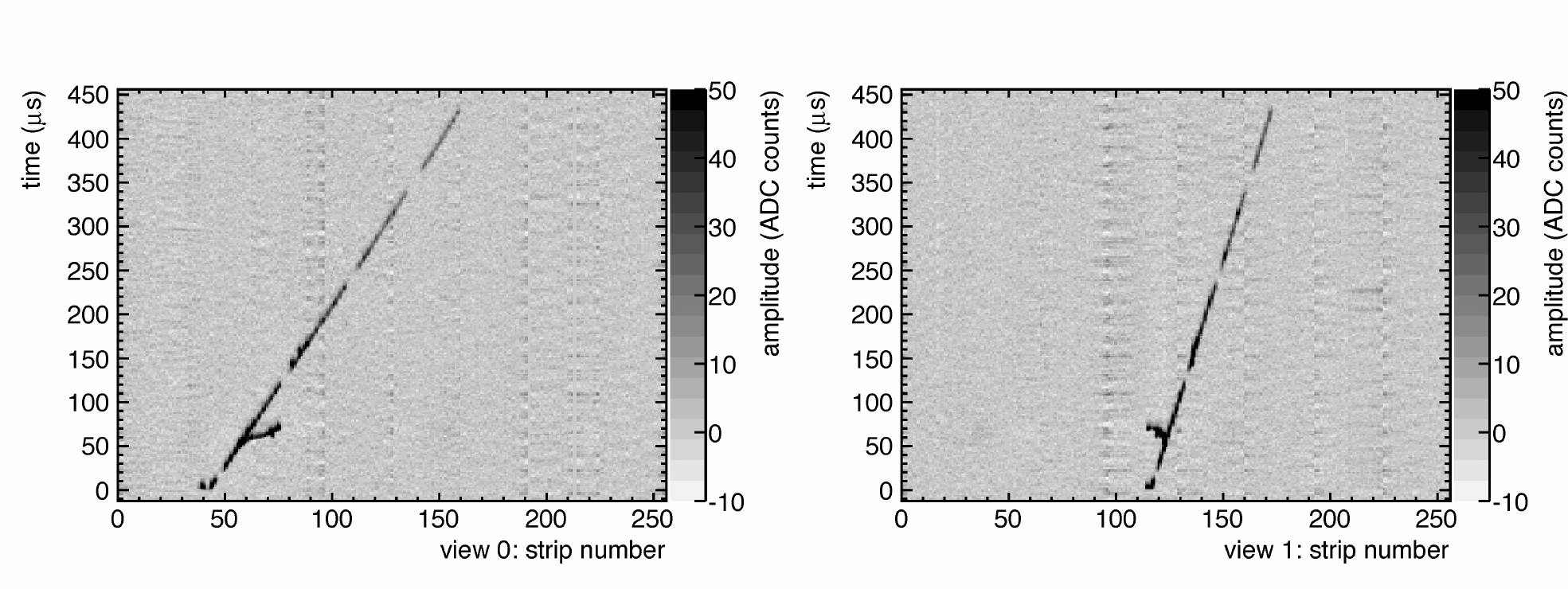}
\includegraphics[width=0.9\textwidth]{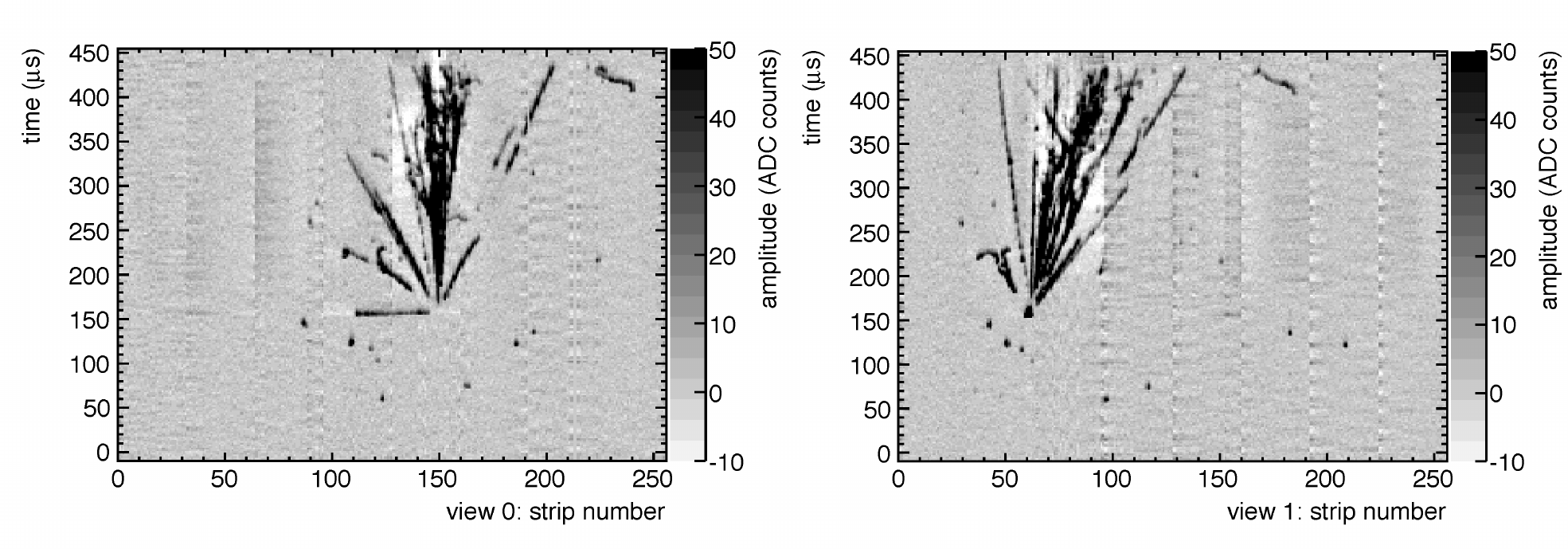}
\includegraphics[width=0.9\textwidth]{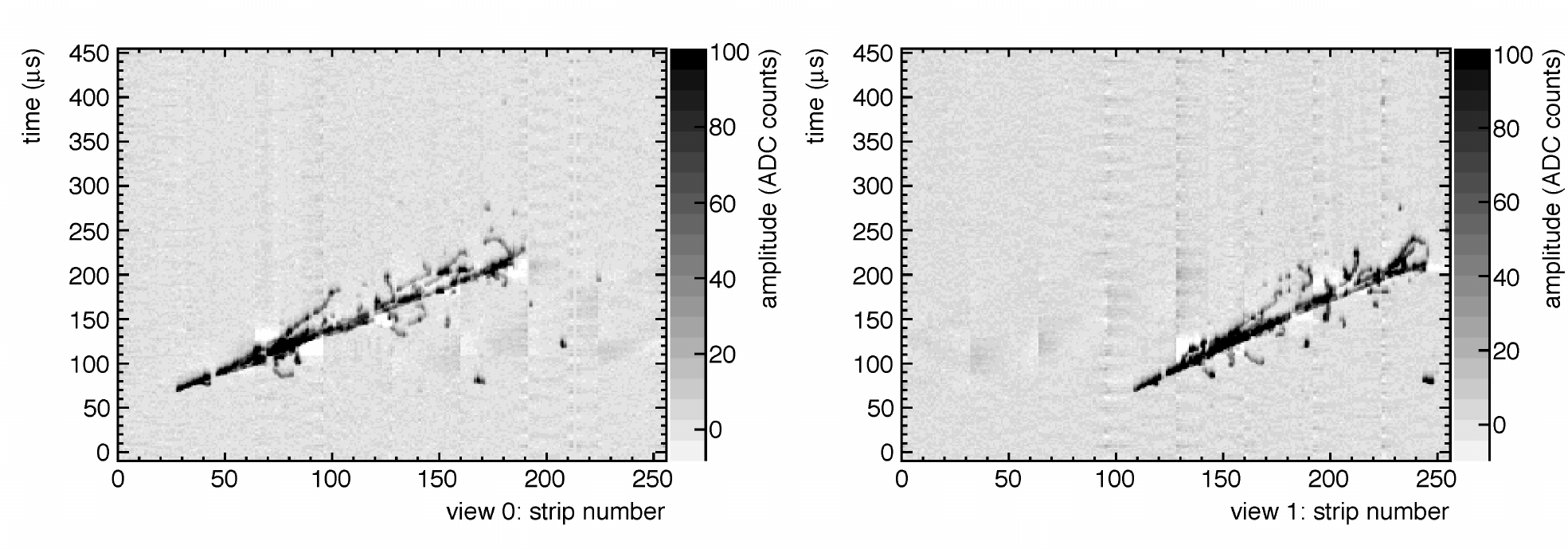}
\caption{Cosmic ray event displays (left: view 0, right view 1),
recorded in double phase conditions with a LEM
  field of 35~kV/cm. The four event displays show from top to bottom
  two muons, a hadronic and an electromagnetic shower candidate. [See
  text for more details.]} 
\label{figure:eventGallery}
\end{figure}

Figure~\ref{figure:eventGallery} shows four typical cosmic ray events, taken from
the highest gain run with an amplification field of 35~kV/cm. From top
to bottom, there are two cosmic muons crossing the detector, a
deep inelastic interaction and an electromagnetic shower candidate. 
These plots show the usual representation of events with the drift 
time as a function of the strip number for both views. The greyscale
represents the signal amplitude for each sample in a linear scale.
It can clearly be seen that both views show symmetric unipolar signals
which are clearly distinguishable from the noise. As a consequence of
the good signal to noise ratio and the spatial resolution, small
structures like knock-on electrons -- also called $\delta$-rays -- can be identified and
reconstructed in three dimensions as demonstrated below. Since cosmic muons
produce straight crossing tracks and the energy deposition is known to
be $\sim$2.1~MeV/cm, the events can be used to characterise the detector
in terms of free electron lifetime and amplification. An obvious limitation of
this charge readout was the inactive area introduced by the
segmentation of the LEM. The 1.6~mm wide gaps between the LEM
electrode segments appear as equally spaced missing charge, as seen in
the two muon events in Figure~\ref{figure:eventGallery}.

In view of the extrapolation of the double phase technique to the
large surfaces foreseen in the DLAr detector further studies were
conducted to simplify the design and reduce the electrical capacitance
of the readouts.  First, as explained in
Ref. \cite{Cantini:2013yba}, we were able to extract the drifting
electrons from the liquid to the gas by means of a single grid placed
just below the liquid surface. In this configuration the LEM is
positioned 50 mm above the liquid surface and the extraction field is
directly provided by the LEM-grid system over the 10 mm distance. This
configuration offers the advantage of 1. simplifying the overall
design, 2. reducing the absolute required high voltage and removing a
high voltage connection in the gas phase and 3. improving the
electrical transparency of the system by avoiding the collection of
the electrons on the top grid.  The typical high voltages applied
across each stage with the single grid configuration are summarised in
Table \ref{tab:efield_distance}. The DLAr detector will follow a
similar high voltage scheme.
\begin{table}[hbtp]
\begin{tabularx}{\textwidth}{@{\extracolsep{\fill}}lccc}
  \hline
  & Distance to the & Typical operating & Field to the stage \\
  & stage above (cm) & potential (kV) & above (kV/cm) \\
  \hline \normalsize
  2D anode & - & 0 & -- \\
  LEM (top electrode) & 0.2 & $-$1 & 5 \\
  LEM (bottom electrode) & 0.1 & $-$4.4 & 34 \\
  Extraction grid  & 1 & $-$6.4 & 2 \\
  \hline
  Cathode & 20.5 & $-$16.7 & 0.5 \\
  \hline
\end{tabularx}
\caption{LAr LEM charge readout system configurations for the
  10$\times$10 cm$^2$ detector equipped with a single extraction grid.}
\label{tab:efield_distance}
\end{table}
In the single grid configuration the anode is kept at ground while the
extraction grid is operated at a negative HV of $\sim$ -7 kV.  An
electric field of 34 kV/cm is applied across the LEM which produces an
effective gain of about 20.

\begin{figure}[h!]
  \centering
  \includegraphics[width=.8
  \textwidth,scale=1]{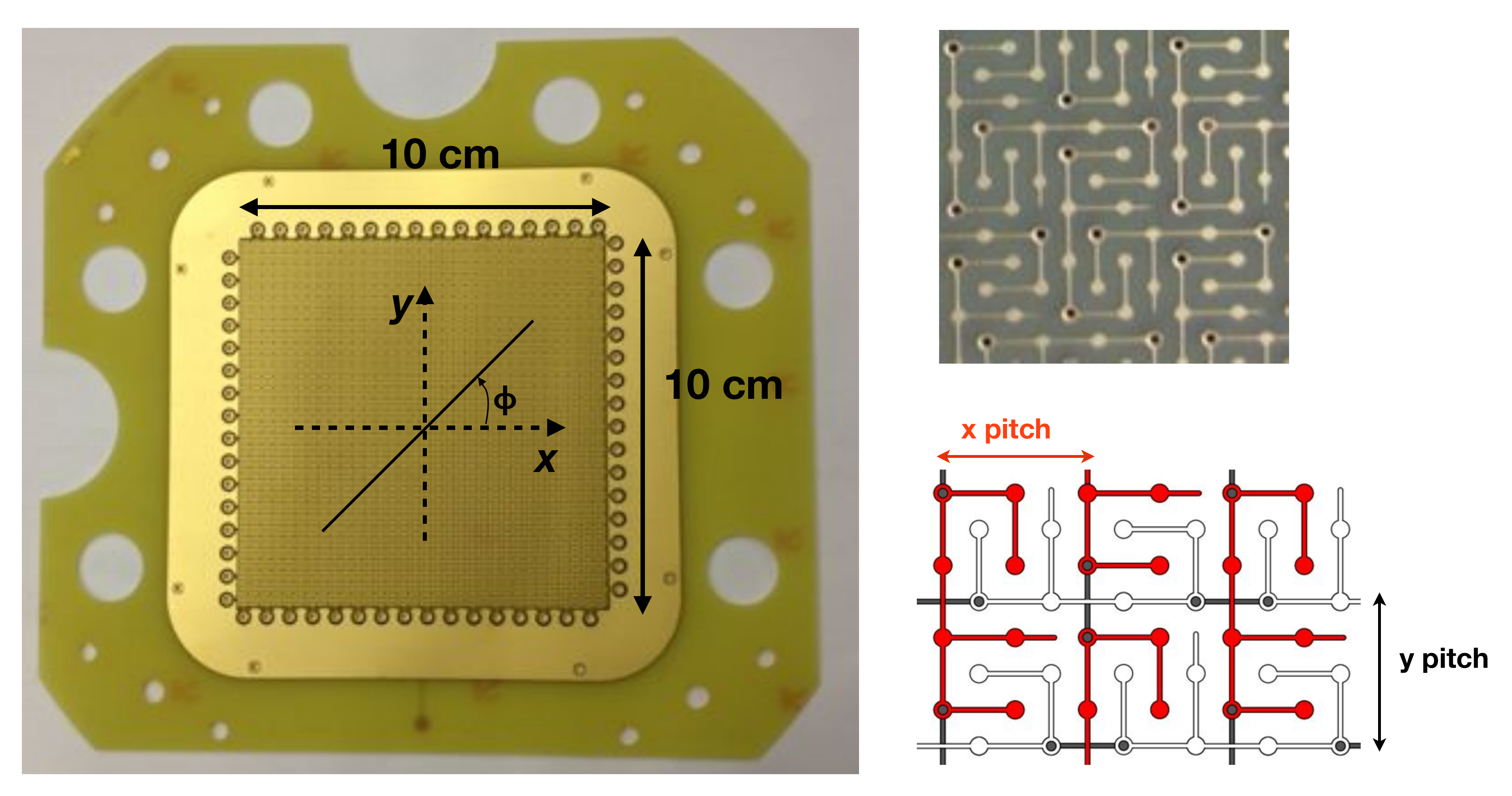}
  \caption{\textit{Left:} Picture of a 10$\times$10 cm$^2$ multilayer
    PCB anode tested in our chamber with dimensions and coordinate
    system super-imposed. \textit{Right:} close up picture of the
    anode showing the copper track pattern that allows a 2 view
    readout on the same circuit board. A schematics explaining the
    interconnections between both views is also shown. The 3 mm
    readout pitches are indicated by arrows. View 0 is filled in red
    and view 1 in white.}
    \label{fig:anodeD}
    \end{figure}

With the aim of easing the manufacturing process for large scale
production and reducing electrical capacitance of the readout we
designed and tested two dimensional projective readout anodes,
manufactured from a single multilayer printed circuit board (PCB). A
picture of the anode used in the 10$\times$10 cm$^2$ setup together
with a close up look and a schematic drawing is shown in
\Cref{fig:anodeD}. Its design parameters are listed
in Table~\ref{tab:anode_para}. 

\begin{figure}[h!]
  \centering
  \includegraphics[width=1.0\textwidth,height=0.22\textheight]{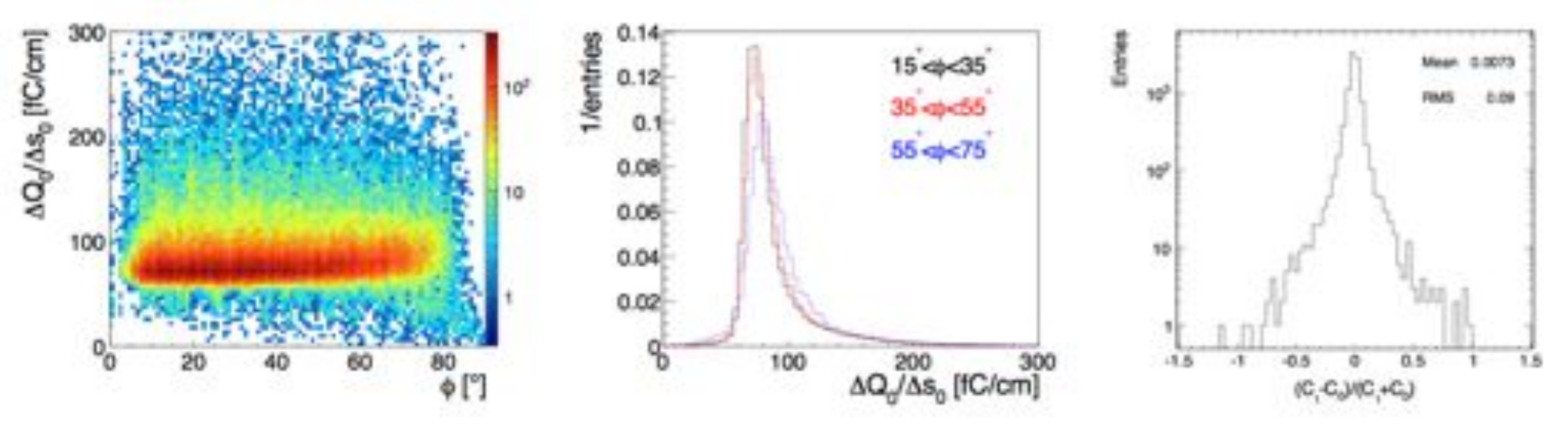}
  \caption{Charge deposition measured on view 0 ($\Delta Q_0/\Delta
    s_{0}$) as a function of the track angle $\phi$ (left) and
    projection of the $\Delta Q_0/\Delta s_{0}$ distribution in three
    $\phi$ intervals (middle). The right plot shows the distribution
    of the difference between the total charge collected on both views
    normalised to their sum.}
  \label{fig:anodeD_prop}
\end{figure}

\begin{table}[hbtp]
\begin{tabularx}{\textwidth}{@{\extracolsep{\fill}}lcccccc}
  \hline
  & Copper & Readout  & Track  & Track & Pad & Via \\
  & thickness ($\mu$m) & pitch (mm) & pitch (mm) & width (mm)&
  diameter (mm) & diameter (mm)\\
  \hline \normalsize
  & 35 & 3.125 & 1.5625 & 0.1 & 0.4 & 0.2\\
  \hline
\end{tabularx}
\caption{Characteristics of the 2D multilayer PCB anode.}
\label{tab:anode_para}
\end{table}
The 3 mm readout pitch of each view is segmented in two interconnected
gold platted copper tracks which are printed on a multilayer PCB.  As
seen in the schematic drawing, the basic pixel contains 8 pads equally
distributed between two views in order to guaranty a fully symmetric
charge sharing. The response of the anode was tested by fitting it to
the 10$\times$10 cm$^2$ prototype and exposing the chamber to cosmic
muons.  We tested that the charge is efficiently shared between both
views which is one of the important requirements of the
anode. Moreover the fine segmentation of the readout strips should
ensure that the response in terms of charge collected per unit length
is independent of the angle at which the track crosses the strips. All
these requirements are demonstrated in \Cref{fig:anodeD_prop}: the
left plot shows the collected charge per unit length on view 0
($\Delta Q_0/\Delta s_{0}$) as a function of the track azimuthal angle
$\phi$ (see \Cref{fig:anodeD} for a definition of $\phi$).
As can be seen from the projections (middle plot) the $\Delta
Q_0/\Delta s_{0}$ distributions are close to a Landau function as
expected from the fluctuations of the collected charge per unit
length. The width and mean value of those distributions are also
similar for all angular intervalls illustrating a uniform response.
To illustrate the charge sharing between both views the right plot
shows the distribution of the difference between the total charge
collected on both views normalised to their sum. The distribution has
an RMS of 9 \% and is centered around 0.7 \% indicating that the anode
is indeed perfectly $x-y$ symmetric and that the charge is equally
shared between both views.  
Thanks to the spacing between the pads and
tracks of both views, the anode has a measured capacitance per unit
length of about 150 pF/m between neighbouring channels.

Compared with the Kapton type anodes used previously in the
$10\times10 $cm$^2$ and $40\times80$ cm$^2$ which had a capacitance
per unit length of $\sim$ 600 pF/m, this anode offers a significant
noise reduction and satisfies our needs in terms of signal-to-noise
ratio for the 3 m long readouts foreseen in the DLAr detector.  Other
anodes with alternate geometries have also been tested. Their designs
and the results are presented in Ref.~\cite{Cantini:2013yba}. The goal
is always to reduce the capacitance per unit length without degrading
the imaging capabilities of the TPC. From that point of view, the one
presented here offered the best performance and was hence chosen as
the final readout option.

\begin{figure}
\centering
\includegraphics[width=.8\textwidth,scale=1]{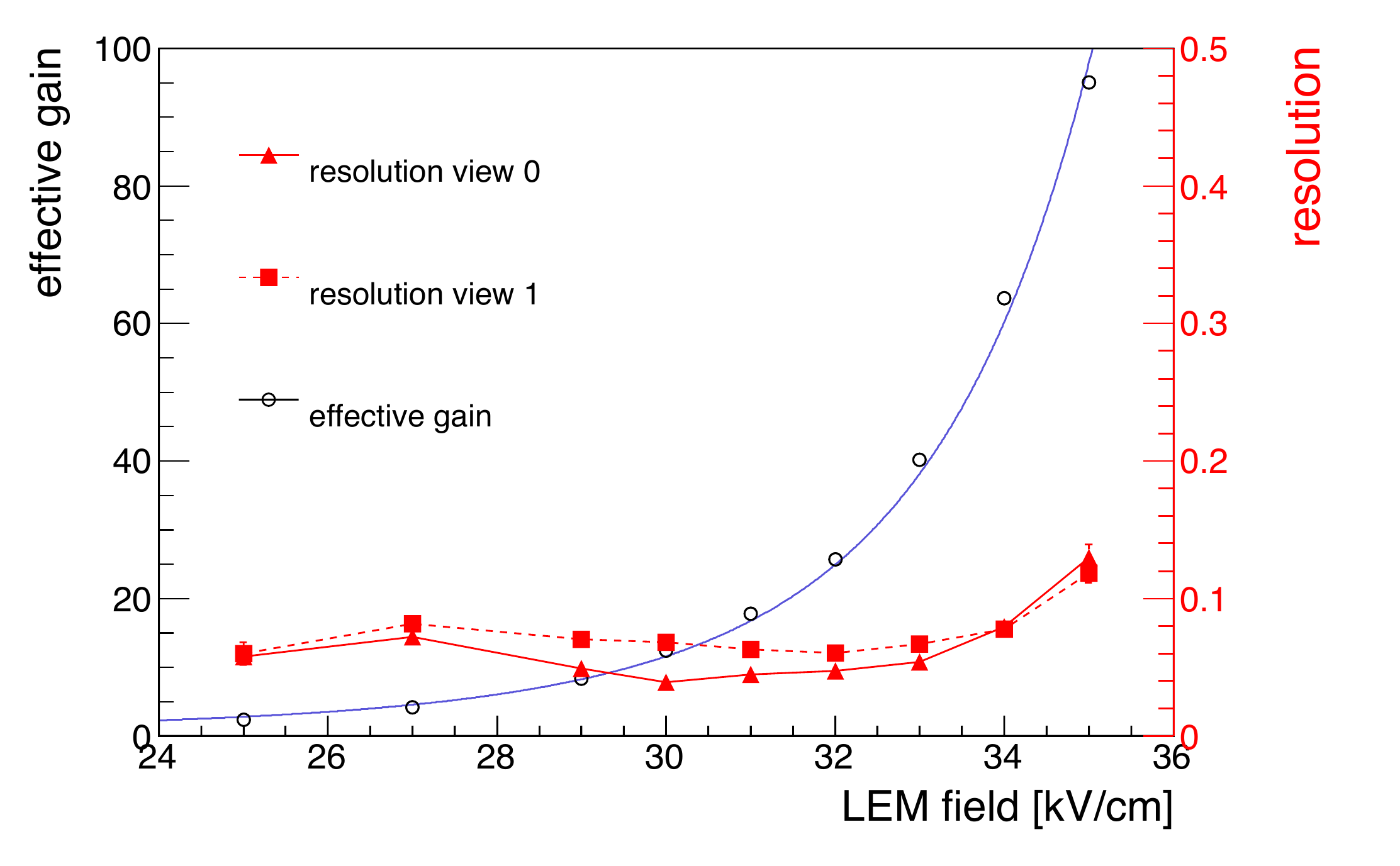}
\caption{Effective gain and resolution of the charge deposition
  measurements on both view as a function of the LEM electric field.}
\label{figure:LEMFieldScan}
\end{figure}

In order to define the optimal high voltage settings across each stage
we systematically tested the response of the chamber in terms of
effective gain and resolution on the deposited charge for different
electric field settings.  For that matter, we define the effective
gain of the device as the ratio of the measured charge (corrected for
the drifting electron lifetime) collected on both views to the
initially produced charge of the ionising particles. Under the
assumption that only MIP events are present in our selected sample,
the average charge deposition along a track, predicted by the
Bethe-Bloch formula and accounting for electron-ion
recombination~\cite{Amoruso:2004dy} is $\langle \Delta Q/\Delta
s\rangle_{MIP} =10$ fC/cm.  The effective gain is hence defined as
\begin{equation}
  G_{eff}=\frac{\langle \dqdxz \rangle +\langle \dqdxo\rangle}{\langle
    \Delta Q/\Delta s\rangle_{MIP}}\label{eq:gain}
\end{equation} 
where the indices correspond to view 0 and 1.  As example we show in
\Cref{figure:LEMFieldScan} the response of the chamber as a function
of the electric field applied across the LEM. On the same figure we
show the effective gain and the resolution on the charge deposition
measurement. The resolution is obtained by fitting the distributions
with a Gaussian convoluted Landau function and is defined as
$\sigma_{gauss}/\langle\dqdxi\rangle$. As can be seen the measured
effective gain exhibits the expected exponential dependence with the
LEM field (see Refs. \cite{Thesis_FilippoResnati,Cantini:2013yba} for
details) and the resolution on both views are similar and stay
constant at around 8\%.
We achieved an effective gain of 90 by setting the amplification field
to 35 kV/cm. The signal-to-noise ratio is then defined as the mean
amplitude of the waveforms produced by the cosmic tracks divided by
the average value of the noise RMS. Given our noise value of about 4-5
adc counts RMS, a gain of 90 corresponds to a signal-to-noise ratio
$S/N$ of about 400 for minimum ionising particles, or $S/N \sim 10$
for an energy deposition of 15 keV on a single readout channel.  

\begin{figure}[htb]
     \centering
     \includegraphics[scale=.35]{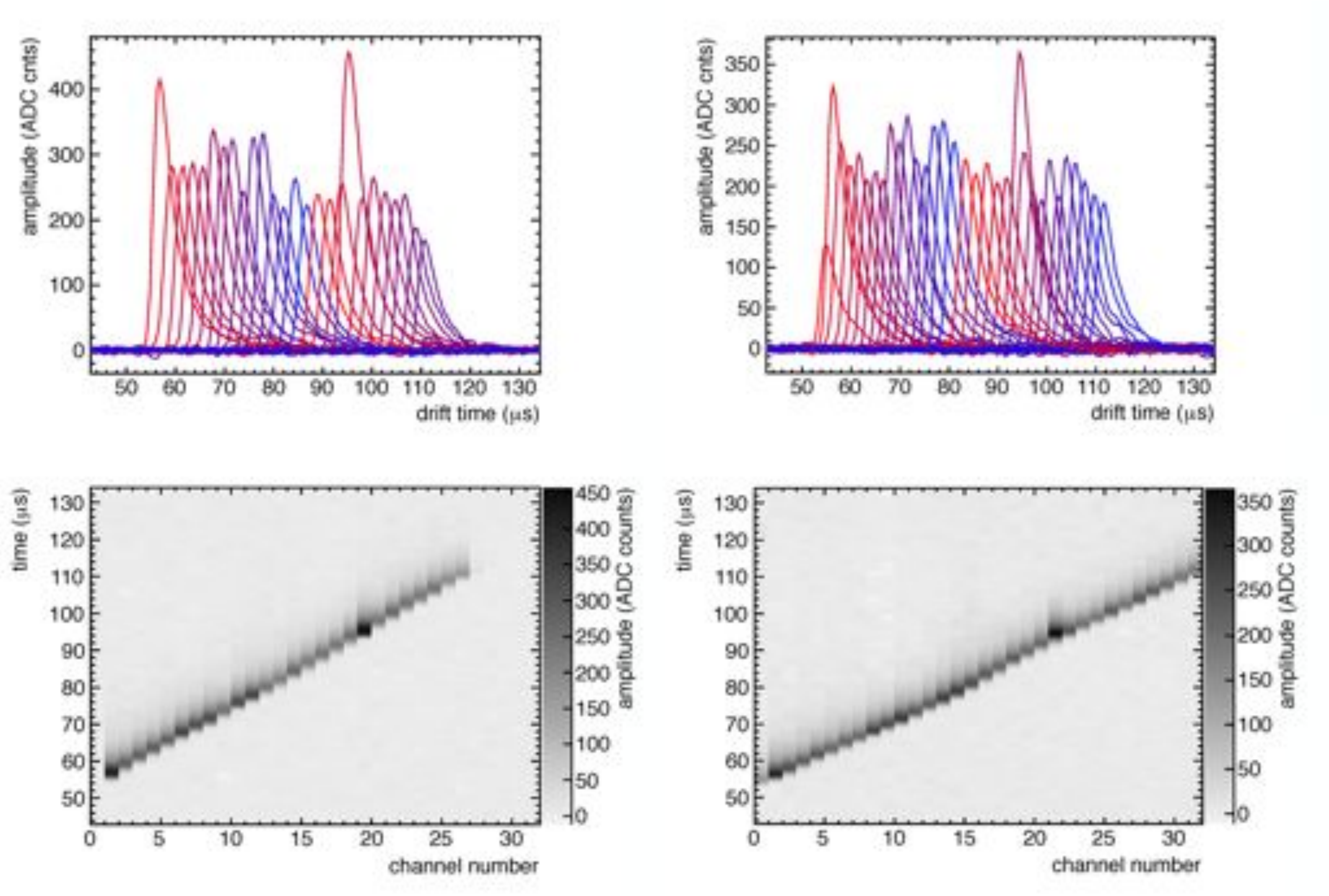}  
     \caption{Event display of a cosmic track with a gain of $\sim$ 15
       obtained in the 3 liter chamber. (top): the raw waveform
       showing the amplitude of the signals on both views. (bottom):
       drift time versus channel number of the reconstructed hits.}
     \label{figure:event_gain_15}
   \end{figure}

As discussed in detail in \cite{Cantini:2013yba} the effective gain
stabilises after an initial decrease of about 1.5 days due to the
charging up of the dielectric of the LEM. Once a state of equilibrium
is reached, the gain is however stable for long periods.  We showed
that with a LEM field of 33kV/cm the detector was in stable operation
for about a month at the gain 15. This means that we can run the
chamber in a stable operation mode with a $S/N$ of at least 60 for
minimum ionising particles. \Cref{figure:event_gain_15} shows a
typical cosmic event acquired with a gain of 15. Further studies will
be performed to check wether the chamber can be continuously operated
at larger LEM fields and if higher gains in stable conditions can be
reached.

   Based on those studies in Table \ref{tab:efield_distance} we
   summarise the typical high voltages applied across each stage of
   the detector equipped with the new anode and operated in the single
   grid configuration.  An electric field of 33 kV/cm is applied
   across the LEM which produces a stable effective gain of about
   15. The DLAr detector will follow a similar high voltage scheme.

\subsubsection{The CRP for the \six}
\label{sec:crpforsix}

\begin{figure}[h!]
      \begin{center}
        \includegraphics[width=.8\textwidth]{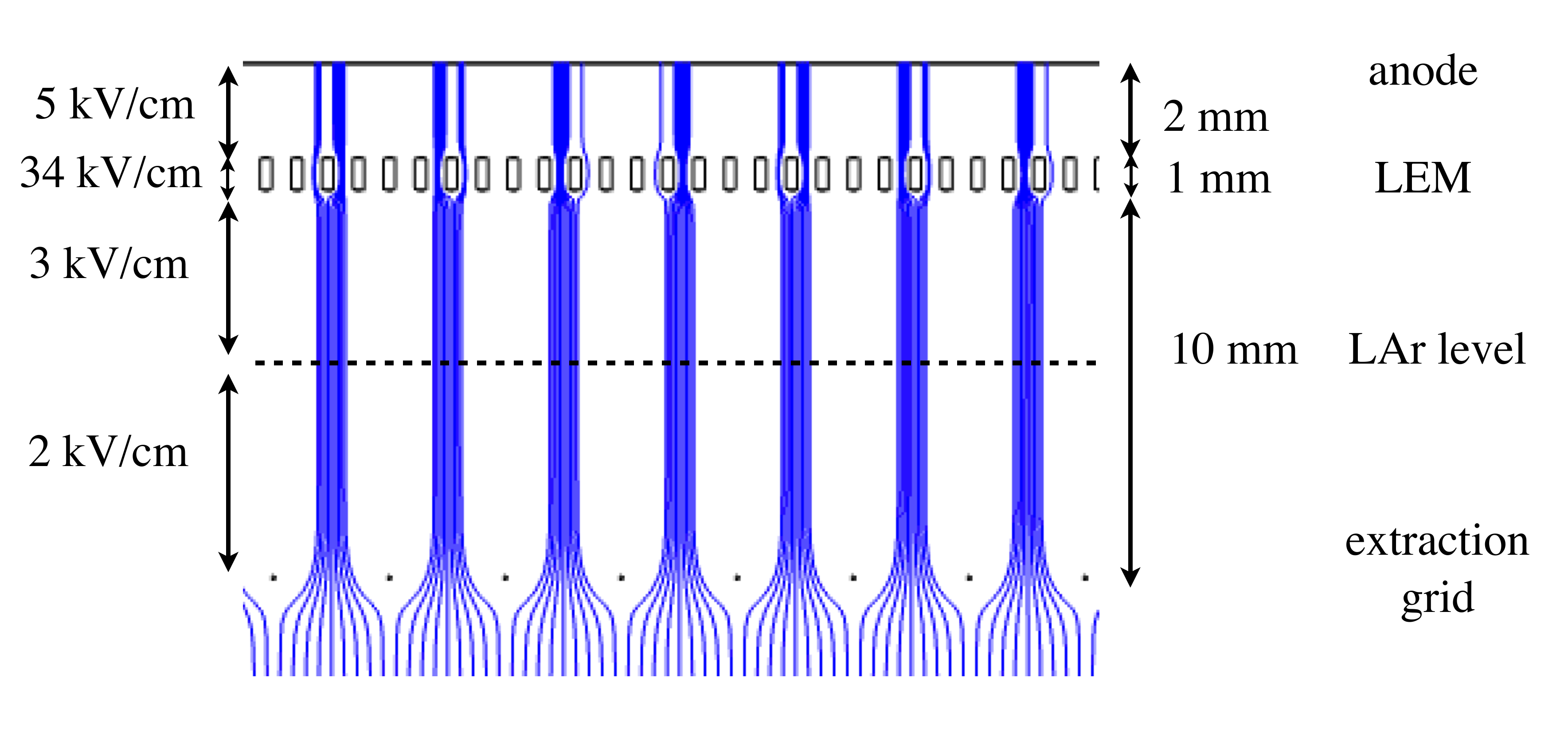}
        \caption{Simulated geometry of the readout showing the
          distances and nominal electric fields. The field lines
          followed by the drifting charges (not taking into account
          diffusion) are also shown.}
        \label{fig:CRP_field_lines}
      \end{center}
    \end{figure}

    The Charge Readout Plane (CRP) module of the \six is a direct
    extrapolation of the successful 10~$\times$~10~cm$^2$ prototype
    with the single extraction grid configuration to an active area
    approximately as large as 600 $\times$ 600 cm$^2$.  In the LAr LEM
    TPC configuration, the CRP consists of the low capacitance 2D
    anode, LEM and extraction grid assembled as a multi-layered
    ``sandwich'' unit with precisely defined inter-stage distances and
    inter-alignment. The distances, tolerances and nominal electric
    fields between each stage are shown in \Cref{fig:CRP_field_lines}
    and summarised in Table \ref{tab:z_tolerance}.  In this
    configuration we successfully operated the 10~$\times$~10~cm$^2$
    prototype at a constant gain of about 15 for around a month with
    very few discharges~\cite{Cantini:2013yba}.

    \begin{table}[htb]
      \renewcommand{\arraystretch}{1.2}
      \begin{center}
        \begin{tabular}{lcccccc}
          \hline 
          &\phantom{ab}& [mm] &\phantom{ab}& electric field [kV/cm]&\phantom{ab}& tolerance [mm]\\
          \hline
          anode-LEM& & 2 && 5&& 0.1 \\
          LEM & & 1 && 34 && 0.01 \\
          LEM-grid& & 10 && 2 && 1 \\
          liquid level& & 5 (from grid)&&-&& 1 \\
          \hline 
        \end{tabular}
      \end{center}
      \caption{\label{tab:z_tolerance} distances and tolerances
        between each stage. The tolerance on the LEM refers to the
        thickness of the LEM insulating material (FR4).}
    \end{table}
   
\begin{figure}[htb]
      \begin{center}
        \includegraphics[width=\textwidth]{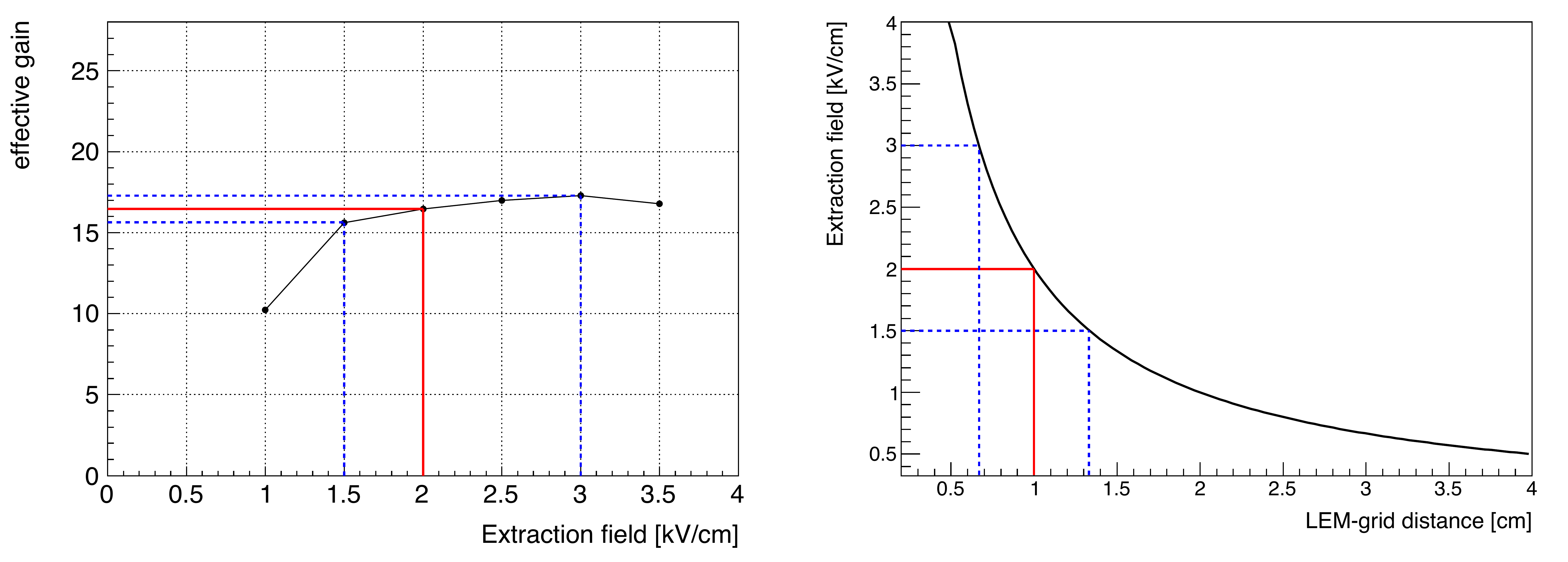}
        \caption{\textit{Left:} measured gain in the
          10~$\times$~10~cm$^2$ prototype as a function of the
          extraction (LEM-grid) field. The red line shows the nominal
          electric field and the dotted blue lines the electric field
          corresponding to a $\pm$5\% variation of the gain
          . \textit{Right:} calculated extraction field as a function
          of the LEM-grid distance. The dotted blue lines indicate the
          variations around the nominal 1 cm distance that keep the
          gain within $\pm$5\%.}
        \label{fig:gain_extr}
      \end{center}
    \end{figure}

    The tolerances are calculated to keep the gain stable within 5\%
    over the entire active area of the readout. We used the
    10~$\times$~10~cm$^2$ prototype to scan the fields applied between
    the different stages and we checked the effect on the gain. As
    example we show in \Cref{fig:gain_extr} the gain for different
    values of the extraction field. The extraction field is defined as
    the electric field across the LEM-grid stage calculated in the
    liquid. As can be seen the gain is rather stable around the chosen
    operating value of 2kV/cm in the liquid and variations of the
    extraction field do not have a dramatic impact on the gain. The
    fact that the gain reaches a stable level indicates that at about
    2kV/cm all the electrons are efficiently extracted from the liquid
    to the gas phase. The allowed variation of the extraction field
    define a region of tolerance for the LEM-grid distance as shown in
    \Cref{fig:gain_extr}-right. The quoted tolerance of 1 mm on the
    grid-LEM distance lies well within the allowed variation of the
    extraction field specified on the figure.

\begin{figure}[h!]
      \begin{center}
        \includegraphics[width=\textwidth]{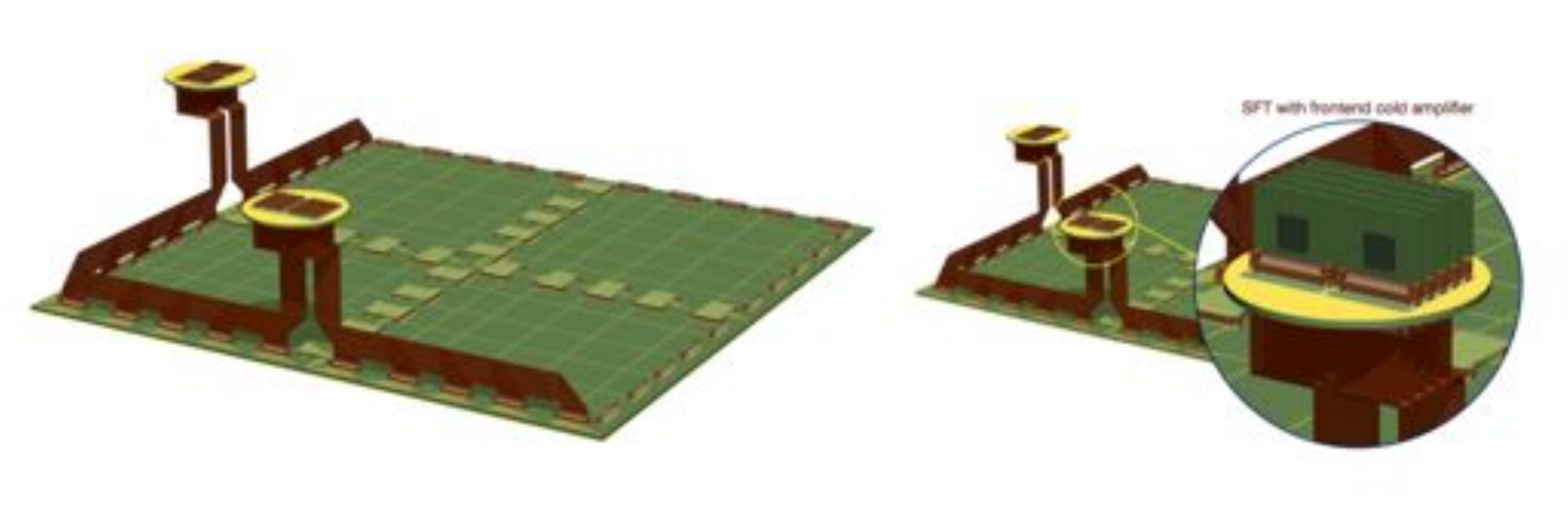}
        \caption{Design concept of the 1$\times$1 m$^2$ CRP: (left)
          overall vue l and zoom on a signal feedthrough (SFT) with
          cold frontend preamplifier; }
        \label{fig:1m2_crp_drawing}
      \end{center}
    \end{figure}

    The design concept of the CRP is illustrated in
    \Cref{fig:1m2_crp_drawing}. The final design of the CRP for the
    \six will be guided by many studies and prototyping. One important
    aspect is to ensure that the various mechanical components have an
    adequate rigidity in order to respect the tolerances reported in
    Table \ref{tab:z_tolerance}. For that matter the mechanical
    structure is being tested on a $1\times1$m$^2$ mockup. In a next
    step we will construct a larger $1\times3$m$^2$ module which will
    be fitted on a LAr LEM TPC of a $3\times3\times1$m$^3$ active
    volume.

\begin{figure}[htb]
      \begin{center}
        \includegraphics[width=\textwidth]{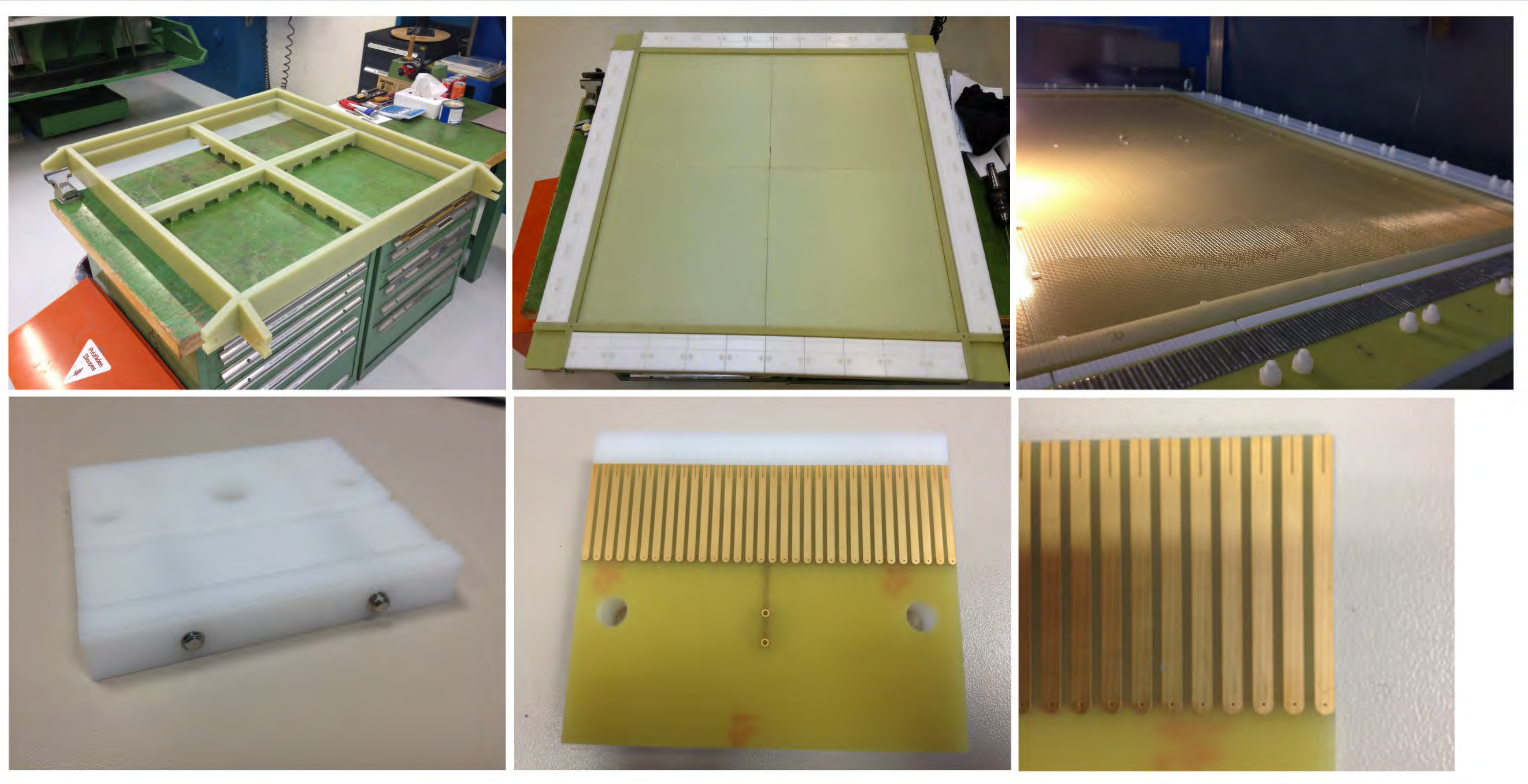}
        \caption{Pictures of the 1 m$^2$ CRP mockup. The top pictures
          show the frame structure made out of G10 which holds the
          four anode-LEM modules and the CRP assembled with a fake LEM
          (1 mm sheet of FR4). The wire tensioning pads made are also
          positioned. The CRP is then equipped with the four
          $50\times50$cm$^2$ anode modules and the tensioned wires of
          the extraction grid. The bottom picture shows the details of
          one wire tensioning pad: the wire holder made out of POM
          (Polyoxymethylene) is equipped with two stainless steel
          screws that provide the tension to the wires by pushing the
          holder against the FR4 frame. A close up picture of the wire
          soldering PCB shows the 200 $\mu$m grooves that allow a
          precise positioning of the wires.}
        \label{fig:CRP_mockup}
      \end{center}
    \end{figure}

    \vspace{.3cm} \textit{1$\times$1 m$^2$ CRP }
    \vspace{.1cm}\\
    In order to test the structural rigidity of the CRP and optimise
    the overall design a mechanical mockup of the CRP module with an
    active area of 1$\times$1 m$^2$ was constructed as shown in
    \Cref{fig:CRP_mockup}.  The mockup consists of a rigid frame
    reinforced by precisely machined external and internal FR4
    bars. In addition to hosting the anode and LEM modules, the frame
    serves as external suspension. The readout is composed of four
    independent modules of $50\times50$~cm$^2$.  Each module is
    composed of a $50\times50$~cm$^2$ anode and a $50\times50$~cm$^2$
    LEM panel. The anode is designed to have 5 connectors at the edges
    which provide the function of either bridging the anodes or
    sending out the signal. The set of connectors in between two
    anodes are used to bridge them together in order to form a fully
    active 1~m$^2$ readout while the others are used to send the
    signal to the front-end electronics. 
    
      \begin{figure}[htb]
      \begin{center}
        \includegraphics[width=\textwidth]{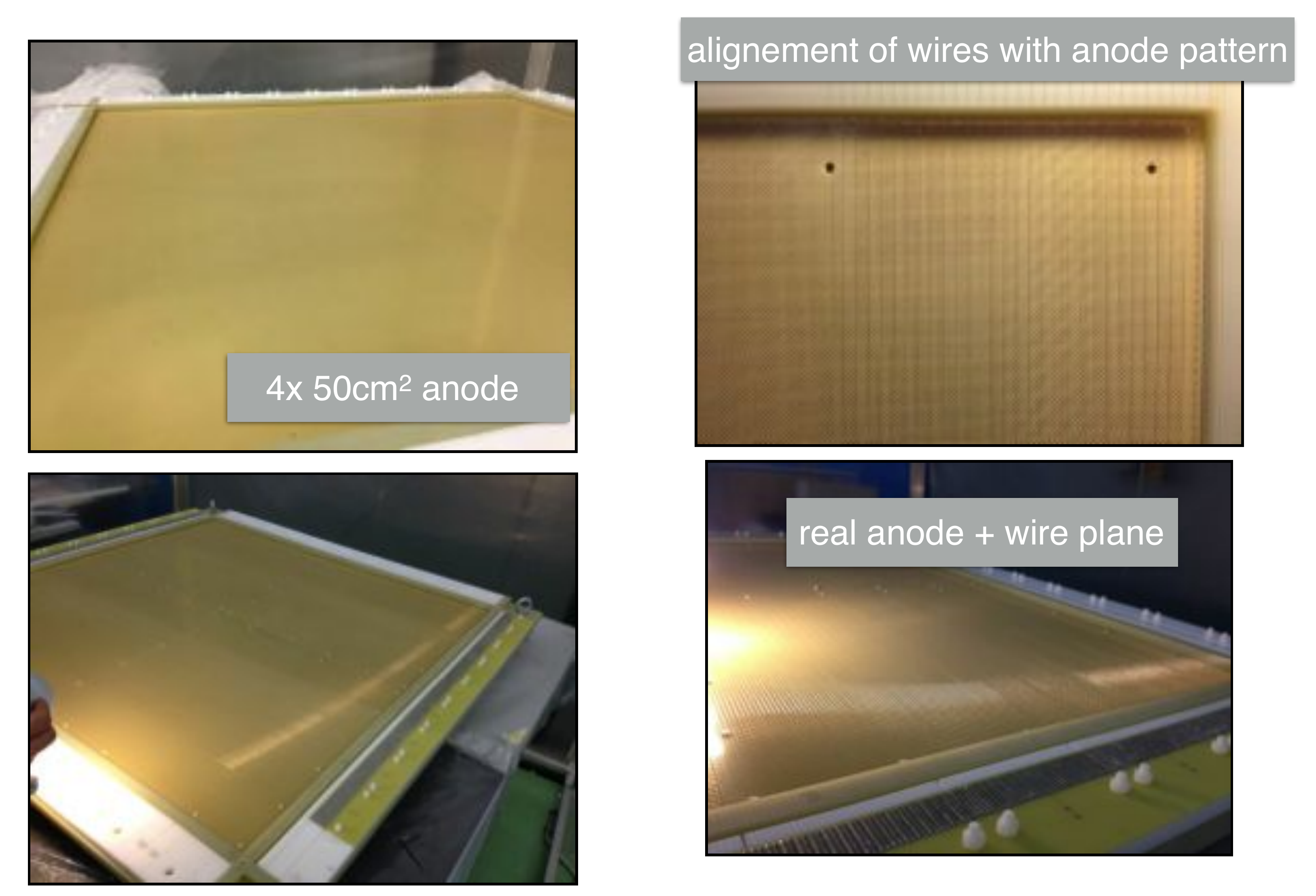}
        \caption{Pictures of the 1 m$^2$ CRP mockup: anode details.}
         \label{fig:CRP_mockup_anode}
      \end{center}
    \end{figure}
      \begin{figure}[htb]
      \begin{center}
        \includegraphics[width=\textwidth]{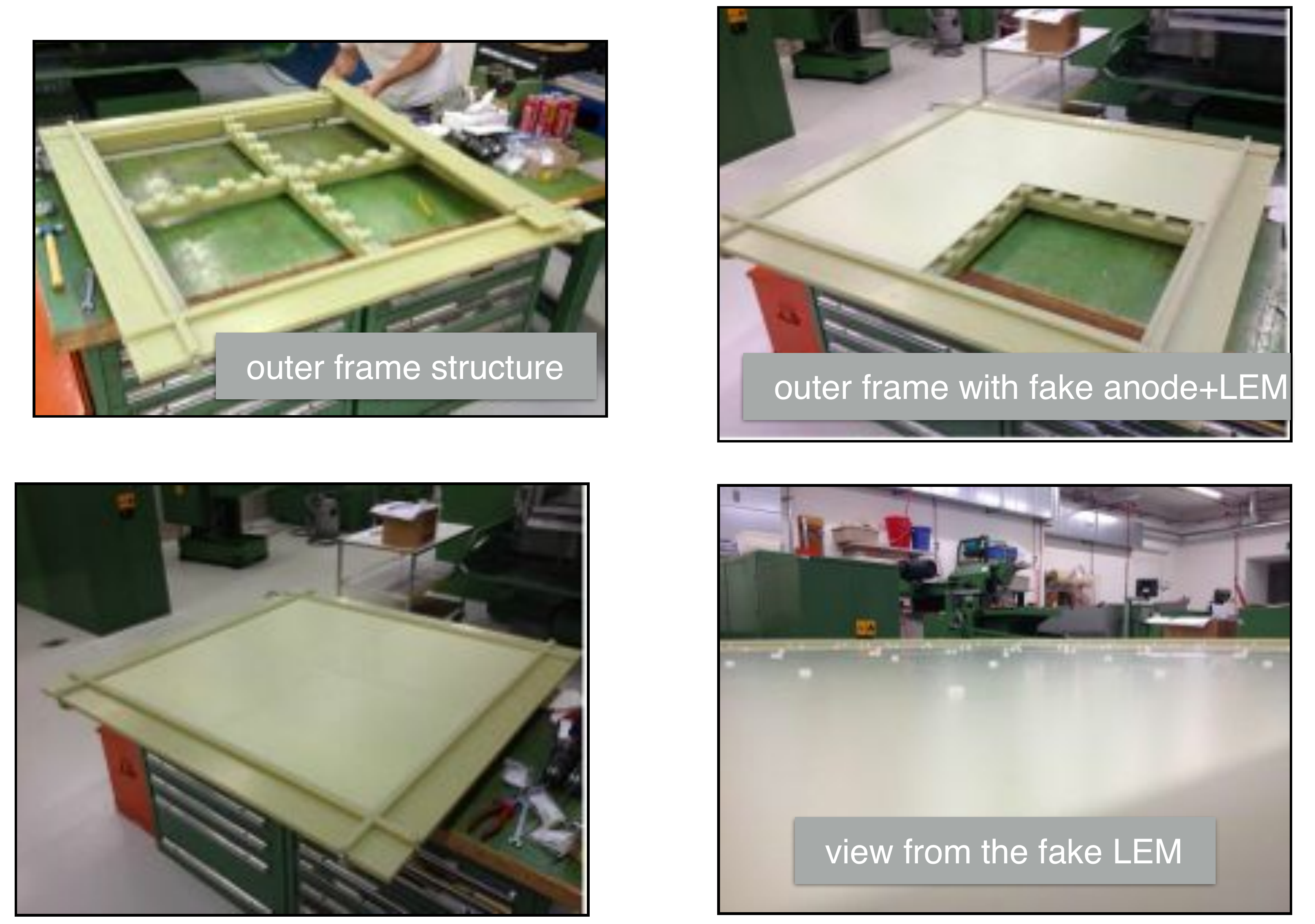}
        \caption{Pictures of the 1 m$^2$ CRP mockup: frame details.}
         \label{fig:CRP_mockup_frame}
      \end{center}
    \end{figure}
      \begin{figure}[htb]
      \begin{center}
        \includegraphics[width=\textwidth]{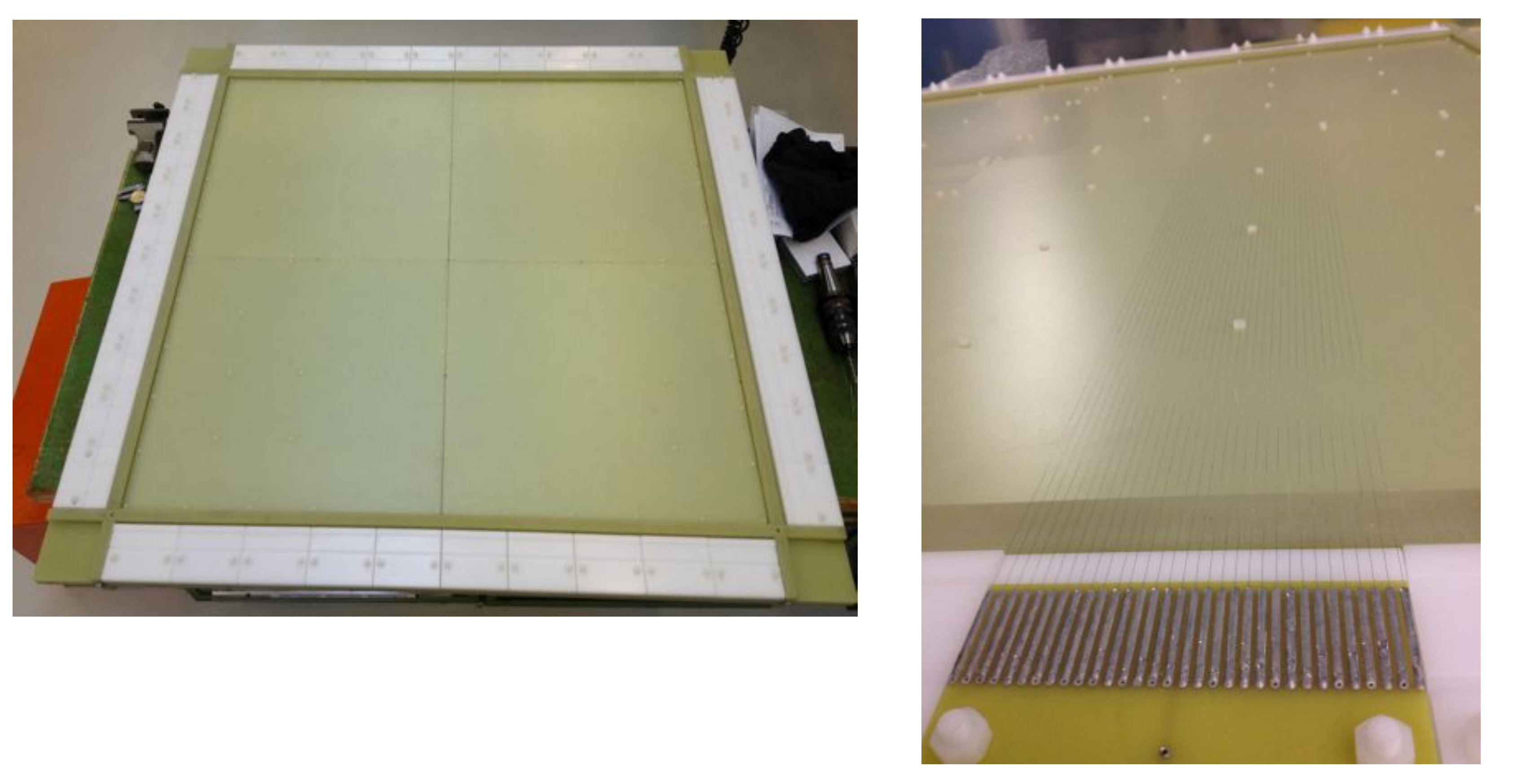}
        \caption{Pictures of the 1 m$^2$ CRP mockup: extraction grid details.}
         \label{fig:CRP_mockup_extractiongrid}
      \end{center}
    \end{figure}

    The extraction grid consists of 100 $\mu$m diameter stainless
    steel wires tensed in both $x$ and $y$ directions. They are
    soldered by group of 32 on independent wire tensioning pads spaced
    on the side of the frame as shown in \Cref{fig:CRP_mockup}. Each
    wire tensioning pad consists of a PCB precisely fixed on a
    mechanical wire holder machined from POM
    (Polyoxymethylene)\footnote{although the wire holders are made out
      of POM for the mechanical mockup, FR4 is contemplated for the
      final design}. The PCB hosts the high voltage connection and has
    32 soldering pads with 200 $\mu$m grooves to precisely position
    the wires. During the wire soldering process each wire is
    tensioned by 150 g lead weights and precisely positioned inside
    the grooves. With this method the precision on the wire pitch,
    measured under the microscope, was better than 50 microns. The PCB
    is then fixed on the wire-holder and the whole system can provide
    precise tension to the group of 32 wires by pushing the holder
    against the CRP FR4 frame with two stainless steel screws (see
    \Cref{fig:CRP_mockup}).  

\begin{figure}[h!]
      \begin{center}
        \includegraphics[width=\textwidth]{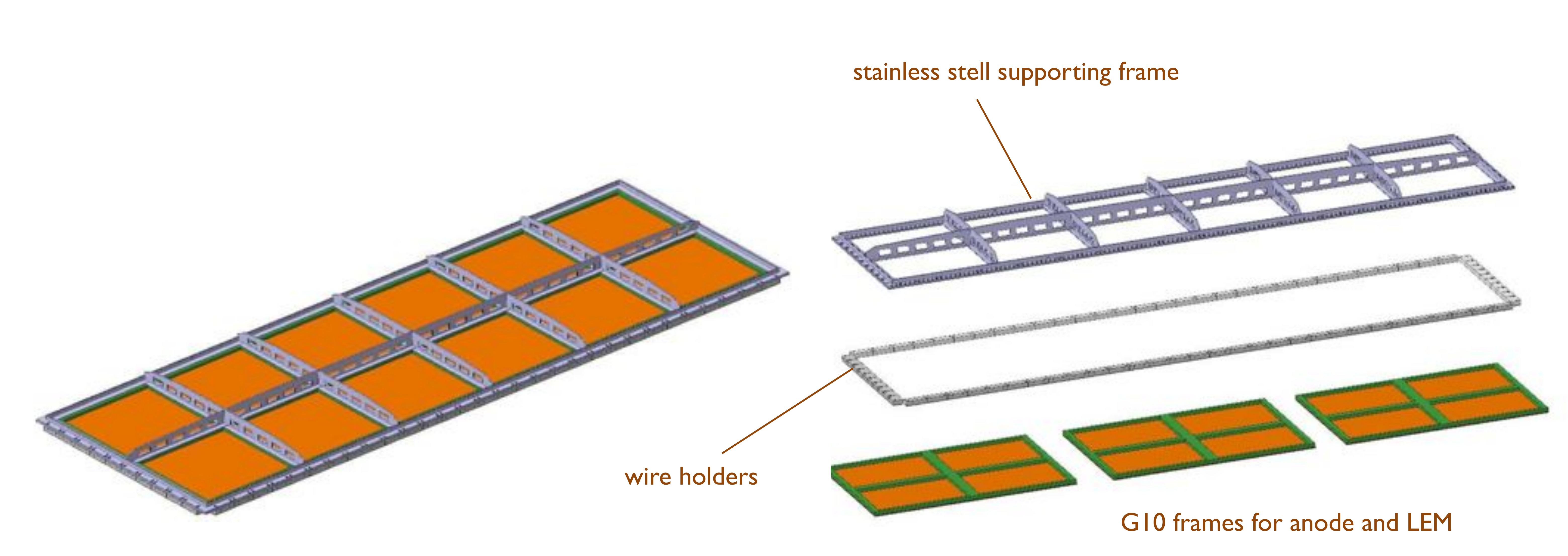}
        \caption{3D drawing of the 1$\times$3 m$^2$ CRP.}
        \label{fig:13_CRP}
      \end{center}
    \end{figure}

    \vspace{.3cm} \textit{1$\times$3 m$^2$ CRP }
    \vspace{.1cm}\\
    A CRP of $1\times3$m$^2$ will soon be assembled and tested on a
    LAr LEM TPC of $3\times1\times1$m$^3$ active volume. The
    mechanical design is shown in Fig~\ref{fig:13_CRP}. It follows the
    basic concept of the 1$\times$1 m$^2$ mockup. The support
    structure is made from stainless steel and houses three G10 frames
    to which the $50\times50$~cm$^2$ anode and LEM modules will be
    attached creating a fully $1\times3$m$^2$ active area. The frame
    also holds on its side the wire holders for the extraction grid.
    
\begin{figure}[htb]
\begin{center}
\includegraphics[width=.8\textwidth]{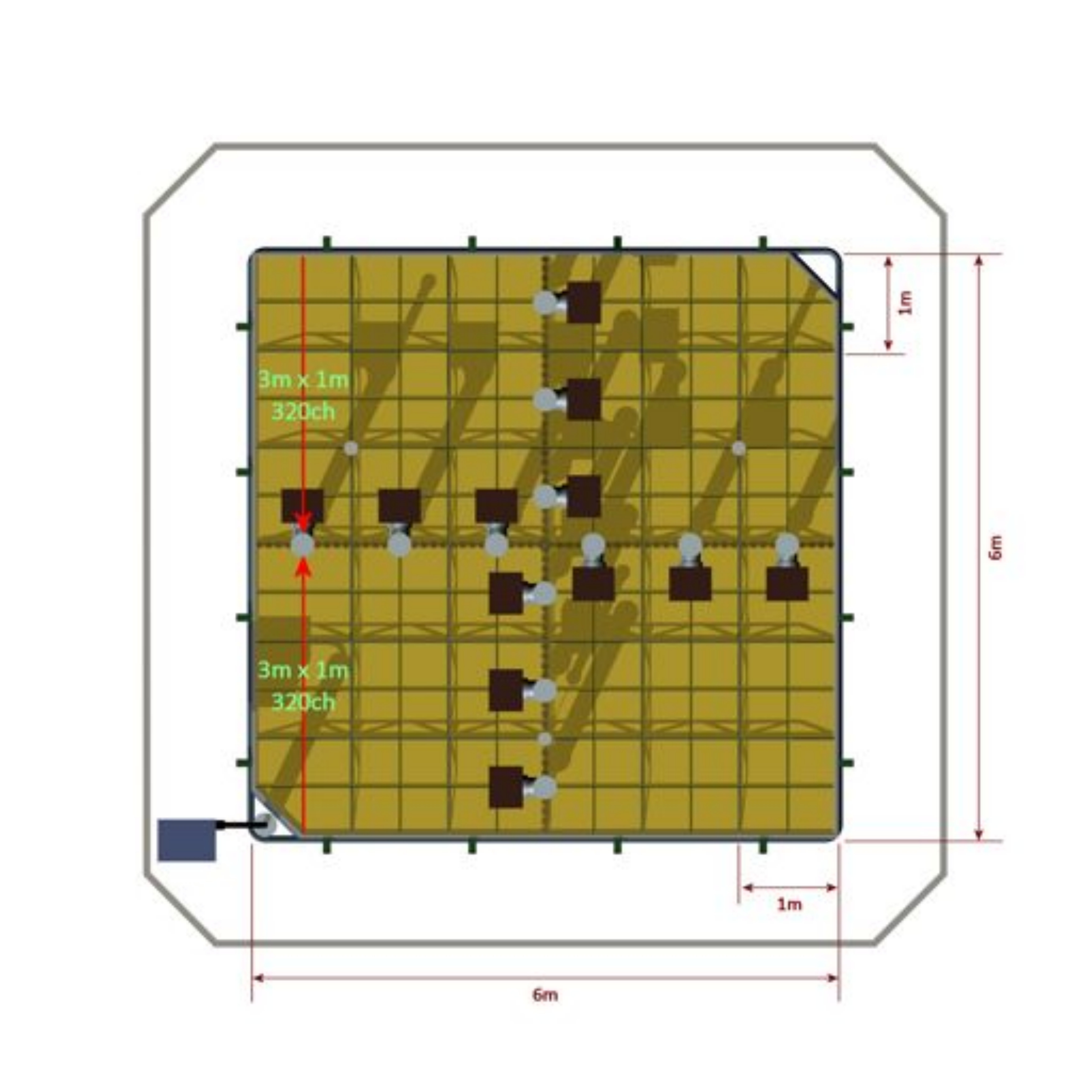}
\vspace{-1cm}
\caption{Top view of the \six CRP.}
\label{fig:CRP_666}
\end{center}
\end{figure}

    \vspace{.3cm} \textit{6$\times$6 m$^2$ CRP }
    \vspace{.1cm}\\
    The 6$\times$6$\times$6 m$^3$ CRP design follows the basic concept
    of the 1$\times$1 m$^2$ mockup and $1\times3$m$^2$ module as shown
    in \Cref{fig:CRP_666}. The frame contains 4 external reinforcement
    bars of 6 meters and 22 internal ones.  The external bars define
    the 6$\times$6 m$^2$ total area and the internal bars divide the
    total area into 144 identical 0.5$\times$0.5 m$^2$ sub-areas.  As
    for the 1$\times$1 m$^2$ CRP, each of the 144 sub-areas has one
    anode and one LEM panel. The extraction grid will be made by
    tensioning and fixing 6 m stainless steel wires to the outer frame
    in both directions. The same method as the 1$\times$1 m$^2$ mockup
    will be used to fix and tension the wires. In total the 7680
    channels are sent to the cold front-end preamplifiers inserted in
    the 12 signal feedthrough chimneys. The signal feedthroughs
    chimneys are positioned to make 4 groups of 3 meter long readout
    strips as explained in \Cref{fig:CRP_666}.  In this configuration,
    adjacent 0.5$\times$0.5 m$^2$ anode panels are bridged by groups
    of 36 to form the maximal readout lengths of 3 meters. The
    suspension feedthrough chimneys provide the function of hanging
    the whole structure from the tank deck and precisely aligning the
    CRP to the liquid argon level. A detailed description of the anode
    deck suspension system is given in Section
    \ref{sec:top_anode_deck}. Table \ref{tab:crp_para} summarises the
    basic components of the \six CRP.

\begin{table}[hbt]
\begin{tabularx}{\textwidth}{@{\extracolsep{\fill}}lccccc}
  \hline
 Component & Anode & LEM  & Signal  & Suspension  \\
           & panel & panel & feed-through & feed-through \\
  \hline \normalsize
  & 144 & 144 & 12 & 3 & \\
  \hline
\end{tabularx}
\caption{Components of the 6$\times$6 m$^2$ CRP.}
\label{tab:crp_para}
\end{table}


\subsubsection{The MICROMEGAS option}
\label{sec:micromegas}

Micro-Pattern Gaseous Detectors (MPGDs) such as GEM/THGEM/LEM 
are well established and high performance devices widely used in particle physics experiments. 
The MICRO-MEsh GASeous Structure or MICROMEGAS is a MPGD invented in 1996 which amplifies electrons
in a typical 100~$\rm{\mu m} $ gap defined by a metallic micromesh placed on top of an anode 
Printed Circuit Board (PCB)\cite{Giomataris199629}. In 2004, a new method to build this detector was
introduced and called the bulk-MICROMEGAS\cite{Giomataris2006405}: a woven micromesh is embedded on
top of the segmented anode plane of the MICROMEGAS by use of standard photolithographic techniques. This 
technology was chosen to instrument the 3 TPCs of the ND280 near detector of the T2K experiment for its 
performance in terms of gas gain uniformity, energy resolution and space point resolution, as well as for its 
capability to efficiently pave large readout surfaces with minimized dead zones. Eighty six 128 $\rm{\mu m}$ 
gap  34$\times$36~$\rm{cm^2} $ bulk-MICROMEGAS modules were produced in 18 months. Eighty of them, for an 
equivalent total surface of 9 $\rm{m^2}$, passed the quality and performance tests. The dispersion of gas 
gain and energy resolution at 5.9~keV  within the whole surface of each module were respectively  2.8\% 
and 6\% r.m.s. The dispersion of mean gain and mean 5.9~keV energy resolution over the eighty modules 
were found to be respectively 8\% and 3\% r.m.s\cite{Delbart2010105}. These facts illustrate the 
maturity of the bulk-MICROMEGAS technology for a high quality mass production at a moderate cost.

One advantage of the MICROMEGAS technique is thus the possibility to easily produce large detectors, with a quality 
suitable for the instrumentation of a large area HEP detector. The industrial production of large 
area detectors is being adopted by the ATLAS Collaboration who will construct about 1200~m$^2$ 
of resistive MICROMEGAS detectors for the upgrade of the first station of the ATLAS muon end-cap 
system (New Small Wheel)
\footnote{\protect\url{http://cds.cern.ch/record/1552862/files/ATLAS-TDR-020.pdf}}.

\begin{figure}[hbt]
\begin{center}
\includegraphics[width=7.5cm]{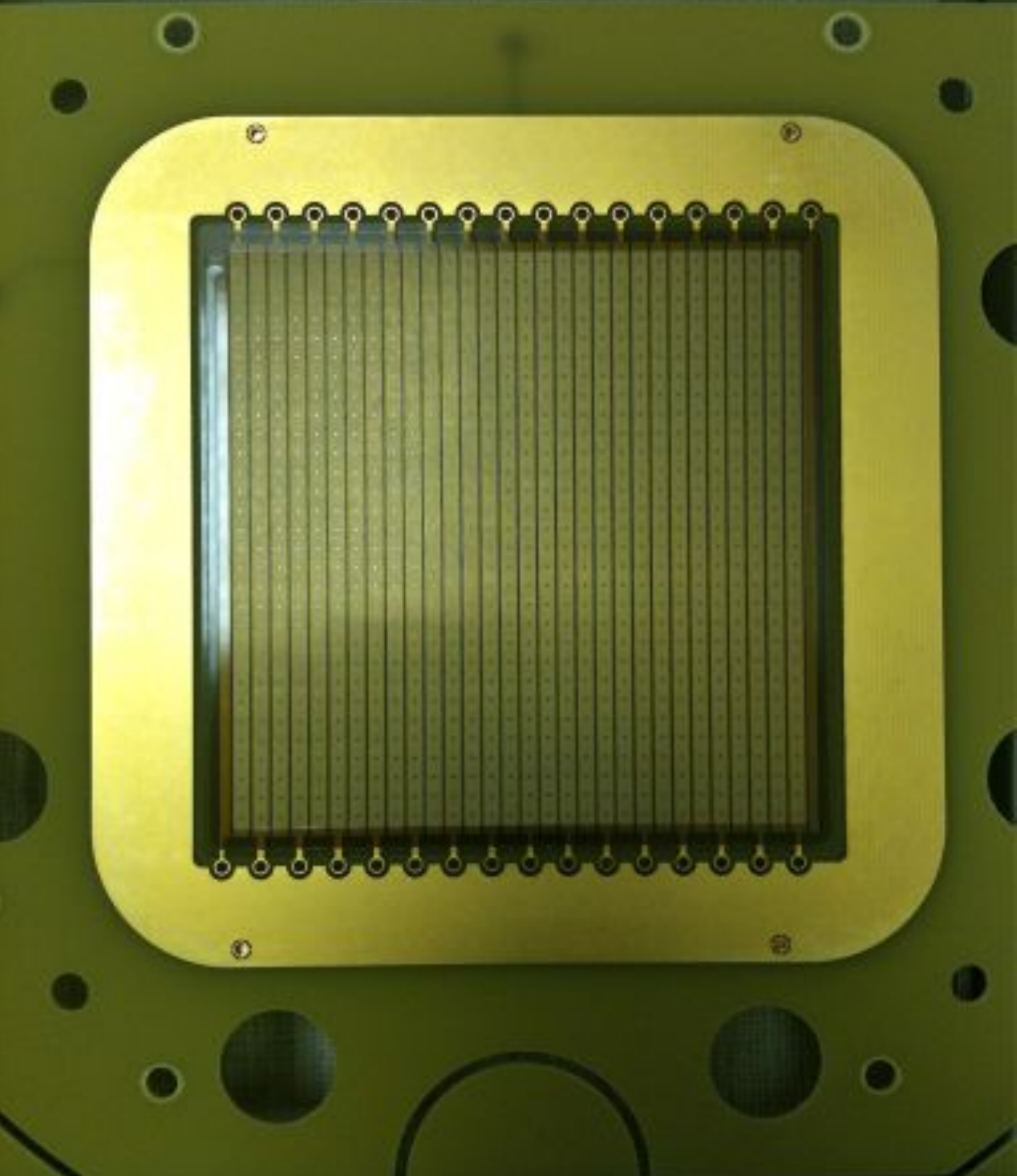}
\caption{The 10$\times$10~cm$^2$ bulk-MICROMEGAS prototype tested in a Liquid Argon TPC. The anode has 32 strips with a 3~mm pitch.}	
\label{fig:micromegas} 
\end{center}
\end{figure}

In 2010, the T2K IRFU group started an R{\&}D project to investigate the feasibility of using 
bulk-MICROMEGAS to instrument a double phase LAr. Gas 
amplification process was first tested tested in argon at room temperature up to a pressure 
of several bars. Then, a 10$\times$10~cm$^2$ bulk-MICROMEGAS with an 
amplification gap of 100~$\rm{\mu m}$ was tested in the 
3L ETHZ Liquid Argon TPC at CERN\cite{Delbart:2011zz} (see \Cref{fig:micromegas}). The anode 
was made of 32 1D strips with a 3~mm pitch. The device operated successfully in 
this cryogenic environment for several days. In particular the following points were demonstrated: 
\begin{itemize}
\item compatibility with the high purity environment; 
\item operation in a cryogenic environment;
\item successful charge readout, with observation of tracks from cosmic rays;
\item charge amplification, up to approximately 5.
\end{itemize}

\begin{figure}[hbt]
\begin{center}
\includegraphics[width=0.7\textwidth]{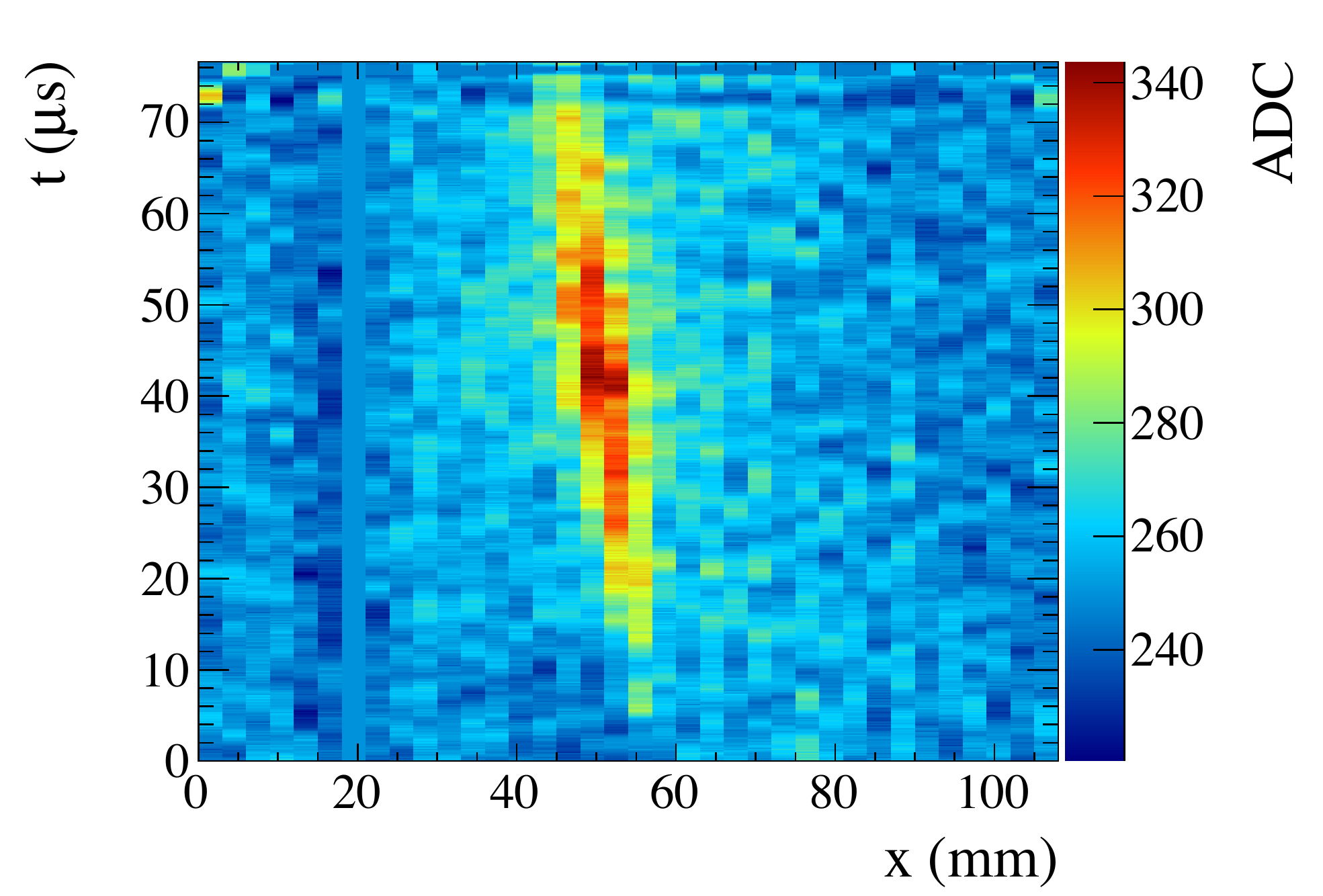}
\caption{Event display of a cosmic ray track observed with a 115 $\rm{\mu m}$ gap bulk-MICROMEGAS detector 
in the Liverpool LAr setup. The horizontal axis gives the position of the collected charge on the strip while the vertical axis indicates the corresponding electron drift time in liquid argon.}	
\label{fig:micromegas_track} 
\end{center}
\end{figure}

In 2013, three bulk-MICROMEGAS prototypes with amplification gaps of 115, 128 and 192 $\rm{\mu m}$ were built 
and tested in a 40 liter LAr cryostat at the University of Liverpool. Each detector had an active 
area of 10.8$\times$10.8~cm$^2$ and the anode subdivided into 36 readout strips with a 3~mm pitch. Tests 
performed at room temperature in pure argon provided a good understanding of the gain dependance 
as a function of the gap size. Using a $^{241}$Am source, gain values were measured 
for applied voltages on the micromesh up to values near the detector breakdown operation 
point. The 115 $\rm{\mu m}$ gap prototype 
was also successfully tested in the Liverpool double phase LAr TPC\cite{Mavrokoridis:2014gka}. Clean cosmic tracks 
could be observed (see \Cref{fig:micromegas_track})
and a gain of about 4, comparable to the one measured during the 2010 test at CERN, was obtained. 

Further R{\&}D activity will be pursued in 2014 in order to optimize the amplification gain of such MICROMEGAS devices.
The main directions of development are: the use of thinner gaps (e.g. 64~$\rm{\mu m}$), tests of resistive 
bulk-MICROMEGAS detectors and the use of a modified anode-micromesh structure to better absorb UV photons produced
in the amplification region.

\subsection{Drift cage}
\graphicspath{{./Section-DetLArComponents/figs/}}

The function of the drift cage is to create a uniform electric field in the inner part of the detector, with a strength between 500 and 1000 V/cm, resulting in a
drift velocity of the quasi-free electrons ranging between 1.6 and 2.0 mm/s.
This mechanical structure supports also the negative HV cathode, towards the bottom of the tank, 1m above the surface of the PMT. 

A full engineering design for the drift cage and the cathode 
has been produced in the context of the LAGUNA-LBNO design study, for detectors of fiducial mass of 20 and 50 kton. For the demonstrator, most of the technical solutions introduced there will be retained.  

The drift cage is composed of 60 equally spaced electrodes, in the form of stainless-steel tubes with a diameter of 69~mm and a pitch of 100~mm. Each tube will be electrically connected to its neighbours through resistors to provide a graded electric field. 
The structure will be supported by 16 FR4 vertical pillars. These pillars might be replaced by insulated links, cut out of FR4 plates. These links could provide a lighter structure and an easier assembly inside the tank. 

The cathode is made of a strong frame with a mesh of tubes. The diameter of the tubes is 60~mm and
their spacing is 500~mm. The mesh is filled with meshed grids with a wire diameter of 5~mm and
pitch 50~mm. The transparency to light is about 80\%, such that scintillation light produced
within the liquid argon fiducial volume can be detected by the PMTs located at the bottom
of the vessel.

\begin{figure}[tbh]
\begin{center}
\includegraphics[width=0.95\textwidth]{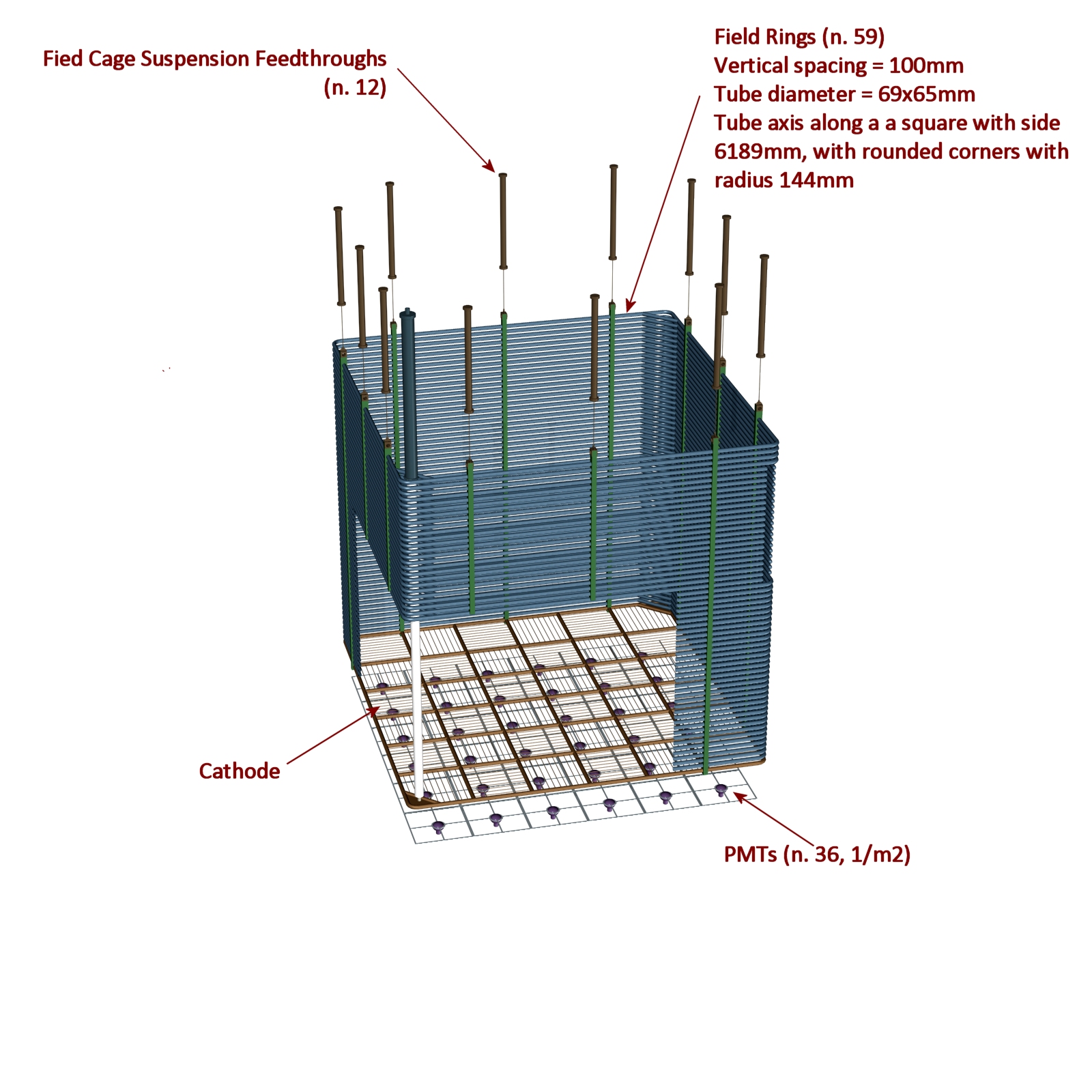}
\vspace{-3cm}
\caption{Conceptual design of the field cage. The high-voltage feed-through is located
in the left corner of the cage.}
\label{fig:fieldcage}
\end{center}
\end{figure}

The whole structure will be hanging from the tank deck with support rods housed in special bellows. As an alternative design option, it could also 
be supported on the floor by feet adapted to the corrugation of the membrane tank.
The drift cage is illustrated in \Cref{fig:fieldcage}. 
The HV feedthrough is placed in a corner of the field-cage. The top of the field cage
is closed by the CRP anode deck. It should be noted that the anode deck is actually
independently hung from the top via the supporting feed-through. The field-cage and
the anode are therefore mechanically independent components, and the anode deck
level is adjusted to the level of the liquid-gas argon interface.

\begin{figure}[tbh]
\begin{center}
\fbox{\includegraphics[width=0.65\textwidth]{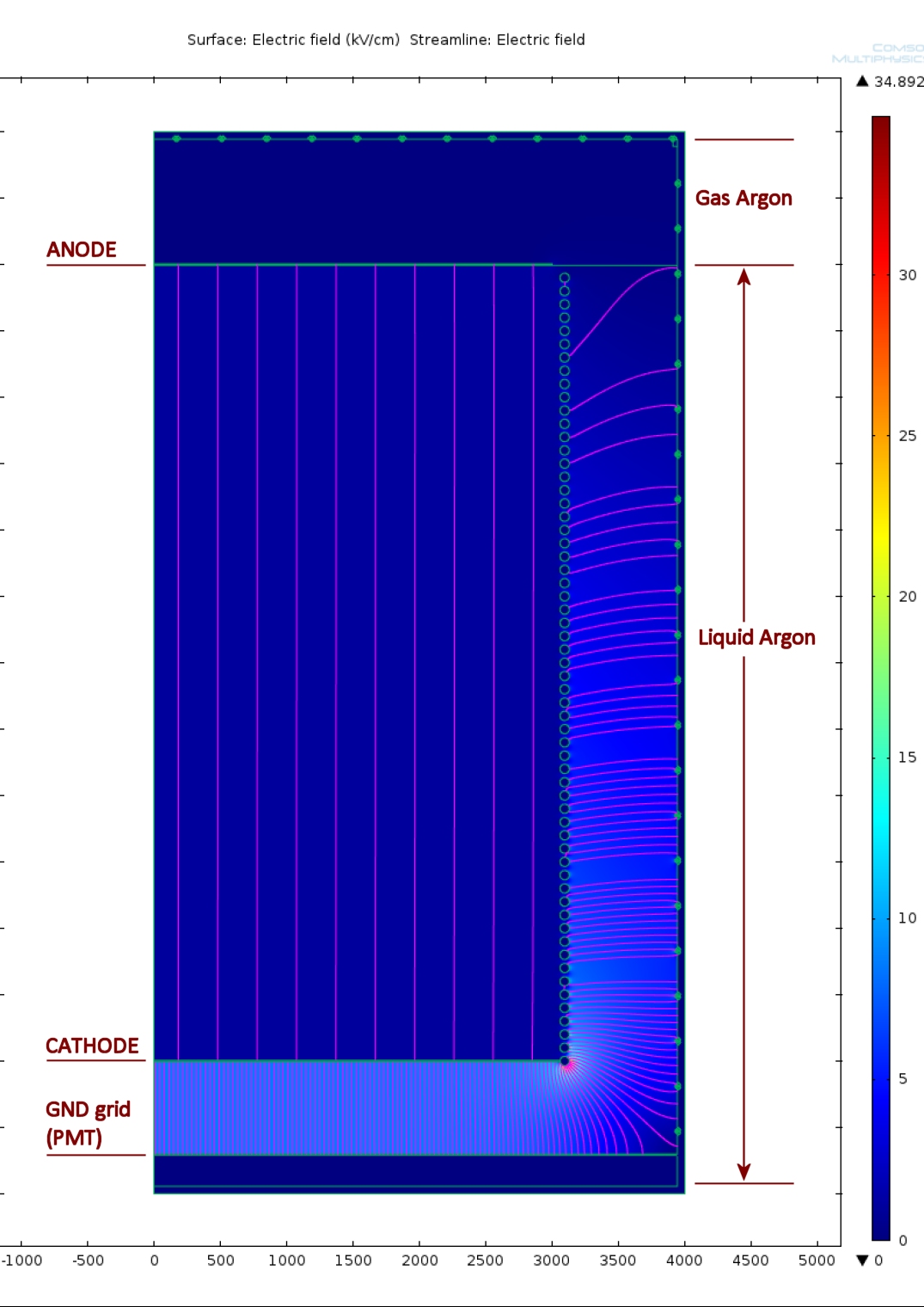}}
\caption{Electrostatic field calculation of the field cage: global view of the field cage embedded in the vessel.
The simulation takes into account the corrugation of the membrane vessel which is display by the green
dots.}
\label{fig:fieldcagecomsol}
\end{center}
\end{figure}

We have performed an electrostatic field calculation to optimise the geometry of the field cage
for a  drift field of 1~kV/cm (equivalent to 600~kV on the cathode).
As can be seen from \Cref{fig:fieldcagecomsol}, the field uniformity is excellent inside the drift volume.
The calculation took also into account the corrugation of the membrane tank (shown as 
green dots on the picture). The zoomed
region around the region of the cathode, where the highest field is expected, is also shown.
It reaches a maximal value of 30~kV/cm nearby the last shaper forming the cathode, a value
which is very acceptable according to our measurements of rigidity in liquid argon
(see \Cref{sec:electricbreakdowninlar}).

\subsection{Drift high voltage}
The intensity of the drift electric field is one of the important design parameters for liquid argon time projection chambers (LAr-TPCs). 
A better collection of ionisation charges is attained by increasing the field intensity because:
(1) more electron-ion recombination is prevented \cite{Amoruso:2004dy}
and 
(2) attenuation of the drift electrons due to their attachment to residual electronegative impurities such as oxygen decreases. 
The dependence of the attachment cross section on the electric field is known to be weak in the practical range of the intensity (0.5--1 kV/cm) \cite{Amoruso:2004ti}. 
The attenuation then is described well by an exponential decrease with drift time, characterised by the drift electron lifetime $\tau$ which is determined dominantly by the impurity concentration. 
The mean drift velocity of the electrons increases with increasing electric field, 
leading to the shorter collection time and consequently to the less attenuation (see~\Cref{c_attachement_impurity}). 
The drift velocity increases by 30\% 
by doubling the electric field intensity from 0.5 to 1 kV/cm, 
and again 30\% from 1 to 2 kV/cm \cite{Walkowiak:2000wf}.
From a technical point of view, however, difficulties increase with increasing field intensity 
which requires higher voltages. 
Therefore, it should be determined by a right compromise between the detector performance and the practicality of the high voltage. 
A field intensity between 0.5 and 1 kV/cm is a reasonable compromise for very long drifts. 

For the \six, a drift field in the range $0.5-1$~kV/cm requires a potential difference 
at the cathode in the range of 300$-$600~kV.
In comparison, the GLACIER drift length of 20 m~\cite{Rubbia:2009md} would require 2~MV. 
Two different approaches have been considered and realised for
 the drift field system for very large LAr-TPCs. 
The first type uses an external HV power supply and feed it into the detector volume using HV
feedthroughs~\cite{Amerio:2004ze}. 
The second type has an internal HV generator directly inside the LAr volume as the Greinacher HV multiplier of the ArDM-1t detector \cite{Badertscher:2012dq,Horikawa:2010bv}. 
Its advantages are: (1) all the HV parts are immersed in LAr which has a large dielectric strength as discussed
above, (2) thus feedthroughs for very HV are not needed, (3) the circuit itself can be used as a voltage divider, so the system needs no resistive load, (4) thus the power dissipation is virtually zero and (5) 
this allows a low frequency (e.g. 50 Hz) of the AC input signal which is fully outside of the bandwidth of the charge amplifiers used for this type of detectors. However, a major drawback is the difficulty to access the device
in case of failure of a component, such as something that could realistically happen in case of discharge given the amount of energy stored in the circuit.

\begin{figure}[htbp]
\centering
\includegraphics[width=0.3\textwidth]{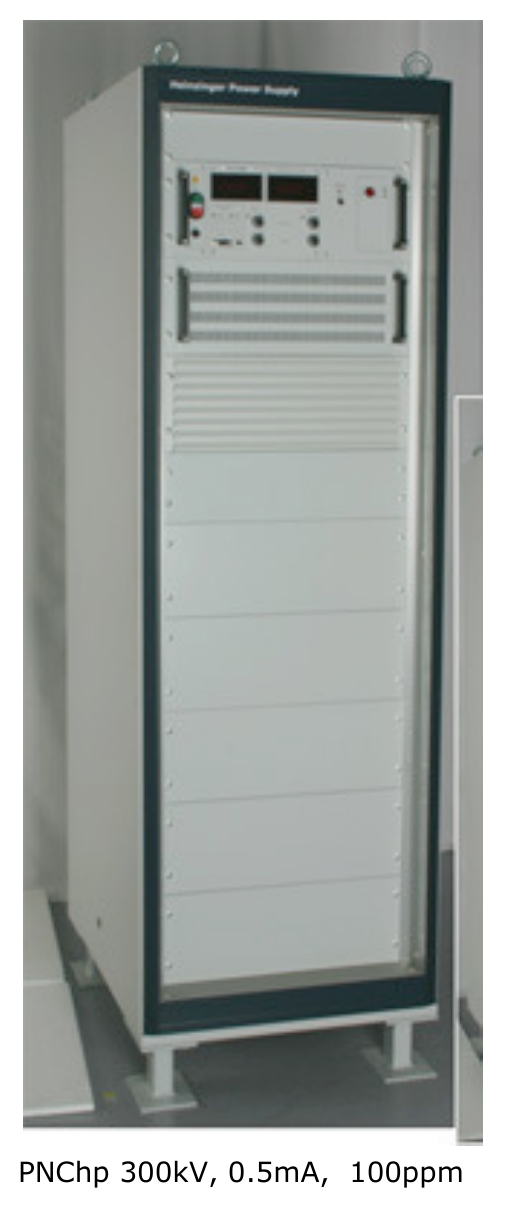}
\caption{The 300~kV power supply from Heinzinger electronic GmbH to be tested in 2014. This power supply
can provide the canonical 500~V/cm drift field over the full 6~m drift distance.}
\label{fig:heinzingerpowersupply}
\end{figure}

A HV feed-through and an external power supply will be adopted for the \six. Power supplies up to 300~kV with the required specifications in terms of stability, noise and low residual ripple, are commercially available in catalog 
(see e.g. Heinzinger electronic GmbH, Rosenheim(D)). We have already ordered the 300~kV (See \Cref{fig:heinzingerpowersupply}) and will perform tests during 2014. The performances for the voltage stabilisation are a reproducibility to $<0.1$\%, a stability of $<0.001$\% over 8 hours,
a ripple of $<0.001$\%pp, and a temperature coefficient $<0.001$\% / K.

Heinziger has also indicated that unique units for 400~kV have been successfully produced and operated. An R\&D phase with industrial partners is considered to develop a 600~kV-able power supply. We note that the 300~kV would be sufficient to operate the \six with the canonical drift field of 500~V/cm. However, in view of the longer drift paths, the \six facility will be used to perform R\&D on higher voltages, with the aim to reach a drift field of 1~kV/cm over 6~m.

As far as the HV feedthrough is concerned, it will be a direct extrapolation of existing design, with an insulating polyethylene thickness of 10~cm. Laboratory tests are been prepared.

\subsection{Front-end and DAQ readout}

\graphicspath{{./Section-FEandreadout/figs/}}

\subsubsection{Requirements for the large scale front-end electronics}

One of the goals of the WA105 demonstrator is to establish the large scale readout systems being developed for the far site LBNO LAr detectors in a configuration as close as possible to their foreseen final architecture. The large scale deployment of the readout systems in the demonstrator will allow testing  their performance with high statistics samples of hadronic interaction showers, their stability and reliability over long time periods and on large data volumes, develop  lossless noise-tolerant zero-suppression schemes on real data, as well as optimizing the full integration in the detector. The large number of charge readout channels, needed for the 20-50 kton LAr detector sizes for LBNO with channel count in the range of 500'000 to 1'000'000,  naturally called  during the last years for  R\&D efforts in view of the development of large scale readout solutions. These are characterized by high-integration levels, significant cost reduction and aims to performance improvement. The  R\&D activities focused on two main axes:

\begin{itemize}
\item  the developments of cold front-end ASIC electronics;
\item  the optimization of the data acquisition system based on modern telecommunication technologies.
\end{itemize}

Both efforts aim to improving the effectiveness and the integration level of the complete readout chain and to cost reductions for the large number of channels to be implemented in the detector.
Shortening of cables needed to bring the analog signals outside the cryostat and reduction of the electronic noise can be achieved using analog amplifiers operating at cryogenic temperatures. The current R\&D on the front-end electronics is based on analog preamplifiers implemented in CMOS ASIC circuits for high integration and large scale affordable production \cite{asicdev}.
 The noise is reduced by exploiting its behaviour as a function of temperature, which has a minimum around 100 K, and thanks to the suppression of the cables used to bring signals outside the cryostat, which otherwise increase the capacitance at the input of the preamplifier. 
In our present baseline the ASIC analog amplifiers can be integrated on the feed-through  flange terminating the chimneys on the roof of the tank, under the insulation layer, in order to be cooled to a temperature near that of liquid argon (see \Cref{sec:coldfeelec}). This solution fully preserves all  the benefits of the cold electronics, as described above, while guaranteeing at the same time accessibility to the amplifiers without affecting the inner volume containing ultra-pure LAr.

For what concerns the DAQ, solutions based on Ethernet capable ``smart sensors" were developed. The ``smart sensors" are Ethernet capable front-end DAQ/processing units acquiring large groups of channels. They are integrated with a time distribution system needed to align the data taken by different units operating independently. Data are output on a Ethernet network and collected with a system of switches to a computing farm which builds the events on the bases of the time stamps associated to the data packets by the different sensors. This concept was further developed for the LAr readout, since the time of its first implementation in the OPERA experiment, with the following technical improvements \cite{girerd-RT2009}:

\begin{enumerate}
\item{  porting it to the Gigabit Ethernet standard;}
\item{  adopting FPGA based virtual processors, in order to achieve cheaper implementation costs and become independent on the market of Ethernet capable front-end processors;}
\item{  integrating the electronics in the micro-TCA form factor, becoming very popular in the world of commercial telecommunication applications;}
\item{  developing a special time distribution system, derived from the Precise Time Protocol standard, integrated in a synchronous Ethernet network. This time distribution scheme achieves a synchronization accuracy among different nodes at better than 1 ns.}
\end{enumerate}

\begin{figure}[h]
 \begin{center}
\includegraphics[scale=0.75]{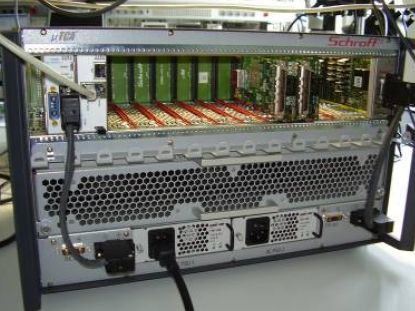}
\end{center}
\caption{ Picture of a micro-TCA crate containing  DAQ boards with 32 ADC channels/board as developed in 2010.}
\label{fig:mutca2010}
\end{figure}

This DAQ scheme allows benefiting of the large-scale integrations developments of the telecommunication industry and decoupling from the market lifetime of commercial processors since it relies on a completely virtual implementation of the processors in the FPGA. A complete setup built out of this R\&D was developed in 2010 for 128 channels (See \Cref{fig:mutca2010}). The proposed DAQ version for the LBNO prototype detector is an evolution of that system, further increasing the channels density and reducing the costs.

\subsubsection{Cold front-end electronics}
\label{sec:coldfeelec}

\begin{figure}[h]
 \begin{center}
\includegraphics[scale=0.7]{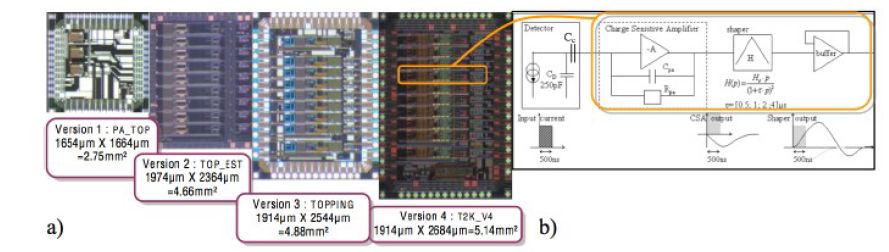}
\end{center}
\caption{ Four ASIC chip die pictures (a) following the development of the
different versions since 2007 and one channel block schematic (b) of the fourth version (V4)}
\label{fig:asics}
\end{figure}

 Since 2006 till 2012, six generations of prototypes of ASIC 0.35 microns CMOS multi-channel preamplifier chips operating at cryogenic temperatures, typically reached in the gas phase above the LAr, were developed. This development was oriented to performance improvement for large size detectors and low cost production. 
See \Cref{fig:asics}.

Each amplifier channel on a ASIC chip includes a complete chain with a pre-amplifier, a shaper and a buffer. Some parameters of the analog chain chain response, as the shaping time, may be controlled for the group of channels in the same chip with a distributed I2C bus.
The ASICs have a power dissipation of 18 mW/channel. The equivalent noise is around 1400 ENC for an input capacitance of 250 pF. The linear dynamic range extends up to  50 m.i.p. The ASIC configuration was mainly optimized for the most critical configuration of a detector with unitary gain, as a LAr TPC with wires readout, and the related possibility of dealing with bipolar signals as from induction planes.

\begin{figure}[h]
\begin{center}
\includegraphics[scale=0.3]{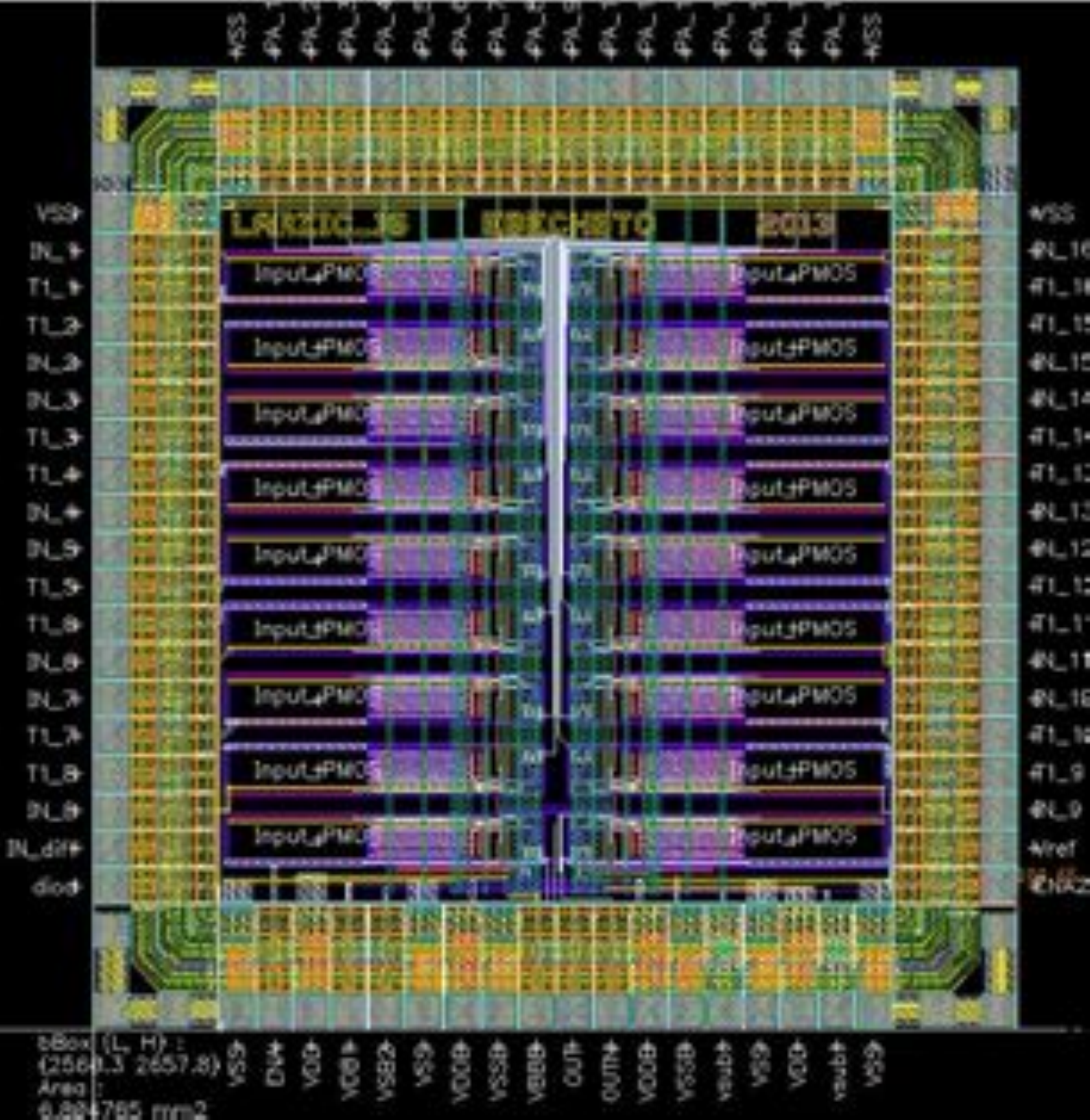}
\label{fig:ASICV7}
\end{center}
 \caption{16 channels version, adapated to the LEM signals, of the ASIC preamplifier(V7)  produced during fall 2013.}
\end{figure}

\begin{figure}[h]
\begin{center}
\includegraphics[scale=0.75]{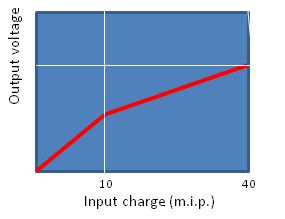}
 \caption{Double slope gain adjustment for an extended dynamic range of the amplifiers.}
\label{figdoubleslope}
\end{center}
\end{figure}

The current ASIC development version (V7), produced in the fall 2013, was upgraded to 16 channels and included as well the adaptation of the pre-amplifier dynamic range to LEM detectors operating at a gain of about 20  and generating unipolar signals from  X and Y anode collection strips with 3~mm pitch (See  \Cref{sec:crpforsix}, \Cref{fig:ASICV7}). The baseline scheme for the production version of the ASIC for the WA105 demonstrator is based on this development version, including its dynamic range adjustment, adapted to cope with the LEM detector gain, with the addition of an optimization of the resolution over the full dynamic range  with a double slope regime  (see Fig.~\ref{figdoubleslope}). The double slope regime is characterized by a high gain region extending up to 10 m.i.p. signals. After 10 m.i.p.  the gain is reduced by a factor 3 in order to overall match a dynamic range of 40 m.i.p. This solution provides the best resolution in the m.i.p. region without limiting the dynamic range for showers, which can still reach  40 m.i.p.
The total strips capacitance at the input of each preamplifier channel, will be around 300 pF, for 3 m long strips. Taking into account 20 m attenuation length of the charges during the drift, the S/N ratio for a m.i.p. signal will vary from 45 to 120 (m.i.p. tracks occurring at the beginning of end of the drift space) for the full size detector of LBNO, and from 90 to 120, for the prototype detector, which has a maximum 6 m drift space instead of 20 m. The front-end electronics will be coupled to the DAQ system, described in the following, based on 12 bits ADC, well matching the needed dynamic range. 

The design of the  \six demonstrator includes a total number of 7680 channels for the charge readout of the X and Y strips views. The strips have a pitch of 3.125~mm and a length of 3 m per channel. Channels are then  arranged in groups of 640 per chimney.

\begin{figure}[h]
\begin{center}
\includegraphics[width=0.8\textwidth]{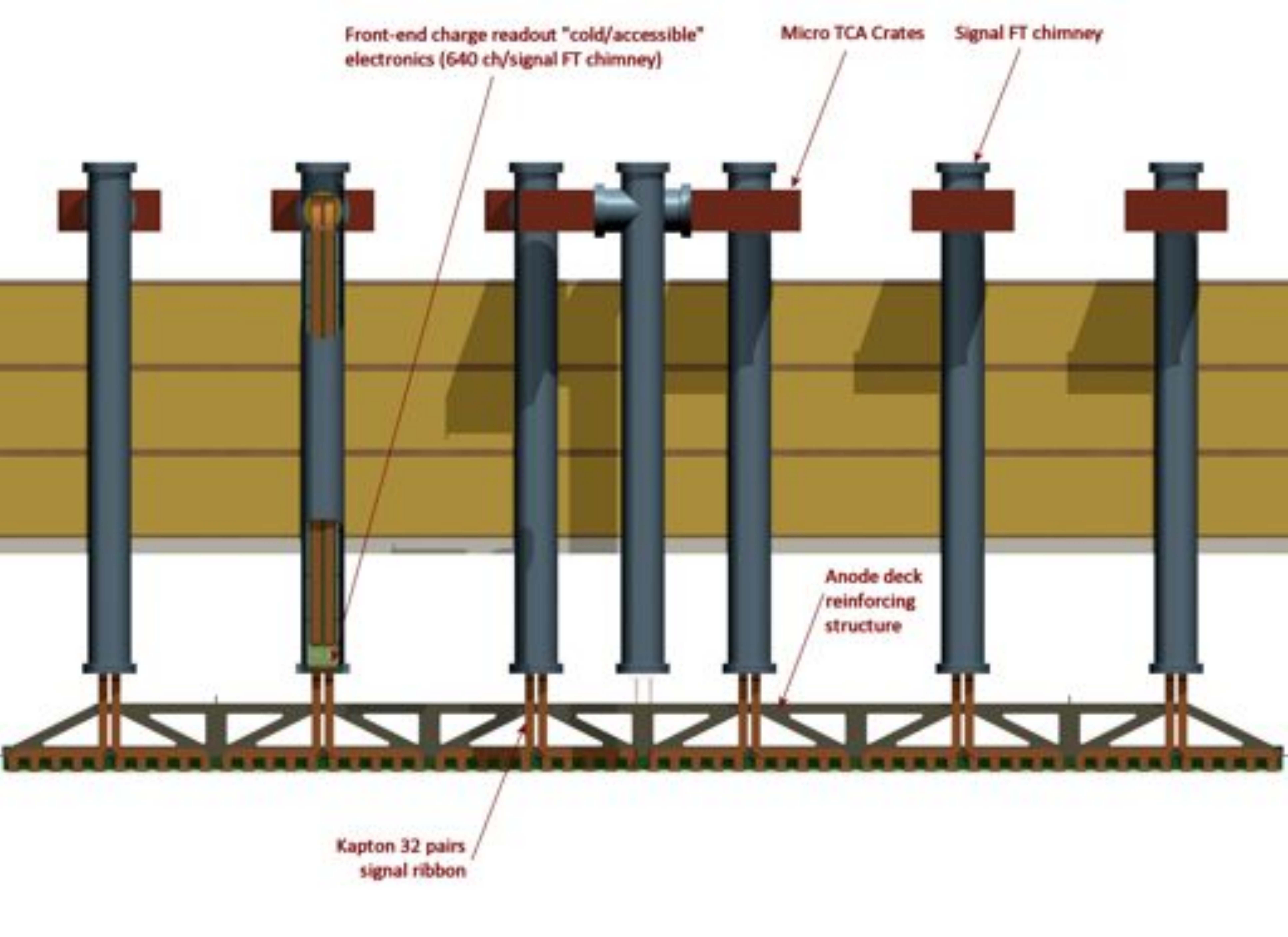}
\vspace{-0.5cm}
 \caption{Location of the FE cards at the bottom of the chimneys.}
\label{figchimney}
\end{center}
\end{figure}

 The 40 ASIC amplifiers needed for the readout of each group of 640 channels will be arranged on cards hosted on the feed-through at the bottom of each chimney  (see \Cref{figchimney}). %
The total power dissipation of the front-end electronics will be of about 11.5 W per chimney. The structure of the chimneys allows for heat dissipation in the cryostat and for  minimising the connection length from the cards hosting the ASIC chips to the double phase detectors down to half a meter. 

The ASIC chips needed to equip the final  LBNO far detectors can be produced in a dedicated run including 6 wafers for a cost going from 0.28 to 0.37 eur/channel, including encapsulation. This cost varies  depending on the  fact that the production version will be based on 16 or 32 channels per ASIC. This dedicated production can handle up to 28k ASIC chips in the 16 channels version (14k in the 32 channels version).   The production volume needed for the  LBNO prototype is much smaller and it will rely on an extended multi-project run with a consequently higher cost (3.91 eur/channel). The total cost per channel for the prototype detector goes up to 6.1 eur per channel when including also the front-end cards and the power supplies. The total cost for the demonstrator front-end electronics will amount to 54.2 keur, including 15\%  spare elements.

\subsubsection{Fine-tuning of the F/E dynamic range}
\begin{figure}[h!]
  \centering
  \includegraphics[width=\textwidth,height=0.32\textheight]{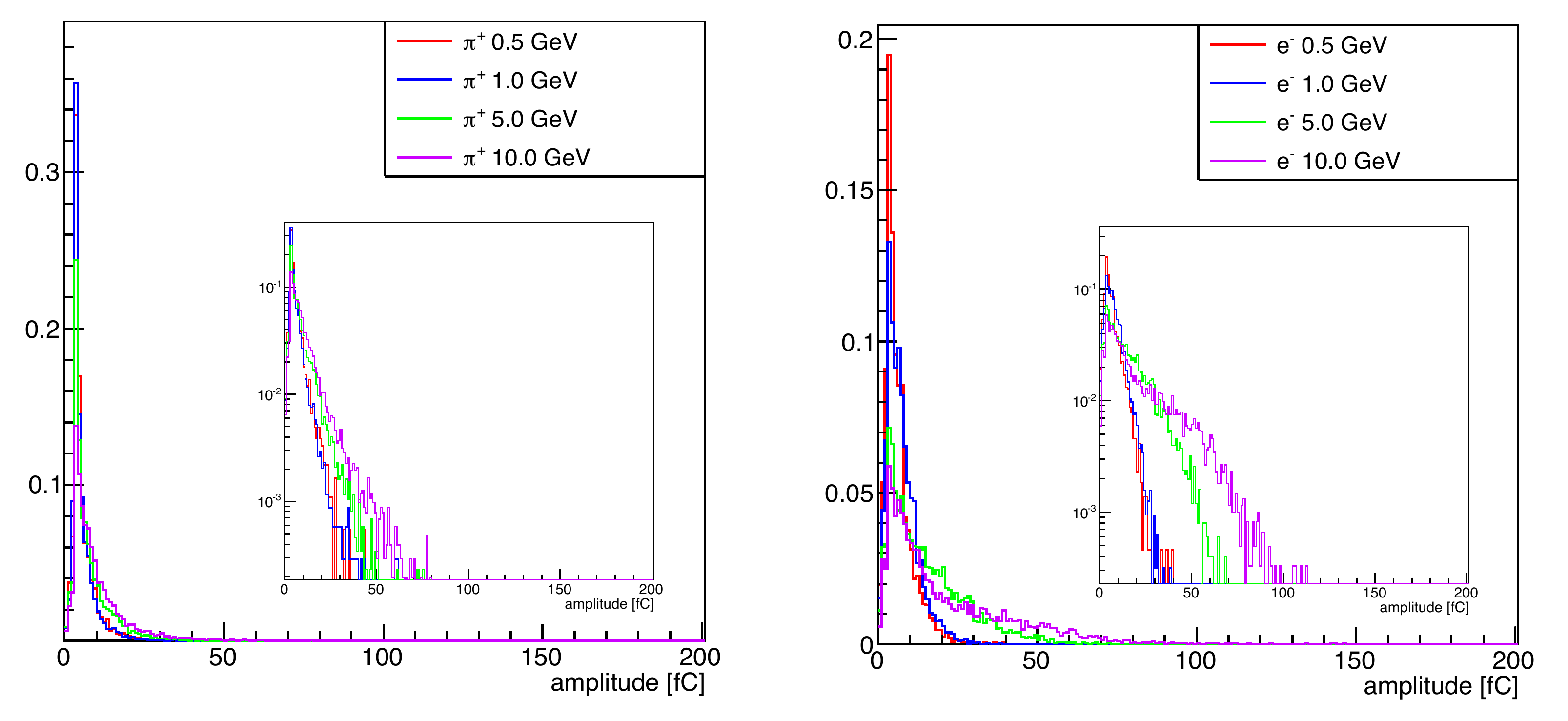}  
  \includegraphics[width=\textwidth,height=0.32\textheight]{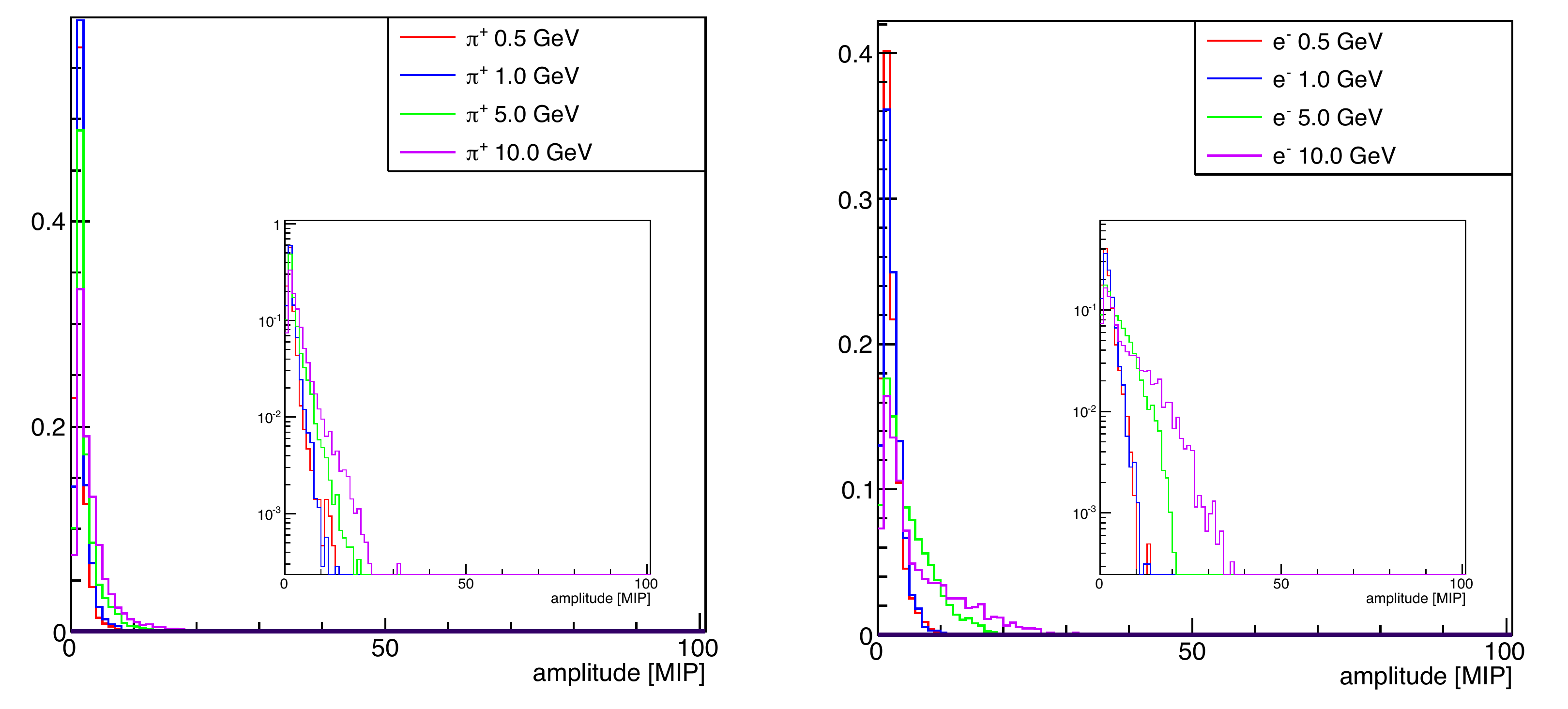}  
  \caption{Distributions of the maximal charge deposited on the strips
    for $\pi^+$ (left) and electron (right) showers of various
    energies. The distributions are given in unit of fC (top) and
    m.i.p (bottom).}
  \label{fig:DynRangeSimu}
 \end{figure}

In order to fine-tune the dynamic range of the preamplifiers the
amount of charge collected on the strips was studied with the
simulated \six geometry. The dynamic range should be large enough to
allow the digitisation of highly ionising events while still providing
good enough resolution for particles that deposit a small amount of
charge on the strips (typically a fraction of a m.i.p). To understand
the amount of charge collected on each strips, pions and electrons of
energies up to 10 GeV were sent through the beam pipe and a
distribution of the maximal charge deposited on each strip was
retrieved. The results are shown in \Cref{fig:DynRangeSimu} for
the strips belonging to one of the views. Since
the beam pipe is orientated at 45 degrees with respect to the strips
the distributions for the other view is identical. Most of the
collected charge is below $\sim$ 10 (resp. $\sim$ 20) fC for pion and
electron showers. The tails extend to about 50 fC and 100
fC for 10 GeV events. The dynamic range proposed in Section 4.4.1 is
therefore suitable as it extends up to signals of 40 m.i.p ($\approx
120$ fC). In addition the double slope configuration of the
preamplifier allows to adjust a range of increased sensitivity. This
range is chosen for signals between 0 and 10 m.i.p ($\approx$30 fC) as it is
the region where most charge is collected.

\subsubsection{Back-end electronics and DAQ global architecture}

\begin{figure}[hbt]
\mbox{\epsfig{file=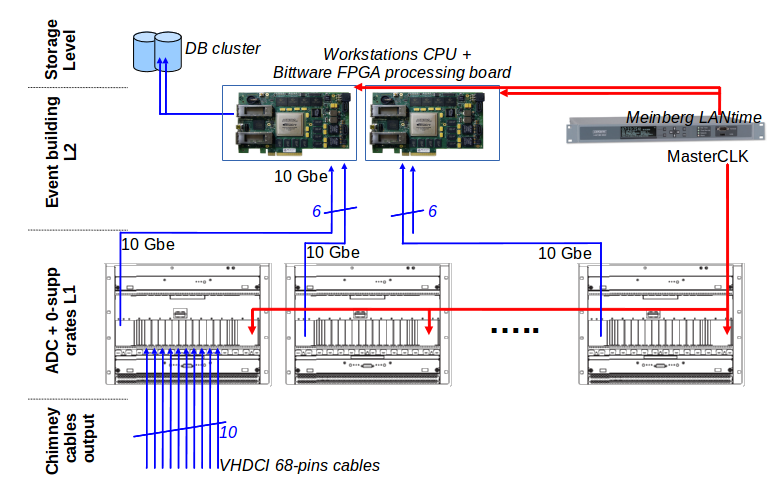,width=0.9\textwidth}}
\caption{\small Global DAQ scheme. At the bottom the DAQ receives the  cables from the F/E electronics. 10 VHDCI cables are connected on each 
rack. One crate is foreseen per chimney, leading to 12 racks in total for the  1st level (L1). Each crate is connected through a 10Gbe uplink
to the next level (L2). The L2 directly connects the racks to the event  builder workstation via a 10Gbe network link on a Bittware FPGA processing 
board. This board may connect up to 8-10Gbe links or 2-40Gbe links. A  lossless transmission scheme is therefore foreseen down to the processing 
board which performs the zero suppression and the event building. A stable common clock is distributed to L1 and L2, as well as the trigger signals (from PMT and  beam). The master clock generator may be derived from a Meinberg LANtime generator.}
\label{figDAQ-scheme}
\end{figure}

The DAQ system uses micro-TCA standards which offers a very compact and easily scalable architecture to manage a large number of channels at low cost.  Those constraints are indeed very close to the one existing in the network telecommunication industry. This has been driving the very first developments based on this type of standards in constant technological evolution and applied to large scale neutrino 
experiments. A generic scheme of the DAQ architecture is displayed in \Cref{figDAQ-scheme}. 

The main component of this DAQ system is an Advanced Mezzanine Board (AMC) reading out the input signals from the front-end electronics and sending the
formatted data through a micro-TCA backplane using a Gigabit or a 10 Gigabit Ethernet link. In the following, details are given on the design of this AMC and
the possible alternatives existing to the Ethernet link on the backplane (e.g. PCI express). The connections from the front-end use the Very High Density
cable interconnect (VHDCI) standard to minimize the number of cables.

A micro-TCA crate (or shelf) interconnects a fixed number of AMC cards through the backplane to collect the data and send them through a standard 10 Gbe 
MicroTCA Carrier Hub (MCH) and to distribute a common clock signal issued from a single, stable, GPS-locked Master Clock (MC).

The clock is made available on the backplane through a dedicated AMC (one per crate) which receives it
from an external connection, together with the trigger signals (internal trigger from the PMT array and external beam trigger). Those triggers are split and
distributed through conventional systems.

\begin{figure}[hbt]
\mbox{\epsfig{file=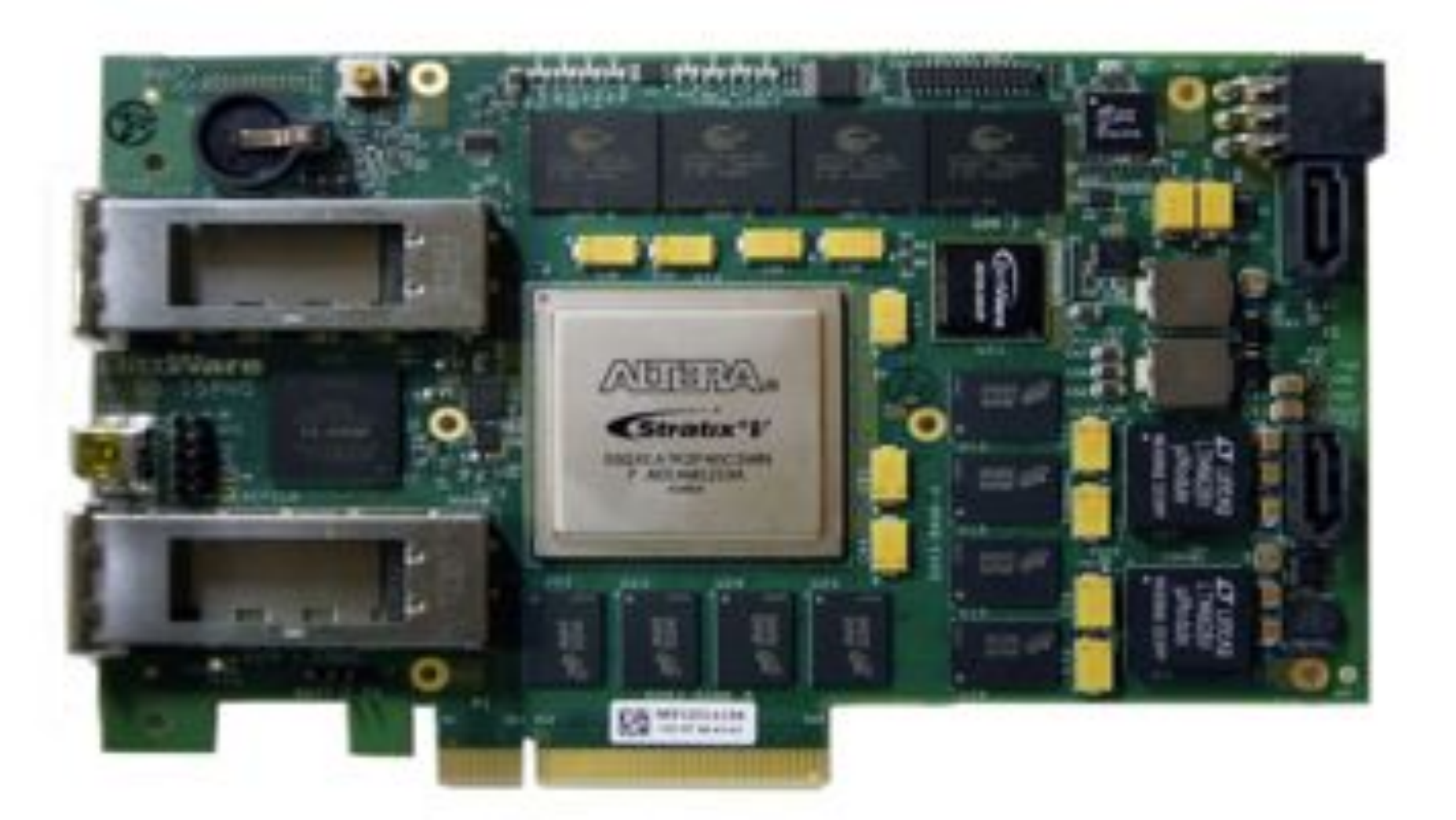,height=4cm,width=6cm}}
\caption{\small FPGA processing board based on Stratix V from Altera. The  board features a dual QSFP+ cages for 40GigE or 10GigE links, 16 GBytes 
DDR3 SDRAM, 72 MBytes QDRII/II+, two SATA connectors and is programmable  via OpenCL.}
\label{fig:Bittware-board}
\end{figure}

A network hierarchical structure is implemented where all crates are interconnected to a dedicated Bittware FPGA processing board (S5-PCIe-HQ, 
Figure \ref{fig:Bittware-board}). This board has two QSFP+ cages to bring  the data direct to the FPGA for lowest possible latency. Up-to 8x10Gbe 
links w/o data loss are available per board.  The board performs further data processing, filtering and transmission to the highest level for 
storage. This type of board is widely used in the massive processing systems and the present generation, based on the Altera Stratix V, will 
evolve to the Aria X and the Stratix X. This version will be probably  available at the time of the construction of the DAQ system.\\

Programming of the processing board is achievable through the OpenCL software suite where a kernel code allows, on top of a host code, to 
program directly the FPGA without a classical VHDL synthesis chain. OpenCL is a high level language for massive parallel processing transparent to its hardware implementation (CPUs, GPUs, FPGAs). Under OpenCL FPGAs provide large computing power for data processing at low power consuption. This highly flexible feature is fully adapted to the requirements of the large  DAQ systems where conditions of filtering, event building etc... may evolve  with time.

The storage is defined as the highest level of the DAQ chain and implies the use of a redundant 2-servers cluster linked to a RAID-5 disks array system.
 
\subsubsection{MicroTCA standard and crates}

MicroTCA offers the possibility to interconnect distributed applications while offering a standard, compact and robust form factor with simplified power 
supply management, cooling and internal clocks distribution. The microTCA backplane is based on high speed serial links arranged in various possible
topologies. Lanes on a microTCA backplane support a large variety of protocols like for example Ethernet 1G or 10G, PCI Express or SRIO. 

\begin{figure}[hbtp]
\mbox{\epsfig{file=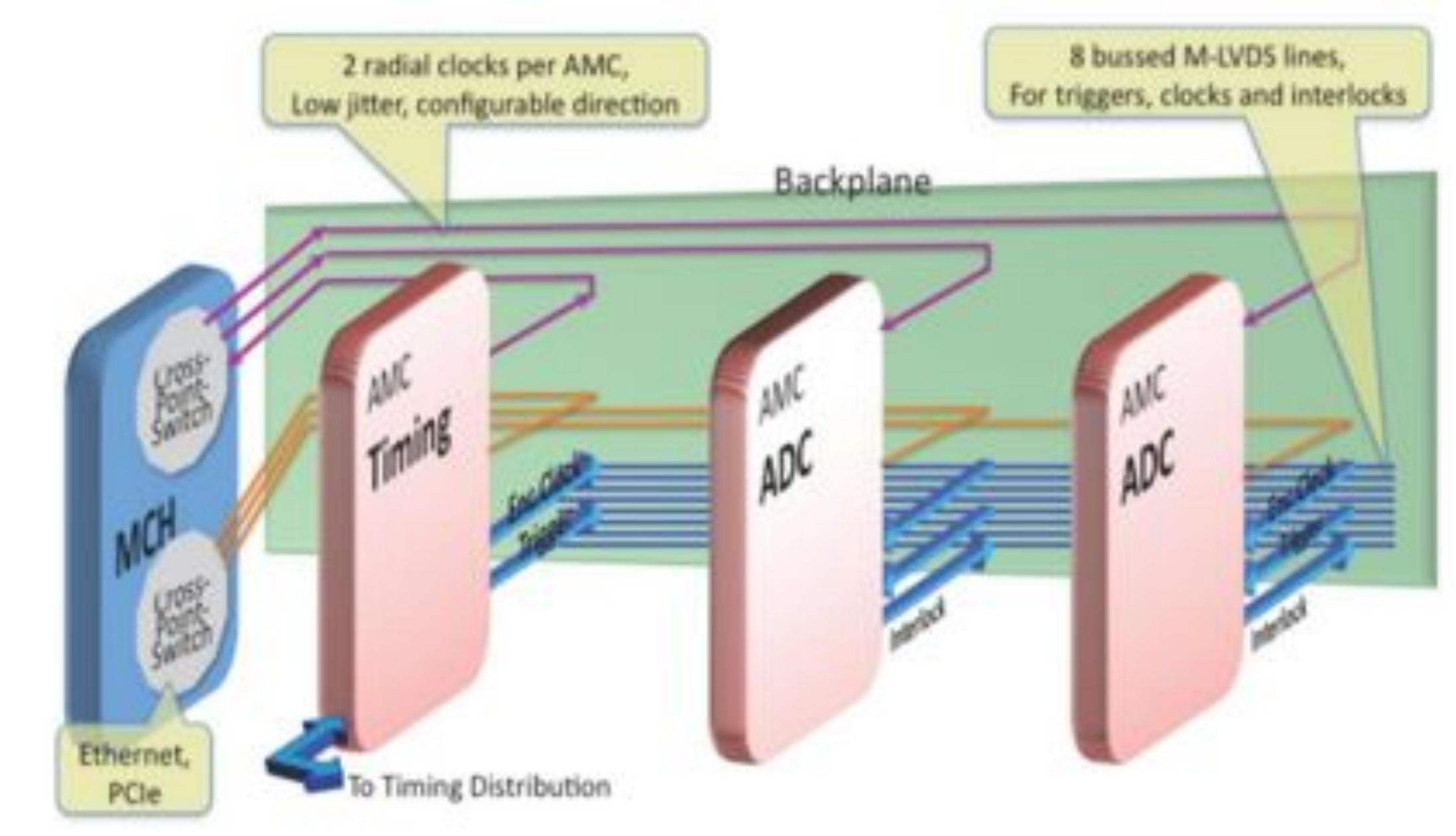,height=6cm,width=8cm}}\hfill
\mbox{\epsfig{file=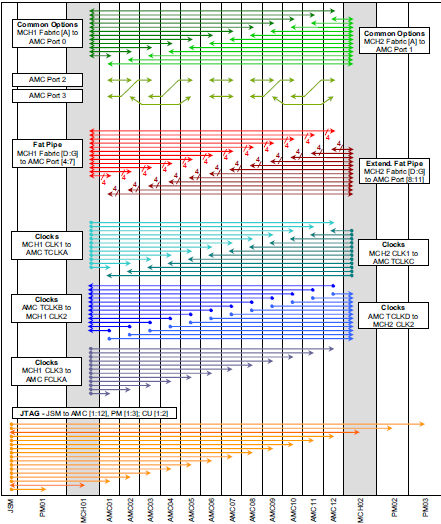,height=6cm,width=8cm}}
\caption{\small Left: global microTCA crate organization. AMC cards (providing basic ADC functions) are connected to the crate controller or MCH which uplinks the  external systems. A dedicated AMC for the clock receives dedicated signals (masterclock, trigger signals) from the timing distribution system and transcript them onto the backplane. Right: backplane technology of the Schroff 11850-015 reference.} 
\label{figmTCA-features}
\end{figure}

The boards plugged into a microTCA shelf are called Advanced Mezzanine Card (AMC)~\cite{picmg-2006}. Each AMC board is connected to one or two MicroTCA Carrier Hub (MCH) through the backplane serial links which provides a central switch function allowing each AMC to communicate with each other or towards external  systems through an uplink access. The backplane also provides the connectivity for the clock distribution allowing the AMC board synchronisation. 
\Cref{figmTCA-features} provides a sketch of the backplane technology and its implementation in one retained shelf reference.

The first developments performed for the LAr TPC readout with the microTCA systems were based on the microTCA.1 standard with connections to the user input signals from the front side only.
In the baseline option we stay as close as possible to this standard although additional standards have emerged like the microTCA.4 offering the possibility to 
enter in a crate both from front and rear sides.
 
\begin{figure}[hbt]
\mbox{\epsfig{file=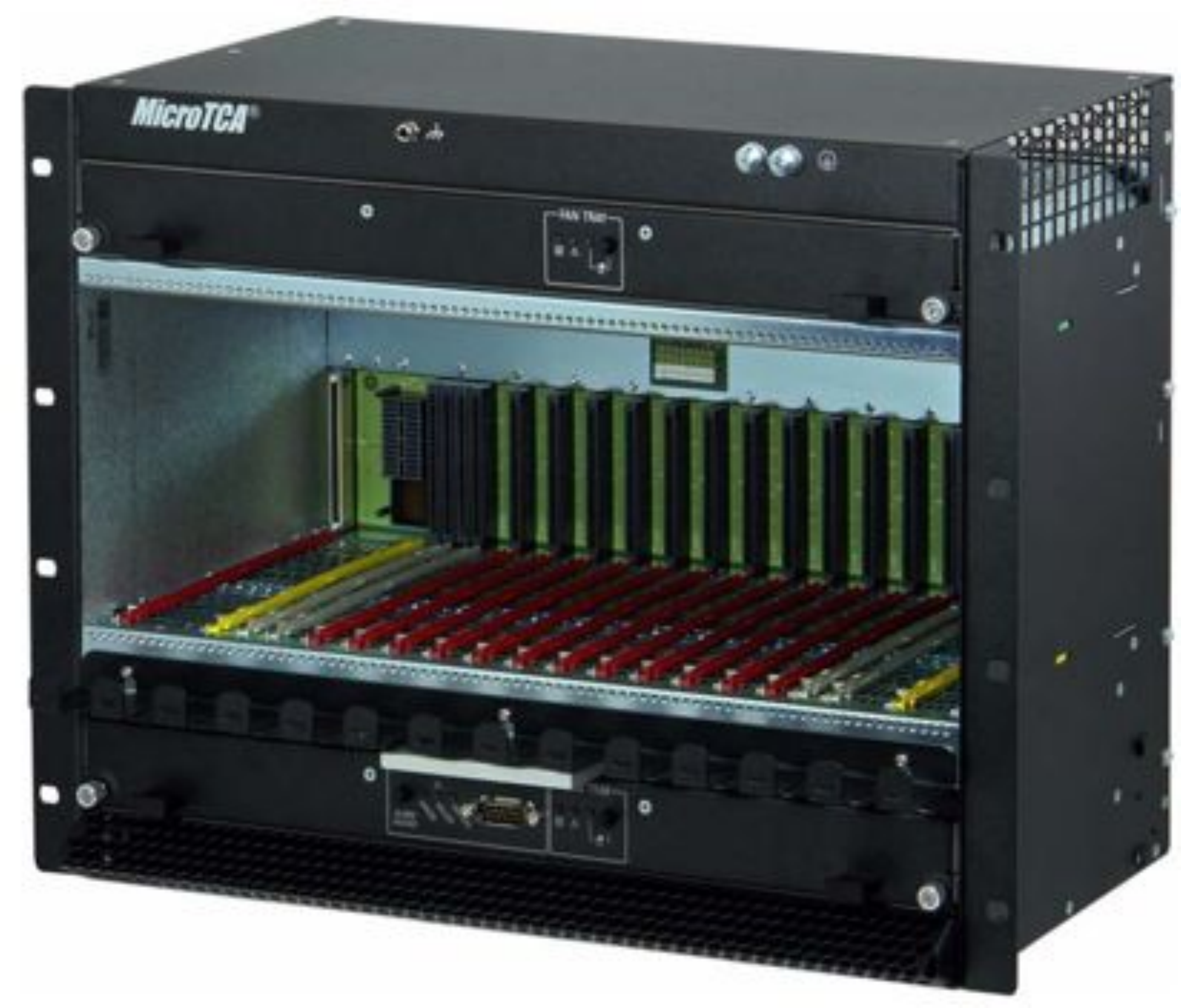,height=6cm,width=8cm}}
\caption{\small Pcture of the Schroff 11850-015 8U shelf.}
\label{figschroff-crate}
\end{figure}

In the microTCA.1 baseline option, one has 1 crate per output chimney, handling 640 channels dispatched over 10 AMC. A candidate crate is the 11850-015 8U shelf from Schroff (\Cref{figschroff-crate}). We are also considering other references from various providers (like the NATIVE-R9 from NAT) in the spirit of
evolving to the microTCA.4 standard. In this case each crate will be located between 2 nearby chimneys.
As shown in \Cref{fig:CRP_666}, the crates are implemented on the top layer of the prototype detector.  Each brown
box features a crate. The inter-chimney distance, of the order of 1 meter, allows this type of crate disposal which minimizes 
the constraints on cables
lengths.

The main features of the 11850-015 Schroff crate references are listed below:
\begin{itemize}
\item 8 U microTCA Shelf, 12+2+3+1 slot for AMC double Mid-size modules,
\item 19" rack mountable,
\item 12 AMC Double Mid-size slots,
\item 2 redundant MicroTCA Carrier Hub (MCH) slots (Double Full-size),
\item 2 Power Module (PM) slots (6 HP Double) at the right side,
\item 1 Power Module (PM) slots (12 (9) HP Double) at the left side,
\item 1 slot for a JTAG module (Double compact),
\item 5 splitting kits to install single module in a double slot.
\end{itemize}

Various MCH references may be used in that design. We selected the NAT  MCH with additional hardware for optical fibres connections 
(NAT-MCH-Base12-GbE, NAT-MCH-XAUIx48, NAT-MCH-UPLNK-SFP+,  NAT-MCH-UPLNK-SFP+850). The optical link will go down to the FPGA 
processing board. This MCH has SRIO (Gen2), PCIe (Gen3), 1GbE and 10GbE  (XAUI), central management up to 13 AMCs, 2 cooling units and 4 power 
modules, e-keying, redundancy and load sharing.

\subsubsection{MicroTCA dedicated AMC}

\begin{figure}[hbtp]
\mbox{\epsfig{file=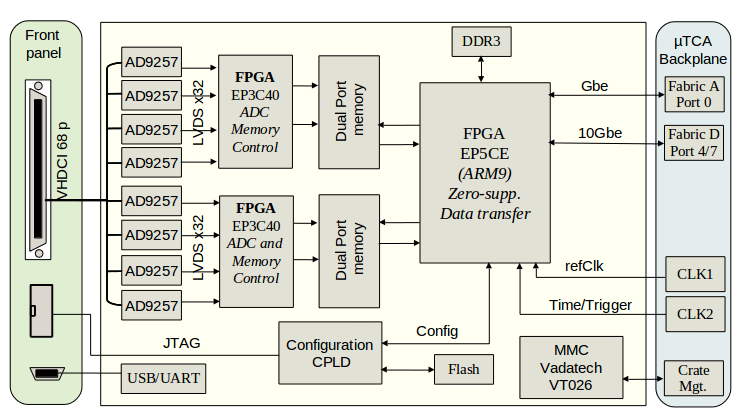,height=6cm,width=9cm}}\hfill
\mbox{\epsfig{file=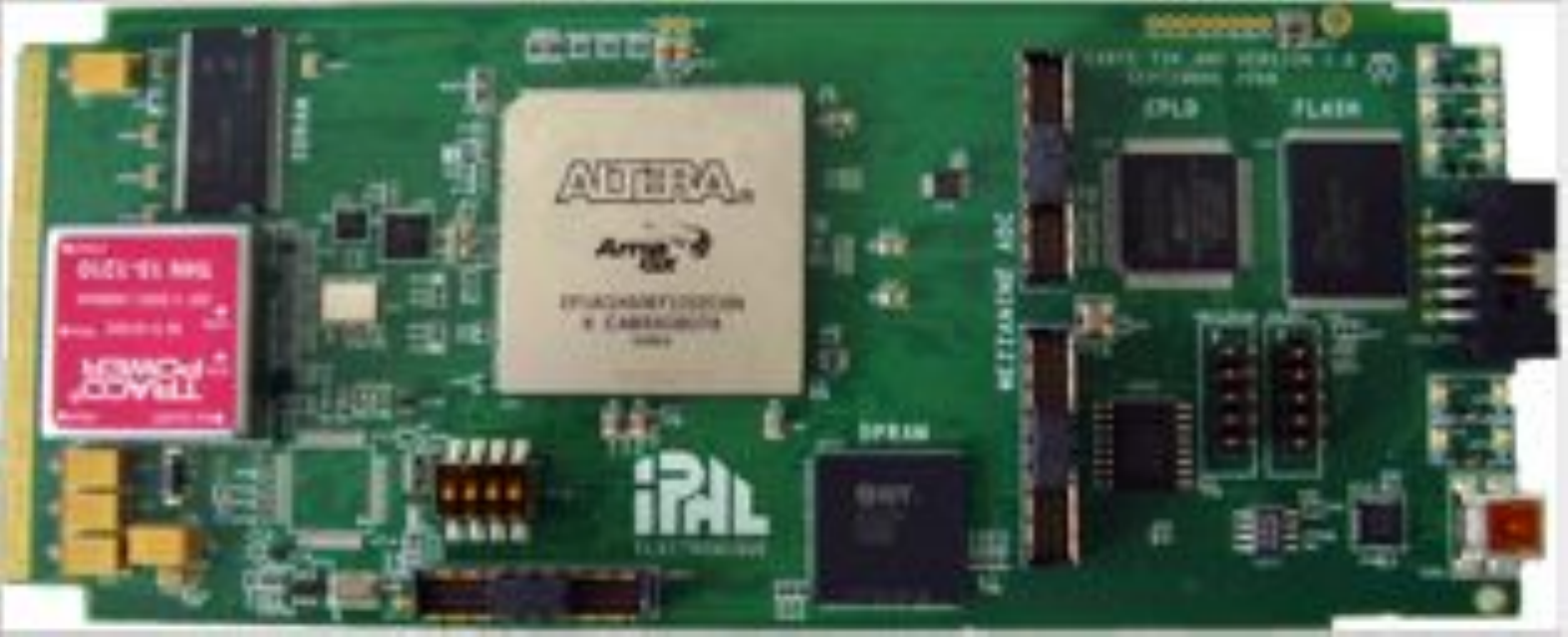,height=5cm,width=7cm}}
\caption{\small Left : AMC bloc diagram scheme. The input stage digitizes the 64 input channels. The ADC readout sequence is controlled via 2 FPGA (one for
32 channels) which handles also the writing into a double bank memory. The management of the trigger input, data output, and link to the backplane if
performed by a third FPGA. The link to the backplane may be direct 10GbE or PCI-e. Right: the first AMC version, single height for 32 channels. Present
design is an extension of this board.} 
\label{figAMC-bloc-diag}
\end{figure}

\paragraph{AMC board design} The idea is to develop only the user AMC offering the desired functions. The generic functional diagram of the AMC is displayed in
\Cref{figAMC-bloc-diag}. The AMC chosen is a double-size module (also compatible with microTCA.4 standard) with a single input connector and a 10GbE or PCIe
link to the backplane. The input stage performs the 64 channels digitization through 8 8-channels 14-bits ADC readout at a 2.5MHz frequency. The ADC readout
sequence is controlled by 2 EP3C40 FPGA from Altera which makes the data available on a double port memory (DPRAM with 9k-samples width from IDT). 
Two banks (B0/B1) are attached to each channel and work in ping pong mode. The write state machine manages the bank address working as a circular buffer. 
If no trigger conditions is detected the samples are written continuously into the memory, the oldest ADC being overwritten by the newest ones. The ADC 
values and the address location in the bank are written into an intermediate FIFO for each bank (B0/B1).
If a trigger occurs, the write state machine keeps on storing the event corresponding to the TPC drift. If the second bank is available, the main state 
machine tells the write state machine to continue to store the samples into this new active bank. The read state machine sets a flag corresponding to the
availability of an event in a bank. All the read state machines are interconnected through a token ring like structure. The zero-suppression algorithm is
applied on all samples available in the bank, the pedestals (mean and width) being computed on the samples themselves. The recorded samples are then
formatted and sent out. These operations are managed by a third FPGA (EP5CE from Altera), which sends the data on the backplane. The baseline option foresees
to send the data directly through a 10GbE link (UDP was used in the first prototype). 
The readout scheme and hardware implementation is very close to the first prototype designed in IPNL
(\Cref{figAMC-bloc-diag}) with slight changes: replacement of the single FPGA by three cost-effective FPGA's and direct implementation of the ADC layer
on the motherboard (in the first version ADC's were on a mezzanine board).

\begin{figure}[hbtp]
\mbox{\epsfig{file=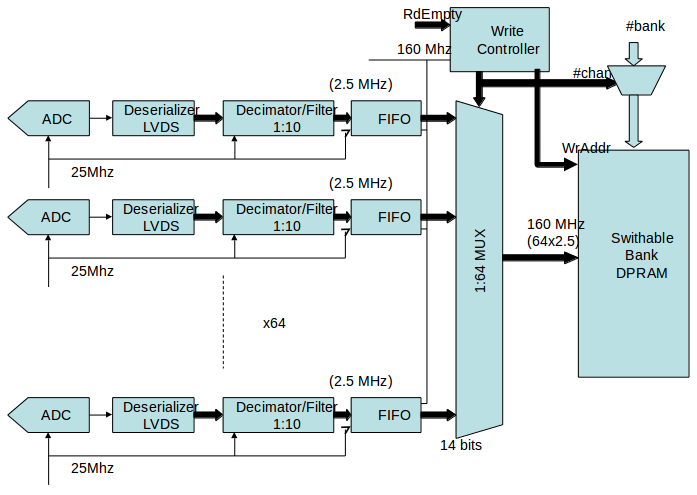,height=7cm,width=11cm}}\hfill
\mbox{\epsfig{file=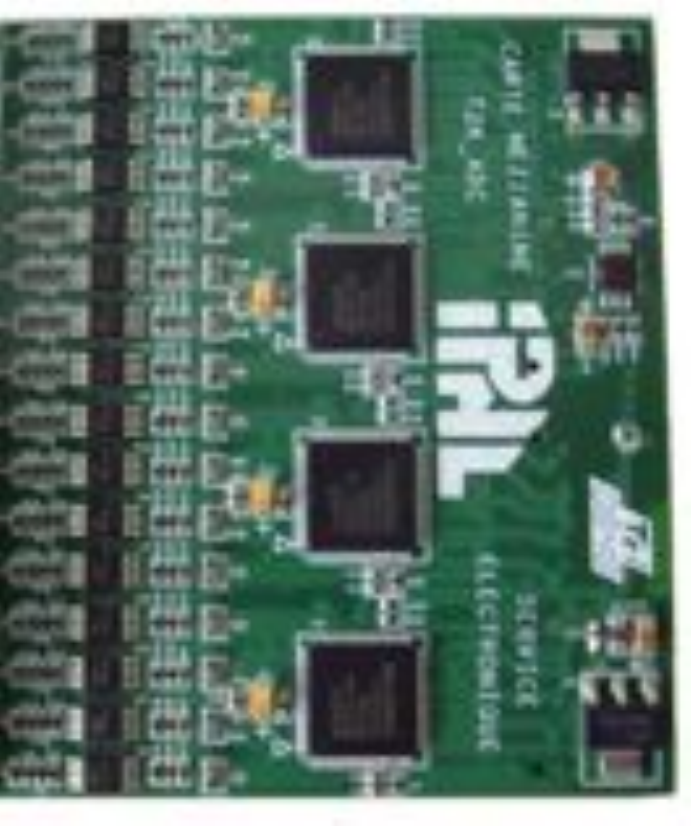,height=5cm,width=5cm}}
\caption{\small Left : ADC readout chain block diagram. The AD amplifier is used to translate the inputs to differential levels. Right: implementation of a
32 channels ADC readout chain in the first prototype version.} 
\label{figAMC-adc-chain}
\end{figure}

\paragraph{ADC readout chain} The analog signals from the F/E ASICS are connected to the AMC front-panel 
through a 68 pins VHDCI connector. The 8$\times$14-bits ADC 8 ch. AD9257 ADC from Analog Device, including a serial LVDS output, has been chosen. The translation
from single ended to differential signals is performed upstream of the ADC. 
The readout scheme is displayed on \Cref{figAMC-adc-chain}. This design offers a high integration level required by the large density of input signals.

\subsubsection{Data rate requirements}

\paragraph{Data reduction} In the present design the useful ADC will be 12-bits resolution and each sample will have 16-bits size. The readout frequency of 2.5 MHz (400ns steps) requires a maximum of 10k samples for the maximal drift time of 4000 microseconds that we want to cover. Given the 6 microseconds shaping of the F/E electronics for a single hit, one obtains: 6($\mu$ s)/400(ns)$\sim$15 useful samples for a single hit over the 10k of a full bank. To be conservative we assume in the following a reduction factor 100 given by the zero suppresion algorithm if implemented  at the AMC level. In the lossless data transmission scheme we will assume the total 10ksamples.

\paragraph{Data rate constraints} In this context we may compute the  saturation limits for one crate and for the full detector. One crate is 
used to readout a full chimney, that is 640 channels (10 AMC). In the case the  zero-suppression is not applied, the maximal affordable data rate to 
saturate the 10GbE uplink of the crate is therefore: 10(GbE)/(16(bits)x640(ch.)x10k(samples))$\simeq$100Hz. Since the crates  are readout in parallel by the FPGA processing boards, this rate  corresponds to the maximal rate for the full detector. This mode is  adapted to a beam-only data readout. In the case the zero suppression is applied we may reach a maximal rate  of 10kHz, compatible with the surface cosmic rays rate ($\simeq$6kHz).  In this mode we may readout beam data and cosmics between the beam cycles.

\subsubsection{Integration of readout electronics}

In the LBNO far detector, 
several hundred thousands channels will need to be routed from the charge readout plane located
inside the vessel to the front-end charge preamplifiers and digitisers to be placed outside
the vessel. Given the large number of channels, it will be necessary to pack as many of them into 
a single flange. 

\begin{figure}[h]
\begin{center}
\fbox{
\includegraphics[width=0.35\textwidth]{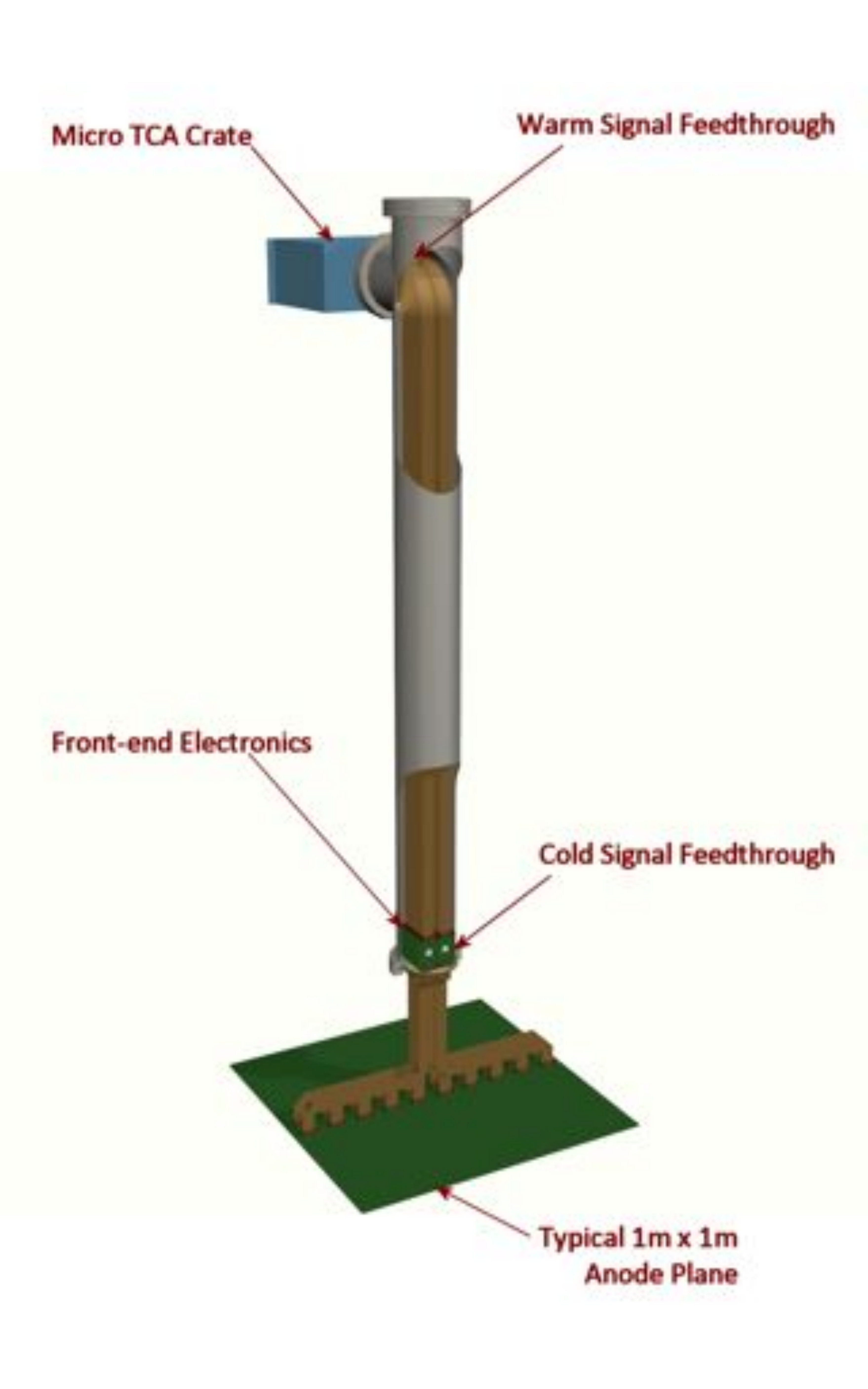}
\includegraphics[width=0.45\textwidth]{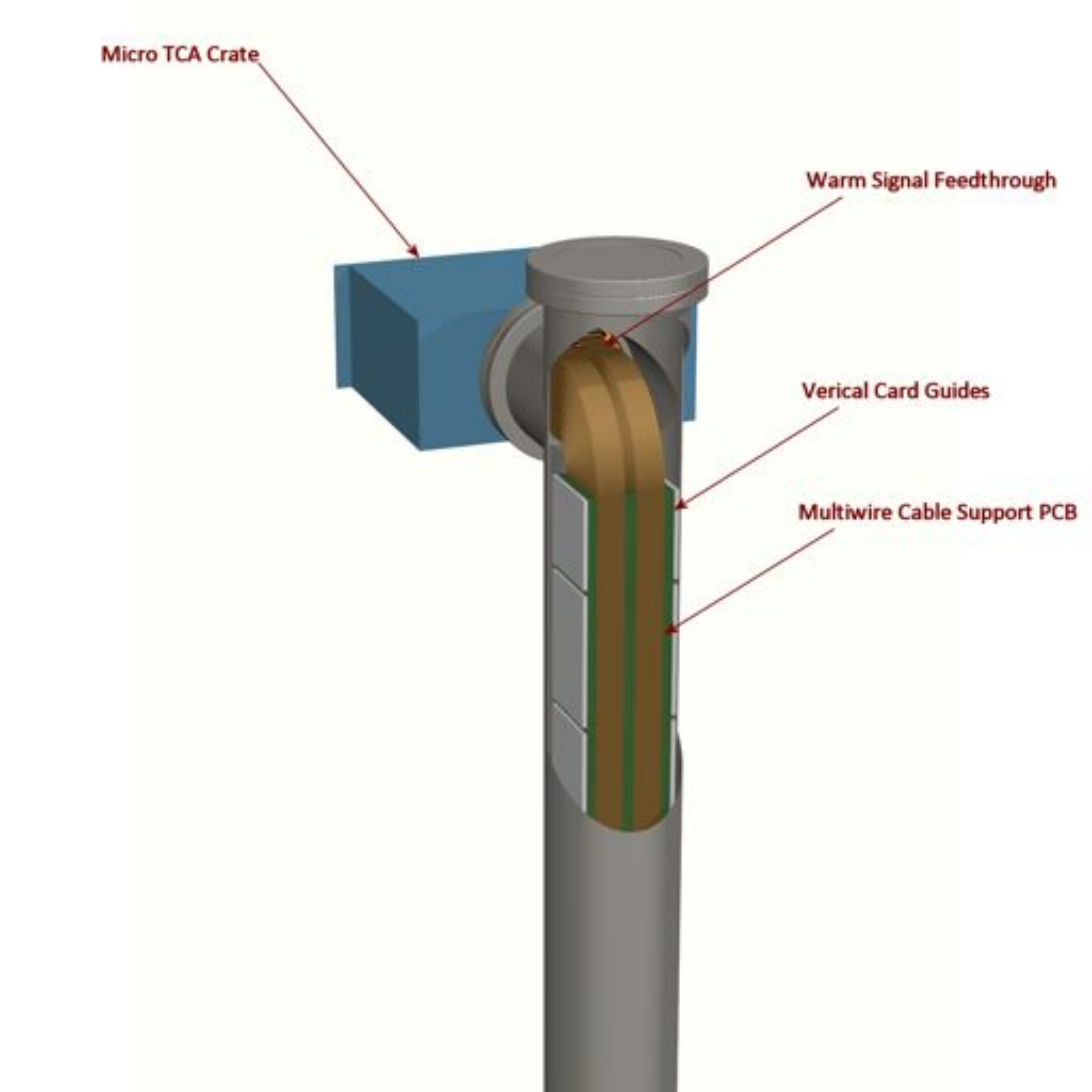}}
\fbox{\includegraphics[width=0.4\textwidth]{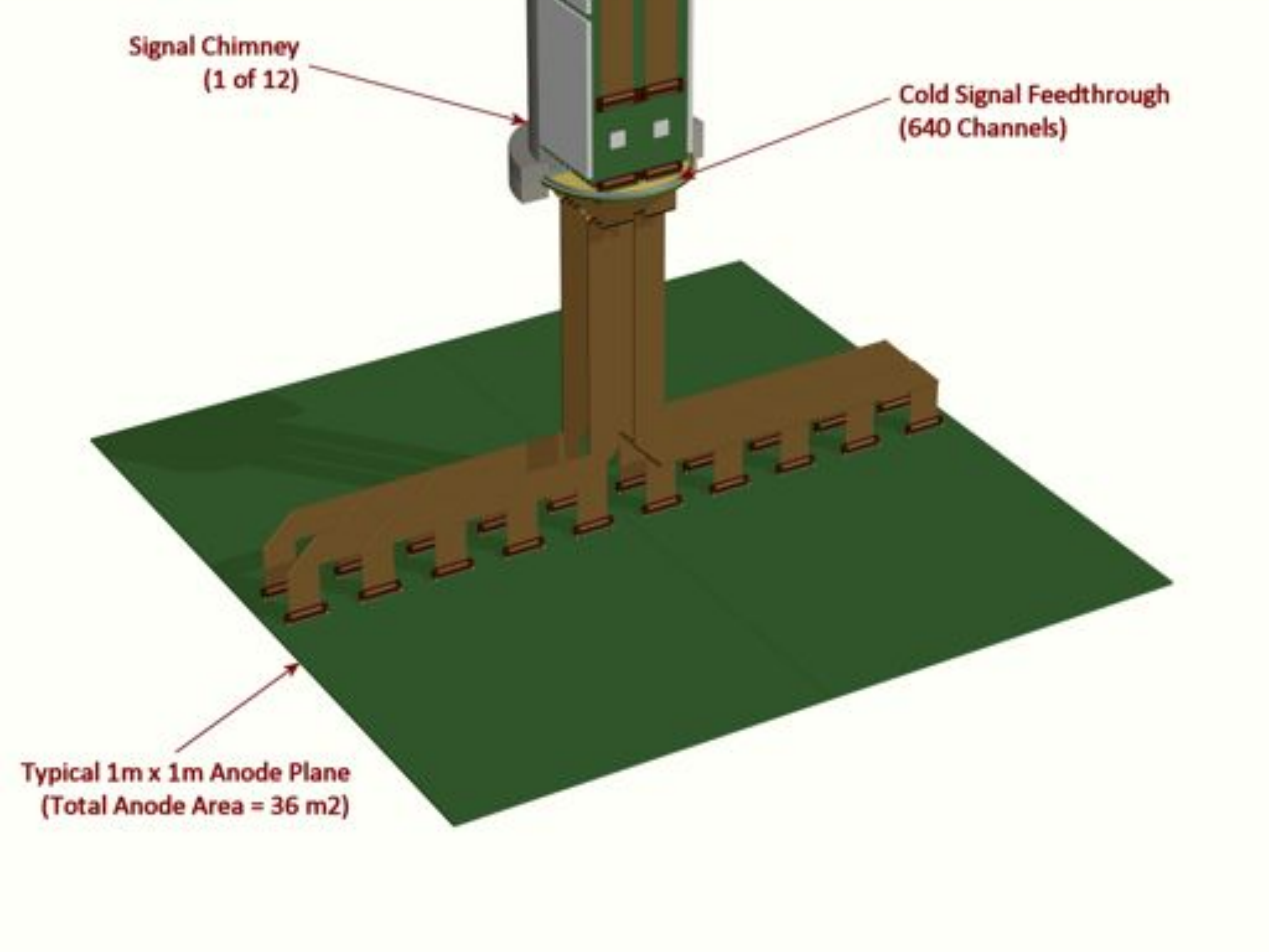} 
\includegraphics[width=0.4\textwidth]{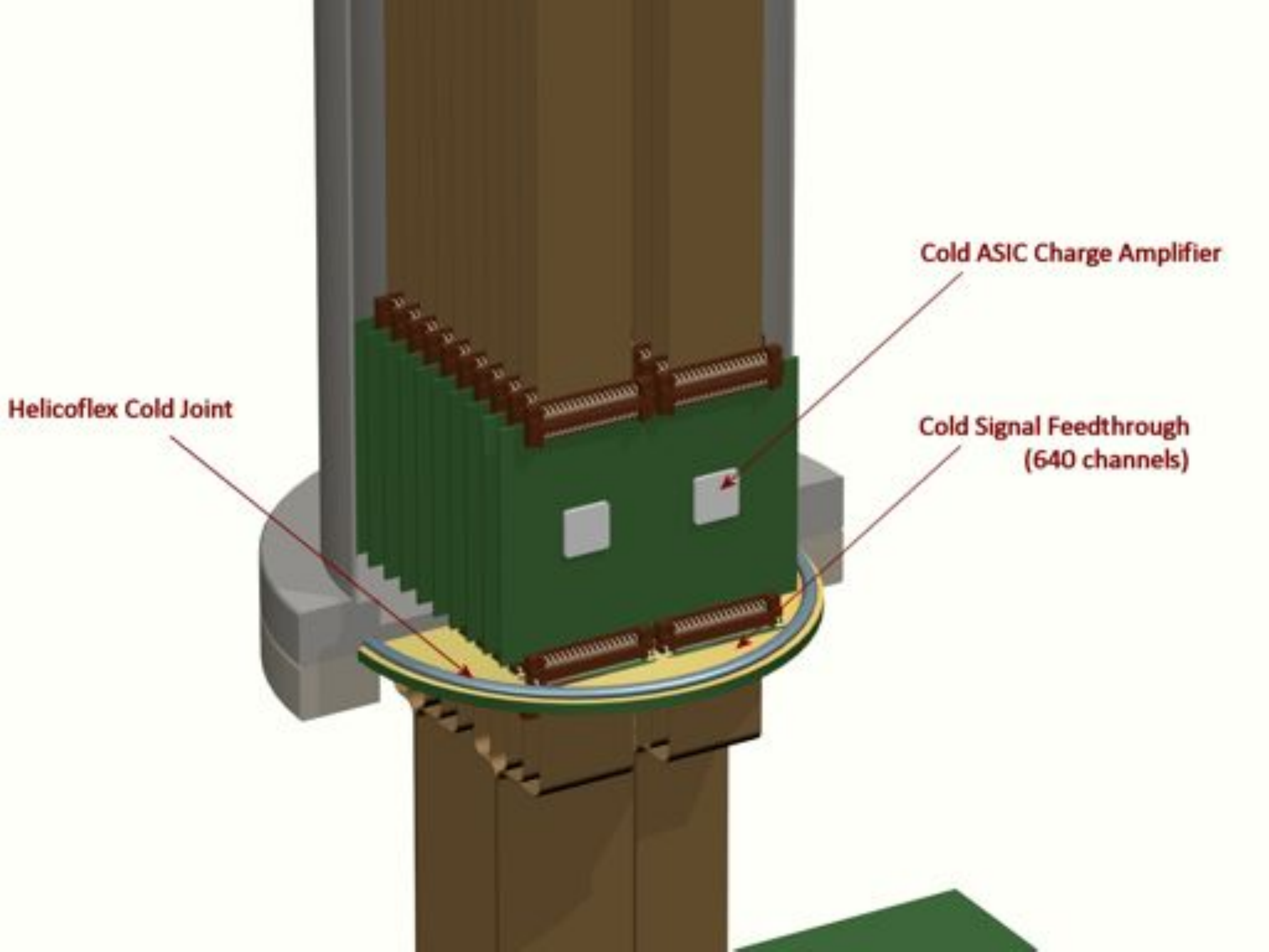} }
 \caption{Details about the chimney, the signal feed-throughs, the inserted F/E electronics cards, and the foreseen cables layout.}
\label{fig:cableslayout}
\end{center}
\end{figure}

Signal cables of the \six 
will be similarly routed towards the top of the detector 
and connected to the signal feed-through flange at the bottom of a chimney. 
Each chimney will be terminated with a feed-through that must
be absolutely ultra-high vacuum leak-tight not to contaminate the pure liquid argon.
\Cref{fig:cableslayout} shows some details about the chimney and the inserted F/E electronic cards,
 and the foreseen cables layout.

\begin{figure}[htb]
\begin{center}
\includegraphics[width=0.4\textwidth]{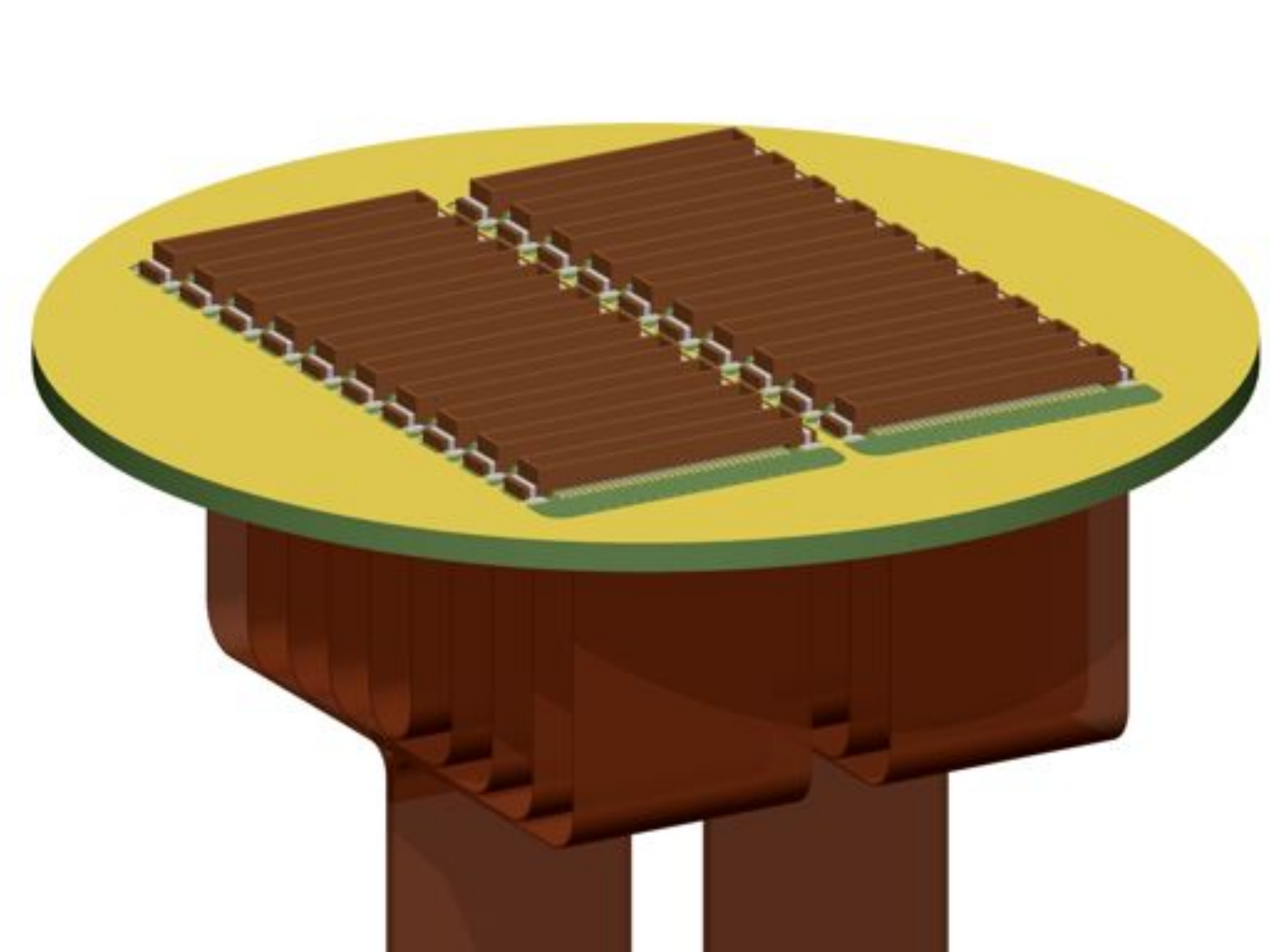} \hfill
\includegraphics[width=0.4\textwidth]{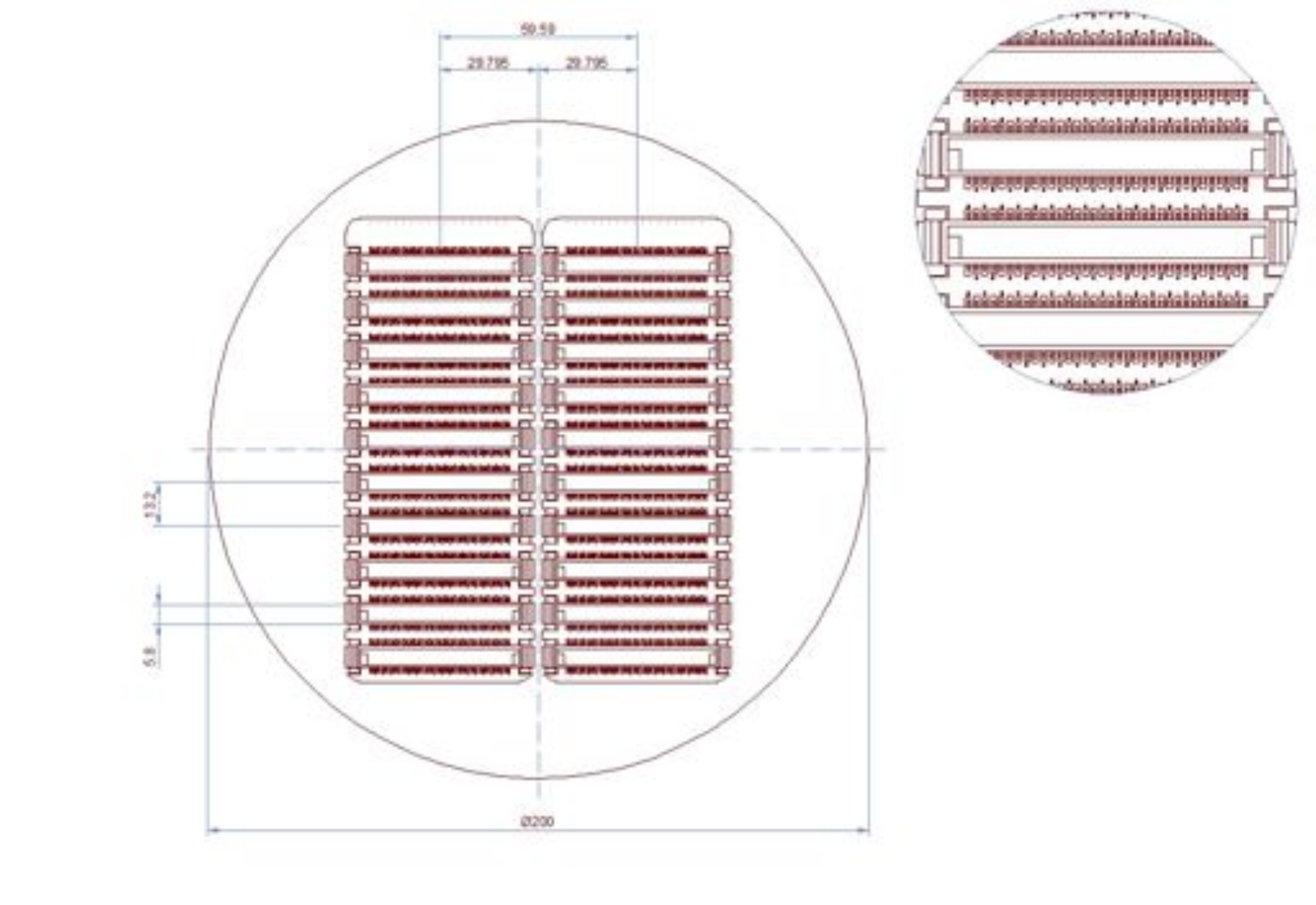} 
\caption{Signal feedthrough and connectors from the CRP (left) 3D view (right) 2D top view.}
\label{fig:flange_666_sft}
\end{center}
\end{figure}

\begin{figure}[h]
\begin{center}
\includegraphics[width=0.95\textwidth]{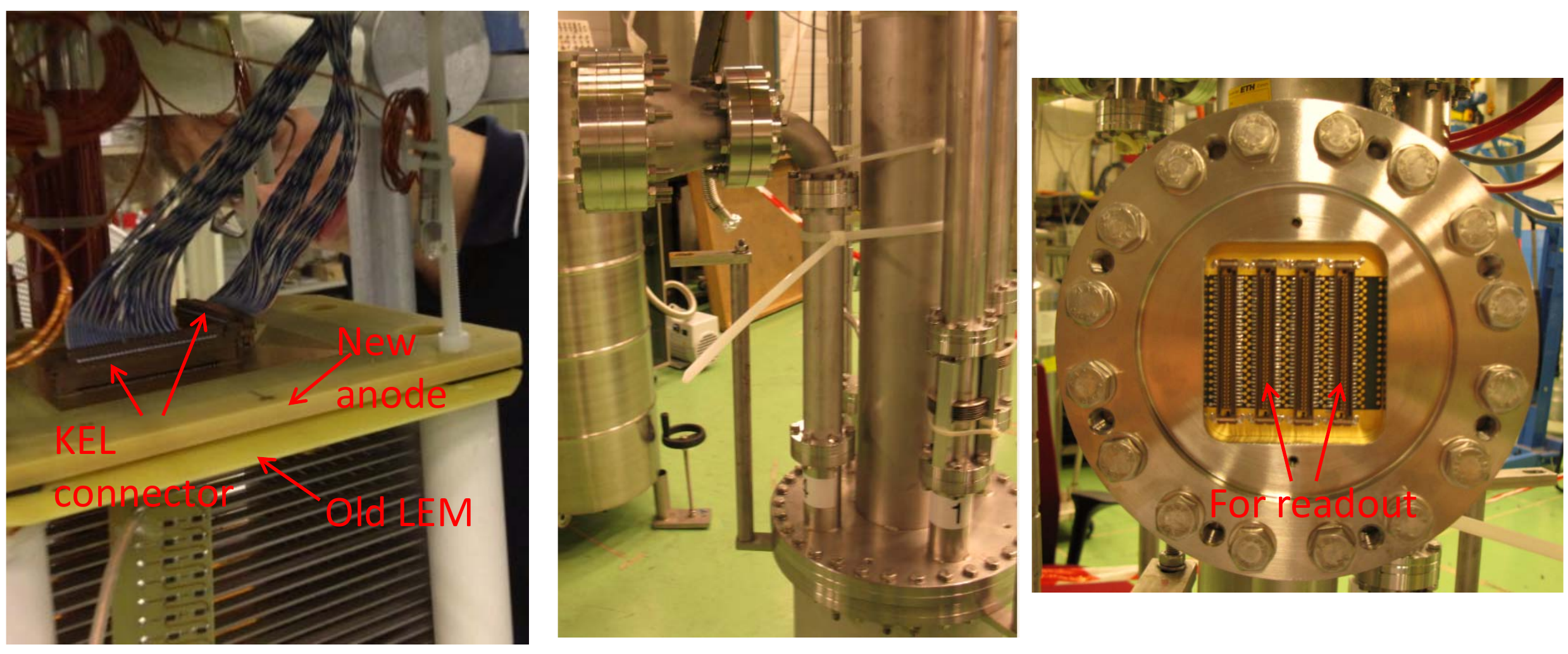} 
\caption{Prototype of the signal feedthrough flange with 4 connectors.}
\label{fig:flange_666_sftproto}
\end{center}
\end{figure}

The dimension of the chosen flange is an ISO CF 200/250
to host 640~channels (20 KEL corporation connectors (\url{http://www.kel.jp/}) with 20$\times$32 channels) as shown in \Cref{fig:flange_666_sft}.
The flange is actually made of a SS ring, in which 
a several mm-thick multilayer PCB is glued or sealed. The multilayer PCB is designed in
order to accommodate connectors on both faces, but internally signals are routed
along a U-shaped path, where the an inner PCB layer is shifted with respect
to the top and bottom layers,  in order not to have passing through holes. 
A prototype with 4 connectors has been built and tested (see \Cref{fig:flange_666_sftproto}).

As mentioned, the signal flanges will be located at the bottom of their chimneys, and below the thermal
insulation panel (See \Cref{figchimney}). 
The temperature of the F/E electronics will be close to that of liquid argon and will
be monitored remotely. The F/E cards with the preamplifiers will be inserted vertically into the
KEL connector. The power dissipation per chimney is 11.5~W which will be taken away to
the main vessel. The chimney will be flushed with dry nitrogen and sealed at the top of the chimney.
This concept will be further developed as to allow access and exchange
of the front-end electronic boards, independently of the main vessel which can remain filled with 
liquid argon. We believe that this is an important requirement for the potential long-term ($>$10~years)
operation of the underground detector.

\subsection{The Light Readout system}
\graphicspath{{./Section-lightreadout/figs/}}

\subsubsection{Scintillation light detection}
Efficient methods to detect the scintillation light in the DUV range have been
studied in the context of ArDM-1t for several years~\cite{Rubbia:2005ge,Boccone:2009kk,Amsler:2010yp}. 
The practical technique for photo-detection based on
large area cryogenic photomultipliers coated with a wavelength-shifter 
is thus well understood. 
For the WLS, we seek a fast response of the light readout system not to distort the pulse shape of the liquid argon scintillation light. Most organic wave shifting materials are well known for their fast optical response caused by the rapid process of radiative recombination of electron hole pairs at the benzene rings in their chemical structure.
The best WLS and  commonly used 
for its well matched emission spectrum to  bialkali photocathodes
is the 
Tetraphenyl-butadiene(1,1,4,4-tetraphenyl-1,3-butadiene or TPB). Different methods for coating the WLS
have been studied in Ref.~\cite{Boccone:2009kk}.
TPB coatings can be made durable with good adherence to the substrate and high resistance to mechanical abrasion.
The light collection efficiency 
is also well reproduced by simulations~\cite{Amsler:2010yp}.

\begin{figure}[htb]
\centering
\includegraphics[width=0.8\textwidth]{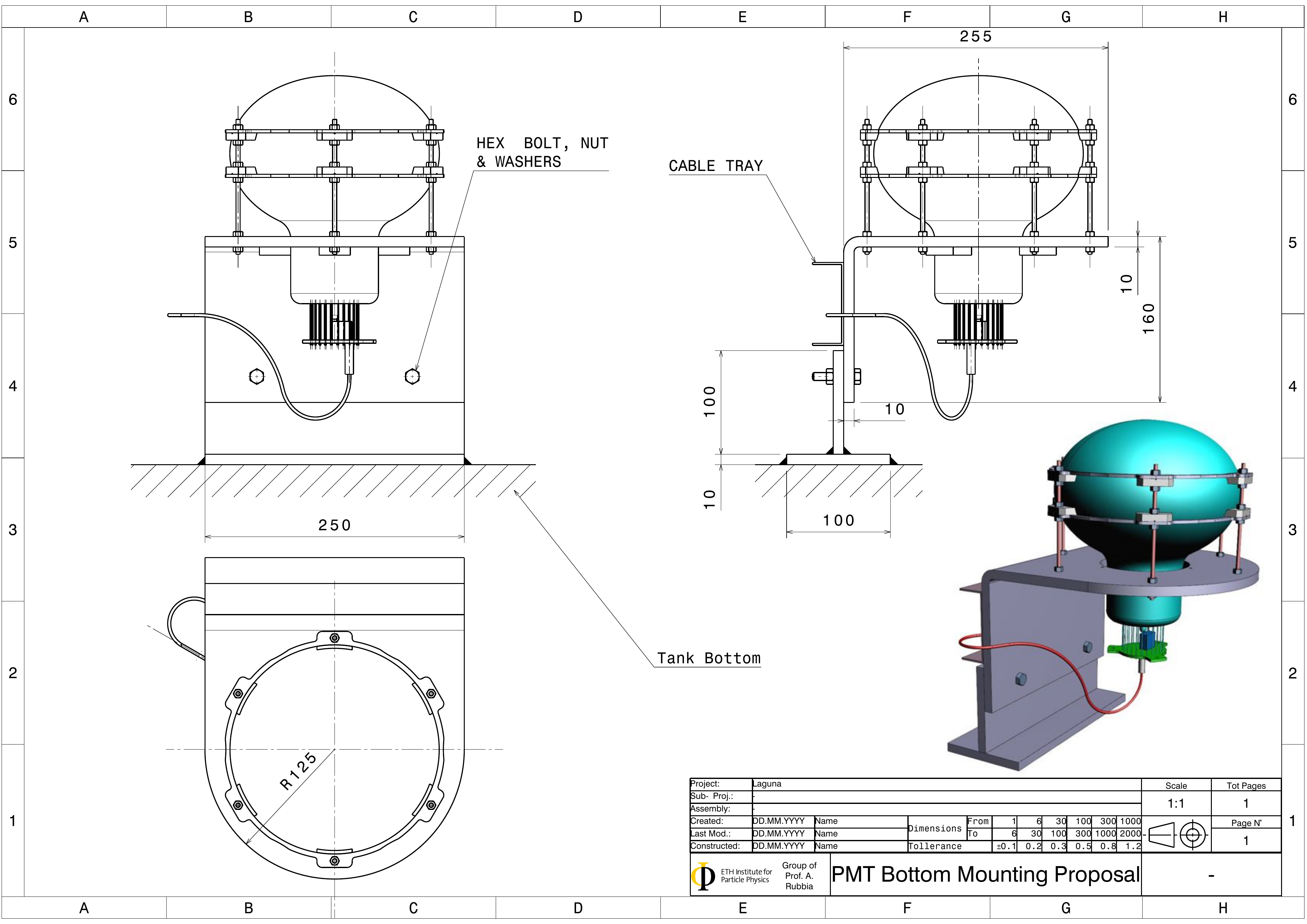}
\caption{PMT mounting arrangement on the bottom of the vessel and its fixation.}
\label{fig:glacier_20kt_pmtattach}
\end{figure}

Based on this expertise, we have developed a baseline design for the light readout of LBNO which is composed
of large area cryogenic PMTs placed uniformly below the (transparent) cathode.
We have estimated that the number of PMT needed to have an efficient light threshold above
several MeV is approximately 1000 for the 20~kton detector, if we adopt 8" photomultipliers with a conservative
quantum efficiency at liquid argon temperature of about 10\%.
The PMT are uniformly located with a coverage of about 1 per square-meter. 
Their buoyancy 
in liquid argon (approx 4~kg) is compensated by appropriately adjusting the weight of
their support. They are anchored at the bottom of the tank, or possibly to the side
of the inner shell wall. See Figure~\ref{fig:glacier_20kt_pmtattach}. The HV cathode is placed
at a height of 2~m above the bottom of the tank and the PMT plane will be distant 
enough from the cathode plane, taking into account the high electrical rigidity of the
liquid argon phase. In order to protect the PMTs an additional mesh plane
will be installed and placed at an identical potential as the PMT photocathode. The
PMT will be negatively polarized to about 1~kV such that the PMT signal can be
readout in DC. We are presently in contact with Hamamatsu Photonics to define
the most optimal configuration of PMT satisfying our requirements, in particular
large area (8" or 12" PMT options are being discussed), cryogenic operation, 
high pressure operation (the PMTs will feel about 2 bar absolute),  and
potentially high QE photocathodes. These solutions will be compared and one will
be adopted for the \six.

\begin{figure}[htb]
\begin{center}
\includegraphics[height=53mm]{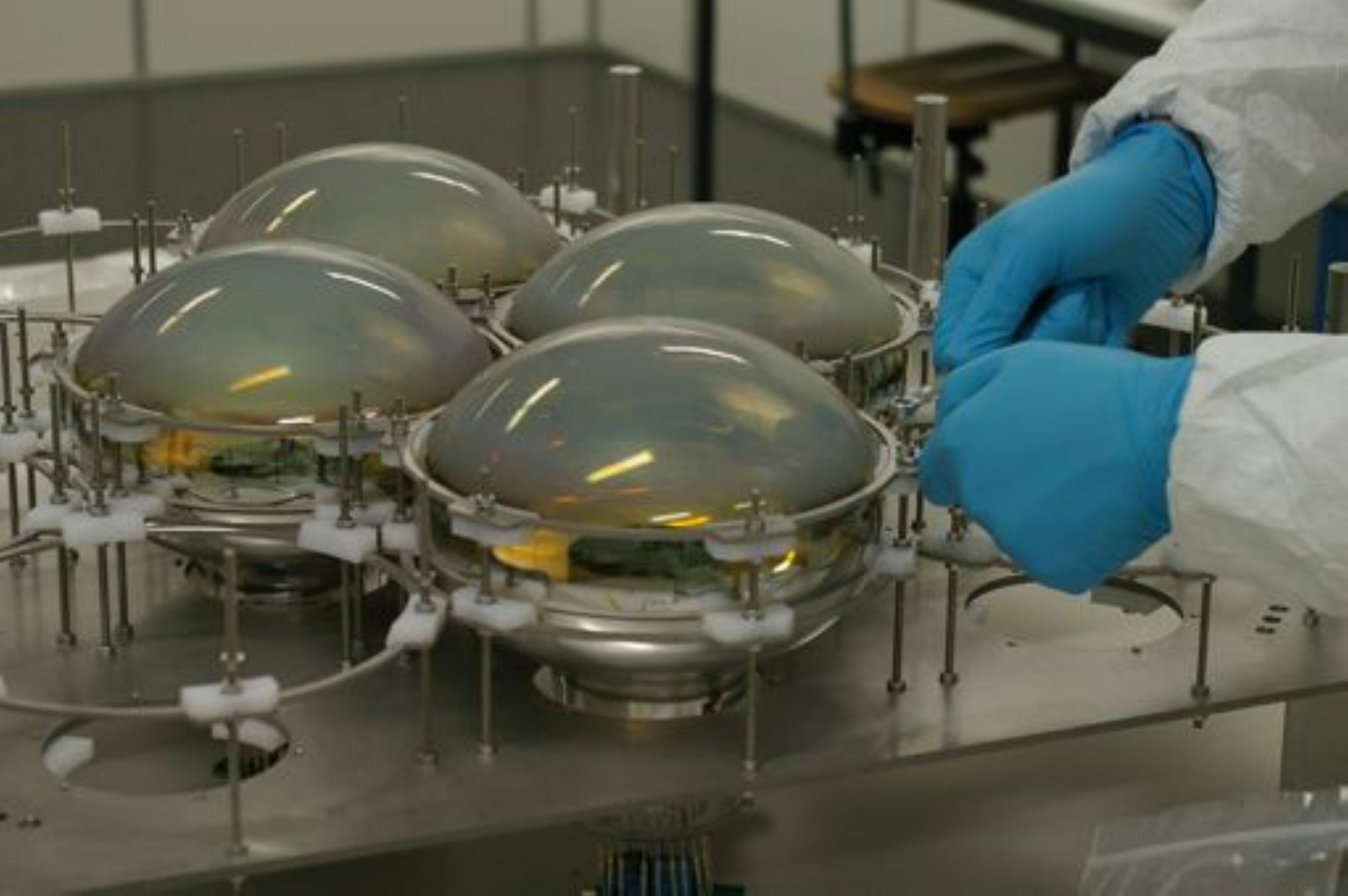}\hspace{5mm}
\includegraphics[trim=18cm 0cm 20cm 16cm, clip=true,height=53mm]{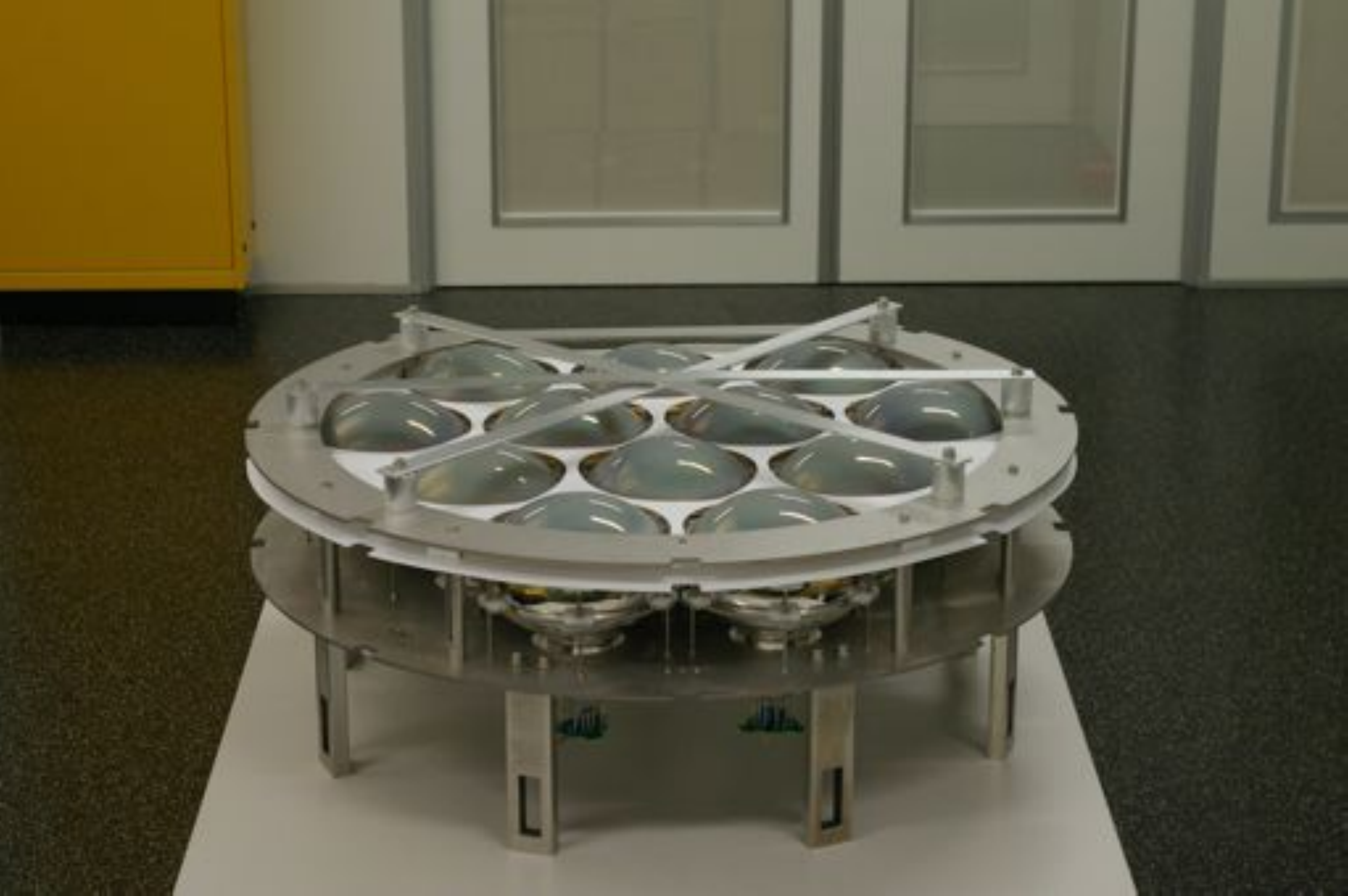}
\caption{The installation and final PMT array for the ArDM experiment~\cite{Marchionni:2010fi,Rubbia:2005ge}
 in the ETHZ clean room. The aluminium crosses over the final array are only for protection during 
 the transport to the Canfranc Underground Laboratory (LSC).}
\label{f_PMT_Assembly}
\end{center}
\end{figure}

A picture of the PMT installation and the final layout of the array developed for the ArDM 
experiment is shown in \Cref{f_PMT_Assembly}.  The PMT were coated with WLS by
TPB evaporation. The aluminium crosses over the final array are only for protection during 
 the transport to the Canfranc Underground Laboratory (LSC).

\subsubsection{Light readout front-end digitiser and DAQ}

In view of the large scale application for the 20~kton detector, the \six prototype will 
allow developing and testing low-cost and high-integration level solutions for the photomultiplier 
signal digitisation on large equipped surfaces. Solutions of this kind have been  studied, in the 
framework of the R\&D programme PMm2~\cite{Campagne1748-0221-6-01-C01081,DiLorenzo:2009rs}, 
for the instrumentation of giant water Cherenkov detectors.

The frond-end electronics for the light readout of the DLAr detector will be based on the solution 
developed within the PMm2 R\&D. The PARISROC ASIC, as described in the following, will be adapted to the 
time structure of the scintillation light produced in the interactions of secondary particles in interactions in LAr. 
The detection of the direct scintillation light is the main purpose of the electronics in order to provide the absolute time.
The system will also be capable to detect the so-called proportional scintillation light produced by the electrons extracted 
and amplified in the gaseous phase, see \Cref{sec:priseclight}.

The solution developed by this R\&D represents an
important handle for costs reduction. The signal digitization is performed by grouping the
photomultipliers in arrays of 16. Each photomultiplier array is then read out by an ASIC
(Application Specific Integrated Circuit) chip in AMS SiGe 0.35 $\mu$m technology. The ASIC,
which is called PARISROC (Photomultiplier ARrray Integrated in Si-Ge Read Out Chip) 
\cite{Campagne1748-0221-6-01-C01081,Genolini2009249}, provides a complete readout system for trigger-less acquisition.

\begin{figure}[htb] 
\begin{center}
\includegraphics[width=0.6\linewidth]{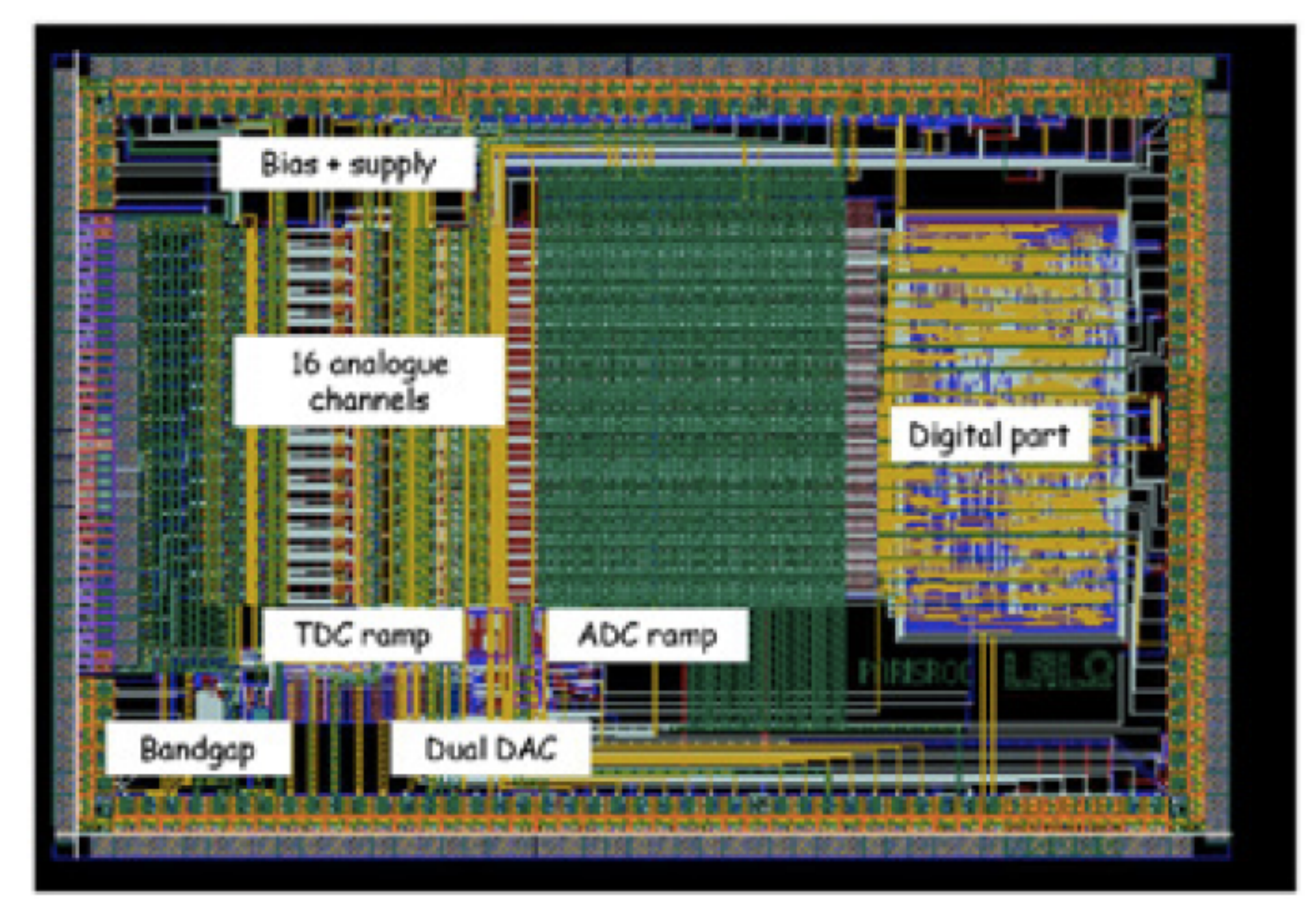}
\caption{Layout of the PARISROC ASIC used for the production of the second iteration of the
chip. (from Nucl.Instrum.Meth. A623 (2010) 492-494).}
\label{fig:asic}
\end{center}
\end{figure}

The PARISROC chip reads the signals of 16 photomultipliers totally independently from each
other. Each analogue channel consists of a low-noise preamplifier with variable and adjustable
gain (8 bits) to compensate the relative photomultiplier gain differences powered by a single high
voltage. The preamplifier is followed by a slow channel for the charge measurement in parallel
with a fast channel for the trigger output. The slow channel includes a variable (50-200 ns) slow
shaper followed by an analogue memory with depth of 2 to provide a linear charge measurement
up to 50 pC; this charge is then converted by a 10-bit Wilkinson ADC. The fast channel is
composed of a fast shaper (15 ns) followed by a low offset discriminator to auto-trigger down to
10 fC. This auto-trigger feature makes the PMT array completely autonomous from the other
PMT arrays. The threshold is loaded by an internal 10-bit DACs common for the 16 channels.

\begin{figure}[htb] 
\begin{center}
\includegraphics[width=\linewidth]{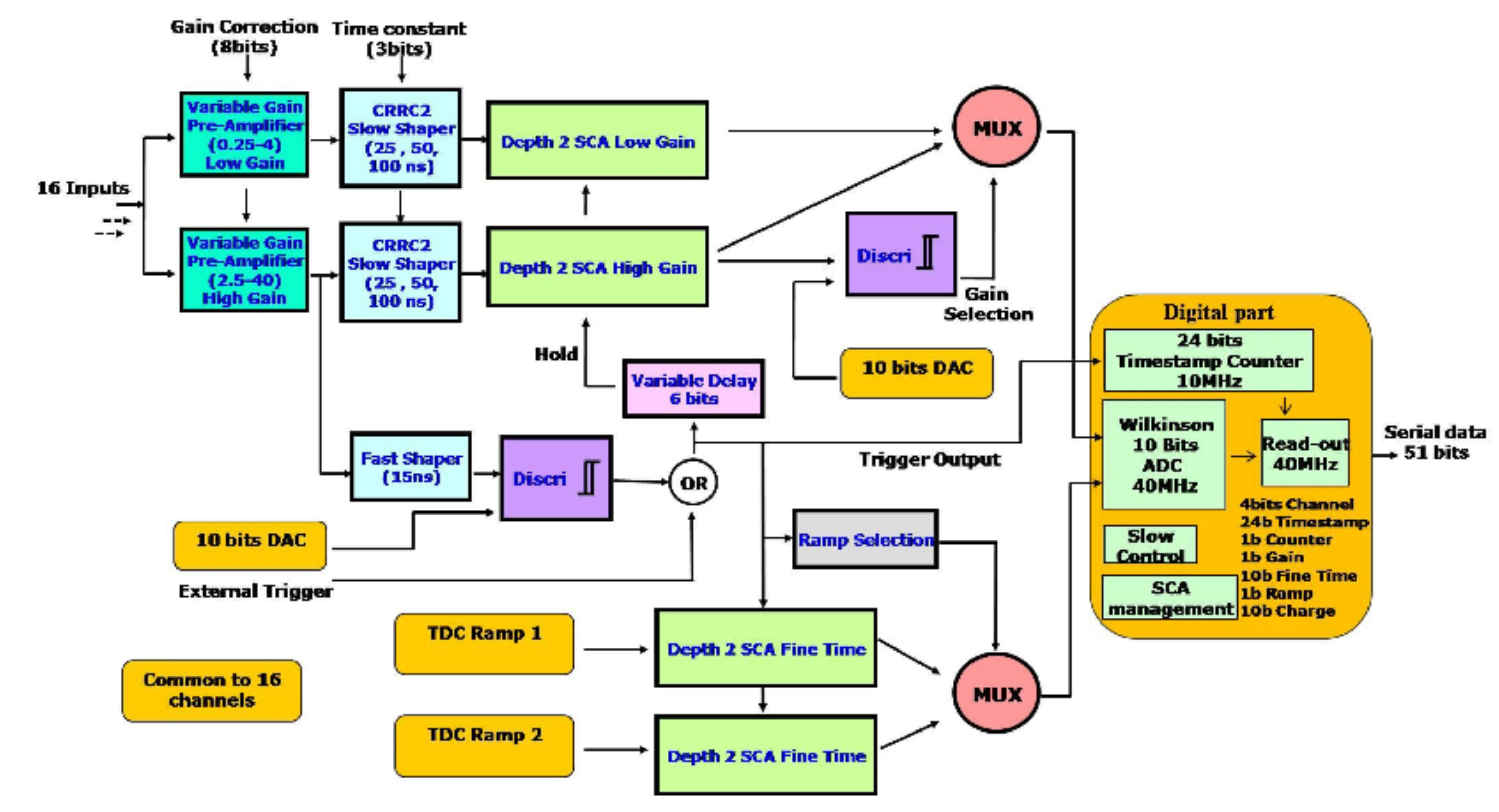}
\caption{Block diagram of the PARISROC ASIC.}
\label{fig:block}
\end{center}
\end{figure}

Each output trigger is latched to hold the state of the response until the end of the clock cycle. It
is also delayed to open the hold switch on the peak output of the slow shaper.
The digital part of PARISROC is built around 4 modules which are acquisition, conversion,
readout and top manager. Currently, PARISROC is based on 2 memories: during acquisition,
analogue signals are stored into the analogue memory and then, during the analogue to digital
conversion, analogue charges and times are converted and stored into a digital memory. At the
end of each cycle, data from this memory are readout to an external system. The analogue to
digital conversion is made via a Wilkinson converter. It is made around a common ramp and
discriminators which sample a digital counter at 40 MHz (for charge and fine time).
The ASIC also provides a complete time stamping solution with a 24-bit coarse time counter at
10 MHz and a fine time measurement with an analogue ramp of 100 ns.
Data are formatted by the ASIC and sent through network cables, which are tight and include also
the clock and slow control connections, to the external data storage. This contributes in
appreciable savings in the connection cables and feedthroughs. The power consumption of the
chip is 15 mW per channel. Figure 6 shows the block diagram of the ASIC.
A traditional digitization scheme with one cable per photomultiplier and external fast digitizers
could be easily setup for the LBNO 6x6x6 m$^3$ prototype, however, in view of the large scale
application for the final detector it is certainly worth to test this high-integration solution which
contributes to reduce the cabling connections and feedthroughs. A mixed configuration can be
envisaged, where different PMT groups can be read out in different configurations, such as
location of the readout cards directly in the argon on the back of the PMTs, or in the evacuation
chimneys on top of the readout plane.

The front end electronics board hosts two main components:
the PARISROC chip coupled to an FPGA that manages the dialog with the surface card.
A series of DC/DC converters
transform the 48 V brought in by the cable to provide the different low voltages for the
electronics and the high voltage power supply common to the 16 PMTs. This reduces by 16 the
cost of the PMT power supply. This high voltage is tunable remotely by a 12 bit DAC. The
PARISROC chip is hosted on a daughter board including its own regulators and supply filters to
avoid noise coupling with the mother board. All the output digital data are read out in serial mode
at a rate of 40 MHz by a FPGA. Special care was taken in the design of the Printed Circuit Board
(PCB) to meet rules of isolation due to the high voltage, rules of filtering, cooling and Electro
Magnetic Compatibility (EMC).

\begin{figure}[htb] 
\begin{center}
\includegraphics[width=0.7\linewidth]{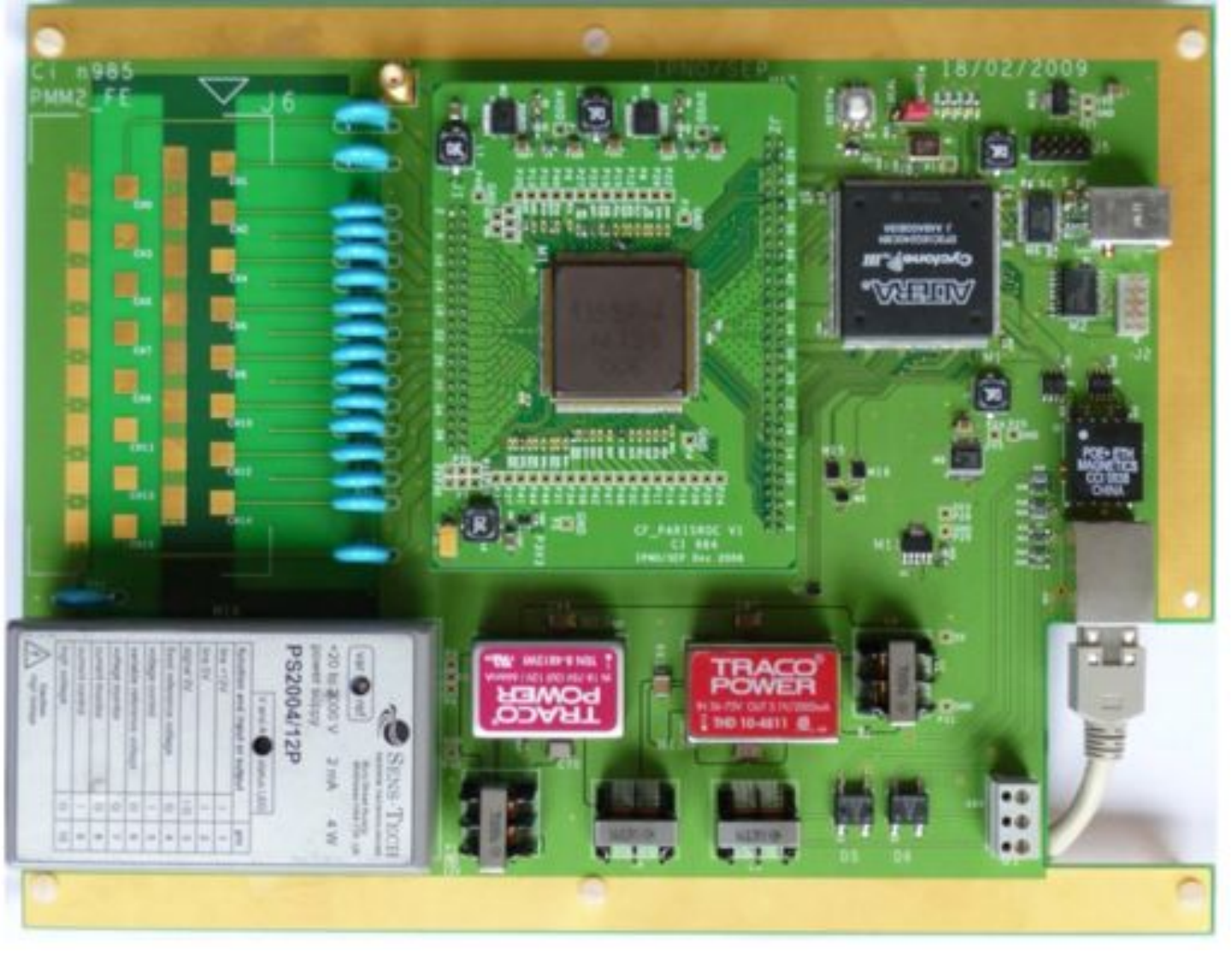}
\caption{Photo of the complete card of the light readout system
under test.}
\label{fig:photo}
\end{center}
\end{figure}

\begin{figure}[htb] 
\begin{center}
\includegraphics[width=\linewidth]{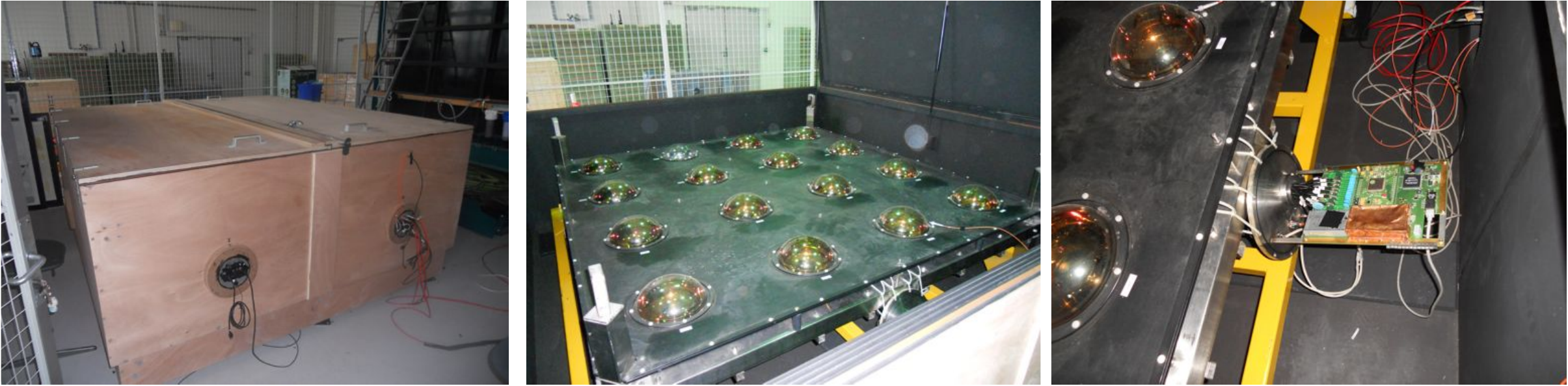}
\caption{The test bench for the PARISROC chip with 16 PMTs. From
left to right: Light-tight box of 2x2x1,5 m$^3$,
matrix of 16 PMTs, front-end card.}
\label{fig:blackbox}
\end{center}
\end{figure}

\begin{figure}[htb] 
\begin{center}
\includegraphics[width=\linewidth]{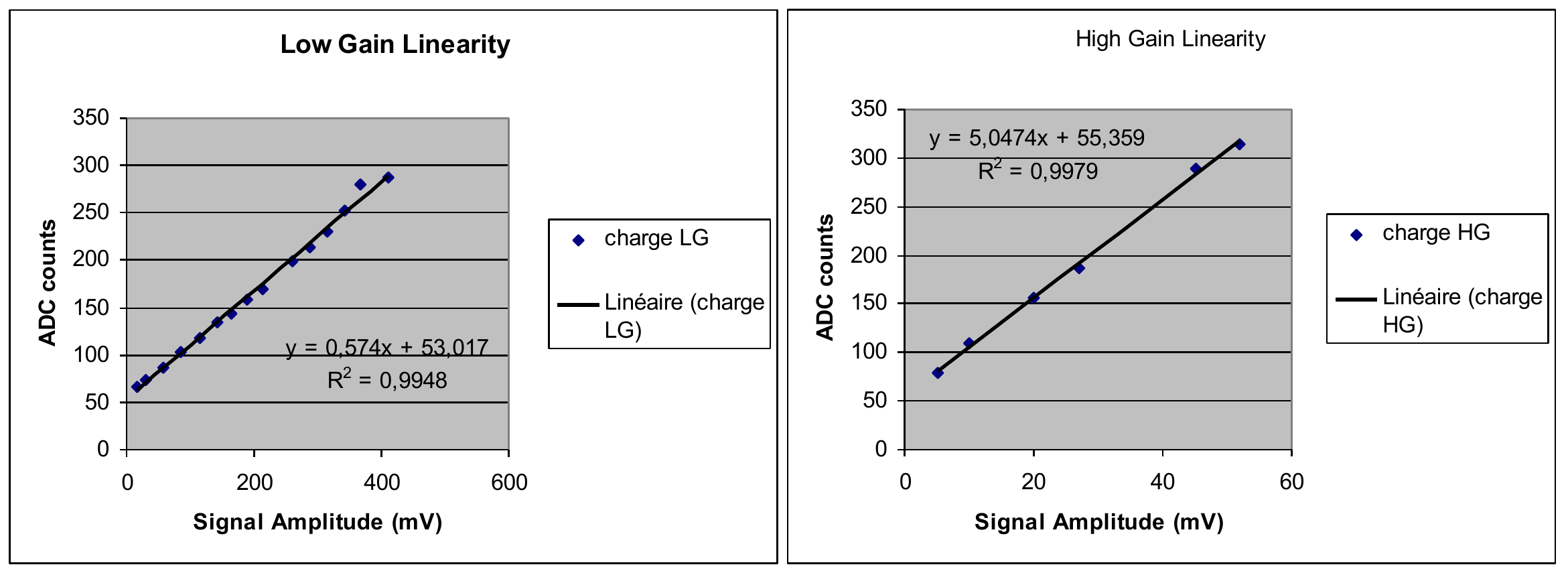}
\caption{Measured gain linearity of the front-end electronic board.}
\label{fig:parisroc_lin}
\end{center}
\end{figure}

The front end card has been successfully tested at room temperature
in a dedicated facility at the APC Laboratory in Paris, shown in Figure~\ref{fig:blackbox}.
The gain linearity has been measured and the result is shown in Figure~\ref{fig:parisroc_lin}.
The PARISROC ASIC can in principle work in liquid argon. An
ongoing development foresees to test and adapt it for this application by possibly increasing the
depth of the Single Channel Analyzer (SCA) and speeding up the digital readout section.

We envisage to study different solutions for the adaptation to operation in liquid argon.
Because of the high electrical rigdity of liquid argon, the system could be brought into the liquid
without a liquid-tight box. The current solution for the HV, common to all (16) PMTs in a matrix,
needs to be adapted: each PMT will receive the HV individually. The heat dissipation in liquid argon and 
the local formation of bubbles around the electronics will be addressed.

\begin{figure}[htb] 
\begin{center}
\includegraphics[width=0.4\linewidth]{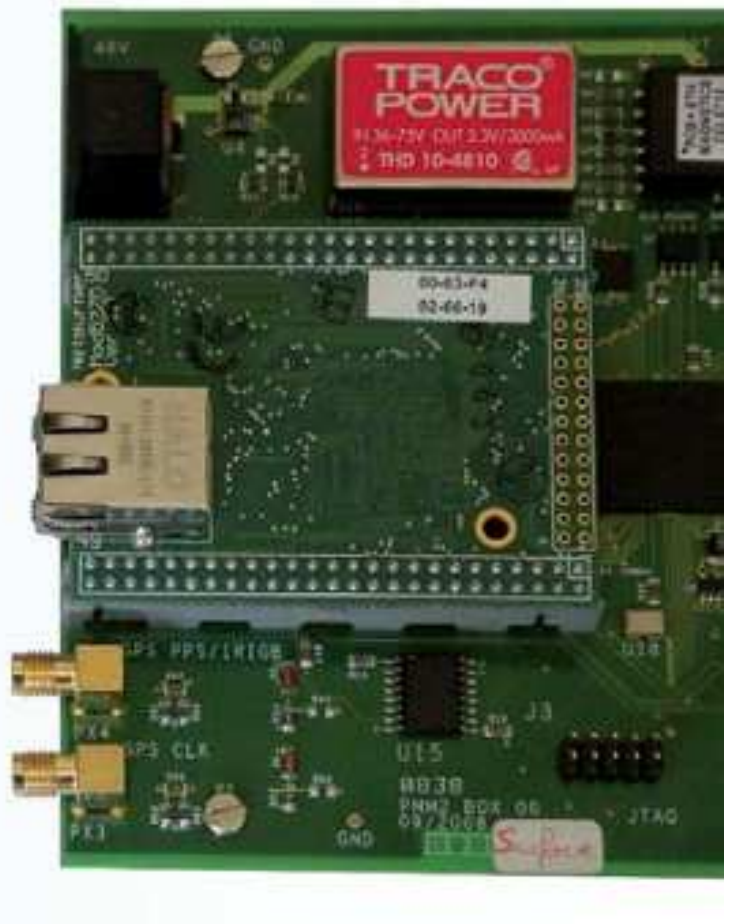}
\caption{Surface card of the light readout system.}
\label{fig:surf_card}
\end{center}
\end{figure}

The second part of the system, situated on surface at room temperature, receives the data
via Ethernet cable. It provides the interface to the global DAQ, the Power Over Ethernet (POV)
for the HV generator, the GPS. This surface card will be redesigned as a dedicated acquisition card imbedded in the microTCA technology used for the general experiment DAQ.
The card is shown in Figure~\ref{fig:surf_card}.

\subsection{Cryogenic vessel}
\graphicspath{{./Section-CryoVessel/figs/}}

The LAGUNA and LAGUNA-LBNO EU FP7 design studies have been focusing on the 
GLACIER concept~\cite{Rubbia:2004tz}, a large double phase Liquid Argon TPC with a long drift and 
charge-sensitive detectors in the gas phase.  According to these studies it is technically feasible to build 
a large (20-50-100 kt) underground tank, based on the LNG industrial technology. 
In a more recent related development, these studies have also considered a novel technique for the tank construction involving 
the membrane technology for the tank. 
This topic has been the subject of several
developments between LAGUNA and industry over many years.
In this technology, the functions of structural support, insulation and liquid containment are all realised by different components, namely an outer concrete structure, specially designed insulating panels and a thin layer of steel plates. These development are very promising for the realisation of a large underground detector, however several areas need to be verified  on a large scale prototype. 

Therefore, the so-called corrugate membrane panels technique (licensed by
GTT/France\footnote{GTT (Gaztransport \& Technigaz), \protect\url{www.gtt.fr}}), has been envisaged as an attractive solution for the
LAGUNA LAr prototype. The inner vessel has a cubic shape with inner dimensions
8.3$\times$8.3$\times$8.3m$^3$.
This volume ensures enough space surrounding the drift cage,
acting as electric insulation ($\sim$1~m of LAr), for safe operation at
HV with up to 300 kV at the cathode (and possibly up to 600~kV, to be tested as
part of the demonstrator).
This volume shall also be used for access and movement inside the
vessel during the construction phase. 
A manhole and a detail-introduction hole are located at the top
face of the vessel.
During the inner detector assembly, additional chimneys are used
to install a controlled air circulation. These additional chimneys are
available for the implementation of the liquid argon process during
normal operation.

\begin{figure}[ptb]
\centering
\includegraphics[width=0.9\textwidth]{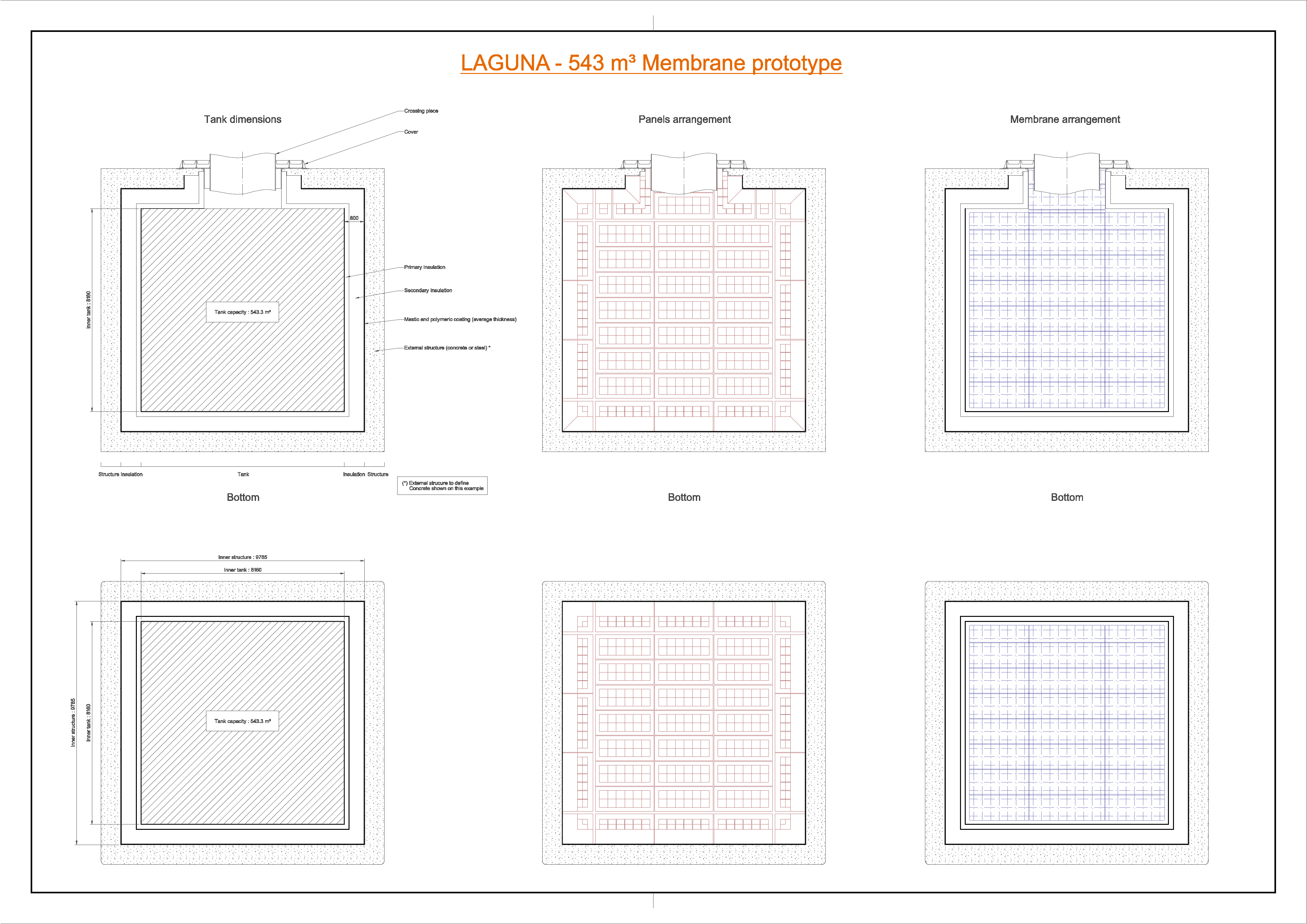}
\caption{Schematic of the 540~m$^3$ membrane vessel developed specifically for 
this proposal by GTT/France. The top cap is not considered at this stage (copyright GTT).}
\label{fig:gttmembrane}
\end{figure}

For practical reasons,
we adopt the panels that come in standard sizes and shapes. A
preliminary panel configuration, developed specifically for us by GTT,
is shown in \Cref{fig:gttmembrane}.
The membrane is 1.2mm thick stainless steel.
The thermal insulation is passive, based on GRPF (glass reinforced
polyurethane foam) layers, interspersed with pressure distributing
layers of plywood. The pressure from the
product is transmitted to a concrete envelope through polyurethane
foam. Its thickness and composition is such to reach a residual heat
input of 5~W/m$^2$ in cold operation.
The total heat input (including the input from the roof and the
cables) in cold operation at LAr temperature is $\sim$2~kW.
The passive insulation is contained in a reinforced concrete
{\textquotedblleft}vessel{\textquotedblright} with $\sim$1.2~m thick
walls. The top outer ceiling is made by a framework reinforced
stainless steel plane, able to support the inner anode and outer
instrumentation (electronics, cryogenics, control). 
A beam pipe (evacuated) for the charged
particles is crossing the concrete outer vessel and the thermal
insulation layers. Its vertical orientation is adapted to the charged
beam vertical axis in its last section. 
A top closed space
({\textquotedblleft}penthouse{\textquotedblright}) is constructed for
hosting the electronic equipment, and various control and monitoring
devices.
A gap space shall remain between the
reinforced concrete vessel and the walls of the pit. The airflow
created by the ventilation and conditioning system of the hall will be
sufficient to keep the temperature of the walls of the pit constant and
avoid condensation on the walls.

\begin{figure}[htb]
\begin{center}
\includegraphics[width=0.9\textwidth]{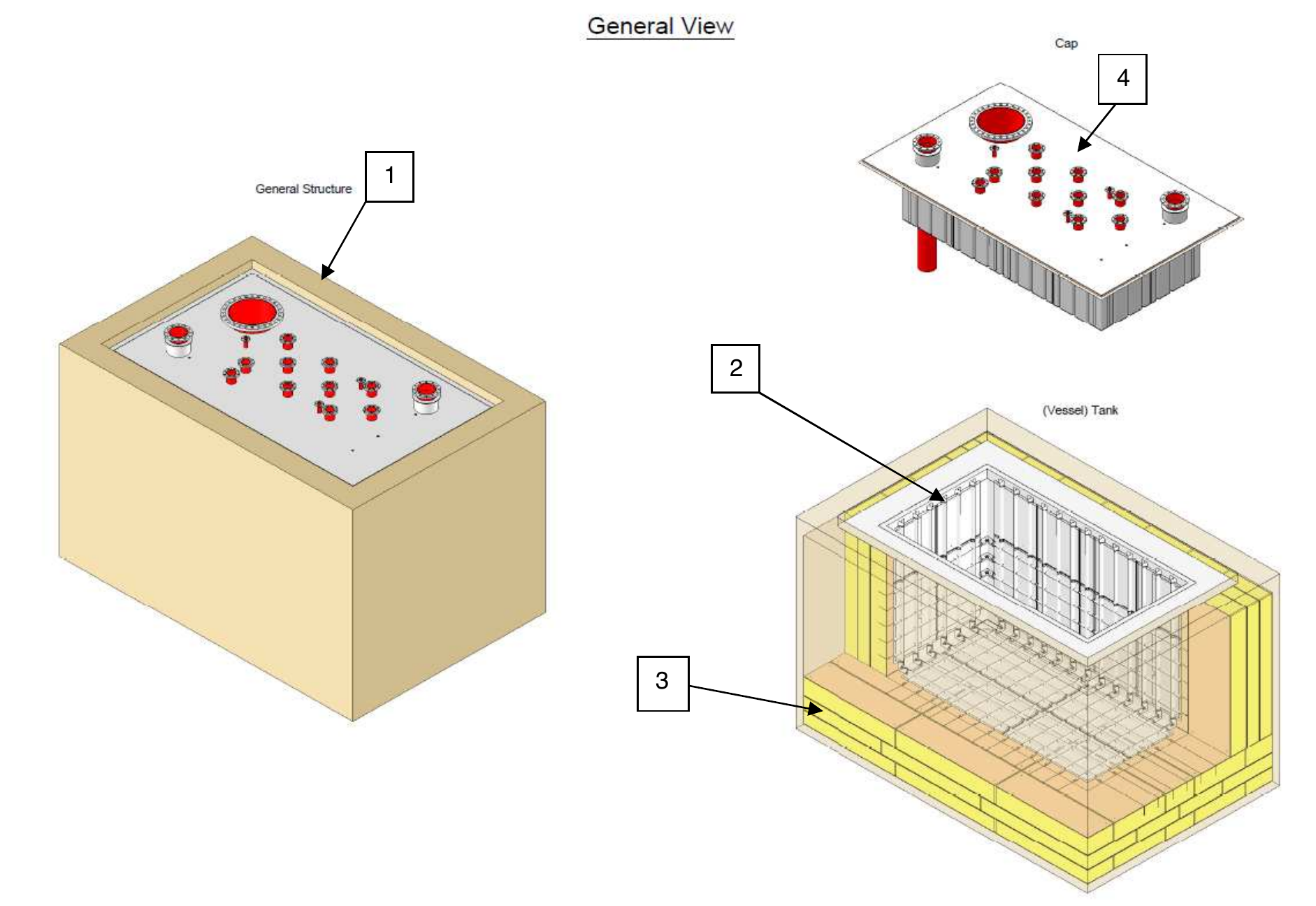} 
\caption{General view of the 17~m$^3$ tank solution presently under construction. Note that the cap
has been specially designed and was the main focus of this development (copyright GTT).}
\label{fig:gttgeneralview}
\end{center}
\end{figure}

In order to understand in more details the use of the corrugated membrane technology,
we have asked GTT to develop the conceptual design of a 17~m$^3$ vessel (See \Cref{fig:gttgeneralview}),
with a central  area of $3 \times 1$~m$^2$, also exploitable for mechanical mockup 
tests of 3 CRP modules, (See Figure 91). The vessel is presently under procurement,
and is planned to be finalised in 2014. A $3 \times 1\times 1$~m$^3$ chamber is under
preparation.
The outer structure (number 1) ensures the structural resistance to contain the LAr, to resist to 
gas pressure and structural weight loads. It will work at ambient temperature and therefore will 
not be submitted to thermal stresses. The current design has considered that this structure will 
be set on pedestal to allow bottom convection.

The corrugated membrane (number 2) acts as primary barrier. It is a liquid and gas tight barrier. The membrane is not a structural component of the system and has a tightness function only. It is a double network of orthogonal corrugations allowing its free contraction/expansion, in two directions, under thermal solicitations.

The insulating structure (number 3) located between the membrane and the outer structure, maintains
the outer tank structure at ambient temperature and keeps the thermal fluxes within the required limit; 
the insulation is said to be "load bearing" in the sense that it transmits the LAr loads from the inner 
containment to the outer structure. This insulation compartment is tight from inside (by the membrane) 
and outside by the outer structure. It is considered as a closed space (insulation space). This space is 
permanently maintained under nitrogen atmosphere, which enables a permanent monitoring. Nitrogen 
breathing allows a control of the integrity of the inner containment tightness.

The Cap (number 4) will ensure the tank tightness from the top and will allow instrumentation to be supported. In addition it has been designed specifically for this Project in order to limit as much as possible the thermal fluxes. The cap will entirely be made out of steel so as to prevent any pollution of LAr and lateral sides will have corrugations in order to compensate for thermal contractions.

\begin{figure}[htb]
\begin{center}
\includegraphics[width=0.6\textwidth]{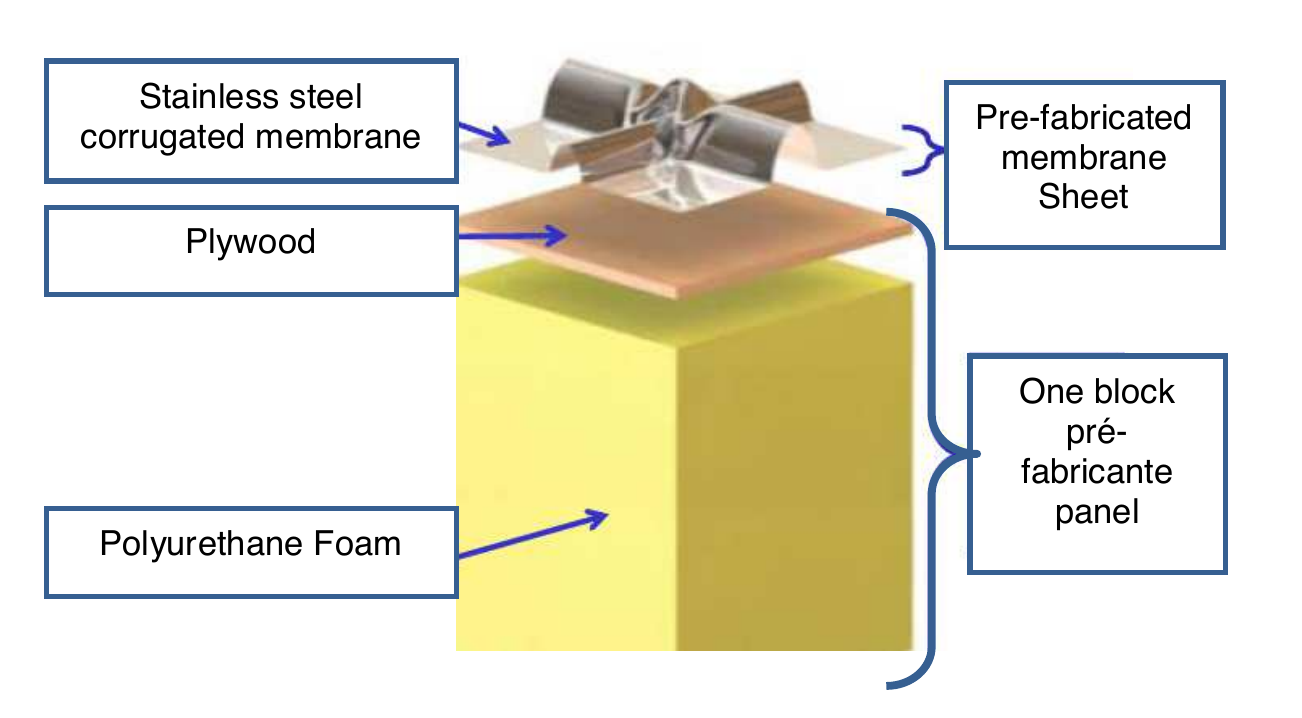} 
\caption{Detail of the GTT insulation panel  (copyright GTT).}
\label{fig:gttinsulationpanel}
\end{center}
\end{figure}

The principle of the membrane system is to uncouple the structural, thermal and tightness functions of the tank. The insulating panels are Òsandwich typeÓ made of rigid closed cells polyurethane foam inserted between two plywood faces bonded to the foam. See \Cref{fig:gttinsulationpanel}.

\begin{figure}[htb]
\begin{center}
\includegraphics[width=0.8\textwidth]{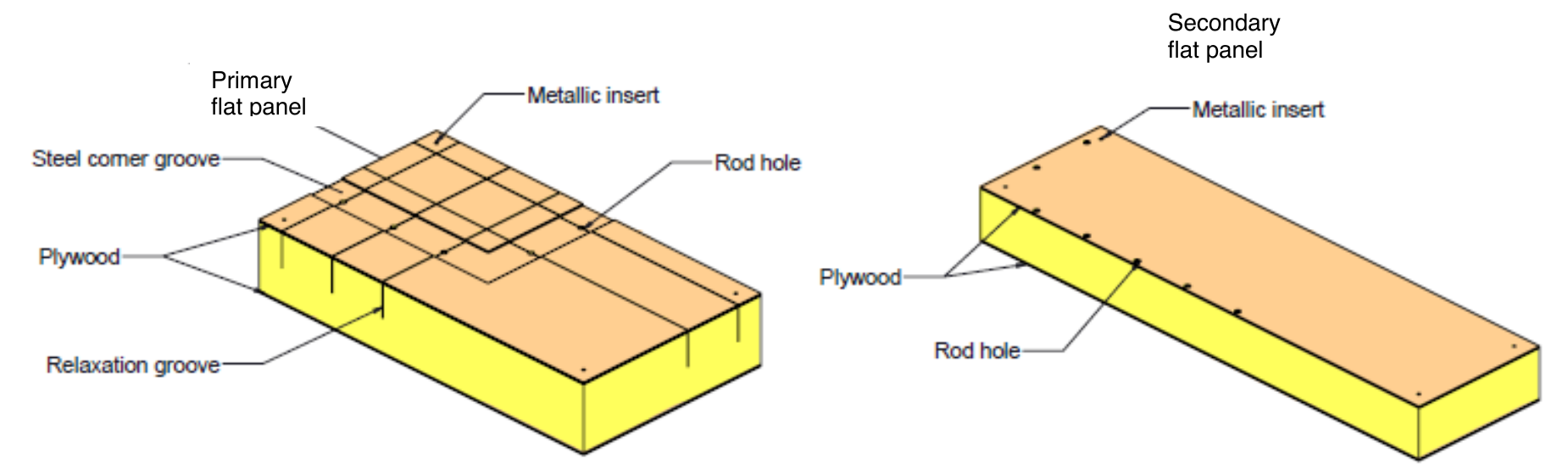} 
\caption{Detail of the GTT insulating panel  (copyright GTT).}
\label{fig:gttinsulatingpanel}
\end{center}
\end{figure}

The insulating structure protects the structural elements from the extreme temperatures. Moreover, it enables to limit ingress of heat and moisture inside the tank. The insulation system is composed of prefabricated insulating panels and flat joints (See \Cref{fig:gttinsulatingpanel}).
The thickness of the insulating panels has been determined upon the required thermal fluxes. The total thickness is 1m divided into three panels (as illustrated: 300 + 300 + 400 mm). The definite specific thicknesses can be adapted to supplierÕs requirement. The size and shape of the panels are optimized to fit easily and properly the whole area of the inner side of the tank. After panel erection, the gap must be filled with glass wool blankets (flat joints) to ensure that no convection effects or thermal bridges occur.

\begin{figure}[htb]
\begin{center}
\includegraphics[width=0.8\textwidth]{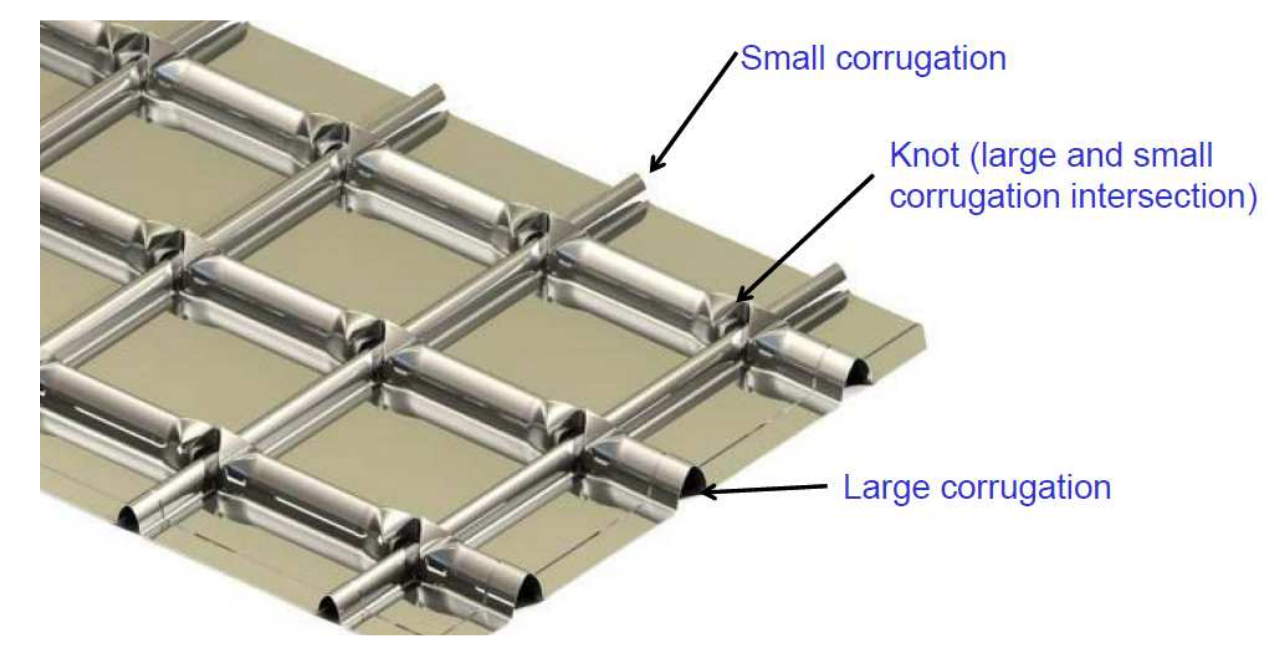} 
\caption{Detail of the GTT metallic membrane panel  (copyright GTT).}
\label{fig:gttmembranel}
\end{center}
\end{figure}

The metallic membrane ensures the containment of the liquid and the tightness of the inner container under normal operating condition. It consists of a stainless steel corrugated membrane on top of the insulating structure all over the tank surfaces (See \Cref{fig:gttmembranel}).
The membrane panels will be displayed on top of the inner containment and lap welded together for liquid and gas tightness.
The corrugations are continuous and cross each other in the form of ÒknotsÓ. The corrugations are formed by dedicated tools using a cold folding process in order to minimize the thickness reduction in the folded parts. One of the families of corrugations is larger than the other one. The sheets shape are optimized to fit easily and properly the whole area of the inner tank.
The metallic membrane consists mainly of flat rectangular sheets. Other shapes are used in the angles of the tank to achieve a proper covering of the corners. When necessary, the corrugations are terminated with dedicated pieces. The junctions between the corrugations of two adjacent walls are achieved by dedicated angle pieces.

After the assembly, the tightness of the vessel will be certified. The standard GTT procedure for testing the tightness is as follows: after the inner container erection is completed, the tightness of all the welds of the membrane is tested by ammonia test. A mixture of nitrogen (75\%) and ammonia (25\%) is introduced in the insulating space, behind the membrane. Welded membrane joints are coated with a reactive paint which changes colour in case of leak in the membrane. All leak test points must be checked to ensure a complete distribution of the $N_2/NH_3$ mixture. Detected leaks must be repaired, leak test points are welded and a new full or global tightness test is carried out. After completion of the test, the insulation space is purged using nitrogen, several vacuum/purging operations are necessary. Finally, the membrane is cleaned and the paint completely removed and evacuated. 
In addition, the tank tightness will be tested with Helium by inserting it the gaps where the insulation is located and operating a helium leak checker inside the main vessel volume.

The cap was given special attention. The crossing pipes (holding the signal and electrical feedthroughs)
were designed according to their function, and their thermal properties and stress have been studied.
Drawings have been produced for the containment system, the cap configuration and the assembly
procedure, along with a construction schedule.
The total inner volume is $3\times 4.8\times 2.4$~m$^3$ = 35~m$^2$. The volume of stored LAr
is 17~m$^2$ with the level of argon at 1.5~m. The residual heat input is 5~W/m$^2$ for a total
BOR of 0.069\%. The insulation thickness is 1~m with a density of 70~kg/m$^3$ (PU Aged HFC245).
The temperature of the LAr will be stored between 86.7 and 87.7~K at a pressure of the tank
$950<P<1050$~mbar.

The instrumentation will be suspended by the cap which has therefore to ensure a mechanical requirement
as it will have to support the weight of the insulation and of the instrumentation. The entire vessel will
also to ensure proper ultra-high vacuum tightness (i.e. He leak rate $<1e-9$~mbar lt/s.
The crossing pipes have been arranged as to provide the optimal path for cables to the readout anode.
Three chimneys are in addition used to hang the anode to the top cap. The cap can be opened
and closed (welded) about four times in the lifetime of the vessel.

The cap will be welded on its periphery to ensure the tightness. In addition all crossing pipes will be welded 
to ensure that no pollution of the Argon occurs. Its total weight will be 3.8~tons.

\begin{figure}[htb]
\begin{center}
\includegraphics[width=0.975\textwidth]{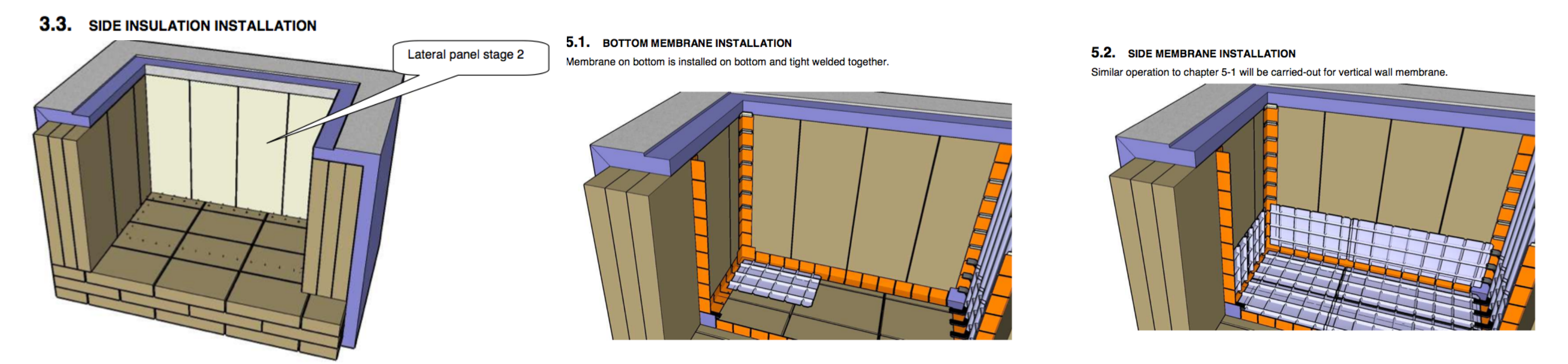} 
\caption{Some steps of the installation sequence  (copyright GTT).}
\label{fig:gttinstallseq}
\end{center}
\end{figure}

The installation sequence has been studied. It consists of several steps for a total execution time of about 6 weeks.
A  concrete or a steel structure can be contemplated for supporting tank and must be assembled first.
Accurate marking will be performed in order to define the exact anchoring system position (rods).
In order to fix the membrane corner elements, rods will be fixed on tank bottom.
These rods will be either welded in case of metallic tank, either screwed in insert in case of concrete structure.
The bottom insulation will be installed next. Lateral rods will be fixed.
Once rods are fixed, vertical insulated panels are installed. As per bottom, insulation is made of 3 layers of insulation.
See \Cref{fig:gttinstallseq}. Steel corners are installed on bottom and vertical corners to anchor the membrane. 
Securing of these corners steel is ensured by the rods. Then the membrane is positioned and welded.
All along corner area membrane will be tighten by means of angle pieces on bottom and vertical area,
overlap on the top part of the vertical area, and end cap on the top part of the tank.


\subsection{Liquid Argon process}
\label{sec:larprocess}
\graphicspath{{./Section-LArprocess/figs/}}


\subsubsection{Sources of impurities from outgassing}\label{c_outgassing}



When pumping a volume from atmospheric pressure, the gas in the volume is removed whereas the speed of evacuation is mainly based on the speed of the vacuum pump. Down to a pressure of $\sim 1 \unit{~mbar}$, this is done with a roughing pump and at lower gas pressures, when the air resistance is low enough, with a turbo molecular pump. The pressure drops exponentially with the time as a function of the pumping speed $S$ and the to be evacuated volume $V$:
\begin{equation}
p(t)\approx p_{0} \, e^{-\frac{S\,t}{V}}
\end{equation}
The gas flow while pumping can be assumed to be laminar. This behavior is true down to about $10^{-3}$ mbar, depending on the used pump. At lower pressures there is no longer any gas flow, but a molecular flow, since single molecules are removed. This is where the turbo molecular pumps become most efficient. In this pressure range, other effects, usually summarized as outgassing, become important. They are complex processes and different mechanisms at different pressure ranges and will be discussed in the following sections:


\paragraph{Zero order desorption}
A first effect, after most of the gas in a volume is removed, is desorption from the surfaces. When there are multiple layers of material sticking to the surface, (e.g. there is water laying in the chamber) the desorption rate is constant. The rate with which molecules are desorbed depends on the latent heat of vaporization $E_{v}$ and the temperature $T$ \cite{Redhead1995_1}:
\begin{equation}
-\frac{dn}{dt}=\alpha \, e^{\frac{-E_{v}}{kT}}
\label{e_desorption}
\end{equation}
where $\alpha$ is a constant depending on the initial pressure.

\paragraph{First order desorption}
When coming to a monolayer of molecules sticking to the surface, the behavior changes. The desorption rate becomes time dependent, according to the molecular density on the surface. To a first approximation it can be assumed that particles are released from the surface, and do not return to it, with a rate proportional to their surface concentration $C$ \cite{OHanlon2003}. This process can be written as:
\begin{equation}
\frac{d\, C(t)}{d\,t}= -K_{1}C(t)=\frac{e^{-E_{d}/N_{A}kT}}{\tau_{0}} \, C(t)
\label{e_FirstOderDesorption}
\end{equation}
where the rate constant $K_{1}$, strongly depends on the activation energy of desorption $E_{d}$ (per mol) and the temperature $T$. $N_{A}$ is the Avogadro number and $\tau_{0}$ the nominal vibration period of an adsorbed molecule; typically of the order of $10^{-13} \unit{~s}$.

By integrating Eq.~\ref{e_FirstOderDesorption} the surface concentration can be written as:
\begin{equation}
C(t)=C_{0} \, e^{-K_{1}t}=C_{0} \, e^{-t/\tau_{r}}
\end{equation}

where $\tau_{r}$ is the average time of a molecule sticking to the surface, i.e. the residence time of a molecule on the surface. It is, using Eq.~\ref{e_FirstOderDesorption}, given by
\begin{equation}
\tau_{r}=\frac{1}{K_{1}}=\tau_{0} \, e^{E_{d}/N_{A}kT}
\label{e_residence_time}
\end{equation}

\paragraph{Second order desorption}
A lot of gas atoms are bound in a diatomic molecular state and they can only be desorbed as a molecule. For example, hydrogen bound to the surface first has to become an H$_{2}$ molecule before being able to be desorbed. 
This process, called second order desorption, is given by a different rate constant $K_{2}$ and the concentration can be written as \cite{OHanlon2003}
\begin{equation}
\frac{d\, C(t)}{d\,t}= \frac{-K_{2} \, C_{0}^{2}}{(1+C_{0} \, K_{2} \, t)^{2}}
\label{f_second_order_desorption}
\end{equation}
$K_{2}$ depends on the desorption energy $E_{d}$ and the temperature. Therefore, it is clear that second order desorption rate, proportional to $1/t^{2}$ is much smaller than the exponential decay of the first order desorption. 

\subsubsection{The desorption from real surfaces}
In a real vessel, gas is not automatically pumped out when desorbed from the surface. A particle will bounce on the walls several times and can be re-adsorbed. With each re-adsorption, it sticks on average for the time $\tau_{r}$ to the surface before being desorbed again. This leads to an experimentally measured outgassing rate $q$ of 
\begin{equation}
q\propto \frac{1}{t^{\alpha}}
\end{equation}
where $\alpha$ is a coefficient approximately equal to 1 \cite{Redhead1995}. This coefficient can be time dependent and varies with the composition of the outgassing components. 
In general the $1/t$ behavior is given for all molecules but for different kind of molecules the time constant of outgassing varies. Therefore, the total desorption is the sum of the desorption of the different molecules. 

Since the residence time (Eq.~\ref{e_residence_time}) is a function of $E_{d}$ and $T$, heating up the walls is increasing the speed of outgassing. For example, for water molecules sticking to a stainless steel surface, the desorption energy $E_{d}$ is $23 \unit{~kcal/mol}=96\cdot 10^{3} \unit{~J/mol}$ and the residence time is $10^{4} \unit{~seconds}$ at room temperature of $22 \unit{~^{\circ}C}$. Increasing the temperature of the chamber walls, gives a strong decrease of the residence time and for $350 \unit{~^{\circ}C}$ it is about $10^{-5} \unit{~s}$. It is therefore favorable to bake out the  components and the inside walls of the vacuum chamber. But doing so, it has to be taken into account that all the chamber must be heated up and not only one part of it. As mentioned before a molecule released is most likely adsorbed again on another surface. Having now, for example, heated up half of the surface to $350 \unit{~^{\circ}C}$ while keeping the other half at $22 \unit{~^{\circ}C}$ gives an average residence time of $\frac{1}{2}(10^{4} \unit{~s}+10^{-5} \unit{~s})\approx 5\cdot10^{3} \unit{~s}$ corresponding to an average temperature of $27.4 \unit{~^{\circ}C}$. 
This shows, that always the coldest surface is dictating the outgassing rate. Thus partial heating of one and later another part of the vessel is not having a significant impact on the overall performance of the vacuum.

\subsubsection{Diffusion}
At a certain pressure most residual gases on the surface of a chamber have been evacuated and diffusion is the major source of gas molecules. This is the transport of molecules inside materials. Gas molecules are moving from inside the detector parts and the walls of the cryostat to the surface where they desorb into the vacuum. 
Because diffusion is on a much slower time scale than desorption, the residence time for releasing the molecules from the surface can be neglected.
The outgassing from diffusion of a solid material for an initial gas concentration $C_{0}$ is
\begin{equation}
\label{eqn_section_vacuumproperties_diffusion}
q=C_{0}\left(\frac{D}{t}\right)^{1/2}\left[1+2\sum_{0}^{\infty} \left( -1 \right) ^n \exp\left( \frac{-n^2 d^2}{D t}\right) \right]
\end{equation}
where $[D]=\unit{m^{2}/s}$ is the diffusion coefficient and $2d$ the thickness of the material. The gas concentration $C_{0}$ is given as residual pressure in Pascal and the diffusion rate $q$ in $[q]=\unit{(Pa \cdot m^{3})/(m^{2}\,s)}$ \cite{Calder1967,OHanlon2003}. 

Instead of solving this equation one has a look at its asymptotic solutions. For short times $(t\simeq0)$, only the first term is of importance and Eq.~\ref{eqn_section_vacuumproperties_diffusion} becomes approximately
\begin{equation}
q\approx C_{0}\left(\frac{D}{t}\right)^{1/2}
\end{equation}
This shows a decrease in the rate proportional to $t^{-1/2}$, which is much slower than the rate given by desorption. For long times, the infinite sum of Eq.~\ref{eqn_section_vacuumproperties_diffusion} has to be taken into account and, by a translation of the mathematical zero of the summation from the center of the volume, at distance $d$ to one of the surfaces, Eq.~\ref{eqn_section_vacuumproperties_diffusion} reduces to a rapidly converging sum and becomes
\begin{equation}
q=\frac{2D \,C_{0}}{d} \, \exp\left( -\frac{\pi^{2}D\,t}{4 \, d^{2}}\right)
\end{equation}

This change in slope is a result of the depletion of gas particles inside the outgassing material \cite{Calder1967,OHanlon2003}.
The diffusion coefficient $D$ not only depends on the diffusing gas and the material it diffuses from, but it is also a function of the thermal activation energy of the gas in the solid. 
\begin{equation}
D=D_{0} \, e^{-E_{D}/kT}
\end{equation}
Therefore, similar as for desorption, heating up the material leads to a much faster outgassing and is preferred. 

\subsubsection{Permeation} \label{c_permeation}A surface facing the atmosphere is never completely depleted but also is adsorbing gas molecules of the atmosphere, which then diffuse through the material (walls) and get desorbed into the vacuum. The total motion can be described as an adsorption followed by diffusion and finally a desorption in the vacuum. Having a combination of the above described processes shows that the rate is much smaller than from the individual processes. This is called permeation. Because of the constant supply with new molecules form the atmosphere, permeation cannot be stopped. The only possibility to further reduce the pressure in a vessel is to seal the chamber wall on the outside with another, less permeable material or, more radically, change the chamber to one made of a different material or with thicker walls. 
For steel surfaces, hydrogen is the main element permeating through. The hydrogen molecule cannot permeate as a molecule but it dissociates on the metal surface into two single atoms that then diffuse through the metal. On the vacuum side of the steel wall the atoms have to recombine to a molecule to be desorbed. The permeation rate is proportional to the square root of the pressure difference \cite{OHanlon2003}.
Different from metals, in plastics, ceramics and glass the whole molecules can permeate through. For glass that is mainly the smallest existing molecule, the helium atom. This is of importance since photomultipliers are glass bulbs. Helium atoms diffusing into the tubes ionize in the strong electric field and destroy the photocathode. 
For this reason, leak testing a chamber, containing PMTs, with a helium leak tester should be omitted. 
Note, that permeation is only observed in ultra high vacuum chambers and was never observed in 
tests performed in smaller LAr TPC by the ETHZ group.

\subsubsection{The $40\times80 \unit{~cm^{2}}$ LEM TPC as an example}
\label{s_250l_purification}

To give a practical example, the test of the $40\times80 \unit{~cm^{2}}$ LEM TPC is considered \cite{Badertscher2012, Badertscher:2013wm}. The detector is mainly made from glass epoxy boards (G10). This material has a lot of epoxy resin and therefore solvents in it and a strong outgassing must be expected. Besides the actual detector, there are also cables, electronic components and the support system for hanging the setup inside the ArDM cryostat. All these components are outgassing and possible sources of impurities. \Cref{t_material_250l} gives an overview of the material budget and the corresponding surface areas. The table also gives an expected outgassing rate per square meter after 24 hours of pumping, derived from several sources.
These rates are average values to get a feeling, what is contributing how much to the outgassing. If there were several values given for the same material, an average has been taken, normalised to 24 hours by assuming an outgassing rate of $R(t)\propto 1/t$.

\begin{table}
\begin{tabular}{ccccl}\hline 
Material & Surface & Outgassing rate & Total outgassing & Reference\\
&$ \left[ \unit{m^{2}} \right] $ & @ $t= 24 \unit{~hr} \left[ \unit{torr \, l / s \, cm^{2}} \right] $ & @ $t= 24 \unit{~hr} \left[ \unit{mbar \, l / s} \right] $ &\\
\hline
Steel & $10.74$ & $2.00 \cdot 10^{-10}$ & $2.86 \cdot 10^{-5}$ & \cite{Edwards1977}\\
G10 & $5.60$ & $1.22 \cdot 10^{-7}$ & $9.11 \cdot 10^{-3}$& \cite{Yamamoto1979}\\
Copper & $1.25$ & $3.01 \cdot 10^{-11}$ & $5.03 \cdot 10^{-7}$& \cite{Koyatsu1996}\\
Nylon & $0.013$ & $2.50 \cdot 10^{-7}$ & $4.20 \cdot 10^{-5}$& \cite{Wong2002, barton1965}\\
Polyethylene & $0.53$ & $6.00 \cdot 10^{-8}$ & $4.25 \cdot 10^{-4}$& \cite{erikson1984}\\
Kapton & $0.30$ & $7.50 \cdot 10^{-8}$ & $3.00 \cdot 10^{-4}$& \cite{Wong2002}\\
Polyolefine & $1.6$ & $5.25 \cdot 10^{-8}$ & $1.12 \cdot 10^{-3}$& \cite{Montanari1998}\\
Teflon & $0.02$ & $ 8.49 \cdot 10^{-9}$ & $2.72 \cdot 10^{-6}$& \cite{Wong2002}\\
\hline
\multicolumn{3}{l}{\textbf{Total outgassing after 1 day}}& $\mathbf{ 1.1 \cdot 10^{-2}}$&\\
\hline \end{tabular}
\caption{Material and calculated outgassing rate of the $40\times80 \unit{~cm^{2}}$ LEM TPC, tested in the ArDM cryostate. All rates are given for $t=24 \unit{~hr}$ from the beginning of the evacuation.\label{t_material_250l}} 
\end{table}

\begin{figure}[th!]
\centering
\includegraphics[height=80mm]{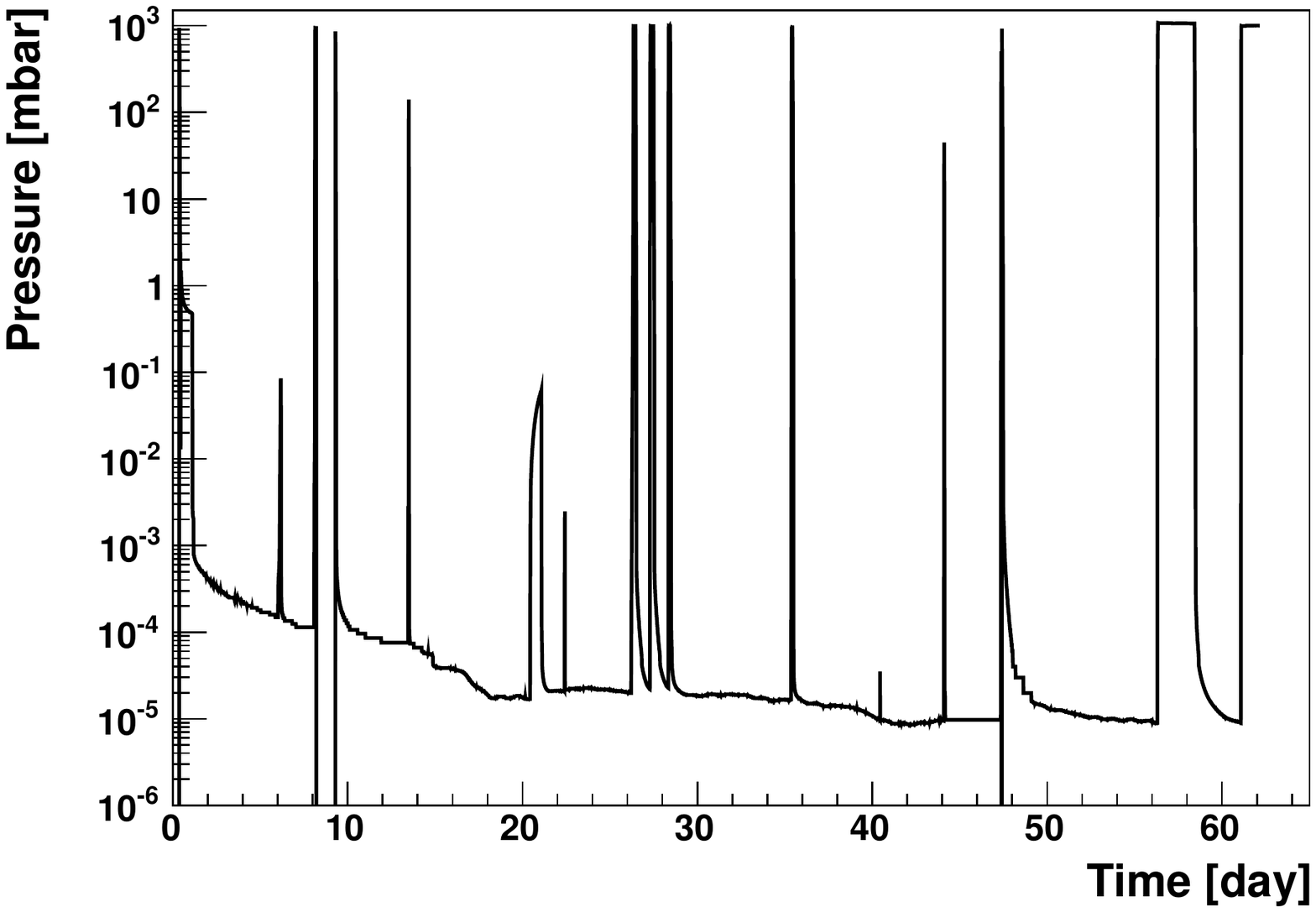}
\caption[Pressure curve as a function of the evacuation time for the $40\times80 \unit{~cm^{2}}$ LEM TPC.]{The pressure as a function of the evacuation time for the $40\times80 \unit{~cm^{2}}$ LEM TPC. The pumping started on August 10, 2011 and lasted for 62 days, before starting to fill the detector with argon. The peaks in the curve are coming from tests with pure argon gas and also from venting the vessel. (See text for details)}
\label{f_TotalVacuumCurve}
\end{figure}

To test our model, outgassing measurements were taken several times during the pumping of the volume. The overall pumping time was around 65 days. \Cref{f_TotalVacuumCurve} shows the pressure inside the cryostat as a function of the days since pumping started. The curve is not smooth
but has several spikes, above as also below the actual pressure curve. 

Values below $10^{-6} \unit{~mbar}$ are indicating the times when the vacuum gauge was switched off and the spikes above the smooth curve have different origins. There are, for example at $\sim 6 \unit{~days}$ and at $\sim 20 \unit{~days}$, measurements of the outgassing with the Rate-Of-Rise method. Other spikes, when reaching more than 1 mbar, are from moments when the detector was filled with argon gas for calibrating the PMTs or to do tests with the charge readout. Also, the cryostat was filled with argon gas when a flange or other components on the top flange had to be replaced.
At $\sim 48 \unit{~days}$ the detector was completely opened again for a few hours, because of a problem with an electrical contact. It can be seen that it took a while to reach the previous vacuum again because of newly installed PCB parts. In the few hours while the detector was exposed to air some water molecules attached to the detector but the main reason for the slow reaching of the previous vacuum value is presumably the water (and other residual gases) in the new glass epoxy parts. 

\begin{figure}[t!]
\centering
\includegraphics[height=80mm]{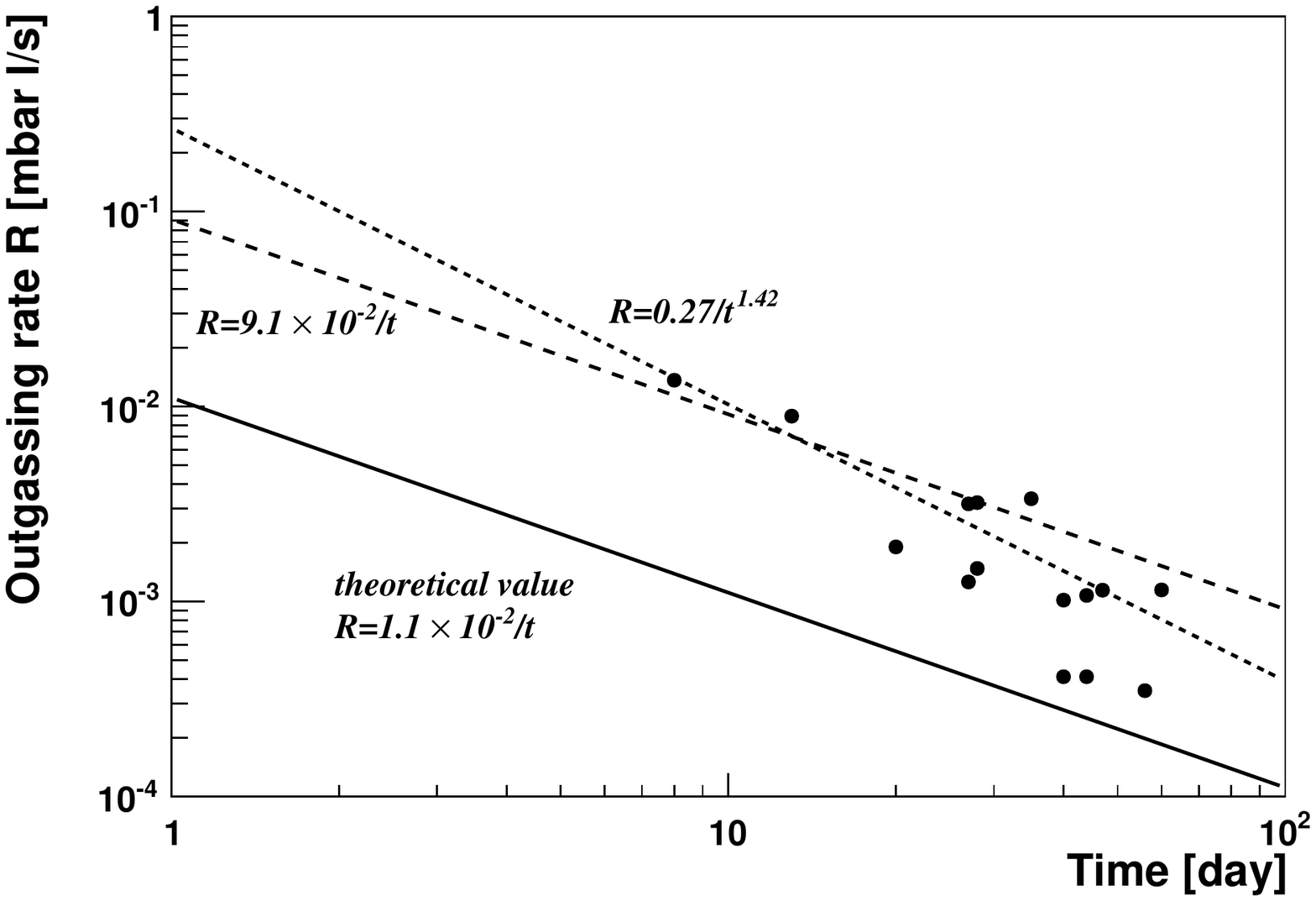}
\caption[Outgassing of the $40\times80 \unit{~cm^{2}}$ LEM TPC]{The outgassing of the $40\times80 \unit{~cm^{2}}$ LEM TPC as a function of the pumping time. The full line indicates the calculated value from \Cref{t_material_250l} and the two dashed lines the fit functions for the measured data.}
\label{f_TotalOutgassingRate}
\end{figure}

The reason for showing \Cref{f_TotalVacuumCurve} in the context of outgassing rates is, that every time the detector is filled with gas, there is a short period of a few minutes when the gate valve to the pump is closed but still there is no gas filled in the detector. During these periods the detector is outgassing and, analyzing the outgassing with the Rate-Of-Rise method, the rate as a function of the time can be measured. The values obtained are presented in \Cref{f_TotalOutgassingRate}. Also the calculated value of $1.1 \times 10^{-2} \unit{~mbar \, l/ s }$, from \Cref{t_material_250l}, is drawn (full line), assuming a simple $1/t$ law. The dashed lines are a fit, strictly proportional to $1/t$ as also to a general power law, giving $R=0.27/t^{1.42}$ with $[t]=\unit{hr}$ and $[R]=\unit{mbar \, l/s}$.
The theoretical outgassing rate and the measured outgassing rate are different by about one order of magnitude.  The actual outgassing is faster than $1/t$. A possible explanation is that, when filling the chamber with argon to atmospheric pressure, water, that is frozen at low pressures, re-liquifies and gets dissolved in the gas.
This effect can be seen in \Cref{f_TotalVacuumCurve} at about 14 days. After a short period of venting the chamber with argon gas, the pressure drops faster, when the pumps are turned on again, compared to the rate before the venting.


\subsubsection{Leaks}\label{c_leaks}
Eventually, the outgassing becomes so small that permeation is the main source of impurities. 
This is the point, where there is no way to improve the vacuum without changing to a bigger, more powerful pump. It is hardly reached with a normal turbo pump and other methods, e.g. using getter pumps, have to be applied. Permeation only gets notable at pressures below $1 \times10^{-8} \unit{~mbar}$. Because of its time independence, the best vacuum reached is 
\begin{equation}
p=\frac{Q_{permeation}}{S}
\label{e_permeation}
\end{equation}
here $S$ is the pumping speed. 

If this absolute possible vacuum is never reached and the pressure becomes constant, this is an indication for a leak. Leaks behave similar to permeation. Also they are letting a constant, time independent, flux of gas into the vacuum. The leak rate, therefore, is given, similar to Eq.~\ref{e_permeation} by
\begin{equation}
Q_{leak}=\frac{\Delta p \cdot V}{\Delta t}=p_{min}\cdot S
\label{e_leakrate}
\end{equation}
where $p_{min}$ is the pressure at the best vacuum obtained. $[Q_{leak}]=\unit{mbar \, l / s}$ the leak rate of the biggest leak. In general each leak has a different ``size'' and therefore the largest has most influence on the final pressure and it is the one to be found first.) Fixing smaller ones will not make any major effect and cannot be noted by looking at the pressure. After the main leak is found, the next smaller is the leading one to be fixed.

Permanent leaks
are actual holes or cracks in the structure, where gas molecules can pass through. Most of the time they occur on connections, where two vacuum parts are flanged together. More rarely they happen to be from production errors like weldings that are not done properly.
A third, very critical place, are connections of different materials like for example feed-throughs. In this case there is a material boundary between the stainless steel flange and the insulator and again between insulator and conductor. The vacuum tightness is done with glue, a soft metal like indium or by mechanical force, just pressing together the two parts.
This kind of leaks sometimes can be cured by having a very liquid epoxy resin. Having vacuum on one side the epoxy can be put on the insulator and, through the leak it is sucked in and blocks the channel where the molecules went in.

Temporary leaks open and close, under certain conditions. Most of the time a temperature effect is the reason. Having the large gradients form room temperature to cryogenic temperatures in ArDM, the contraction of different materials becomes important. Normally plastics have a thermal contraction of a few percent for a $\Delta T$ of $\sim 200 \unit{~^{\circ}C}$. Metals on the other hand contract about ten times less.
The connection between two different materials can be tight at room temperature but by cooling it down it becomes leaky. 
\begin{figure}[t!]
\centering
\includegraphics[height=80mm]{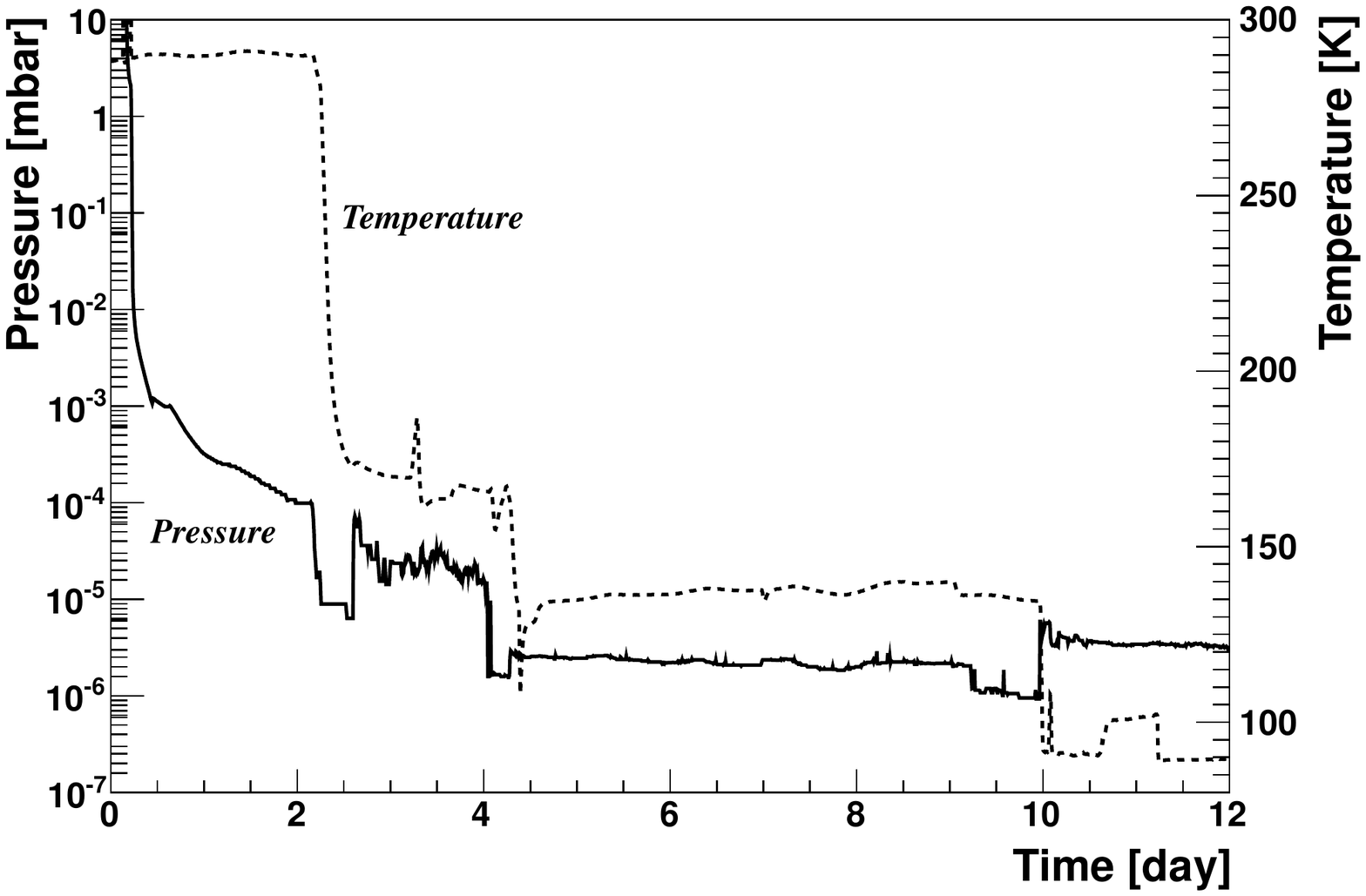}
\caption[Cold Leak]{The correlation between the pressure of the vacuum insulation (full line) and the temperature of a cold line inside it (dotted line) shows the behavior of cold leaks. Most obvious is it at $T= 10 \unit{~days}$ where a temperature drop from gaseous to liquid argon immediately gives an increase in the pressure. }
\label{f_ColdLeak}
\end{figure}
\Cref{f_ColdLeak} shows the correlation between the pressure in the vacuum insulation of ArDM, and the temperature of tubes containing the argon inside it. In the beginning the normal behaviour of evacuation is visible. With the start of cooling down the detector, at $T\sim 2 \unit{~days}$, the vacuum improves because of the cold surfaces. Later, it goes back on its original curve with a lot of fluctuations, which already indicates the release of a gas from a leaky connection. When cooling down more, to $\sim100 \unit{~K}$, the leak seems to close again and the pressure drops again. At $T= 10 \unit{~days}$ liquid was filled in the cryostat (what can be seen by the sudden temperature drop to 89 K). In exactly the same moment, the pressure is rising what is a clear indication for a cold leak.
This example shows that the cold temperature created a small leak in one of the joints that only opens when the temperature falls below 150 K. 


\subsubsection{Gas recirculation}
The best way to reduce the amount of impurities is to pump the cryostat for a reasonably long time, if possible for several months, before filling with argon. But even after a long evacuation phase there is outgassing. After stopping of the pump the impurities can diffuse into the volume without being pumped away. The argon has to be purified constantly.
To make an example, taking the lowest value for outgassing of $3.4 \times 10^{-4} \unit{~mbar \, l / s}$ from \Cref{f_TotalOutgassingRate} and the total volume of 1700 l of the system, with a total pressure at the moment when the gate valve is being closed of $\sim 2 \times 10^{-7} \unit{~mbar}$. 15 minutes later, it is only $\sim 1.8 \times 10^{-4} \unit{~mbar}$, which corresponds to 180 ppb of impurities when assuming the volume filled with 1 bar of argon gas at room temperature. 

One possibility to slow down the outgassing process is to cool down the vessel while still pumping it. The time $t(T)$ for the outgassing of stainless steel at temperature $T$ can be found by fitting the values given in \cite{Calder1967}, with an exponential function:
\begin{equation}
t(T)=\alpha \cdot \,e^{-0.0164 \cdot T}
\end{equation}
The 15 minutes at 300 K therefore become $\sim 400 \unit{~min}$ at 100 K. Further, the cold gas is 3 times more dense than the argon at room temperature. Therefore, filling with pure argon and cooling it to $\sim100 \unit{~K}$, reduces the ratio of impurities by a factor 3 compared to warm argon gas. When filling with pure liquid argon, it is reduced even by a factor of $\sim 800$ with respect to argon gas at $20\unit{~^{\circ}C}$.

This example shows that an initial cool down, with vacuum inside the chamber has a positive effect on the concentration of impurities diffused into the volume as also the ratio is improved. The disadvantage is, that all molecules that did not desorb are still on the surfaces and eventually some are dissolved in the liquid. A constant recirculation and purification of the liquid is needed. 

Another approach is, to actually fill the chamber with pure argon gas, while it is still warm. The outgassing in the gas is large but, contrary to cryogenic liquids, purification of the warm gas is not difficult and commercial cartridges can be used. 
Having a given gas flow $S$ ($[S]= \unit{l/s}$) and an outgassing rate $R$ ($[R]= \unit{mbar \, l / s}$) the total amount of impurities $N(t)$ as a function of time can be estimated by the following idea:

The change of impurities is given by 
\begin{equation}
\frac{dN(t)}{dt}=\frac{dN_{in}(t)}{dt}-\frac{dN_{out}(t)}{dt}
\end{equation}
Where the total outgassing rate can be assumed to be constant and the total molecules outgassing into the vessel is given by
\begin{equation}
N_{in}=N_{0}+\frac{R}{k\cdot T}\cdot t
\end{equation}
with $N_{0}$ being the number of molecules still inside the volume after pumping. They are given by the general gas equation
\begin{equation}
N_{0}=\frac{p\cdot V_{tot}}{k\cdot T}
\end{equation}
with $p$ the pressure before closing the gate valve. The differential outgassing rate into the vessel is given by
\begin{equation}
\frac{dN_{in}}{dt}=\frac{R}{k\cdot T}
\end{equation}

The change of impurities filtered out depends on the pump speed and the total impurity concentration inside the vessel. It is given by
\begin{equation}
\frac{N_{out}}{dt}=\frac{S}{V_{tot}}\cdot N(t)
\end{equation}
where the efficiency of the filter has been assumed to be $100\%$, i.e. all impurities passing through it are filtered out. 

Combining the two parts leads to the following differential equation:
\begin{equation}
\frac{N}{dt}=\frac{R}{k\cdot T}-\frac{S}{V_{tot}}\cdot N(t)=A-B\cdot N(t)
\end{equation}
and solved to
\begin{equation}
N(t)=\frac{A}{B}+C \cdot e^{-B\cdot t}\label{e_recirculation}
=\frac{R\cdot V_{tot}}{k\cdot T\cdot S}+C e^{-\frac{St}{V_{tot}}}
\end{equation}
The constant $C$ depends on the initial amount of impurities $N_{0}$ in the vessel.

Taking a total volume $V_{tot}=1700 \unit{~l}$ for the dewar, an outgassing rate of $R=3.4 \cdot 10^{-4} \unit{~mbar \, l/s}$ and an actual gas flow of $S=80 \unit{~l/min}$ for the gas recirculation, the equilibrium is reached when the rate of impurities, that can be trapped, is equal to the rate of new impurities produced by outgassing, i.e. $N(t \rightarrow \infty)$ in Eq.~\ref{e_recirculation}:
\begin{eqnarray}
N&=&\frac{R}{k\cdot T}\cdot \frac{V_{tot}}{S}\notag \\
& = &\frac{3.4\cdot 10^{-4} \unit{~mbar \, l/s}}{k \cdot 297 \unit{~K}}\cdot \frac{1700 \unit{~l}}{80 \unit{~l/min}}=1.05 \cdot 10^{19} \unit{~particles}\\
&= &\,\sim255 \unit{~ppb} \mbox{ @ SATP}\notag
\end{eqnarray}

\begin{figure}[t!]
\centering
\includegraphics[height=80mm]{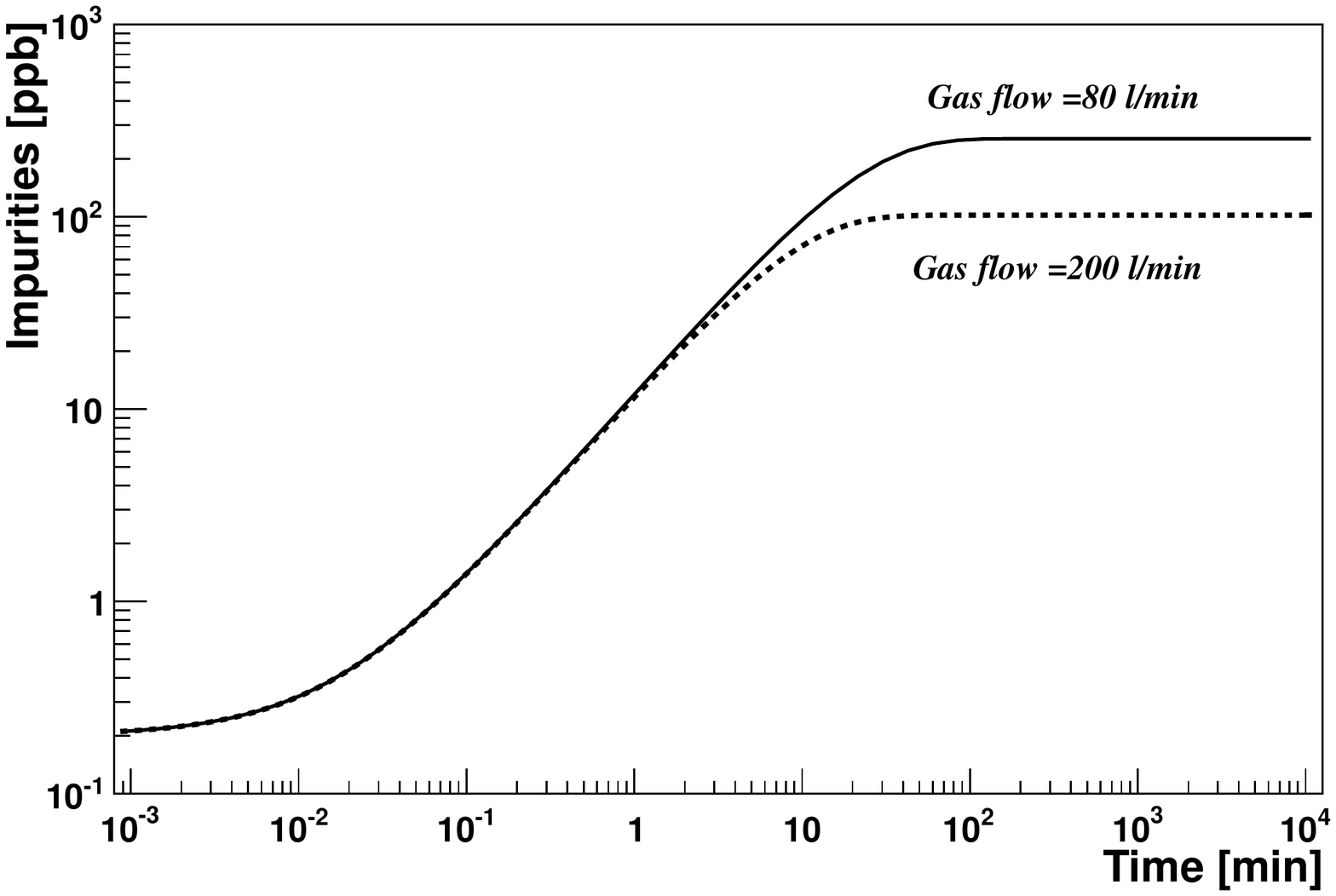}
\caption[Outgassing of impurities after stopping the evacuation.]{The contamination of the argon gas at SATP conditions with impurities as a function of the time, from the moment on when the gate valve has been closed and pure argon gas was filled in the system. The curve is calculated according to Eq.~\ref{e_recirculation} for the values discussed in the text for the $40\times80 \unit{~cm^{2}}$ LEM TPC in the ArDM dewar.}
\label{f_Gasrecirculation}
\end{figure}

Using Eq.~\ref{e_recirculation} the contamination of the argon gas with impurities (in ppb) is plotted as a function of time in \Cref{f_Gasrecirculation}. For comparison, there is the actual gas flow of the ArDM system of $80 \unit{~l/min}$ as also an increased gas flow of $200 \unit{~l/min}$ shown. 
Analyzing these curves shows that it is most important to reduce the outgassing rate as much as possible. Once stopping to pump pollutes the volume in a few minutes with more than $100 \unit{~ppb}$ of impurities. Also it can be seen that the maximum contamination with impurities is given by the gas flow through the filter and not the initially reached vacuum.


\subsubsection{Liquid argon purification}\label{c_LArPurification}

The initial purity of commercial liquid argon is in the order of ppm of oxygen equivalent impurities. It is not sufficiently clean for the experiment and has to be purified in situ. Different techniques, hereafter discussed, are used to filter out different molecules. 

For large setups evacuation is not possible and the majority of the air inside the dewar is removed by flushing the setup with argon gas. A big help is that argon (40 g/mol) is heavier than the other molecules like O$_{2}$ (16 g/mol), N$_{2}$ (14 g/mol) and H$_{2}$O (10 g/mol). Filling up a volume with argon presses out the other gases and after a few exchanges of the whole volume the contamination in a vessel can be reduced to the level of ppm, equivalent to the initial purity of the liquid argon.
An example for a large, non evacuable volume is a dewar of about $6 \unit{~m^{3}}$. After ten volume exchanges the oxygen concentration was measured to be $\sim 4 \unit{~ppm}$ \cite{Curioni2011}.

Having reached the initial purity of the liquid argon, the only way of improvement is to actually purify the argon. Depending on the kind of molecule to be removed different filters are used. Also an important role plays the argon state in which the purification takes place, i.e. whether the argon is purified as gas or liquid.
In general impurities in the liquid argon are filtered out by binding them to a surface. This can be a cold trap (freezing impurities to a cold surface), physical adsorption (van der Waals force) or chemical adsorption (reaction with other atoms). 

\subsubsection{Molecular sieve}
For filtering out water molecules, in general physical adsorption is used. This can be silica gel, aluminum oxide or a molecular sieve. While the silica gel has the biggest capacity for holding back water, the molecular sieves are the strongest adsorbents. In general the power of adsorption is given at the temperature of the dew-point. For silica gel it is $\sim -5 \unit{~^{\circ}C}$ and for a molecular sieve up to $\sim -100 \unit{~^{\circ}C}$ \cite{Trent1992}. Assuming a vapor pressure of $1 \times 10^{-4} \unit{~mbar}$ for water at $-100 \unit{~^{\circ}C}$, this gives a ratio of $\sim100 \unit{~ppb}$ for gaseous argon at room temperature and $\sim0.1 \unit{~ppb} $ of water for liquid argon. This increase in purity for liquid is due to the $800$ times higher density of argon atoms in liquid compared the gas/vapor at room temperature. For the water molecules there is no such increase assumed since the amount is given by the partial pressure of $10^{-4} \unit{~mbar}$ and independent of the other molecules in the mixture. 

\begin{figure}[t!]
\centering
\includegraphics[scale=1]{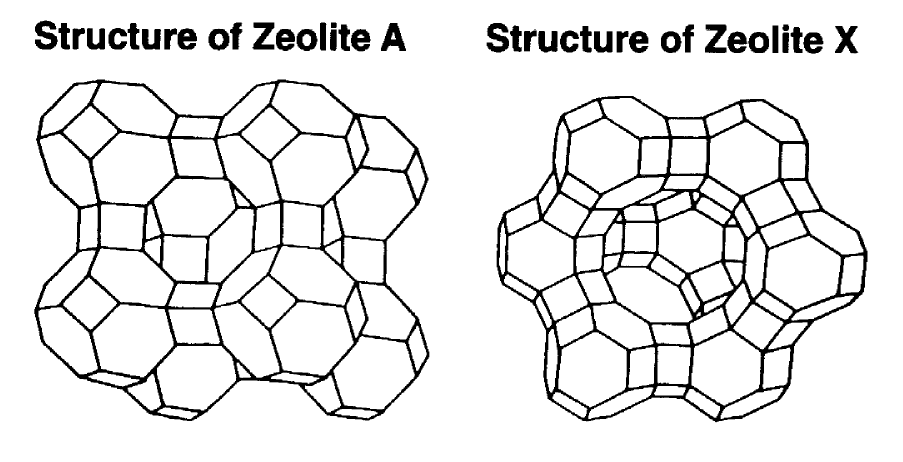}
\caption{Molecular structure of the two most common types of zeolites. The pore size in the center depends on the element with which some of the sodium atoms are replaced. It is between $\sim 3 \unit{~\AA}$ and $\sim 5 \unit{~\AA}$ in diameter for type A and $\sim 10 \unit{~\AA}$ for type X.}
\label{f_Zeolite}
\end{figure}

Usually a molecular sieve consists of porous aluminosilicates, so called zeolites. They are able to adsorb water and when heated up to release it. For commercial use these kind of crystals are synthetically produced and specially designed to have a uniform and precisely defined pore size. \Cref{f_Zeolite} shows the crystalline structure of the two most common types of zeolite. 
Each ``ball'' of the structure (in the Zeolite A structure in the eight corners) has a silicon or an aluminum atom in each corner of the square surfaces, interconnected with oxygen atoms. Because the aluminum atom is trivalent the structure as a whole becomes an anion. To become electrically inert there are sodium, calcium or potassium cations attached to the molecule. Depending on the ratio and the type of these exchangeable cations the properties of the molecule can be influenced.

A crystal with sodium atoms has a pore size of $\sim 4 \unit{~\AA}$, while replacing a sodium atom with a potassium atom shrinks the diameter to $\sim 3 \unit{~\AA}$. Replacing it with a calcium atom increases the diameter to $\sim 5 \unit{~\AA}$.
In such a crystal, molecules smaller than the diameter of the pore can enter, larger ones are blocked. Capillary condensation takes place with the molecules entering to a much larger scale than to others since most of the surface of the crystal is inside the structure. So molecules, smaller than a certain size can be attached, while larger ones flow around the crystal. 
Considering only this property, also the argon atoms should be filtered out and the sieve immediately would saturate. This is not the case since, besides the capillary condensation, there is a second effect The polarity of the sieve material makes molecules stick to the surface.
Because of the crystal structure made out of electrically unbalanced charged parts at fixed positions, the crystal has a polar surface. Therefore, polar molecules like water are attracted and stick to the surface while apolar molecules only are held very weakly. The attachment of argon and nitrogen atoms, can be found in \cite{Barrer1959}. With this property water can be trapped up to $\sim 28 \%$ of the crystal weight and these kind of crystals really act as a ``sieve'' for filtering out certain molecules. 
The amount of trapped molecules depends on the temperature of the crystal since at higher temperature molecules are easier detached again. This behavior is also used to regenerate the material. By heating it up to $\sim 250 \unit{~^{\circ}C}$ and flushing it with dry nitrogen/argon gas the trapped water is released and the zeolite can be used again.

Commercially, there are different molecular sieves available, depending on the diameter of the pore, they are called 3A, 4A or 5A, where ``A'' stands for Angstrom. Zeolites of type X in general have bigger pores of about $10 \unit{~\AA}$.

\begin{figure}[t!]
\centering
\includegraphics[width=100mm]{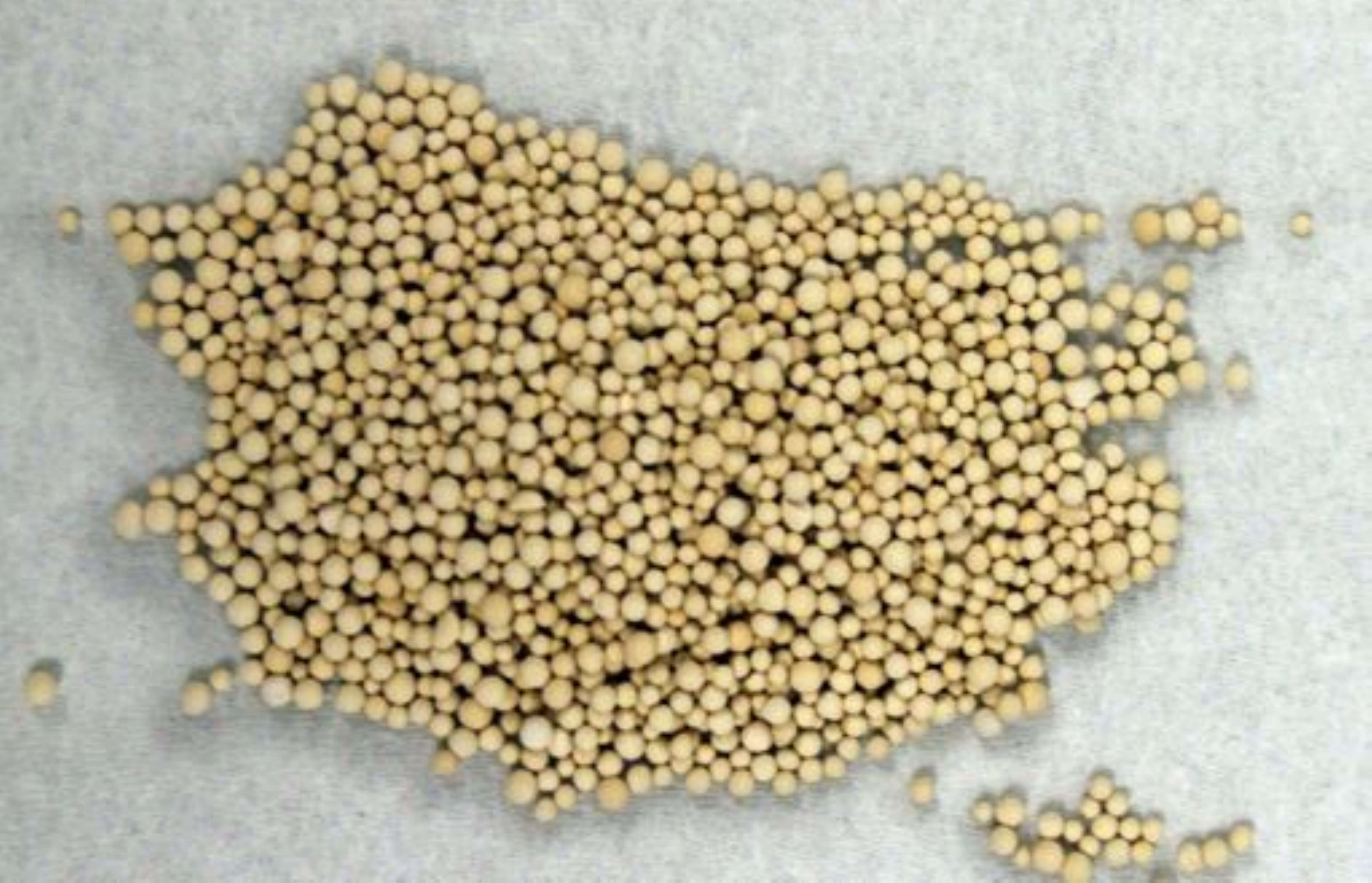}
\caption{Molecular sieve \textit{ZEOCHEM 3A}$^{\tiny{\copyright}}$ grains before inserting them in the regeneration cartridge. Each grain has a diameter of about $2\unit{~mm}$.}
\label{f_molecular_sieve}
\end{figure}

In ArDM, a molecular sieve is not included in the standard recirculation circuit. Water is removed as much as possible from the detector by evacuating the volume for a long period before filling with argon. The remaining water, if attached to a surface, is sticking to it very strongly due to the low temperatures. If it is dissolved in the liquid argon, it (partly) sticks to the oxygen filter described in \Cref{c_oxysorb}. 
Therefore, the only way for water to enter the detector volume is to be already dissolved in the liquid argon, used for filling the detector. For this reason a purification cartridge, partly filled with a 3A molecular sieve is placed before the inlet to the detector. It is filtering out the water from the commercial liquid argon. \Cref{f_molecular_sieve} shows the 3A crystals with a diameter of $\sim 2 \unit{~mm}$ each. In total the cartridge contains $0.7 \unit{~l}$ of molecular sieve.

\subsubsection{Oxygen filters}\label{c_oxysorb}
As mentioned before, the most critical impurity for drifting electrons in liquid argon is oxygen. A constant recirculation and purification of the argon is mandatory to keep the contamination low. 
Oxygen is entering the detector volume through micro leaks in the detector shell or, more dramatically, through cold leaks, discussed in \Cref{c_leaks}.
Also not to be neglected is the outgassing of the detector material. By evacuating the detector, a lot of the residual gases are pumped out but, by lowering the partial pressure of the vessel, impurities might condense or freeze and can no longer be pumped. Breaking the vacuum with pure argon gas and re-pumping it helps to reduce them. 

\begin{figure}[t!]
\centering
\includegraphics[height=80mm]{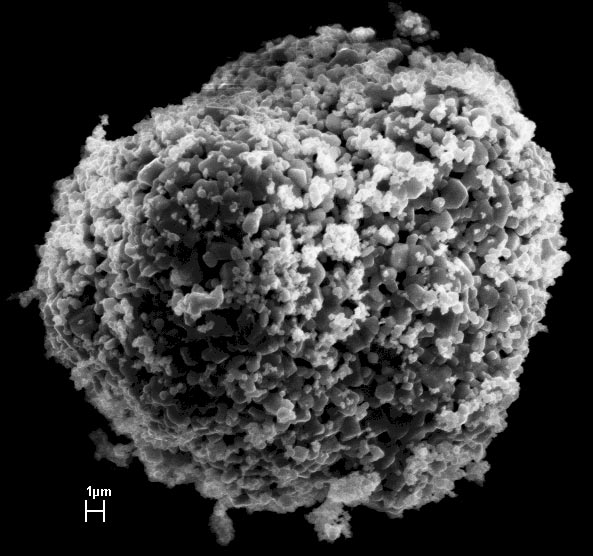}
\caption{The copper grains, sintered together from smaller grains, have a diameter of a few $10 \unit{~\mu m}$. This technique grants a large ratio of surface to volume.}
\label{f_sintered_copper}
\end{figure}

For filtering out oxygen from argon gas or liquid, chemical adsorption is used. More specifically, the argon oxygen mixture is guided over a surface of a reduced material. It oxidizes and only argon is left. 
The challenge is to find a material with a low reaction potential for an efficient oxidation and a large surface. Elements in question are easy to be oxidized metals. We have successfully developed filters containing pure copper powder. Oxidation is an exothermal reaction and takes place according to
\begin{equation}
\unit{Ar + O_{2} + 2 \, Cu \rightarrow Ar + 2\, CuO}
\end{equation}
The heat of adsorption is $\sim 82\unit{~kcal/mole}$ on the true oxidation and goes down to $\sim 55\unit{~kcal/mole}$ for the partly oxidized copper grains \cite{Dell1953}.
This reaction is a surface process and takes place in a monolayer of the copper. In fact the thickness of the layer is increasing with the temperature. At $87\unit{~K}$ it is $\sim 4\unit{~\AA}$ while at $300\unit{~K}$ it is $\sim 17\unit{~\AA}$ \cite{Rhodin1950}. Since the temperature is given by the liquid argon, it is important to have a maximum possible surface per volume. 
It can be reached by the use of sintered copper pellets\footnote{The product in use is \textit{FLUKA Copper(II)oxide
purum, ³98.0\% (RT); 61202} from Sigma-Aldrich Chemie GmbH}. Each grain of copper has a diameter of a few $10 \unit{~\mu m}$. It is not a single junk of copper but sintered together from smaller grains in the nanometer scale. \Cref{f_sintered_copper} shows an image of such a grain taken with an electron microscope.
Because of its exothermal reaction with oxygen, copper is sold already oxidized to copper(II)oxide ($\unit{CuO}$), not to be confused with copper(I)oxide ($\unit{Cu_{2}O}$).

\begin{figure}[t!]
\centering
\includegraphics[height=90mm]{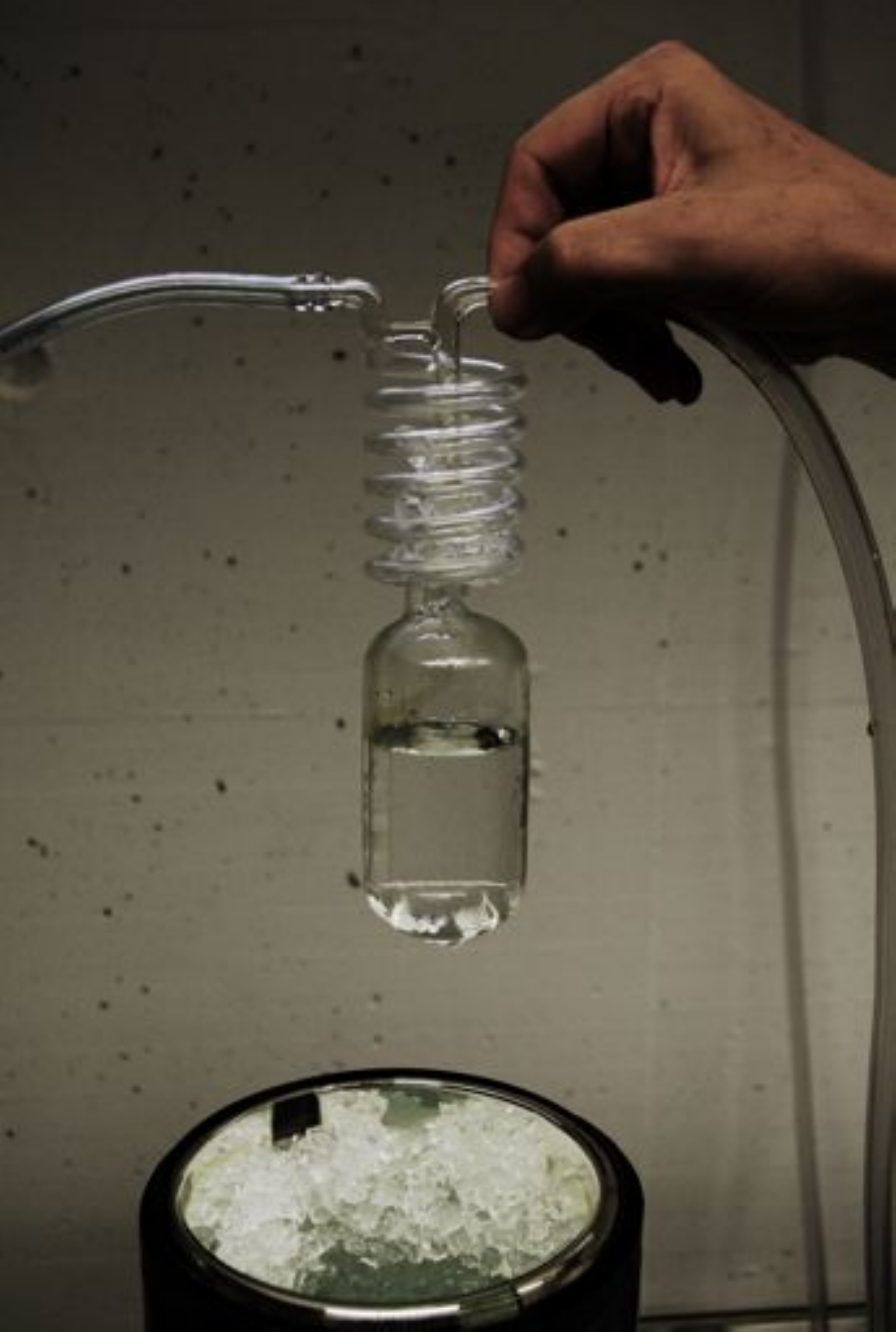}
\caption{Collected water from regeneration of a copper cartridge.}
\label{f_watercondensation}
\end{figure}

Before being used as a filter it has to be regenerated. This happens by another exothermal reaction where the copper(II)oxide is flushed with hydrogen. The oxide is reduced to pure copper and water according to 
\begin{equation}
\unit{CuO + H_{2} \rightarrow Cu + H_{2}O}
\end{equation}
Since the reaction is exothermal, the temperature has to be monitored by measuring the temperature of the exhaust gas from the cartridge in oder not to harm the pellets. The temperature is controlled by the amount of hydrogen entering the cartridge.
To initiate the reaction, a temperature of $\sim 150\unit{~^{\circ}C}$ of the copper(II)oxide is needed, which is achieved with heating bands wound around the purification cartridge. By the reaction the temperature rises and, as a clear indication of a complete regeneration, the temperature drops back to the value of the heater. Another indication for the cartridge to be regenerated is to see how much water is extracted from the cartridge by cooling the output gas. The water condenses and can be measured as seen in \Cref{f_watercondensation}.

Hydrogen is a dangerous gas and the concentration must stay below 3\% in order not to take the risk of an explosion. For this reason the regeneration does not take place with pure hydrogen but with a mixture of hydrogen and argon. Also at the end, when the copper is regenerated, the cartridge is flushed with pure argon gas, at $\sim 150\unit{~^{\circ}C}$, to clean out all the water vapor trapped in the sintered material.
For transporting and storing, the regenerated cartridge is filled with pure argon gas to an overpressure of $\sim 0.5 \unit{~bar}$. 


\subsubsection{Design of the ArDM LAr purification system}

\begin{figure}[htb]
\begin{center}
\includegraphics[width=0.9\textwidth]{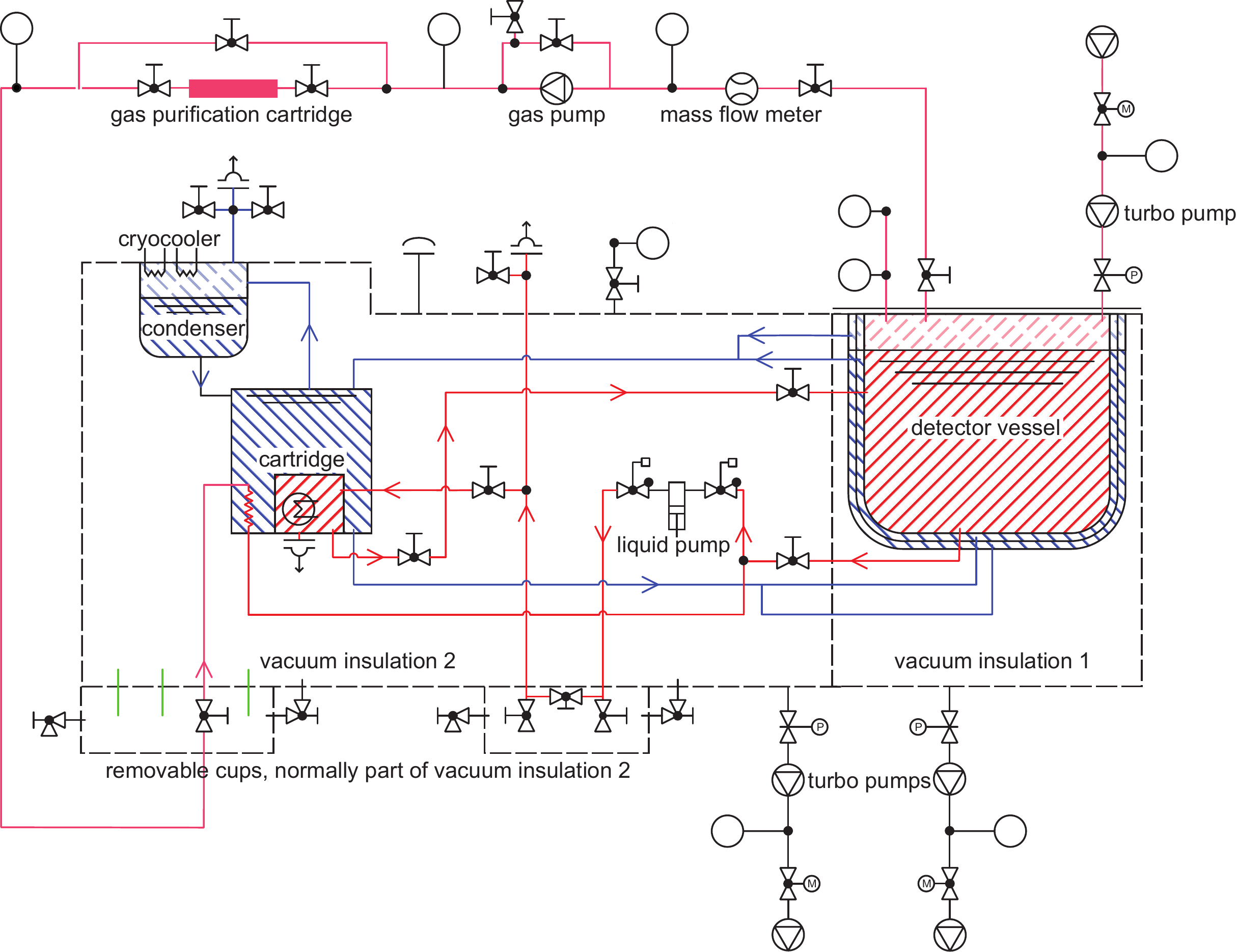}
\caption{Schematic view of the cryogenic installation of the ArDM detector. The actual dewar is situated on the right side and all the cryogenic installations containing the re-condensing and the purification of the argon are on the left side.}
\label{f_SchemaArDM}
\end{center}
\end{figure}

A schematic view of the cryogenic installation of ArDM is shown in \Cref{f_SchemaArDM}. The dewar, hosting the detector can be seen on the right side
of the schematic. All cryogenic services, including the purification cartridge and the liquid argon pump are situated on the left of the drawing. On the top left, above the cartridge the re-condenser unit, with the two cryocoolers, is situated. All these parts are inside two separate vacuum insulations, one around the dewar (\textit{vacuum insulation 1}) and the other for the cryogenic facilities (\textit{vacuum insulation 2}). 

As indicated by the different colours, there are two independent argon circuits. One containing the highly purified argon used as detector target (red) and a second circuit of argon to cool the primary circuit (blue). The primary argon from the dewar is cleaned from electro negative molecules in a purification cartridge as described in \Cref{c_LArPurification}. It is pumped by a bellows pump that is completely inside the vacuum insulation. 

Besides the liquid purification circuit, also a gas purification system is installed. It is indicated on top in \Cref{f_SchemaArDM}.
The recirculation of the primary liquid argon circuit is done by extracting the liquid from the botton of the dewar, pressing it through the filter and then to re-inject it to the detector volume from the top. The recirculation pump, maintaining the constant flow of argon, is situated in the lowest position of the experiment. This gives the maximum  \textit{net positive suction head} (NPSH) to press the liquid through the check valve into the pump body. 
Shortly before the pump, there is a T-connection for the entry of the re-condensed gas, in case the gas recirculation system is used.

Between pump and filter, the liquid argon has to pass through a valve, indicated in the lower middle part of \Cref{f_SchemaArDM}. This configuration of valves is normally inside the \textit{vacuum insulation 2} and the central bypass valve is open to let the liquid argon flow from the pump to the purification cartridge.
The other two valves are for filling, and for emptying, the dewar. In case of their use, the volume around the valves is separated from the \textit{vacuum insulation 2} and the vacuum is broken. 
For filling the dewar, the bypass valve from the pump is closed and argon is fed in the system. It first is pressed through the filter to be pure when entering the detector volume. When the dewar is full, the input is closed and the bypass opened again.
For emptying the system, the bypass is closed and the other valve is opened. The argon from the dewar now flows through the pump (check valves open automatically by the pressure difference between the dewar and the atmosphere) and leaves the detector. By using the pump, the emptying process can be speeded up. 
The purification cartridge for the liquid argon is, as indicated in \Cref{f_SchemaArDM}, immersed in a bath of liquid argon for keeping it cold and to re-condense possible argon vapor. 

\subsubsection{Liquid argon purification in the 40$\times$80~cm$^2$ operation at CERN}
As described in the previous section,
the cryogenic system of the 40$\times$80~cm$^2$~\cite{Badertscher:2013wm} developed
for ArDM, was
equipped with two separate purification systems. During the run period
both systems were operated, thus it was important to monitor the
liquid argon purity. In order to measure the charge attenuation as a
function of the drift time, we have used the reconstructed cosmic muon
tracks.  
\begin{figure*}[t]
  \centering
  \includegraphics[width=0.49\textwidth]{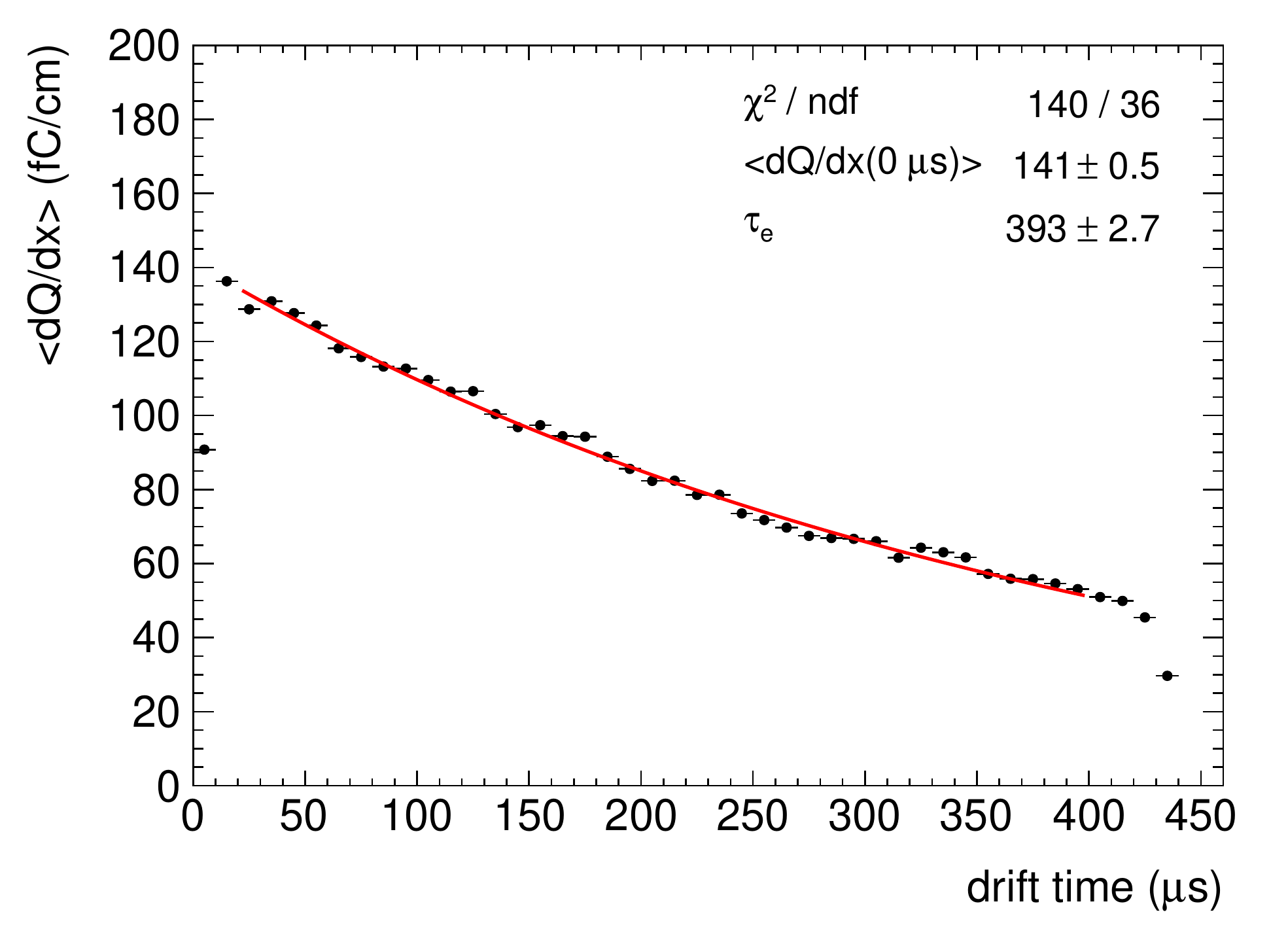}
  \includegraphics[width=0.49\textwidth]{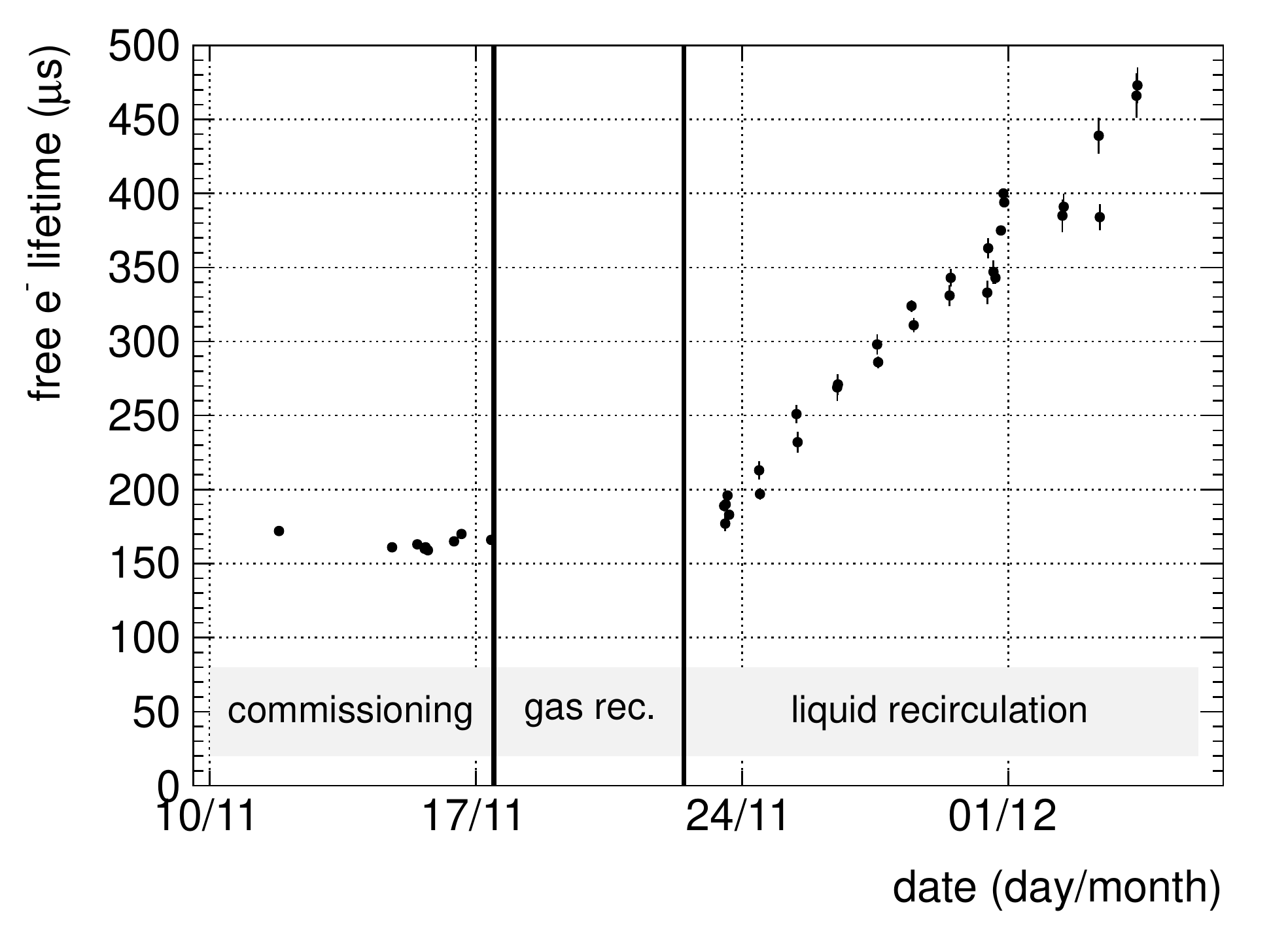}
  \caption{Left: $<dQ/dx>$ as a function of the drift time
    superimposed with an exponential function with the decay time constant
    $\tau_e=393\pm3$~$\mu$s fitted to the data points.  Right:
    evolution of the free electron lifetime during the run period.} 
  \label{fig:lifetime}
\end{figure*}
The data points in Figure~\ref{fig:lifetime} show the mean $dQ/dx$
values of all reconstructed muons for different drift times. It can be
seen that the exponential function $e^{-t_{drift}/\tau_e}$ with the free electron
lifetime $\tau_e$ perfectly fits the obtained distribution. 
The maximum electron drift time is about 430~$\mu$s.
The plot
on the right side shows the lifetime 
evolution during the last three weeks of operation. Starting from a
lifetime of about 170~$\mu$s we first did initial tests with the
liquid recirculation. During this commissioning period the free electron lifetime
remained constant. In a second phase, the gas recirculation was turned
on. Due to changes in pressure and liquid argon level we did not
acquire data with the double phase readout. However, after stopping
the gas recirculation, the measured free electron lifetime was not
significantly improved although was maintained at constant value. 
During
the last phase we have  operated the liquid argon recirculation
system at a flow rate of about 5 to 7~lt LAr/h. 
Since we saw a steady increase of the
free electron lifetime until we stopped the run, we conclude that the liquid argon
recirculation system was efficiently working. When we stopped the run
a lifetime of 470~$\mu$s was measured, but still growing linearly with time. 
This corresponds to an oxygen
equivalent impurity concentration of about 0.64~ppb.


\subsubsection{Concept for the gas and liquid purification of the $6\times 6\times 6$~m$^3$ prototype}
The liquid argon process and the performance of its sub-units are a critical item for the successful
operation of the \six prototype. The basic parameters are summarised in \Cref{tab:larprocparam}.
The design is based on the extensive experience developed during several years of tests and R\&D.
As purity requirements are so stringent  welded connections or ISO CF flanges with metal seals will be used.
The system will be scaled to reach the required purification volume given the size of the \six vessel.
\begin{table}[htdp]
\caption{Overview of the parameters for the liquid argon process}
\begin{center}
\begin{tabular}{|p{7cm}| c|}
\hline
\multicolumn{2}{|c|}{General properties} \\
\hline
Total inner-vessel volume & $\unit[600]{m^{3}}$\\
Membrane surface in inner-vessel & $ \unit[430]{m^{2}}$\\
Liquid level above floor&$\unit[7]{m}$\\
Amount of liquid&$\approx \unit[530]{m^{3}}$\\
Amount of gas&$\approx \unit[70]{m^{3}}$\\
Instrumented volume & $ \unit[6 \times 6 \times 6]{m^{3}}=\unit[216]{m^{3}}$\\
\hline\hline
\multicolumn{2}{|c|}{Gas purification} \\
\hline
\multirow{2}{*}{With 2 pumps of \unit[500]{l/min} each}&$\approx\unit[10]{hr/vol.}$ for warm gas (\unit[300]{K})\\
&$\approx \unit[30]{hr/vol.}$ for cold gas (\unit[100]{K})\\
\hline\hline
\multicolumn{2}{|c|}{Boil-off} \\
\hline
Heat input (vessel)&$\unit[2150]{W}$ ($\unit[5]{W/m^{2}}$)\\
Evaporation rate&$ \unit[450]{l_{gas}/min}$\\
\hline\hline
\multicolumn{2}{|c|}{Liquid purification} \\
\hline
Purification speed for $\tau=\unit[1]{day}$&$\unit[22]{m^{3}/hr}\approx \unit[400]{l/min}$\\
\hline
\end{tabular}
\end{center}
\label{tab:larprocparam}
\end{table}%
The inner-vessel volume which contains all equipment and the argon is about 600~m$^3$. It must
be a totally sealed volume with no leaks towards the outside in order to avoid the penetration
of any contaminant or the loss of pure argon. 

The refrigeration component will be provided by external cryocooler with separate loops,
thermally joined with heat exchangers.

During construction and assembly phase, the inner-vessel volume will be maintained
in a controlled air environment. Starting from that state, the volume will be first replaced
with argon. In \cite{Curioni:2010gd} we have experimentally demonstrated on a volume
of 6~m$^3$ that the replacement of air with argon gas
down to ppm impurities concentration is achievable through flushing.

\begin{figure}[ptb]
\centering
\includegraphics[width=0.6\textwidth]{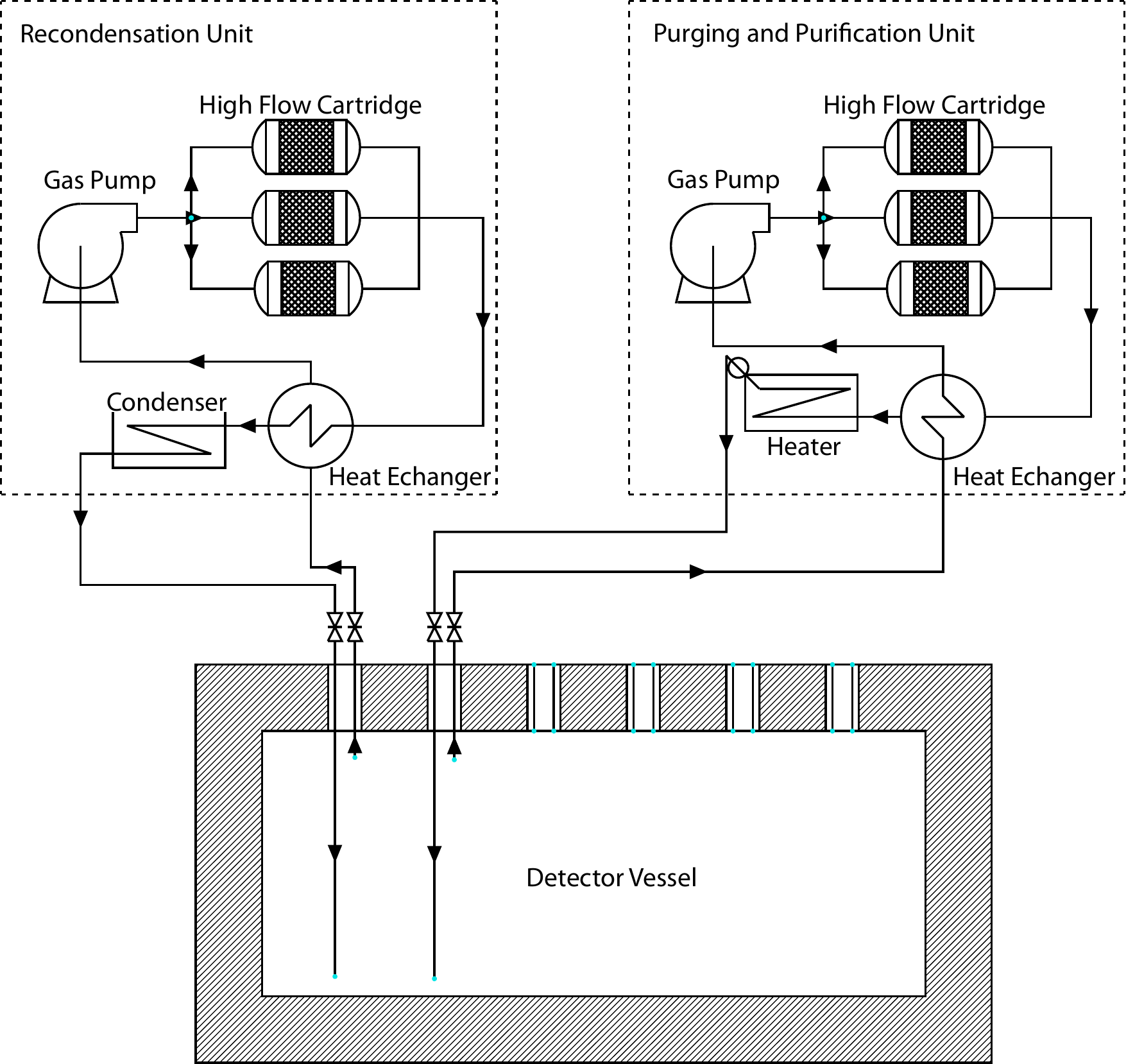}
\caption{Schematic diagram for the gas recirculation system.}
\label{fig:gasrecirc}
\end{figure}

A fully-submersed centrifugal cryogenic pump (e.g. AC-32 from ACD Cryogenic Industries\footnote{ACD CRYO, Gutenbergstrasse 1 CH-4142 Muenchenstein Switzerland, \protect\url{www.acdcryo.com}}) will be operated
in order to create the liquid argon flow through the purification cartridges. The pump has been
chosen at FNAL for the LAPD and 35~ton membrane test~\cite{Montanari:2013/06/13aqa}.
Our pumps will be hanging inside a dedicated chimney and will hang from the top, hence will be lower
from the top before usage, or lifted to the top in case of removal when maintenance is required. 
The pump and the motor are fully flooded with liquid, minimizing start-up and downtime and 
guaranteeing quick, responsive pumping. The multi-frequency motor provides an efficient and 
broad range of operation and power, ranging from 15 to 1514 lpm. The pump head is above requirement,
and causes no problem. The electrical motor and bearing life is extended being cooled by the cryogen it is pumping, 
the heat input is minimal and carried off by discharging liquid. The power is in the range 2-14~kW.
\begin{figure}[ptb]
\centering
\includegraphics[width=0.6\textwidth]{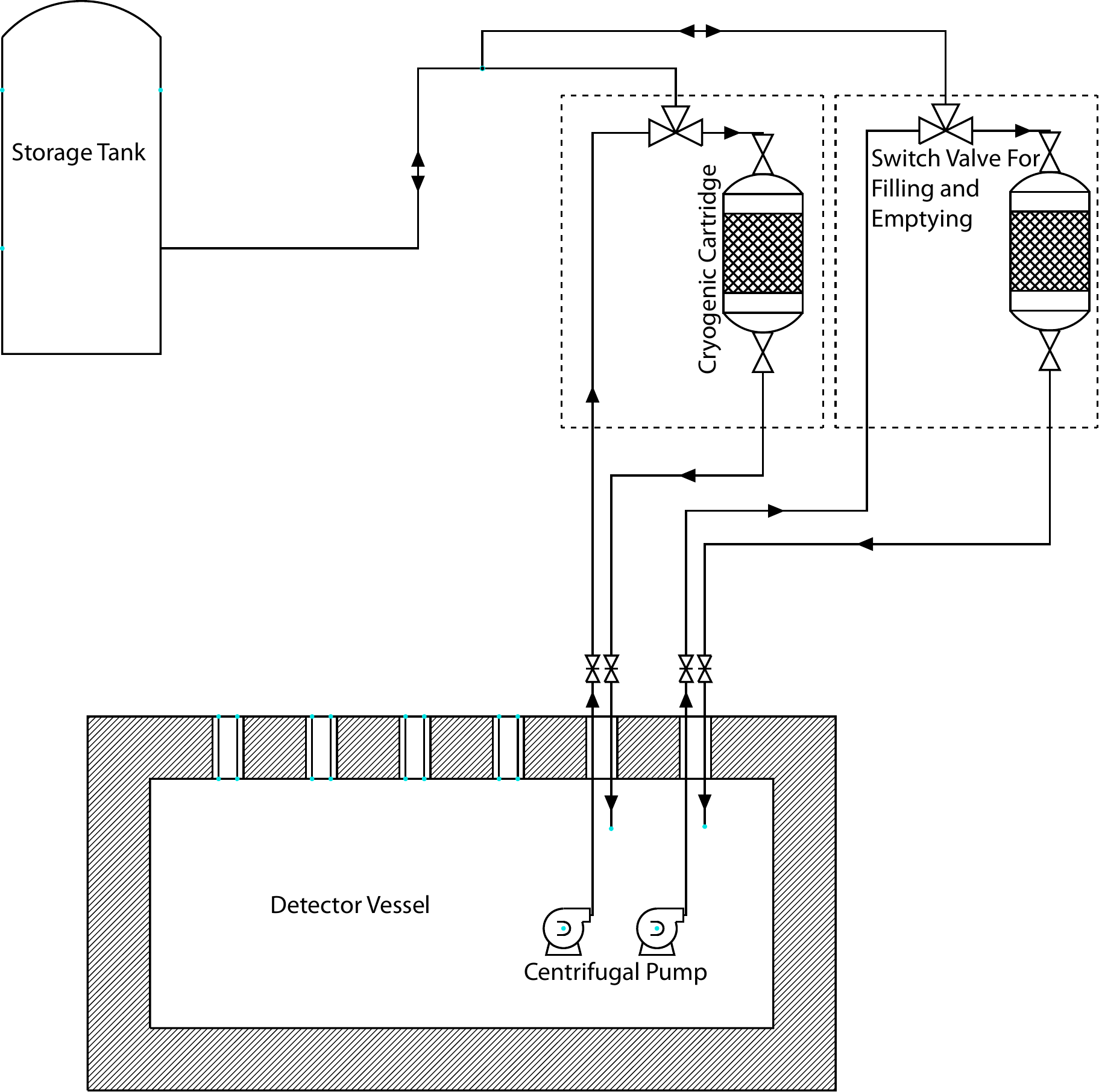}
\caption{Schematic diagram for the liquid phase purification system.}
\label{fig:larrecirc}
\end{figure}

\begin{figure}[htb]
\centering
\includegraphics[width=0.3\textwidth]{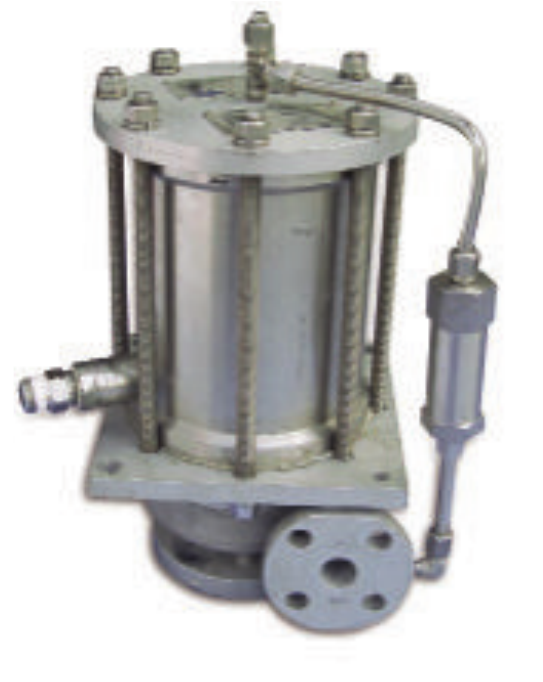}
\includegraphics[width=0.65\textwidth]{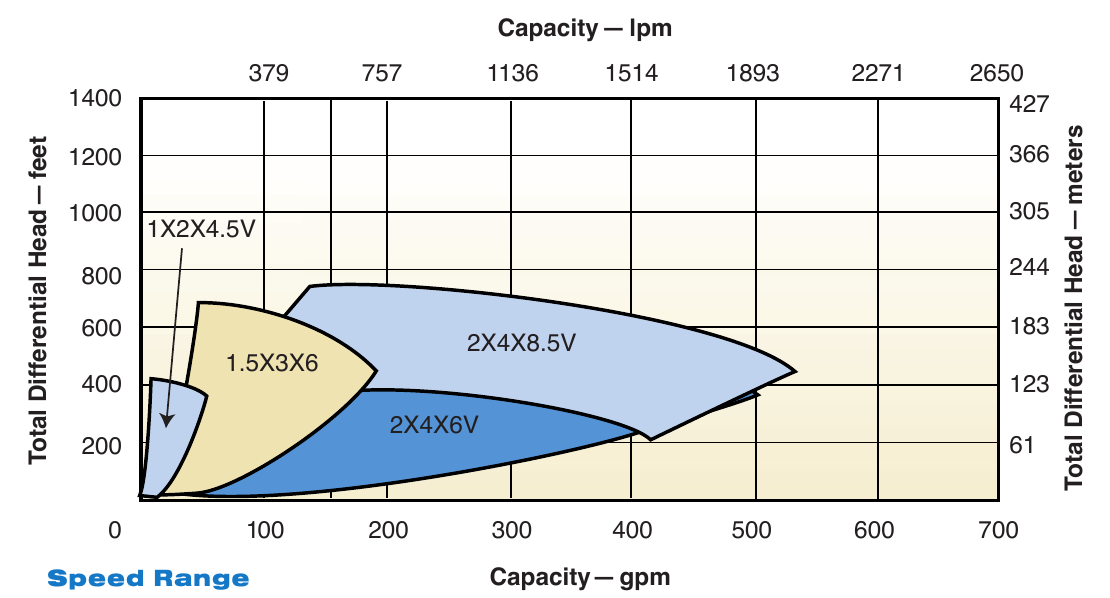}
\caption{Specifications of the sealess cryogenic pump from ACD Cryogenic industries.}
\label{fig:acdcryopump}
\end{figure}

\begin{figure}[ptb]
\centering
\includegraphics[width=0.6\textwidth]{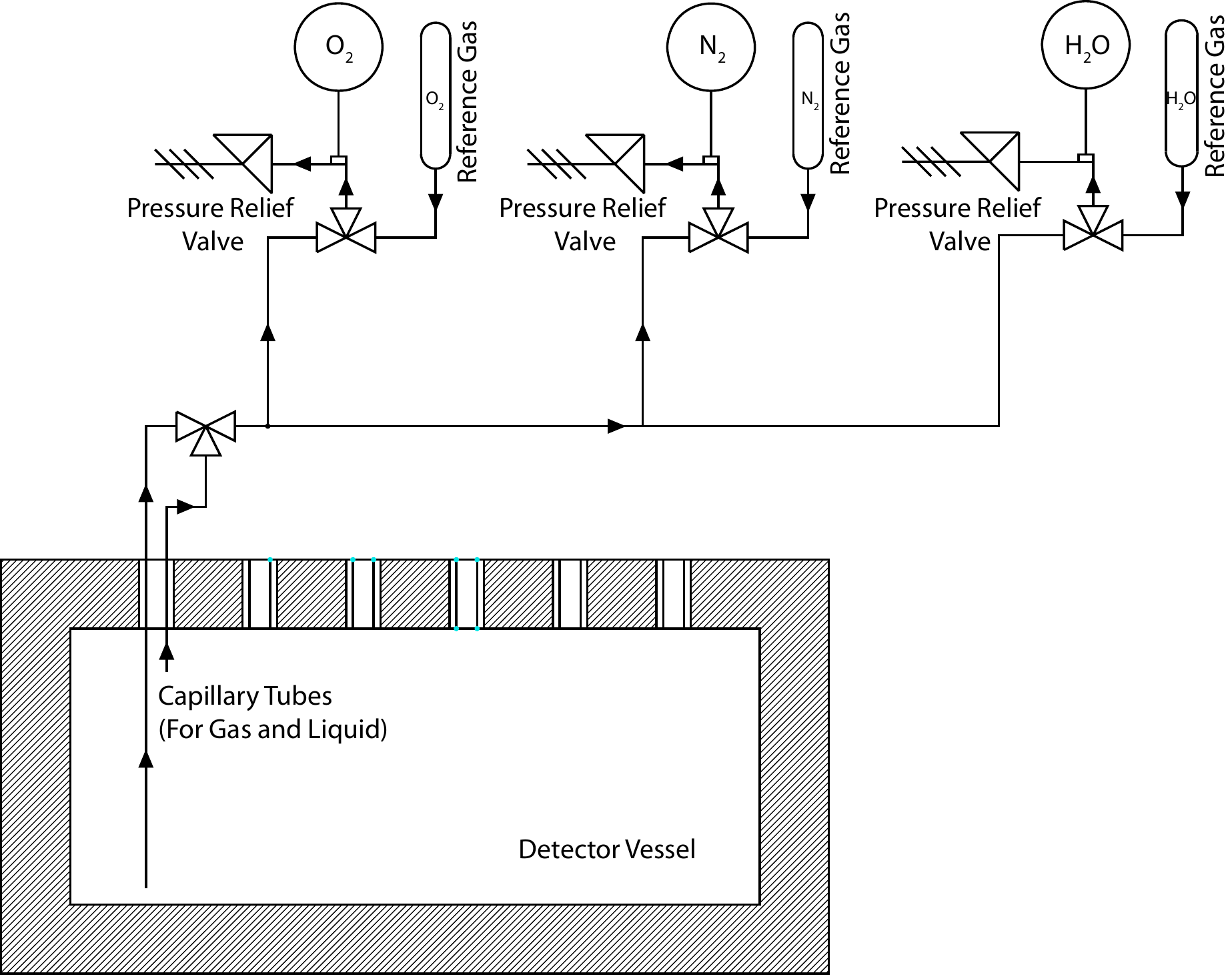}
\caption{Schematic diagram for the liquid and gas phase impurity analyser systems.}
\label{fig:larrecircgasanalyse}
\end{figure}

\subsection{Process control and monitoring}

The main function of the process control and monitoring system is the monitoring of the condition of the prototype and the regulation of the cryogenics and high voltage systems in order to allow a safe operation of the detector. In addition, the control system is a safety monitor that alerts people in case of an imminent danger, such as oxygen deficiency or fire.
The  control system allows for process control and is based on a programmable logic controller (PLC). A PLC is a computer specially designed for process automation with a large number of inputs and outputs for sensors and actuators. PLCs possess a high reliability due to specialised hard- and software and therefore the control system serves as a safety system.

\begin{figure}[htb]
\centering
\includegraphics[width=0.9\textwidth]{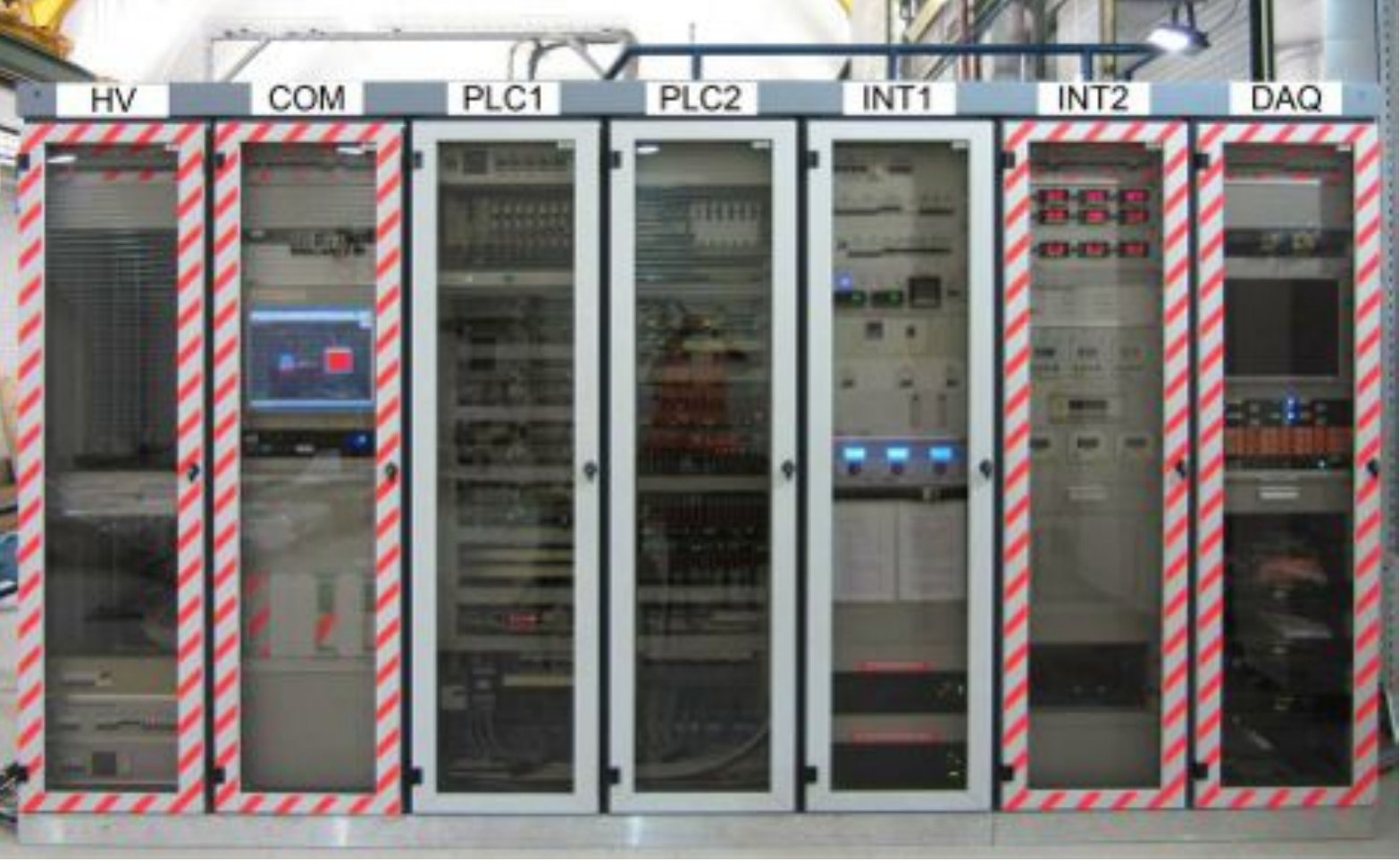}
\caption{The seven standard 19Ó racks containing the ArDM (CERN RE18) control system (COM, PLC1, PLC2, INT1 and INT2 rack), the high voltage supply of the PMTs (HV rack) and the data acquisition system (DAQ rack).
The system was built in Collaboration with the CERN/PH-DT group.}
\label{fig:ardmplc}
\end{figure}

The control system developed for the ArDM (RE18) experiment built in Collaboration with the CERN/PH-DT group, is shown in \Cref{fig:ardmplc}.
The control system for ArDM is located in seven standard 19Ó racks next to the detector together with the high voltage supply for the PMTs (HV rack in \Cref{fig:larrecirc}) and the data acquisition system (DAQ rack in \Cref{fig:larrecirc}). The PLC modules are installed in the PLC1 and PLC2 racks. The processor module and the analog input and output modules are in the PLC1 rack, the digital input and output modules are in the PLC2 rack. The INT1 and INT2 racks contain the interfaces for the sensors and actuators and displays for pressure and vacuum sensors whereas
the COM rack contains the supervisory computer and its accessories.
For safety reasons every rack contains its own power distribution unit that provides the power for the entire rack and allows to cut the power of the complete rack with one single circuit breaker. The main power line of ArDM (three-phase, 400 V, 80 A per phase) arrives to the power distribution unit in the INT1 rack. From there all the power distribution units in the other racks, with the exception of the PLC1 and PLC2 racks, are supplied with single-phase current (230 V, 20 A). The PLC1 and PLC2 racks are supplied with 24 VDC which are delivered by a rectifier in the COM rack. Also all the sensors and actuators installed in the ArDM experiment are powered from the racks of the control system. To protect the  control system from short power cuts, all the equipment, with exception of the devices supplied by the INT1 rack, are connected to two UPSs (uninterruptible power supplies), UPS 12 and UPS 23 , with a total power of 6300 W. The two UPSs can cover the power consumption of the connected equipment for about 15 min. In case of a power cut, the UPS 1 supplies the equipment in the INT2, the DAQ and the HV rack, the UPS 2 supplies the COM rack and thus also the PLC1 and PLC2 racks.
The control software is based on PVSS (Prozessvisualisierungs- und Steuerungssystem, process visualisation and control system) used by CERN.

Although larger in scale, we expect that the process control and monitoring system for the \six prototype will be functionally very similar to the one developed and successfully operated for ArDM. We therefore expect that it will be rather straight-forward to develop and implement a new dedicated process control and monitoring system.

\clearpage
\section{DLAr offline requirements and software}
\label{chap:reconstruction}

\graphicspath{{./Section-Reconstruction/figs/}}

\subsection{Overview}
LAr TPCs record the complete electronic image of an ionizing
event. Due to the optimal electron transport properties of the LAr
medium (high electron mobility and small diffusion), large volumes can
be instrumented with a spacial resolution down to the
millimeter-scale. Ionization electrons drift over the distance 
$z=v\cdot t_{drift}$ towards the charge readout passing from liquid
to gas phase, where they induce
signals on different sets of readout electrodes, hereafter referred to as
\textit{views}. The readout views provide at
each sampling time two or more projections of the $(x,y)$
image of the event. In addition to the event topology, the amplitude of the recorded
signals give a direct measure of the produced ionization charge along
each track,  providing calorimetry and particle specific local energy loss
information. 

Track reconstruction aims at the extraction of
the relevant information, such as the types and the momenta of the detected
particles. In the simplest case, where 
particles stop inside the detector, the stopping power along the
produced track allows to identify the
particle~\cite{Bueno:2007um,Arneodo:2006aa}. In case the particle 
does not stop inside the fiducial volume, an
estimation of the momentum can be obtained by measuring the multiple
scattering of the traversing particle~\cite{Ankowski:2006aa}. Besides
the reconstruction of tracks, it
is possible to reconstruct showers: in this case, it is not
needed to reconstruct each single hit in three
dimensions. A strategy is to reconstruct global event
parameters like the total charge, the shower direction, the charge
profile and the vertex as well as the initial part of the shower. This
latter is used 
to discriminate $e^-/e^+$ from $\gamma/\pi^0$ showers~\cite{Epi0Note2003}.

\subsection{The Qscan software}
\label{sec:reco-qscan}

Qscan is a multipurpose software framework to simulate,
reconstruct and visualize events from LAr TPC detectors.  
It was originally created and developed
 by the ETHZ group for the ICARUS experiment and 
 was then adapted to other detectors. It was employed 
 for all analyses the ICARUS data collected in 
 Pavia~\cite{Amoruso:2003sw,thesis_JavierRico,Amoruso:2004ti,Amoruso:2004dy} 
 as well as to reconstruct quasi-elastic
neutrino events in the 50~L ICARUS TPC that was exposed to the CERN West Area
Neutrino Facility WANF~\cite{Arneodo:2006aa,Thesis_AlbertoMartinez}. 
Qscan 
allows to simulate, reconstruct and visualize events for various
detectors. It is able to read raw-data of all our hardware setups developed since 1997.
Due to the development of several new double phase LAr TPC prototypes,
including the 3~L~\cite{Badertscher:2010zg}, the 120~L at J-PARC 
and the 200~L detector at CERN~\cite{Badertscher:2012dq,Badertscher:2013wm}, a set of new
algorithms and a new
ROOT\footnote{http://root.cern.ch}~\cite{brun199781} based graphical
user interface have been implemented. 

The Qscan software framework
comes along with three main functionalities: 
\begin{itemize}
\item It provides a set of tools to decode, store and
  reconstruct data from all our hardware setups. 
\item Being interfaced to VMC  (and currently GEANT3 and
  GEANT4) to propagate particles through any detector geometry, it
  allows to produce fully digitized MC events that can be post-processed
  similar to real data from a detector.  
  \item A batch-mode
\item An interactive mode with graphical user interface that allows to scan raw data
  online or offline. Moreover, Qscan has a 3D event display to visualize MC
  truth events as well as the reconstructed event for both real
  and MC generated data inside the detector geometry. 
\end{itemize}

\begin{figure*}[t]
\centering
\includegraphics[width=0.7\textwidth]{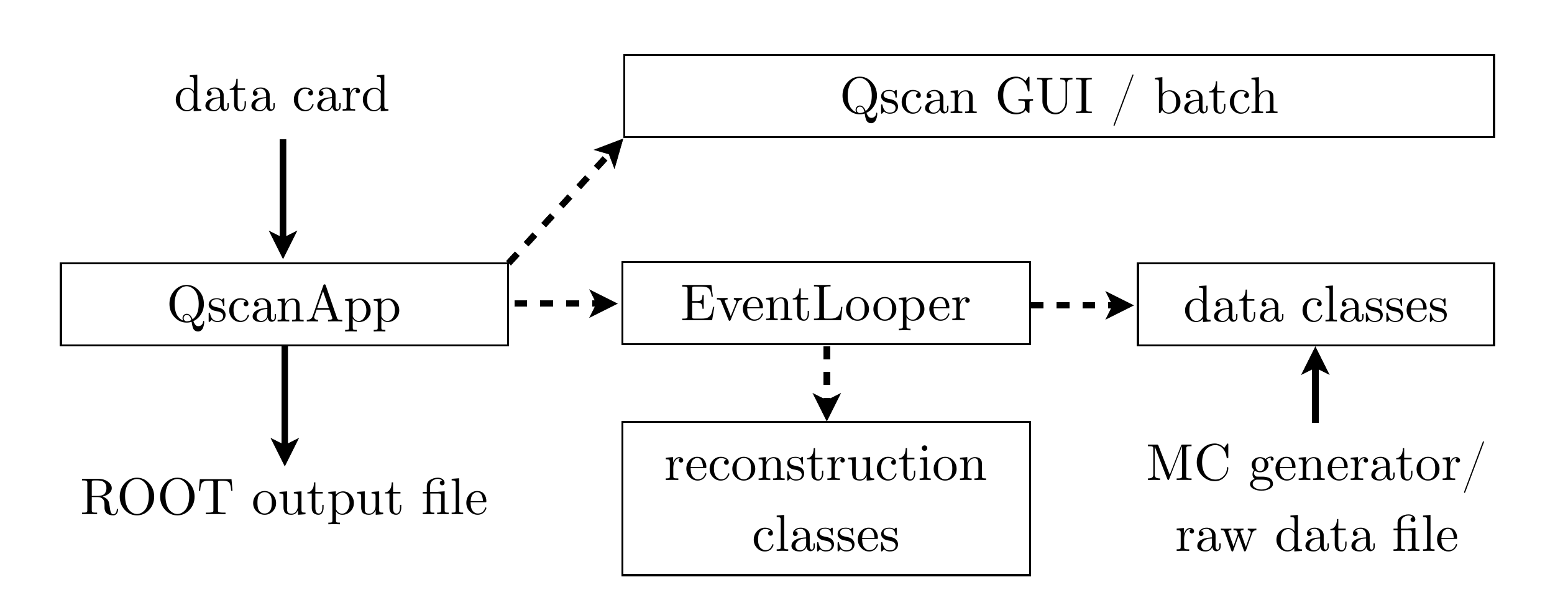}
\caption{Basic organisation of the Qscan software: boxes without frame
are calibration, in and output files, while framed boxes name
different classes.}
\label{fig:qscan-organisation}
\end{figure*}
In order to have a well structured software package that allows to add
new detectors with minimal efforts, Qscan is written in the
object oriented C++ language. As shown in the diagram in
\Cref{fig:qscan-organisation}, the main program \textit{QscanApp}
can either run in batch mode in order to process a large amount of data, or
it initiates the graphical user interface of Qscan. Depending on
which detector and which kind of reconstruction is asked by the user in the so called \textit{data
  cards} file, the proper data class with the detector geometry is
initialized as well as the reconstruction menu. After the
initialization, the \textit{EventLooper} goes from event to event,
either reading an external data file or generating artificial events via MC
simulation. Once all the signal waveforms are stored in the data
class, the event is processed by the reconstruction algorithms. The
resulting reconstructed objects are finally written into a file with
the common ROOT format that can later be used for a further
analysis. 
More details on the software layout are
presented in~\cite{Lussi:2010aa}.

\subsection{Event simulation}
\label{sec:reco-mc-sim}

To optimally utilise the fact that LAr TPCs ideally provide a high resolution image of
any ionizing event, MC simulations are an important tool to validate
reconstruction algorithms and to study the detector
performance. Moreover, a working MC simulation of a
detector allows to study different effects like non-uniform drift
fields, different noise levels, impurities, \textit{etc}. This is in
particular important for physics performance studies of giant LAr
detectors for future long baseline neutrino oscillation
experiments~\cite{Stahl:2012exa}. 
All the MC simulation code and in particular
the digitization procedures that are presented hereafter,  have been
validated with both testbeam and cosmic ray data, as reported in~\cite{Lussi:2010aa}. 

\subsubsection{Particle propagation in detector geometries}
\label{sub:reco-geant}

\begin{figure*}[t]
\centering
\includegraphics[width=1\textwidth]{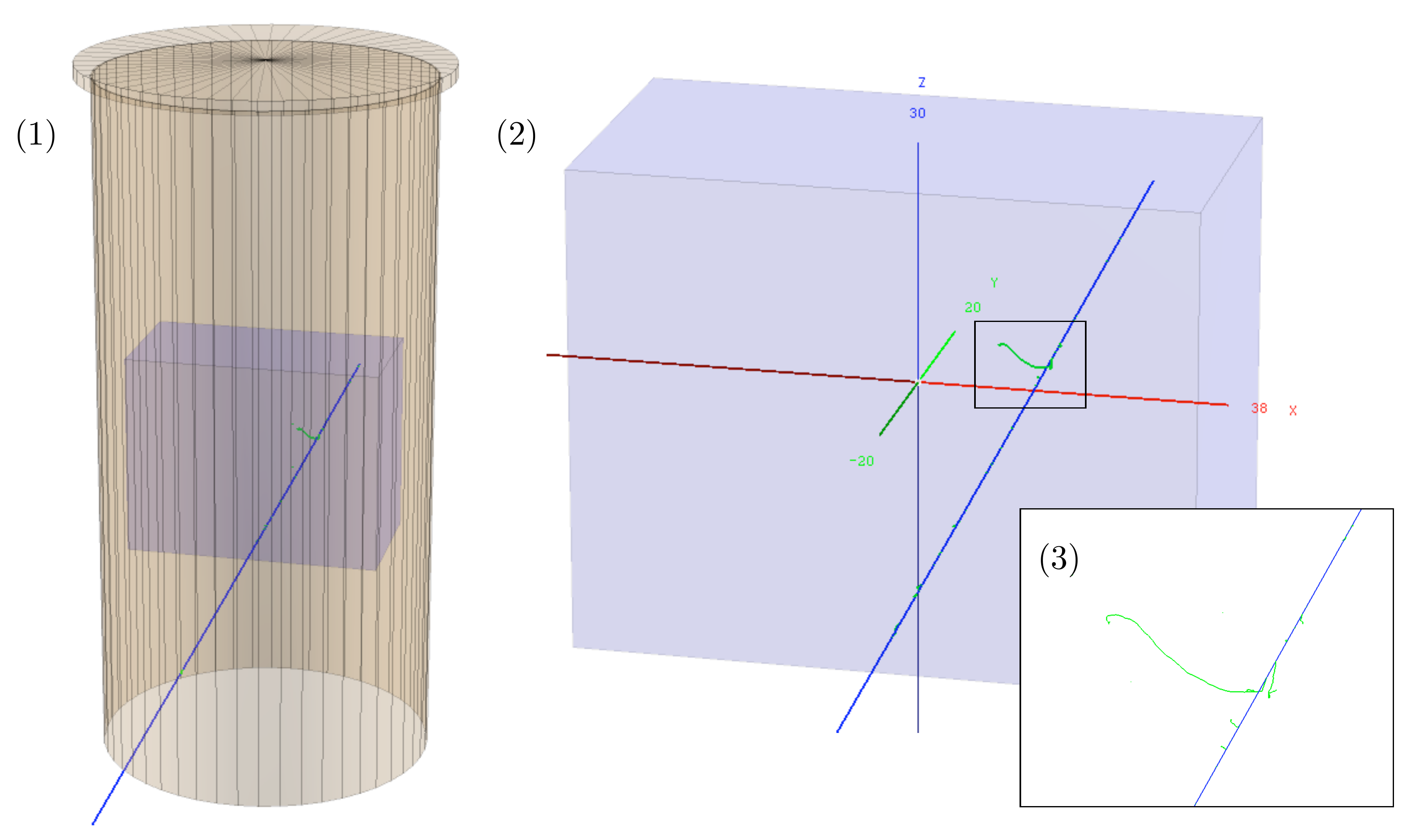}
\caption{Three dimensional event display of Qscan, showing the
  ROOT geometry of the 200~L detector and $\mu^-$, tracked through the
  geometry by \textsc{Geant4}. The figure shows an overview
  of the geometry~(1), the extracted fiducial volume together with its local
  coordinate system~(2) and a closeup of the $\mu^-$ track with
  several $\delta$-electrons~(3).}
\label{fig:mc-truth-event}
\end{figure*}

Qscan uses the Virtual Monte Carlo package
VMC\footnote{http://root.cern.ch/drupal/content/vmc}
(see~\cite{Hrivnacova:2012gz} and references therein) to interface
with \textsc{Geant4}\footnote{http://geant4.cern.ch}, a toolkit for the
passage of particles through matter~\cite{Agostinelli2003250}. 
Since the VMC interface loads the MC libraries at runtime, Qscan is
completely independent from the MC simulation code, allowing for
instance to use \textsc{Geant3} instead of \textsc{Geant4}. The second
advantage is that the ROOT geometry package \textit{TGeo} can be used
to define the geometry of the experimental setup. This feature is important
since the same geometry definition can be used to track particles in
\textsc{Geant3} and \textsc{Geant4}. Moreover, TGeo is also the
preferred geometry format to display the detector together with its
reconstructed event objects in the three dimensional, ROOT based,
event display. An example of a MC simulated muon event in the 200~L
detector~\cite{Badertscher:2012dq,Badertscher:2013wm} is displayed in
\Cref{fig:mc-truth-event}. (1) shows the overview of the geometry,
consisting of the cylindrical cryostat and the cuboidal detector. The
isolated TPC volume with the MC truth tracks are shown in (2) and (3)
shows a closeup of the $\mu^-$ track in blue with some $\delta$-ray
electrons in green. 
 
\textsc{Geant4} provides a large set of different physics
processes. Depending on the required level of details and
physics models, it is possible to enable or disable different electromagnetic
and hadronic processes via physics lists. Typically we are using the
QGSP\_BERT physics list that uses the Quark Gluon String model
for simulating high energy hadronic interactions. To provide accuracy
at energies below 10~GeV, the list is extended with the Bertini cascade
model~\cite{Apostolakis:2008aa}. Concerning the tracking
configuration, depending on the readout 
pitch of the simulated detector, one usually has to define an upper
limit for the stepping size in order not to see any descrete
effects. Typically the maximum step length is fixed to about 10\% of
the readout pitch. Unlike in the case of typical \textsc{Geant4}
simulations, particle propagation cut-offs are defined as kinetic
energies, rather than ranges. In order to properly implement charge
quenching effects, one wants to track secondary electrons, being produced
along ionizing tracks, down to very low energies. As discussed
elsewhere~\cite{Amoruso:2004dy,Araoka:2011pw},  an electron tracking cut-off of
10~keV is a good compromise between microscopic accuracy and computing
speed. 

\begin{figure}[h!]
  \centering
  \includegraphics[width=.49\textwidth,height=0.38\textheight]{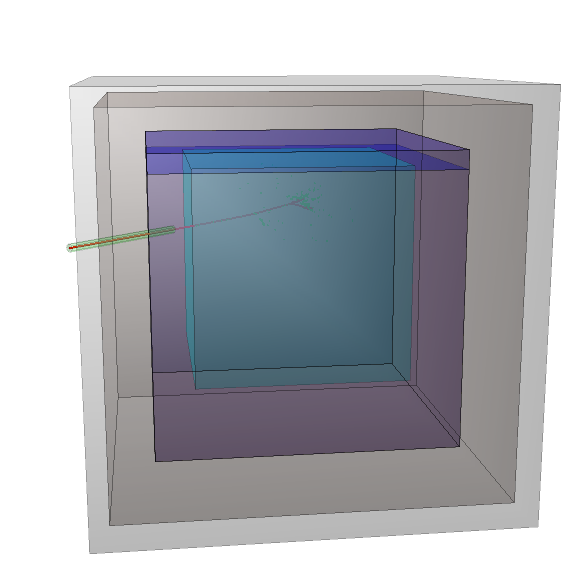}  
  \includegraphics[width=.49\textwidth,height=0.38\textheight]{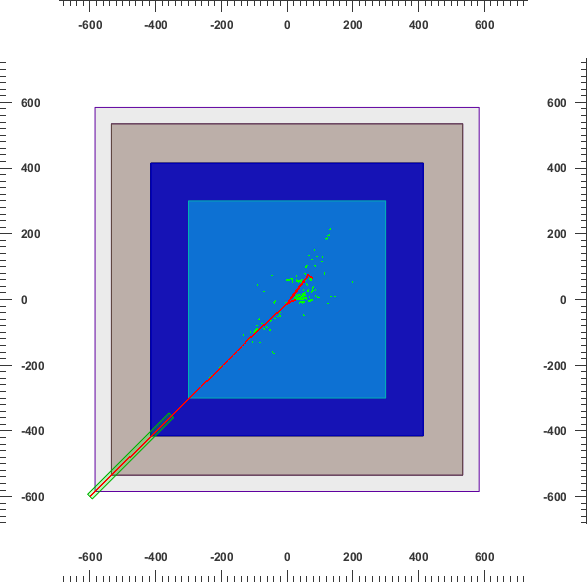}  
  \caption{Simulation of a 5 GeV/c $\pi^+$ event in the $6\times6\times6$ m$^3$
    detector. 3 dimensional view (left) and top projection of the detector (right).}
    \label{fig:qscan_piplus_666}
    \end{figure}

A first version of the $6\times6\times6$ m$^3$ detector geometry
has been implemented in Qscan.  The geometry is described with the
ROOT $TGeo$ package and the particle tracking through the detector can
be performed with either Geant4 or Geant3 using the Virtual Monte
Carlo interface, as discussed above. The
materials and their densities used in the simulation are listed in
Table \ref{tab:666_material}. Figure \ref{fig:qscan_piplus_666} shows
a simulated event of a 5 GeV/c $\pi^+$ shot trough (a preliminary version of) the beam pipe and
initiating a shower in the LAr fiducial volume. Most parts of
the detectors can be seen in the figure: the fiducial (cyan) and LAr (blue)
volumes, the corrugated membrane (too thin to be visible), the
insulating polyurethane foam (dark gray) the concrete outer vessel
(light gray) and the preliminary beam pipe (green). The latter consists of a
horizontal vacuum cylinder placed at 1.5 m from the top of the liquid
argon level. The beam axis is oriented at 45 degrees with respect to
the readout strips of both views.
\begin{table}[htb]
\renewcommand{\arraystretch}{1.5}
\renewcommand{\tabcolsep}{4mm}
\centering
    \begin{tabular}{lll}
      \toprule
      Detector component&Material  & Density[g/cm$^3$]  \\
      \hline
      tracking medium&liquid argon  & 1.396 \\
      membrane& stainless steel  & 8.030\\
      thermal insulator& polyurethane foam & 0.100\\
      outer vessel& concrete  & 2.500\\
      \hline
        \end{tabular}
        \caption[Simulated decay channels and branching ratios.]{
          \label{tab:666_material} Materials and densities for
          the simulated $6\times6\times6$ m$^3$ detector.
        }
\end{table}

\subsubsection{Waveform generation}
\label{sub:reco-digit}

The goal of the waveform generation is to convert the MC truth information to
LAr TPC data, as it would be recorded in a real experiment.
Based
on the energy deposit $\Delta E_{dep}$ and the length $\Delta x$, given
at each step of every tracked particle in the MC
simulation, the readout signals are produced as explained in the following:

\begin{enumerate}
\item The deposited energy $\Delta E_{dep}$ is converted into the
  amount of produced ionization charge ${\Delta Q_0=e\,(\Delta
E_{dep}/W_{ion})}$ with the elementary charge $e$ and the
ionisation energy of ${W_{ion}=23.6}$~eV, as given in
\Cref{t_argon_specifications}. Due to electron ion
recombination, which depends on the ionization density as well as on
the applied electric field, a fraction of the produced electrons
recombine. Since Birks' approximation gives the best fit to the
reconstructed data~\cite{Amoruso:2004dy}, it is
reasonable to implement quenching at MC level by multiplying the
charge with $\mathcal{R}_{Birks}(\Delta E_{dep}, \Delta E_{dep}/\Delta
x)$. The two parameters
of the model $A_{MC}=0.8$ and
${k_{MC}=0.05}$~${(\text{kV/cm})\left(\frac{\text{MeV}}{\text{g/cm}^2}\right)}$ 
provide good agreement with the data. 

\item The remaining charge is then transported up to the collection
  plane. After choosing the coordinate system such that the anode
  plane is parallel to the $(x,y)$ plane, the charge, being initially
  produced at $(x,y,z)$, is drifted along the electric field lines
  until it leaves the fiducial volume or it reaches the
  readout plane. In case of a constant field $(0,0,\mathcal{E})$, the
  final charge position equals $(x'=x,y'=y,z'=z_{anode})$ and
  the corresponding drift time is
  $t_{drift}=(z_{anode}-z)/v_{drift}(\mathcal{E})$. Otherwise, in case
  of an inhomogeneous electric field, an external look-up table, providing
  the final coordinate as well as the drift time for any point within
  the fiducial volume, is used. After including also charge
  attenuation effects due to an imposed finite free electron lifetime
  $\tau_{MC}$, the charge is given by
  \begin{equation}
    \Delta Q=\frac{\Delta E_{dep}}{W_{ion}} \cdot e \cdot
    \frac{A_{MC}}{1+k_{MC}/\mathcal{E}\cdot \Delta E_{dep}/\Delta x}
    \cdot e^{-t_{drift}/\tau_{MC}}.
  \label{eq:mcCharge}
  \end{equation}
Longitudinal diffusion effects are included by convoluting the charge with a Gaussian
of width given by Eq.~\ref{e_transversal_diffusion_gas}.

\item Depending on the arrival position $(x',y',z'=z_{anode})$ of the drift
  charge $\Delta Q$, currents are induced on the corresponding readout
  electrodes of the different views. These currents are then further
  processed by a charge sensitive preamplifier. The finally recorded
  signal $V_{out}(t)$ is a convolution of the induced current $I(t)$
  and the response of the preamplifier $h(t)$:
  \begin{equation}
    V_{out}(t)=I*h(t)=\int^t_{t_0} I(t')\cdot h(t-t') dt'.
    \label{eq:induction}
\end{equation}
  Obviously both the preamplifier response and the induced currents depend
  on the readout type and the used electronics.
  The  simplest case is the LEM readout with a 
  2D anode: due to the fast
  electron drift in gas and the short induction gap between LEM and 2D
  anode, the induced current $I(t)$ approaches a $\delta$-function and
  the signal is directly given by the fast response of the
  preamplifier. 
\end{enumerate}
After finishing this procedure for each step, noise of a given
amplitude can be added on top of the signal waveforms. Besides the
generation of white noise, it is also possible to
use a specific frequency spectrum that has e.g. been taken
from real data. The final step is the digitization of the generated
waveforms. 
\begin{figure*}[t]
\centering
\includegraphics[width=1\textwidth]{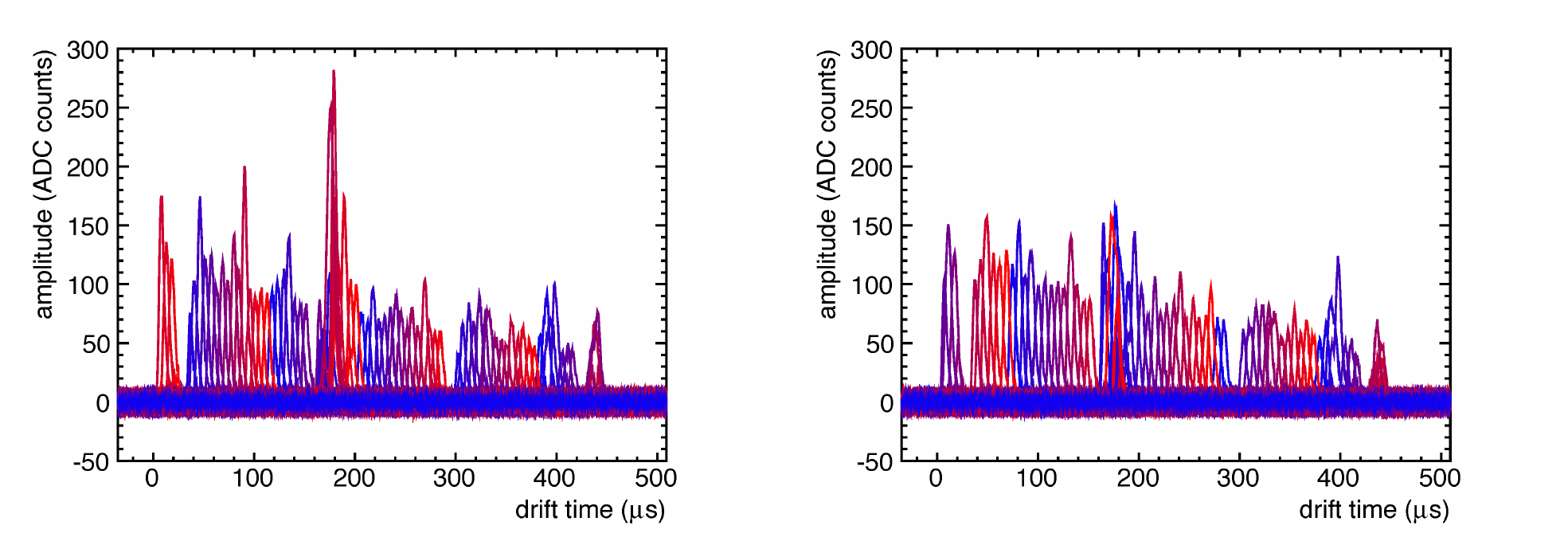}
\includegraphics[width=1\textwidth]{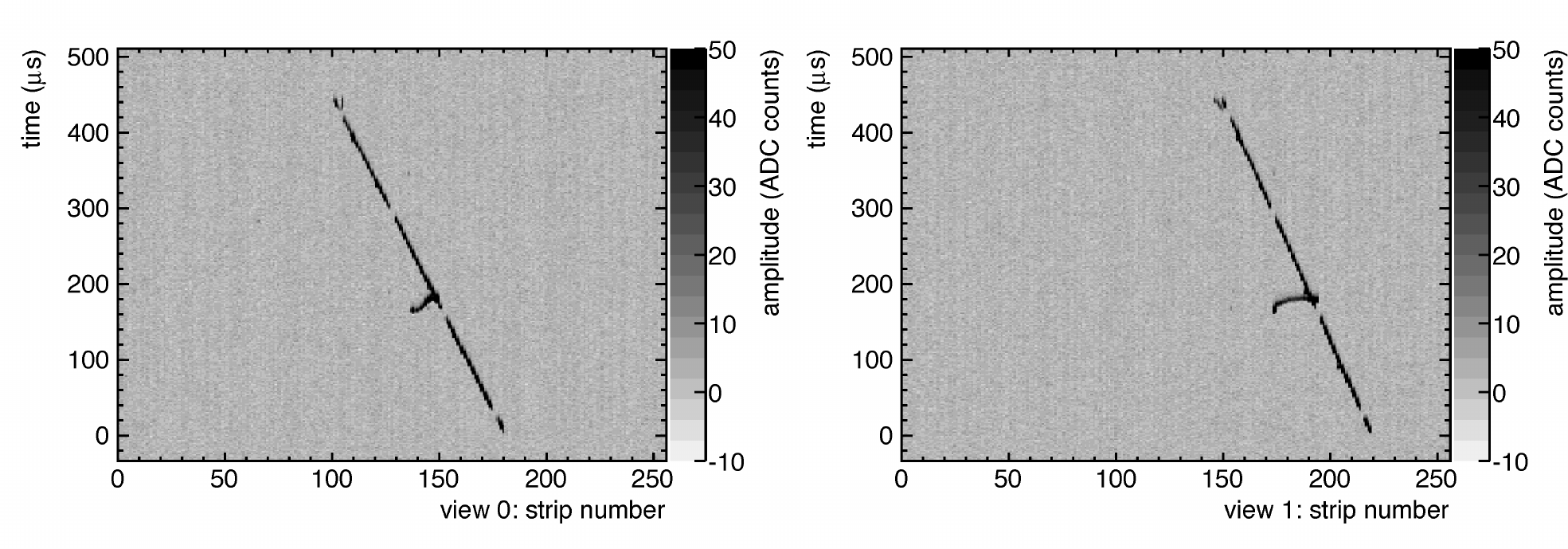}
\caption{Example of a fully MC generated $\mu^-$
  event in the 200~L detector. The MC steps of the same event are
  shown in \Cref{fig:mc-truth-event}. The waveforms for the two views
  (left and right) have been generated, using the known ETHZ
  preamplifier response, and Gaussian
  noise with an RMS value of 3~ADC counts. Top: waveforms $V_{out}(t)$ of
  all the readout channels are shown with different colors. Bottom:
  typical event display, showing the drift time vs strip number. The
  greyscale is proportional to the amplitude. }
\label{fig:digitized-event}
\end{figure*}
\Cref{fig:digitized-event} shows on top the generated waveforms of the
MC $\mu^-$ from \Cref{fig:mc-truth-event} and on the bottom the
corresponding event display.

\subsection{Event Reconstruction}
\label{sec:event-reco}

After recording data from the detector, the data
is processed by several reconstruction algorithms. Starting from the local,
channel-by-channel signal discrimination, the final goal is to
reconstruct physical objects like tracks, showers and event
vertices. As there can be large differences among detectors and event
types, Qscan provides various methods to reconstruct events. 
The reconstruction basically goes through the following steps:
\begin{enumerate}
\item The raw waveforms are processed: this involves noise reduction as
  well as the subtraction of the baseline (see \Cref{sub:reco-signal-processing}). 
\item Hits, defined as signals
  that are discriminated from the noise, are identified and reconstructed (see \Cref{sub:reco-hit}). 
\item Clusters are formed by grouping close hits together (see \Cref{sub:reco-clustering}).
\item Tracks are identified for each view (see \Cref{sub:reco-tracking}).
\item Finally, the three dimensional track reconstruction is done by matching 
  coincident tracks from different views (see \Cref{sub:reco-3dtracking}). 
\end{enumerate}

\subsubsection{Signal processing}
\label{sub:reco-signal-processing}

The data acquisition system records the amplified, shaped and digitized
output voltage~$V_{out}(t)$ for each electrode with a discrete time
sampling. In order
to extract physical signals efficiently and accurately, a minimal signal to
noise ratio of about~10 is required. Although the
electronics is in principle designed to fulfill this requirement,
experimental data can be distorted by external noise sources and an
imperfect shielding of the detector and the signal lines. In order to suppress
noise without affecting the signal component too much, hence improving
the signal to noise ratio, two different
algorithms are used: the \textit{Fast Fourier Transform (FFT) filter},
and the \textit{coherent noise subtraction} algorithm. 

The \textit{FFT filter} makes use of the fact that
induced noise, being produced by external sources like switching power
supplies, computers, \textit{etc}, is often dominated by a few specific
frequencies. After transforming the waveforms $V_{out}(t)$ to the
frequency space, the Fourier transformed waveforms 
$\hat{V}_{out}(\omega)$ are processed in order to reduce the
noise components. The final, noise suppressed waveforms
$V_{out,filt}(t)$ are then obtained by applying an inverse FFT to
$\hat{V}_{out,filt}(\omega)$. Due to the fact that the waveforms are
discrete and the total number of samples is an integer power of two,
the computing time is minimized by using the FFT implementation
from~\cite{recipes}. 

\begin{figure*}[t]
\centering
\includegraphics[width=0.49\textwidth]{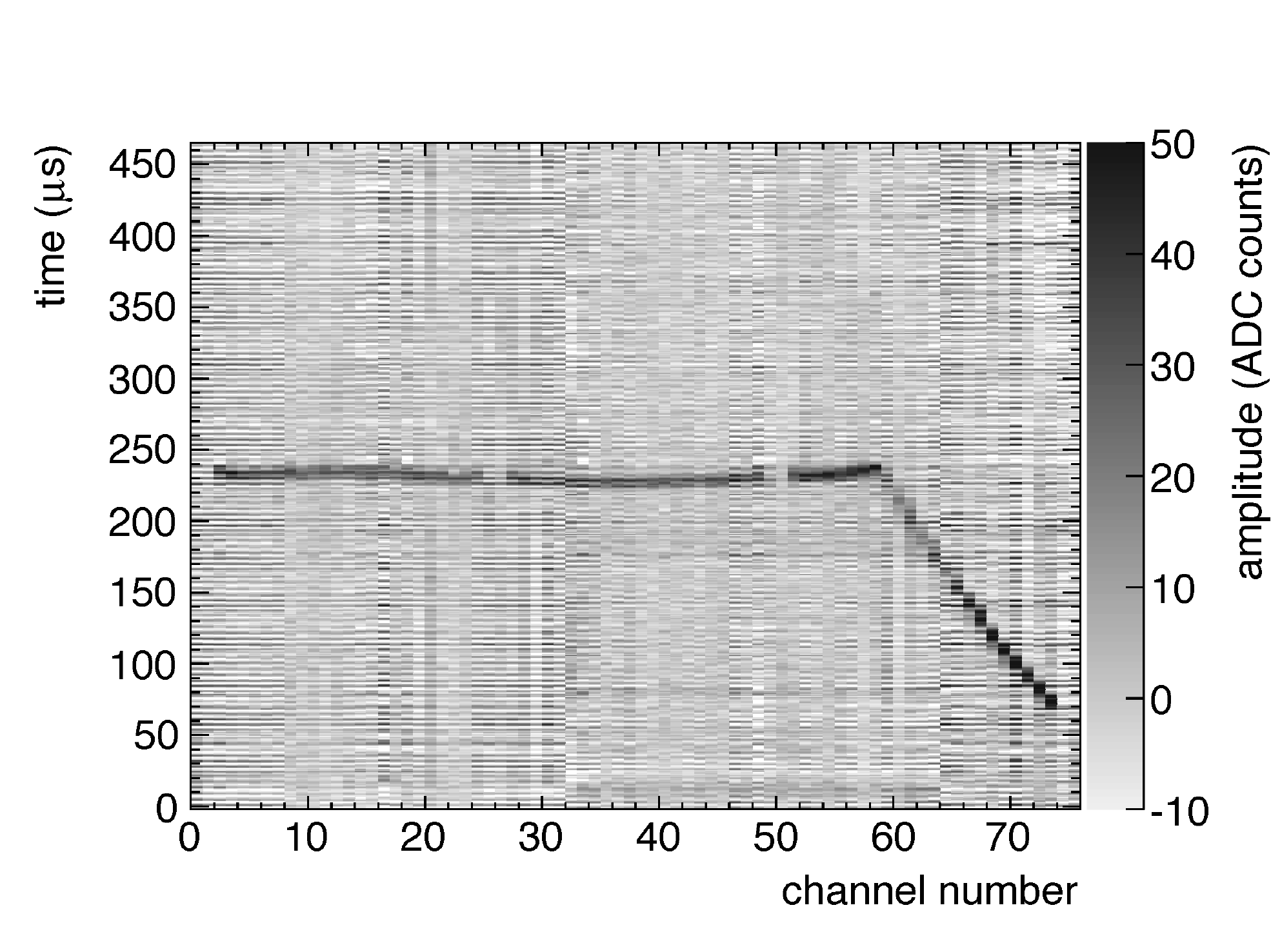}
\includegraphics[width=0.49\textwidth]{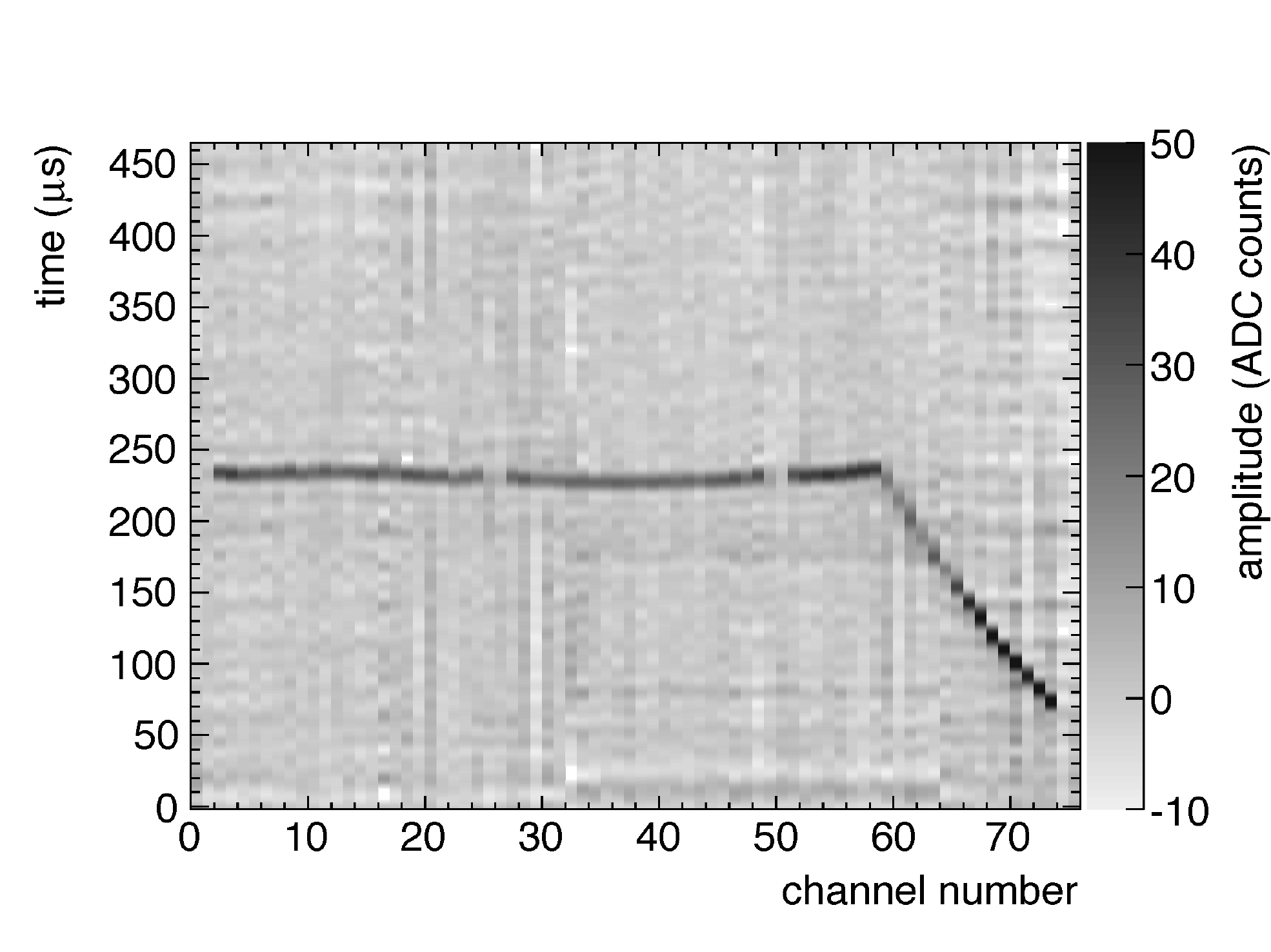}
\caption{Raw (left) and FFT filtered (right) stopping K$^+$ event from
  the 120~L single phase LAr TPC in a beam test at J-PARC (\cite{Araoka:2011pw}). The noise
pattern, shown on the raw event on the left, is removed by the low pass
filtering with a smoothened cut-off at the frequency of 80~kHz.}
\label{fig:raw-and-fftfiltered-event}
\end{figure*}
\begin{figure*}[t]
\centering
\includegraphics[width=0.49\textwidth]{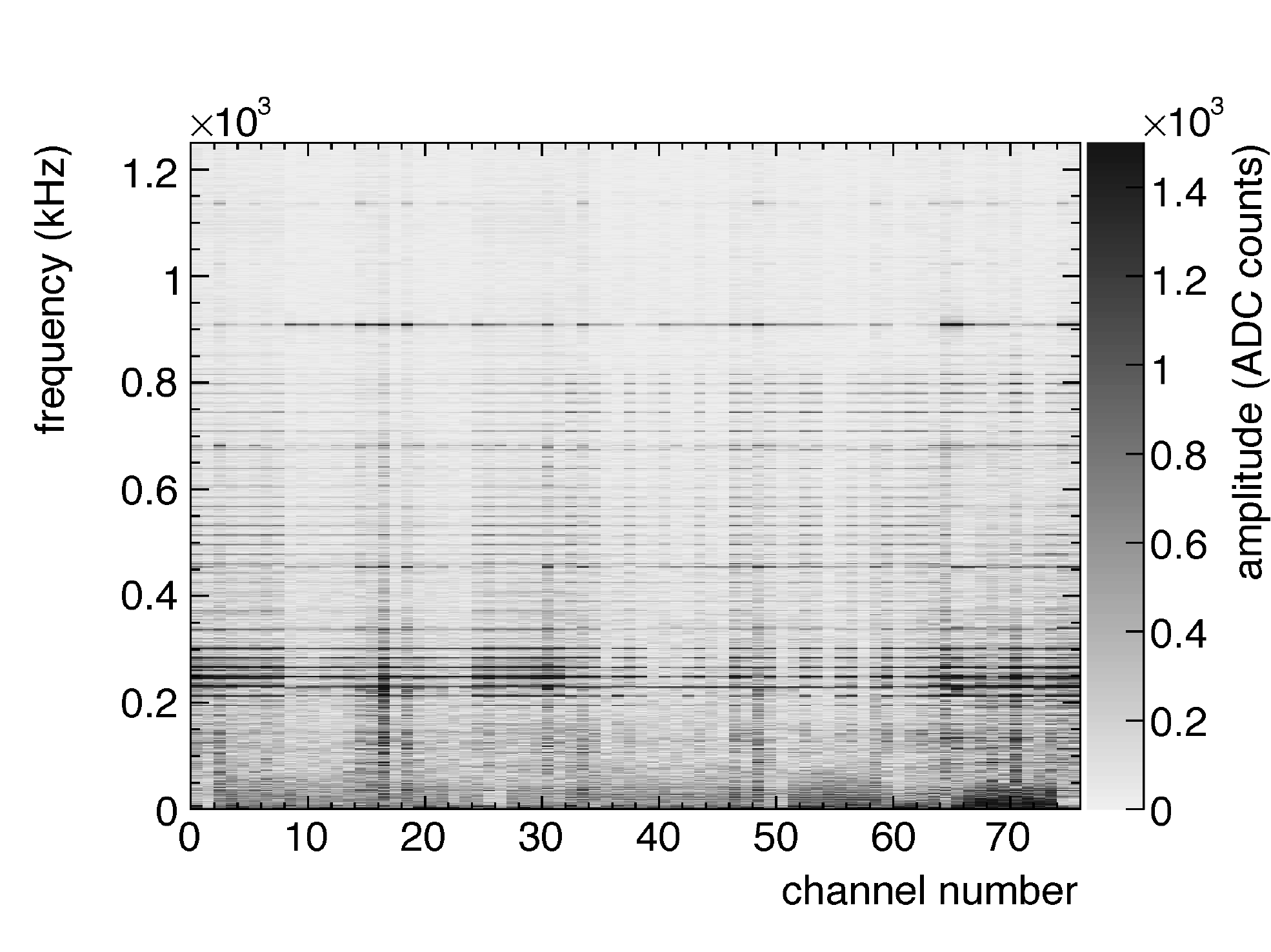}
\includegraphics[width=0.49\textwidth]{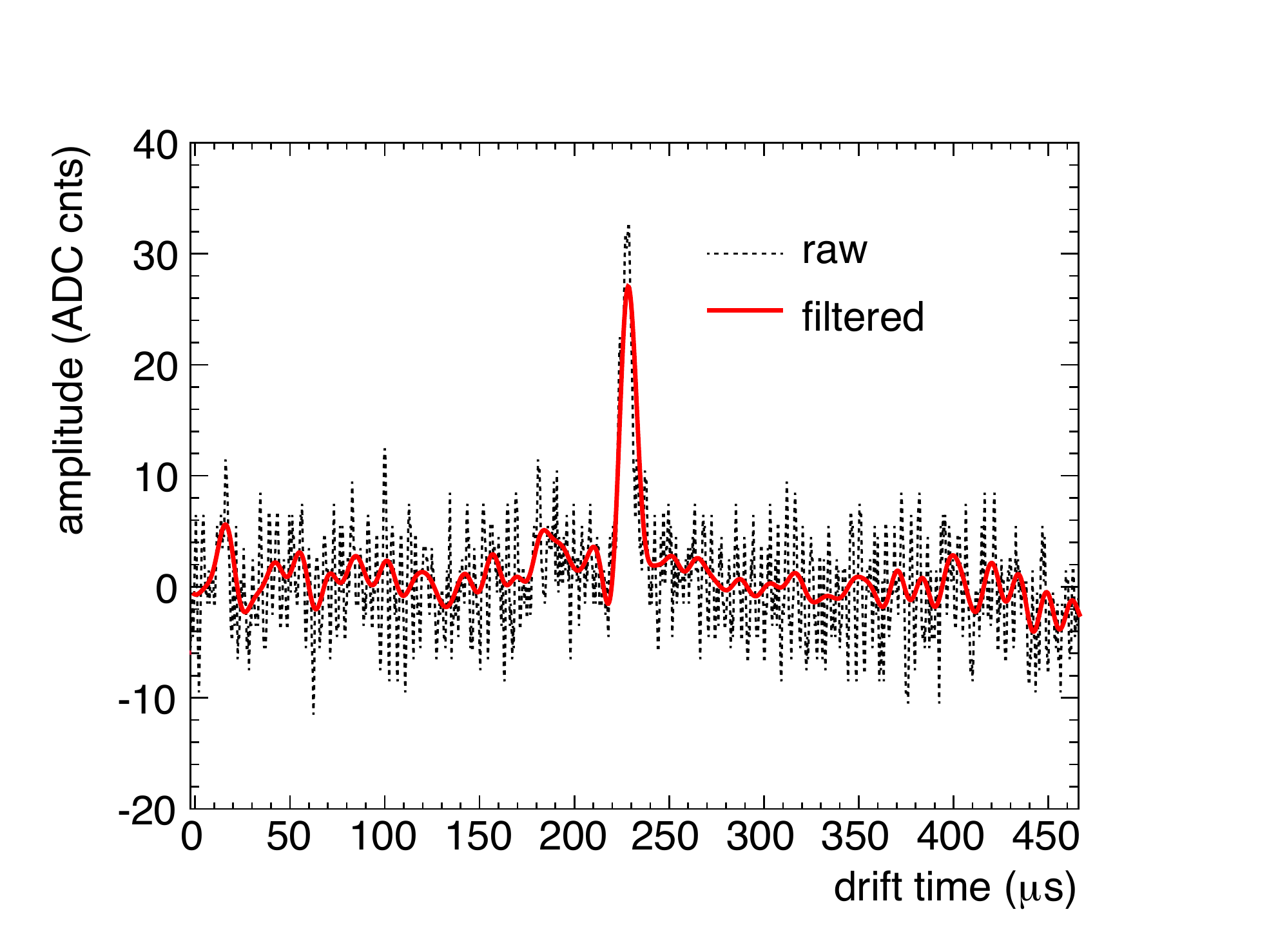}
\caption{Left: power spectrum of the raw event from
  \Cref{fig:raw-and-fftfiltered-event}. The two dimensional histogram
  shows the amplitude for each readout channel and frequency
  bin. Common noise frequency on all the channels are visible well
  above the 80~kHz cut-off. Right: single readout channel raw (dashed
  black line) and the filtered waveform (solid red line).}
\label{fig:noise-power-spectrum}
\end{figure*}
A concrete example of the FFT algorithm is presented in
\Cref{fig:raw-and-fftfiltered-event,fig:noise-power-spectrum}: the
K$^+$ beam event, recorded with the 120~L single phase LAr TPC in 
the beam test at J-PARC~\cite{Araoka:2011pw}, demonstrates the effect of the filter. The high
frequency noise, seen in the raw event on the left of
\Cref{fig:raw-and-fftfiltered-event}, is efficiently removed in the filtered
event on the right. In this case the best way to improve the signal to
noise ratio was to suppress frequencies above 80~kHz. The left plot of
\Cref{fig:noise-power-spectrum} shows the amplitude for each channel
and frequency. In this plot it can be seen that a continuous background is
superimposed with discrete frequency lines. While the continuous
component is due to the signals and the intrinsic preamplifier noise, the sharp
lines correspond to discrete noise frequencies that are induced by
external sources. Due to the fact that most of the
noise frequencies are above 100~kHz, a smooth cut-off, implemented
with a Fermi potential of width 3~kHz, efficiently suppresses the noise,
while keeping the signals, besides a reduction of the bandwidth,
unaffected. Unlike a smooth cut-off, a sharp
cut in the frequency spectrum introduces artefacts in the time
domain. \Cref{fig:noise-power-spectrum} 
shows the effect on a single waveform before (dashed black line) and
after filtering (solid red line). 

The \textit{coherent noise filter} is implemented to remove identical
noise patterns that are seen on larger sets of readout
channels. Unlike in the case of the FFT filter, which directly suppresses
the frequencies of single channels and thus reducing the signal
bandwidth, the coherent noise filter ideally subtracts only the 
noise while keeping the signals unchanged. During the operation of
detectors it is often observed that all the readout channels, being hosted
on the same readout board, have exactly the same noise in terms of
frequency, phase and amplitude.
\begin{figure*}[t]
\centering
\includegraphics[width=0.49\textwidth]{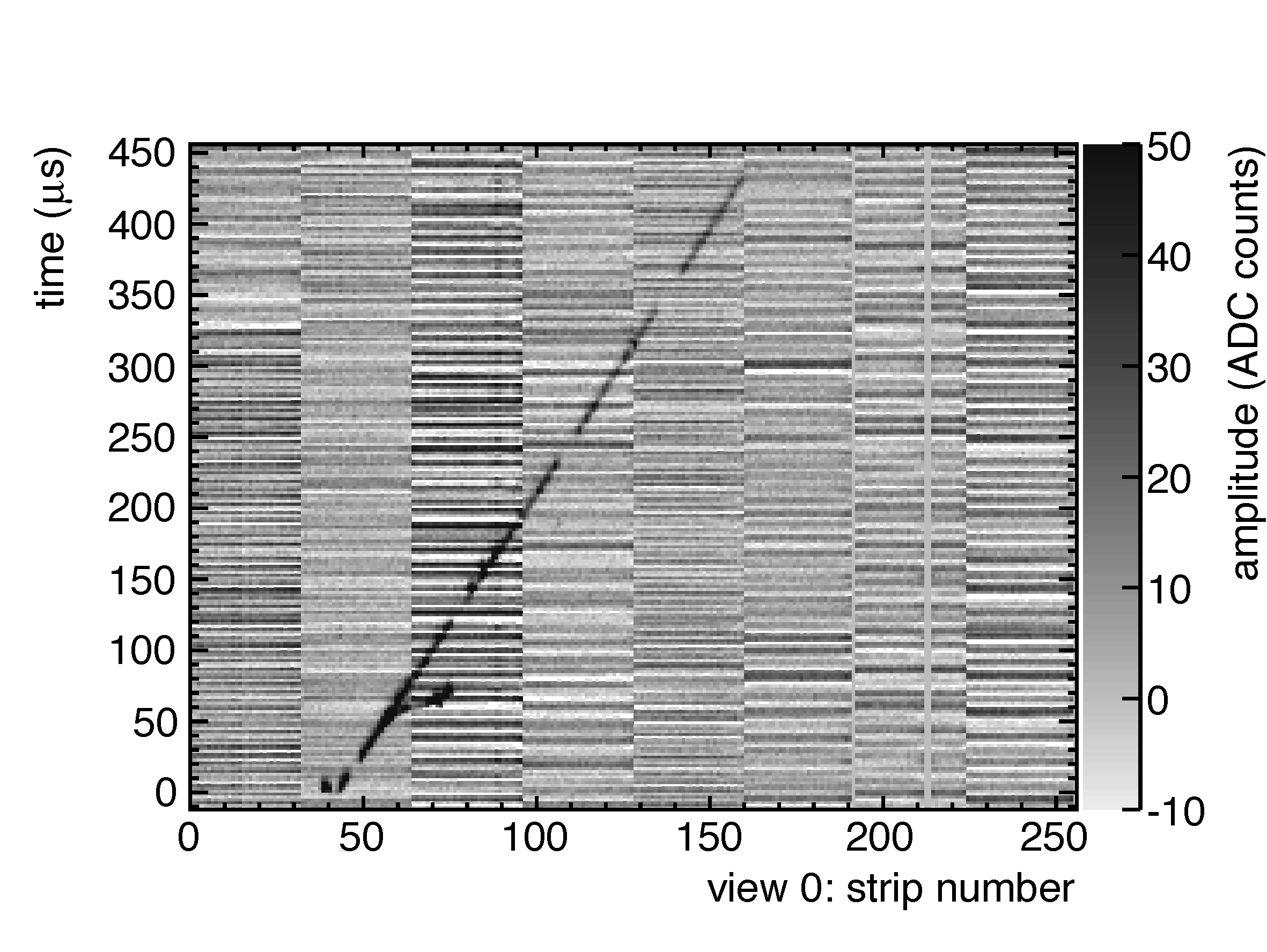}
\includegraphics[width=0.49\textwidth]{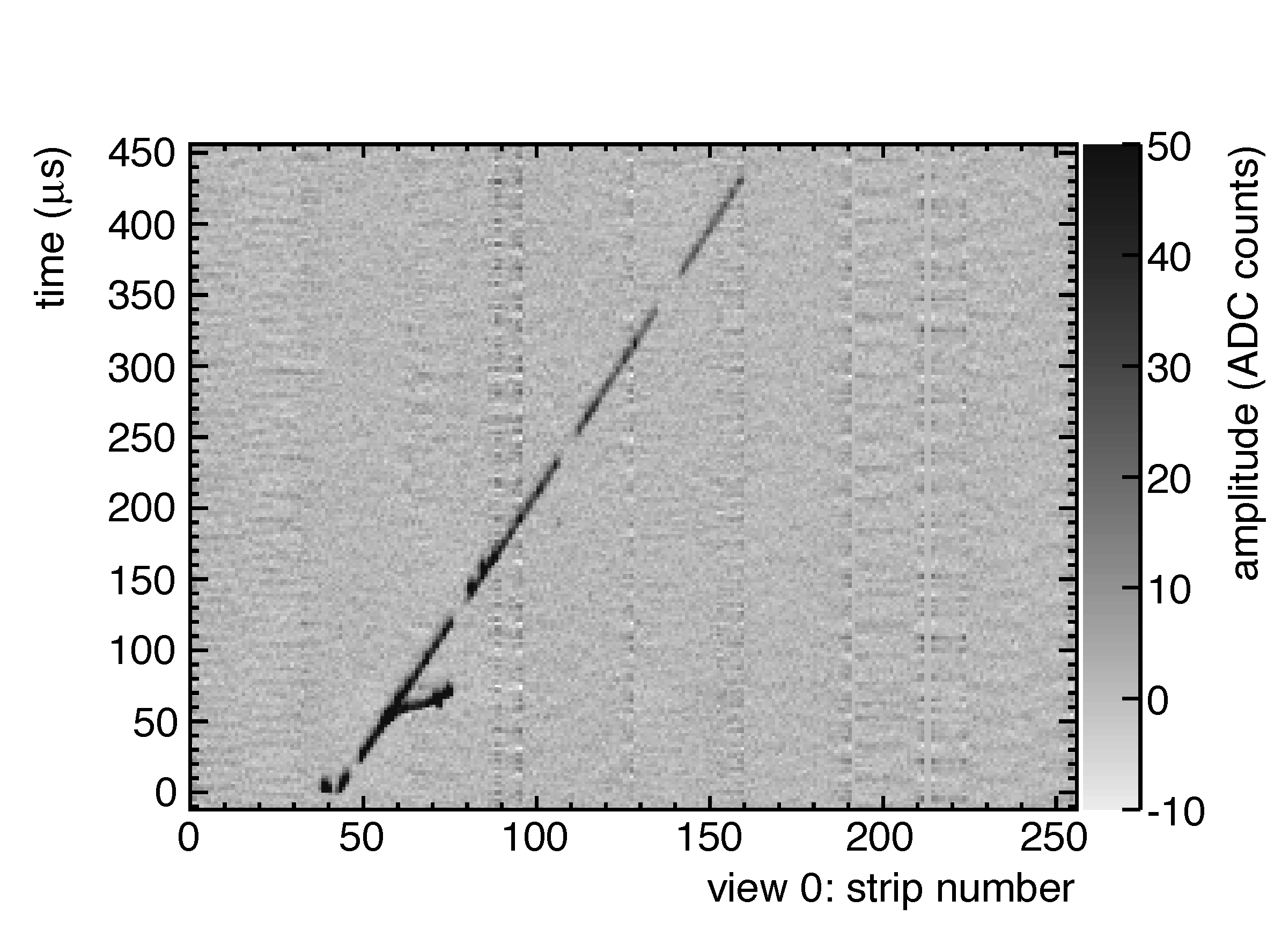}
\caption{Cosmic ray event from the $40\times 80$~cm$^2$ LAr LEM-TPC. Left: raw event that shows a
  characteristic coherent noise pattern. Right: final event, after
  applying the coherent noise filter. }
\label{fig:noise-coherenceFilt}
\end{figure*}
 An example of such an event, recorded with the 200~L double phase LAr
 LEM-TPC~\cite{Badertscher:2012dq,Badertscher:2013wm}, is given in
 \Cref{fig:noise-coherenceFilt}. Since the noise of every channel that
 belongs to the same acquisition board is almost identical, it can be
 considered as disturbance of the baseline. This time-dependent
 baseline first has to be computed for each single time sample, including all the channels on a physical
 readout board. Finally, similar to the subtraction of a constant
 pedestal, the baseline is subtracted from each sample. The difficulty
 of the calculation of the time-varying baseline is to select only the
 channels without a signal. Since signals are amplitude fluctuations
 with respect to the baseline, a good way to compute the baseline
 value for a given time sample is to use only the $N$ channels with
 the smallest voltage values. Depending on the needs, $N$ can be
 chosen between 1~and~32: in the case of $N=1$, the baseline is defined at
 each time sample $t$ by the minimum Voltage $V_{out}(t)$ of all the
 32 readout channels per acquisition board. On the other hand, in the
 case of $N=32$, signals are discarded and all the channels are used
 to calculate the baseline. The choice $N\approx16$ allows that
 signals are not affected, even in case they are present in half of
 the readout channels. 
 \begin{figure*}[t]
\centering
\includegraphics[width=0.8\textwidth]{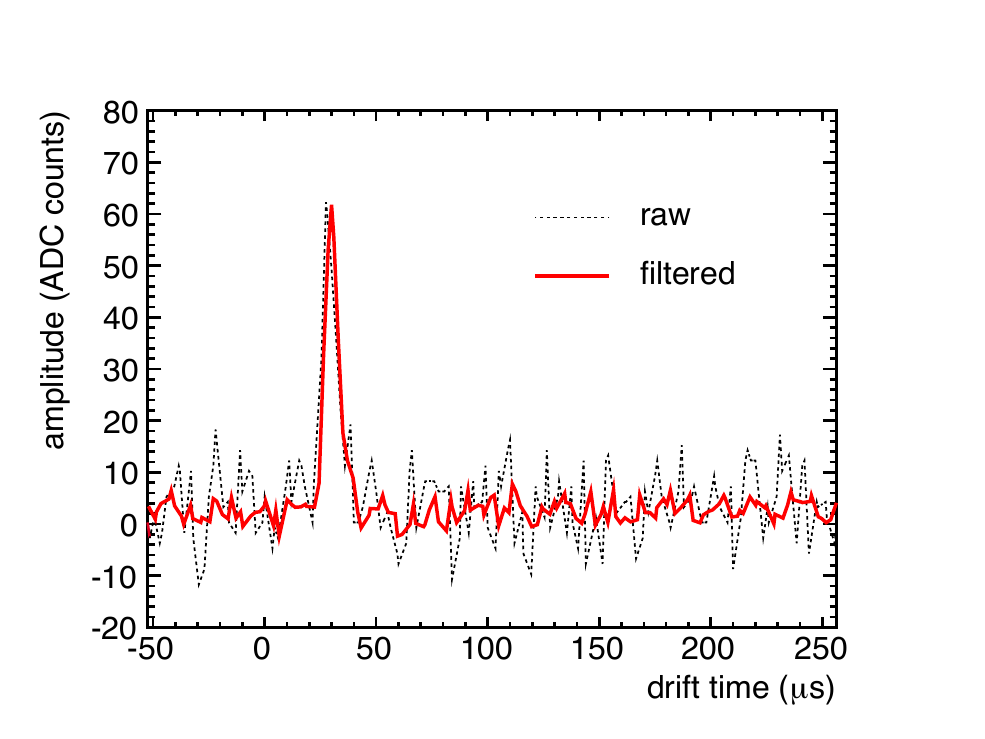}
\caption{Single waveform, extracted from the event shown in
  \Cref{fig:noise-coherenceFilt}. Both the raw (dashed black line) and
coherent noise filtered (solid red line) waveform are shown. }
\label{fig:noise-coherenceFilt-sig}
\end{figure*}
\Cref{fig:noise-coherenceFilt-sig} shows a single readout channel, taken
from the same event that is presented in
\Cref{fig:noise-coherenceFilt}. The noise fluctuations that are present
in the unfiltered event (dashed black line) are significantly suppressed
after applying the filter (solid red line). It can also be seen that
the algorithm does not affect the signals, despite its significant 
effect on the noise.

After suppressing the noise, the (constant) pedestal of each waveform
has to be computed and subtracted from each sample. In order to avoid
any bias due to physical signals, only 
pre-trigger samples are used to compute the mean value. Another
possibility is to use the most probable value since it does not depend
on tails, which are due to signals. 

\subsubsection{Hit identification and reconstruction}
\label{sub:reco-hit}
The smallest sub-unit of the reconstructed event is the
\textit{hit}. Physically, a hit corresponds to a track segment below
a readout strip. The charge of this
segment is then drifted towards the strip, where it induces a signal. 
The information that is attributed to a hit is the deposited
charge~$\Delta Q$, the three-dimensional length $\Delta x$ of the
track segment, the drift time, which is equivalent to the drift
coordinate~$z$, the readout view and the electrode strip number. 
Hits have to be extracted from the signal
waveforms by means of a standard threshold discrimination. Due to
changing noise conditions, the threshold is defined in relation
to the measured RMS noise  value $\sigma$, which is measured for each
event and readout strip, using the pre-trigger samples. A typical value for the threshold is $V_{thresh}=3\sigma$.  
\begin{figure*}[t]
\centering
\includegraphics[width=0.8\textwidth]{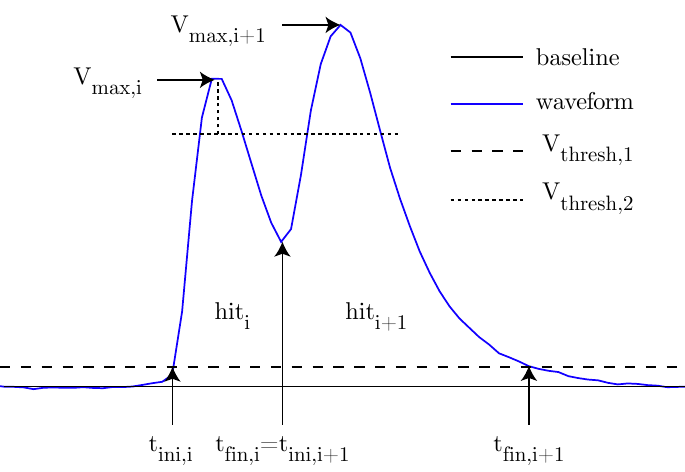}
\caption{Waveform (blue line) from a cosmic ray event that was recorded with the 3~L
  detector, showing a double hit structure (hit$_i$ and
  hit$_{i+1}$). It defines the hit variables $t_{ini}$, $t_{fin}$ and
  $V_{max}$ as well as the \textit{baseline} (black line) and the two
  thresholds $V_{thresh,1}$ and $V_{thresh,2}$.}
\label{fig:hit-finding}
\end{figure*} 
In the general case there can be several close, or overlapping tracks
per event, producing a superposition of several signals/hits on a single
readout strip. The shaping
time constants of the preamplifiers are chosen such, that double
tracks being separated by a few $\mu$s can be resolved. In order to
explain the working principle of the \textit{hit-finding} algorithm, \Cref{fig:hit-finding}
shows an example of two subsequent hits in a single readout
channel. The waveform was recorded 
with the $10\times10$~cm$^2$ LAr LEM TPC prototype~\cite{Badertscher:2010zg} and
shows a single waveform from a cosmic ray track with an emitted
knock on electron. 

Moving from the left to the right, 
a hit candidate $hit_i$ is initiated in case the signal waveform (blue) exceeds
a pre-defined threshold $V_{thres,1}$ (dashed line) and terminated, when it
either goes again below the same threshold or in case a new,
subsequent hit candidate $hit_{i+1}$ is triggered. The imposed trigger condition for subsequent
hits is that the minimum voltage between the two hit candidates goes
below the value
${min(V_{max,i},V_{max,i+1})-V_{thres,2}}$. $V_{thres,2}$ (dotted line) is a secondary
pre-defined threshold, responsible for the re-triggering of subsequent
hits. Besides increasing the hit finding
threshold, the number of fake hits due to noise can be reduced by imposing a
minimum time over threshold $t_{fin}-t_{ini}>\Delta T$.
Initial and final drift times ($t_{ini}$ and $t_{fin}$), as well
as the hit amplitudes $V_{max}$ are defined as indicated in
\Cref{fig:hit-finding}. In order to calculate the integral of each
hit, the waveform is integrated from the initial to the final time
sample. To reduce any bias, coming from the height of the threshold,
the window, in which the integral is computed, can be extended in case
there is no other hit attached. 

The main parameters of the hits that need to be reconstructed are the
\textit{hit time} and the  \textit{hit integral}: together with the location of the
corresponding readout channel, the hit time directly provides the
information of the hit location in the considered view (projection),
whereas the hit integral is related to the produced ionization charge
and therefore provides the calorimetric information. In order to
improve the accuracy of these two evaluated parameters, the signal
waveforms can be fitted with a pre-defined function. 
\begin{figure*}[t]
\centering
\includegraphics[width=0.49\textwidth]{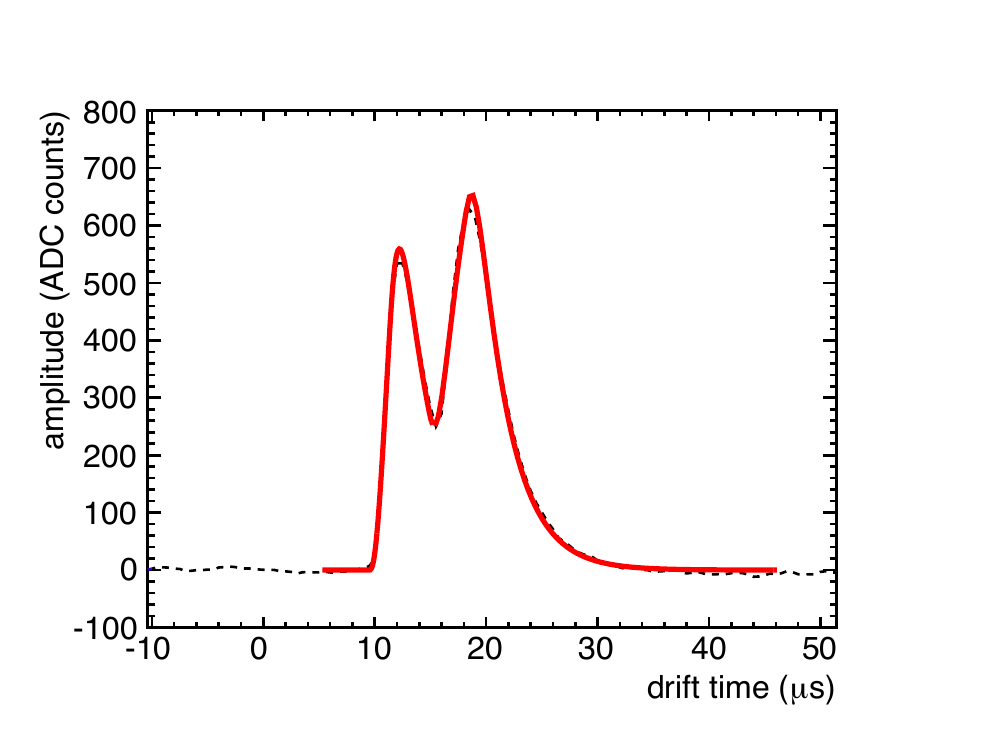}
\includegraphics[width=0.49\textwidth]{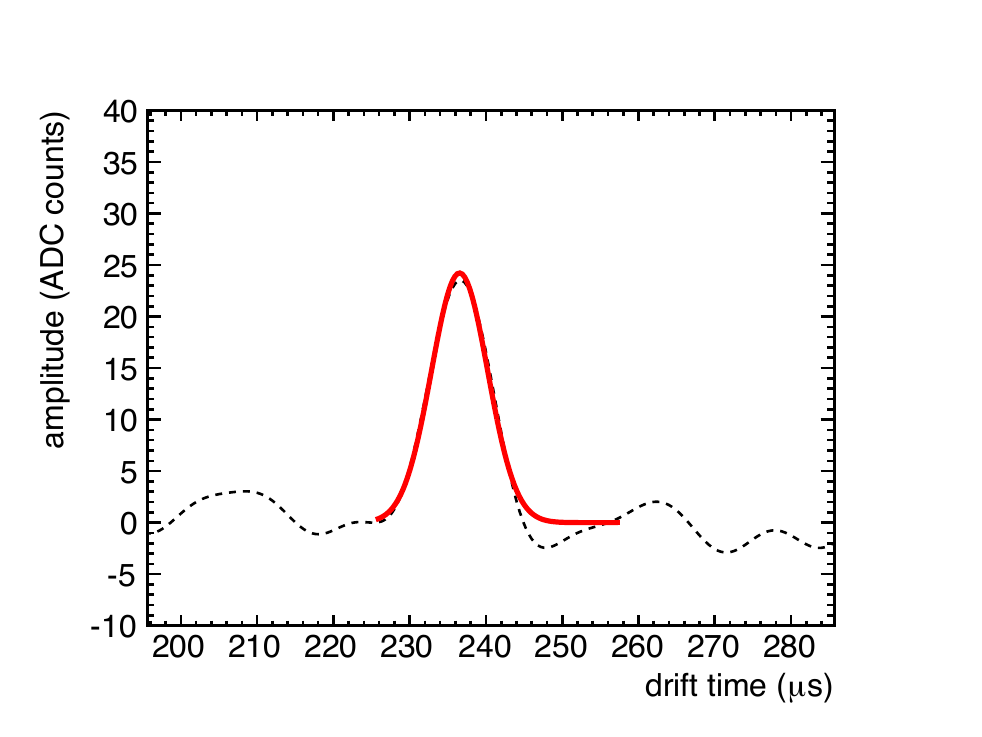}
\caption{Waveforms (dashed black line) from two different detectors,
  superimposed with the 
  fitted functions (solid red line). Left: the cosmic ray event from the 3~L double
  phase LAr LEM-TPC~\cite{Badertscher:2010zg} is, due to the fast response, fitted with a
  function that is based on the preamplifier
  response function. Right: beam event from the 120~L single phase LAr
  TPC~\cite{Araoka:2011pw}, due to its slow signals fitted with a Gaussian.}
\label{fig:hitfinding-fit}
\end{figure*}
Two examples of waveform fits are shown in \Cref{fig:hitfinding-fit}:
the left plot shows the waveforms of a double hit from a cosmic
ray event, recorded with the 3L double phase LAr LEM-TPC~\cite{Badertscher:2010zg}, whereas the
right plot is a beam event, recorded with the 120~L single phase LAr
TPC~\cite{Araoka:2011pw}. As the LEM-TPC has a faster signal induction, the used fitting
function is a convolution of a constant current ($\theta(t)$ is the step function):
\begin{equation} 
I(t):=I_0\cdot\theta(t-t_0)\cdot\theta(t_0+\Delta t - t)
\label{fitting-function-current}
\end{equation}
of duration $\Delta t$ and integral $I_0$ and the normalised response of the ETHZ 
preamplifier $h(t)$~\cite{Badertscher:2013wm}:
\begin{equation}
V(t)=I*h(t)=\int_{t_0}^{\infty} I(t')h(t-t') dt'.
\label{fitting-function}
\end{equation}
The function is analytically computed and since the response with
the integration and differential time constants is already known, the
only three fitting parameters are the integral $I_0$, the time $t_0$
and the signal width $\Delta t$. 

\subsubsection{Cluster finding}
\label{sub:reco-clustering}

After a successful hit-finding, adjacent hits of different readout
channels are grouped together. Since physical objects like tracks or
showers are extended, the \textit{hit clustering} is the first step
towards a more global reconstruction of the event. The second
advantage of clustering is that single hits 
can easily be suppressed by applying a cut on the
cluster size. Besides the improvement of the purity, also the hit
finding efficiency can be increased, since it is possible to search for
more hits around the borders of clusters with a lowered threshold.
The clustering algorithm is based on the search of directly adjacent hits: 
starting from a single hit, the \textit{nearest neighbour} algorithm
(\textit{NN}) iteratively expands the cluster by adding close
hits. Looping over all hits in the cluster, it searches for unclustered
hits within a pre-defined time and readout channel range around the 
current hit. Once a new hit is found, it is added to the cluster. The
algorithm continues looping, until no more hits are found in the vicinity of the
cluster. Like in the case of the hit-finding algorithm, the
parameters of the algorithm depend on the detector as well as on the event type. Since
the NN clustering is optimised to reconstruct connected objects, like
tracks, it was perfectly suitable for the reconstruction of the 3~L,
120~L and 200~L TPC data. A more detailed description of the
implementation of the NN clustering algorithm is presented
in~\cite{thesis_JavierRico}.


\subsubsection{Track reconstruction}
\label{sub:reco-tracking}

Starting from general clusters that can assume any topology, the tracking
algorithm, which was used for the reconstruction of the 200~L LAr
LEM-TPC data~\cite{Badertscher:2012dq,Badertscher:2013wm}, aims at the identification of straight tracks. The
straight track assumption is justified, since the multiple scattering is
negligible for the maximally 60~cm long cosmic ray tracks. The method
being described here can be easily generalized to the case of bent tracks. 

In the following, we first describe a Hough-transform
based tracking algorithm for the identification of the main,
through-going cosmic rays and then a second algorithm to
reconstruct knock-on electrons, appearing as secondary tracks.
The basic idea of the track identification, being described here, 
is to convert the problem of finding aligned hits into the trivial
problem of finding a maximum. The \textit{Hough Transform
  (HT)}~\cite{Hough:1962} transforms hits with coordinates $(x,z)$,
where $x$ is given by the readout strip number and $z$ by the drift
time, into the parameter space of straight 
lines, also called \textit{Hough space}. The basic procedure of the 
HT is to evaluate the parameters of all the straight lines, going
through each hit in the $(x,z)$ plane. If a subset of points are
aligned, there is a single line going through all of them and its
parameter pair (angle $\theta$ and offset $r(\theta)$) will appear as
a maximum in the two dimensional, binned Hough space. Looping over
each cluster, the HT based track identification algorithm goes through
the following steps:  
\begin{enumerate}
\item Each hit coordinate pair $(x,z)$ is transformed via HT to the
  parameter space. As shown in \Cref{fig:reco-htparam}, straight
  lines are parametrized by an angle $\theta$ and the minimal distance
  to the origin of the coordinate system $r$.
  \begin{figure*}[t]
    \centering
    \includegraphics[width=0.35\textwidth]{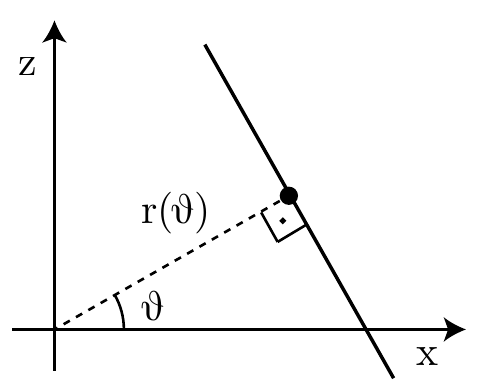}
    \caption{Straight line parametrization, used for the Hough transformation.}
    \label{fig:reco-htparam}
  \end{figure*}
  The transformation to the Hough space is then given by
  \begin{equation}
     r(\theta)=x\cdot\cos{\theta}+z\cdot\sin{\theta}.
    \label{eq-ht}
  \end{equation} 
  After defining the binning of the Hough space with the variables
  $(r,\theta)$, the algorithm computes $r(\theta)$ for each
  angle $\theta$ and adds the point with weight~1 to the two dimensional
  histogram. This step is then repeated for each hit of the cluster. 

\item The bin with the largest number of entries
  $(\theta_{max},r_{max})$ gives then directly the parameters of the
  straight line that crosses most of the hits. In order to avoid fake
  tracks, a minimum threshold of typically $N_{min}=4$~hits is required. The two
  dimensional track is then defined by all the hits that are close to
  the found straight line. 

\item In case it is needed to find other straight tracks, step
  number~2. can be repeated for the residual hits that belong
  to the cluster but not to a track, until the number of hits in a
  track candidate goes below the user-defined threshold
  $N_{min}$. Finally, the main track is again fitted with a straight line.

\begin{figure*}[t]
\centering
\includegraphics[width=1\textwidth]{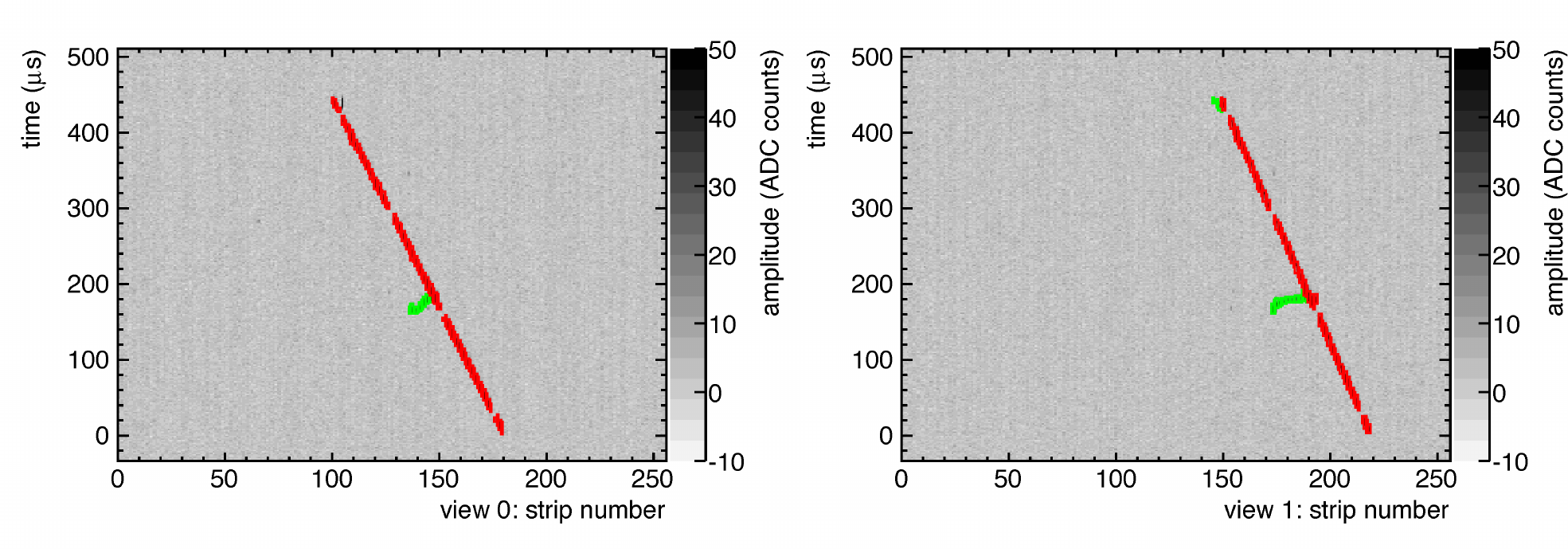}
\caption{Two views of a fully reconstructed MC generated muon with a knock-on
  electron in the TGeo geometry of the 200~L double phase 
  LAr LEM-TPC~\cite{Badertscher:2012dq,Badertscher:2013wm}. Hits belonging to the
  main track are highlighted in red; hits belonging to secondary tracks
that are attached to the main tracks are green.}
\label{fig:mc2D-reco}
\end{figure*}
\item After finding the main track, all residual hits that belong to the
  cluster but not to the track, are tagged as $\delta$-ray hit
  candidates. Then, the algorithm groups and tags nearby $\delta$-ray
  hit candidates as $\delta$-rays. Since single hits can easily be
  produced by noise, $\delta$-rays require at least two consecutive
  hits. \Cref{fig:mc2D-reco} shows the MC generated $\mu^-$ with a
  $\delta$-ray. The event display for both views~0 (left) and~1
  (right) show in red the hits that belong to the main track and in
  green the hits that belong to $\delta$-rays.  
\end{enumerate}

\subsubsection{Three dimensional track reconstruction}
\label{sub:reco-3dtracking}

The most important information, obtained from the hits, is the charge
$\Delta Q$ that is related to the signal integral and the three
dimensional track length $\Delta x$. The information of the two
dimensional tracks from complementary views have to be combined in
order to reconstruct $\Delta x$  as well as the complete three
dimensional image of the event. Due to intrinsic
ambiguities coming from the projection technique, this task is
in general complex. However, the fact that only straight tracks
are considered, simplifies the three dimensional reconstruction a
lot, since the problem is reduced to the simple matching of
two-dimensional tracks: comparing the drift times of the
first hits of tracks, an algorithm loops over all two dimensional
track candidates until a unique pair of tracks from both views is
found. In case of the event shown in \Cref{fig:mc2D-reco}, the
algorithm links the two red tracks, since the first hits from view~0
and~1 both have a similar drift time $t_{drift}\approx 0$ (the $\mu$
is entering from the top of the chamber through the anode).
To reconstruct the three dimensional hit coordinates
$(x_0,y_0,z_0)$ and $(x_1,y_1,z_1)$ for the views 0~and~1,
the straight line approximation is used: first, the tracks of both
views (view~0 and view~1) are fitted with the linear equation 
\begin{equation}
z(x)=a_0\cdot x+b_0 \,\text{ and }\, z(y)=a_1\cdot y+b_1 
\label{eq:linear}
\end{equation}
with the slopes $a_{0,1}$, the offsets $b_{0,1}$ and the common drift
coordinate $z=t_{drift}\cdot v_{drift}$. Eq.~\ref{eq:linear} makes use of
the fact that the strips of the two readout views are perpendicular
and the coordinate system is chosen such that view~0 (1) 
directly provides the $x$ ($y$) coordinate. Besides the naturally
provided readout coordinate $z_0$ ($z_1$) and $x_0$ ($y_1$) for
view~0 (view~1), inverting Eq.~\ref{eq:linear}, the missing coordinates
$y_0$ ($x_1$) are given by
\begin{equation}
y_0=\frac{z_0-b_1}{a_1} \,\text{ and }\,
x_1=\frac{z_1-b_0}{a_0}.
\label{eq:linear2}
\end{equation}
Besides the absolute position of the hits of view~0 and view~1, the
three dimensional track length
\begin{equation}
\Delta r_{0,1}=\sqrt{\Delta x_{0,1}^2+\Delta
  y_{0,1}^2+\Delta z_{0,1}^2}
\label{eq:dr}
\end{equation}
has to be computed for both views independently. In the case of
view~0 (view~1), $\Delta x_0$ ($\Delta y_1$) is equal to the readout
pitch that is typically for both views equal to 3~mm. Further,
$\Delta y_0$ and $\Delta x_1$ can be computed according to
Eq.~\ref{eq:linear2} and $\Delta z_{0,1}$ is given by
Eq.~\ref{eq:linear}. As final result for the three dimensional track
pitch we get
\begin{equation}
  \Delta r_0=\Delta x_0 \sqrt{1+a_0^2/a_1^2+a_0^2} \,\text{ and }\, 
  \Delta r_1=\Delta y_1 \sqrt{1+a_1^2/a_0^2+a_1^2}.
\label{eq:drFinal}
\end{equation}
Finally, using the proper charge calibration of the readout and
$\Delta r_{0,1}$ from Eq.~\ref{eq:drFinal}, the expression $\Delta
Q_{0,1}/\Delta r_{0,1}$ can be computed for both views. For reasons of
simplicity, hereafter the same expression is often renamed to $\Delta
Q/\Delta x_{0,1}$ or $dQ/dx_{0,1}$. 

\begin{figure*}[t]
\centering
\includegraphics[width=0.8\textwidth]{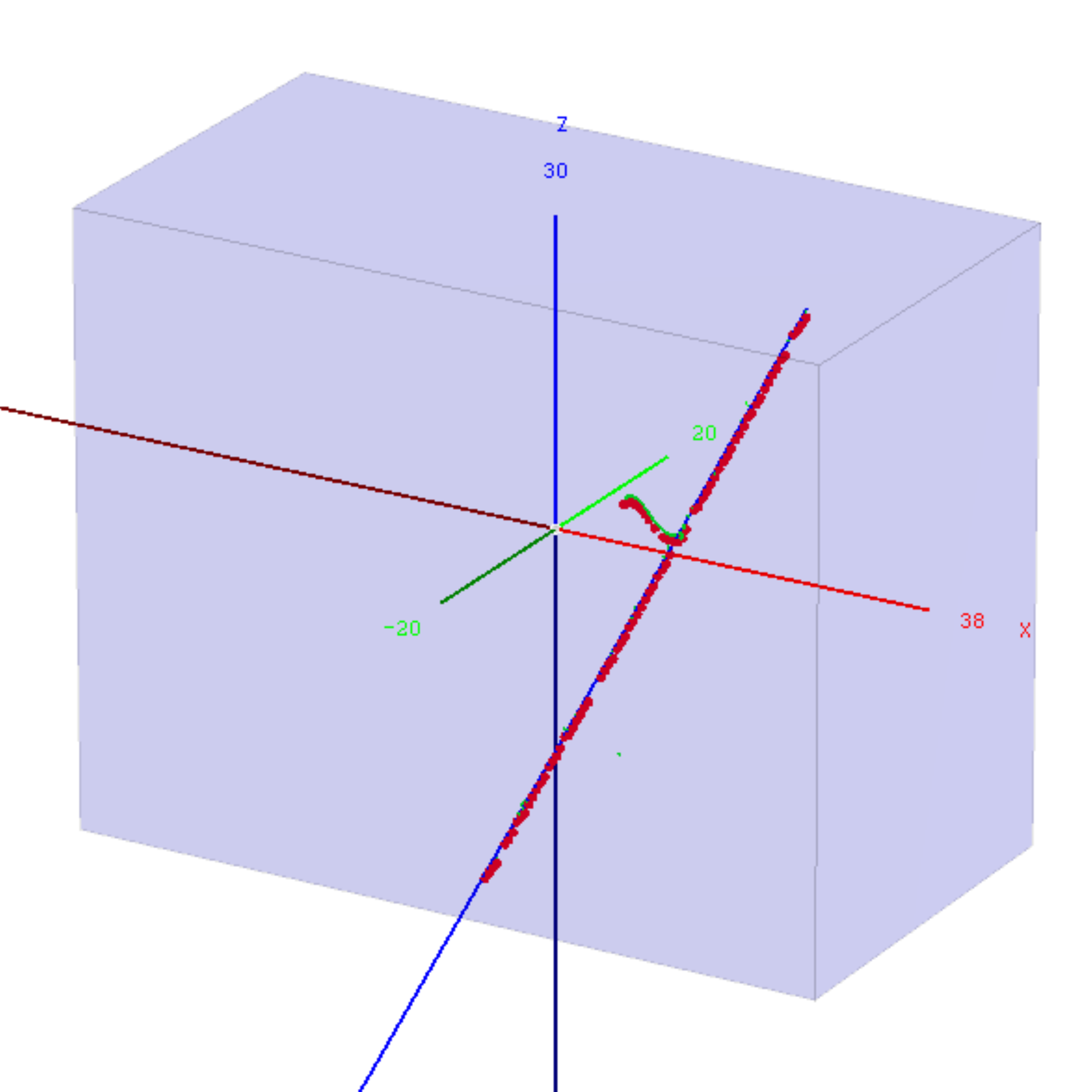}
\caption{Full reconstruction of view~0 hits (red dots) of the MC
  generated cosmic ray event from
  \Cref{fig:mc-truth-event,fig:mc2D-reco}. The MC truth $\mu^-$ track
  is shown in blue, the emitted $\delta$-ray in green.}
\label{fig:mc3D-reco}
\end{figure*}

The hits that have been tagged as $\delta$-ray hits (see
\Cref{sub:reco-tracking}), do not necessarily form straight tracks. Therefore,
the best way to reconstruct those hits is to find for each
$\delta$-ray hit of view~0 a coincident $\delta$-ray hit of view~1.
This means, that the resulting coordinate of each hit on view~0 are
defined as ${(x_0,y_1,z_0)}$, where $y_1$ is taken from a hit on
view~1 with a similar drift time,
i.e. $|t_{drift,1}-t_{drift,0}|<\epsilon$. In case no coincident
$\delta$-ray hit is found on the other view, it is matched with the
main track, since in such a case it can be assumed that $\delta$-ray
and $\mu$ are superimposed.
An example for the final, three dimensionally reconstructed MC
generated event from \Cref{fig:mc-truth-event,fig:mc2D-reco} is shown
in \Cref{fig:mc3D-reco}: the coordinates of all reconstructed hits of
view~0 are shown in the 3D display, superimposed with the MC
tracks. It can be seen that the through-going $\mu^-$, as well as the
$\delta$-ray in green match with the reconstructed hits in red.

\subsubsection{Particle flow - the PANDORA methodology}
\label{sec:pandora}
Alternative ways of reconstructing events starting from Qscan hits are
being developed, in an effort to create a 
fully-automatic reconstruction package capable of handling all
expected neutrino interaction topologies.

The PANDORA framework~\cite{Thomson200925} was originally written for particle flow
calorimetry in heterogeneous collider detectors, specifically the ILC
and CLIC, by a group at the University of Cambridge led by Mark
Thomson. This method of reconstruction attempts to identify all
primary final-state particles in an event, and then associate all
energy deposits in the detector with a specific primary particle, thus
allowing to estimate the energies of all final-state particles through
calorimetric methods. 

The original PANDORA algorithms are not immediately suitable for
reconstruction of neutrino events in LAr TPCs. For example, in a
non-collider environment, there is no a priori knowledge of the vertex
location from the beam geometry, and so the primary vertex must be
reconstructed from event data. In addition, the original algorithms
use the fact that different particle species deposit their energy in
different parts of a heterogeneous detector (e.g. tracker, ECAL, HCAL,
MRD) -- this is not the case in a homogeneous LAr detector.  

For this reason, the PANDORA developers have produced a suite of
algorithms designed specifically for LAr
reconstruction~\cite{Thomson200925_2}. Essentially, the algorithms do the following: 
\begin{itemize}
\item Split data into separate 2D projections.
\item Perform first-pass hit clustering, by
 \begin{itemize}
 \item clustering individual hits using a nearest-neighbour method; then
 \item associating subclusters based on proximity and direction.
 \end{itemize}
\item Split clusters using a kink search algorithm.
\item Find the primary vertex based on presumed beam direction and cluster positions/directions.
\item Identify primary final-state particles (``seed clusters''), based on cluster length and proximity to vertex.
\item Association of each remaining, â``non-seed'' cluster, with ``seed''
clusters, so that all energy deposits from a single final-state
particle are grouped together. 
\end{itemize}

In practice, this process is rather involved and each step is split
into several discrete algorithms, each performing a tightly specified
operation on the existing list of clusters; the philosophy followed is
that each step should be as conservative as possible, since it is much
easier to combine objects later than to separate objects which have
already been combined. The present version of the PANDORA
reconstruction reconstructs hits in 2D projections rather than in 3D;
an algorithm for matching the views to produce a 3D event does not yet
exist, but is in development. 

\begin{figure}[htb]
  \centering
  \includegraphics[width=\textwidth]{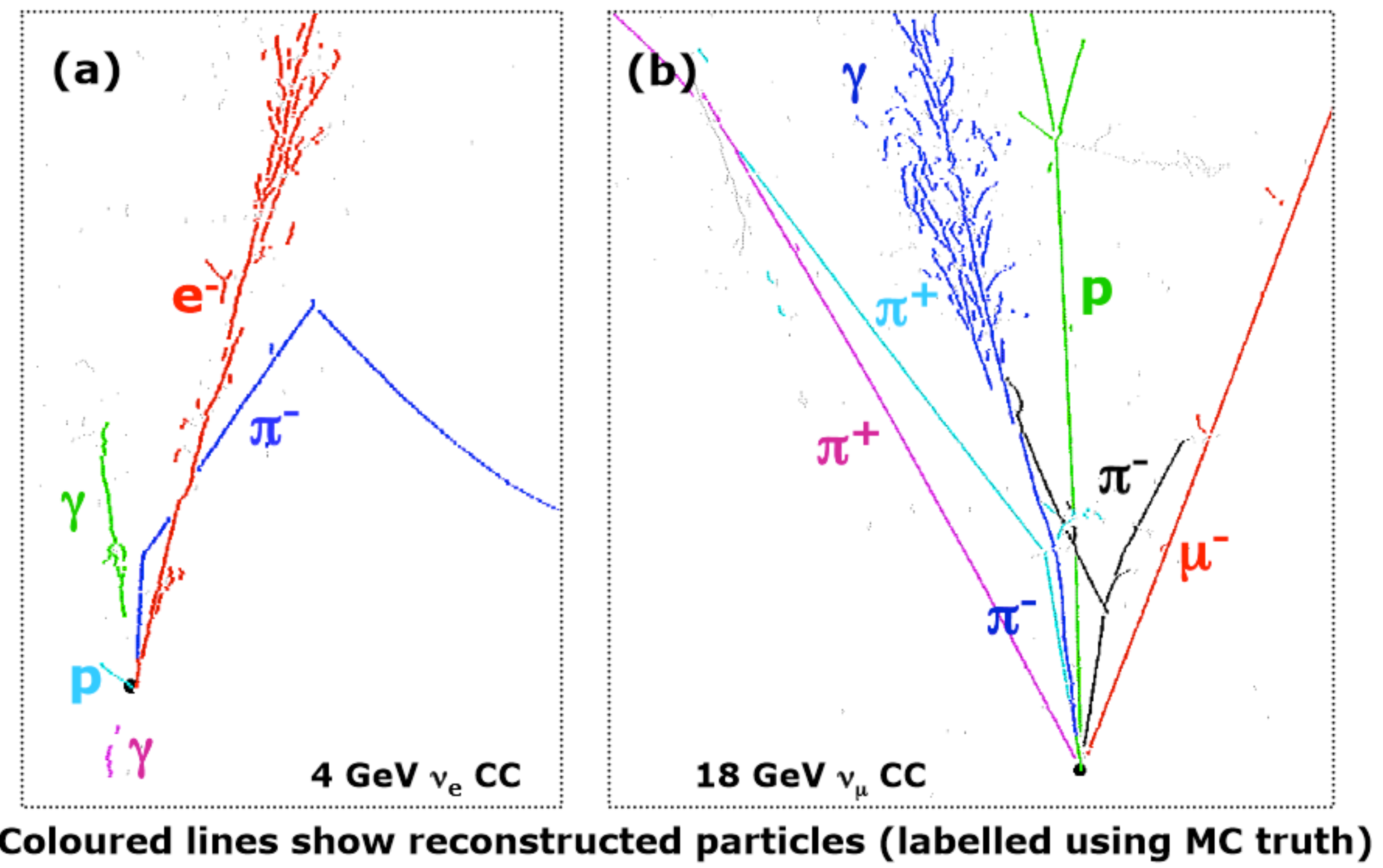}  
  \caption{Examples of successfully reconstructed Monte Carlo events
    in LAr, produced by the PANDORA developers (taken
    from~\cite{Thomson200925_2}). Black circles indicate the reconstructed
    primary vertex.} 
    \label{fig:pandoraexample}
    \end{figure}

Examples of successfully reconstructed Monte Carlo events in liquid
argon can be seen in Figure~\ref{fig:pandoraexample}.

\clearpage
\section{MIND500: North Area MIND}
\graphicspath{{./Section-MIND500/}}

\subsection{MIND500 overview}
MIND500, a ~500 ton Magnetized Iron Neutrino Detector, will operate downstream of the DLar in the North Area. Some characteristics of MIND detectors are poorly known, either because they have not been tested or because the conditions under which tests were carried out were not fully representative of the operational environment (e.g. no B-field). Improvements in technologies and the evolution of software tools such as reconstruction algorithms call for a comparison to experimental data for validation. There is a strong incentive to study in particular:
\begin{itemize}
\item Muon charge identification, for wrong sign muon signature of a neutrino oscillation event: golden channel at a neutrino factory: requires correct sign background rejection of 1 in $10^4$, 0.8 $\rightarrow$ 5 GeV/c.
\item Hadronic shower reconstruction for identification of charged current neutrino interactions and rejection of neutral current neutrino interactions. Test beam: protons/pions 0.5 $\rightarrow$ 9 GeV/c.
\end{itemize}

Dimensions for the MIND500 are defined by the requirement for having an adequate acceptance for events originating in the DLar.

\begin{table}[h]
\centering
\caption{\em MIND500 parameters.}
\begin{tabular}{lccccccc}
\toprule
\textbf{Parameter} &	\textbf{Symbol}  & 	\textbf{Unit}  &	\textbf{Nominal value} & \textbf{Range Min.} &	\textbf{Range Max.}\\
\hline
\multicolumn{5}{l}{Detector global dimensions}\\
\hline
Detector width	&	$w_{det}$	&	m	&	5.0	&	2.5	&	6.0 \\	
Detector height	&	$h_{det}$	&	m	&	5.0	&	2.5&	6.0 \\	
Detector depth	&	$d_{det}$	&	m	&	3.2	&	2.0	&	5.0 \\	
\hline
\multicolumn{5}{l}{Iron plates}\\
\hline
Material grade &	- &	-&	AISI1010 &	-	&	ARMCO \\
Number of plates &	$n_{iron}$ &	-	&	70 &	-	&	- \\
Iron width &	$w_{iron}$ &	m	&	5.0 &	2.5	&	6.0 \\
Iron height &	$h_{iron}$ &	m	&	5.0 &	2.5	&	6.0 \\
Iron thickness &	$t_{iron}$ &	cm	&	3.0 &	1.0	&	5.0 \\
Total iron mass &	$m_{iron}$	&	tons	&	412 &	-	&	-	\\
Total iron area &	$a_{iron}$ &	m2	&	1750 &	-	&	- \\
Number of slots for coil &	$n_{slots}$ &	-	&	2 &	2	&	4 \\
Slot for coil: width &	$w_{slot}$ &	cm	&	10.0 &	-	&	-	\\
Slot for coil: height &	$h_{slot}$ &	cm	&	20.0 &	-	&	- \\
Support structure &	- &	-	&	TBD &	-	&	- \\
\hline
\multicolumn{5}{l}{Gap between iron plates}\\
\hline
Number of gaps	&	$n_{gaps}$	&	-	&	70	&	-	&	- \\	
Gap thickness	&	$t_{gap}$	&	cm	&	1.5	&	1.5	&	2.0 \\	
Material	&	-	&	-	&	air + plastic	&	-	&	- \\	
\hline
\multicolumn{5}{l}{Plastic scintillator}\\
\hline
Material &	- &	-&	Polysterene &	-	&	-	& \\
Number of planes per module (xy or uv)	&	$n_{module}$ &	-	&	2.0 &	-	&	- \\
Gap between planes within module &	- &	cm	&	0 &	0	&	0.05 \\
Module envelope thickness &	$t_{env}$	&	cm	&	0.05 &	0	&	0.05	\\
Scintillator bar length &	$l_{sci}$ &	cm	&	500.0 &	100.0	&	500.0 \\
Scintillator bar width &	$w_{sci}$ &	cm	&	3.0 &	1.0	&	5.0 \\
Scintillator bar height &	$h_{sci}$ &	cm	&	0.7 &	0.6	&	1.0	\\
Bars per plane &	$n_{bars,pla}$ &	-	&	167 &	-	&	- \\
Bars per module &	$n_{bars,mod}$ &	-	&	334 &	-	&	- \\
Total number of bars &	$n_{bars,tot}$ &	-	&	23380 &	-	&	-	\\			
\hline
\multicolumn{5}{l}{Light readout and conversion}\\
\hline	
Light readout optical fibres &	- &	-	&	WLS &	C.F.	& -	\\	
Total length of fibre &	fibre &	m	&	120000 & 	-	&	- \\
Readout device &- &	-	&	SiPM &	- &	-	\\
Readout channels per bar &- &	-	&	2 &	1&	2	\\
\bottomrule
\end{tabular}
\label{mind500}
\end{table}

\begin{figure}[hbt]
\centering
\includegraphics*[width=0.30\textwidth]{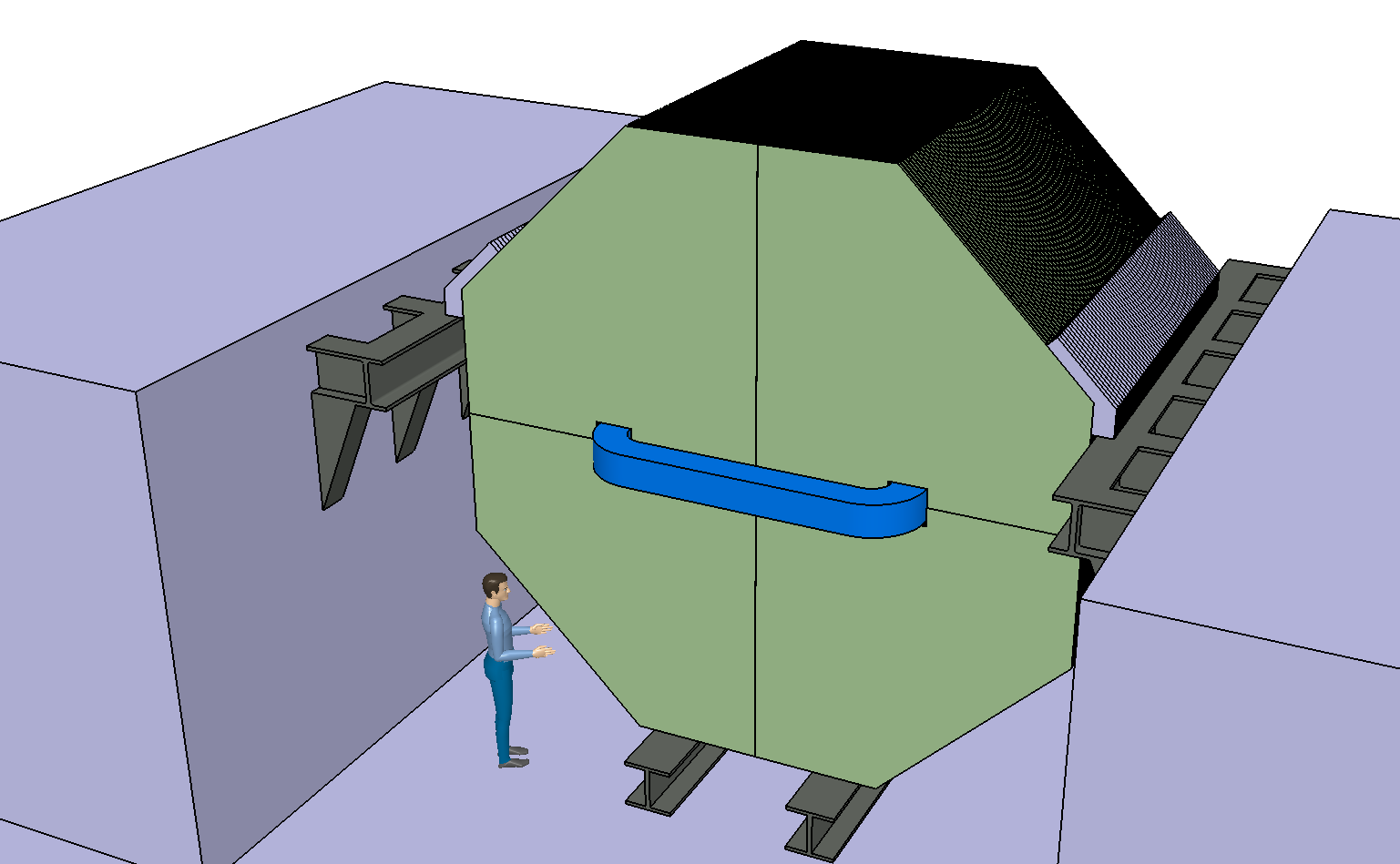}
\includegraphics*[width=0.30\textwidth]{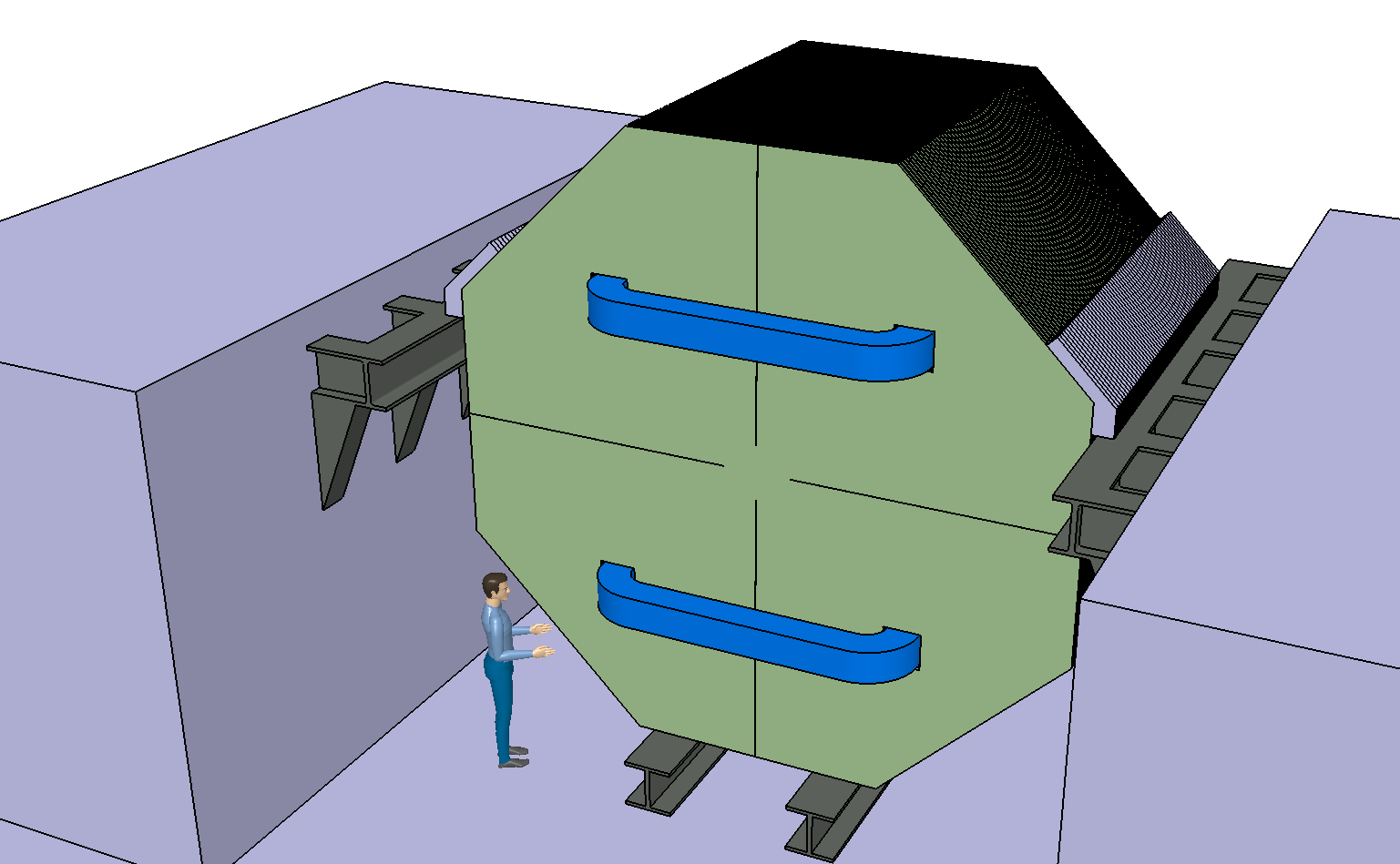}
\includegraphics*[width=0.30\textwidth]{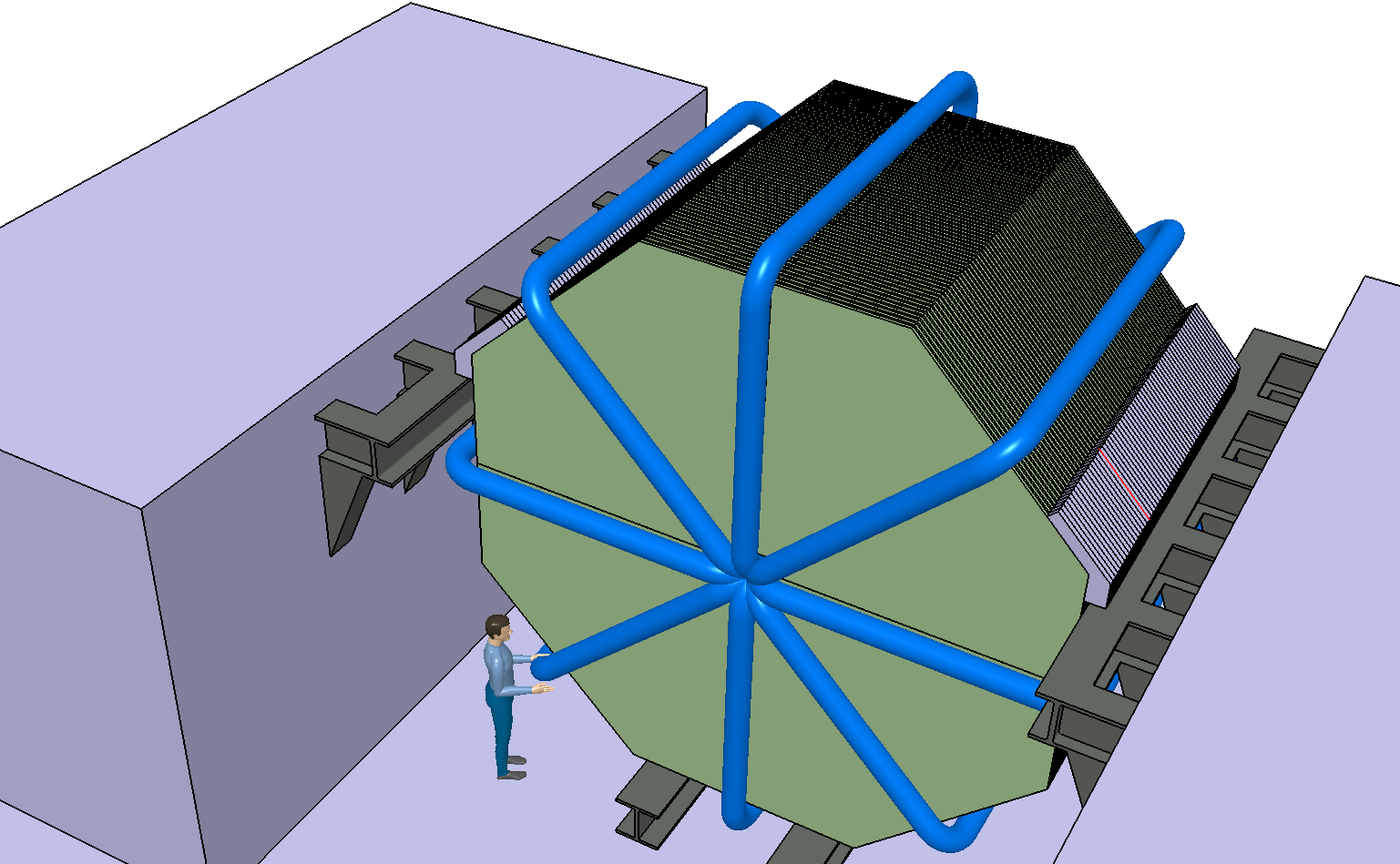}
\caption{Three different magnetization schemes for the iron are being considered. The scheme on the left depicts a 1-coil configuration which would provide a dipole magnetic field. The scheme in the center is a 2-coil configuration that leads to a reduction of the $B_y$ component of the field in the central zone of the detector. The scheme on the right leads to a toroidal magnetic field. The cradle support structure is not shown here.}
\label{MIND500-magnet}
\end{figure}

The three magnetization configurations under study are shown in Figure \ref{MIND500-magnet}:
\begin{itemize}
\item{1-coil configuration;}
\item{2-coil "Helmholtz" configuration generating a dipole field, with a much smaller $B_y$ component compared to the 1-coil configuration;}
\item{8-coil configuration generating a toroidal field.}
\end{itemize}

The toroidal field option is the one which is currently favoured. A MIND with a toroidal field configuration has been studied for the Neutrino Factory, showing equivalent $\delta_{CP}$ reach to a MIND with a dipole field \cite{Bross:2013oua}. The 100 kTon Neutrino Factory MIND has an octagonal cross-section of 14 m $\times$ 14 m, and consists of 3 cm plates of iron interleaved with a 2 cm-thick XY lattice of plastic scintillator bars. Another MIND with a toroidal field, the Super B Iron Neutrino Detector (SuperBIND), has been proposed for nuSTORM. SuperBIND has a circular cross-section of diameter 6 m, a length of 16 m, and consists of 1.5 cm plates of iron interleaved with a 2.0 cm-thick XY lattice of plastic scintillator bars.

\begin{figure}[hbt]
\centering
\includegraphics*[width=0.70\textwidth]{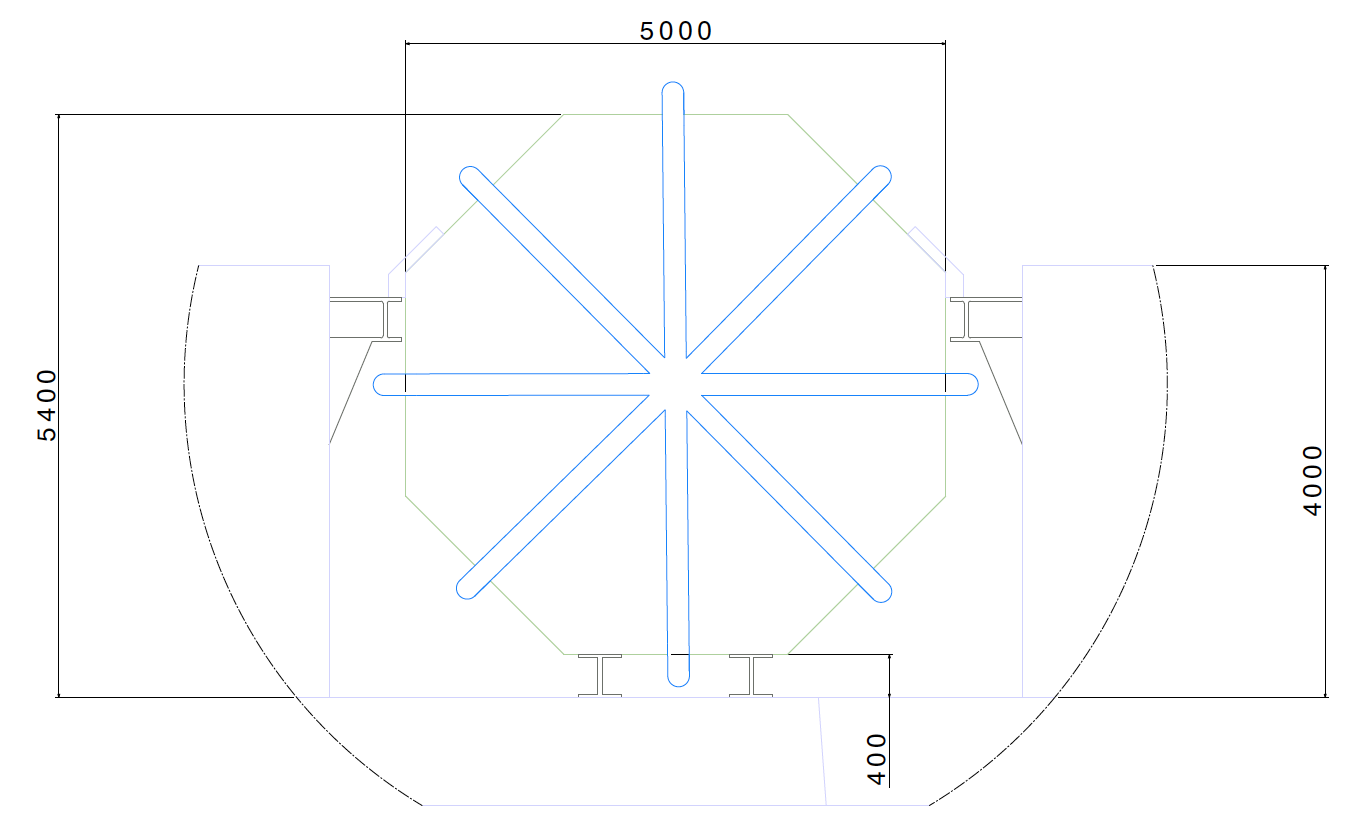}
\caption{Dimensions of the North Area MIND.}
\label{MIND500-dimensions}
\end{figure}

\begin{figure}[hbt]
\centering
\includegraphics*[width=0.70\textwidth]{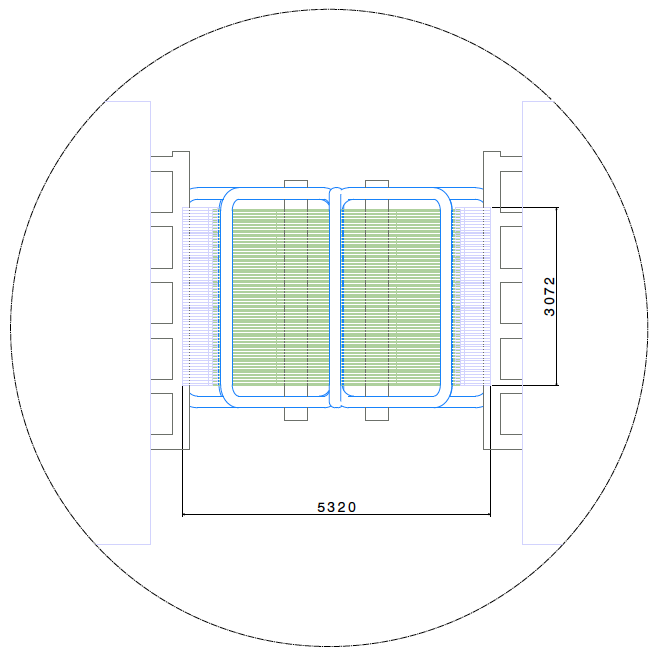}
\caption{Top view of the coil configuration generating a toroidal magnetic field in the North Area MIND.}
\label{MIND500-top}
\end{figure}

\subsection{MIND reconstruction efficiencies}
Preliminary simulations in Geant4 of a MIND prototype are reported here. Further optimisation is ongoing, particularly concerning the geometry, definition of the magnetic field and reconstruction algorithms for pions. Assumptions taken for the geometry are octagonal plates 2 m $\times$ 1 m, two transmission lines, with a 2.8 m detector depth (along the beam axis). 
Four different scenarios were tested to validate the steel thickness and for an indication of whether the choice of scintillator geometry is acceptable (0.7 cm high rectangular bars vs 1.7 cm high triangular bars), scintillator pitch = 1.0 cm in all scenarios:
\begin{itemize}
\item 3 cm steel plate, 1.5 cm scintillator module (i.e. 0.75 cm X plane + 0.75 cm Y plane),
\item 2 cm steel plate, 1.5 cm scintillator module,
\item 3 cm steel plate, 3.5 cm scintillator module,
\item 2 cm steel plate, 3.5 cm scintillator module.
\end{itemize}

In assessing the various efficiencies from simulations for a small MIND prototype exposed to a charged particle beam, the following were taken into consideration:
\begin{itemize}
\item{a) The total number of tracks (or simulated particles in the detector);}
\item{b) The number of tracks reconstructed (using the Kalman filter);}
\item{c) The number of successful tracks (where successful means that the correct charge is identified).}
\end{itemize}

The efficiencies are then defined as:
\begin{itemize}
\item{1) The reconstruction efficiency is b/a.}
\item{2) The charge identification efficiency is c/a.}
\end{itemize}

\subsubsection{Muon reconstruction efficiencies}
\begin{figure}[hbt]
\centering
\includegraphics*[width=0.24\textwidth]{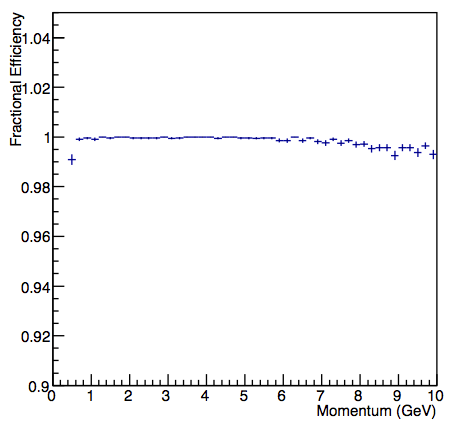}
\hfill
\includegraphics*[width=0.24\textwidth]{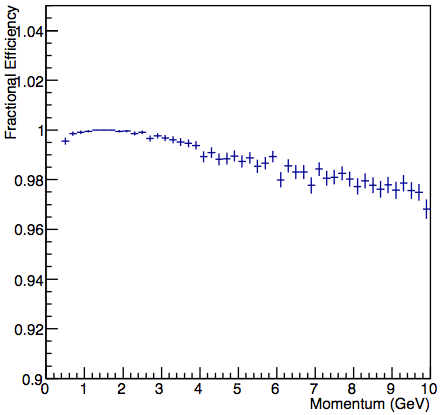}
\hfill
\includegraphics*[width=0.24\textwidth]{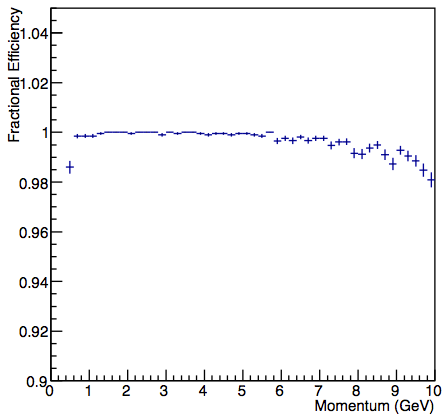}
\hfill
\includegraphics*[width=0.24\textwidth]{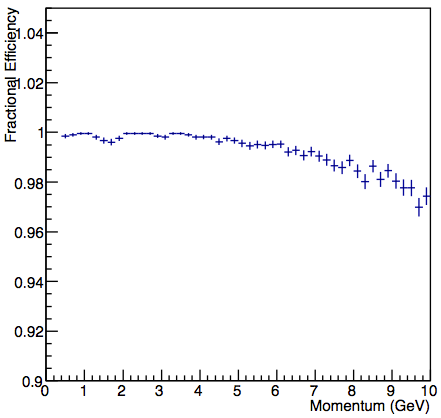}
\caption{Reconstruction efficiencies for $\mu$$^+$ for different steel and scintillator combinations, clockwise from top left: a) 3 cm steel, 1.5 cm scintillator. b) 2 cm steel, 1.5 cm scintillator. c) 3 cm steel, 3.5 cm scintillator. d) 2 cm steel, 3.5 cm scintillator.}
\label{mindmuonefficiency}
\end{figure}

Muon reconstruction efficiencies are shown in Figure \ref{mindmuonefficiency}. All four combinations of steel and scintillator thicknesses show good efficiencies at low momenta $<$2 GeV/c. Above 1 GeV/c, the combination showing the best performance over the widest momentum range is 3.0 cm of steel and 1.5 cm of scintillator, with efficiencies close to 100\% up to 6 GeV/c, staying above 99\% up to 10 GeV/c. These efficiencies remain good for the other scenarios, dropping to 97\% at high momenta.

A small fraction of events are not successfully reconstructed for the small MIND50 prototype. Track reconstruction using the Kalman filter from RecPack is based on a number of criteria, the most important being the number of hits along the track and the apparent curvature of the track (i.e. is the curvature well enough defined as to determine the momentum). The Kalman filter algorithm allows for the propagation of track parameters back through successive detector planes using a helix model that considers multiple scattering and energy loss. In a neutrino detector where a $\nu_\mu$ charged current interaction leads to hadronic activity at the interaction vertex in addition to an outgoing muon, the filter works well when the muon travels much further than particles related to the hadronic interactions. By ranking hits as a function of distance from the interaction vertex, it is possible to distinguish hits furthest away from this vertex as due only to a muon. Those hits then act as a seed for the Kalman filter. A typical reconstruction analysis will consider a number of planes (e.g. five) furthest downstream where hits are due to muons only. At high momenta, the technique works well. It has however limitations at high $Q^2$ or low neutrino energy, when the muon range is comparable to the range of the hadronic activity and identification of the muon becomes more challenging. When the Kalman filter fails, a cellular automaton method can be applied. It forms possible trajectories from a ranking of hits using a neighbourhood function. 

In a charged particle beam scenario with a reduced detector depth along the charged particle beam axis, the Kalman filter can fail to reconstruct a track at low momentum if the number of hits is insufficient, or at high momentum if the curvature of the track is insufficient. Analyses for the much larger Neutrino Factory MIND or nuSTORM SuperBIND with true neutrino interactions yield efficiencies which are much closer to 100\%, with a few events ($<<1\%$) failing reconstruction due to significant scattering affecting track curvature. For these detectors, pattern recognition and event reconstruction are more complex, and merit some comparison with test beam data.

\begin{figure}[hbt]
\centering
\includegraphics*[width=0.24\textwidth]{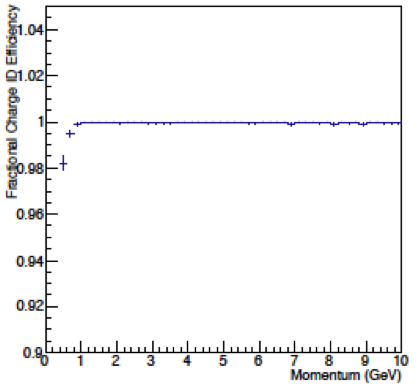}
\hfill
\includegraphics*[width=0.24\textwidth]{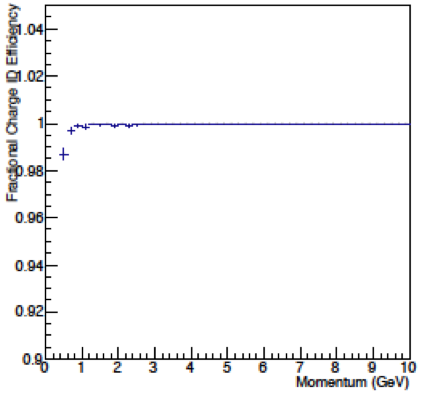}
\hfill
\includegraphics*[width=0.24\textwidth]{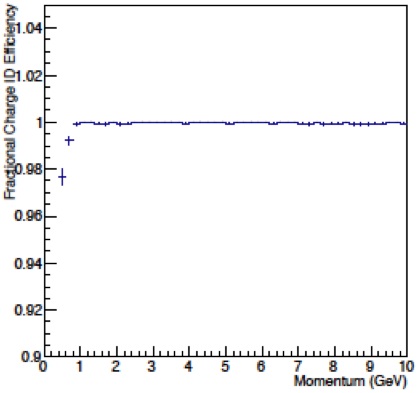}
\hfill
\includegraphics*[width=0.24\textwidth]{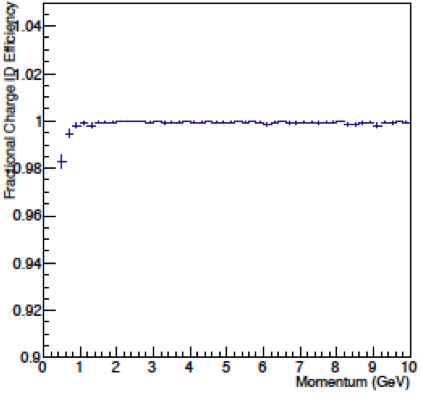}
\caption{Charge identification efficiencies for $\mu$$^+$ for different steel and scintillator combinations, clockwise from top left: a) 3 cm steel, 1.5 cm scintillator. b) 2 cm steel, 1.5 cm scintillator. c) 3 cm steel, 3.5 cm scintillator. d) 2 cm steel, 3.5 cm scintillator.}
\label{mindmuchargeID}
\end{figure}

Charge identification efficiencies are similar for all scenarios, close to 100 \% for all momenta above 1 GeV/c, Figure \ref{mindmuchargeID}. Detailed simulations were not carried out below 1 GeV/c. It is to be expected that the thickness of steel and scintillator start playing a role because of multiple scattering. A hint of the fall in charge identification efficiencies can be seen in Figure \ref{mindmuchargeID} with a couple of data points below 1 GeV/c where the thinnest steel and plastic scintillator combination has marginally better efficiency.

\subsubsection{Pion reconstruction efficiencies}

\begin{figure}[hbt]
\centering
\includegraphics*[width=0.24\textwidth]{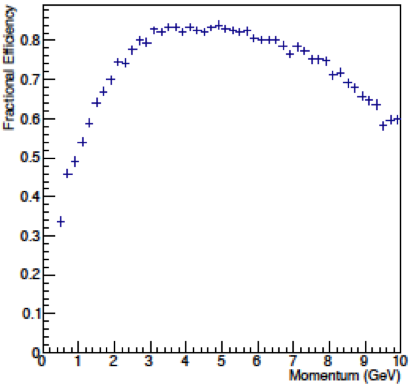}
\hfill
\includegraphics*[width=0.24\textwidth]{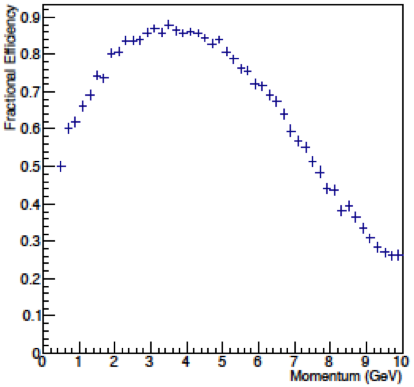}
\hfill
\includegraphics*[width=0.24\textwidth]{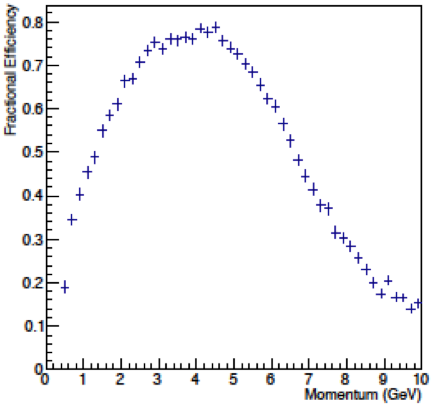}
\hfill
\includegraphics*[width=0.24\textwidth]{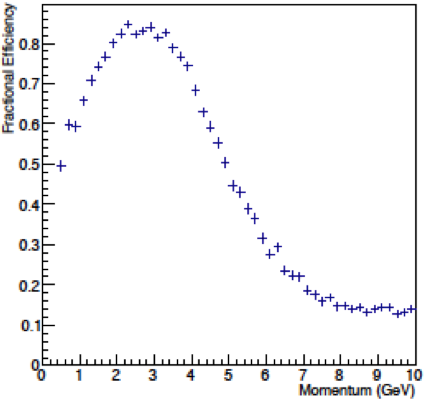}
\caption{Reconstruction efficiencies for $\pi$$^+$, clockwise from top left: a) 3 cm steel, 1.5 cm scintillator. b) 2 cm steel, 1.5 cm scintillator. c) 3 cm steel, 3.5 cm scintillator. d) 2 cm steel, 3.5 cm scintillator.}
\label{mindpionreconeff}
\end{figure}

\begin{figure}[hbt]
\centering
\includegraphics*[width=0.24\textwidth]{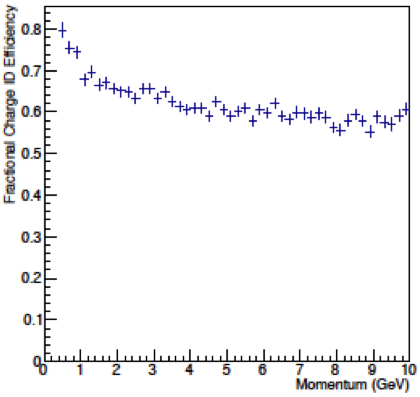}
\hfill
\includegraphics*[width=0.24\textwidth]{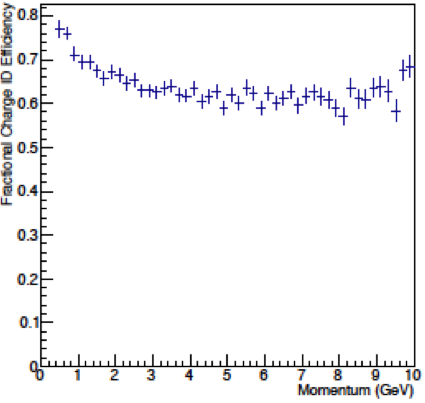}
\hfill
\includegraphics*[width=0.24\textwidth]{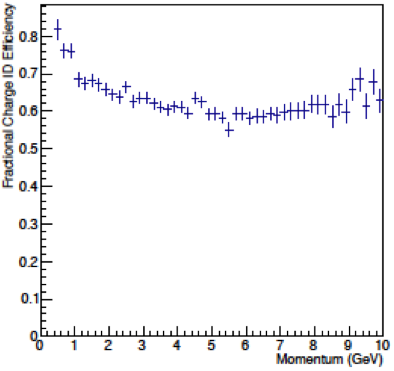}
\hfill
\includegraphics*[width=0.24\textwidth]{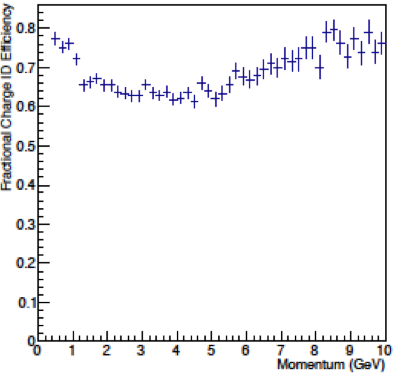}
\caption{Charge identification efficiencies for $\pi$$^+$, clockwise from top left: a) 3 cm steel, 1.5 cm scintillator. b) 2 cm steel, 1.5 cm scintillator. c) 3 cm steel, 3.5 cm scintillator. d) 2 cm steel, 3.5 cm scintillator.}
\label{mindpionchargeID}
\end{figure}

Pion track reconstruction is more challenging because of the hadronic shower development. The pion reconstructed momenta show that individual tracks are poorly reconstructed, Figure \ref{mindpionreconeff}. The input momentum distribution was uniform in $p_z$ between 0.3 GeV/c and 10 GeV/c. The reconstructed momentum distribution is peaked at low momenta. Comparison of $\pi$$^+$ and $\pi$$^-$ shows some discrimination of charge although further analysis is required.
Charge identification efficiencies are poor for pions, Figure \ref{mindpionchargeID}. Single track charge identification is unreliable and requires further work.

\subsection{Detector modules}
MIND500 detector module design will be based on developments carried out for the MICE-EMR and MIND50 detectors. The major difference compared to those detectors where scintillator slabs were typically $\sim$ 1 m long, comes from the factor $\times$ 5 greater length with associated significant attenuation losses for the light.

\begin{figure}[hbt]
\centering
\includegraphics*[width=0.32\textwidth, height=0.20\textwidth]{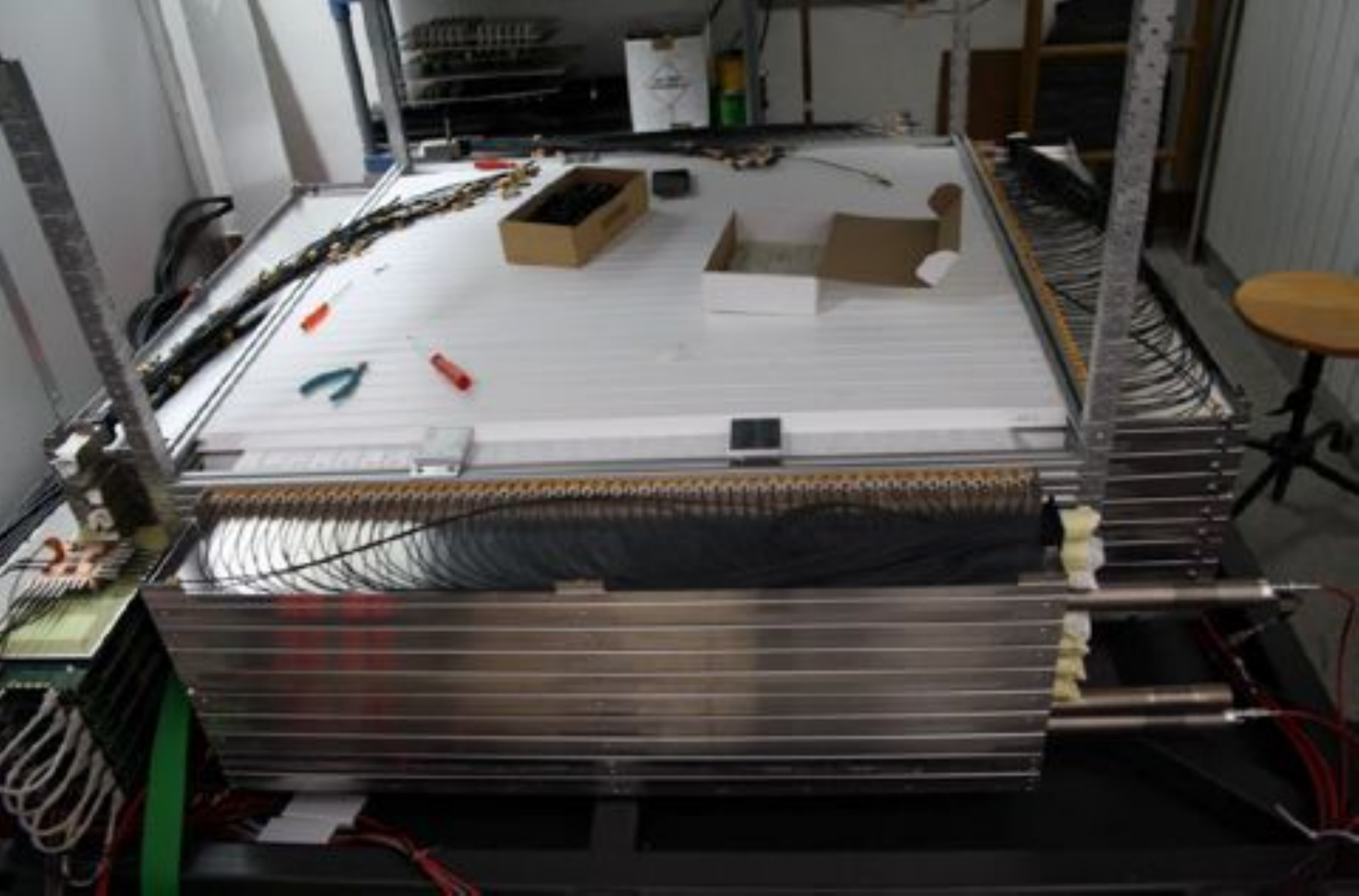}
\hfill
\includegraphics*[width=0.32\textwidth, height=0.20\textwidth]{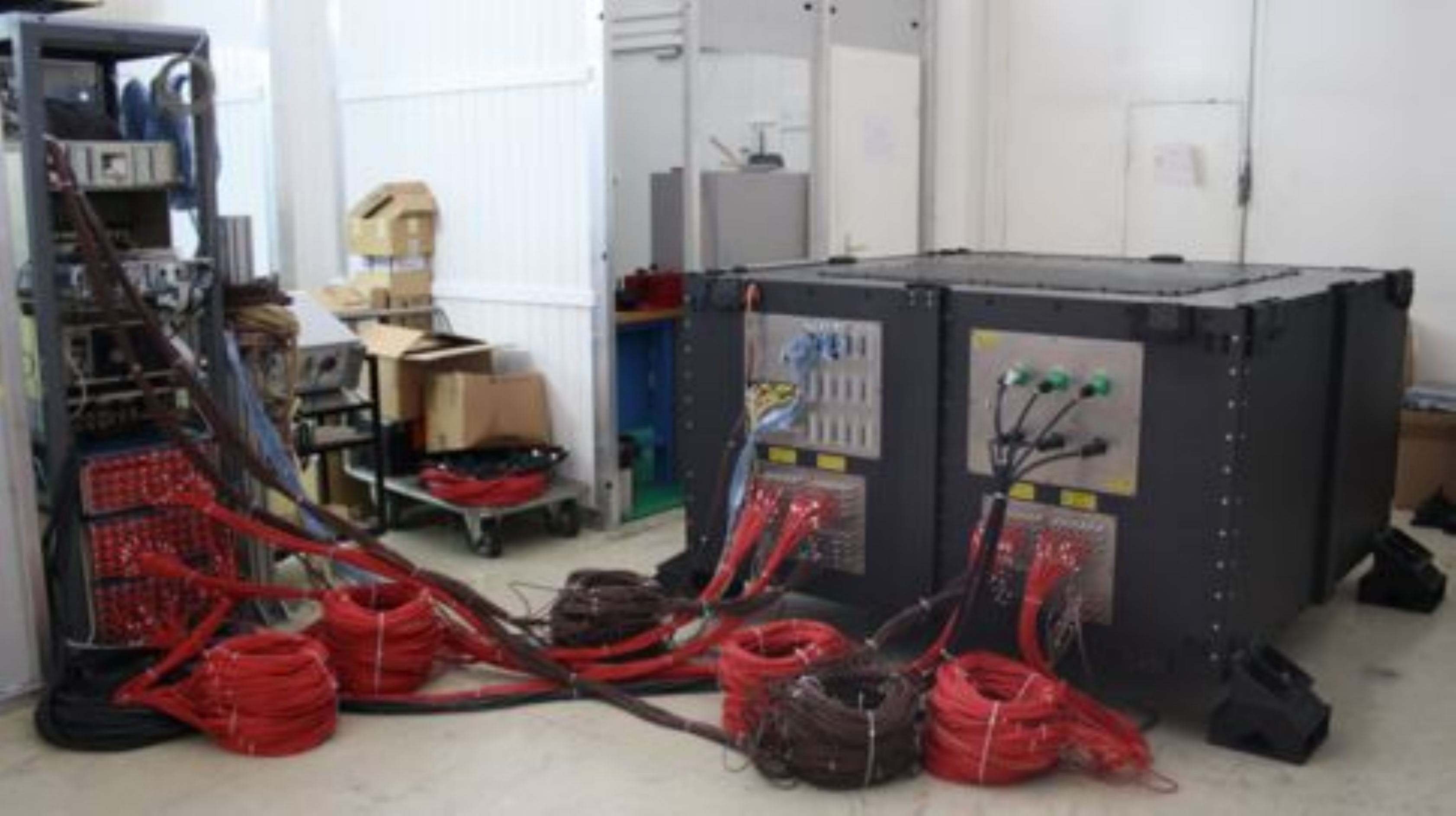}
\hfill
\includegraphics*[width=0.32\textwidth, height=0.20\textwidth]{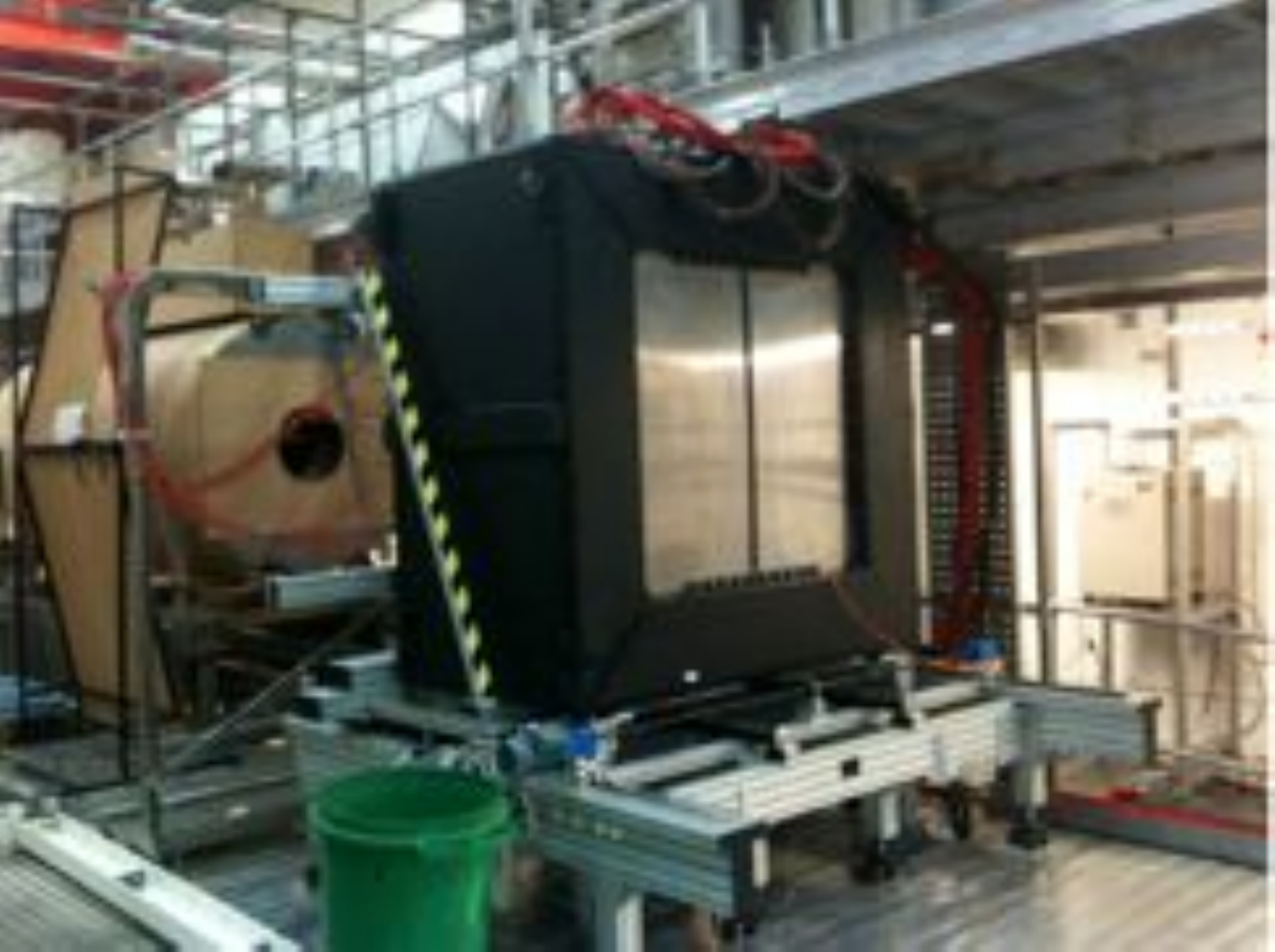}
\caption{Commissioning of the MICE Electron Muon Ranger (EMR) in 2013: left) Assembly of the MICE-EMR detector at the University of Geneva, photo taken in June 2013, middle) Completed MICE-EMR, photo taken mid-September 2013, right) The MICE-EMR installed at the end of the MICE beamline at the Rutherford Appleton Laboratory (RAL) in the UK, photo taken 27th September 2013.}
\label{MICE-EMR}
\end{figure}

\begin{figure}[hbt]
\centering
\includegraphics*[width=0.5\textwidth]{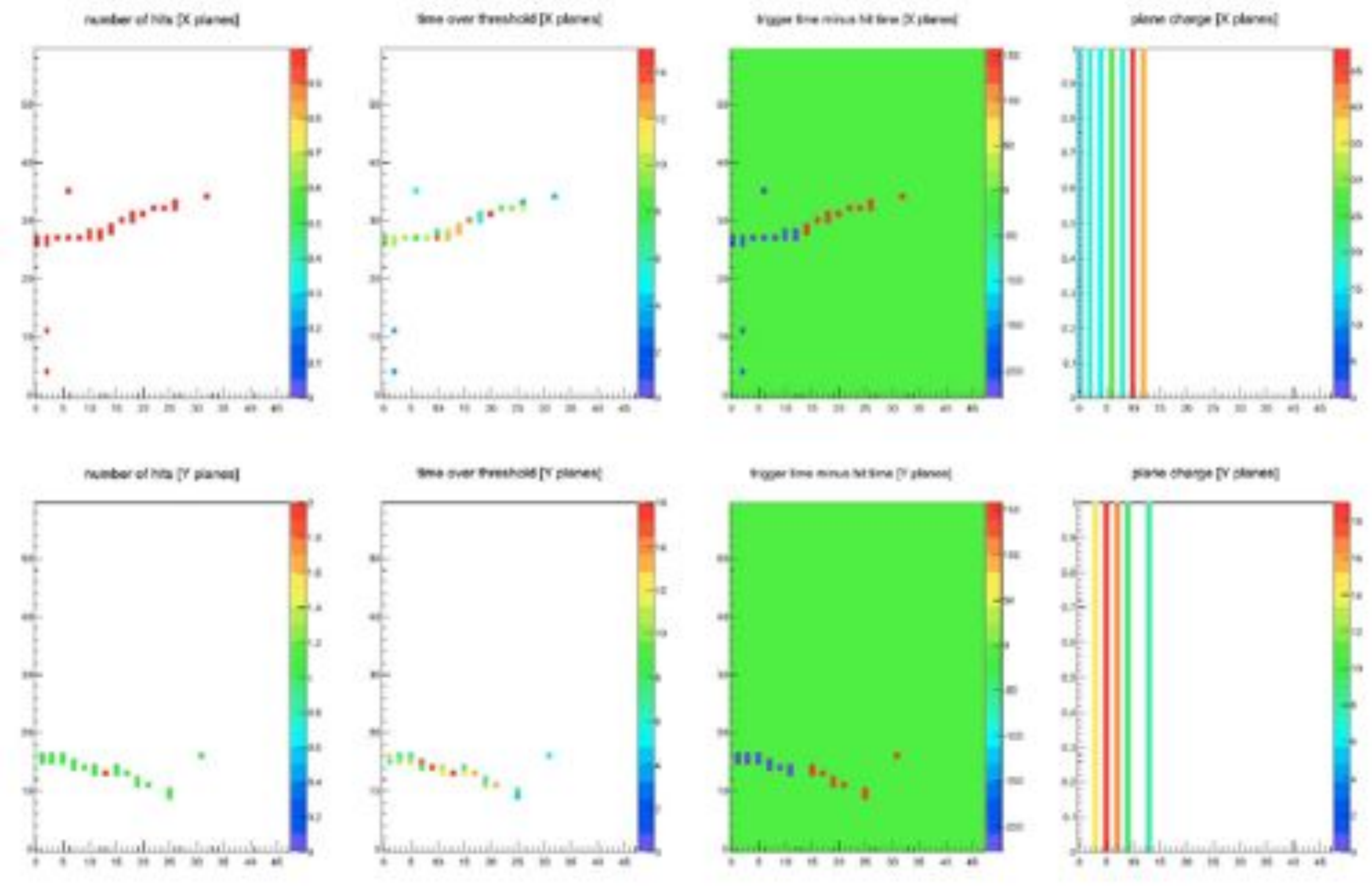}
\caption{MICE-EMR event display showing a 140 MeV/c negative muon entering the detector from the left, recorded during test beams on the MICE beamline in October 2013.}
\label{EMR-event}
\end{figure}

The detector modules consist of bars of plastic scintillators with wavelength shifting fibers and silicon photo-multipliers as light sensors. Considerable experience has been gained at the University of Geneva with all stages of the realization of such detectors. Recently, the MICE-EMR detector with similar modules was designed and constructed. It was commissioned in summer 2013, and shipped to the Rutherford Appleton Laboratory (RAL) in the UK where it was installed at the end of the Muon Ionization Cooling Experiment (MICE) beamline on 27th September 2013, Figure \ref{MICE-EMR}.
First online test beam runs were carried out in October 2013, showing excellent particle identification capabilities of the TASD for low momentum ($< 400$ MeV/c) muons, pions and electrons. Data analysis is ongoing, an event display is shown in Figure \ref{EMR-event}.


\subsubsection{Detector module mechanics}

\begin{figure}[hbt]
\centering
\includegraphics*[width=0.45\textwidth, height=0.25\textwidth]{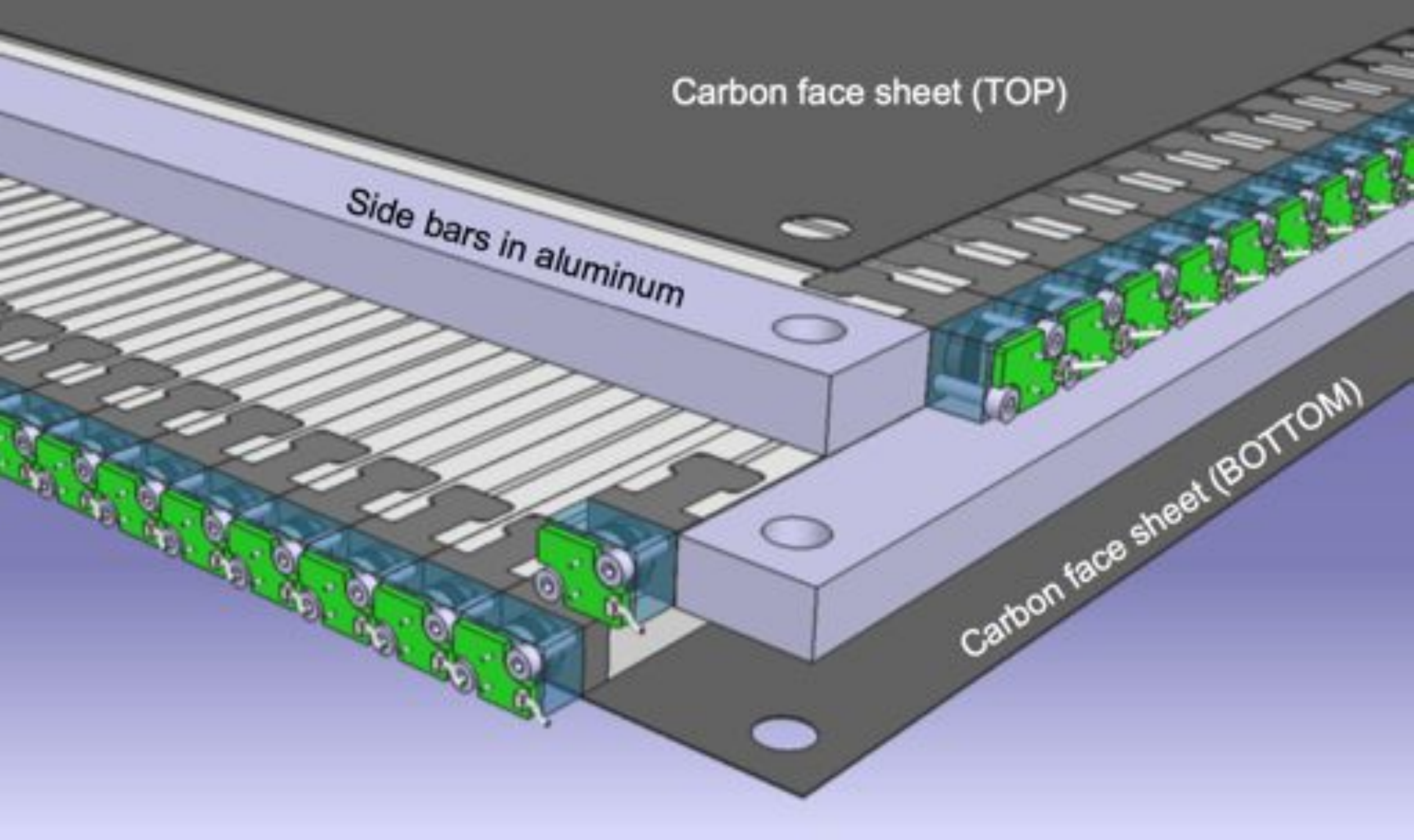}
\hfill
\includegraphics*[width=0.49\textwidth, height=0.25\textwidth]{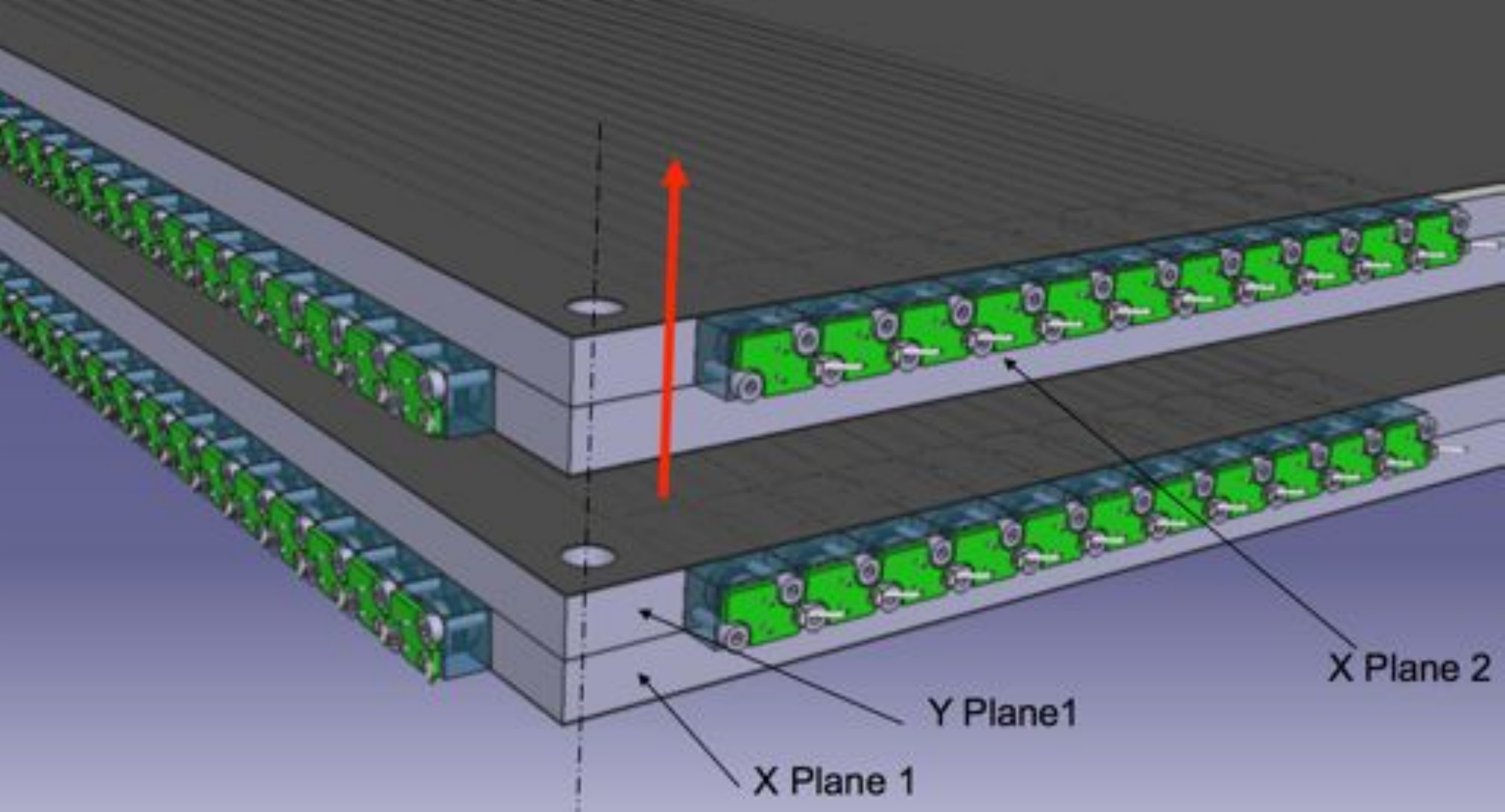}
\caption{MIND detector module: assembly.}
\label{mechanics}
\end{figure}

The detector module mechanics proposal is shown in Figure \ref{mechanics}. The concept is to have independent modules that can be inserted as cassettes between the steel plates in a MIND.

A module consists of one X plane with a number of scintillator bars glued at either end onto an aluminium support bar and positioned onto one Y plane with an equivalent number of scintillator bars also glued at either end onto a second pair of aluminium support bars. The X and Y planes are sandwiched between two carbon or kapton sheets, of thickness $200\, \mathrm{{\mu}m}$. These sheets provide mechanical support, reinforcing the planar alignment of the scintillator bars, and also provide some degree of light tightness. The carbon sheets will not affect the physics capabilities of the MIND500 detector, since their thickness is negligible with respect to that of the steel plates.

\subsubsection{Scintillators}
\begin{figure}[hbt]
\centering
\includegraphics*[width=0.4\textwidth]{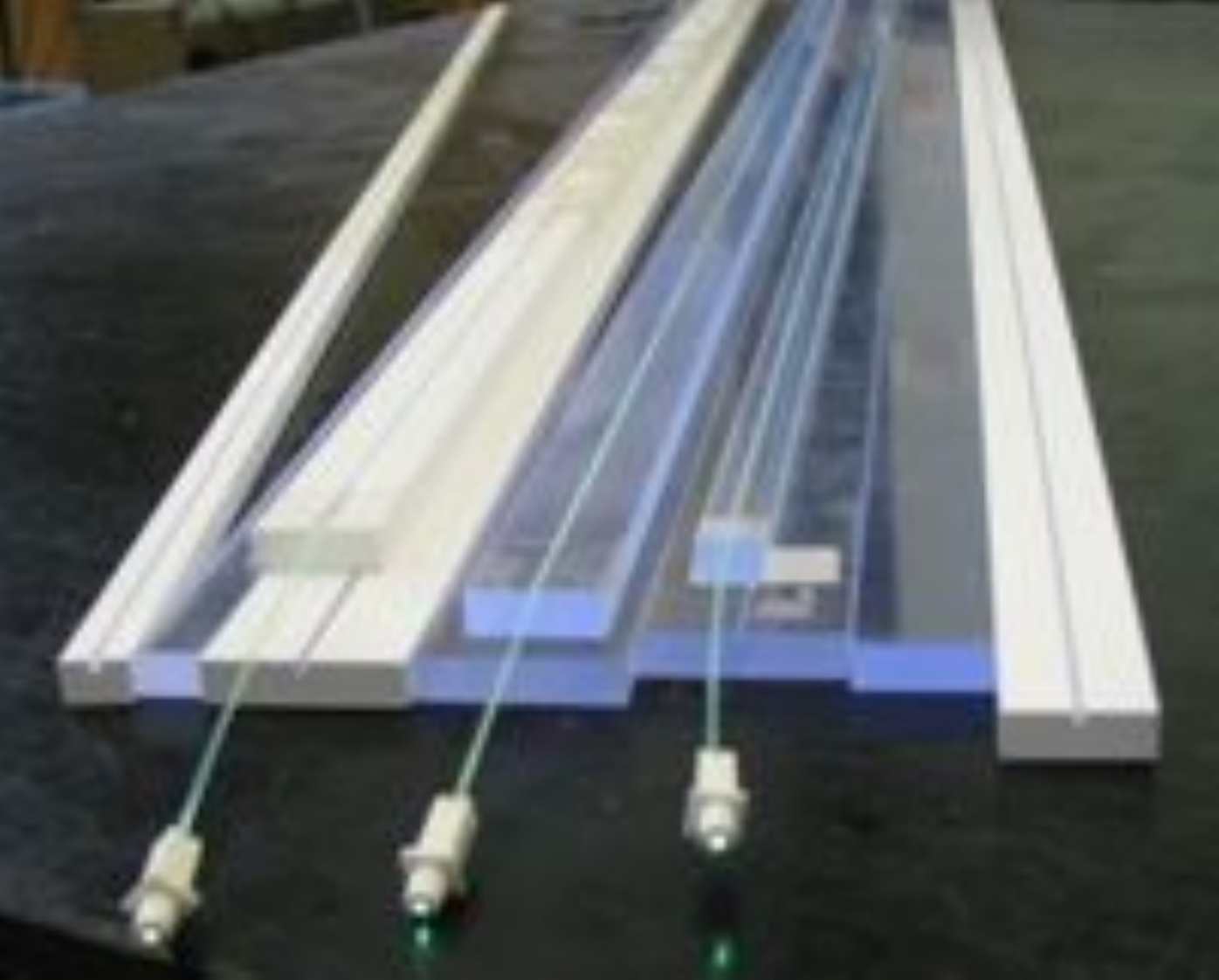}
\hfill
\includegraphics*[width=0.5\textwidth]{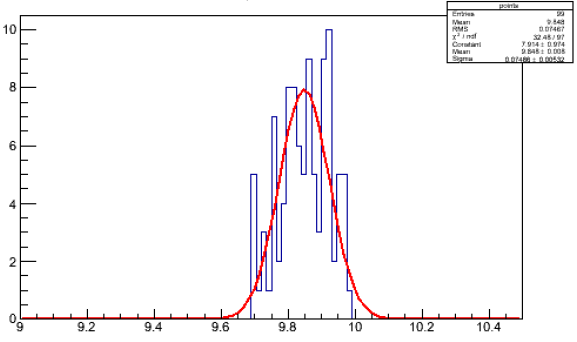}
\caption{Left: Prototype plastic scintillators of different dimensions produced by Uniplast. Right: Distribution of measured width for 100 randomly selected plastic scintillator bars for the chosen width of 10 mm, the X scale is [mm].}
\label{scintillator}
\end{figure}

The plastic scintillator bars will be supplied by the Institute for Nuclear Research (INR) of the Russian Academy of Sciences. The nominal parameters for the geometry are bars of 90 cm long, 0.7 cm in height and 1.0 cm in width, examples are shown in Figure \ref{scintillator}. A small batch of prototypes has been manufactured by Uniplast based in Vladimir (Russia). These extruded scintillator slabs are polystyrene-based with 1.5\% of paraterphenyl (PTP) and 0.01\% of POPOP, similar to the plastics used for the T2K SMRD detector counters. The surface is etched with a chemical agent (Uniplast) to create a 30-100 $\mu$m layer acting as a diffusive reflector. Slabs of three different sizes have been manufactured (895 $\times$ 7 $\times$ 10 mm$^3$, 895 $\times$ 7 $\times$ 20 mm$^3$, 895 $\times$ 7 $\times$ 30 mm$^3$) with 2 mm deep grooves  of different widths (1.1 mm, 1.3 mm or 1.7 mm) to embed optical fibres of different diameters.

\begin{figure}[hbt]
\centering
\includegraphics*[width=0.4\textwidth]{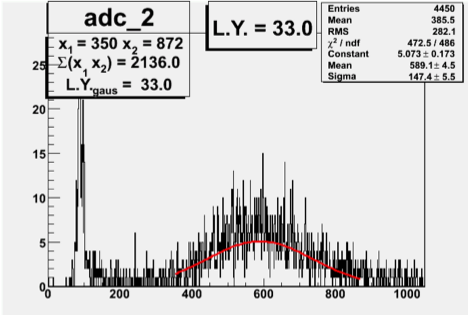}
\hfill
\includegraphics*[width=0.4\textwidth]{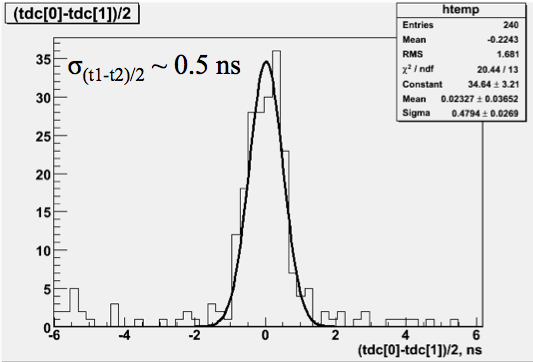}
\caption{Response of a scintillator slab and silicon photomultiplier to a minimum ionising particle: a) light yield and b) timing properties.}
\label{scintillatormip}
\end{figure}

\begin{figure}[hbt]
\centering
\includegraphics*[width=0.35\textwidth]{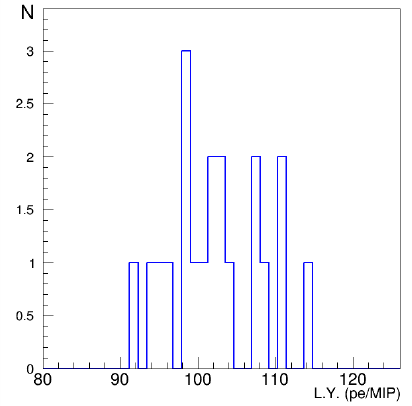}
\includegraphics*[width=0.45\textwidth]{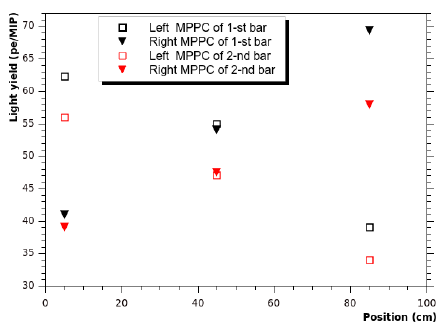}
\caption{Light yield from a 0.7 cm-thick plastic scintillator bar, readout from both ends. The plot on the left shows the sum of the two photosensors. The plot on the right shows the individual contribution of each of the two photosensors to the light yield as a function of position along the bar. The photosensors used for these tests were of the T2K-type, 1.3 $\times$ 1.3 mm$^2$ sensitive area.}
\label{lightyield}
\end{figure}

Tests were carried out at INR to determine basic light yield and timing properties. A wavelength shifting fiber (WLS) from Kuraray (200 ppm, S-type) of d = 1.0 mm was embedded into the 1.1 mm wide groove with a silicon grease (TSF451-50M) to improve optical contact between the scintillator groove surface and the fiber. Hamamatsu MPPC photosensors (1.3 $\times$ 1.3 mm$^2$, 667 pixels, 50 $\times$ 50 $\mu$m$^2$,  gain = 7.5 $\times$ 10$^5$ @$25\,^{\circ}\mathrm{C}$) were connected to both ends of the $\sim$1m long WLS fibers. A cosmic telescope was set up with two trigger counters. Measurements were made at the center of the scintillator slabs. The temperature during testing was 25-$28\,^{\circ}\mathrm{C}$.
Results are summarised in Table \ref{lightyields}. Typical response to a minimum ionising particle is shown in Figure \ref{scintillatormip}. Results show good light yield for all bar thicknesses, the highest light yield was obtained with the narrowest 10 mm width with 83 p.e. Comparisons with/without chemical reflector show an increase of light yield of a factor 2.5 when the chemical reflector is present. The effect of the silicon grease is close to 60\%. For the final assembly, the silicon grease would be replaced by glue, which is expected to have roughly the same effect. An additional Tyvek reflector provides a 20\% increase in light yield, though this reflector is not currently planned for the prototype detectors.
The light yield was measured to be 50 p.e./MIP on average for one photosensor, Figure \ref{lightyield}, in the case of the previous generation Hamamatsu MPPC 1.3 $\times$ 1.3 mm$^2$. Very preliminary results of the latest generation process from Hamamatsu show an average of 58.5 p.e./MIP for smaller 1.0 $\times$ 1.0 mm$^2$ MPPC.
Timing properties were studied for the two-sided readout, combining both ends with the result: $\sigma$($(t_1-t_2)$/2) = 0.5 ns. The timing is mostly determined by the fiber decay constant, $\tau$$_{fiber}$ $\backsim$ 12 ns.

\begin{table}[h]
\centering
\caption{\em Light yields from cosmic tests with prototype scintillator bars of different widths, unit is the photoelectron.}
\begin{tabular}{cccccccc}
\toprule
\textbf{Bar width [mm]} &	\textbf{MPPC 1 [p.e.]}  & 	\textbf{MPPC 2 [p.e.]}  &	\textbf{Sum [p.e.]}\\
\hline
\multicolumn{4}{l}{Bar with no chemical reflector}\\
\hline
10	&	15.7	 &	15.8	&	31.5 \\		
20	&	15.5	 &	13.6	&	29.1 \\	
30	&	12.8	 &	11.5	&	24.3 \\
20 + Tyvek reflector 100-200 $\mu$m		&	41.8	 &	34.8	&	76.6 \\
\hline
\multicolumn{4}{l}{Bar with chemical reflector}\\
\hline		
10	&	46.0	 &	36.8	&	82.8 \\	
20 (1) w/o grease	&	25.7	 &	22.1	&	47.8 \\	
20 (1)	&	39.7	 &	35.7	&	75.4 \\	
20 (1) + Tyvek reflector	&	49.3	 &	44.0	&	93.3 \\	
20 (2)	&	32.6	 &	28.2	&	60.8 \\		
30	&	31.2	 &	26.6	&	57.8 \\	
\bottomrule
\end{tabular}
\label{lightyields}
\end{table}


\subsubsection{Scintillator and fiber connectors}
A good geometrical interface between the SiPM sensitive area and the fiber is a crucial step in achieving good signal transmission efficiency and signal quality. Experience gained with the design of the MICE EMR optical connectors at the University of Geneva has proved valuable in designing the photosensor connectors. Connector prototypes were manufactured with 3D lithography. The final connector design and mass production using plastic injection moulding is to be carried out by the INR. Particular attention will be paid to fiber polishing and assembly stages, where quality assurance must be guaranteed and costs and schedule controlled.

\begin{figure}[hbt]
\centering
\includegraphics*[width=0.45\textwidth]{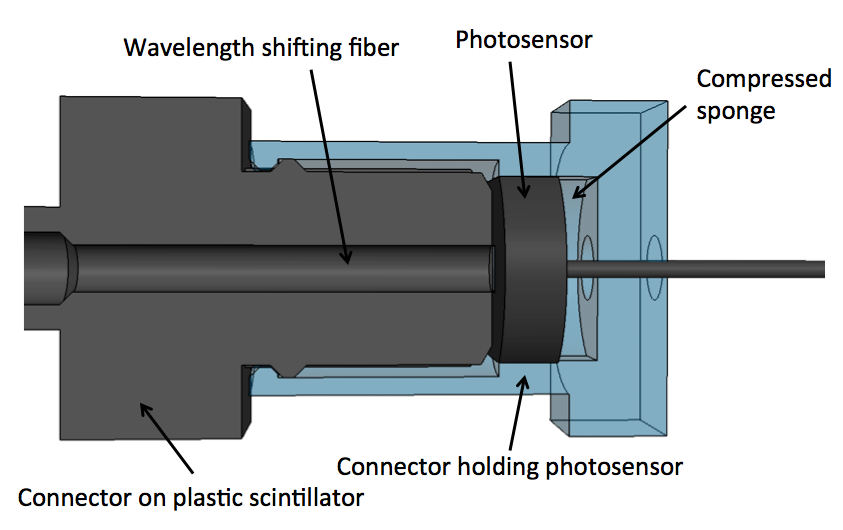}
\hfill
\includegraphics*[width=0.45\textwidth]{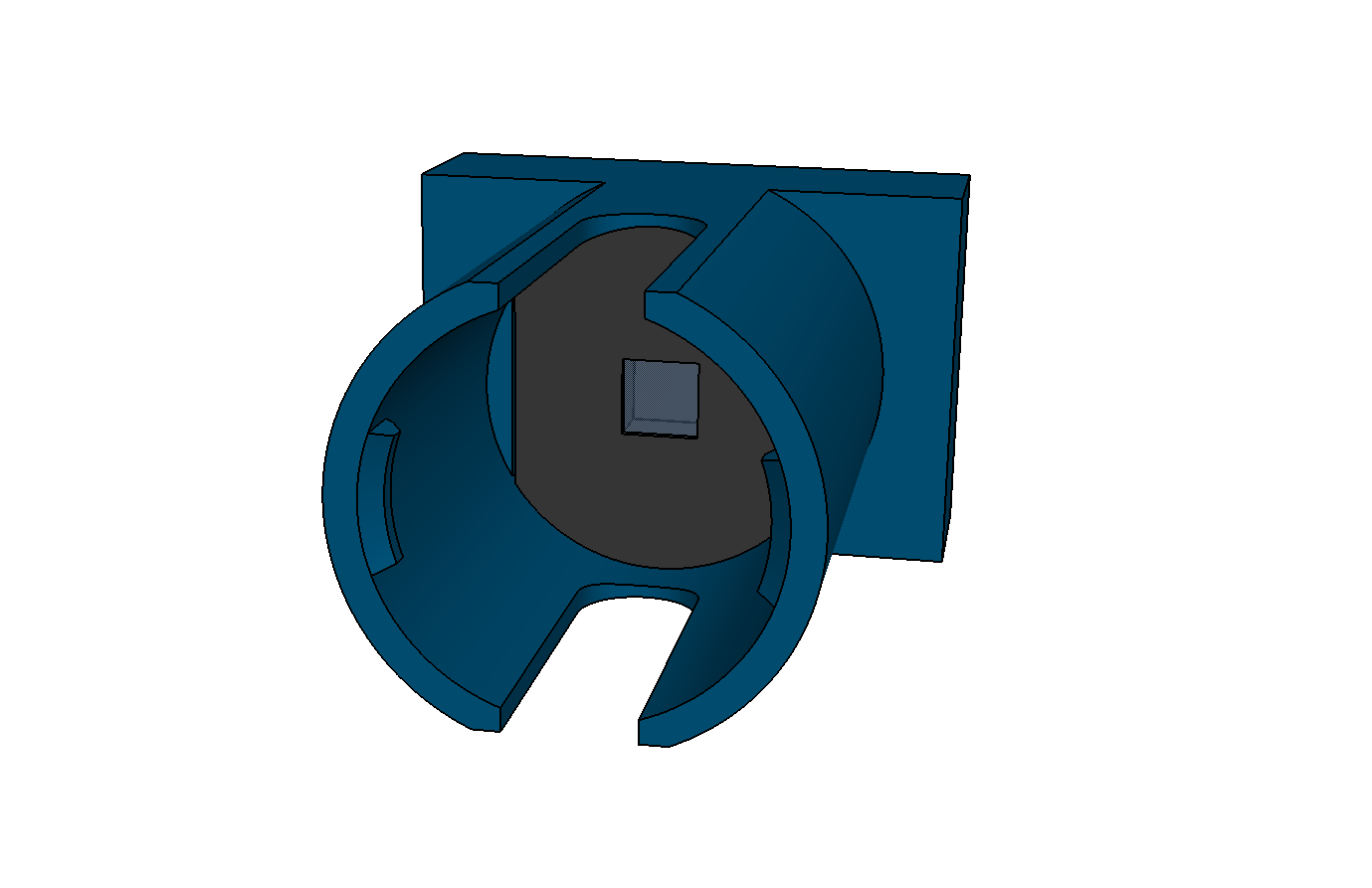}
\hfill
\caption{Photosensor connector concept. The design has evolved from these initial sketches, to include constraints imposed by the plastic injection moulding process.}
\label{connector}
\end{figure}

The concept for a connector system to ensure optimal coupling between the wavelength shifting fibre and photosensor surface is shown in Figure \ref{connector}. It consists of the following main components:
\begin{itemize}
\item The plastic scintillator slab;
\item The wavelength shifting fibre;
\item Connector A, which ensures the WLS fibre is centered with respect to the plastic scintillator slab;
\item The photosensor (MPPC);
\item A component with a spring effect (sponge);
\item Connector B, which holds the photosensor in place;
\item A miniature pcb, which couples to the photosensor pins.
\end{itemize}
The first connector A once glued onto the plastic scintillator bar provides a support enabling the polishing of the fiber ends. The second connector B is designed to house the photosensor. Some parameters for the above components have been fixed, some choices are still possible. The production of plastic scintillator slabs has started, and in this context, the connector concept was designed to allow for production to proceed on the elements that affect the plastic scintillator production (the WLS, connector A).

The connector concept proposed here is based on a connector for the Hamamatsu MPPC devised for the T2K experiment. It is therefore based on the dimensions of physical objects, such as the MPPC, the pcb for electrical connections which hosts a mini-coaxial connector etc... Many of the geometrical constraints are very similar, modifications were brought to the design to account for the smaller dimensions of the connector. 

\begin{figure}[hbt]
\centering
\includegraphics*[width=0.45\textwidth]{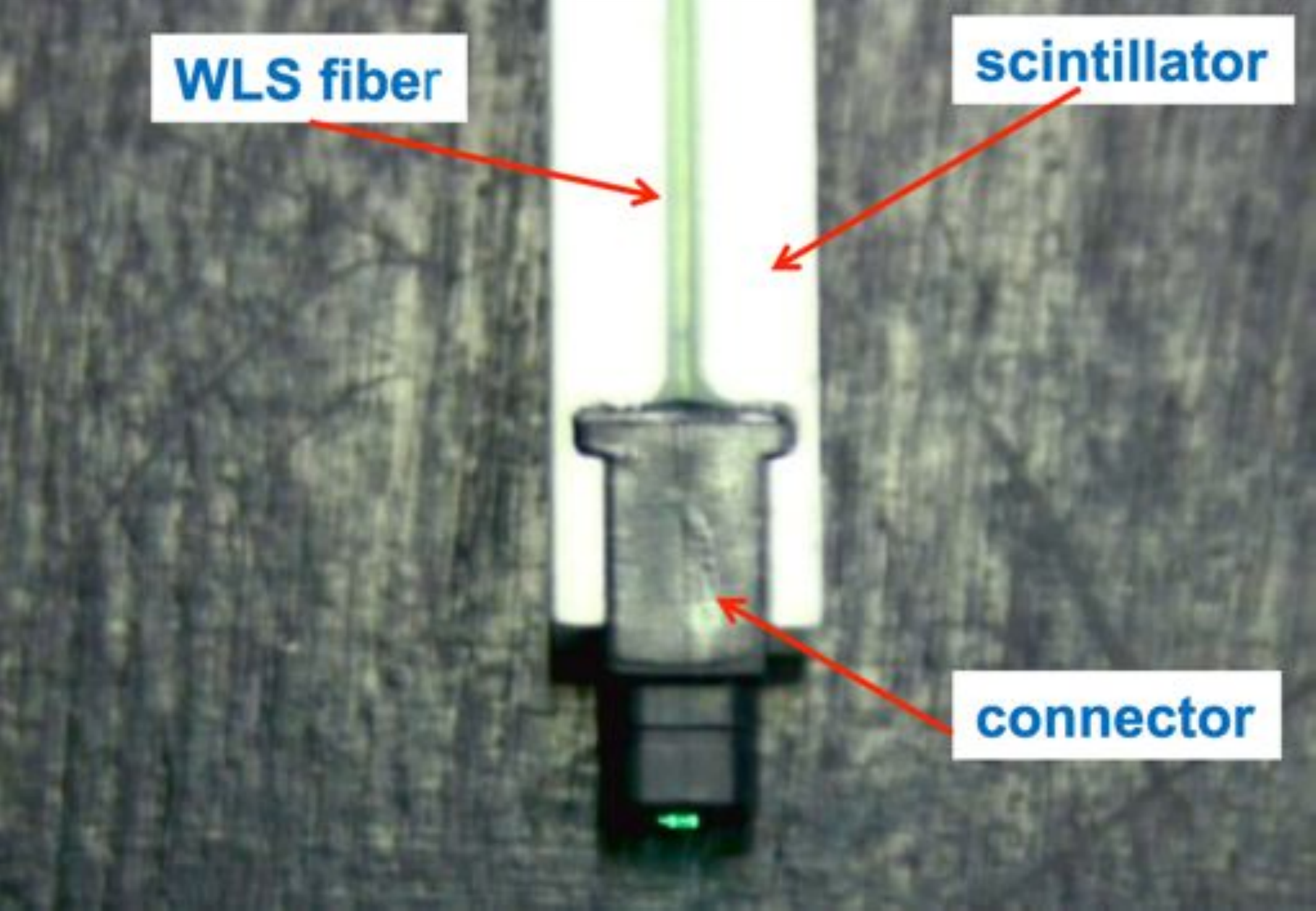}
\includegraphics*[width=0.45\textwidth]{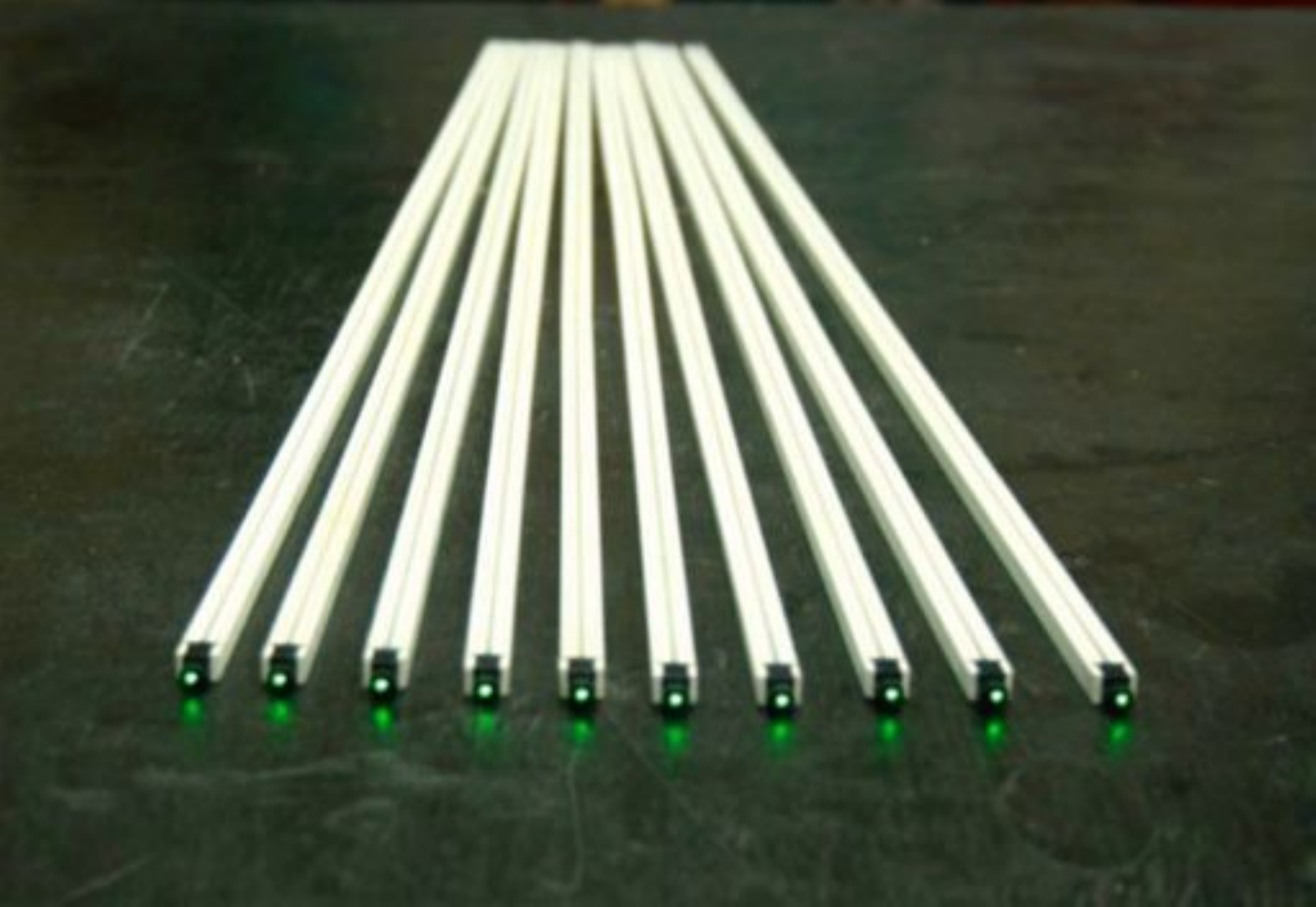}
\caption{Plastic scintillator bars with their connectors, as planned for the MIND modules.}
\end{figure}

\subsubsection{Choice of photosensor}

The first silicon photomultiplier device used on a large scale in a physics experiment is the Hamamatsu MPPC S10362-13-050C instrumenting all scintillator detectors at the ND280 near detector complex of the T2K experiment, \cite{Amaudruz:2012esa}. It was derived from a commercial device, with the sensitive area increased from 1 $\times$ 1 mm${^2}$ to 1.3 $\times$ 1.3 mm${^2}$ to provide better acceptance for the light emitted from the 1.0 mm diameter wavelength shifting fiber from Kuraray. This MPPC consists of 667 pixels, each working in limited Geiger mode with an applied voltage slightly above the breakdown voltage ($V_{bd}=70V$). With the production of a photo-electron in a pixel, a Geiger avalanche is generated, which is then passively quenched by a built-in resistor in each pixel. The induced charge is independent of the number of photo-electrons produced, and is directly proportional to the over voltage which is therefore a crucial parameter: $Q=C(V-V_{bd})$ or $Q=C{\delta}V$. In order to operate in the linear regime, where the MPPC output charge is directly proportional to the incoming photons, it is crucial that the total number of photo-electrons remains below the number of pixels in the device.

The thorough work done in the selection and characterization of photosensors for the T2K experiment serves as a very good basis for the selection of photosensors for the AIDA prototypes. Photosensor design is a fast evolving field. Several manufacturers offer a variety of products.  Table \ref{photosensorcomparison} lists characteristics and measured performance for a selection of devices from different manufacturers tested at the INR in Russia in 2013.

Further tests are planned of the latest generation devices from Hamamatsu. Specifications from this manufacturer indicate improved dark noise, lower after pulse and higher PDE. Smaller cell sizes are also available, increasing the dynamic range for our application. Hamamatsu will also make available so-called precision measurement devices with significantly less cross-talk between adjacent pixels which is achieved by surrounding each pixel by a trench. These devices will be available from April 2014.

The full electronics chain must be taken into consideration before deciding on the photosensor. In order to determine several of the key operational parameters, a test stand was setup with the final configuration of the plastic scintillator bar, a photosensor connector allowing for tests of various photosensors and an evaluation board for the EASIROC chip designed by Omega micro electronics. The test board is described in more detail in another section. Charge information for the photosensors is shown in Figure... The exact working point for the photosensors, which can be resumed to the applied overvoltage, is still to be determined from more detailed studies. A large overvoltage leads to a large signal from the photosensor, which in turn leads to high ADC counts per photo-electron once the charge information is digitized by the 12-bit ADCs which are external to the chip. It also leads to a higher signal-to-noise ratio. The disadvantages of a large over voltage are the significantly higher cross-talk, and the smaller dynamic range, which is ultimately limited by the ADC. The dynamic range can be extended to some extent by using the low gain analogue signal path, though it is less straightforward to calibrate, due to the much lower gain, which limits the resolution of single photo-electron peaks.

The following points are therefore relevant in setting the operating point of the photosensors and EASIROC chip, from the perspective of the readout electronics; i.e. the over voltage of the photosensor, and the pre-amp and shaper settings of the EASIROC chip:
\begin{itemize}
	\item{Calibration of signals: peak-to-peak resolution:}
	\begin{itemize}
		\item{should be able to calibrate high gain signal path;}
		\item{should be able to calibrate low gain signal path;}
		\item{this puts stringent requirements on the cross-talk;}
		\item{calibration can be performed at higher pre-amp gain if linear.}
	\end{itemize}
	\item{ADC counts per photo-electron peak:}
	\begin{itemize}
		\item{high enough to resolve individual photo-electron peaks;}
		\item{high enough to provide good signal/noise: noise $\sigma=5 ADC$;}
		\item{required noise level is 0.2 p.e. Pk-to-pk should be $> 25 ADC counts$.}
	\end{itemize}
\end{itemize} 

As an example, setting the chain to obtain 25 ADC/p.e. on the high gain path, with a 12-bit ADC giving a range of 4096 ADC and a baseline around 1000 ADC (could be improved), the full range of the high gain path would be 120 p.e. Using the same assumptions for the low gain signal path but this time with 5 ADC/p.e., the full range of the low gain path would be 620 p.e. This example shows that the system can be optimized for both:
\begin{itemize}
	\item{the most likely events with the high gain signal path (one MIP would yield 100 p.e.);}
	\item{events with high energy deposition such as stopping particles (estimated at 500 p.e.).}
\end{itemize}

\begin{table}[h]
\centering
\caption{\em Comparison of specifications and performance of different photosensors from a range of manufacturers, measured under conditions representative of the AIDA detector modules.}
\begin{tabular}{cccccccc}
\toprule
\textbf{Parameter} & \textbf{Unit} & \textbf{MPPC-T2K} &	\textbf{ASD-40}  & 	\textbf{KETEK}  &	\textbf{SensL}\\
\hline
\multicolumn{4}{l}{Manufacturer reported specifications}\\
\hline
Pixel size & ${\mu}m$ & 50	&	40	 &	50	&	20 \\		
Number of pixels	&	& 667	 &	600	&	400 & 848 \\	
Sensitive area	& mm$^2$	& 1.3 $\times$ 1.3	 &	dia 1.2 & 1.0 $\times$ 1.0	&	1.0 $\times$1.0 \\
Gain	& 	& $7.5\times10^5$	 &	$1.6\times10^6$ & -	&	-\\
Dark rate & MHz 	& $\leq1$	 &	$\sim3$ & $\leq2$	&	$\leq2$\\
Bias voltage	& V	& $\sim70$	 &	30-50 & 33-50	&	30\\
\hline
\multicolumn{4}{l}{Performance}\\
\hline
Overvoltage	& V	& $\sim$1.4	 &	3.6 & 4.5	&	2.7\\
Dark rate & kHz 	& 900	 &	3630 & 1250	& 1960\\
Crosstalk & \% 	& 10	 &	13.4 & 35	& 9.7\\
Pulse shape& - 	& good	 &	good & long tails& good\\	
Peak separation& - 	& good	 &	good & bad& bad\\	
PDE& \% 	& 25.6	 &	11 & 26.4& 14.2\\	
\bottomrule
\end{tabular}
\label{photosensorcomparison}
\end{table}

\begin{figure}[hbt]
\centering
\includegraphics*[width=0.45\textwidth]{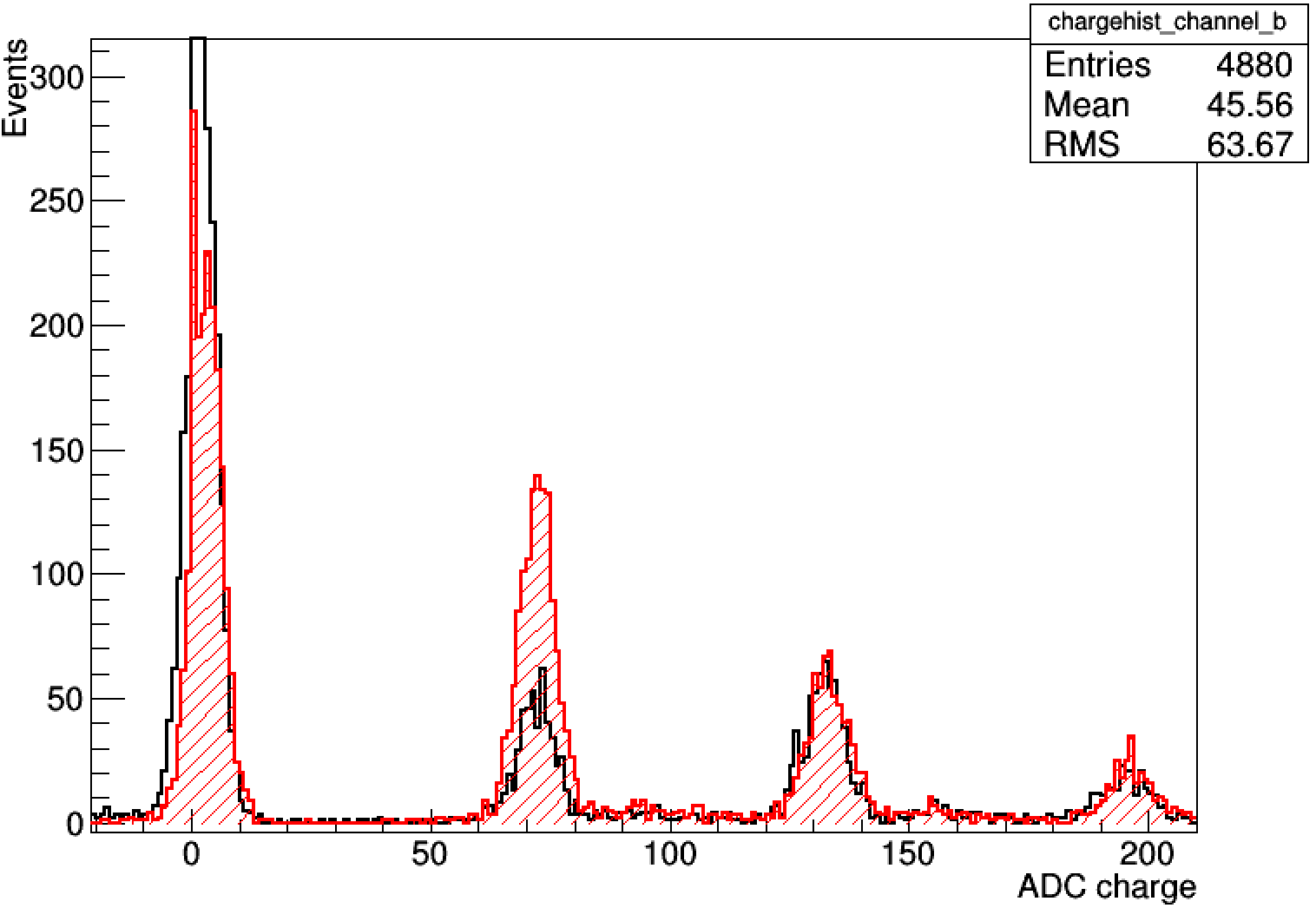}
\includegraphics*[width=0.45\textwidth]{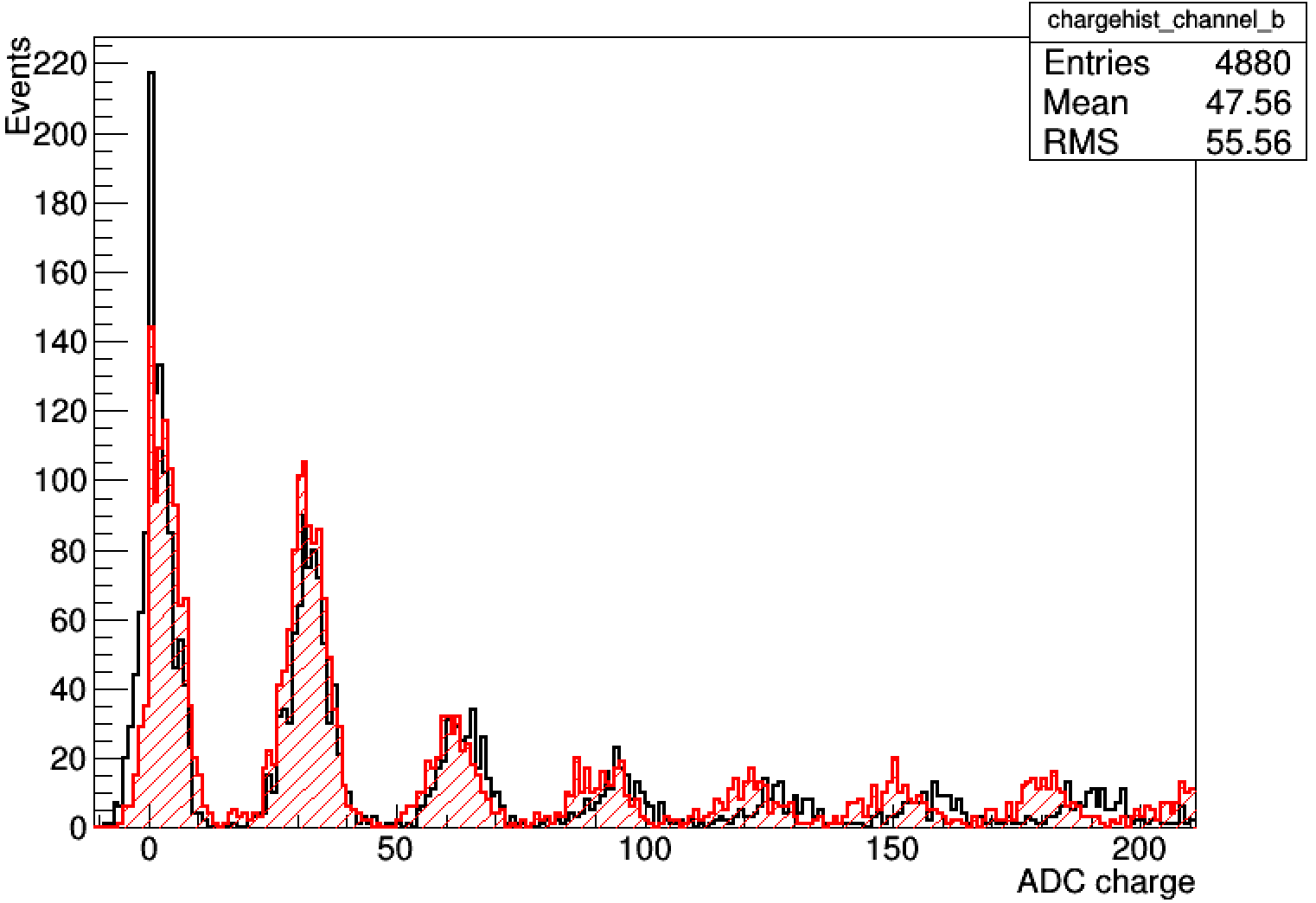}
\caption{Charge spectra for the MPPC S12571-050C, a 50-micron cell size, 1 $\times$ 1 mm$^2$ device. Analogue data from the high gain signal path from the EASIROC chip, digitized with a 12-bit ADC, demonstrates the excellent photo-electron peak-to-peak separation. The EASIROC pre-amp feedback capacitance is set to 100fF, the shaper time constant is set to 50ns. Left) high over voltage leading to $\sim$ 65 ADC/p.e. Right) low over voltage leading to $\sim$ 30 ADC/p.e. Difference in over voltage between left and right acquisitions is 1.75 V.}
\label{50-micron-MPPC}
\end{figure}

\begin{figure}[hbt]
\centering
\includegraphics*[width=0.45\textwidth]{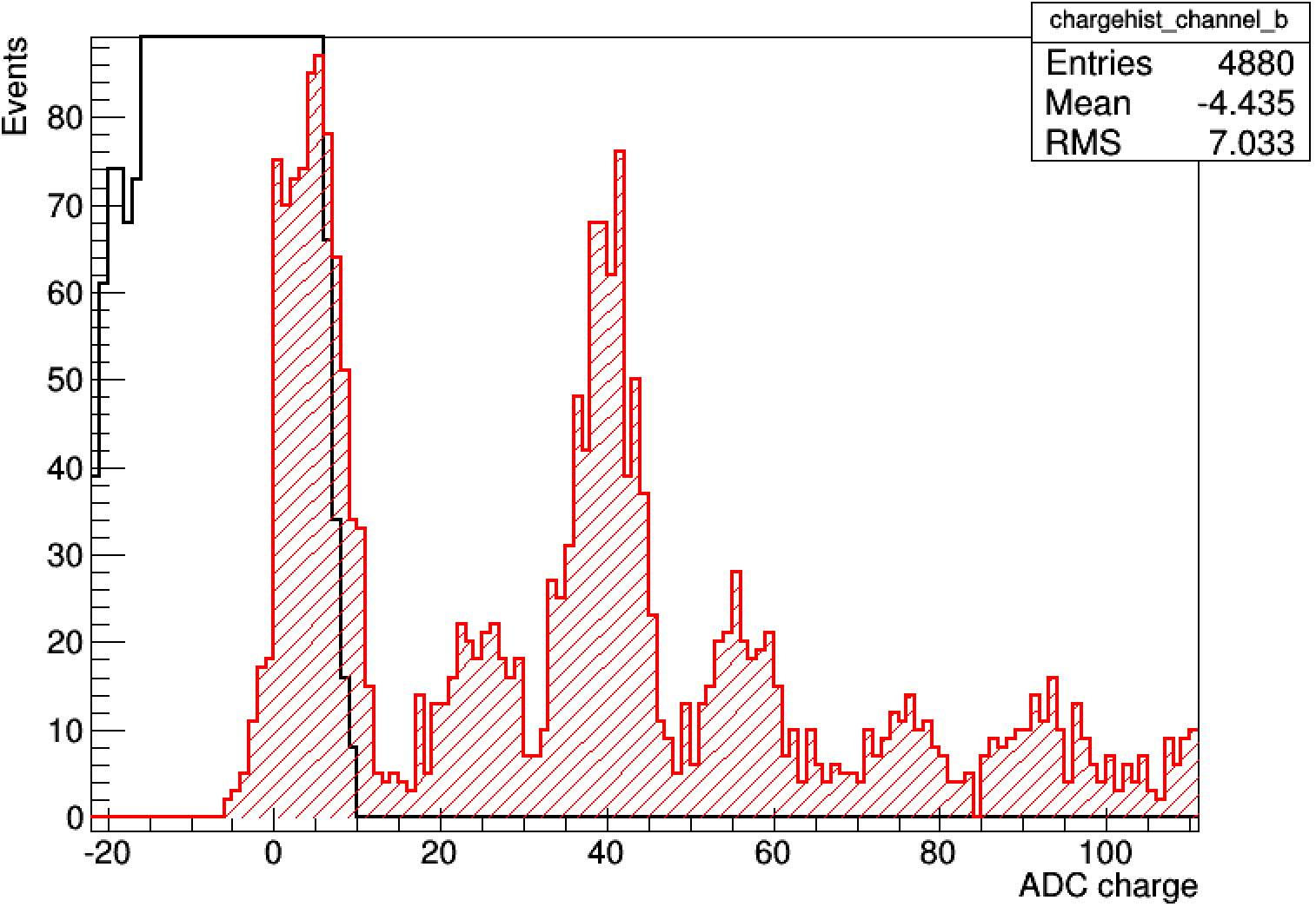}
\includegraphics*[width=0.45\textwidth]{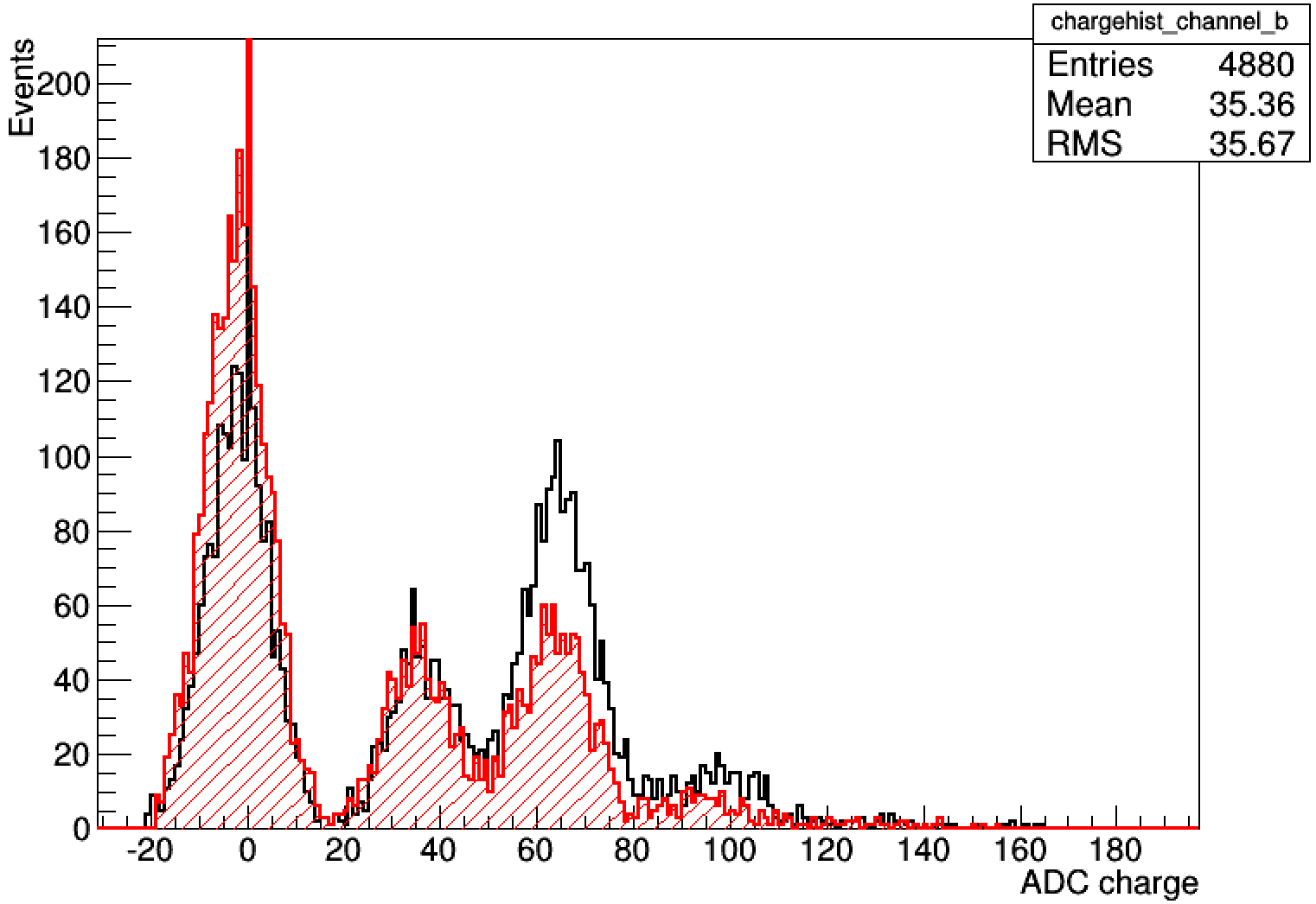}
\caption{Charge spectra for the MPPC S12571-025C, a 25-micron cell size, 1 $\times$ 1 mm$^2$ device. Analogue data from the high gain signal path from the EASIROC chip, digitized with a 12-bit ADC. The EASIROC pre-amp feedback capacitance is set to 100fF for the left plot, and 0fF for the right plot. The shaper time constant is set to 50ns. A relatively low over voltage was used here, it could be increased to provide higher gain.}
\label{25-micron-MPPC}
\end{figure}

\subsection{Electronics and DAQ}
Emphasis will be placed on the electronics options which offer the best opportunity for further development to cover medium-term foreseeable requirements for neutrino detectors and related applications.
The following options have been studied:
\begin{itemize}
\item{DRS4: potential with long term perspectives but currently expensive,}
\item{EASIROC: 3kHz readout rate demonstrated, architecture close to the MICE EMR,}
\item{T2K ND280 TRIP-t option: will not be considered for this test beam.}
\end{itemize}

The data acquisition system will be adapted from the MICE EMR DAQ. The EASIROC readout chip is the baseline solution. R\&D on the DRS4 chip is ongoing at the University of Geneva within the framework of upgrades to the NA61 experiment. Depending on progress, a few modules could be equipped with DRS4 readout boards.

 We plan to adopt a readout system where we readout separately the slower low and high gain analogue signal paths at 1 kHz providing charge information, and the faster hit-only digital triggers representing 4000 samples for every particle trigger running at 400 MHz, covering 10 $\mu$s after each event, with an average readout rate of 100 kHz. The faster digital triggers should allow tagging of delayed signals related to each event (e.g. muon decay to electron). 


\subsubsection{Outline of electronics chain}
The planned electronics chain is based on the electronics chain developed for the MICE-EMR detector, installed on the MICE beamline at the Rutherford Appleton Laboratory in September 2013. It is to be adapted taking into consideration:
\begin{itemize}
	\item{the beam characteristics of the beamline at CERN;}
	\item{the different photosensors, going from PMTs to silicon photomultipliers;}
	\item{the different readout chip, from MAROC to EASIROC.}
\end{itemize}

In particular, a new front-end board must be designed. Coupling of the photosensors to the EASIROC chip can be done using the scheme implemented on the EASIROC evaluation board.

Concepts to be re-used from the MICE-EMR detector include the use of a Digitizer Buffer that acts as a TDC, assigning a time stamp to every event. The VME Readout Board (VRB) will also be re-used. It collects information from the front-end memory buffers and transmits it to the PC.

\begin{figure}[hbt]
\centering
\includegraphics*[width=0.7\textwidth]{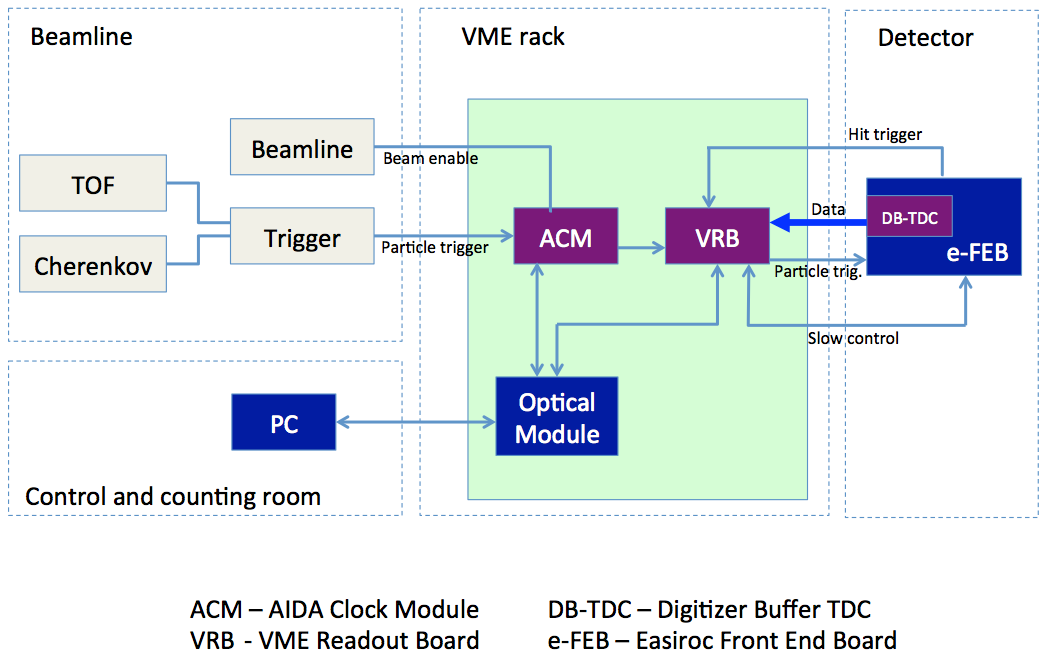}
\caption{Overview of the electronics chain for the AIDA neutrino detector prototypes.}
\label{electronics-scheme}
\end{figure}

\subsubsection{Beam considerations for electronics}
\begin{figure}[hbt]
\centering
\includegraphics*[width=0.8\textwidth]{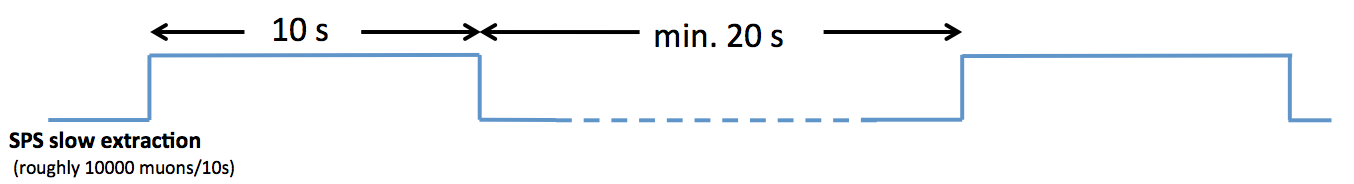}
\caption{SPS slow extraction beam time structure. These assumptions are taken in the design of the AIDA neutrino detector electronics.}
\label{spsslowextraction}
\end{figure}

The planned set of runs in the North Area foresees the implementation of a very low energy beam line, delivering protons, pions, electrons, muons in the energy range 0.5 to 9 GeV/c, Table \ref{particlerates}. The beam is a slow extracted beam with a pulse length of up to 10 s, delivering particles at a rate of 1 kHz or so, Figure \ref{spsslowextraction}. The pulse repetition frequency is maximum 0.03 Hz. 
 
\begin{figure}[hbt]
\centering
\includegraphics*[width=0.8\textwidth]{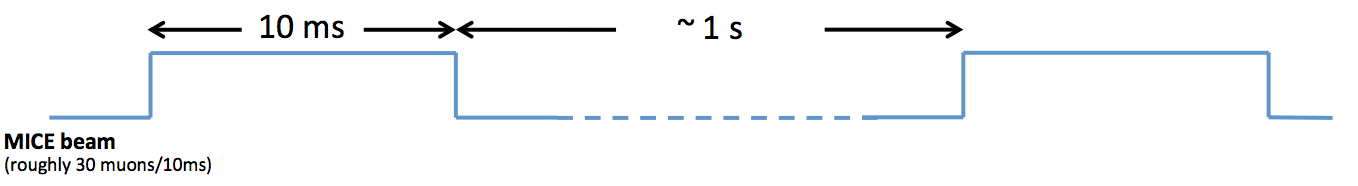}
\caption{MICE beam structure. The downstream components of the MICE electronics chain will be adapted for use in the AIDA readout electronics chain.}
\label{spsslowextractiontwo}
\end{figure}

\begin{figure}[hbt]
\centering
\includegraphics*[width=0.8\textwidth]{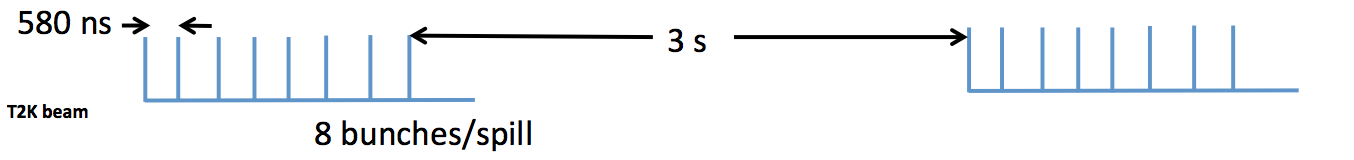}
\caption{T2K beam structure. The Trip-t electronics for T2K were initially chosen as the baseline but are no longer being considered for AIDA.}
\label{spsslowextractionthree}
\end{figure}

The background is expected to be mainly from high energy muons created from the interaction of the primary beam with the first target. It should be low enough that triggering on interesting events will be achievable with high efficiencies.


\subsubsection{EASIROC chip}
The EASIROC (Extended Analogue SI-pm ReadOut Chip) chip is designed in 0.35 $\mu$m SiGe technology, with first versions available since 2010 and used in a variety of experiments such as PEBS, MuRAY, at JPARC and in medical imaging. It is a 32 channel fully analogue front-end ASIC dedicated to readout of SiPM photosensors. It is derived from the SPIROC chip which was developed for hadron calorimetry foreseen for the International Linear Collider.

The chip integrates a 4.5 V (2.5 V) range 8-bit DAC for individual SiPM gain adjustment. A multiplexed charge measurement is available from 160 fC to 320 pC with 2 analogue outputs. These charge paths are made of 2 variable gain preamplifiers followed by 2 tunable shapers and a track and hold. 

The analogue core is sensitive to positive SiPM signals. For each channel, two parallel AC coupled voltage preamplifiers ensure the read out of the charge from 160 fC to 320 pC (ie. 1 to 2000 photoelectrons with SiPM Gain = $10^6$, with a photoelectron to noise ratio of 10). Two variable shapers are used to reduce noise; each of them has an adjustable peaking time from 25 to 175 ns to allow the user to minimize the noise depending on the final application. A trigger line is available from the high gain preamplifier. It is composed of a 15 ns peaking time fast shaper followed by a discriminator. The threshold is set by an internal 10-bit DAC and is common to the 32 channels. The 32 triggers are available as a 32-bit output bus and can be either latched or directly outputted.

In addition to the charge output, timing measurements are possible via a trigger path, that integrates a fast shaper followed by a discriminator. It's threshold can be set via a common 10-bit DAC. The 32 trigger outputs are complemented by an OR32 output. 

Power consumption is 5 mW/channel and unused features can be powered OFF. 

\begin{figure}[hbt]
\centering
\includegraphics*[width=0.30\textwidth]{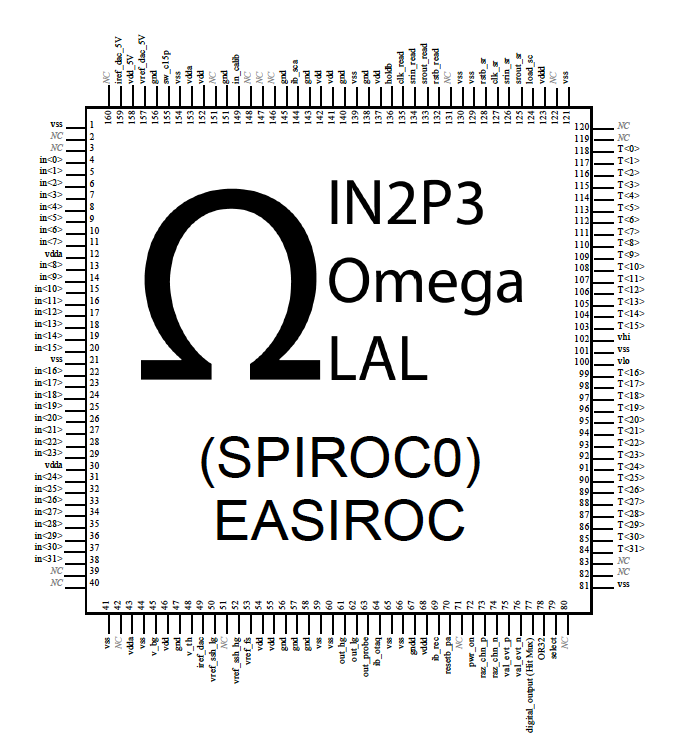}
\includegraphics*[width=0.55\textwidth]{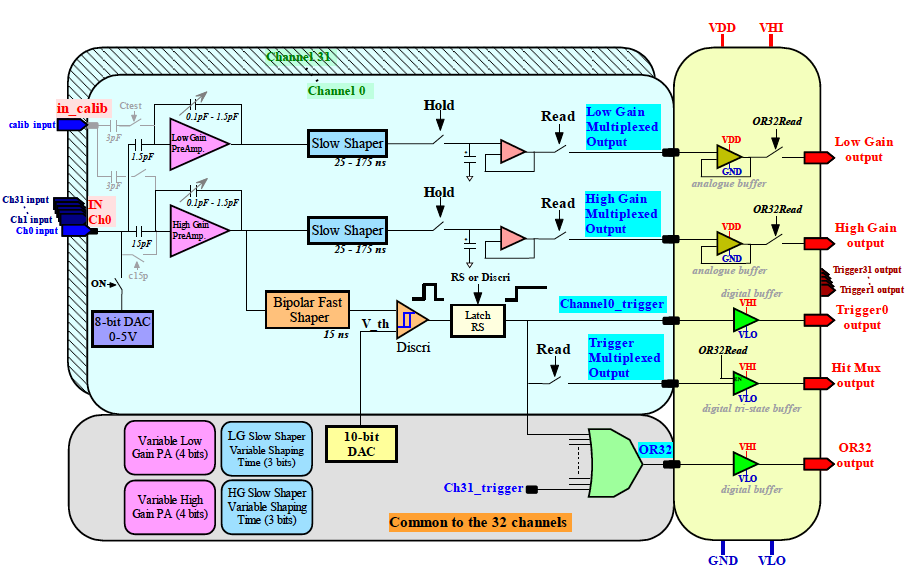}
\caption{Pinout and schematic of the EASIROC chip.}
\label{easiroc-pins}
\end{figure}


\subsubsection{Readout modes}
As described above, two readout modes from the EASIROC are possible:
\begin{itemize}
	\item{Mode A: slow readout of the analogue multiplexed signals;}
	\item{Mode B: Fast readout of the 32 triggers.}
\end{itemize}

\begin{figure}[hbt]
\centering
\includegraphics*[width=0.6\textwidth]{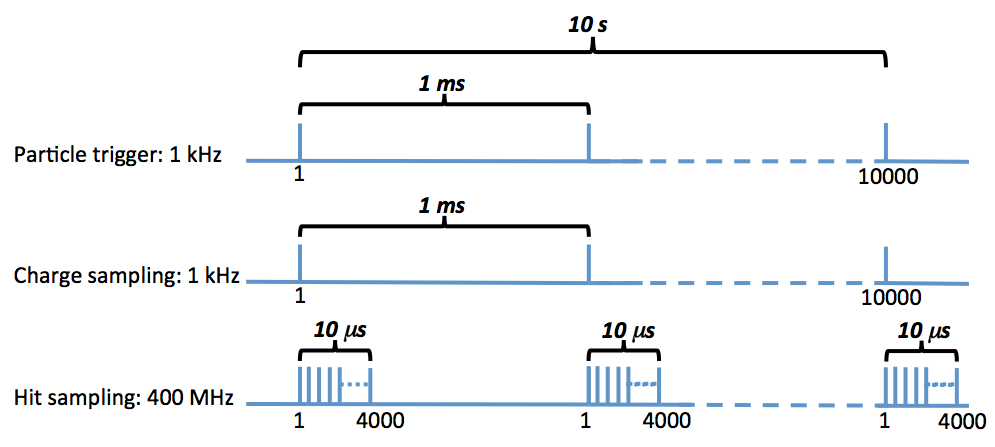}
\caption{Particle triggers, charge sampling and hit sampling within a 10s slow extraction spill from the SPS.}
\label{sampling-rates}
\end{figure}

Used in combination, it is planned to operate with an external particle trigger provided by a system upstream, for example a Time of Flight detector, at rates of few $\sim$kHz. Mode A would then provide one charge sample per particle trigger. Mode B can be operated at 400 MHz, recording 10 $\mu$s-worth of digital hits occurring immediately after the primary event, such as decay products, i.e. 4000 samples per particle trigger, see Figure \ref{sampling-rates}.

The Front-End Board will be designed with sufficient memory buffers to record all data within a spill, lasting 10s in the case of a slow extracted beam at the SPS. The stored data will be transmitted between spills to the VME Readout Board (VRB), and then onto a PC. The minimum length between spills is 20s.


\subsubsection{Data stream}
 It is planned to have one Front End Board per plane, i.e. 90 ch/FEB. However, detailed studies are required, including costing, before committing to a 1 FEB/plane scheme. Another possibility is to have one FEB/EASIROC chip, i.e. 30 ch/FEB, 3 FEB/plane. Since only roughly 1/3 of channels will be instrumented in phase 1, it is to be decided whether the corresponding fraction of each module will be instrumented, or whether full modules will be instrumented, but only 1/3 of the total number of modules will be used.

\begin{figure}[hbt]
\centering
\includegraphics*[width=0.7\textwidth]{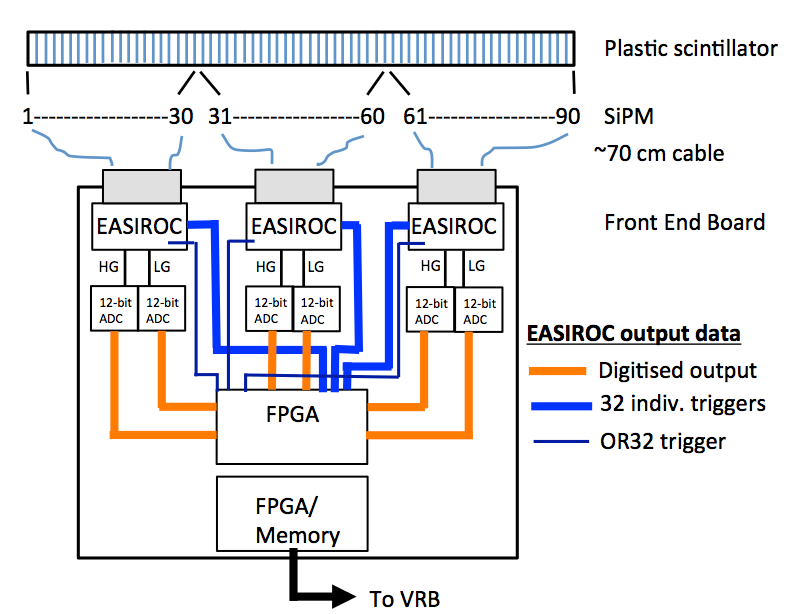}
\caption{Readout scheme for one plane.}
\label{Electronics-sketch}
\end{figure}

The bit allocation scheme is reported in Table \ref{bit-allocation}. The scheme was devised to allow flexibility in the recording of time, with several configurations possible by combining the 16-bit \mbox{\bf{Hit time measurement}} with the 20-bit \mbox{\bf{Event time tag}}. 

For timing under normal operation at the SPS with a slow extraction spill, the hit time measurement should cover 10 $\mu$s with sampling every 2.5 ns: 12 bits are sufficient.
The event time tag should reset every 10 $\mu$s, a rate of 100 kHz, and cover 10s which is the spill length: 20 bits are required.
\begin{table}[h]
\centering
\caption{\em Allocated bits for different parameters in the recorded data.}
\begin{tabular}{cccccccc}
\toprule
\textbf{Parameter} & \textbf{Allocated bits} & \textbf{Full range} &	\textbf{Requirement} 	\\
\hline
Word type & 4 & 16	&	-	 \\
Board ID & 10 & 1024	&	100	 \\
Spill number & 16 & 65536	&	-	 \\
Event count (Ext. particle trigger) & 28 & 268435456	&	1e8	 \\
Event ID (TBC) & - & -	&	-	 \\
Channel ID & 7 & 128	&	90	 \\
Hit ID & 5 & 32	&	-	 \\
Hit time measurement & 16 & 65536	 &	4000	 \\
Amplitude status & 4 & 16	&	2	 \\
Hit amplitude measurement & 12 & 4096	&	4096	 \\
Hit count within event & 6 & 64	&	-	 \\
Event time tag & 20 & 1048576	&	1e6	 \\
Spill width (TBC) & 22 & 4194304	&	-	 \\
\bottomrule
\end{tabular}
\label{bit-allocation}
\end{table}
 
\begin{figure}[hbt]
\centering
\includegraphics*[width=1.0\textwidth]{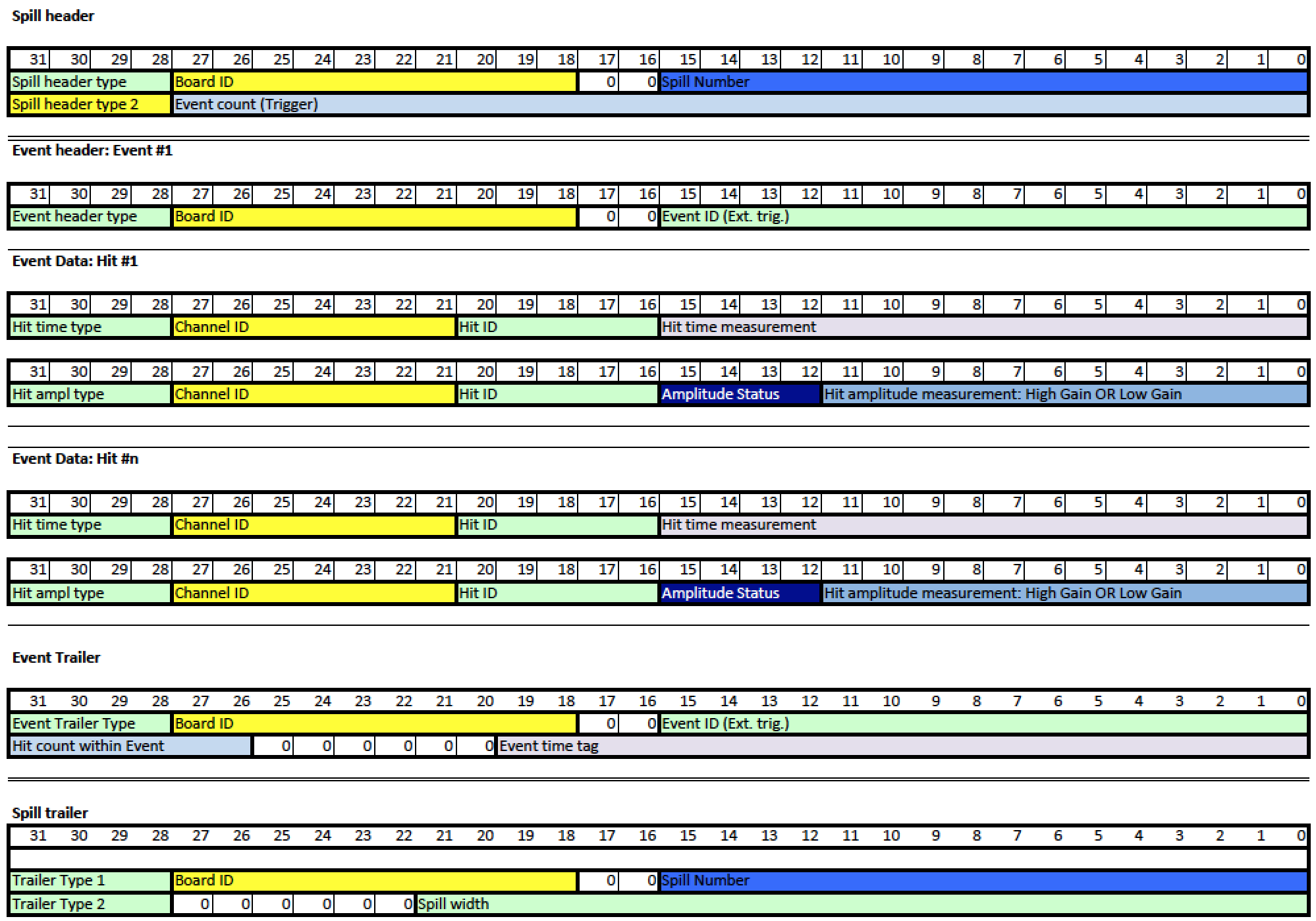}
\caption{Data structure for the MIND neutrino detector prototypes.}
\label{detectormodules}
\end{figure}

The estimation of data rate is critical in defining the buffer size for memory on the Front End Board. Several assumptions are made, these are usually conservative. Experience with the MICE detector would suggest that during normal data taking, the buffers are over-dimensioned by a large factor. However, some events, such as electromagnetic or hadronic showers, lead to a large number of hits, which must be recorded.

Table \ref{data-rates} lists the expected data rates for runs under the assumptions of a slow extracted beam at the SPS outlined previously. In summary, we expect \mbox{\bf{1 MB/plane/spill}}, where a spill is 10s. The VRB memory buffer of the existing MICE-EMR detector is 8 MB. With one FEB/plane, it is relatively straightforward to reconfigure the firmware of the VRBs so that each VRB serves 8 planes, or 4 modules.
A fraction of the 9000 plastic scintillator bars will be instrumented in the first phase. 3000 channels will be instrumented, we would therefore need 34 FEBs, 5 VRBs, 1 VME crate. The transfer rate from VRB to PC was measured to be 80 $\mu$s per kB, it should be re-measured with the configuration described here. It would lead to transfer times of 2.5s/event for the instrumented part of the detector, i.e. 3000 channels.

\begin{table}[h]
\centering
\caption{\em Data rates estimated during operation of the AIDA neutrino detector prototypes. Note that in the first phase, only a fraction of the whole detector, 3000 of 9000 bars, will be instrumented.}
\begin{tabular}{cccccccc}
\toprule
\textbf{Parameter}  & 	\textbf{Per plane}  &	\textbf{Instrumented detector (Whole)}\\
\hline
\multicolumn{3}{l}{Items per plane}\\
\hline
\# channels		 &	90	&	3000 (9000) \\		
\# FEB	 &	1	&	34 (100) 	 \\
\# EASIROC	 &	3	&	102 (300) 	 \\
\hline
\multicolumn{3}{l}{Charge and hits per particle trigger(event)}\\
\hline			
\# Charge output/Particle trig/plane &	10	&	-  \\
\# Hit output/Particle trig/plane &	10	&	-  \\
\hline
\multicolumn{3}{l}{Stored information}\\
\hline	
Stored Bytes/event	&	92.0	&	 \\		
Particle triggers/spill	 	&	10000 &	 \\
Stored Bytes/spill [MB]	&	0.920 & 31.28	 \\			
\bottomrule
\end{tabular}
\label{data-rates}
\end{table}

\subsubsection{Slow control}
It is planned to implement the slow control via the VRB chain, i.e. one link for both data and control.

\subsection{Iron plates}
\subsubsection{Plate dimensions}
The iron plates for the MIND500 will be octagonal, with a height of 5 m and a width of 5 m. Their thickness is currently set to 3 cm, with a potential range of 1.5 - 5 cm. Whether the iron thickness between detector modules is made up of 1 or 2 layers of iron will be determined from optimization of the mechanical engineering of the MIND. A cradle configuration is being explored, adapted from MINOS, where the iron plates are hung on ears. The main advantage of a cradle configuration is that detector positioning and alignment are independent of the building, particularly important given that it is planned to move the MIND500 within the North Area building onto several beamlines. The displacement  of the MIND will have to be studied in detail, since it is unlikely that modularity will be possible, given that the 8 superconducting coils will have to be spliced together once the detector is assembled. A system on rails might be possible, though the floor levels for different beam lines in the extension to the North Area building are planned to be different.

\begin{figure}[hbt]
\centering
\includegraphics*[width=0.70\textwidth]{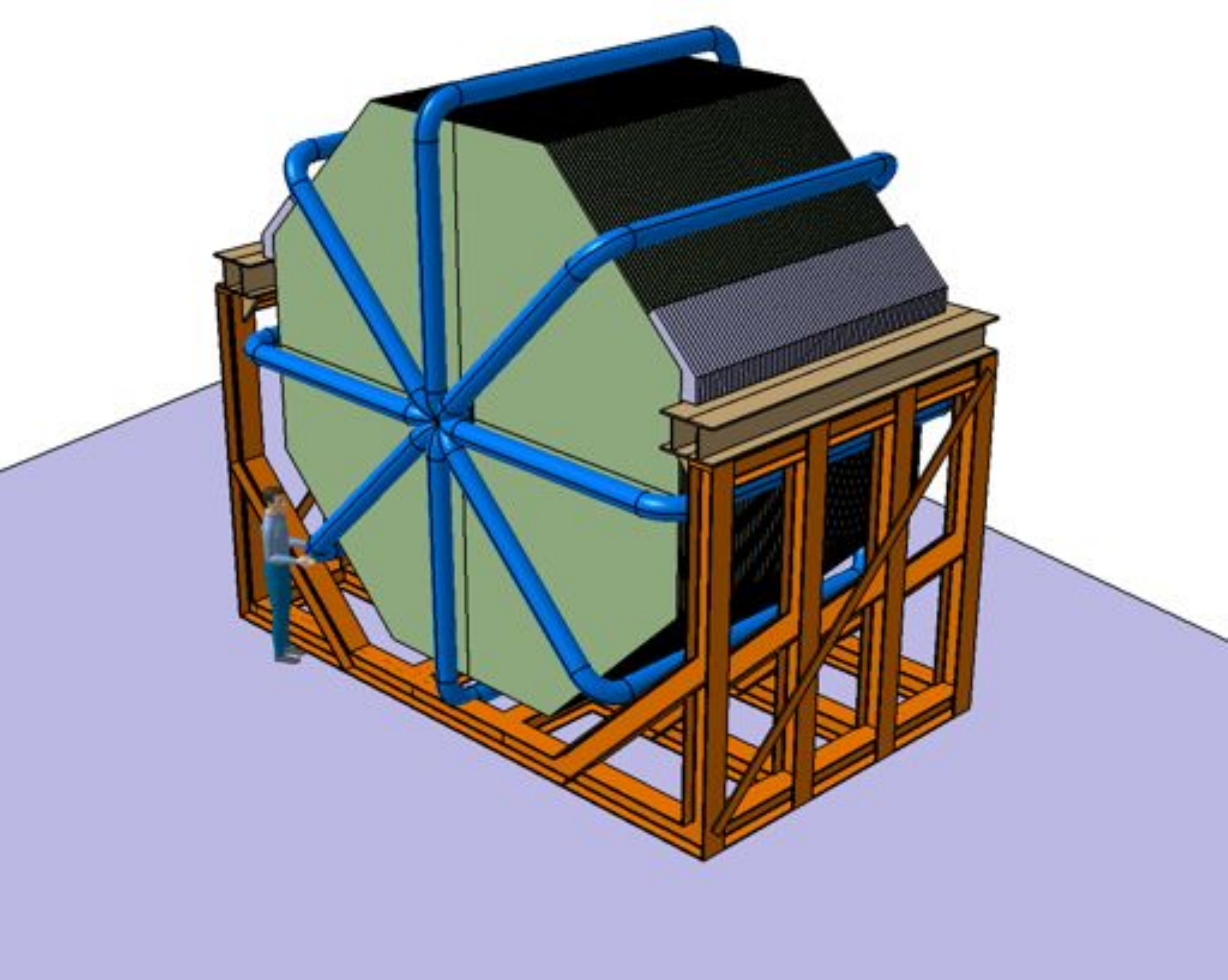}
\caption{The 8-coil configuration for the MIND500 with supporting cradle.}
\label{MIND500-coils}
\end{figure}

Each plate is fabricated from 2 half-octagons which are skip-welded together. The 2.5 m high by 5 m wide half-plates are transportable via standard means from the point of fabrication. The ears on the side would be welded and screwed onto one of the two half-octagons. They serve the purpose of supporting the load from the weight of the fully assembled plate, and will be used for alignment and to block the plates in position on the craddle.

\subsubsection{Material selection}
Selection of the iron is a compromise between adequate mechanical properties (tensile strength), and good magnetic properties (high magnetic permeability). Given the cost of pure iron such as ARMCO, we plan to select a more affordable construction steel with low carbon content such as the construction steel AISI1006. Our specifications are derived from previous experience with the MINOS and BaBar magnet steels. Both those experiments used AISI1006. Similar construction steels such as AISI1010 (DIN CK10, 1.1121) could also be considered.
An example specification is the American Society of Testing and Materials (ASTM)  "Standard Specification for General Requirements for Rolled Steel Plates, Shapes, Sheet Piling and Bars for structural use." ASTM Designation A6.

Table \ref{plate-dimensions} shows the flatness tolerances which are critical for this type of detector. The flatness is taken to be for steel plates lying on a perfectly flat horizontal surface such as a large marble slab. The ASTM specifications for flatness for these steel plate dimensions are a flatness of 18 mm. This is too large a value for the MIND steel plates, where we request a flatness of 1.0 mm.

Tolerances for plate thickness are taken from the standard ASTM Designation A6. The thickness is measured 10.0 to 20.0 mm from the longitudinal edge of the plates.
\begin{table}[h]
\centering
\caption{\em Dimensions and tolerances for the MIND steel plates.}
\begin{tabular}{cc}
\toprule
\textbf{Dimension} &	\textbf{Value [mm]}\\
\hline
{Plate thickness}	&	30.0\\		
{Thickness tolerance}	&	+1.5, -0.3\\
{Plate length}	&	5000.0\\		
{Length tolerance}	&	25.0\\
{Plate width}	&	2500.0\\		
{Width tolerance}	&	16.0\\
{Flatness tolerance over 5.0 m}	&	$\pm1.0$\\
\bottomrule
\end{tabular}
\label{plate-dimensions}
\end{table}

\subsubsection{Chemical and mechanical properties}
The plate manufacturer will be requested to demonstrate that analysis of the chemical composition can be made for each heat used in the production of all steel plates. The chemical composition of AISI1006 is given in Table \ref{chemical-composition} as an example. The most important element to control is Carbon, since it affects the magnetic properties. Other elements should be monitored in case reported values could lead to significant differences with respect to the mechanical properties.
\begin{table}[h]
\centering
\caption{\em Chemical composition (\% by weight) for the MIND steel plates (from MINOS TDR).}
\begin{tabular}{cc}
\toprule
\textbf{Property} &	\textbf{Specification}\\
\hline		
{Carbon}	&	0.04\% to 0.06\%\\
{Manganese}	&	0.40\% (Max.)\\
{Silicon}	&	0.40\% (Max.)\\
{Sulfur}	&	0.01\% (Max)\\
{Phosphorus}	&	0.07\% (Max)\\
{Nitrogen}	&	0.008\% (Max)\\
{Aluminium}	&	0.05\% (Max)\\
{Chromium}	&	0.05\% (Max)\\
{Copper}	&	0.06\% (Max)\\
{Nickel}	&	0.06\% (Max)\\
{Molybdenum}	&	0.01\% (Max)\\
{Vanadium}	&	0.01\% (Max)\\
{Niobium}	&	0.01\% (Max)\\	
\bottomrule
\end{tabular}
\label{chemical-composition}
\end{table}

The mechanical requirements in Table \ref{mechanical-requirements} are given as an indication. Full finite element analysis for the particular design of the MIND detector prototype could call for a revision of these numbers.
\begin{table}[h]
\centering
\caption{\em Mechanical specifications for tensile properties of the MIND steel plates (from MINOS TDR).}
\begin{tabular}{cc}
\toprule
\textbf{Property} &	\textbf{Specification}\\
\hline
{Ultimate tensile strength}	&	275 MPa minimum\\		
{Yield strength}	&	140 MPa minimum\\	
{Elongation of 5 cm}	&	22\% minimum\\
\bottomrule
\end{tabular}
\label{mechanical-requirements}
\end{table}

\subsection{Magnetization of MIND500}
\subsubsection{The Superconducting Transmission Line}
Concerning the magnetization of the MIND prototype, a low carbon steel will be selected. There are no particular radiation or environmental constraints (corrosion/humidity). The magnetization will be set by passing a current through one or more conductor coils. Specifications for the field are the following:
\begin{itemize}
\item field value: 1.5 T ± 20\%,
\item knowledge of field in volume of interest to a precision of $10^{-4}$, especially $B_x$ and $B_y$ components,
\item field uniformity within steel along projection of plastic scintillator volume: 10\%,
\item field value outside MIND volume: maximum = 10 mT.
\end{itemize}
The assumption made concerning power supplies is that one can be provided by CERN. The coil design will therefore be made as a function of available power supplies.
Initial studies were carried out to optimise the uniformity of the $B_y$ component of the field whilst minimising the $B_x$ component. Although this does not represent the current consensus on MIND-type detector design, this approach was meant to minimise uncertainties in the knowledge of the B-field. The resulting geometry led to a considerable height increase for the steel plates (factor $\times$2), in order to have a return path for the field lines well away from the detector plane area. The results of the optimisation of field lines is an impressive constraining of the $B_x$ component of the field, 115 Gauss (0.7\% of $B_y$) for the two coil configuration, compared to 8210 Gauss (110\% of $B_y$) for the one coil configuration. However, the doubling of the cost of steel and the introduction of large "empty" slots are disadvantages which drive the adoption of the more affordable and adequate one coil configuration in the case of the MIND50.

\begin{figure}[hbt]
\centering
\includegraphics*[width=0.7\textwidth]{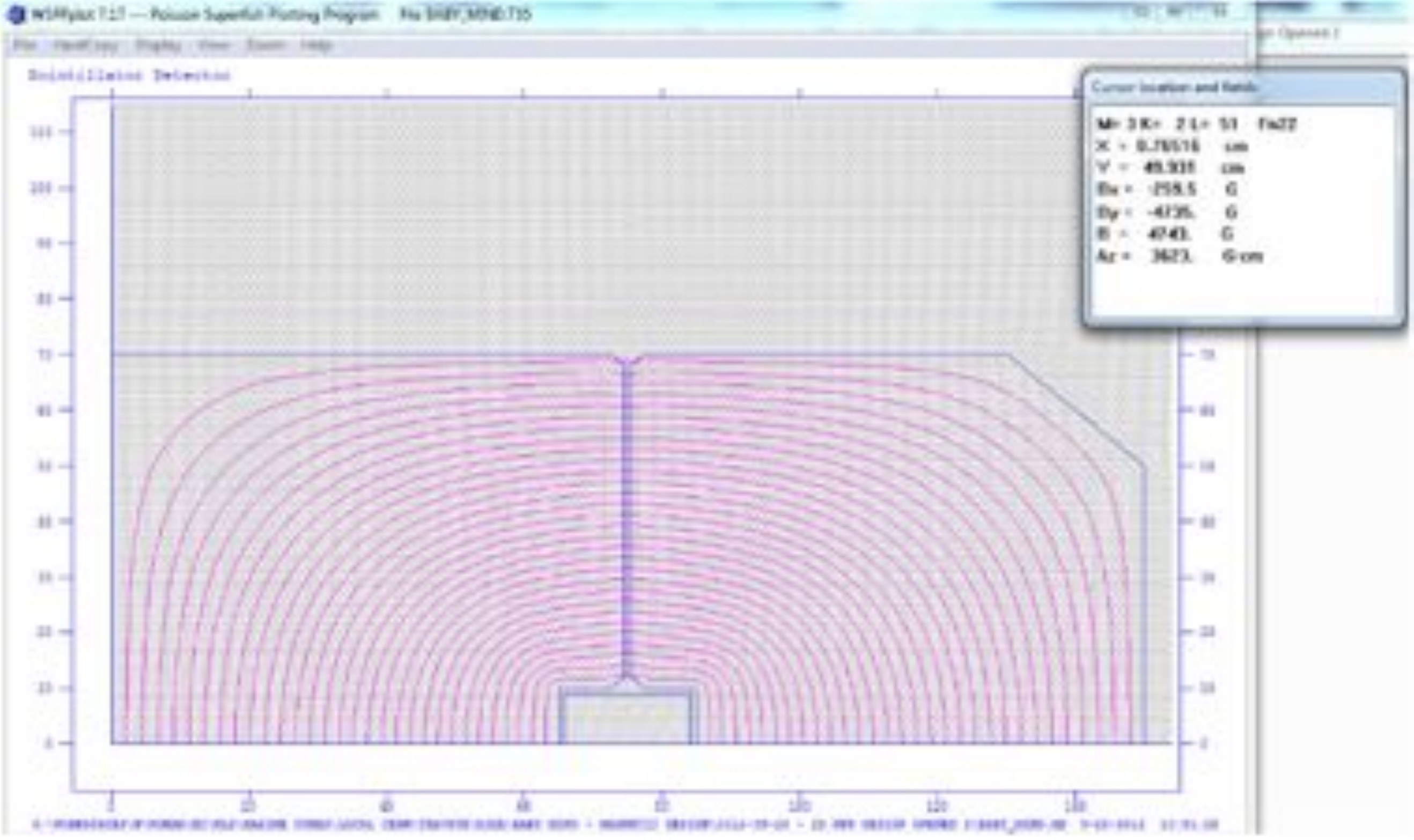}
\caption{Simulation of the magnetic field lines for the MIND50 prototype with a vertical gap for the insertion of a measuring device. This gap can be seen at $\sim$ 75 cm on the horizontal axis. One possible method of measuring the B field is to insert a probe embedded in a non-magnetisable material into this vertical gap. Field lines in the detector module area between 0 and +45cm on both vertical and horizontal axes can be inferred from measurements made with the probe along the gap length.}
\label{mindfieldgap}
\end{figure}

Having chosen the one-coil configuration, attention is now turning to the challenge of determining with accuracy the value of the field, not simply the total field but separate knowledge of the $B_x$ and $B_y$ components of the field. One solution that has been proposed and which is currently being investigated is to create a slot away from the detector planes from the coil to the outer edge of the steel plate, into which a non-magnetised material such as a stainless steel or aluminium is inserted with an embedded magnetic field sensor. By displacing this sensor along the entire length of the gap, it is possible to create a map of the field in the gap and thus infer the field lines in the area of the detector planes, Figure \ref{mindfieldgap}. This technique could have relevance for the much larger MIND-type detectors planned for future facilities.
Further optimisation work is required to produce a design for the MIND that includes detailed maps of the B- field, thorough mechanical design and integration of a B-field measurement system.

\begin{figure}[hbt]
\centering
\includegraphics*[width=0.70\textwidth]{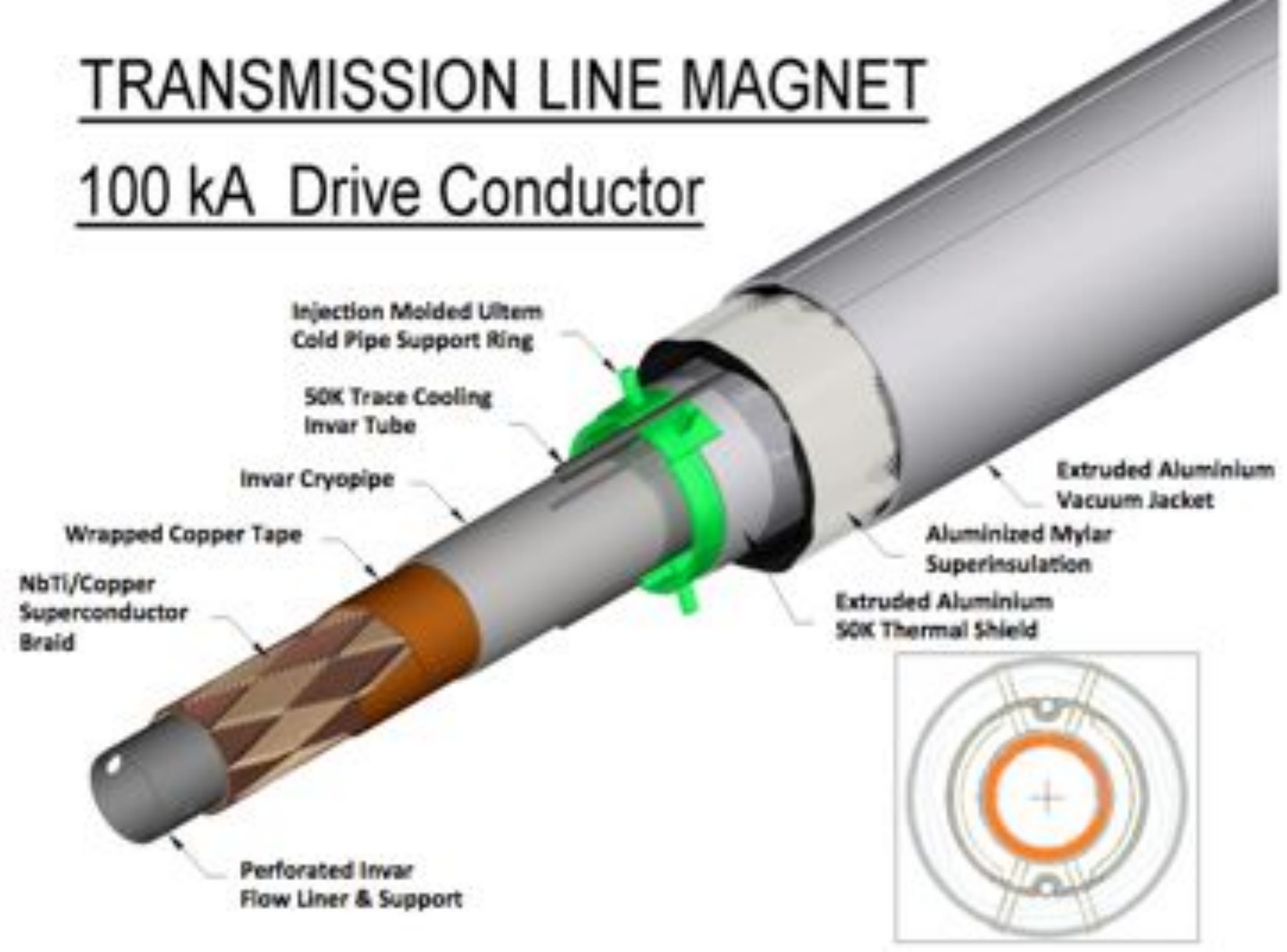}
\caption{The Superconducting Transmission Line (STL) developed for the VLHC.}
\label{fig:stl}
\end{figure}

The excitation of the MIND iron plates can be achieved with the Superconducting Transmission Line concept (STL) developed during the design study for a Staged Very Large Hadron Collider, see \Cref{fig:stl}. The excitation current required is likely to be higher than the 40 kA-turns of the MINOS near detector. This is definitely the case if the concept for in-situ measurement of the B-field is adopted, where a much higher excitation current is required to establish a homogeneous field through the gap in the iron plate.

The STL is based on a NbTi superconductor cooled by forced flow of LHe. This approach proposed by G. W. Foster was successfully tested for the VLHC, carrying up to 100 kA and generating up to 2 T in the double aperture dipole magnet of the VLHC \cite{Foster1018363,Foster795658}. The compactness of the line (80 mm OD) is cost effective since it eliminates the requirement for a large and expensive cryostat.  The total current required for the magnetization of the MIND will be split evenly between the 8 coils, keeping the maximum current in each coil at an acceptable level to reduce Lorentz forces on the superconductor.

\subsubsection{Dimensioning the STL for MIND500}
The nuSTORM proposal outlines in detail characteristics of the STL for SuperBIND, reported here \cite{Adey:2013pio}. The peak effective magnetic field in the iron is 2.5 T which drops to 1.87 T at a radius of 2.5 m. The peak field on the superconductor is 0.83 T on the coil inner radius. For SuperBIND, 30 kA of current in each of the 8 coils is required to generate those magnetic fields, with a total coil length of 320 m. The total coil length for the MIND500 is 104 m. The total current is likely just as high in the MIND500, though the total stored energy will scale with the coil length. Radial Lorentz forces on the inner and outer conductor for SuperBIND are -271 kN and 40 kN on inner and outer conductors respectively. The longitudinal force component is 2.4 kN. These loads are within the capability of the VLHC STL design where critical elements are the Ultem rings supporting the cold cable mass within the vacuum shell. A stainless steel slotted tube is required in the center to sustain the large radial forces directed to the center experienced by the inner coil. 

The inner coil assembly cold mass has a diameter of 120 mm and is cooled by forced helium flow through a 40 mm-diameter central hole. Eight straight NbTi bars are mounted with epoxy resin between the central tube and the outer aluminium tube. During cool down, the aluminium shell provides a pre-stress for the coil, which increases the superconducting coil mechanical stability. Aluminized Mylar super insulation is wrapped around the cold mass and the nitrogen shield. The cold mass is mounted inside a stainless steel vacuum vessel of 200 mm diameter and includes a nitrogen shield. 
\begin{figure}[hbt]
\centering
\includegraphics*[width=0.50\textwidth]{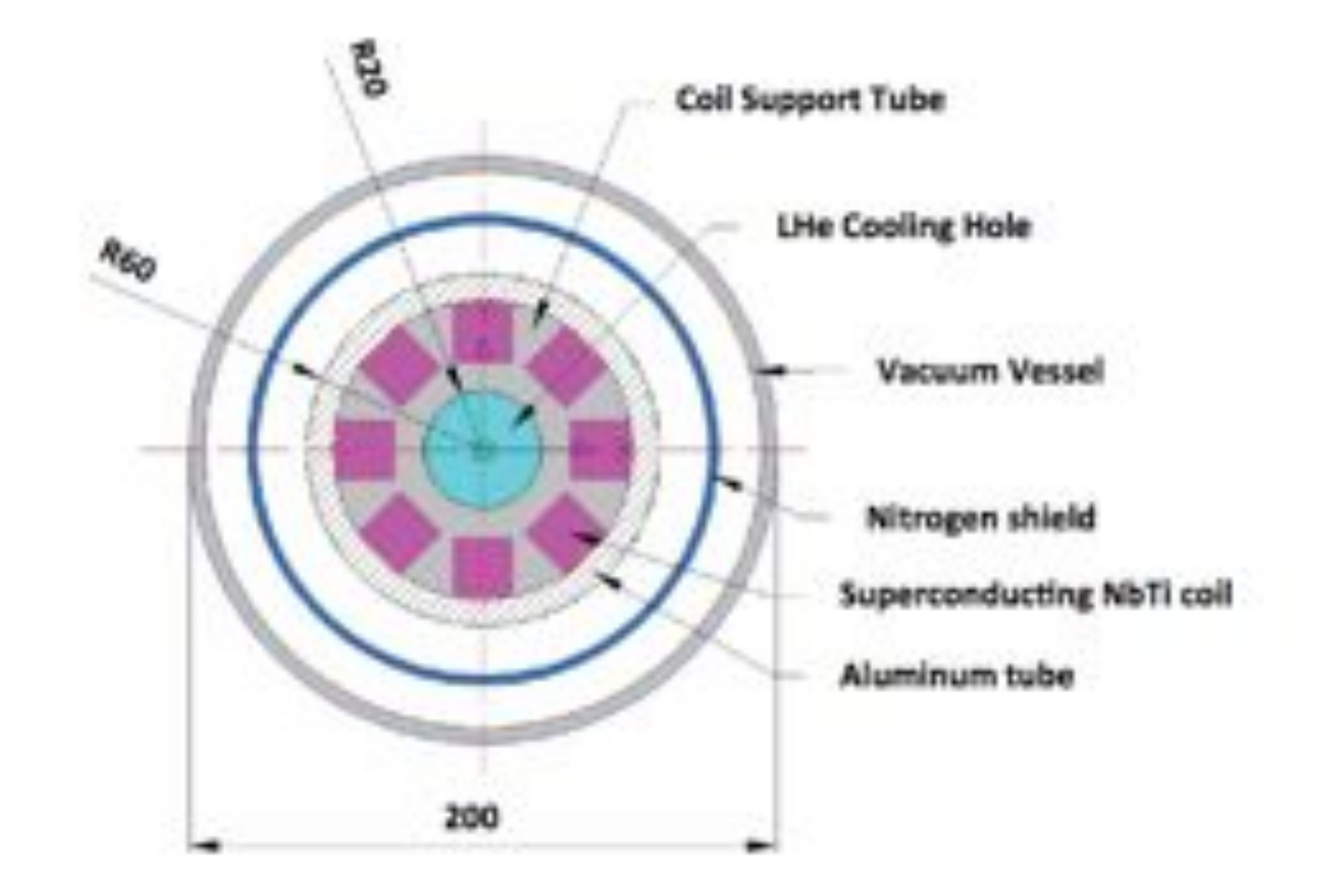}
\caption{Cross-section of the STL coil inner assembly.}
\label{fig:stlass}
\end{figure}

\subsubsection{The STL superconducting coil}
A couple of options are available for the assembly of the NbTi superconducting coil:
\begin{itemize}
\item{the SSC-type Rutherford cables used in the original STL \cite{Foster1018363,Foster795658}; }
\item{the ITER-type Cable in a Conduit Conductor (CICC) \cite{Devred6151053}.}
\end{itemize} 
The latter option is widely used for ITER correction coils, poloidal coils, buses and manifold cables. Production technology and facilities are developed and these cables have been tested in Europe, Russia and China. CCIC has direct cooling through the cooling hole for large currents, or cooling between cable strands. The transmission line is one of the cost drivers for large superconducting magnet systems. The STL for the VLHC was estimated at \$500/m including cost of vacuum and shield tubes, Ultem spacers and super-insulation. The ITER project evaluated correction coil costs at \$220/m. For nuSTORM, the STL VLHC estimate runs at \$160k for 320 m. Scaling for the MIND500 leads to an estimate of \$52k.

\clearpage
\section{Overall layout and space requirements}

\graphicspath{{./Section-Layout/figs/}}
\label{sec:genrequirements}

\subsection{General Requirements}

The Experimental Hall EHN1, being at
present largely occupied by other fixed target experiments, must be
extended to host the LAGUNA/LBNO prototypes. This choice allows not
perturbing the existing experimental program in EHN1. Services (i.e.
crane, electricity, water, etc{\dots}) could still be derived from the
existing area{, provided that the experiment
does not suffer or generates perturbations related to the operation of
other experiments or activities in the Hall.

The space requirements can be summarised as such:
\begin{itemize}
\item {{\textgreater}256
m}{\textsuperscript{2}}{
recessed floor space: }The  DLAr prototype is located in a
recessed floor region (a pit of ${\geq}$16m large, ${\geq}$16m long,
$\sim$7m deep) of the extension of the EHN1 experimental hall (EHN1-X).
\item {5m$\times$16m =
80m}{\textsuperscript{2}}{
clean assembly + control room}
\item {5m$\times$16m =
80m}{\textsuperscript{2}}{
unloading area}
\item 10m$\times$10m should be left open for installing the MIND.
\end{itemize}

In the baseline option, the detectors are located in a recessed floor region (${\geq}$16m
large, ${\geq}$16m long, $\sim$7m deep) of the EHN1 extension,
as illustrated in \Cref{fig:laguna_lardetector_illustration}.
 \begin{figure}[htb]
\begin{center}
\includegraphics[width=0.75\textwidth]{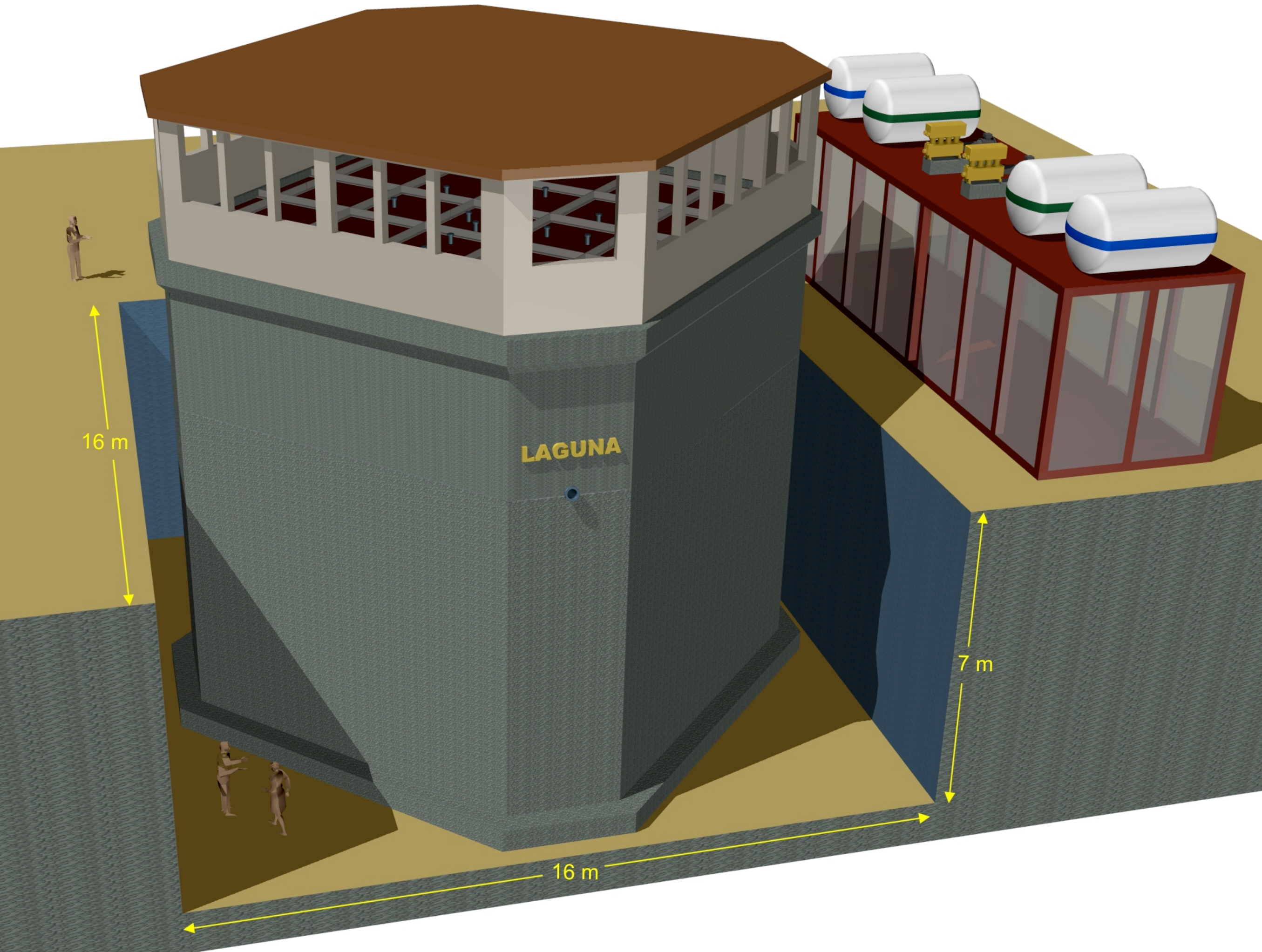} 
\caption{Illustration of the \six DLAr demonstrator in the trench.}
\label{fig:laguna_lardetector_illustration}
\end{center}
\end{figure}

The pit can house the detectors in such a way that they
can receive a beam of charged particles.
The geometry and tolerance for the direction of the incoming
charge particles are presented in \Cref{sec:testbeamrequirements}.
The floor of the pit should be designed
to support a weight of 800 tons.
The detector should be standing on
pillars whose number and size must be appropriately chosen. The
supporting pillars act as thermal insulation (to avoid freezing of the
floor) and as anti-seismic shock absorbers.

The requirements for the MIND detector will be defined at a later stage.

\subsection{Control room}
\label{sec:controlroom}
The control room hosts the computers
responsible for the data acquisition and the local storage.
The control room hosts several screens
to monitor the various functions and statuses of the experiment.
The control room is the location where
physicists take shifts (24 hours per day, 7 days per week during beam time).
The noise level should be, according to
safety rules, such that extended period of stays are possible by
staff.
The ventilation of the control room
should be dimensioned to dissipate up to 50 kW.

\subsection{Clean-room}
During assembly of the inner detector, the inner vessel shall
become a clean room and be connected to an external clean room.
The external clean room shall have a section with an external door
to bring in preassembled and sealed components. The external clean room
is used to open the shipped sealed packages and install individual
components inside the inner vessel by means of a 10T crane (see
\Cref{sec:cranesrequirement}).
The environment in these volumes should be a class
1:100{\textquoteright}000 (ISO Class 8).

\subsection{Access to EHN1-X area}
Equipment needed during the various phases of the installation and
assembly will be transported by truck, truck-containers or with
internal CERN vehicles (likely from the Meyrin site e.g. Blg 182).
Unloading from the trucks will be done with the cranes or with
forklifts.
A convenient and unobstructed road access and a truck-unloading
zone accessible from the cranes and with the forklifts should be
available.
During the initial construction phase of the vessel and the
installation of the cryogenic facilities (see \Cref{sec:larfillingemptying,sec:larboiloff,sec:larfiltration}), 
the 40T crane will be used.
During the inner detector construction within the clean room,
trucks will be unloaded with a forklift. The 10T crane to be located
within the clean tent will be used to insert elements into the vessel
from the top roof openings.

\subsection{Liquid Argon filling and emptying}
\label{sec:larfillingemptying}

A liquid argon receiving and transmitting station will be used.
The total liquid argon volume is $\sim$500 m\textsuperscript{3} or
approximately the equivalent amount transported by 25-30 trucks, with a
maximum rate of several trucks per day giving a filling or emptying
time of $\sim$10 days.
The delivered Liquid Argon shall be of
the Welding Grade with following maximum contaminants at the arrival
(contaminants at departure + contamination during loading and by truck
itself):
\begin{itemize}
\item {Purity\ \ \ \ \ \ {\textgreater}\ \ 
99.99\% molar }
\item {Oxygen\ \ \ \ {\textless}\ \  \ 4 ppm
molar}
\item {Nitrogen\ \ \ \ {\textless}\ \  \ 2 ppm
molar}
\item {Water\ \ \ \ \ \ {\textless} \ \  \ \ 1 \ ppm molar}
\item {Hydrocarbons\ \ \ \ {\textless}\ \ 
\ 0.5 ppm molar}
\end{itemize}
In order to reduce nuisance and
interference caused by the many trucks in the experimental hall, the
LAr receiving and transmitting zone should be located externally,
outside the experimental hall, and connected with the prototype
cryogenic facilities via appropriate vacuum-insulated cryogenic
transfer lines.
An immersed LAr pump is necessary to
empty the vessel (See Section \Cref{sec:larfiltration}).
Heaters will be located inside the
vessel to warm up the detector after emptying the liquid argon.
Alternatively purging warm argon gas inside the tank can be considered
to warm up the inner detector (See \Cref{sec:garpurgepur}).
In order to test the aging of the membrane weldings and the
potential creation of micro-leaks, which would deteriorate the liquid
argon purity even after complete outgassing of the inner detector
materials, the vessel will be purposely cycled by filling
and emptying several times.
For an efficient filling and emptying, a local external storage
with a capacity of 500~m$^3$ LAr should be available, using e.g. nr. 10
cryogenic standard tanks Linde LITS 2 of net capacity 58,500 lt /
each\footnote{
http://www.linde-engineering.com/en/plant\_components/cryogenic\_tanks/index.html}.

\subsection{Liquid argon boiloff recondensation}
\label{sec:larboiloff}

See \Cref{sec:larprocess}.
With the insulation thickness chosen, the average heat input is
\~{}5 W/m$^2$ and the total heat input (including also the roof input) is
about \~{}4 kW at LAr temperature. The corresponding boiloff rate is
\~{}2~ton~LAr/day (\~{}0.3\%/day).
A cryogenic plant must re-condense the boiloff rate in order to
keep the pressure and the level of liquid argon constant.
The absolute pressure of the argon remains constant {\textpm}1
mbar, and can differ {\textpm}75 mbar relative to the atmospheric
pressure.
In order to avoid contamination of the pure liquid argon, the
boiloff will be re-condensed via heat exchangers connected to a
nitrogen loop. The re-condensed argon will be purified (see next
section) before it is returned to the main vessel.
The LAr condensation can be made using
liquid nitrogen at 1.5 barg and -187.1$^\circ$C in a cross current heat
exchanger. Liquid nitrogen is introduced at bottom of exchanger and
exits on top in gaseous state. Gaseous argon goes at top of exchanger
and exits at bottom in liquid form. Gaseous
nitrogen is compressed and expanded in the thermodynamic cycle.
Preliminary calculations indicate a
thermodynamic efficiency of about 10\%, so an electrical power of 40kW
is needed to re-condense the boiloff argon.

\subsection{Gas phase argon purging and purification}
\label{sec:garpurgepur}

See \Cref{sec:larprocess}.
Gas purging in warm phase has been shown to be a very effective
method to remove air from the vessel and to reduce the impurities to
\textit{ppm} level~\cite{Curioni:2010gd}.
A forced gas recirculation and filtration is implemented to remove
molecules (mostly water) outgassed from materials at warm temperature,
before the detector is cooled down.
Flushing gaseous argon at a temperature of
$\sim$40-50$^\circ$C improves the outgassing, by warming up the
surfaces of the detector components.

\subsection{Liquid Argon filtration}
\label{sec:larfiltration}

See \Cref{sec:larprocess}.
The free-electron lifetime in liquid
must be more than 10~ms in order to drift over 6~m without significant
charge amplitude degradation.
The impurities in the liquid argon
should be less than 0.030 ppb = 30 ppt O$_2$ equivalent.
A filtration system removes the
impurities from the commercially delivered bulk argon (see Section
\Cref{sec:larfillingemptying}) to the
required level. Filtration is done by flowing liquid argon through
custom-made purification cartridges, made of sections of molecular
sieves followed by oxygen-reduced copper oxide powder, such as e.g.
those successfully operated in CERN Blg182 in the context of ArDM (CERN
RE18).
The total volume of liquid argon in the
vessel is continuously extracted by an immersed pump, and forced to
flow through purification cartridges to filtrate residual impurities
and those arising from outgassing materials inside the vessel. The flow
of recirculation is at least 20~m$^3$ LAr/hr.
The heat input into the LAr caused by
the liquid argon filtration is about 4 kW.

\subsection{Cranes}
\label{sec:cranesrequirement}
A 40T crane to be used during the initial phases of construction
(e.g. tank assembly) and to install the cryogenic and liquid argon
filtration plants.
A dedicated 10T crane is located within the clean room tent (see
\Cref{sec:controlroom}) and is used during the phase of installation of
the inner detector.

\subsection{Ventilation requirements}
The temperature and humidity in the hall are controlled by the
ventilation and air conditioning system.
An exhaust pipe from the recessed floor region is necessary.
Dimension and location will be specified in agreement with the safety
review.
An exchange of the air volume in the recessed pit in which the
LBNO LAr prototype is mandatory in order to mitigate the risk of
accumulation of argon at the bottom of the recessed pit. The flow is to
be defined in agreement with safety regulations.
An increased air flow (two- or three-fold) which should be
activated in case of emergency, might be required by safety
regulations.

\subsection{Cryorefrigeration requirements}
The cryorefrigerator has a total of 10 kW cooling power at the LAr
temperature.
The operation at the full capacity dissipates about 100-150 kW.
A totally redundant system (i.e. 20 kW) might be required to
satisfy safety regulations.
External buffers of LN2 ($\sim$30~m$^3$
LN2/each) are used as cold source to ensure boil-off re-condensation in
case of major event on the cryorefrigerator system. The total storage
volume of the external buffers will be defined after the safety review.
An area of 10x10~m$^2$ must be foreseen to install the
cryo-refrigerators. 

\subsection{HVAC requirements}
The total heat dissipation is dominated by the cryo-refrigerators.
The heat generated by the cryo-compressors is evacuated via
water-cooling (see \Cref{sec:coolingwater}).
The heat dissipated by the rest of the
equipment (electronics, power supplies, pumps, control systems,
computers, {\dots}) can be removed by air-cooling via an appropriate
air conditioning system.
The temperature and humidity of the hall
should be controlled and stable. The dew point in the hall should be
$\sim$10$^\circ$~C.

\subsection{Cooling water requirements}
\label{sec:coolingwater}
The heat generated by the cryo-compressors is evacuated via
water-cooling, using {a flow of fresh
demineralised water is foreseen with standard CERN pressure (4-6 bar)
and modest flow ($\sim$1m$^3$/h).
The temperature of the inflow water is
in the range
6\~{}10}{\textsuperscript{o}}{C.}

\subsection{Electrical requirements}
The area should be equipped with an arrival and a switchboard with
protected lines dedicated. In the first phase, the electrical power
consumptions are:
\begin{itemize}
\item 200 kW 400V/3P for the cryocoolers
\item 100 kW 400V/3P for the cryogenic system, pumps, etc.
\item 100 kW 400V/3P, 220V/1P for the detector electronics, power
supplies, slow control, computers, etc.
\item power for the magnetisation of MIND (tbd).
\end{itemize}
The available electric power might be upgraded 
to accommodate the MIND magnetised detector.

\subsection{Additional laboratory space at CERN}
The chain for the assembly, Q\&A and
test area to assemble the 36 m$^2$ for the charge readout components
requires $30\times20$~m$^2$ laboratory space. 
The amount of space and the already
existing infrastructure used by CERN RE18 (ArDM experiment) is adequate
and can be in large part reused/upgraded.

\subsection{Data storage and computing requirements}
Gigabit Ethernet connections to the CERN
backbone are expected. The needed storage of raw data in CASTOR of the
order of 100TB per year of running.


\clearpage
\section{Test beam requirements}
\label{sec:testbeamrequirements}


In order to characterise the response of the detectors in terms of tracking, electromagnetic and hadronic calorimetry, and
the physics of secondary interactions the detectors will be exposed to charged particle beams. 
The charged particles momentum should be
selectable and cover the range 0.5-20~GeV/c with a bite of $\Delta p/p\approx5\%$.
The beam composition should be pions,
muons, electrons,{\dots} and sign selected.
Its composition is being studied with beam line simulations that takes
into account the geometry of the various magnet elements.
The rate of particles should be in the range 200-1000 Hz.
Preliminary calculations assuming primary pion beam in the 60-80~GeV/c range
show that this is feasible down to the desired momenta below 1~GeV/c.
At 1~GeV/c the pion-to-muon ratio is estimated to be about 10:1. Below 1 GeV/c
the fraction of muons increases due to upstream decays. Above 1~GeV/c, the fraction
of muons decreases and reaches about 100:1 at 10~GeV/c.
External instrumentation should include
counters and chambers and early trigger signals.

Several configuration will be considered: (1) DLAr only, (2) MIND only, and (3) DLAr+MIND where
the MIND500 measures particles exiting the DLAr.
One of the more significant elements to be checked in the simulations will be the digitisation, 
which requires hardware efficiencies that are measured online. 

A close collaboration between detector  and beam group is required 
because a number of parameters such as particle rate, beam size and angle of incidence on the 
detector affects the detector design, e.g. readout electronics buffers.
For the DLAr, the incoming direction should be at
45{\textpm}15{\textsuperscript{o}}
w.r.t. the two perpendicular readout views orientations of the TPC, and
vertically point to the center of the active volume
{\textpm}15\textsuperscript{o}. A beam window is foreseen in order
not to spoil the momentum of the incoming particles and to minimise
interactions upstream of the active volume. 
The top view of the present layout is shown in \Cref{fig:EHN1_top_view},
and the side view in \Cref{fig:EHN1_side_view}. Because of the
recessed floor, the H2 line will be bent towards in order to hit 
the center of the DLAr detector.

\begin{figure}[htb]
\begin{center}
\includegraphics[width=0.85\textwidth]{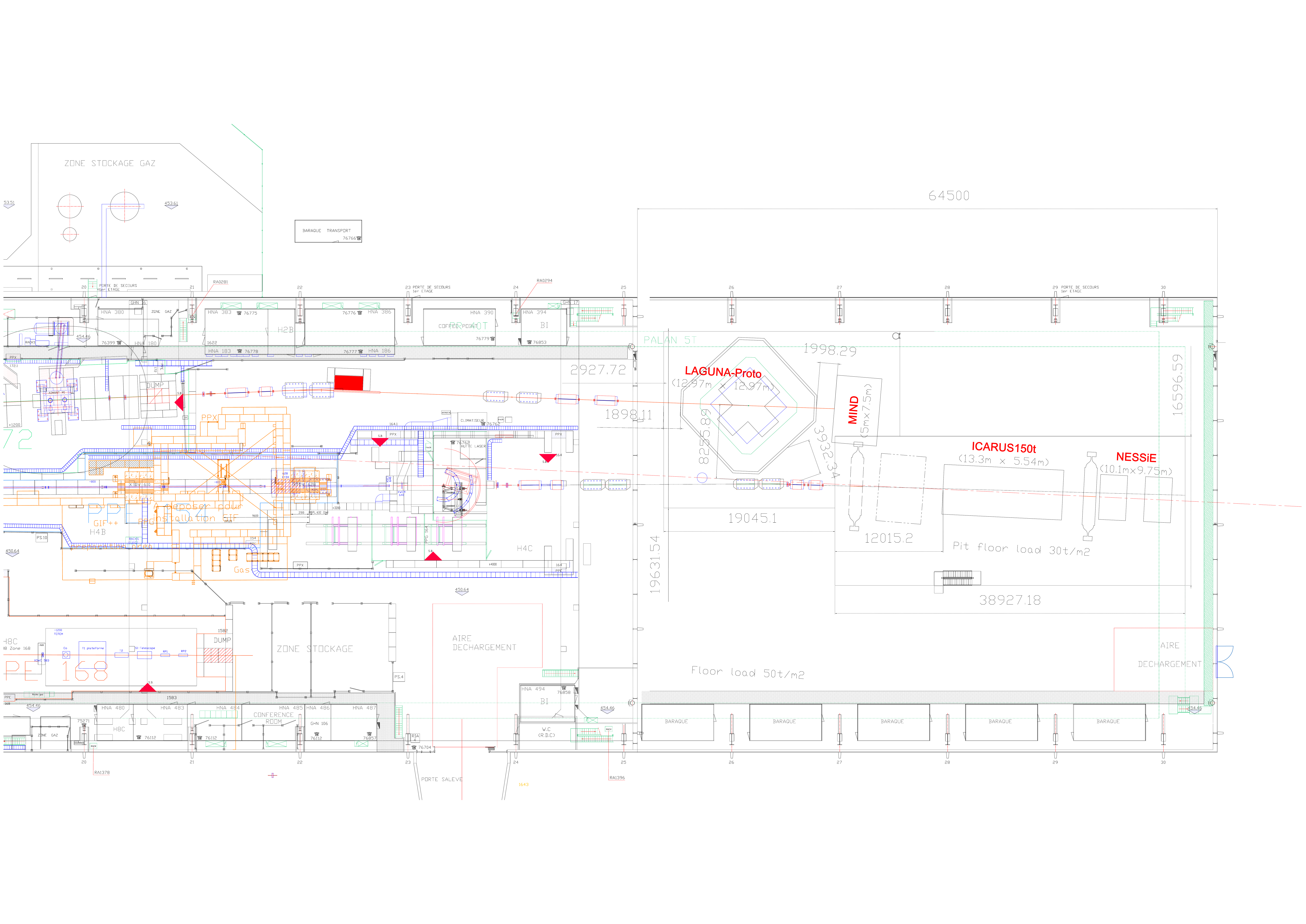} 
\caption{ Top view of the layout of LBNO-DEMO (LAGUNA-Proto and MIND) in the EHN1 building and its extension.}
\label{fig:EHN1_top_view}
\end{center}
\end{figure}

\begin{figure}[htb]
\begin{center}
\includegraphics[width=0.85\textwidth]{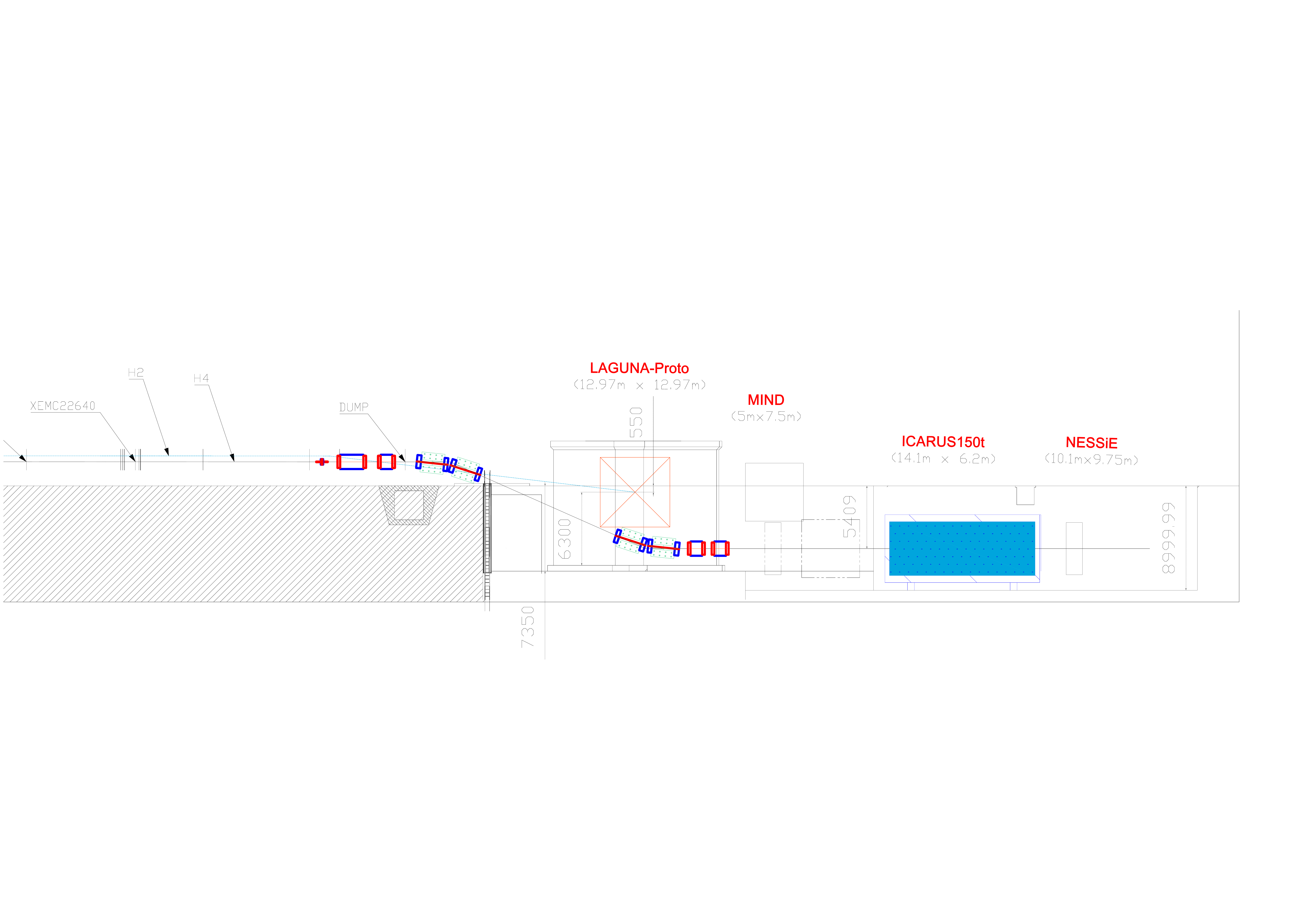} 
\caption{ Side view of the layout of the LBNO-DEMO (LAGUNA-Proto and MIND) in the EHN1 building and its extension.}
\label{fig:EHN1_side_view}
\end{center}
\end{figure}

Particle rates presented in Table \ref{particlerates} are given as an indication, in order to have a rough idea of the estimated beam time for sufficient statistics. The rate assumed is in the range of 100~Hz -- 1 kHz of particles per spill.  With a sample of $2\times 10^6$ particles for each momentum bin and configuration,
same overall test beam time online, estimated to be 6 weeks if we make the assumption of 175 $\times$ $10^6$ triggers in total 
at 100~Hz with 50\% running efficiency.

\begin{table}[h]
\centering
\caption{\em Requirements for particles and their momenta. The particle rate here is the rate within a spill, regardless of the spill length, slow extraction is assumed.}
\begin{tabular}{cccccccc}
\toprule
\textbf{Type} &	\textbf{Momentum [GeV/c]}  & 	\textbf{Rate [kHz]}  &	\textbf{Total} & \textbf{Time est. [hrs]}\\
\hline
\multicolumn{5}{l}{Muon tracks}\\
\hline			
$\mu$$^{+/-}$&	0.8, 1.0, 1.5, 2.0, 5.0, 10.0, 20.0 &	0.1	&	$5\times 10^6$$\times$14 &	200 \\
\hline
\multicolumn{5}{l}{Shower reconstruction}\\
\hline	
$\pi$$^{+/-}$	&	0.5, 0.7, 1.0, 2.0, 5.0, 10.0, 20.0	 &	0.1	&	$5\times 10^6$$\times$14 &	200 \\		
$e$	&	0.5, 0.7, 1.0, 2.0, 5.0, 10., 20.0	 &	0.1	&	$5\times 10^6$$\times$7 &	100 \\			
\bottomrule
\end{tabular}
\label{particlerates}
\end{table}


\clearpage
\section{Organization, cost estimate, schedule and risk assessment}
\subsection{Organisation}
The tentative organisation of the Collaboration (see  \Cref{fig:organigramm}) is based on typical schemes commonly successfully adopted in
high-energy experiments. It will be finalised and formally voted in the coming months.

The Collaboration Board (CB) is the top-level decision-making and arbitration body. 
It has one representative with voting rights from institution involved in the project. 
It will review the progress of the project at the annual meetings, and, where necessary, 
will decide on changes in the work plan and budget allocation for the next period. 

The Executive Board (EB) assures the day-to-day follow-up of the program and it is formed by the spokesperson, 
the deputy-spokesperson, the technical and scientific board leaders plus the administration responsible members. It will be responsible for the co-ordination and harmonization of all WA105 actions and particularly for the administrative and co-operative support of all transnational research activities. It will follow up all important horizontal issues and will prepare the IB meetings of the Collaboration. It will also be responsible for the official contact to CERN and its various committees, for public relation issues and for the contents of the WA105 websites. It will meet at least every two months, and decisions will be taken on a unanimity basis. On exceptional cases differences may be resolved by qualified majority rule (2/3 of the members) 
or can be directed to an exceptional CB meeting.

The Technical Board (TB), coordinated by the Technical Coordinator, manages the technical follow-up of the project. 

The Scientific board (SB), coordinated by the Scientific Chair, manages the scientific follow-up of the project.

The  Finance Review Committee will be composed of representatives from the various national and international funding agencies and research entities and will be setup to provide recommendations to the Project and guidance to the IB from the point of view of the Funding Agencies. The FRC shall usually meet every year. It will make recommendations on the use of resources and financing for the various implementation phases of the project. 

\begin{figure}[htb]
\begin{center}
\includegraphics[width=0.85\textwidth]{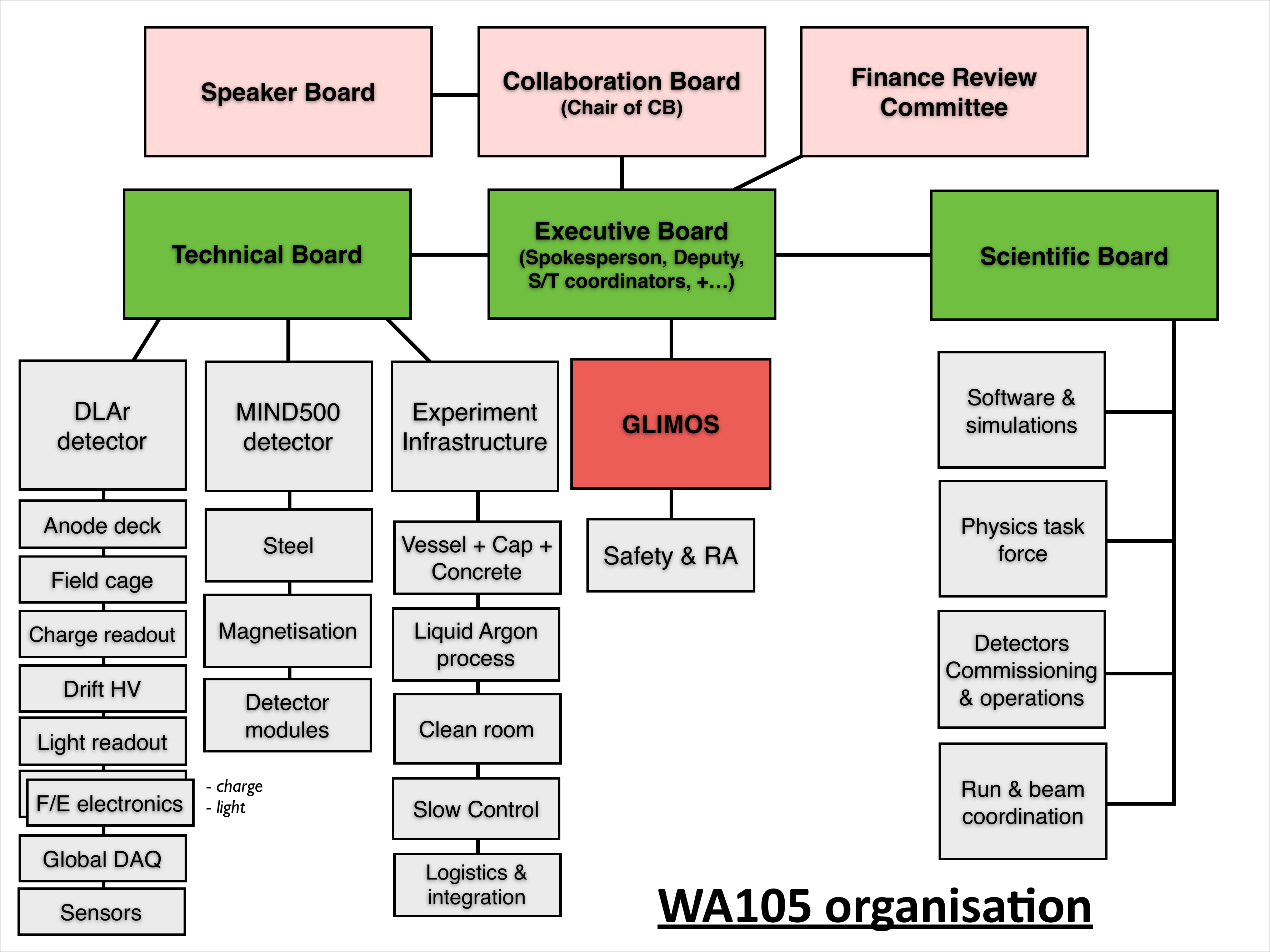} 
\caption{Tentative organisation of the Collaboration.}
\label{fig:organigramm}
\end{center}
\end{figure}

The definition of the Work breakdown structure (WBS) down to level 3 with the contributing institutes is shown in 
\Cref{tab:wpdef}.

\begin{table}[htdp]
\centering
\begin{tabular}{|llc|}
\hline
\hline
& Work package & Contributing institutes \\
\hline
WBS 1 & Management & -- \\
\hline
WBS 2 & Technical Board &  \\
\hline
\,\,\,WBS 2.1 & DLAr detector &  \\
\,\,\,\,\,\,WBS 2.1.1 & Anode deck &  LAPP \\
\,\,\,\,\,\,WBS 2.1.2 & Field cage &  CEA, ETHZ \\
\,\,\,\,\,\,WBS 2.1.3 & Charge readout &  CEA, ETHZ, UCL \\
\,\,\,\,\,\,WBS 2.1.4 & Drift HV &  LPNHE, ETHZ \\
\,\,\,\,\,\,WBS 2.1.4 & Light readout &  APC, Barcelona, CIEMAT, KEK \\
\,\,\,\,\,\,WBS 2.1.5 & F/E readout (charge) &  IPNL, ETHZ \\
\,\,\,\,\,\,WBS 2.1.6 & F/E readout (light) &  APC, LAPP, KEK \\
\,\,\,\,\,\,WBS 2.1.7 & Global DAQ and trigger &  UCL, IPNL, APC, KEK, Jyv\"askyl\"a \\
\,\,\,\,\,\,WBS 2.1.8 & Sensors &  KEK, ETHZ, Oulu, IFIN-HH \\
\hline
\,\,\,WBS 2.2 & MIND500 detector &  \\
\,\,\,\,\,\,WBS 2.2.1 & Steel &  INR, UniGE \\
\,\,\,\,\,\,WBS 2.2.2 & Magnetisation &  Glasgow, INR, UniGE \\
\,\,\,\,\,\,WBS 2.2.3 & Detector modules & Glasgow, INR, Sofia, UniGE \\
\hline
\,\,\,WBS 2.3 & Experiment infrastructure &  \\
\,\,\,\,\,\,WBS 2.3.1 & Vessel + cap + concrete & CERN, ETHZ \\
\,\,\,\,\,\,WBS 2.3.2 & Liquid Argon process &  CERN, ETHZ \\
\,\,\,\,\,\,WBS 2.3.3 & Clean room &  CERN, ETHZ \\
\,\,\,\,\,\,WBS 2.3.4 & Slow control (DCS) &  CERN, ETHZ \\
\,\,\,\,\,\,WBS 2.3.5 & Logistics and integration &  CERN, ETHZ \\
\hline
WBS 3 & Safety \& GLIMOS &   \\
\hline
WBS 4 & Scientific Board &  \\
\,\,\,WBS 4.1 & Software \& simulations &  all \\
\,\,\,WBS 4.2 & Physics task force &  all \\
\,\,\,WBS 4.3 & Detector commissioning \& operations & all  \\
\,\,\,WBS 4.4 & Run \& beam coordination &  all \\
\hline
\hline
\end{tabular}
\caption{Definition of the Work breakdown structure (WBS) down to level 3.}
\label{tab:wpdef}
\end{table}%
%
%

\subsection{Cost estimate}
The total estimated cost of the LAGUNA $6\times 6\times 6$~m$^3$ LAr demonstrator
is presented and detailed in Table~\ref{tab:lagoodetailed}. The figures based
on preliminary discussion with industrial partners and on extensive experience
of the groups on smaller scale prototypes are assumed
to have an error at the level of 20-30\%. 

\begin{table}[htdp]
\caption{Total estimated cost of the  DLAr demonstrator}
\centering
\small
\begin{tabular}{|l|l|c|c|}
\hline
\hline
Nr. & Item &  Cost (kCHF) & Remark \\
\hline
\hline
1 & Anode charge readout 				& 800 & \\
2 & Hanging anode frame + movement		& 100 & \\
3 & Drift cage							& 500 & \\
4 & Light readout						& 300 & \\
5 & Power supplies						& 300 & \\
6 & Drift field supply						& 300 & \\
7 & F/E electronics						& 350 & \\
8 & DAQ								& 300 & \\
9 & Supporting infrastructure 				& 100 & \\
10 & Cryogenic vessel					& 2300 & (market survey)\\
11 & Cryogenic plant						& 1400 & (market survey)\\
12 & Detector DCS						& 360 & \\
13 & Clean room						& 430 & \\
14 & Logistics							& 450 & \\
\hline
& {\bf Total} & {\bf 7'990} & preliminary \\
\hline
\end{tabular}
\label{tab:lagoodetailed}
\end{table}%

\begin{table}[h]
\caption{Preliminary cost estimate of the MIND500 detector}
\centering
\small
\begin{tabular}{|l|l|c|c|}
\hline
\hline
Nr. & Item &  Cost (kCHF) & Remark \\
\hline
\hline
1 & Steel  					& 800 & \\
2 & Mechanics + coils  		& 200 & \\
3 & Power supply 			& 300 & \\
\hline
4 & Plastic scintillator 		& 500 & \\
5 & WLS fiber				& 400 & \\
6 & Optical connectors 		& 137 & \\
7 & Photosensors 			& 480 & \\
8 & Mechanics 				& 100 & \\
9 & Electronics + DAQ 		& 720 & \\
\hline
& {\bf Total} & {\bf 3'637} & preliminary \\
\hline
\end{tabular}
\label{tab:mindcostdetailed}
\end{table}%


\subsection{Schedule and milestones}
Pending a prompt approval of the activities presented in this document,
the schedule and milestones are summarised below:
\begin{itemize}
\item Beneficial occupancy EHN1: September 2015
\item Vessel constructed: March 2016
\item Inner-detector constructed: Jan 2017
\item Detector start commissioning: Mar 2017
\item Beginning test-beam data-taking: Spring 2017
\end{itemize}
The construction of the infrastructure and vessel for the DLAr must be done on site.
At this stage, it therefore appears that the readiness of the EHN1 extension is on the critical path.
Once the vessel is constructed, the inner detector will be assembled inside, using a side
opening defined as the TCO (Temporary Construction Opening). The assembly sequence
will follow as closely as possible the one developed with industrial support
for the LAGUNA/LBNO 20 and 50 kton detectors.

\subsection{Risk assessment}

A set of measures will be implemented to
ensure H\&S of people, protect equipment, and to mitigate risks
associated to the storage and process of large amounts of cryogenic
liquids. Additional equipment might be necessary as the results of
detailed risk analyses and CERN specific safety regulations.
Fire hazards are expected to be
predominantly located in specific areas of the facility such as (1) the
electronic racks located on top of the detector and (2) the high
current cryogenic facilities. Measures to limit fire propagation from
those areas might be required.
The cryogenic infrastructure should be
separated and protected from the regions of potential fire hazard.
CERN is expected to contribute to the infrastructure resulting
from the CERN safety rules (see e.g. CERN
Safety Instruction IS 47 - The use of
cryogenics fluids, \url{http://edms.cern.ch/file/335812/LAST_RELEASED/IS47_E.pdf}).
CERN provides and installs remote
cabinets, which are connected to appropriately located ODH and fire
detectors, and which the CERN fire brigade remotely monitors. The
building extension is equipped with fire detectors and evacuation
alarms.
CERN takes the responsibility for the
measures needed to mitigate risk interference between the LBNO
prototypes and potential other experiments or equipment to be installed
in the EHN1 extension hall.
The present design concept might be affected by so-far unforeseen
and additional requirements set by safety regulations and identified by
safety reviews. CERN will take possible measures to implement these
additional requirements.


\section{Conclusions}

For over a decade (some) members of this collaboration 
have worked towards the realisation of a giant underground detector 
for next-generation  long-baseline experiments, neutrino astrophysics and proton decay.
Such experiments require concrete and large-scale prototyping for their far detectors -- best realised on surface
in laboratories with extensive expertise and competence such as CERN -- as well as an
accompanying campaign of measurements in precisely-defined charged particle beams
aimed at assessing the tracking and calorimetric response of the detectors, the sophisticated
energy-flow and other reconstruction algorithms, and the phenomenology of secondary hadronic interactions. 
All these parameters and their
systematic errors
will  crucially affect the sensitivity of future long-baseline physics programmes, in particular
for what concerns their ability to discover CP-violation in the leptonic sector.

In this context, following our submission of an Expression of Interest for a long-baseline experiment (LBNO),
and the SPS Committee having reviewed positively the technology choices, it was requested to submit
a technical proposal describing the R\&D programme to be led at CERN.
We have defined as priority the construction and operation of a \six (active volume) double-phase liquid
argon demonstrator (DLAr) and in parallel the development of the technologies necessary for large
magnetised MIND detectors.

Successful operation of these detectors will provide very important feedback and will represent in
many areas milestones that have so far never been achieved. The detectors concepts symbolise
concrete steps towards the extrapolation of the technologies to very large masses, with the goal to 
reach tens of kilotons in underground environments. The proven technical choices deployed for the demonstrators will be
directly scalable and their components mass-produceable in view of a potential large-scale implementation.

The programme and the results that will be obtained will be quite general and be useful for 
several neutrino programmes being presently contemplated by the community. 
Specifically, the detectors described in this document and their
successful operation could, if performed in a timely fashion, have an important impact on the
design of LBNE and be the seed for a significant European contribution to the US project.
Similarly, the technologies could be considered in the context of the near detectors of the
Japanese HyperKamiokande project. 

Last but not least, the deployment and successful operation of
these detectors at CERN could naturally be revisited in the context of a short
baseline neutrino programme, based either on the CENF conventional neutrino beam,
the NUSTORM beam from a muon storage ring, or on the SHIP beam dump for which
prompt neutrinos would need to be studied. The goal of these programmes (in particular
the first two) would be to complement the long-baseline efforts, with measurements
aimed at improving significantly our knowledge on neutrino and antineutrino cross-sections 
and reducing their systematic errors to the required level for discovery of CP-violation in
the leptonic sector.

Hence, in the coming years, we intend to focus on the realisation and operation of the
demonstrators, whose technical designs were presented in this document. We will reconsider
potential future steps after that milestone has been reached.

We look forward to a prompt support of the SPSC towards ensuring a timely implementation of the North 
Area facility and the exposure to charged particle beams before the second LHC long shutdown (LS2).


\section*{Acknowledgements}
We acknowledge the support from the FP7 Research Infrastructure ÒDesign StudiesÓ LAGUNA
(Grant Agreement No. 212343 FP7-INFRA-2007-1),
LAGUNA-LBNO (Grant Agreement No. 284518 FP7-INFRA-2011-1),
and project AIDA (Grant Agreement No. 262025).

We are also grateful to the CERN Management for their encouragements,
recognising the importance of CERN  in the coherent definition
of a potential future program for the European neutrino community.

\appendix

\clearpage
\section{Appendix: Event rates with neutrino beams}
\label{sec:appendixA}
The proposed prototype will collect a large number of neutrino
interactions if a neutrino beam will be built in the North Area.

These interactions will be used to measure neutrino cross section on
Argon in the GeV region. The lack
of a precise knowledge of the neutrino-nucleus interactions at this
energy region is one of
the main sources of systematic uncertainties in long baseline
neutrino oscillation experiments. The measurements of $\nu_{\mu}$ cross section still have
large uncertainties while no measurements exist
of $\nu_e$ cross section
at this energy. The situation is even worth for antineutrinos, where cross-sections
presently have uncertainties above 10-20\%. This is an issue for
experiments relying  on the ratio of neutrinos
to antineutrinos to measure the CP violation phase.

LBNO presently assumed systematic errors for its MH and CPV sensitivity studies
at the level of 5\% for the signal~\cite{Stahl:2012exa}. The predicted sensitivities of LBNO could be improved
if one could show that these systematic errors could be reduced~\cite{agarwalla:2013kaa}.
Other future experiments, looking for CP violation in the leptonic sector,
aim for systematic uncertainties of the order of (1-2)\% and this will
only be possible if precise measurements of electron and muon flavours
cross sections for both neutrinos and antineutrinos will be available.

In this appendix we have computed the expected event rates in the
LBNO prototype for two different beams: a conventional neutrino beam
(CENF)~\cite{cenf} in which mainly $\nu_{\mu}$ are produced by the decay of pions
and kaons with a small intrinsic component of $\nu_e$ and the NuSTORM
beam~\cite{Kyberd:2012iz} in which $\bar{\nu}_e$ and $\nu_{\mu}$ ($\nu_e$
and $\bar{\nu}_{\mu}$) are produced from
the decay of $\mu^-$ ($\mu^+$). The event rates are obtained with the
GENIE neutrino simulation program~\cite{Andreopoulos:2009rq} in which we simulated
the proposed prototype and the two different beam configurations.

\subsection{CENF conventional beam}
For the conventional neutrino beam we used the CENF fluxes that are
produced by the SPS accelerating protons to 100 GeV/c~\cite{cenfflux2014}. To avoid large
pile-up the beam is produced off-axis with respect to the
detector. The center of the prototype is at a radial distance from the
axis of the beam of $\sim9$~m.

The detector is rotated by $45$ degrees with respect to the beam
direction to maximize the sharing on the two readout views of the
ionization produced by particles coming from neutrino interactions.

The CENF neutrino fluxes are available for $\nu_{\mu}$ and $\nu_{e}$ for
the positive and negative horn polarities and for
different radial bins $R_T$ with $R_T$ going from 0 to 25 meters. An
example of the fluxes with the positive horn polarity is shown in
\Cref{fig:cenf_flux}.

\begin{figure} [h]
\includegraphics[width=7.cm]{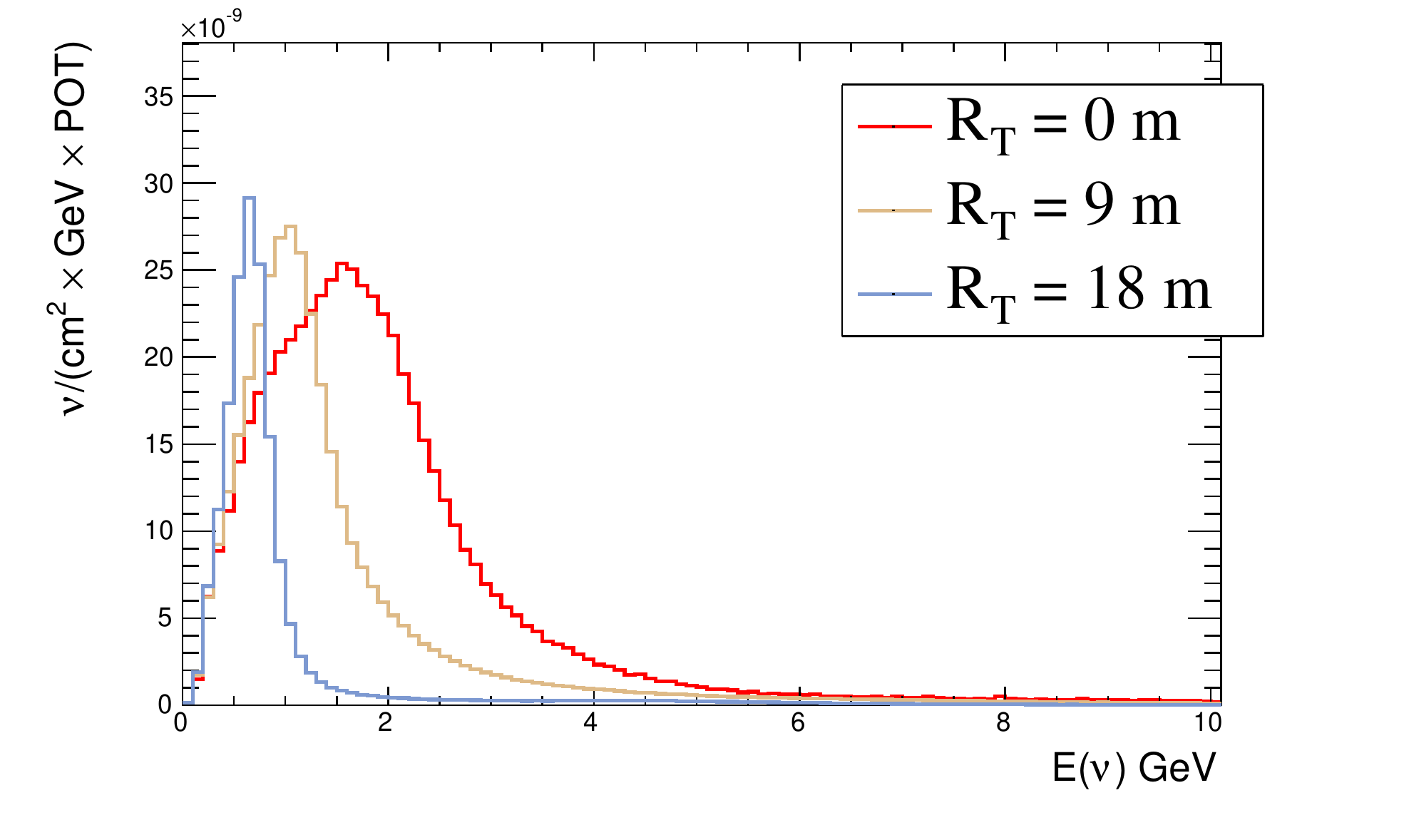}
\includegraphics[width=7.cm]{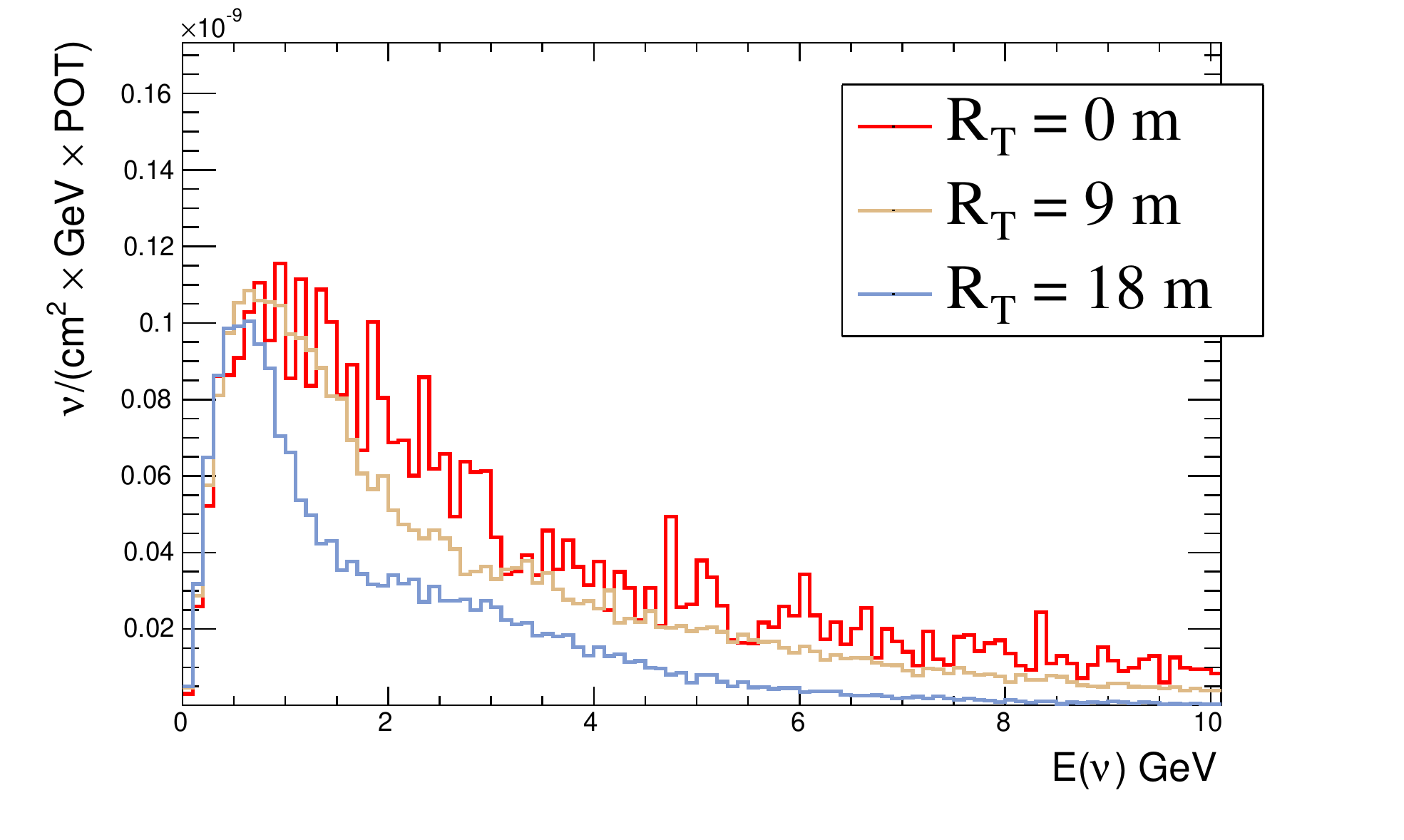}
\caption{\label{fig:cenf_flux}Expected CENF fluxes for $\nu_{\mu}$ (left) and
$\nu_e$ (right) for different off-axis distances.}
\end{figure}

The neutrino fluxes are used to obtain the event rates with the GENIE neutrino
event generator. The geometry of the prototype is simulated
and the neutrino cross section for the different materials
that compose the detector are computed according to the predefined
models included in GENIE.

We computed the event rate for an exposure of $10^{19}$ POT. We
only counted neutrino interactions occurring inside the active volume of
the prototype corresponding to 300 tons of Liquid Argon.

The number of neutrino interactions inside the active volume is shown in
\Cref{tab:cenfnu} for $\nu_{\mu}$ and
$\nu_e$. The energy spectrum of the interacting neutrinos is shown in
\Cref{fig:cenfeventrate}.

\begin{table}
 \begin{tabular}{|c|c|c|c|c|c|c|}
\hline
Int. & $N_{exp}$ & Fraction & Mean $E(\nu_{\mu})$ &
$N_{exp}$ & Fraction & Mean $E(\nu_e)$ \\ 
Type & $\nu_{\mu}$ & (\%) & (GeV) & $\nu_{e}$ & (\%) & (GeV) \\ \hline
 Inclusive & 559168 & 100 & 2.44 & 9734 & 100 & 3.86\\\hline 
CC & 411004 & 73.5 & 2.47 & 7312 & 75.1 & 3.87\\\hline 
 CCQE & 235786 & 42.2 & 1.37 & 2448 & 25.1 & 2.44\\\hline 
\end{tabular}
 \caption{\label{tab:cenfnu} Expected $\nu_{\mu}$ and $\nu_e$ event rates and
  mean neutrino energy for different neutrino interaction types, in
  the Liquid Argon active volume, for an exposure of $10^{19}$ POT and
  for $E(\nu)<10$~GeV.}
 \end{table}

\begin{figure}
\includegraphics[width=7.cm]{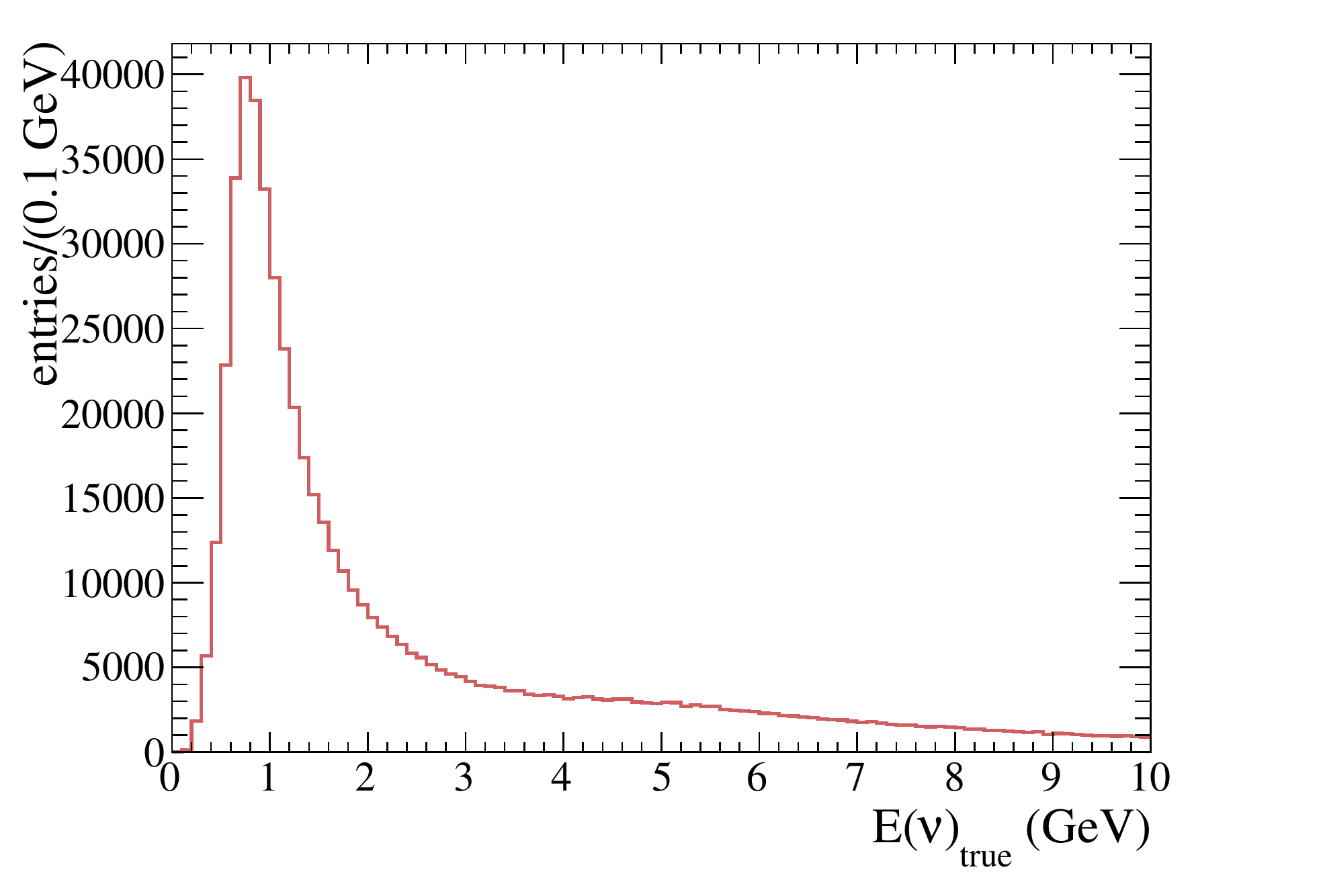}
\includegraphics[width=7.cm]{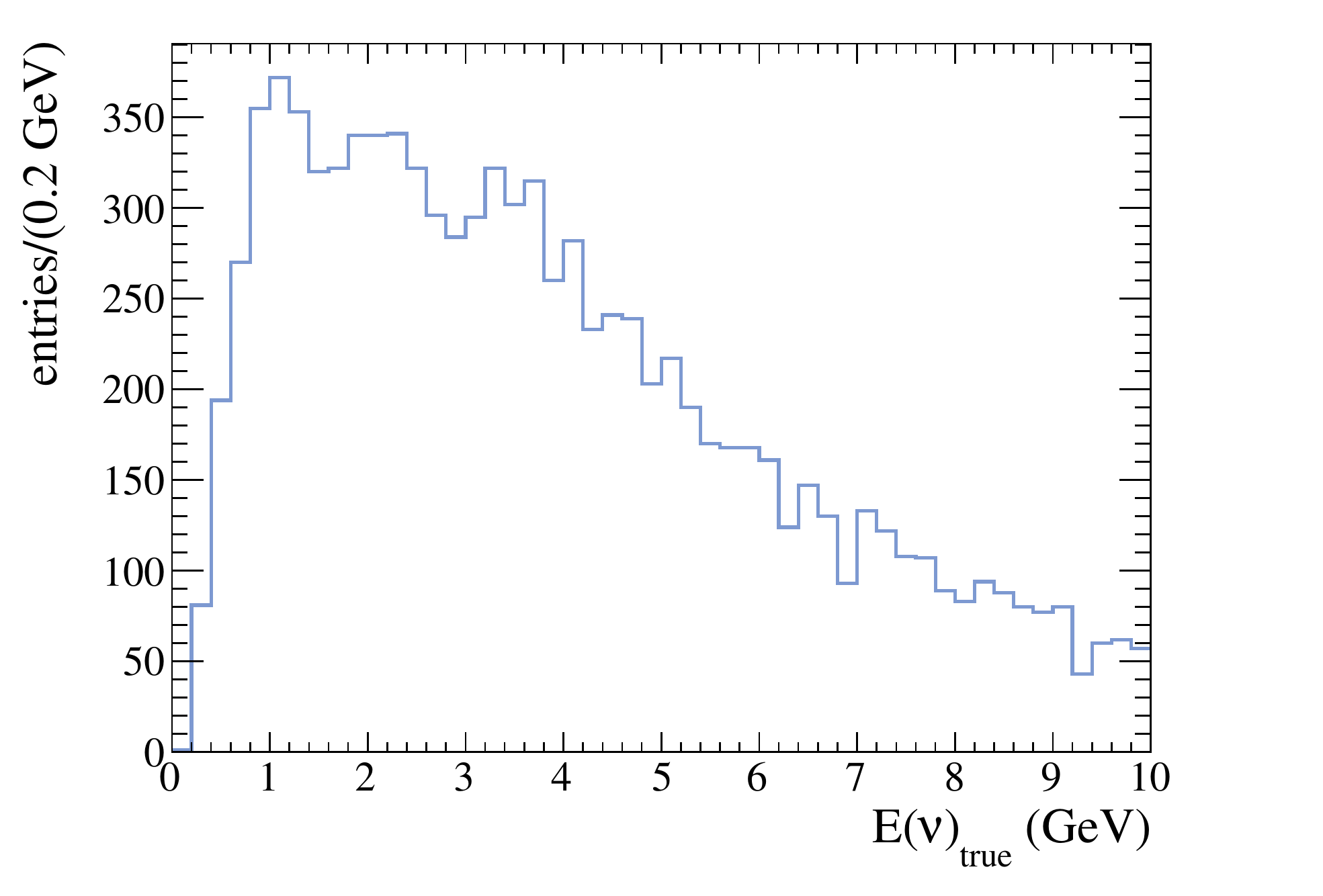}
\caption{\label{fig:cenfeventrate}True neutrino energy for the inclusive
sample of neutrino interactions occurring inside the prototype active
volume for $10^{19}$ POT for $\nu_{\mu}$ (left) and $\nu_e$ (right) interactions.}
\end{figure}

This large data sample will allow to perform precise measurements
of $\nu_{\mu}$ cross sections on Argon in different
exclusive channels. As far as the $\nu_e$ cross section are concerned,
the expected fraction of $\nu_e$ interactions inside the active volume
is 1.7\% but the large mass of the prototype will allow to select a
sizable data sample to measure also $\nu_e$ cross section.

For this off-axis configuration the event pile-up is
expected to be small: by assuming that in each spill there will be
$10^{13}$ protons, we expect $\sim0.6$ neutrino interactions per spill
in the active volume.

We have also performed the simulation in the case of horns' opposite
polarity in which mainly $\bar{\nu}_{\mu}$ and $\bar{\nu}_e$ are
produced. The expected event rates for $\bar{\nu}_{\mu}$ and
$\bar{\nu}_e$ in the active volume of the prototype for an exposure of
$10^{19}$ POT is shown in \Cref{tab:cenfnupos}.

\begin{table}
 \begin{tabular}{|c|c|c|c|c|c|c|}
\hline
Int. & $N_{exp}$ & Fraction & Mean $E(\bar{\nu}_{\mu})$ &
$N_{exp}$ & Fraction & Mean $E(\bar{\nu}_e)$ \\ 
Type & $\bar{\nu}_{\mu}$ & (\%) & (GeV) & $\bar{\nu}_e$ & (\%) & (GeV) \\ \hline
 Inclusive & 167837 & 100 & 2.47 & 2999 & 100 & 3.96 \\\hline 
 CC & 108831 & 64.8 & 2.59 & 2059 & 68.7 & 4.04 \\\hline 
 CCQE & 74339 & 44.3 & 1.59 & 887 & 29.6 & 2.79 \\\hline 
\end{tabular}
 \caption{\label{tab:cenfnupos} Expected $\bar{\nu}_{\mu}$ and $\bar{\nu}_e$ event rates and
  mean neutrino energy for different neutrino interaction types, in
  the Liquid Argon active volume, for an exposure of $10^{19}$ POT and
  for $E(\nu)<10$~GeV.}
 \end{table}

\subsection{Muon storage ring - NuSTORM}
We have also computed the expected event rates using neutrinos
produced by the decays of stored muons (NuSTORM proposal).
In this case we have assumed muon decays in a straight line of 226~m
with the end point of the straight line at a distance of 55~m from the
center of the detector. The neutrino fluxes have been simulated using a software
developed by the NuSTORM collaboration.

We have simulated $10^{16}$ muon decays that, assuming a 1/1000 ratio
between the number of useful muons and the number of protons would roughly correspond
to an exposure of $10^{19}$ POT (the same exposure used for the event
rates with the CENF). The spectrum of $\nu_{\mu}$ and $\nu_e$ produced by
muons decays in NuSTORM and arriving to a disk with a 3~m radius at a distance of 50~m
from the end of the straight line is shown in \Cref{fig:nustorm_flux}.

\begin{figure} [h]
\includegraphics[width=7.cm]{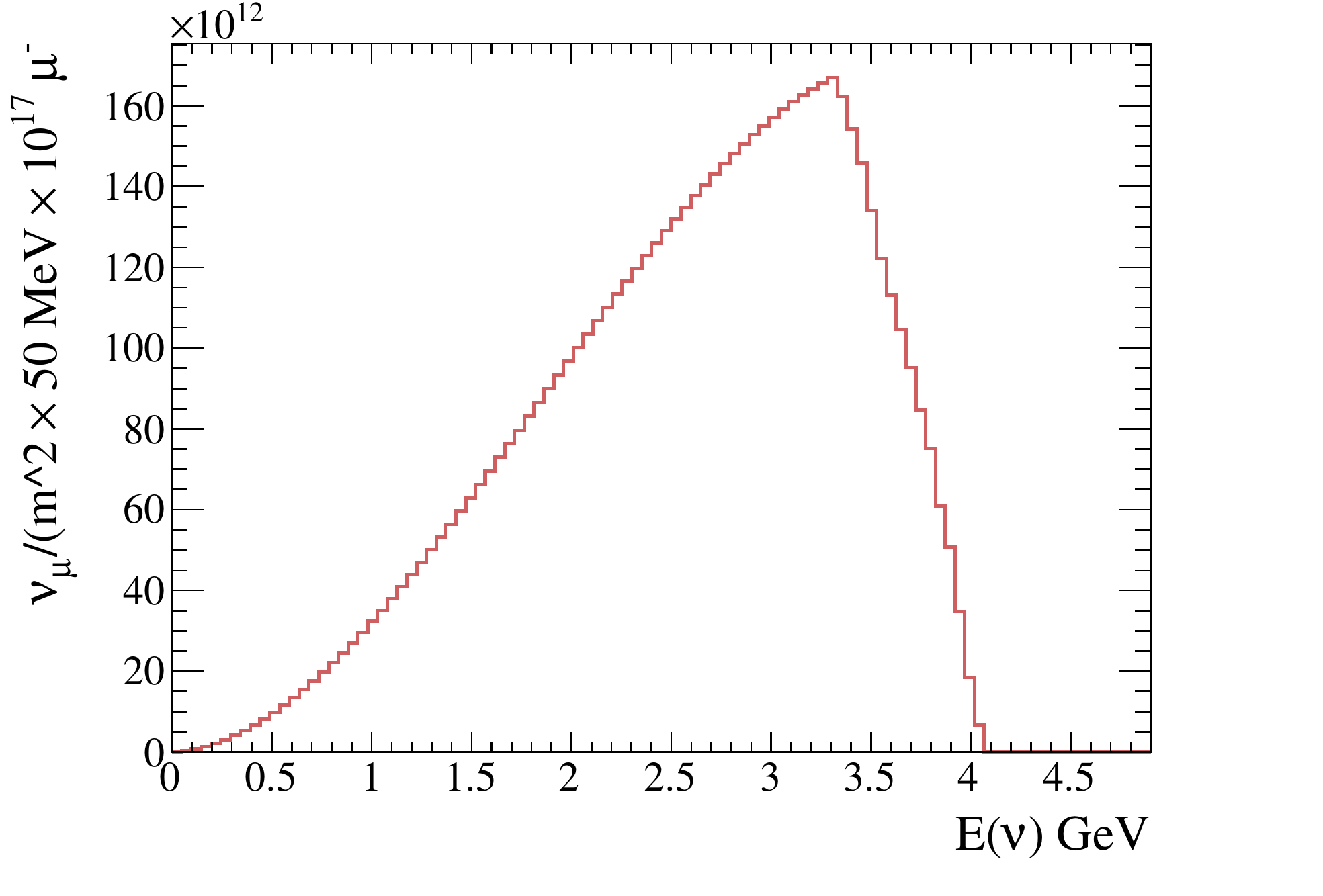}
\includegraphics[width=7.cm]{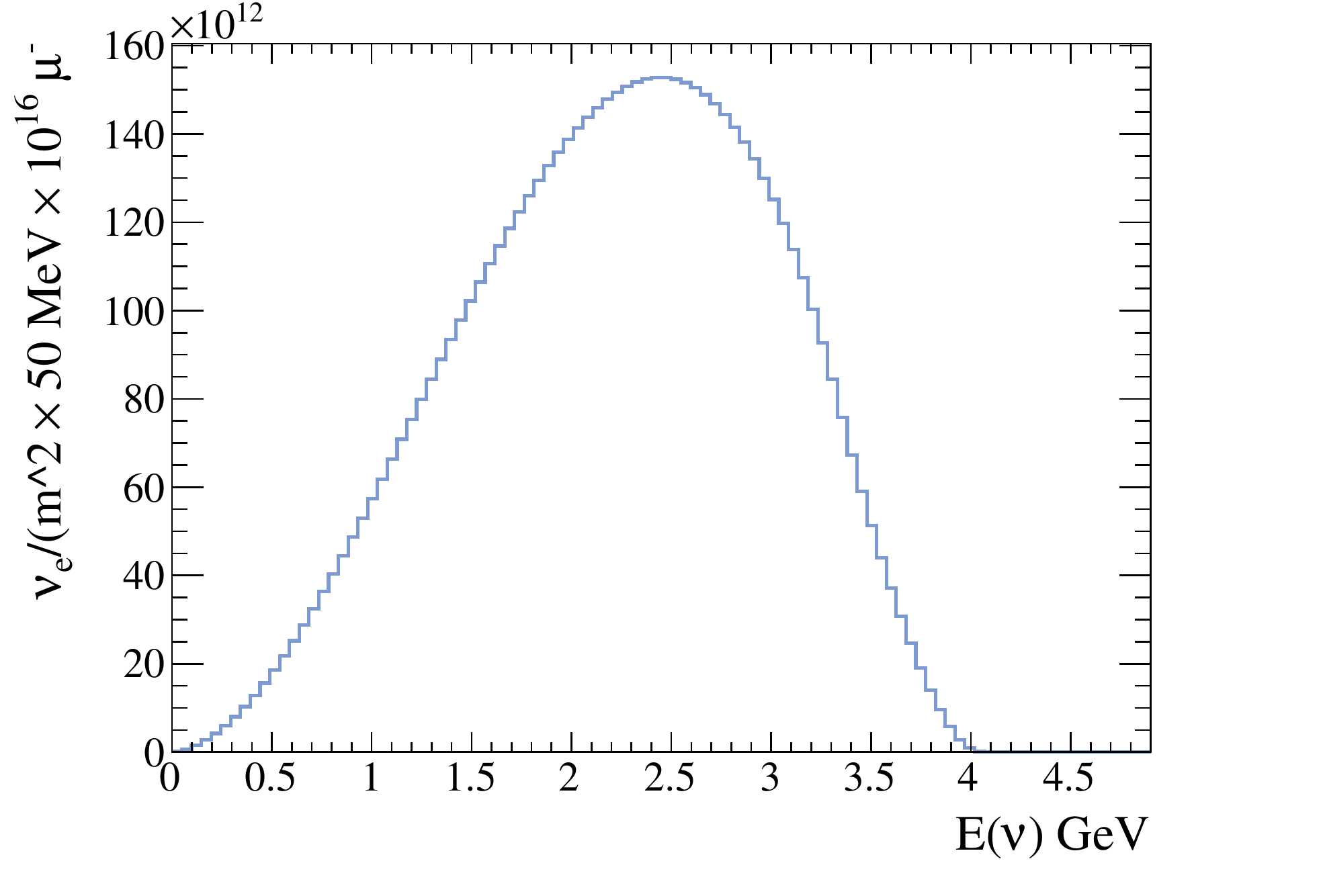}
\caption{\label{fig:nustorm_flux}Expected NuSTORM fluxes for $\nu_{\mu}$ (left) and
$\bar{\nu}_e$ (right) ($\bar{\nu}_{\mu}$ and $\nu_e$) produced in
$\mu^-$ ($\mu^+$) decays.}
\end{figure}

The expected number of events in the active volume of the prototype for
the neutrino fluxes of \Cref{fig:nustorm_flux} are shown in
\Cref{tab:nustormmum,tab:nustormmup} for the $\mu^-$
and $\mu^+$ decays respectively. The energy spectra of the
interacting neutrinos for the
inclusive samples are shown in \Cref{fig:nustormeventmum,fig:nustormeventmup}.

\begin{table}[h]
 \begin{tabular}{|c|c|c|c|c|c|c|}
\hline
Int. & $N_{exp}$ & Fraction & Mean $E(\nu_{\mu})$ &
$N_{exp}$ & Fraction & Mean $E(\bar{\nu}_e)$ \\ 
Type & $\nu_{\mu}$ & (\%) & (GeV) & $\bar{\nu}_{e}$ & (\%) & (GeV) \\ \hline
 Inclusive & 97251 & 100 & 2.82 & 34794 & 100 & 2.51\\\hline 
 CC & 72213 & 74.2 & 2.82 & 23645 & 67.9 & 2.52\\\hline 
 CCQE & 23284 & 24.0 & 2.60 & 11123 & 32.0 & 2.36\\\hline 
\end{tabular}
 \caption{\label{tab:nustormmum} Expected $\nu_{\mu}$ and $\bar{\nu}_e$ event rates and
  mean neutrino energy for different neutrino interaction types and
  for $10^{16}$ $\mu^-$ decays.}
 \end{table}

\begin{figure} [h]
\includegraphics[width=7.cm]{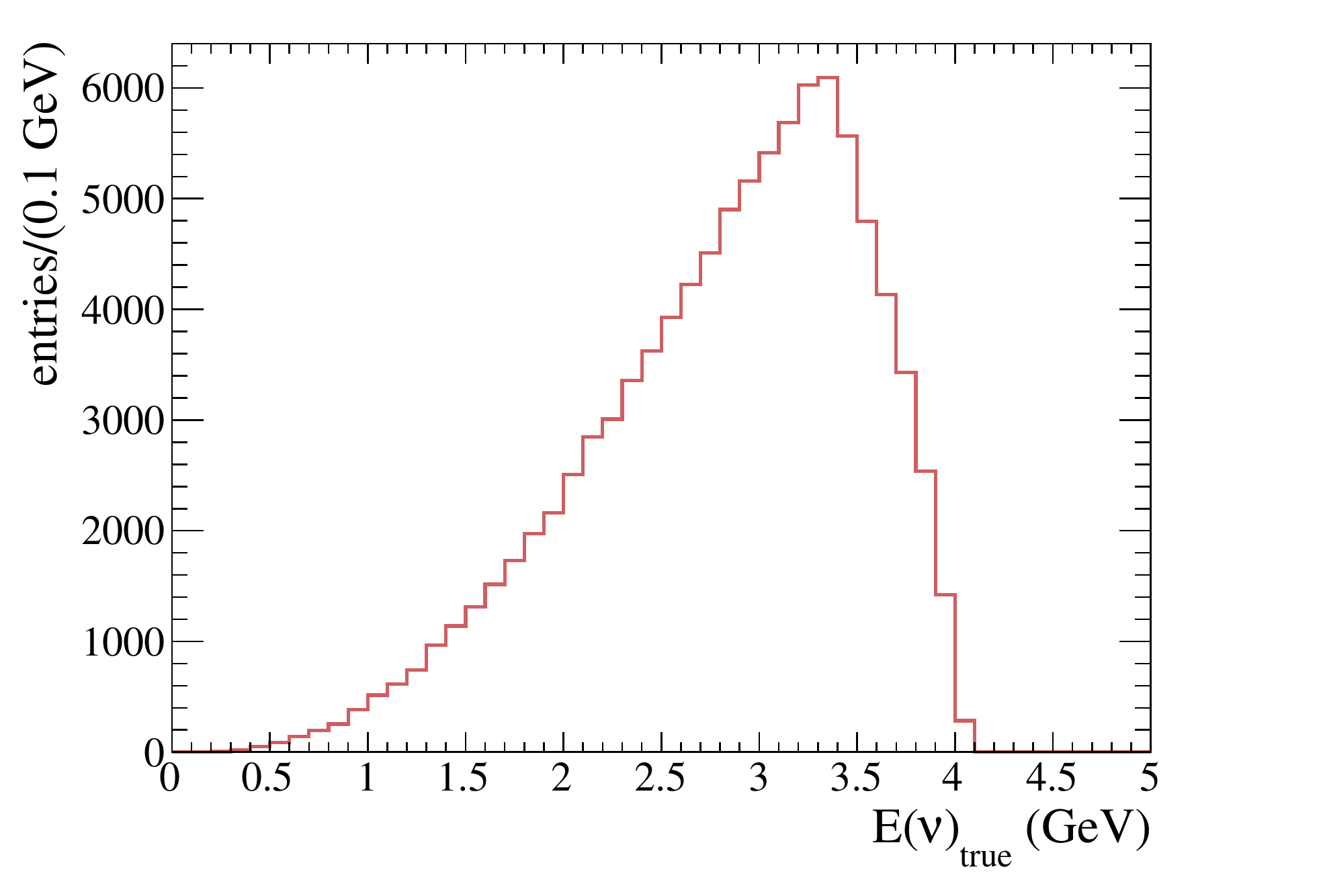}
\includegraphics[width=7.cm]{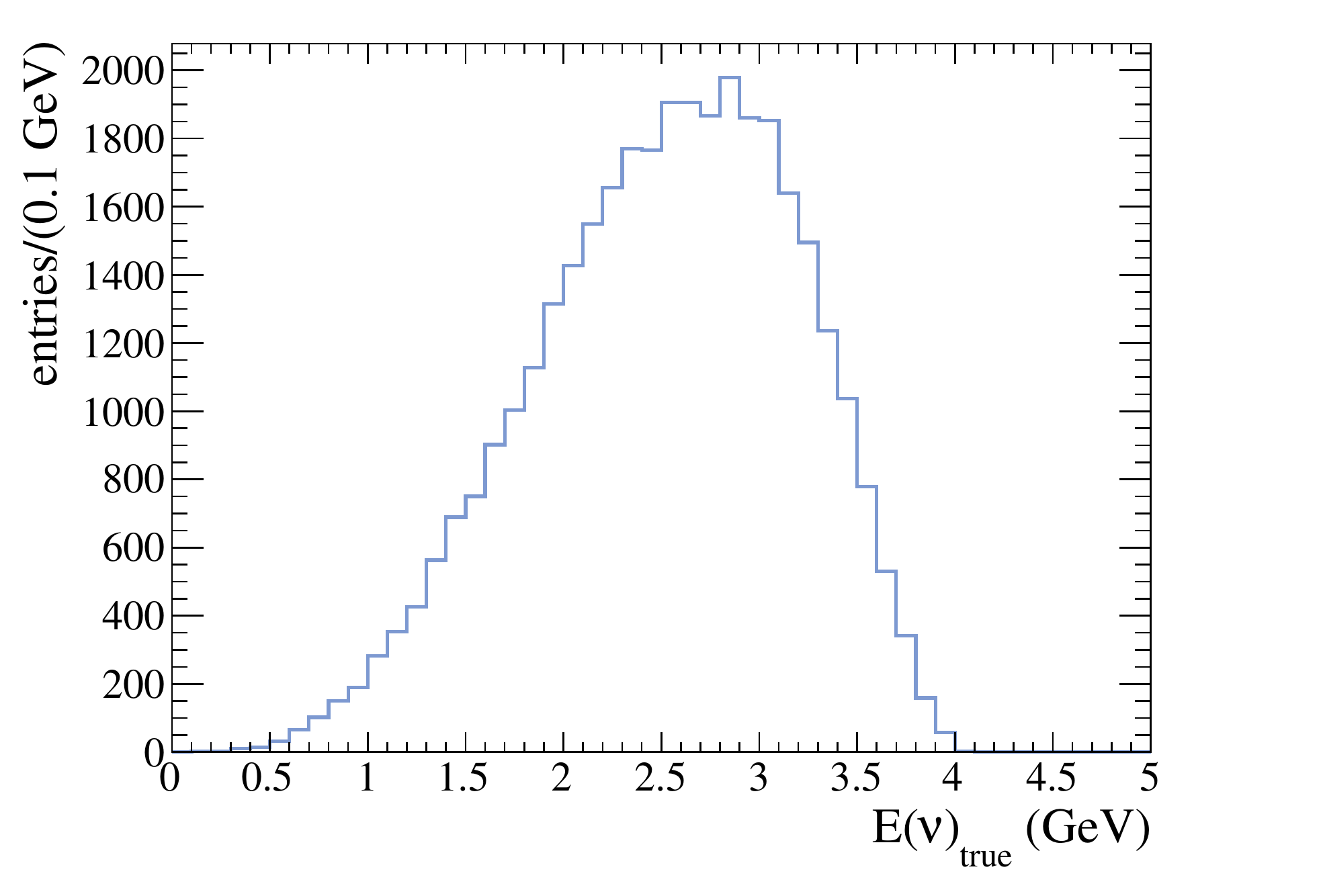}
\caption{\label{fig:nustormeventmup}True neutrino energy for the inclusive
sample of neutrino interactions occurring inside the prototype active
volume for $10^{16}$ $\mu^-$ decays for $\nu_{\mu}$ (left) and $\bar{\nu}_e$ (right) interactions.}
\end{figure}

\begin{table}[h]
 \begin{tabular}{|c|c|c|c|c|c|c|}
\hline
Int. & $N_{exp}$ & Fraction & Mean $E(\bar{\nu}_{\mu})$ &
$N_{exp}$ & Fraction & Mean $E(\nu_e)$ \\ 
Type & $\bar{\nu}_{\mu}$ & (\%) & (GeV) & $\nu_{e}$ & (\%) & (GeV) \\ \hline
 Inclusive & 40328 & 100 & 2.86 & 86074 & 100 & 2.46\\\hline 
 CC & 27538 & 68.3 & 2.87 & 64071 & 74.4 & 2.46\\\hline 
 CCQE & 11991 & 29.7 & 2.73 & 23288 & 27.1 & 2.25\\\hline 
\end{tabular}
 \caption{\label{tab:nustormmup} Expected $\bar{\nu}_{\mu}$ and $\nu_e$ event rates and
  mean neutrino energy for different neutrino interaction types and
  for $10^{16}$ $\mu^+$ decays.}
 \end{table}

\begin{figure}[h]
\includegraphics[width=7.cm]{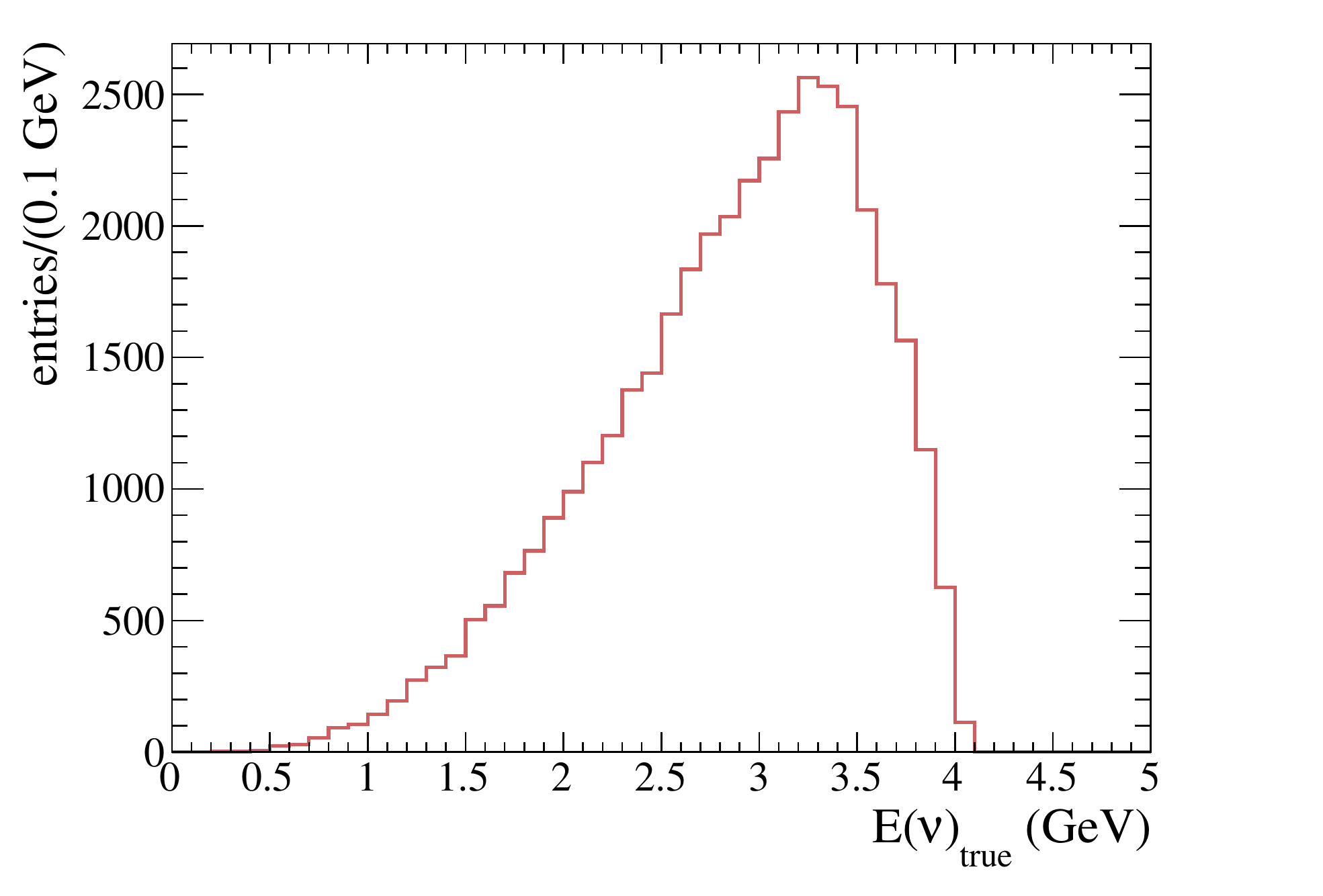}
\includegraphics[width=7.cm]{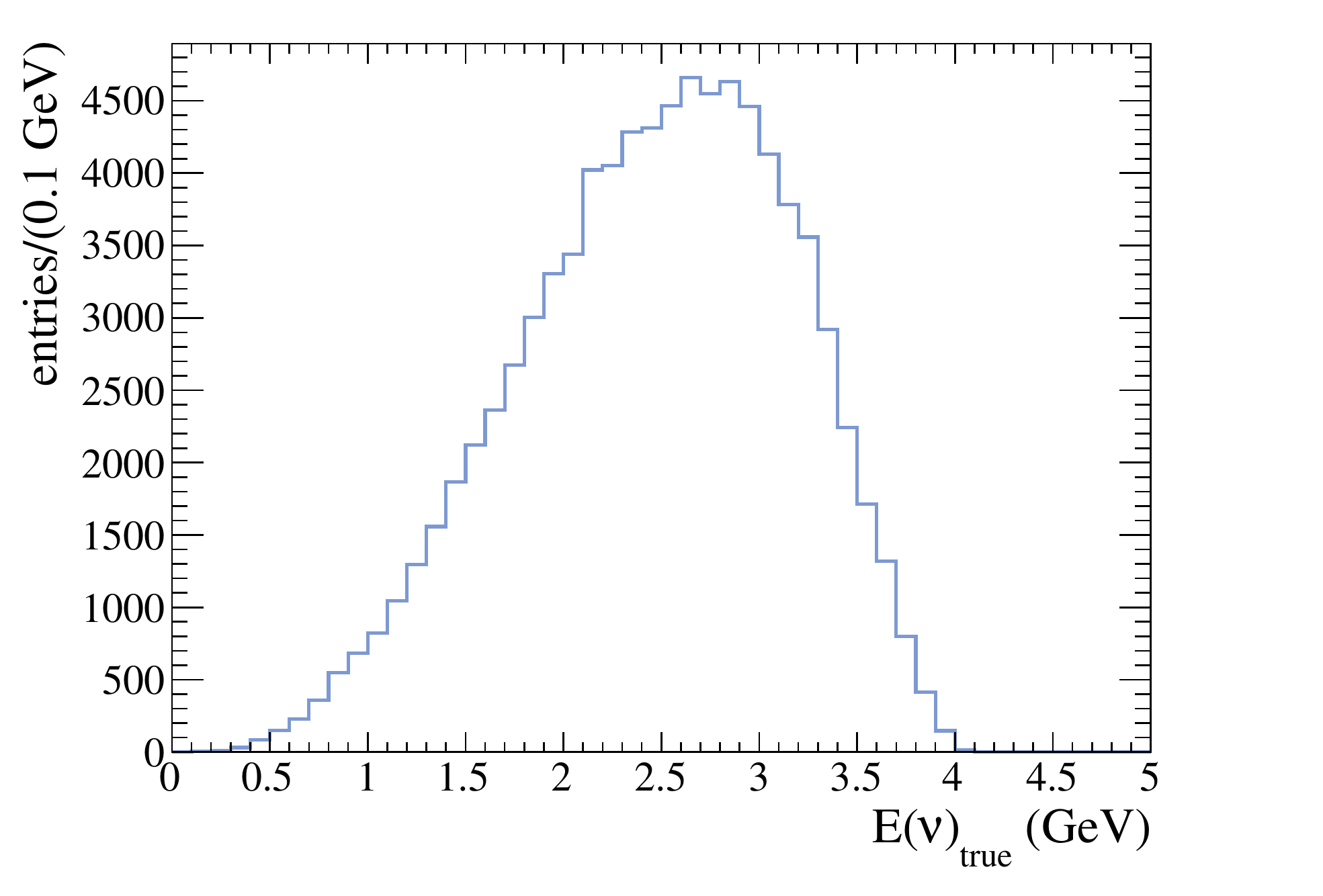}
\caption{\label{fig:nustormeventmum}True neutrino energy for the inclusive
sample of neutrino interactions occurring inside the prototype active
volume for $10^{16}$ $\mu^+$ decays for $\bar{\nu}_{\mu}$ (left) and $\nu_e$ (right) interactions.}
\end{figure}

One of main advantage of the NuSTORM beam with respect to the CENF one will be to
produce a large statistical sample of $\nu_e$ and $\bar{\nu}_e$ interactions, while in the CENF beam, $\nu_e$
are expected to be only 1.7\% of the total flux. This will allow
to perform precise measurements of $\nu_e$ cross section that are
currently completed unknown.  The systematic error on the absolute flux is also expected
to be much smaller for NuSTORM compared to CENF, since neutrinos are produced
in very well-defined muon decay processes. One should also stress that the ratio of neutrinos
to antineutrinos is very well defined. The energy of the beam is very well defined
by the ring energy and could in principle be changed to study systematics associated
to the energy scale.

With this data sample it will be possible to precisely investigate
cross section differences between $\nu_e$ and $\nu_{\mu}$
and this will be of fundamental importance for future experiments
looking for CP violation and the mass hierarchy in the leptonic sector. All these experiments
aim to measure CP and the mass hierarchy by observing $\nu_e$ ($\bar{\nu}_e$) appearance in $\nu_{\mu}$ ($\bar{\nu}_{\mu}$)
 beams. In such experiments, systematic uncertainties are constrained mainly by observing the unoscillated $\nu_{\mu}$ flux at the
 Near Detector, while after the oscillation $\nu_e$ interactions are
 observed at the Far Detector. As a consequence, if $\nu_e/\nu_{\mu}$
 cross section differences will not be precisely measured, they will have a large impact on the
 systematic uncertainties and on the discovery potential of the
 experiments.

\clearpage
\section{Appendix: Test of full-scale APA's for LBNE}
In Collaboration with LBNE colleagues, we have started to investigate the possibility to insert in the cryostat for the \six (see \Cref{fig:laguna_innerdet_cryo}) the APA prototype detector developed at FNAL for the LBNE project.  This prototype instruments a parallelepiped volume of $\sim$267 m$^3$ (374 ton LAr), with dimensions H = 7.20m, W = 7.48m, L = 5.16~m, vertically split into two drift volumes by a double face vertical wire chamber (APA) positioned at half-width, with 2 vertical cathode planes on its right and on its left. The drift length results of 3.74m, requiring a negative high voltage in the range 150 kV - 380 kV on the 2 cathodes. 

It appears possible to insert the LBNE APA detector in the LAGUNA/LBNO DLAr cryostat, although the dimensions of the vessel will likely have to be increased slightly compared to the ones presented in this document. To cope with its vertical dimension and with the position of the signal, HV and suspension feedthroughs, a dedicated top thermal insulation/support cap will be required (see \Cref{fig:appBlbnepersp,fig:appBlbneside}, with adjustable dimensions). This poses no issue as the cap will be welded and removable without interference with the overall structure of the storage vessel. A detailed design is being developed and will be ready by the time the vessel needs to be procured.

\begin{figure}[hbt]
\centering
\includegraphics*[width=0.8\textwidth]{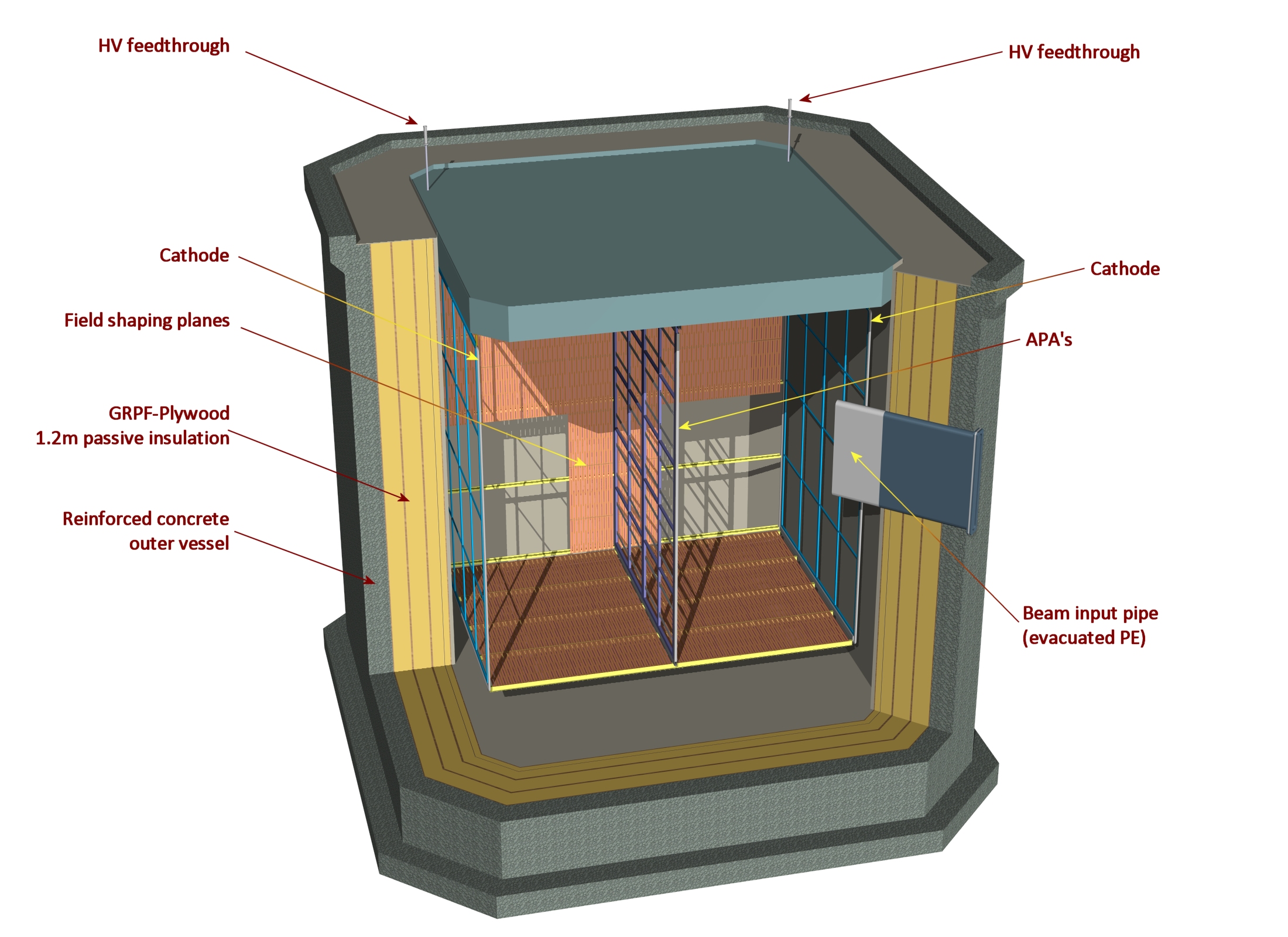}
\caption{Perspective view of the \six vessel and the inserted APA + cathode chamber based on the LBNE design.}
\label{fig:appBlbnepersp}
\end{figure}

\begin{figure}[hbt]
\centering
\includegraphics*[width=0.8\textwidth]{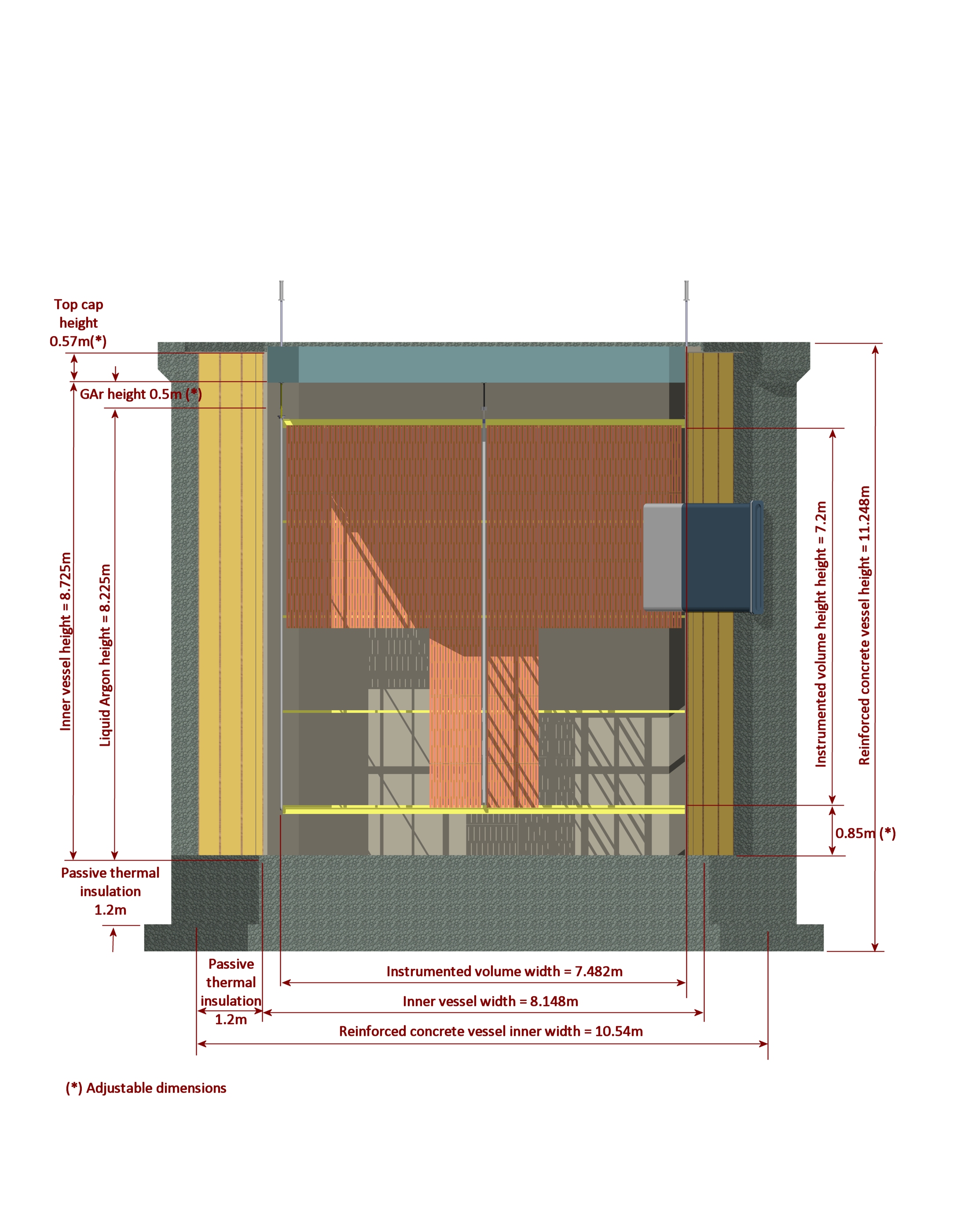}
\caption{Side view of the \six vessel and the inserted APA + cathode chamber based on the LBNE design.}
\label{fig:appBlbneside}
\end{figure}

\clearpage

\bibliographystyle{hieeetr}
\bibliography{mybibliography}
  
\end{document}